\documentclass[12pt]{cernrep}
 
\usepackage{epsfig, url, psfrag, color}
\usepackage{fancyhdr, multind}
\usepackage[square,comma,numbers]{natbib}
\usepackage{dcolumn, bm, mathrsfs, xspace, axodraw, graphicx,subfigure}
\usepackage{array,amsfonts,amssymb,amsmath,longtable}

\usepackage{landscape}

\newcommand{\PT}{\ensuremath{P_T}\yspace}
\newcommand{\mtop}{$m_t$}
\begin{document}

\begin{titlepage}

\vspace*{-2cm}
\hbox to\hsize{\null}

\noindent \hspace*{-2mm} \today \\ [-15mm]
\begin{flushright} FERMILAB-CONF-06-284-T
\end{flushright}

\vspace*{0.1cm}

\begin{center}\hspace*{-.5cm}
{\Large \bf Tevatron-for-LHC Report: Preparations for Discoveries}

\vspace*{0.6cm}

\addtolength{\baselineskip}{3mm}

{\normalsize
Salavat~Abdullin$^1$,
Darin~Acosta$^2$,
Shoji~Asai$^3$,
Oleksiy~Atramentov$^{4,5}$, 
Howard~Baer$^5$,
Csaba~Balazs$^{10}$,
Paolo~Bartalini$^2$,
Alexander~Belyaev$^{16}$,
Ralf~Bernhard$^8$,
Andreas~Birkedal$^{19}$,
Volker~Buescher$^{6\, {\large  \P{}} }$,
Richard~Cavanaugh$^2$,
Marcela~Carena$^{1\, \dagger}$,
Mu-Chun~Chen$^1$,
Christophe~Cl\'{e}ment$^{11}$,
AseshKrishna Datta$^{9}$,
Ytsen~R.~de~Boer$^{12}$,
Albert~De~Roeck$^{13}$,
Bogdan~A.~Dobrescu$^{1\, {\large  \P{}}}$,
Alexey~Drozdetskiy$^2$,
Yuri~Gershtein$^5$, 
Doug~Glenzinski$^1$,
Craig~Group$^2$,
Sven~Heinemeyer$^{30}$,
Michael~Heldmann$^6$,
Jay~Hubisz$^1$,
Martin~Karlsson$^{14}$,
Kyoungchul~Kong$^2$,
Andrey~Korytov$^2$, 
Sabine~Kraml$^{13}$,
Tadas~Krupovnickas$^{15}$,
Remi~Lafaye$^{13}$, 
Kenneth~Lane$^7$,
Fabienne~Ledroit$^{17}$,
Frank~Lehner$^8$,
Cheng-Ju~Lin$^1$,
Cosmin~Macesanu$^{18}$,
Konstantin~T.~Matchev$^2$,
Arjun Menon$^{10,29}$,
David~Milstead$^{11}$,
Guenakh~Mitselmakher$^2$, 
Julien~Morel$^{17}$,
David~Morrissey$^{20}$,
Steve~Mrenna$^{1\, \dagger}$,
Jorge~O'Farrill$^5$,
Yuriy~Pakhotin$^2$, 
Maxim~Perelstein$^{21}$,
Tilman~Plehn$^{22}$,
David~Rainwater$^{23\, {\large  \P{}}}$,
Are~Raklev$^{13,24}$,
Michael~Schmitt$^{25\, \P{}}$, 
Bobby~Scurlock$^2$, 
Alexander~Sherstnev$^{26}$,
Peter~Skands$^1$,
Zack~Sullivan$^{10}$,
Tim~M.P.~Tait$^{10}$,
Xerxes~Tata$^{27}$,
Ingo~Torchiani$^6$,
Benjamin Trocm\'e$^{17}$,
Carlos~Wagner$^{10,29}$,
Georg~Weiglein$^{31}$,
Dirk~Zerwas$^{28}$

\setcounter{footnote}{5}
\footnotesep=5mm
\renewcommand{\thefootnote}{\fnsymbol{footnote}}
\footnotetext{\normalsize \ Convenors of the Physics Landscapes working group \\ [.1mm]
\hspace*{3mm} $^\dagger$ \hspace*{0.7mm}Organizers of the TeV4LHC Workshop}

}

\vspace{6mm}

\addtolength{\baselineskip}{-1mm}
{\small \em \noindent
\hspace*{-0.15in}
\parbox[t]{1.8in}{
$^1$ Fermilab \\ [-0.1mm]
$^2$ University of Florida \\ [-0.1mm]
$^3$ University of Tokyo \\ [-0.1mm]
$^4$ Iowa State University \\ [-0.1mm]
$^5$ Florida State University \\ [-0.1mm]
$^6$ Universit\"at Freiburg \\ [-0.1mm]
$^7$ Boston University \\ [-0.1mm]
$^8$ University of Z\"urich \\ [-0.1mm]
$^9$ Harish-Chandra \\ [-0.1mm] \hspace*{2mm} 
Research Institute \\ [-0.1mm]
$^{10}$ Argonne National Lab. \\ [-0.1mm]
}\hspace*{0.1in} 
\parbox[t]{2.6in}{
$^{11}$ Stockholm University \\ [-0.1mm]
$^{12}$ ITEP Moscow \hspace*{-1mm}/\hspace*{-1mm}
Univ. of Twente \\ [-0.1mm]
$^{13}$ CERN \\ [-0.1mm]
$^{14}$ Lund University\\ [-0.1mm]
$^{15}$ Brookhaven National Laboratory \\ [-0.1mm]
$^{16}$ Michigan State University \\ [-0.1mm]
$^{17}$ LPSC, Grenoble \\ [-0.1mm]
$^{18}$ Syracuse University \\ [-0.1mm]
$^{19}$ Univ. of California, Santa Cruz \\ [-0.1mm]
$^{20}$ University of Michigan \\ [-0.1mm]
$^{21}$ Cornell University \\ [-0.1mm]
}\hspace*{0.01in} 
\parbox[t]{1.8in}{
$^{22}$ MPI Munich\\ [-0.1mm]
$^{23}$ University of Rochester \\ [-0.1mm] 
$^{24}$ University of Bergen \\ [-0.1mm]
$^{25}$ Northwestern University \\ [-0.1mm] 
$^{26}$ Moscow State University \\ [-0.1mm]
$^{27}$ University of Hawaii \\ [-0.1mm]
$^{28}$ Universit\'e de Paris-Sud \\ [-0.1mm]
$^{29}$ University of Chicago \\ [-0.1mm]
$^{30}$ IFCA, Spain \\ [-0.1mm]
$^{31}$ IPPP, Univ. of Durham \\ [11mm] 
}}

\vspace{-5mm}

\begin{abstract}
This is the ``TeV4LHC'' report of the ``Physics Landscapes'' Working Group,
focused on facilitating the start-up of physics explorations at the LHC 
by using the experience gained at the Tevatron.
We present experimental and theoretical results
that can be employed to probe various scenarios for physics beyond the 
Standard Model.
\end{abstract}
\end{center}

\end{titlepage}

\pagestyle{plain}
\setcounter{page}{2}
\setcounter{tocdepth}{2}
\tableofcontents

\clearpage
\section{Introduction and Overview}
\label{sec:introduction}

The direct exploration of the energy frontier currently performed by
the D0 and CDF experiments using $p\bar{p}$ collisions at
$\sqrt{s}=2$~TeV provided by the Tevatron is possible as a result of a
long process of development involving many people.  The existing sets
of data, of over 1.5~fb$^{-1}$ each, already contain information that
will advance the understanding of the basic laws of physics.  Within
the next three years, many new aspects of physics beyond the Standard
Model will be probed with four-times-larger data sets.

The capability of exploring the energy frontier will make a huge leap
forward with the ATLAS and CMS experiments using $pp$ collisions at
$\sqrt{s}=14$~TeV at the LHC, planned to start operating in 2008, only
two years from now.  This is a daunting endeavor, behooving the whole
community of high-energy physicists to prepare.

The purpose of the TeV4LHC series of workshops, held at Fermilab,
Brookhaven and CERN since 2004, is to facilitate the start-up of the
physics explorations at the LHC by using experience built up at the
Tevatron.  This report describes the activities of the Physics
Landscapes working group, focused on physics beyond the Standard
Model.  Three other TeV4LHC working groups, dealing with Higgs, QCD
and electroweak/top physics, will summarize their activities in
separate reports.

There are various experimental issues at the LHC that can be addressed
using experience gained at the Tevatron.  Furthermore, there are
solutions to analysis problems for searches at CDF and D0 that can be
transferred to CMS and ATLAS.  In particular, many of the tools
developed to facilitate Tevatron searches for new particles may be
used at the LHC.

One should keep in mind though that the LHC is not a scaled-up
Tevatron.  Its $pp$ collisions, as opposed to $p\bar{p}$ at the
Tevatron, change the nature of the underlying processes.  In fact, in
certain cases there is a complementarity between the machines.  For
example, if a $Z^\prime$ boson exists such that a resonance will be
discovered at the Tevatron in the dilepton invariant mass
distribution, only a certain combination of $Z^\prime$ couplings to
quarks may be measured.  Observation of the same resonance at the LHC
would provide a measurement of a different combination of $Z^\prime$
couplings to quarks.  Putting together the two measurements would then
allow the determination of the $Z^\prime$ couplings to up and down
quarks separately.
	
In addition, it should be emphasized that the LHC environment will be
much more challenging, with huge backgrounds and more stringent
triggers.  There are possible scenarios for physics beyond the
Standard Model in which the Tevatron has a better capability than the
LHC to discover certain new particles.  For example, a weakly-coupled
$s$-channel resonance that decays predominantly to $b$ jets could be
observable at the Tevatron if it is light enough, but may be too hard
to distinguish from background at the LHC.  Nevertheless, the much
higher center-of-mass energy of the LHC leads to a truly impressive
discovery potential.  This report is intended to be a small step
toward optimizing that potential.

A generic hurdle in assesing and optimizing the discovery potential of
the LHC, as well as of the Tevatron, is that it is impossible to
reliably predict how physics looks at the TeV scale.  Progress in
theoretical high-energy physics has shown that the range of
possibilities for physics at the TeV scale is much broader than was
contemplated a decade ago.  The only robust piece of information comes
from the computation of the amplitude for longitudinal $WW$
scattering~\cite{Lee:1977eg}, which shows that perturbative unitarity
is violated unless certain new particles exist at the TeV scale. More
concretely, at least one of the following statements must be true:
\\[2mm]
$i)$ There is a Higgs boson with mass below about 700~GeV.  This
possibility is analyzed in the TeV4LHC report of the Higgs working
group.
\\[2mm]
$ii)$ There is no Higgs boson, but instead there are several spin-1
particles that couple to $WW$. These may be strongly coupled, as in
the case of Technicolor (see Section~\ref{sec:TC}), or weakly coupled,
as in the case of the so-called Higgsless models (see
Sec.~\ref{sec:higgsless}).  Note that unlike theories that are
extensions of the Standard Model, which reduce to the Standard Model
in some decoupling limit, the absence of the Higgs boson would imply
that the electroweak symmetry breaking sector of the Standard Model is
not realized in nature.
\\[2mm]
$iii)$ Our current ability to compute cross sections breaks down at
the TeV scale, either because of the complicated nature of some
strongly-coupled field theory, or because quantum field theory is no
longer a good description of nature at that scale. Evidently, either
case would imply a most intriguing development for physics. Given that
further progress in this direction would likely be data-driven, we
will not discuss this possibility further.

Beyond the problem of unitarity in longitudinal $WW$ scattering, there
is little to guide us regarding what the ATLAS and CMS might
observe.  There are many well-motivated models that predict new
particles which may be tested at the LHC, and it would be useful to
analyze as many of them as possible in order to make sure that the
triggers are well-chosen and that the physics analyses cover
sufficient ground.  Fortunately, ATLAS and CMS are multi-purpose
discovery instruments, able to measure many different parameters in
large classes of models.  Moreover, any observation at CDF or D0 of
physics beyond the Standard Model, as well as tighter limits on
parameters in extensions of the Standard Model, would help the LHC
experiments to focus on signatures likely to disentangle the correct
description of nature at the TeV scale.

The next three Sections collect several self-contained contributions
from individual authors.  Each of these three sections starts with an
introduction describing the connections between various contributions.
Section 2 is focused on experimental aspects, such as the
identification of simple and compound objects.  Sec.~3 deals with
experimental signatures associated with the cases where a single new
particle will be accessible in the beginning at the LHC or Tevatron.
Although this might sound like a simplistic scenario, it is realized
in large regions of parameter space of many interesting models.
Sec.~4 covers the more complicated cases of models where several new
particles will be revealed at once.  An important question tackled
there is how to differentiate between models that lead to similar
collider signatures, even though they have completely different
origins.  A classic example is pair production of heavy colored
particles followed by cascade decays, which occurs in supersymmetric
models with $R$-parity, in models with universal extra dimensions, and
in little Higgs models with $T$-parity.  We conclude the report, in
Section 5, with a brief summary of some of the striking results
presented in Secs.~2, 3 and 4.

\clearpage\setcounter{equation}{0}\setcounter{figure}{0}\setcounter{table}{0}
\section{Experimental Aspects}
\label{sec:exp}

After collecting an integrated luminosity of more than 1~fb$^{-1}$,
searches at both Tevatron experiments CDF and D\O\ currently explore
new territory beyond existing limits.  At peak luminosities that are
now reaching 2$\times 10^{32}$~cm$^{-2}$s$^{-1}$, this requires the
detectors, trigger systems and reconstruction algorithms to handle
events at high rates and with high occupancy.  With beam crossings
producing up to $O$(10) simultanuous interactions, these challenges
are quite similar to what the ATLAS and CMS experiments will face at
the LHC.

A lot of the knowledge and experience gained at Run~II of the Tevatron
can therefore serve as a basis for a quick startup of searches at the
LHC.  In this chapter, a number of examples with relevance to searches
analyses are discussed to show how experimental techniques developed
and refined at the Tevatron can be transferred to the LHC.  This
includes the reconstruction of leptons, jets and event quantities in a
busy hadronic environment, the separation of new physics from huge jet
backgrounds, and the modelling of backgrounds using data-driven
methods.

Sections \ref{sec:em} through \ref{sec:jetmet} discuss aspects of the
reconstruction and identification of electrons, photons, muons,
tau-leptons as well as jets and missing transverse energy.  The
discovery of any of the signals discussed in
chapters~\ref{sec:particle} and~\ref{sec:models} relies on the
capability to reconstruct these objects efficiently.  In addition it is
crucial to model their efficiency and background correctly, which is a
non-trivial challenge in the complex hadron-collider environment.
Techniques for measuring efficiencies and energy scales of electrons,
photons and muons are presented in sections~\ref{sec:em} and
\ref{sec:muon}.  Sec.~\ref{sec:tau} shows how Tevatron data can be
used to predict background rates to tau lepton reconstruction at the
LHC.  The modelling of jets and missing transverse energy is discussed
in Section~\ref{sec:jetmet}.  In particular for jets, the Tevatron
experiments play an important role in testing new generators that will
be essential to model background from jet radiation at the LHC.

Finally, in the last section an example of an indirect search for new
physics is summarized in full detail, including a discussion of the
provisions necessary to trigger on the signal and study the various
background processes.

\clearpage\setcounter{equation}{0}\setcounter{figure}{0}\setcounter{table}{0}
\subsection{Electron and Photon Reconstruction and Identification}
\label{sec:em}

Yuri~Gershtein$^1$, Oleksiy~Atramentov$^2$ \\ [2mm]
{\em
$^{1}$ Florida State University \\ [-0.1mm]
$^2$ Iowa State University 
}

\subsubsection{Overview}
\label{sec:em:intro}

Photon reconstruction at the Tevatron starts with finding clusters of
energy in the electromagnetic calorimeter.  For electrons, in addition
to the calorimeter-seeded algorithms, a track-seeded algorithm exists,
although it is used primarily for reconstruction of non-isolated
electrons.  For both reconstruction algorithms, however, the idea
behind it is similar.

The main background to electrons and photons comes from jets.  Also, a
photon can be misidentified as an electron and vice versa.  For
example, $W\gamma$ events form a major background to multi-lepton SUSY
searches when the $W$ decays semi-leptonically and the photon
undergoes convertion in the tracker~\cite{Abazov:2005ku}, and to
di-photon SUSY searches when the $W$ decays into an electron and its
track is not reconstructed~\cite{Abazov:2004jx}.  For CDF's study on
the exact composition of the electron fakes see
Ref.~\cite{Culbertson}.

Silicon trackers have revolutionized heavy flavor tagging at hadron
colliders.  However, the price one must pay for being able to tag
heavy flavors is the large amount of material that electrons and
photons must transverse before reaching the calorimeter.  This
introduces a significant problem for the Tevatron detectors, and will
be an even bigger problem at the LHC, since both CMS and ATLAS
detectors have much more material in the tracker (up to $\sim$ 1.6
radiation lengths).  We will discuss this in more detail in
Sec.~\ref{sec:em:material}.

Having more than one radiation length of material in front of the
calorimeter is already challenging, but experience at the Tevatron
shows that the amount of material included in the Monte-Carlo (MC)
simulation of the detector is significantly smaller than it is in
reality.  This and other effects lead to a substantial disagreement
between the data and the MC at start-up.

{\it In situ} measurement of the material distribution and tuning of
MC parameters is a long and elaborate process.  However, analyses of
the first data cannot wait for a perfect MC.  It is therefore of
utmost importance to develop algorithms to extract everything needed
for the analyses (reconstruction and identification efficiency, energy
scale and resolution, {\it etc...}) from the data itself.  A lot of
experience in this has been accumulated for electrons coming from $Z$,
$J/\psi$, and $\Upsilon$ decays.  Photons, on the other hand, present
more of a challenge, since there is no clean and abundant resonant
production of isolated photons at the Tevatron.  One of the
achievements of this series of workshops is the realization that at
the LHC the $\mu\mu\gamma$ final state provides such a source.  These
issues are discussed in Sec.~\ref{sec:em:data}

\subsubsection{Effects of the Tracker Material}
\label{sec:em:material}

The large (up to 1.6 radiation lengths) amount of the material in
front of the ECAL has a significant negative effect on reconstruction
of electrons and photons.  Electrons lose their energy by
bremsstrahlung in material while curving in the magnetic field, which
turns usually narrow EM showers into azimuthally wide sprays.  This
leads to a certain energy loss due to imperfect clustering.  Even more
important is the fact that the bremsstrahlung photons convert, and the
resultant electrons curl in the magnetic field and do not reach the
calorimeter.  The combination of these two effects results in a
non-linear energy scale for electrons that depends on the material
distribution in front of the ECAL, and therefore on rapidity and (to a
lesser extent) azimuth.

\begin{figure}
\begin{center}
\includegraphics[width=12cm]{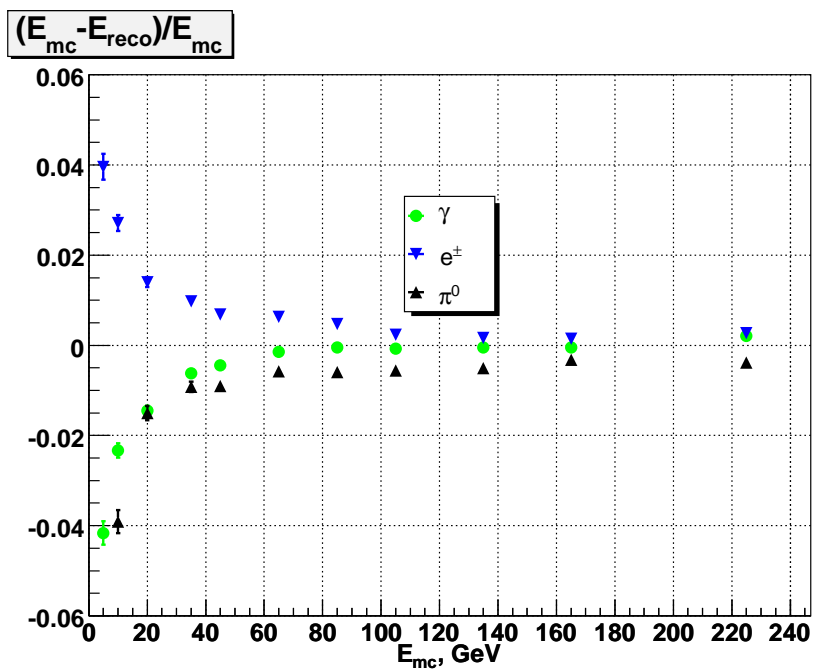}
\end{center}
\caption{Simulation of the linearity of the response of D\O\ detector
to single electrons, photons, and neutral pions.}
\label{fig:em:mat1}
\end{figure}

Photons, as opposed to electrons, propagate in the material in a
different way.  They stay totally intact until the first conversion.
Therefore, for unconverted photons the material-induced non-linearity
is not an issue.  However, when a photon converts, its energy is
shared between two electrons and the effect of the material is
effectively doubled.  As a result, the electron and photon energy
scales are different and non-linear (see Fig.~\ref{fig:em:mat1} for
simulation of the D\O\ detector response).

%

\subsubsection{Extraction of Efficiency and Energy Scale from Data}
\label{sec:em:data}

The experience of previous experiments, including the most recent from
CDF and D\O\ at the Tevatron, is such that the amount of material
included in the GEANT description of the detector is severely
underestimated by start-up time.  First, the as-built detector is not
the same as the as-drafted.  Second, because the tracking system is
complex and comprises so many elements, some of them end up
inadequately implemented in the MC.  The magnitude of the disagreement
can be as large as a factor of two.  For CDF, for example, in which
three silicon detectors were build during Runs~I and~II, the amount of
unaccounted material in the MC implementation of the last one at the
start-up was only about $50\%$ of the actual amount.

The above consideration makes it too risky to rely solely on the MC
simulation for a proper description of electrons and photons in first
analyses at the LHC.  The plan for start-up should therefore be
two-pronged:
\begin{itemize}
\item[{\bf A.}] Measure the amount of material in the tracker 
{\it in situ} using a combination of several methods (converted photon
yields and distributions, mass of low-lying resonances and measurement
of transverse momentum variation from the beginning to the end of
electron tracks).  The end result of this activity would be a MC
simulation that describes the real detector.
\item[{\bf B.}] In parallel to the work described in {\bf A}, 
efficiencies, resolutions, and energy scales of electrons and photons
should be measured for different detector regions and for different ID
cuts.
\end{itemize}
\begin{figure}[t]
\begin{center}
\includegraphics[width=12cm]{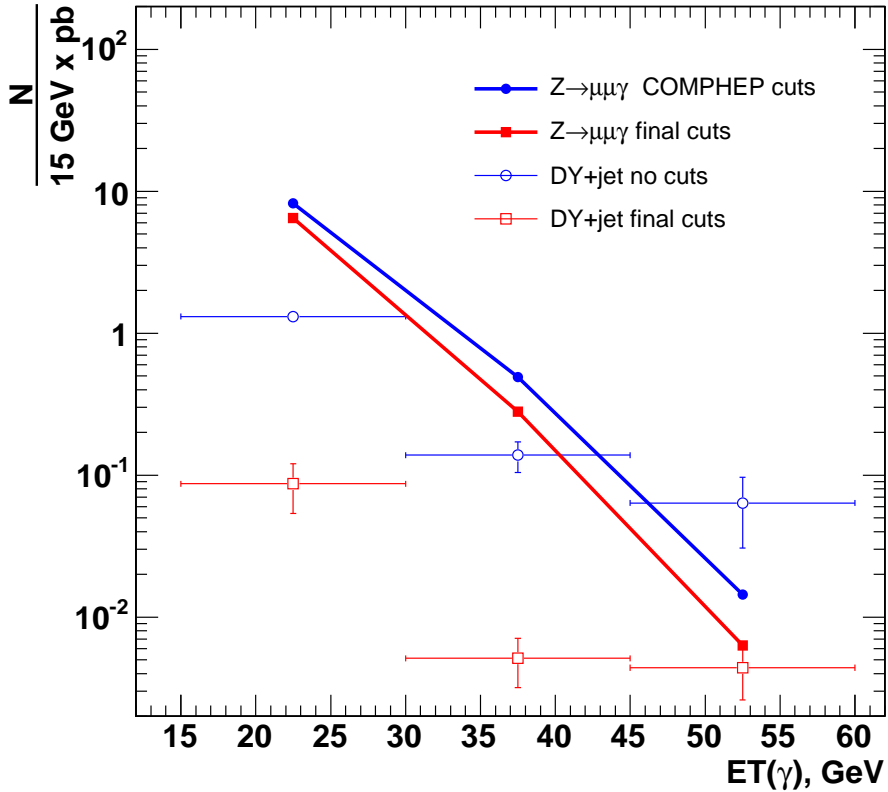}
\end{center}
\caption{Radiative $Z$ decay signal and background yields before and 
after the cuts on event kinematics.}
\label{fig:em:uug1}
\end{figure}

The Tevatron experiments followed this strategy, using $Z$,
$\Upsilon$, and $J/\phi$ decays to calibrate electrons.  At the LHC,
both the center of mass energy and luminosity are high enough to
provide a source of clean and isolated photons from radiative $Z$
decays.  A study using the detailed simulation of CMS detector
showed~\cite{CMS-PTDR-Vol1} that using simple kinematic cuts on
dilepton mass ($40<m_{\mu\mu}<80$~GeV) and photon-lepton separation
($\sqrt{\Delta\phi_{\mu\gamma}^2+\Delta\eta_{\mu\gamma}^2}<0.8$) a
reasonable signal-to-background ratio can be achieved (see
Fig.~\ref{fig:em:uug1}).

When extracting detector performance from data, one should be wary of
possible biases.  For example, D\O\ measures the electron
identification efficiency in $Z\to e^+e^-$ events using the
``tag-and-probe'' method.  In this method one of the electrons is
required to pass stringent identification criteria to improve purity
of the sample while the other -- the probe -- is used to measure the
efficiency.  Here, the biases arise from correlations between the tag
and the probe electrons.  For example, in the early stages of electron
identification, the efficiency turned out to be dependent on the
primary vertex position, and since the selection of the tag biased the
vertex distribution, the electron identification efficiency, obtained
from the probe, was found to be shifted toward higher values.
Although the full D\O\ MC simulation did not reproduce the effect
exactly, it was enough to suggest a corrective action, i.e. it was
chosen for the short term to parametrize efficiency as a function of
both rapidity and vertex position while developing a new version of
electron identification that did not have such a strong vertex
dependence.

\begin{figure}
\begin{center}
\includegraphics[width=12cm]{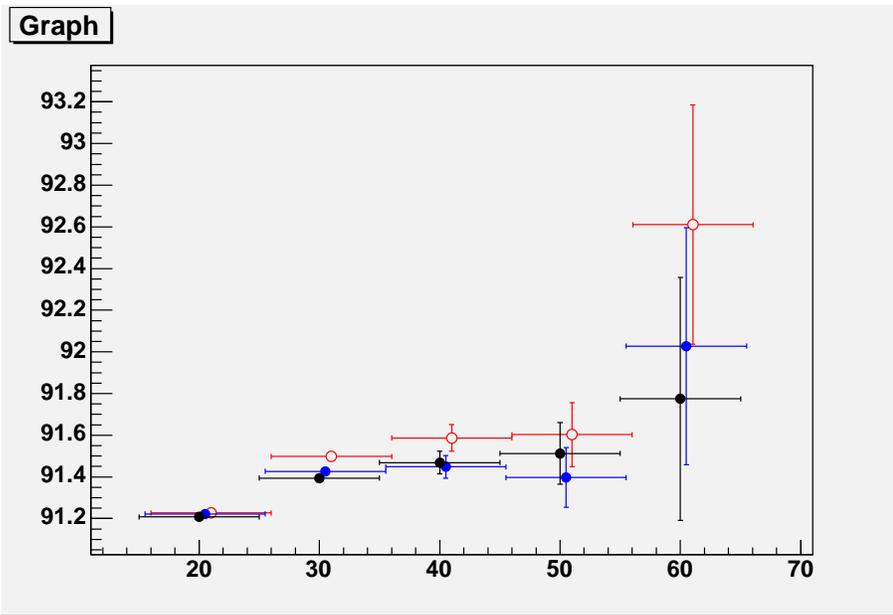}
\end{center}
\caption{Fitted value of the $Z$ mass v. the photon $E_T$ in GeV. 
See text for details.}
\label{fig:em:uug2}
\end{figure}

For energy scale measurements, the biases can arise from both
instrumental and physics effects.  As an example, let us consider
photon energy scale measurement with $Z\to\mu\mu\gamma$ events.  The
instrumental effect comes from the photon energy resolution.  The
photon $E_T$ spectrum in radiative $Z$ decays falls sharply.
Therefore, a sample of $\mu\mu\gamma$ events with large photon $E_T$
will be enriched by events where the photon energy has been
mis-measured, and the $Z$ peak would shift toward larger masses.  The
second effect arises from the large natural width of the $Z$.  The
importance of both effects can be estimated using a simple
parametrized MC simulation (see Fig.~\ref{fig:em:uug2}).  We fit the
three body mass distribution in bins of photon transverse energy,
first using generator level information (black points), and then
smearing the generator information by the best energy resolution that
one might expect at the LHC detector~\footnote{$\frac{\sigma_E}{E}=
\frac{0.027}{\sqrt{E}}\oplus\frac{0.155}{E}\oplus 0.0055$} (blue
points).  The red points correspond to the case in which we add an
extra $2\%$ constant term to the resolution function. The fitted
values of the $Z$ mass can be shifted by almost 0.4~GeV, which
corresponds to a photon energy scale shift of $2\%$.

\clearpage\setcounter{equation}{0}\setcounter{figure}{0}\setcounter{table}{0}

\subsection{Sensitivity of the Muon Isolation Cut Effic. to the
Underlying Event Uncertainties}
\label{sec:muon}

    S.~Abdullin$^1$, D.~Acosta$^2$, P.~Bartalini$^2$,
    R.~Cavanaugh$^2$, A.~Drozdetskiy$^2$, A.~Korytov$^2$,
    G.~Mitselmakher$^2$, Yu.~Pakhotin$^2$, B.~Scurlock$^2$,
    A.~Sherstnev$^3$ \\ [3mm]
{\em  $^1$ Fermi National Laboratory, Batavia, Illinois, USA
  \\$^2$ University of Florida, Gainesville, Florida, USA
  \\$^3$ Moscow State University, Moscow, Russia} \\

{\em Uncertainties in predicting the muon isolation cut efficiency are
    studied by varying the PYTHIA parameters responsible for
    simulation of the underlying event. Study is performed on the
    example of the Standard Model Higgs search. The following processes are
    considered: $H \rightarrow ZZ \rightarrow 4 \mu$, $\rm ZZ
    \rightarrow 4 \mu$, and $t\bar{t} \rightarrow 4\mu + X$. 
    We show that an inclusive Z data sample will allow for a direct
    experimental measurement of the 4-muon isolation cut efficiencies
    associated with $H \rightarrow ZZ \rightarrow 4 \mu$ and
    $ZZ \rightarrow 4 \mu$ events with a systematic uncertainty
    of less than 2\%.}

\subsubsection{Introduction}

In future searches for the Higgs boson at the LHC via its 4-muon decay
channel, $H \rightarrow ZZ \rightarrow 4 \mu$, the muon
isolation cut plays a key role in suppressing many otherwise
dominating backgrounds where all or some muons originate from hadronic
decays (${\rm t\bar{t} }$ and ${\rm Zb\bar{b} }$ are the most
important processes in this category). Having reduced the ${\rm
t\bar{t} }$ and ${\rm Zb\bar{b} }$ backgrounds to a negligible level,
we also suppressing the ZZ background and signal. Therefore, one must
worry about the efficiency of the muon isolation cut with respect to
the ZZ background and Higgs boson signal and, even more, about the
sensitivity of this efficiency to large theoretical uncertainties
associated with a poor understanding of the underlying event (UE)
physics. The UE is defined as \cite{PAOLO} all the remnant activity
from the same proton-proton interaction.  

The goal of the studies presented here was not to optimize the 
muon isolation cut in order to maximize the signal-over-background 
significance, but rather to understand how well can we predict 
the isolation cut efficiency using the current Monte Carlo generators, 
and to develop a method of measuring the isolation cut efficiency 
using the experimental data themselves. The proposed technique of 
evaluating the isolation cut efficiency for ZZ events is based on 
sampling energy flow in cones of random directions in inclusive 
${\rm Z \rightarrow 2 \mu }$ data sample. At Tevatron, Z-boson 
di-muon data samples and random/complimentary cones are widely
used for various calibration purposes, which was an original 
inspiration for us in developing the method we present further below.
In these generator-level studies, we looked only at the tracker-based
isolation cut.

The analysis presented in this subsection is done in accordance with
official guidelines described in \cite{PAOLO} for UE for a particular
Monte Carlo generator with a particular set of model parameters. Only
effects of the first order influencing UE in this model are
considered.


\subsubsection{Event Generation Parameters for PYTHIA}

Higgs boson, ${\rm t\bar{t} }$ and Z-inclusive data samples were
generated with PYTHIA 6.223 \cite{Sjostrand:2000wi}. The ZZ data
sample was generated at the matrix-element level with CompHEP
\cite{Boos:2004kh} and, then, PYTHIA was used to complete the event
simulation (parton shower development, UE, hadronization, and particle
decays). The PYTHIA parameters that drive the UE simulation were
consistently chosen to match those selected for the Data Challenge
2005 (DC05) CMS official production (see
Table~\ref{tab:DC05}). Detailed discussion of the associated
phenomenology and the corresponding references can be found elsewhere
\cite{PAOLO}.

\begin{table}[htb]
  \label{tab:DC05}
  \begin{center} \small \renewcommand{\arraystretch}{1.4}
    \begin{tabular}{||c|c|c|c|c|c||} \hline
      parameter & CDF & ATLAS & \parbox[t]{3.1em}{ \ CMS \\ (DC04) \\ [-5mm] } & \parbox[t]{3.1em}{ \ CMS \\ (DC05)} & comment \\ \hline
      PARP(82) & 2    & 1.8  & 1.9 & 2.9 & \parbox[t]{19.1em}{regularization scale of PT spectrum for MI } \\ \hline
      PARP(84) & 0.4  & 0.5  & 0.4 & 0.4 & \parbox[t]{19.1em}{parameter of matter distribution inside hadrons\\ [-3mm]} \\ \hline
      PARP(85) & 0.9  & 0.33 & 0.33 & 0.33 & \parbox[t]{19.1em}{probability in MI for two gluons with color connections\\ [-3mm]} \\ \hline
      PARP(86) & 0.95 & 0.66 & 0.66 & 0.66 & \parbox[t]{19.1em}{probability in MI for two gluons (as a closed loop)\\ [-3mm]} \\ \hline
      PARP(89) & 1800 & 1000 & 1000 & 14000 & \parbox[t]{19.1em}{reference energy scale} \\ \hline
      PARP(90) & 0.25 & 0.16 & 0.16 & 0.16 & \parbox[t]{19.1em}{power of the energy-rescaling term} \\ \hline
      ${\rm pt_{\textrm{cut-off}} }$ & 3.34 & 2.75 & 2.90 & 2.90 & final ${\rm pt_{\textrm{cut-off}} }$ \\ \hline
    \end{tabular} 
\caption{Parameters in PYTHIA for multi-parton interactions (MI) and
  UE for CDF, ATLAS and CMS.}
  \end{center}
\end{table}

The most critical parameter affecting the UE activity is ${\rm
pt_{\textrm{cut-off}} }$, the lowest PT allowed for multi-parton
interactions.  The smaller ${\rm pt_{\textrm{cut-off}} }$ is, the
larger is the number of tracks associated with the underlying
event. The ${\rm pt_{\textrm{cut-off}} }$ value and its evolution with
the center of mass energy of proton-proton collisions are defined via
the following formula:

\begin{center}
${\rm pt_{\textrm{cut-off}} = PARP(82)*(14000/PARP(89))^{PARP(90)} }$
\end{center}

The three parameters, PARP(82,89,90), have meaning only in this
combination.  The parameters PARP(89) and PARP(90) are fixed at 14,000
and 0.16, correspondingly. We decided to vary ${\rm
pt_{\textrm{cut-off}} }$ by ${\rm \pm 3\sigma }$, or ${\rm \pm 0.5
~GeV }$, which seems to be a sensible estimation of theoretical
uncertainties arising from UE modeling \cite{Nason:1999ta}. Note that
${\rm pt_{\textrm{cut-off}}=3.34 ~GeV }$, as extracted from CDF's Tune
A of PYTHIA MI parameters, differs from the default values used by
ATLAS (${\rm 2.75 ~GeV }$) and CMS (${\rm 2.9 ~GeV }$) by ${\rm \sim
0.5 ~GeV }$ because it was done using a different PYTHIA parameter
tuning model and is listed for completeness only in
Table~\ref{tab:DC05}.


\subsubsection{Monte Carlo sample production}

Processes used in these studies were: ${\rm t\bar{t} }$ (PYTHIA
parameter MSEL = 6); Higgs boson signal (${\rm m_H = 150 }$ GeV,
PYTHIA parameters MSEL = 0, MSUB(102,123,124) = 1 with H allowed to
decay to ${\rm Z/\gamma* }$ only, ${\rm Z/\gamma* }$ allowed to decay
to ${\rm e/\mu/\tau }$ pair only and ${\rm \tau }$ allowed to decay to
${\rm e/\mu }$ only); ZZ (PYTHIA parameters MSEL = 0, MSUB(1) = 22
with ${\rm Z/\gamma* }$ allowed to decay to ${\rm e/\mu/\tau }$ pair
only and ${\rm \tau }$ allowed to decay to ${\rm e/\mu }$ only);
Z-inclusive (PYTHIA parameters MSEL = 0, MSUB(1) = 1 with Z allowed to
decay to muon pair only). For Higgs boson signal, we used PHOTOS as a
generator of bremsstrahlung photons.

Generator-level cuts:

\begin{itemize}

\item ${\rm t\bar{t} }$: at least four muons with ${\rm PT > 7 ~GeV }$ and ${\rm |\eta| <
  2.4 }$;

\item Higgs boson signal: at least four muons with ${\rm PT > 7 ~GeV
}$ and ${\rm |\eta| < 2.4 }$; \break ${\rm 5 < M_{inv}(\mu^+\mu^-) < 150 ~GeV
}$ for 2 intermediate resonances (${\rm Z/\gamma* }$);

\item ZZ-sample: same as for signal;

\item Z-inclusive: no user defined cuts.

\end{itemize}

\subsubsection{Event selection}

Event-selection cuts were further imposed on the produced Monte Carlo samples.
These cuts were chosen to mimic those optimized for the future data analysis.
There are two distinct sets of such cuts.

First, only ''good muons'' were selected. A muon was considered to be
''good'' if it had ${\rm PT > 7 }$ GeV in the barrel region (${\rm |\eta| < 1.1 }$)
or ${\rm P > 9 }$ GeV in the endcaps (${\rm 1.1 < |\eta| < 2.4 }$). This ensures
that the muon reconstruction efficiencies are at their plateau, which
helps minimize systematic uncertainties on the muon reconstruction
efficiency.

Then, event-selection cuts similar to the full analysis cuts were
applied. They are:

\begin{itemize}

\item At least 2 opposite sign muon pairs with invariant masses for
   all ${\rm \mu^+\mu^- }$ pair permutations being greater than 12 GeV
   (this cut suppresses heavy-quark resonances).

\item PT of all four selected muons must be greater than 10 GeV
   (signal-over-background optimization).

\item invariant mass of the four muons must be greater than 110 GeV
   and less than 700 GeV (Higgs boson with ${\rm M<114.4 ~GeV }$ is
   excluded at LEP, Higgs boson with mass over 700 GeV is strongly
   disfavored by theory and, also, would have too low a production
   cross section).

\item ${\rm ISOL = \sum{PT_i} }$ (PT with respect to the beam
   direction) should be less or equal to 0, 0, 1, 2 GeV for the four
   muons when the muons are sorted by the ISOL parameter. The sum runs
   over only charged particle tracks with PT greater then 0.8 GeV and
   inside a cone of radius ${\rm R = \sqrt{(\Delta\phi)^2 +
   (\Delta\eta)^2}=0.3 }$ in the azimuth-pseudorapidity space. A PT
   threshold of 0.8 GeV roughly corresponds to the PT for which tracks
   start looping inside the CMS Tracker. Muon tracks were not included
   in the calculation of the ISOL parameter.

\end{itemize}

\subsubsection{Tracker-based muon isolation cut efficiency}

\begin{figure}[htb]
    \begin{tabular}{p{.47\textwidth}p{.47\textwidth}} 
      \resizebox{\linewidth}{0.65\linewidth}{\includegraphics{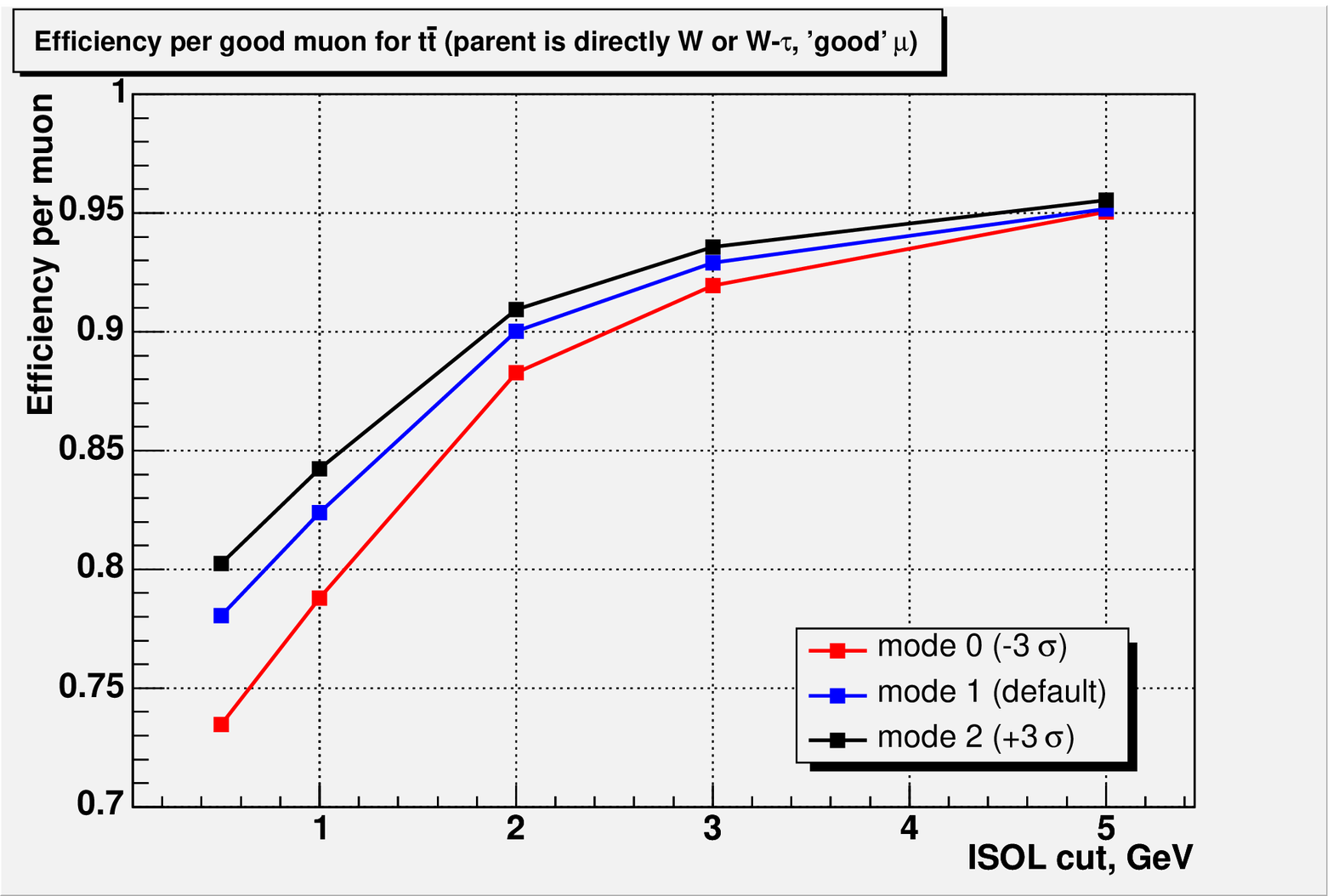}} &
      \resizebox{\linewidth}{0.65\linewidth}{\includegraphics{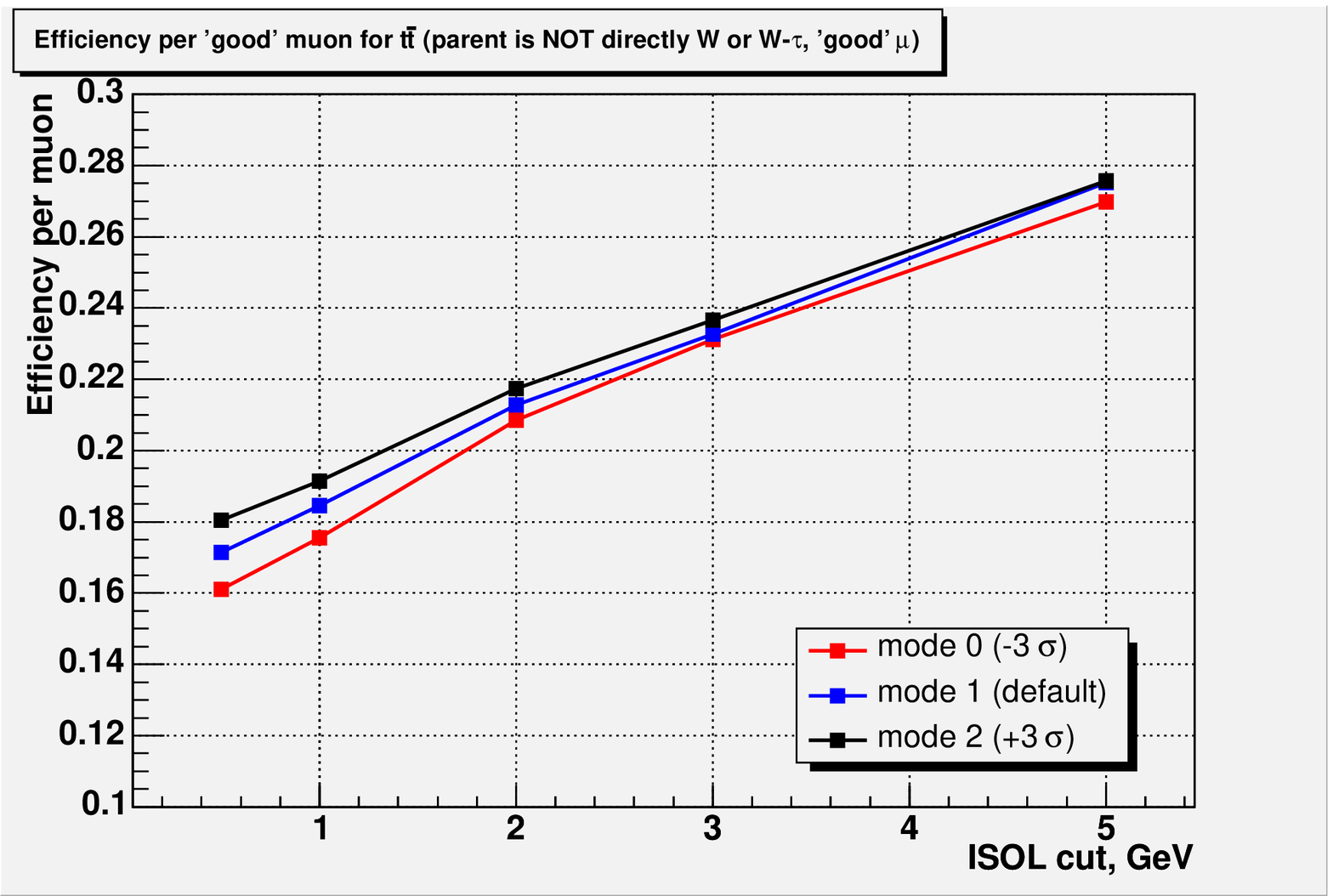}} \\
      \caption{Muon isolation cut efficiency averaged over selected muons
	whose parents are W bosons (${\rm t\bar{t} }$ events). The blue middle line
	is for the default MI ${\rm pt_{\textrm{cut-off}} }$, the black upper line is for
	downward ${\rm -3\sigma }$ variation of ${\rm pt_{\textrm{cut-off}} }$ value, the red lower
	line is for upward ${\rm +3\sigma }$ variation.}
      \label{fig:GoodBckgW} &
      \caption{Similar to Fig.~\ref{fig:GoodBckgW} for muons from hadronic
	decays (${\rm t\bar{t}}$ events).}
      \label{fig:GoodBckgB} \\
      \resizebox{\linewidth}{0.65\linewidth}{\includegraphics{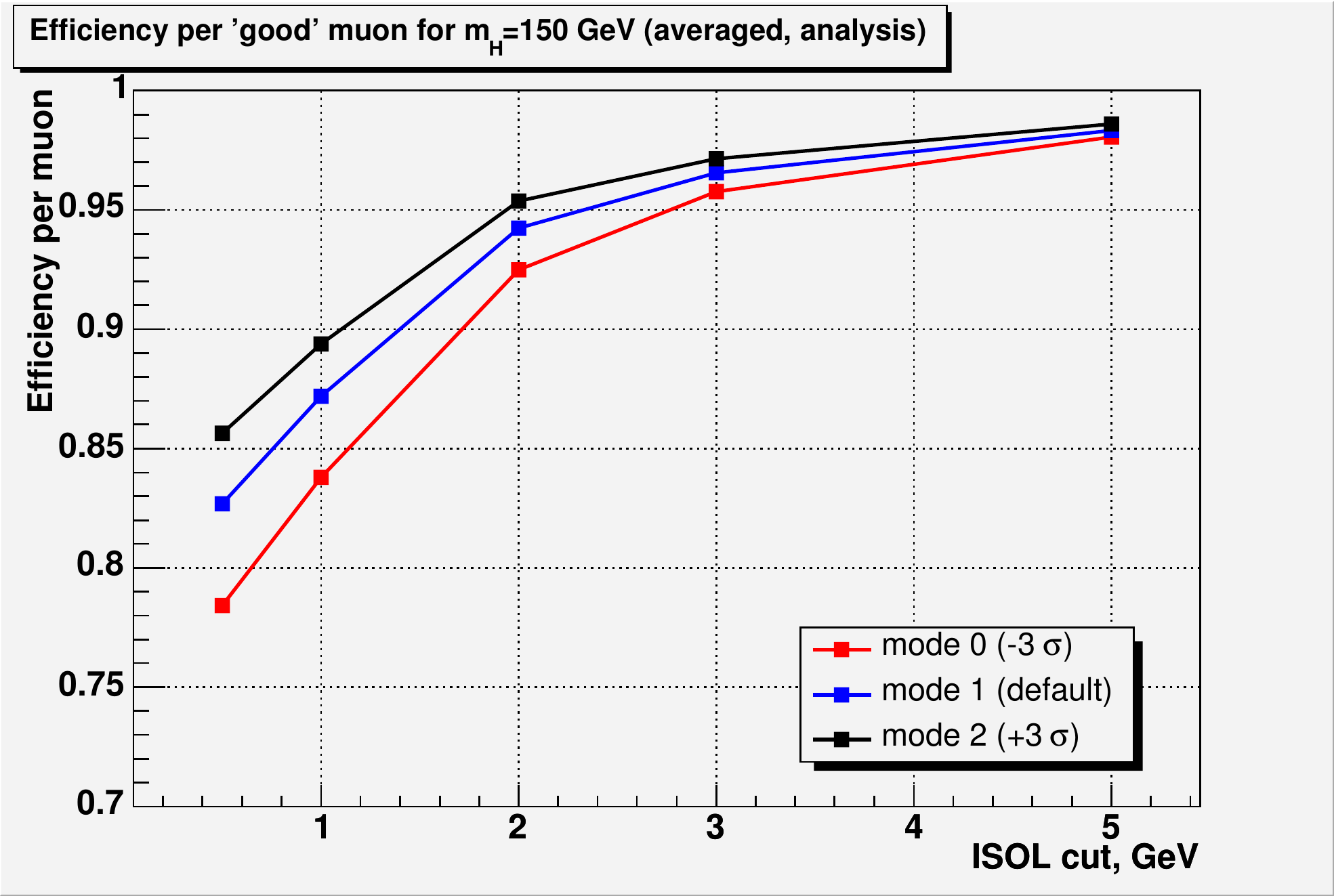}} &
      \resizebox{\linewidth}{0.65\linewidth}{\includegraphics{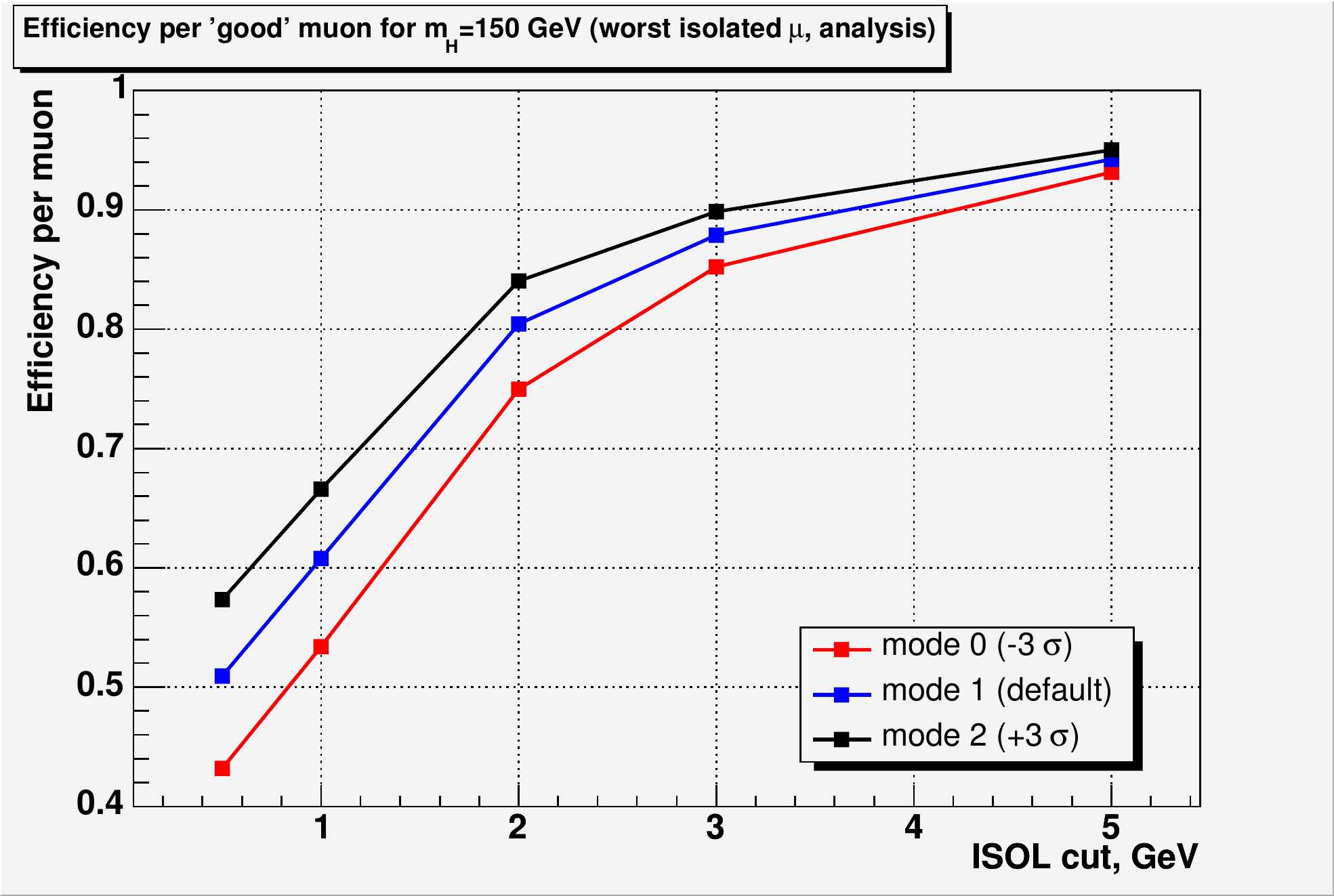}} \\
      \caption{Similar to Fig.~\ref{fig:GoodBckgW} for Higgs boson events.}
      \label{fig:GoodSig} &
      \caption{Muon isolation cut efficiency for the least isolated muon
	from 4 selected ones in Higgs boson events.}
      \label{fig:SigWorst} \\
    \end{tabular}
\end{figure}

Figures \ref{fig:GoodBckgW},~\ref{fig:GoodBckgB} and \ref{fig:GoodSig}
show the muon isolation cut efficiency averaged over all "good" muons
(see section 3.2) for the ${\rm t\bar{t} }$ sample and the Higgs
boson. For ${\rm t\bar{t} }$ background, we show two plots: one for
muons originating from ${\rm W \rightarrow \mu\nu }$ and ${\rm W
\rightarrow \tau\nu \rightarrow \mu\nu\nu\nu }$ decays and the other
for muons originating from hadronic decays (typically, the former
would tend to be isolated and the latter non-isolated). The average
isolation efficiency per "good" muon is calculated as the ratio of the
number of "good" muons with the isolation parameter ISOL below a
particular threshold to the total number of "good" muons.  Figure
\ref{fig:SigWorst} shows the isolation cut efficiency for the least
isolated muon out of four (Higgs boson sample). We use a cut at ISOL=2
GeV for such muons. One can see that this cut alone will have ${\rm
\sim 80\% }$ efficiency with ${\rm \pm 5\% }$ uncertainty in a
considered UE model.
    
\begin{figure}[t]
  \begin{tabular}{p{.47\textwidth}p{.47\textwidth}} 
    \resizebox{\linewidth}{0.65\linewidth}{\includegraphics{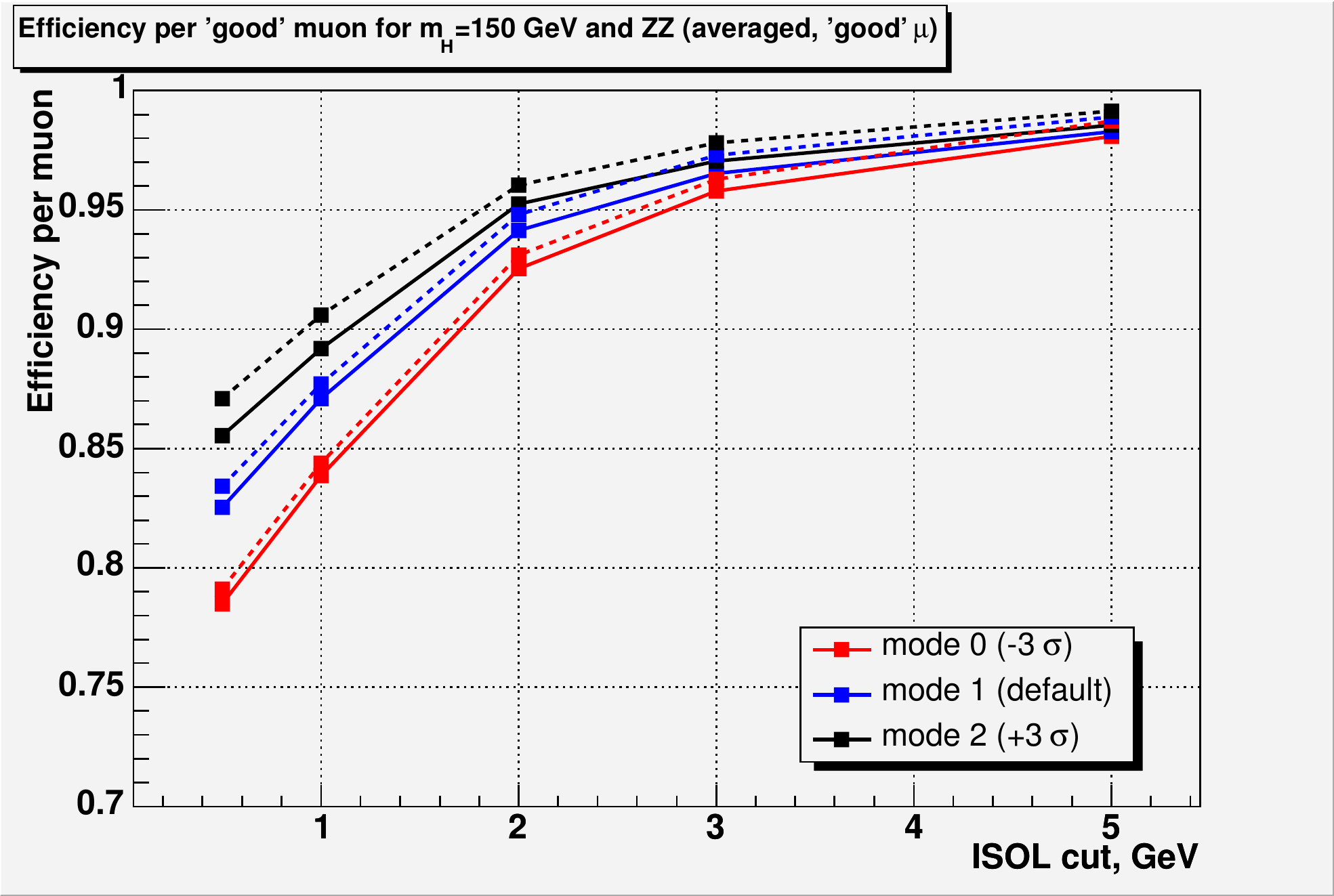}} &
    \resizebox{\linewidth}{0.65\linewidth}{\includegraphics{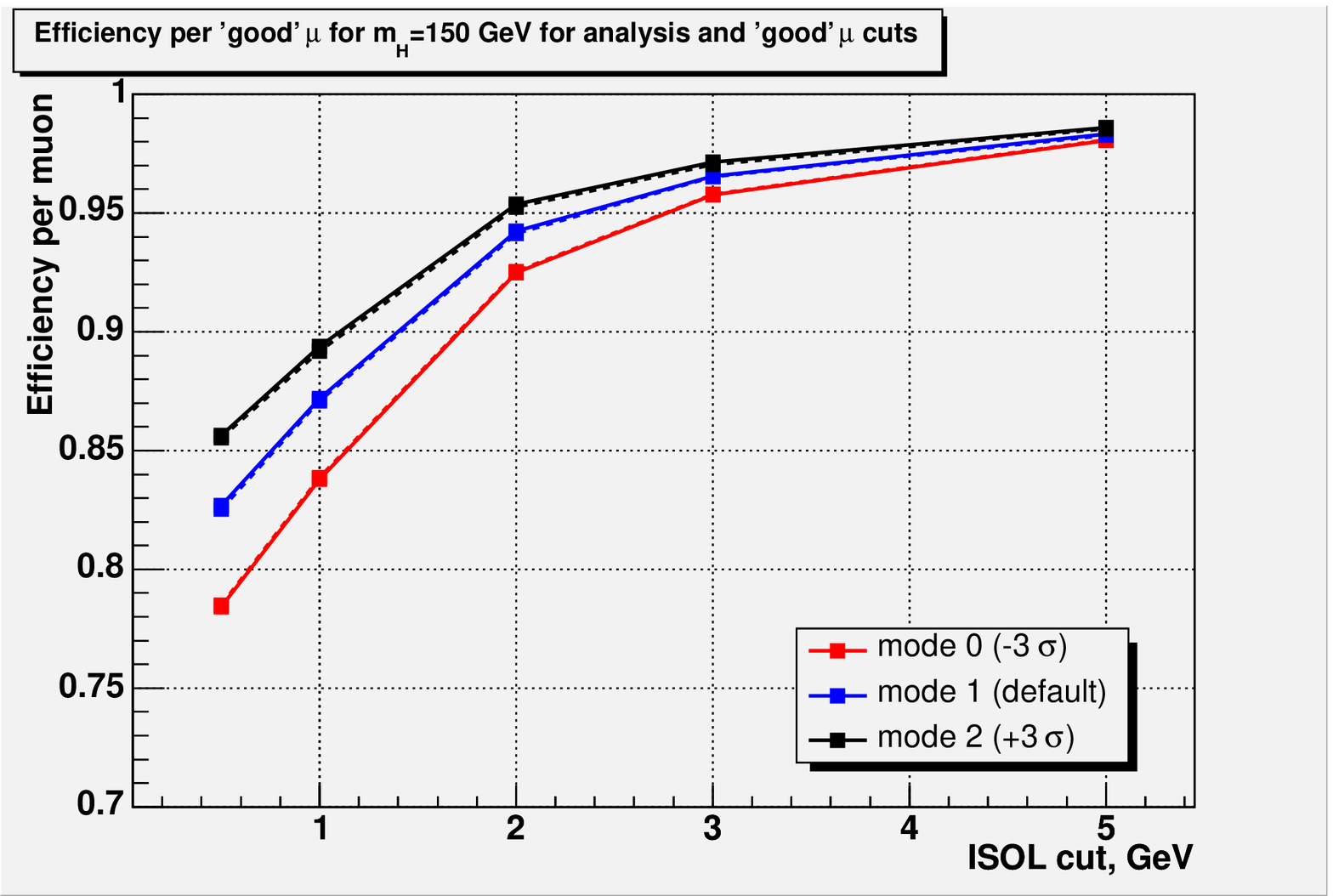}} \\
    \caption{Muon isolation cut efficiency averaged over 4 selected
      muons for signal events (solid lines, Fig.~\ref{fig:GoodSig})
      and ZZ background (dashed lines). The blue middle line is for
      the default MI ${\rm pt_{\textrm{cut-off}} }$, the black upper line is for
      downward ${\rm -3\sigma }$ variation of ${\rm pt_{\textrm{cut-off}} }$ value, the red
      lower line is for upward ${\rm +3\sigma }$ variation.}
    \label{fig:IsolZZsig} &
    \caption{Muon isolation cut efficiency averaged over 4 selected
      muons for signal events. Solid lines are for good muons from
      events after analysis cuts (see Fig.~\ref{fig:GoodSig});
      dashed lines are for good muons from events before analysis
      cuts. There is no difference at statistical precision level for
      two graph sets. The blue middle line is for the default MI ${\rm
	pt_{\textrm{cut-off}} }$, the black upper line is for downward
      ${\rm -3\sigma }$ variation of ${\rm pt_{\textrm{cut-off}} }$
      value, the red lower line is for upward ${\rm +3\sigma }$
      variation.}
    \label{fig:GoodBeforeAfter} \\
    \resizebox{\linewidth}{0.65\linewidth}{\includegraphics{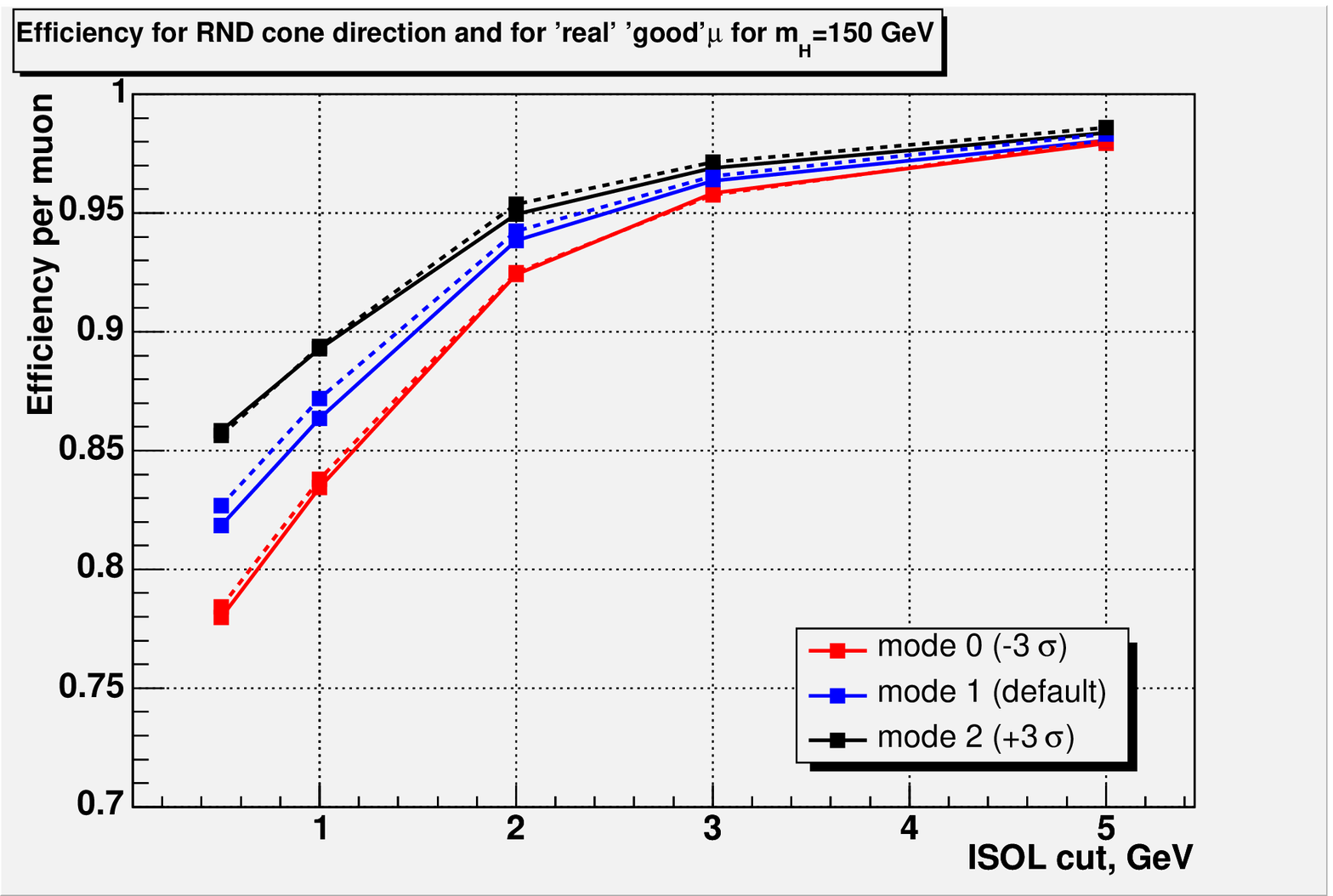}} &
    \resizebox{\linewidth}{0.65\linewidth}{\includegraphics{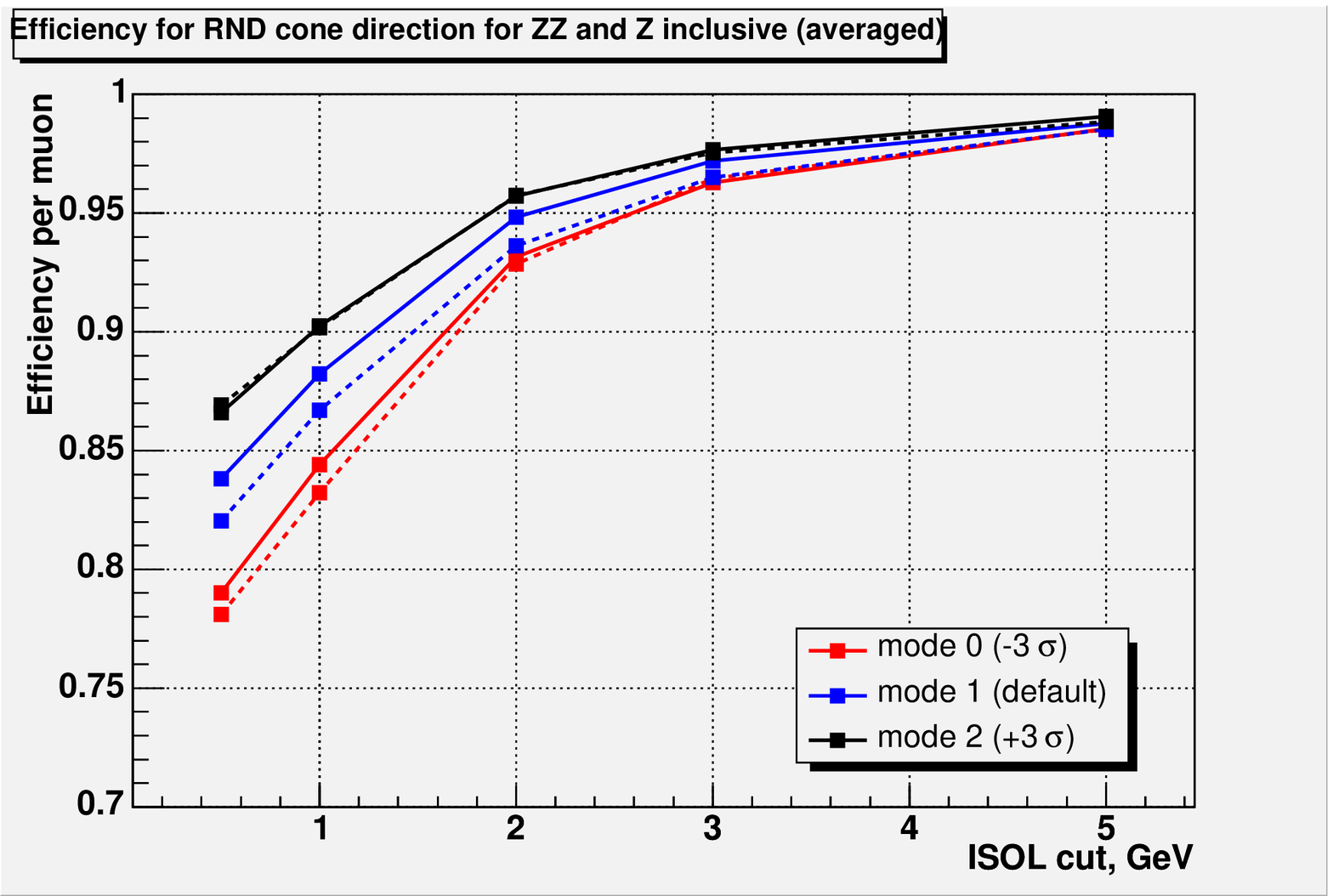}} \\
    \caption{Muon isolation cut efficiency for random-cone directions
      (solid lines) and for muons (dashed lines) for signal events. The
      blue middle lines are for the default MI ${\rm pt_{\textrm{cut-off}}
      }$, the black upper lines are for downward ${\rm -3\sigma }$
      variation of ${\rm pt_{\textrm{cut-off}} }$ value, the red lower
      lines are for upward ${\rm +3\sigma }$ variation.}
    \label{fig:SigRND} &
    \caption{Muon isolation cut efficiency for random-cone directions for
      Z-inclusive (dashed lines) and ZZ (solid lines) events. The blue
      middle lines are for the default MI ${\rm pt_{\textrm{cut-off}} }$, the black upper
      lines are for downward ${\rm -3\sigma }$ variation of ${\rm pt_{\textrm{cut-off}} }$ value,
      the red lower lines are for upward ${\rm +3\sigma }$ variation.}
    \label{fig:ZincRND} 
  \end{tabular}
\end{figure}

Figure \ref{fig:IsolZZsig} compares the muon isolation cut efficiency
curves for the main irreducible ZZ background and for the Higgs boson
events. Clearly, these efficiencies are very similar.

\paragraph{Sensitivity to kinematical cuts}

Figure \ref{fig:GoodBeforeAfter} demonstrates another very important
feature of the tracker-based muon isolation cut: its efficiency is not
very sensitive to the kinematical analysis cuts. The figure has two
sets of efficiency curves: one is obtained for "good" muons and
another for "good" muons passing further event selection cuts as
described in section 3.2. One can hardly see any
difference. Therefore, the conclusions of this analysis will not
depend on the choice of the final event selection cuts.

\paragraph{Evaluation of the muon isolation cut efficiency from data using random-cone directions}

Figure \ref{fig:SigRND} shows the isolation cut efficiency as
calculated for random directions uniformly distributed in ${\rm
\eta-\phi }$ space (${\rm |\eta|<2.4 }$). The algorithm of the ISOL
parameter calculation is the same as for ``real'' MC muons, except
that now the ISOL parameter takes into account the sum of PT for
tracks around random directions in the acceptance region. The Higgs
boson Monte Carlo sample was used to make these plots. We see that the
graphs obtained for the random cone (solid lines) and for ``real''
muons (dashed line; identical to Figures \ref{fig:GoodSig} and
\ref{fig:GoodBeforeAfter}) look very similar. In fact, they agree
within statistical uncertainties.  This observation motivated us to
investigate whether we can measure the isolation cut efficiency by
using some distinct reference data sample and applying the random-cone
technique. The reference data sample must have a large cross section
(to provide good statistics), be relatively clean from backgrounds,
and have a similar underlying structure to ZZ events. Inclusive ${\rm
Z \rightarrow \mu\mu }$ seems to be just what we need. The cross
section is ${\rm \sim1.6 }$ nb, ${\rm Z \rightarrow \mu\mu }$ has a
very clean signature.

Figure \ref{fig:ZincRND} shows the isolation cut efficiencies computed
for random-cone directions in Z-inclusive Monte Carlo sample.  One can
see that the isolation cut efficiencies for muons in the ZZ sample
are very well mimicked by the efficiencies calculated for random cones
in the Z-inclusive sample. The variations in the UE
${\rm pt_{\textrm{cut-off}} }$ have nearly identical effects on both data
samples.

\paragraph{${\rm 4\mu}$ Isolation cut efficiency per event}

Efficiencies per event are listed in Table~\ref{tab:perEvnt}.  We
observe that the values for Signal, ZZ-background, and Z-inclusive
using random-cone technique samples are in agreement with each other
for all three tested UE scenarios. The range of efficiencies for the
ZZ-background spans from ${\rm \sim0.72 }$ to ${\rm \sim0.84 }$. This
range of ${\rm \pm 6\%}$ absolute of the central value can be
associated with the uncertainties on the 4-muon isolation cut
efficiency arising from theoretical uncertainties on considered UE
parameters in PYTHIA.

\begin{table}[htb]
  \small
  \label{tab:perEvnt}
  \begin{center} \renewcommand{\arraystretch}{1.3}
    \begin{tabular}{||c|c|c|c|c||} \hline
      process/case & efficiency (default) & efficiency (${\rm -3\sigma }$) & efficiency (${\rm +3\sigma }$) \\ \hline
      signal, ${\rm m_H = 150 }$ GeV         & $0.775 \pm 0.004$ & $0.707 \pm 0.005$ & $0.812 \pm 0.004$ \\ \hline
      ZZ background                   & $0.780 \pm 0.004$ & $0.721 \pm 0.005$ & $0.838 \pm 0.004$ \\ \hline
      4 RND muons, Z-inclusive events & $0.762 \pm 0.007$ & $0.706 \pm 0.007$ & $0.821 \pm 0.006$ \\ \hline
      ${\rm t\bar{t} }$ background           & $0.016 \pm 0.001$ & $0.013 \pm 0.001$ & $0.015 \pm 0.001$ \\ \hline
    \end{tabular}
  \caption{Efficiency per event using different events samples: Higgs
  boson signal with ${\rm m_H=150 }$ GeV, ZZ background, Z-inclusive
  (4 RND muons), ${\rm t\bar{t} }$ background. ``4 RND muons'' means
  that for a particular process in each event 4 random cone directions
  were used to calculate the ISOL parameter and the corresponding
  values were treated as ones for ``real'' muons.}
  \end{center}
\end{table}

On the other hand, it appears possible to use the Z-inclusive sample
to gauge the UE activity and evaluate the 4-muon isolation cut
efficiency experimentally.  There might be a small systematic shift of
the order of ${\rm \sim 2\% }$ in efficiencies between the ZZ and
Z-inclusive samples, and this is a shift for calibration from data
technique, which makes the result to a large degree independent from a
particular UE scenario which would be actually realized in nature.
For the three different UE simulations we used in these studies, we
obtain the following offsets: $0.018 \pm 0.008$, $0.015 \pm 0.009$,
$0.017 \pm 0.007$. Much larger Monte Carlo samples would be needed to
pin it down more accurately. Meanwhile, conservatively, one may just
ignore this correction and assign a $2\%$ systematic uncertainty on
the Z-sample-based estimate of the 4-muon isolation cut efficiency for
ZZ-background and Higgs boson signal events. This uncertainty is
already much smaller in comparison to the other systematics such as
experimental uncertainties on the muon reconstruction efficiency,
theoretical uncertainties associated with the choice of PDF's and QCD
scale, etc.

The efficiency for accepting ${\rm t\bar{t} }$-events is of the order of
0.015 $\pm$ 0.001. Its sensitivity to the UE could not be studied due to
lack of statistics, but it is not expected to be too large as it is
dominated by the jet activity. In fact, if the reducible ${\rm t\bar{t} }$-
and ${\rm Zb\bar{b} }$-backgrounds could not be suppressed well below the
ZZ-background, one would need to study their sensitivity to the UE
physics, as well as to the jet fragmentation modeling.


\subsubsection{SUMMARY}

The isolation cut efficiency per muon due to uncertainties in the
considered UE models can vary as much as $\pm 5$\% (the efficiency
itself and its uncertainty strongly depend on how tight the ISOL cut
is). The 4-muon isolation cut efficiency per event for ${\rm ZZ
\rightarrow 4 \mu }$ background is measured to be $\sim (78 \pm 6)\%$.

To decrease these large uncertainties to a negligible level with
respect to other systematic uncertainties, one can calibrate the
isolation cut efficiency from data using Z-inclusive events (${\rm Z
\rightarrow 2 \mu }$) and the random-cone technique. We show that this
indeed significantly decreases uncertainties associated with the poor
understanding of the UE physics. There might be $\sim 2\%$ systematic
shift in the 4-muon isolation cut efficiencies obtained this way. In
principle, one can correct for this shift, but it does not appear to
be necessary as this uncertainty is already very small.

The results and described techniques in this letter may be of interest
for all analyses relying on lepton isolation cuts.

\subsubsection{ACKNOWLEDGMENTS}

We would like to thank 
M.~Aldaya, 
P.~Arce, 
J.~Caballero, 
B.~Cruz, 
G.~Dissertori,
T.~Ferguson,
U.~Gasparini,
P.~Garcia,
J.~Hernandez,
I.~Josa, 
M.~Konecki,
P.~Moisenz, 
E.R.~Morales, 
N.~Neumeister,
A.~Nikitenko,
F.~Palla and
I.~Vorobiev
for their active participation in the analysis discussions and
comments on this letter.


\clearpage\setcounter{equation}{0}\setcounter{figure}{0}\setcounter{table}{0}
\newcommand{\g}{$\gamma$ }
\newcommand{\co}{$ {}^{60}Co $}
\newcommand{\na}{$ {}^{22}Na $}
\newcommand{\revident}{\hspace{-0.52cm}}

\newcommand{\ah}{$A^0/H^0$}
\newcommand{\aintautau}{$bbA^0/H^0 \rightarrow bb\tau \tau $~}
\newcommand{\zintautau}{$Z^0 \rightarrow \tau \tau $~}
\newcommand{\hchatau}{$H^{\pm} \rightarrow \tau^{\pm} + \nu_{\tau}$~}
\newcommand{\hchaw}{$pp\rightarrow t\bar{t} \rightarrow H^{\pm} b W^{\pm} \bar{b}$~}
\newcommand{\dR}{\ensuremath{\Delta \rm{R}} }
\newcommand{\Eta}{\ensuremath{\eta}}
\newcommand{\dRf}{\ensuremath{\Delta \rm{R}=\sqrt{\Delta ^{2} \eta  + \Delta ^{2} \phi }}}

\newcommand{\Rem}{\ensuremath{\rm{R}_{]\rm{em}}}~}
\newcommand{\Isofrac}{\ensuremath{\Delta E_T^{12}}~}
\newcommand{\Stripwidth}{\ensuremath{\Delta\eta}~}
\newcommand{\Etawidth}{\Stripwidth}
\newcommand{\Ntr}{\ensuremath{\rm{N}_{\rm{Tr}}}~}
\newcommand{\DEot}{\ensuremath{ \Delta E^{12}}~}
\newcommand{\Etopt}{\ensuremath{\et / \rm{p}_{\rm{T},1}}}
\newcommand{\Sipsig}{\ensuremath{ d_0 / \sigma_{d_0}}~}
\newcommand{\degr}{^\circ}
\newcommand{\Degr}{$^\circ$}
\newcommand{\Nstrip}{\ensuremath{\rm{N}_{\eta-hits}}~}
\newcommand{\ta}{\ensuremath{\tau}}
\newcommand{\Wmunu}{\ensuremath{\rm{W} \rightarrow \mu \nu}~}
\newcommand{\tjet}{$\tau$ jet}
\newcommand{\ttype}{\mbox{$\tau$--type}}
\newcommand{\ttrack}{\mbox{$\tau$--track}}

%
%
%




\subsection{\ta-Identification, from D\O\ to ATLAS}

Michael Heldmann, Ingo Torchiani \\ [2mm]
{\em Freiburg University, Germany}

\subsubsection{Introduction}
\label{sec:tau}

Excellent reconstruction and identification of all lepton species is
crucial at the Large Hadron Collider (LHC). Tau leptons are the most difficult ones in this
respect, since they produce neutrinos and hadrons among their decay
products. Many different physics channels contain \ta\ leptons in their
final states. In particular the heaviest Higgs bosons in the Minimal
Super Symmetric Model (MSSM) can be observed through their decays to 
\ta\ leptons (\aintautau, \hchatau) and also the Standard Model (SM) Higgs produced 
through Vector Boson Fusion (VBF) can be observed when it decays 
to a \ta\ pair. Additionally \ta\ leptons can be an
important signature for SUSY.
Since it will not be possible to discriminate between prompt light leptons
(electrons and muons) and leptons from \ta-decays, the hadronic decay
modes will have to be explored.
Various characteristics of a \ta-decay allow a discrimination against
 jets from QCD-events (QCD-jets) or heavy quark-jets (as from \ttbar). 
This discrimination will be called \ta -identification in the following.
\par
For physics at the LHC many studies have been undertaken to evaluate the discovery potential in these channels. Since data from the LHC will not be available for at least one more year these studies had to rely on Monte Carlo simulation. 
Though a lot of work has been invested to provide a detailed description of the physics and detector effects, the potential uncertainty for such a complex variable like a multivariate discriminator between \ta-jets and other jets might be substantial.

In terms of the similarities in physics environment and detector design, the Tevatron and D\O\ specifically is the best available tool to investigate the reliability of such Monte Carlo techniques to predict the performance of a given algorithm to separate between \ta-jets and other jets.

Therefore an attempt has been made to estimate the uncertainty on the performance of the \ta-identification algorithm used in ATLAS.

To accomplish this we try to establish a chain of understanding composed of the following steps:
\begin{itemize}
\item D\O\ Algorithm on D\O\ data
\item D\O\ Algorithm on D\O\ MC
\item D\O\ Algorithm on ATLAS MC
\item ATLAS Algorithm on ATLAS MC
\item Prediction for ATLAS Algorithm on ATLAS data
\end{itemize}

To implement this chain the strategy will be to select a signal and a background sample in D\O\ data and study the \ta\ preselection as described below on these samples. After the selection we investigate a simple cut based \ta-identification using three key variables, which have been implemented in ATLAS. Afterwards it will be shown to what extent these results are transferable to ATLAS.  



\subsubsection{\ta-reconstruction and identification in ATLAS}

The reconstruction of \ta-candidates in ATLAS is done by a package called ``tauRec'' \cite{TAUREC}.
The seeds for the building of tau candidates are provided by a sliding window cluster algorithm. It runs on  $ \Delta \eta \times \Delta \phi = 0.1 \times 2 \pi / 64 $ calorimeter towers. 
Only clusters with $ \rm{E}_{\rm{T}} > 15 ~\rm{GeV} $ are considered.

\begin{figure}[htbp]
  \centering
  \begin{minipage}[t]{6.9 cm}
    \includegraphics[width=6.9cm]{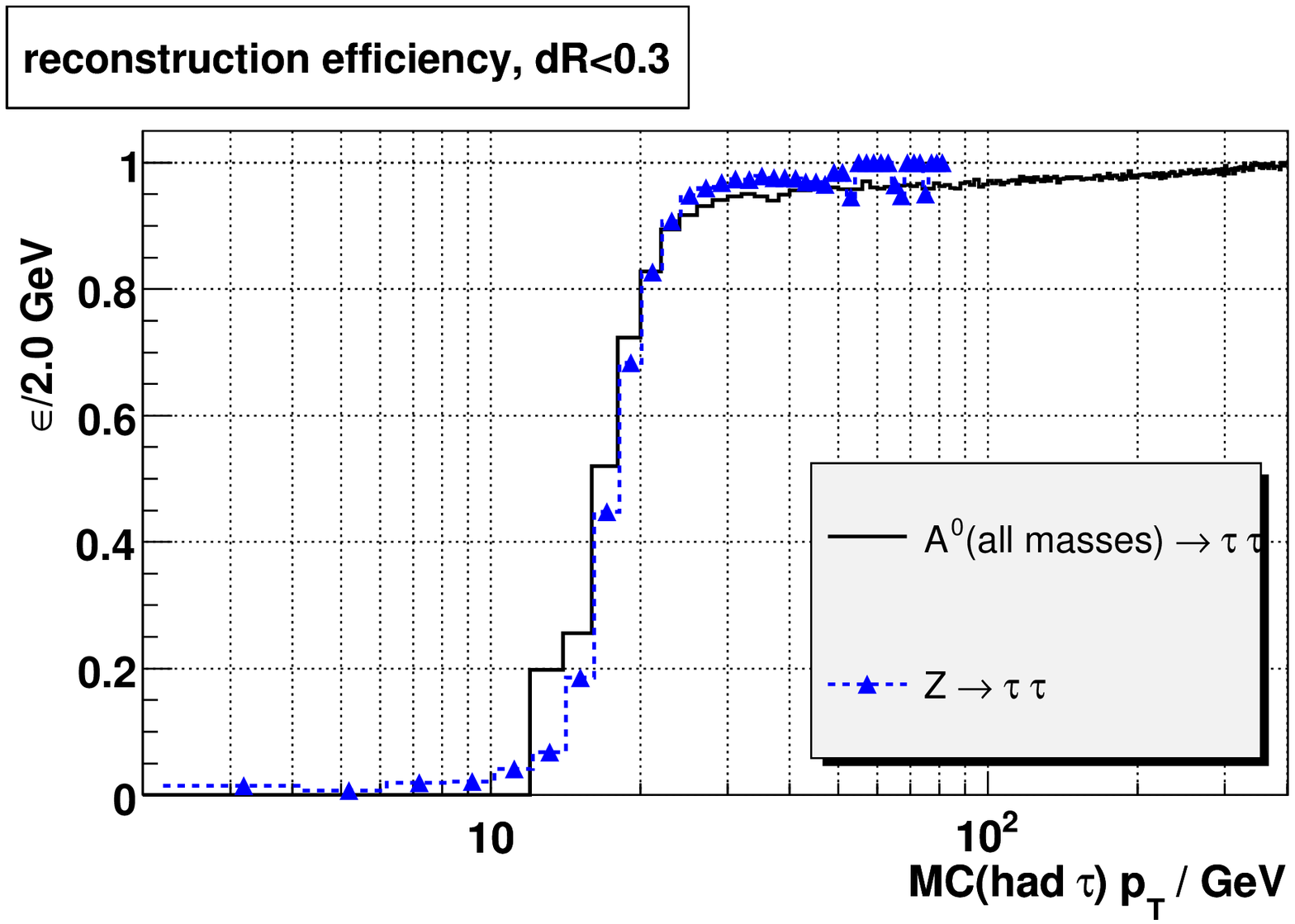}  
    \caption[]{{\small Reconstruction efficiency for \ta-jets as a function of \ET. Black is an average over signal samplesfrom \aintautau events, dotted is only a \zintautau sample.}}
    \label{RECOEFF}
  \end{minipage}
  \hspace{1cm}
  \begin{minipage}[t]{6.9 cm}
    \includegraphics[width=6.9cm]{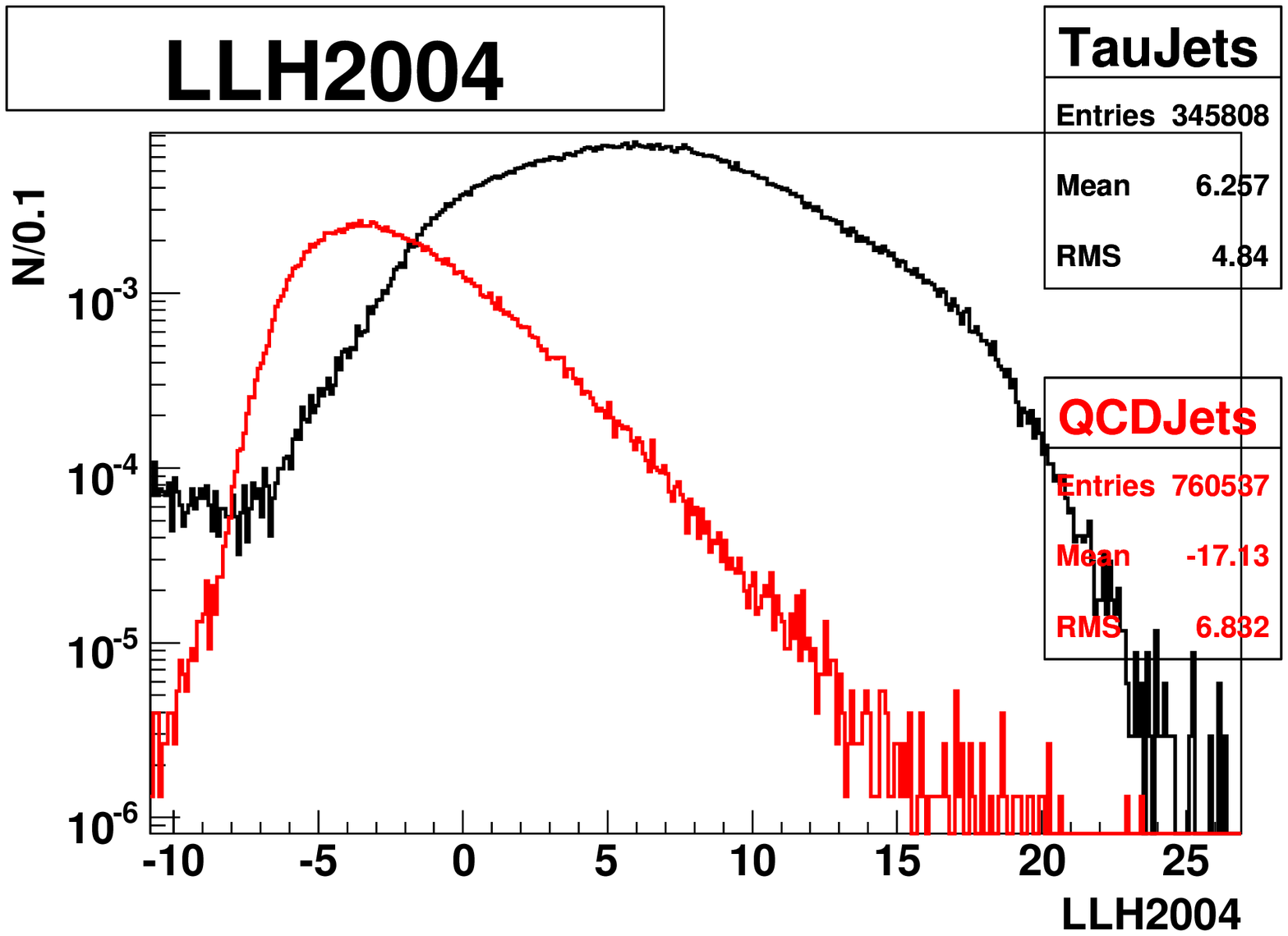} 
    \caption[]{{\small LLH2004 distribution for \ta-jets (black) and QCD-jets (red). Candidates with a LLH $< -10 $ had variables outside the boundaries of histograms used when obtaining the PDFs. The LLH is applied after a preselection of $1 \le \Ntr \le 3$. Due to statistical limitations, only one-dimensional distributions have been used. Nevertheless a good separation between \ta-jets and QCD-jets is achieved.}}
    \label{LLH2004ALLG3NoNoise}
  \end{minipage}
\end{figure}

In Figure~\ref{RECOEFF} the probability is shown for a true \ta-jet
within $|\eta|<2.5$ to be reconstructed as a \ta -candidate as a
function of \ET for two different samples. \ET represents the
transverse energy of the visible decay products. A true \ta-jet is called reconstructed
if a cluster is found with a barycenter within $\Delta \rm{R} < 0.3$
around the \ta-jet direction. The efficiency rises from 20~\% at
15~GeV over 88~\% at 20~GeV and saturates at 98~\% for
$\rm{E}_{\rm{T}}>30\ \rm{GeV}$.

The reconstruction is followed by a step called identification. For
the purpose of separating \ta-jets from other jets a set of variables
is calculated for each \ta -candidate. The three most important of
these variables are \Rem, \Isofrac and \Ntr.

\begin{itemize}
\item \Ntr: number of tracks, extrapolated to the calorimeter, within $ \Delta R < 0.2 $
around the cluster barycenter, with $P_T > 2\GeV $
\item \Rem: transverse energy radius in the EM calorimeter layers
\begin{equation}
\label{EQUEMRADIUS} 
R_{em} = \frac{\sum_{i=1}^{n} E_{Ti} \sqrt{ \left( \eta_i - \eta_{cluster} \right)^2 + \left( \phi_i - \phi_{cluster} \right)^2 } } {\sum_{i=1}^{n}E_{Ti}}
\end{equation}

\hspace{0.5cm} {\sf \small i runs over all electromagnetic calorimeters cells in the cluster with $\Delta \rm{R} < 0.4$,}

\item \Isofrac: transverse energy isolation
\begin{equation}
\label{EQUISOFRAC}
\Delta \rm{E}_{\rm{T}}^{12} = \frac{\sum_{j=1}^{n'}E_{Tj}}{\sum_{i=1}^{n}E_{Ti}}
\end{equation}

\hspace{0.5cm} {\sf \small j runs over all electromagnetic calorimeters cells in the cluster with $0.1 < \Delta \rm{R} < 0.2$,}

\hspace{0.5cm} {\sf \small n' denotes their number, $\rm{E}_{\rm{T}j}$ is the transverse energy in cell j}

\hspace{0.5cm} {\sf \small i runs over all electromagnetic calorimeters cells in the cluster with $\Delta \rm{R} < 0.4$,}
 
\hspace{0.5cm} {\sf \small n denotes their number, $\rm{E}_{\rm{T}i}$ is the transverse energy in cell i}

\end{itemize}

These three along with five other variables are combined into one discriminant using a likelihood ratio method, shown in Figure~\ref{LLH2004ALLG3NoNoise}. A good separation between \ta-jets and light jets can be obtained by tuning a cut on this single variable LLH2004 to the desired efficiency.

\subsubsection{\ta-reconstruction and identification in D\O}
\label{d0tauid}

Reconstruction and identification of hadronically decaying \ta-leptons has been used successfully at the \dzero experiment in various analyses, e.g. a measurement of the $Z/\gamma^*\rightarrow\tau\tau$ cross section \cite{Abazov:2004vd}. The search for \ta-candidates at D\O\ starts with the reconstruction of energy clusters in the calorimeter using a $0.3$-cone algorithm seeded either by a calorimeter tower with a transverse energy of more then 1~GeV or a track with a transverse momentum of more than $5$~GeV. For being considered as a \ta-candidate, the total transverse energy in the cluster is required to be larger than $4$~GeV or $2$~GeV, respectively. In order to find energy deposits from neutral pions, a nearest-neighbor algorithm is used to reconstruct energy clusters in the electromagnetic calorimeter only. If the transverse energy of such an EM cluster is above $0.8$~GeV it is regarded as a $\pi^0$ candidate. 

In a second step tracks which are reconstructed in the central tracking system are associated to the calorimeter cluster. Up to three tracks can be assigned to a \ta-candidate. Tracks are processed in the order of decreasing transverse momentum and have to fulfill the following requirements for being associated to the \ta-candidate:

\begin{itemize}
\item The transverse momentum has to be larger than 1.5~GeV.
\item The distance at the point of closest approach between first, second and third track has to be smaller than 2~cm.
\item The invariant mass calculated from first and second track is smaller than 1.1~GeV.
\item The invariant mass calculated from first, second and third track is smaller than 1.7~GeV.
\item The charge of all tracks adds up to $\pm 1$.
\end{itemize}

After this reconstruction process the tau candidates are classified into three types:

\begin{itemize}
\item {\bf \ttype\ 1}: $\tau^\pm \rightarrow h^\pm\nu_\tau$ ($\pi$-like). The \ta-candidate consists of a calorimeter cluster and one track, without any reconstructed EM cluster.
\item {\bf \ttype\ 2}: $\tau^\pm \rightarrow h^\pm\nu_\tau+ (n\pi^0), n > 0$ ($\rho$-like). The \ta-candidate consists of a calorimeter cluster, one track and at least one EM cluster.
\item {\bf \ttype\ 3}: $\tau^\pm \rightarrow h^\pm h^\mp h^\pm \nu_\tau + (n\pi^0), n \ge 0$ (3-prong). The \ta-candidate consists of a calorimeter cluster and two or three tracks.
\end{itemize}

\begin{figure}[t]
\begin{center}
\includegraphics[width=7.4cm]{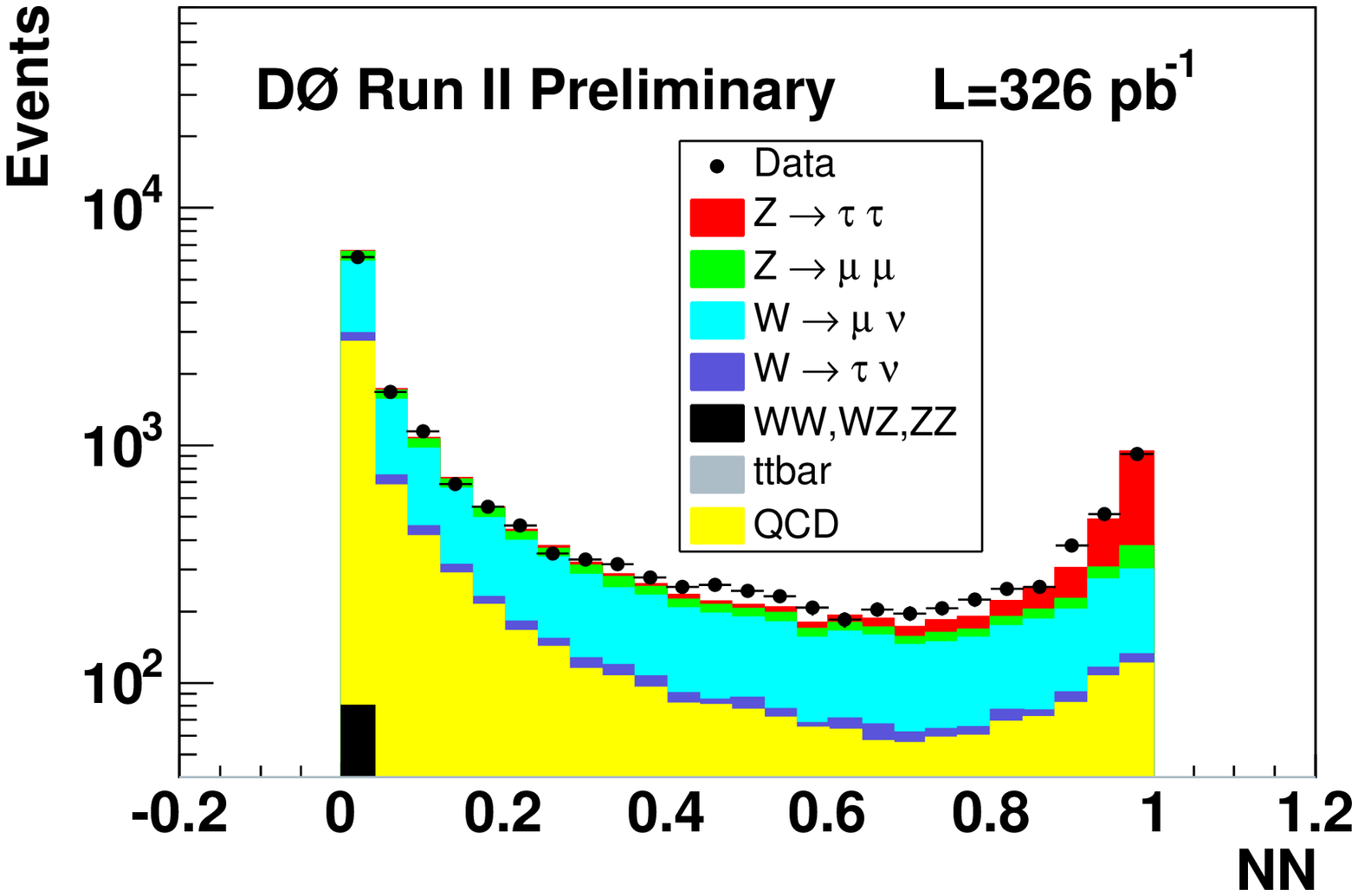}
\includegraphics[width=7.4cm]{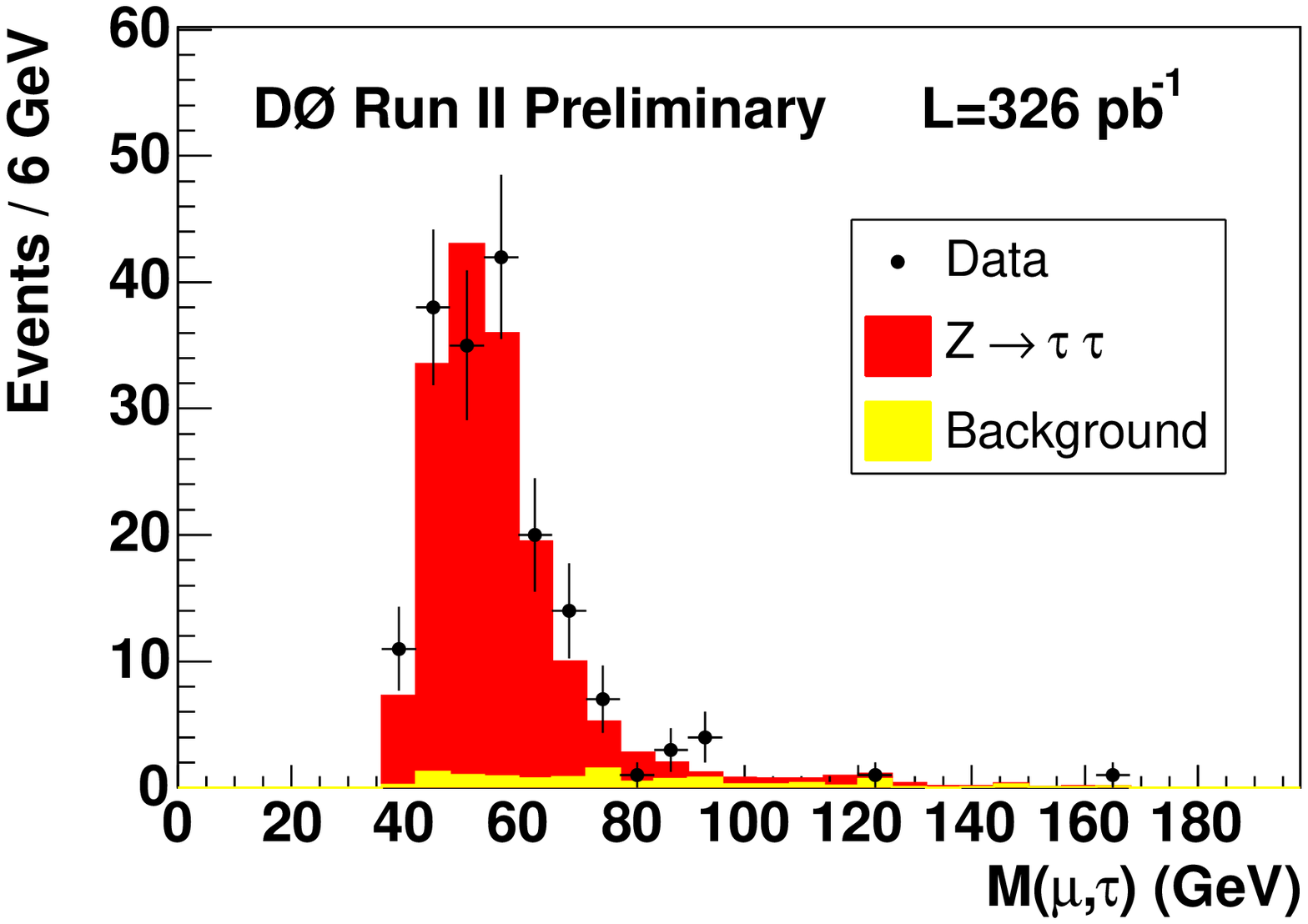}
\includegraphics[width=7.4cm]{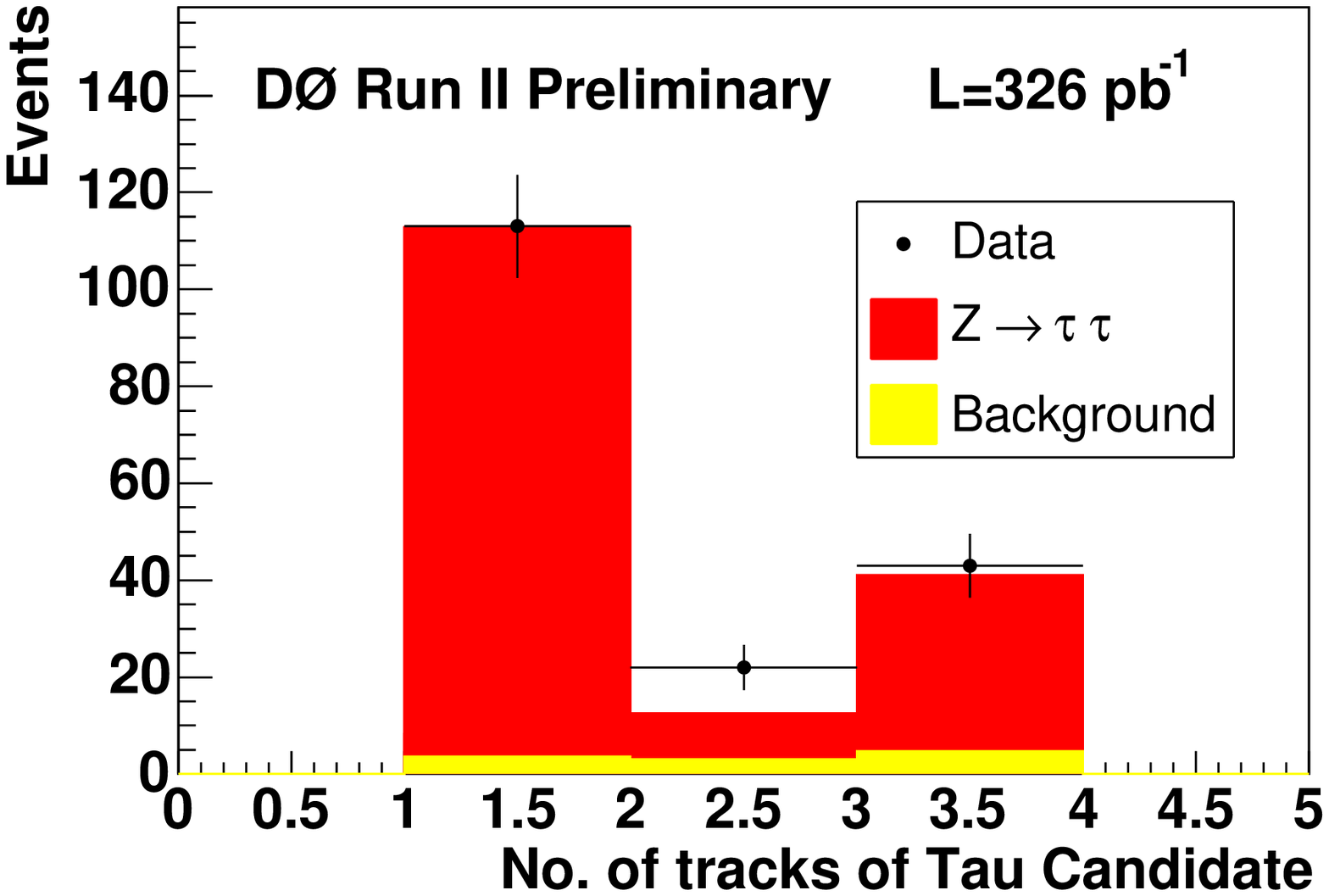}
\includegraphics[width=7.4cm]{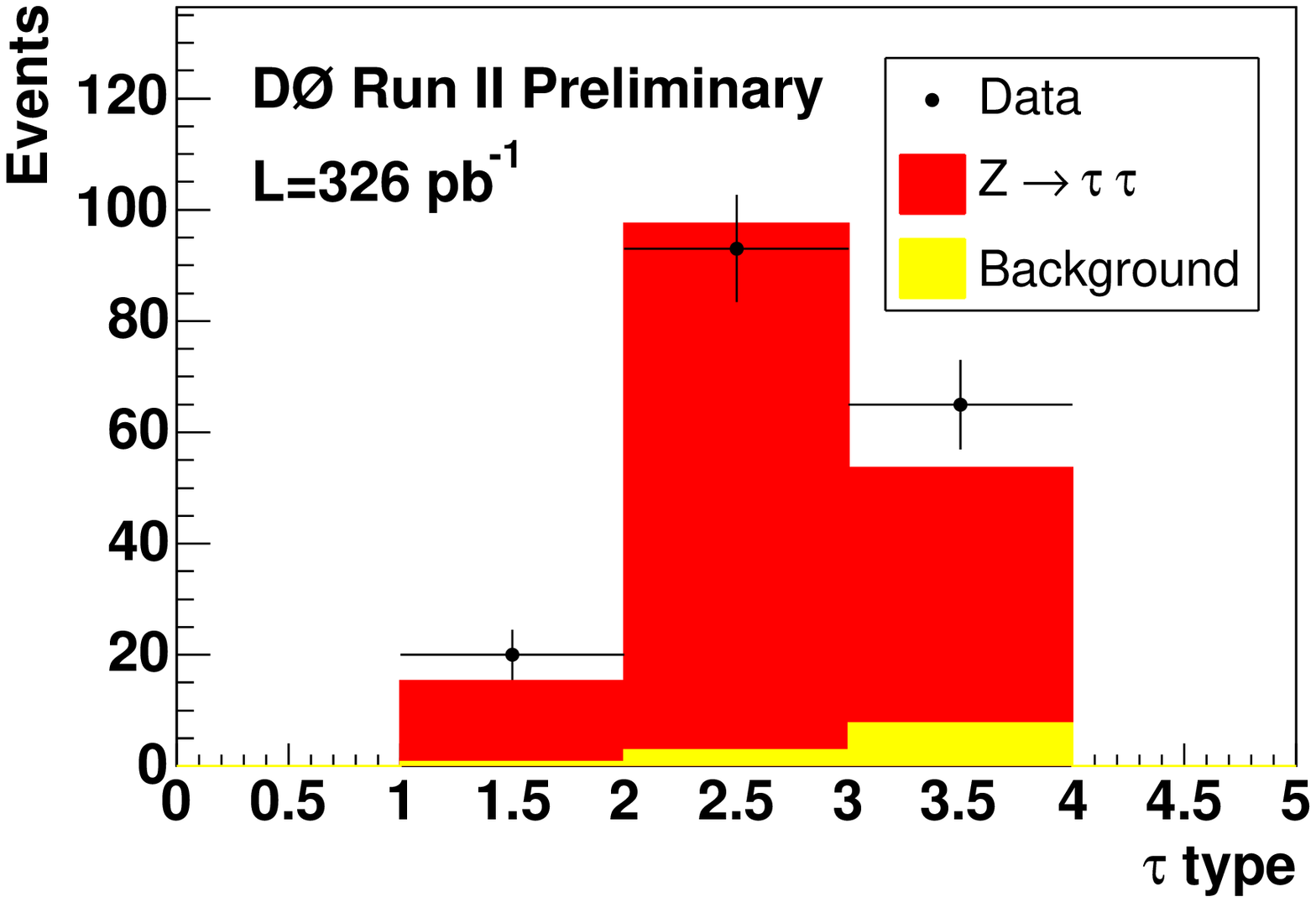}
\caption[]{{\small Distribution of the output of the neural net used for the \ta-identification at D\O\ taken from a $\mu+\tau_{had}$-selection for $p_T^\tau > 7$~GeV and distributions of the $\mu$-$\tau_{had}$ invariant mass, track multiplicity and \ttype\ for $p_T^\tau > 25$~GeV after a selection optimized for $Z/\gamma^*\rightarrow\tau\tau$ \cite{D0note4742CONF}.}}
\label{D0NNOUTPUT}
\end{center}
\end{figure}

\ta-candidates to which no track could be matched will not be discussed, since they are currently not used for physics analyses. After this classification the identification of hadronically decaying tau leptons and the rejection against background from QCD jets is performed using three neural networks, one for each \ttype . The neural nets consist of one input, one hidden and one output layer. The input and hidden layer contain as many nodes as input variables are used, while the output layer holds only a single node. Only the three most important input variables are discussed here, since they are comparable to variables used by ATLAS:

\begin{itemize}
\item {\bf Profile:} $(E_T^1+E_T^2)/E_T(0.5)$, $E_T^i$ represents the transverse energy in the calorimeter tower with the highest and second highest transverse energy and $E_T(0.5)$ the transverse energy in a 0.5 cone around the \ta-candidate.
\item {\bf Isolation:} $(E_T(0.5)-E_T(0.3))/E_T(0.3)$, where $E_T(x)$ represents the transverse energy in a $x$-cone around the \ta-candidate.
\item {\bf Track isolation:} Scalar sum of $p_T$ of tracks which are not associated to the tau, divided by the scalar sum of $p_T$ of all tracks in a 0.5-cone around the \ta-candidate.
\end{itemize}
A distribution of the output of the neural net is presented in Figure~\ref{D0NNOUTPUT}. It also shows $Z/\gamma^*\rightarrow\tau\tau$ signal, \ttype\ and track multiplicity, after a $\mu +\tau_{\rm{had}}$ selection, which uses the neural networks and is optimized for $Z/\gamma^*\rightarrow\tau\tau$ \cite{D0note4742CONF}.

\subsubsection{Signal and background selection}
As a signal sample \zintautau was chosen to provide true \ta-jets. As a background sample \Wmunu is used to provide light jets. \Wmunu is used because it is an important background to many channels with \ta\ final states and because it allows to obtain an unbiased jet-sample using a single $\mu$ trigger, down to rather low \PT.

\begin{figure}[htbp]
  \centering
    \includegraphics[width=9cm]{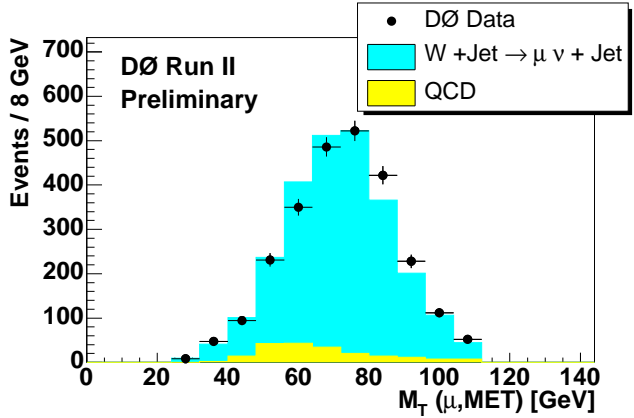} 
    \caption[]{{\small Distribution of the transverse mass between the leading muon and \met\ after cuts described in the text. Shown are data (black dots), \Wmunu (red) and QCD background (yellow).}}
    \label{D0WMUNUMTMUMET}
\end{figure}

\paragraph{\Wmunu sample}
The selection cuts used to obtain the \Wmunu sample were:

\begin{itemize}
\item $\PT(\mu) > 25\GeV,\ |\eta(\mu)| < 1.5$
\item $\PT(jet) > 15\GeV,\ |\eta(jet)| < 1.0$
\item $\met > 20\GeV,\ \rm{m}_{\rm{T}} > 30\GeV$
\item $\rm{m}(\mu,track) < 60\GeV$
\item $\Delta \phi(\met,jet)>0.4$
\end{itemize}
Figure~\ref{D0WMUNUMTMUMET} shows the distribution of the transverse mass between \met\ and the muon after the selection for D\O\ data and Monte Carlo, where the dominating process is \Wmunu.
 
\begin{figure}[t]
\begin{center}
\begin{tabular}{rl}
\includegraphics[width=7.4cm]{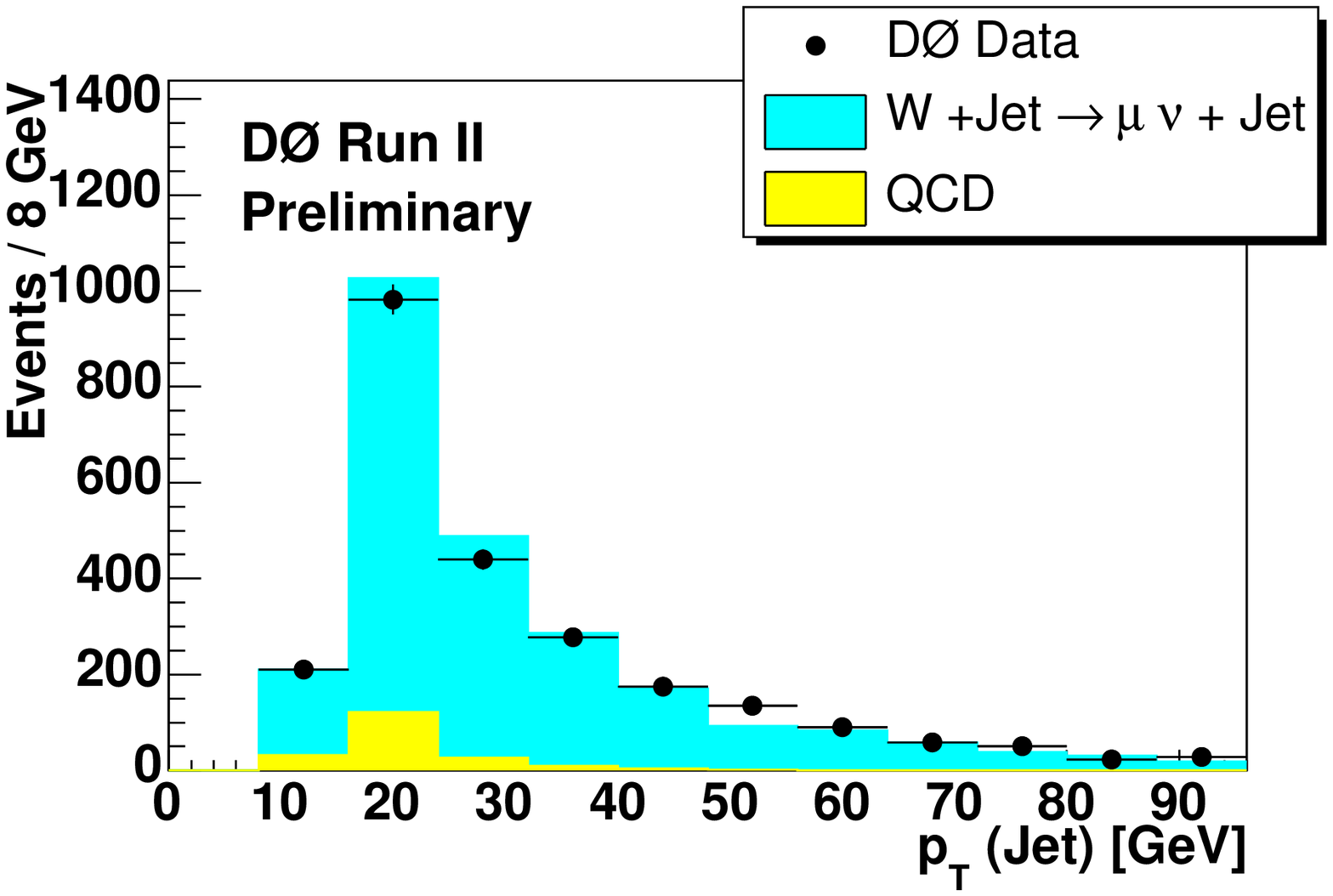} 
\includegraphics[width=7.4cm]{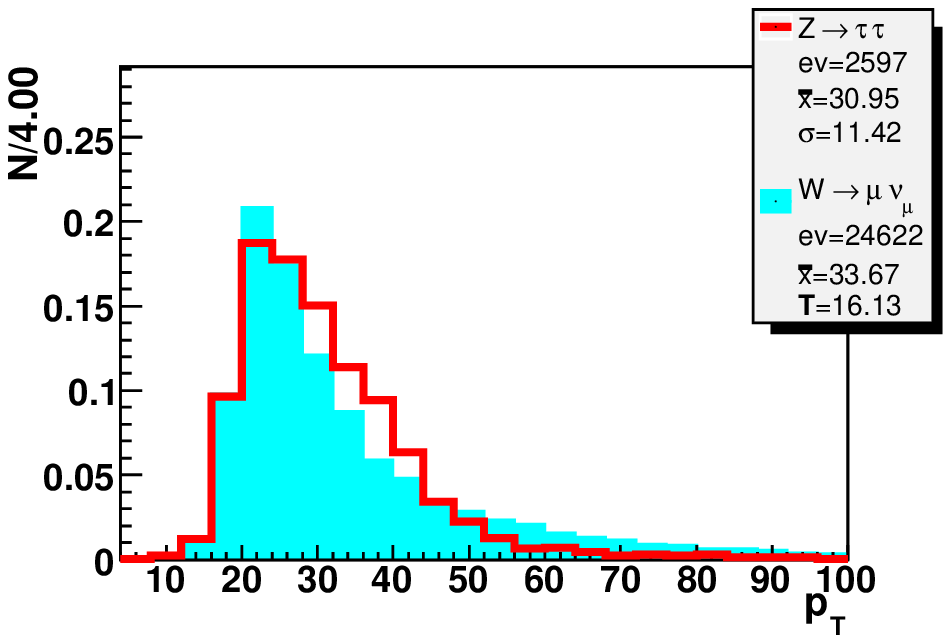} \\ 
\includegraphics[width=7.4cm]{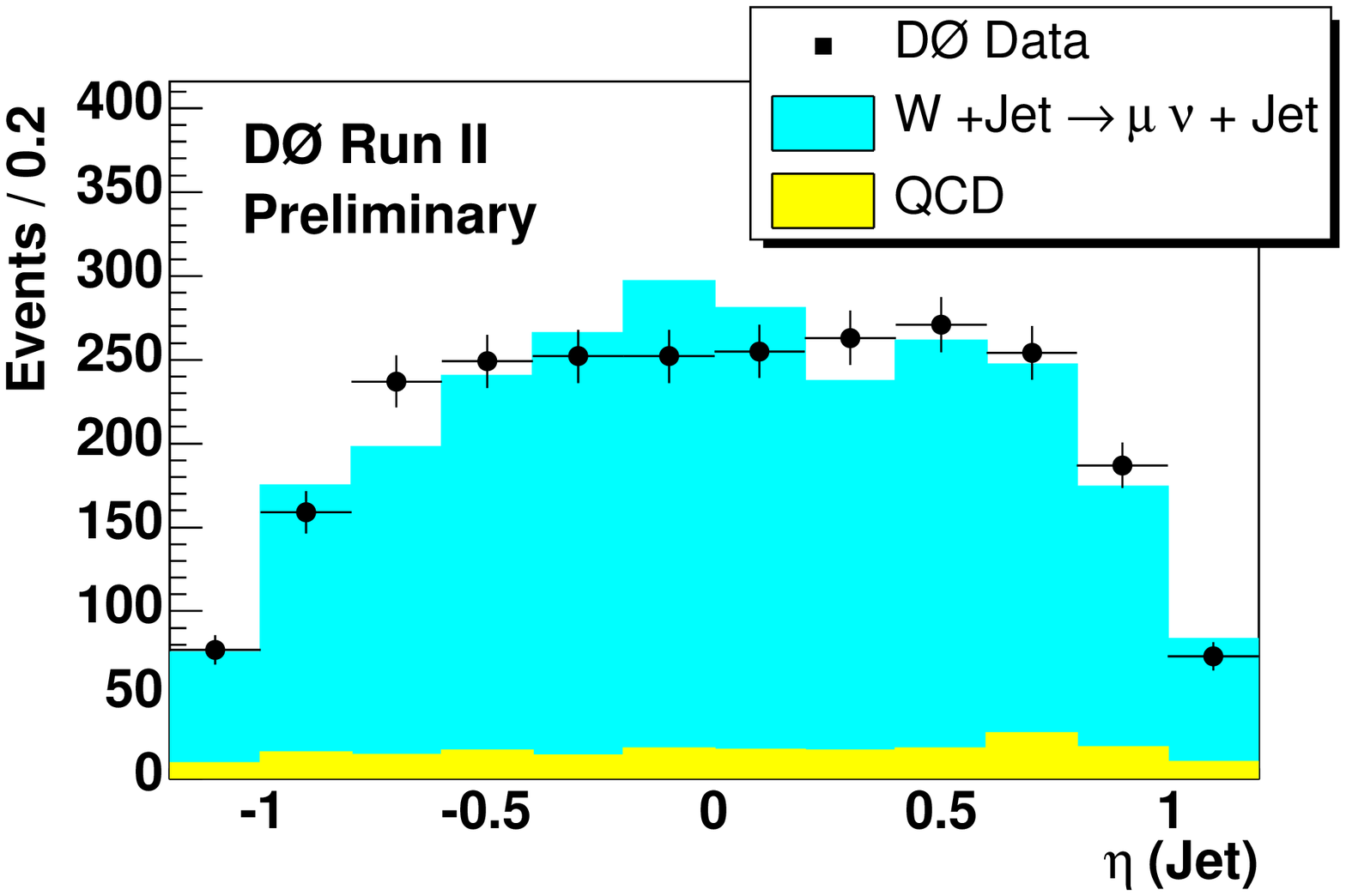} 
\includegraphics[width=7.4cm]{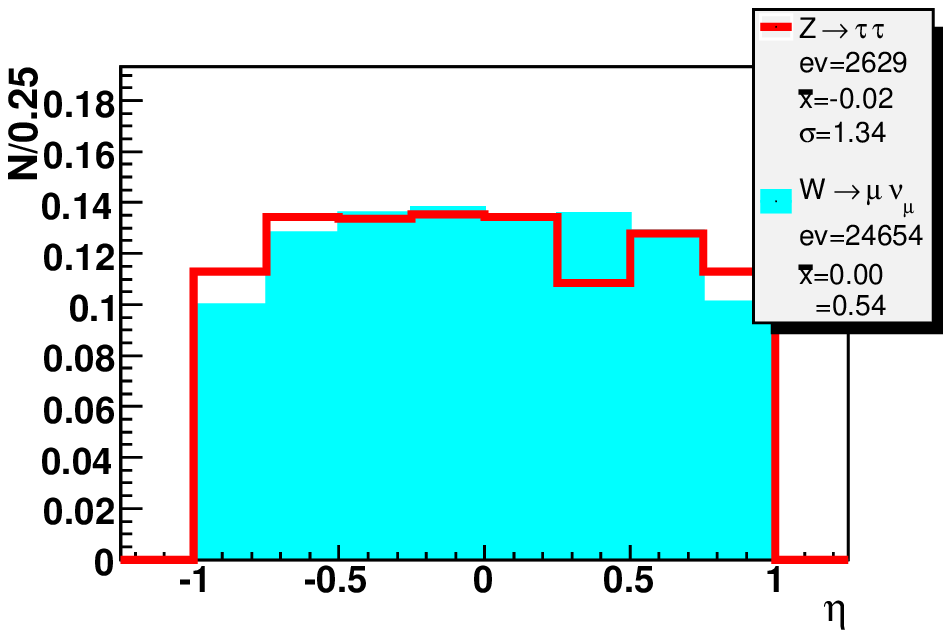} 
\end{tabular}
\caption[]{{\small Distribution of \PT and $\eta$ for the leading jet for D\O\ (left) and ATLAS (right) in the $W\rightarrow\mu\nu$ sample. Distributions are given for D\O\ data, \Wmunu (cyan), QCD (yellow). For ATLAS \PT\ and $\eta$ of the leading \ta-jet are shown in red for \zintautau.}}
\label{D0WMUNUJETPTETA}
\end{center}
\end{figure}

Since the rejection against jets will depend strongly on the kinematic  variables (\ET and \Eta) of the jets these are important. Figure~\ref{D0WMUNUJETPTETA} shows the comparison between data and Monte Carlo for the leading jet \PT and \Eta\ after the \Wmunu selection. Data and Monte Carlo agree within statistical errors.

Similar samples were produced for ATLAS using the same generator (PYTHIA). The same selection cuts were applied to make the samples as comparable as possible, see Figure~\ref{D0WMUNUJETPTETA}.

\paragraph{\zintautau sample}
The \zintautau$\rightarrow\mu+{\rm had}$ sample was selected in D\O\ Monte Carlo by requiring:
\begin{itemize}
\item \PT$(\mu) > 14\GeV,\ |\eta(\mu)| < 1.5$
\item \PT$(\tau) > 15\GeV,\ |\eta(jet)| < 1.0$
\end{itemize}

\subsubsection{Jet properties}

\begin{figure}[t]
\begin{center}
\begin{tabular}{rl}
\includegraphics[width=7.4cm]{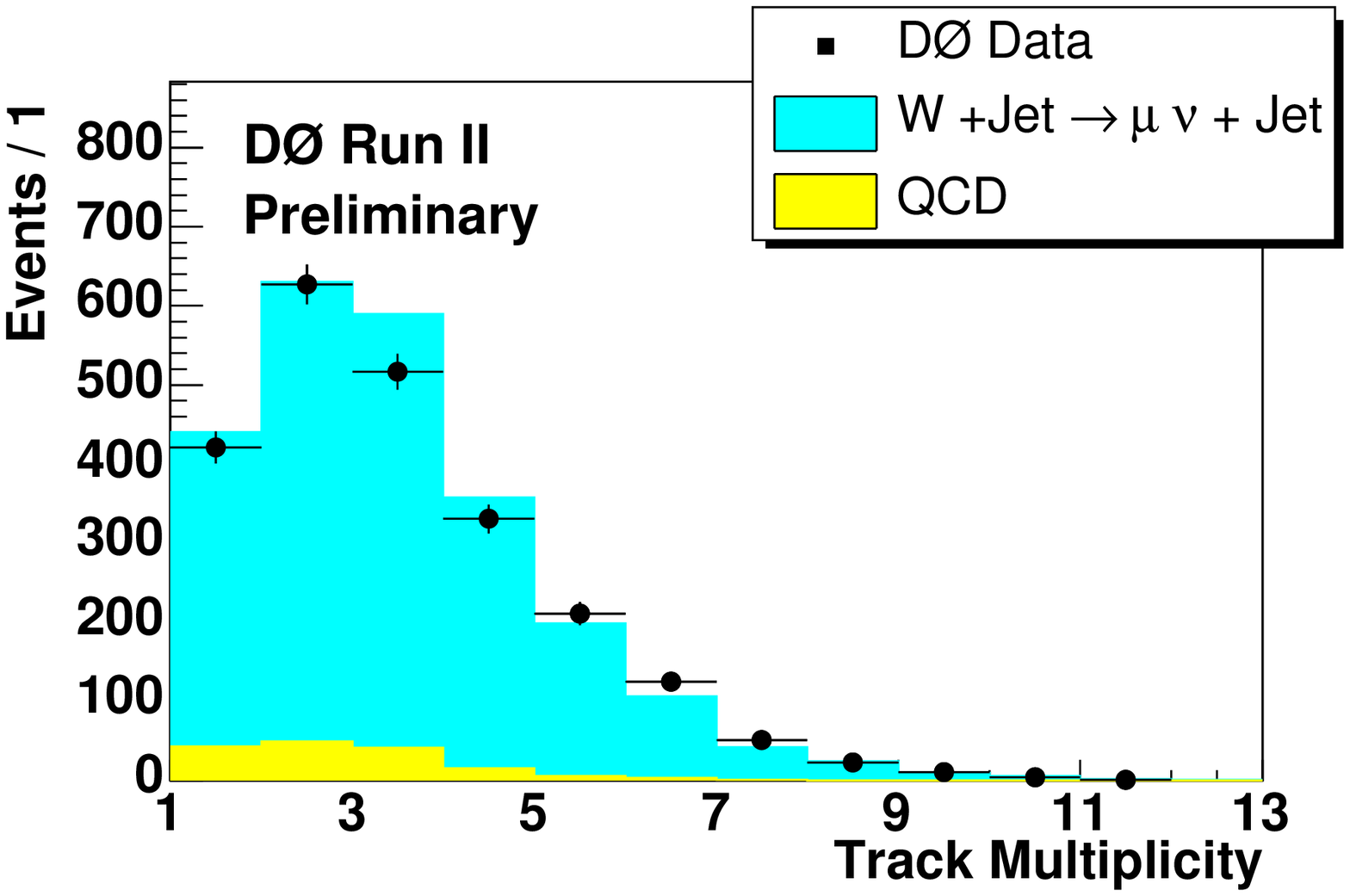}
\includegraphics[width=6.5cm]{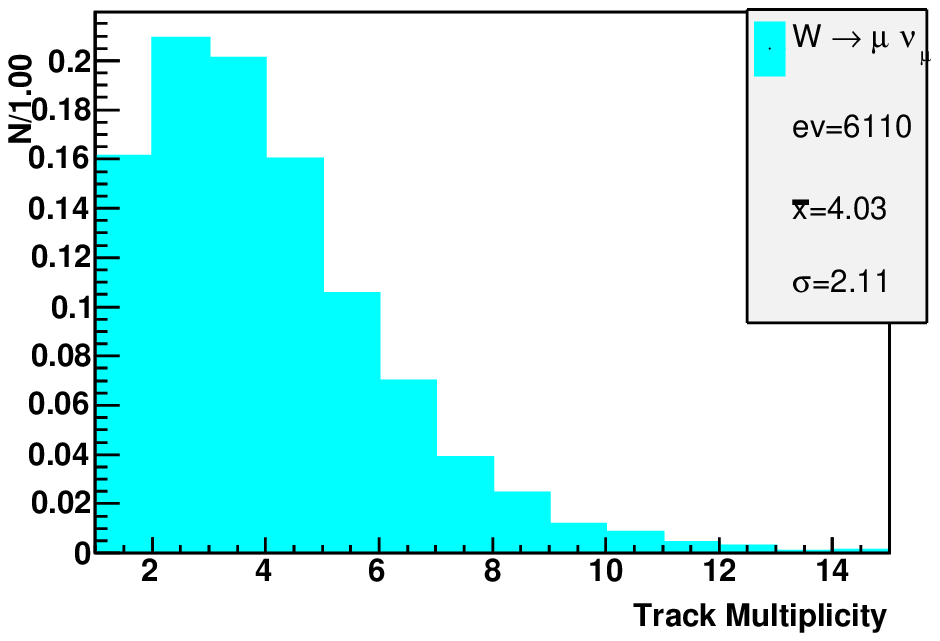} \\
\end{tabular}
\caption[]{{\small Distribution of the track multiplicity for jets for D\O\ (left) and ATLAS (right). Distributions are given for D\O\ data, \Wmunu (cyan), QCD (yellow).}}
\label{ATLASWMUNUJETTRACKSCALARSUM}
\end{center}
\end{figure}

A \ta -selection, in one form or the other, makes use of the well known basic features of a hadronically decaying tau lepton: One (, two), or three tracks near the calorimeter cluster center and a corresponding energy deposition in the calorimeter. 

Figure~\ref{ATLASWMUNUJETTRACKSCALARSUM} shows the number of tracks within $\dR < 0.3$ around the jet axis, for \dzero and ATLAS. Every jet with $\pt > 15~\gev$ and $|\eta| < 1.0$ from the W+mu selection enters the plot. The track multiplicity is shown for tracks with $\PT > 1\GeV$. The comparison for \dzero between Monte Carlo and data shows good agreement. Taus are, in most cases, selected only within jets having between one and three tracks. 

The comparison with distributions for the same quantities obtained with simulated ATLAS data shows that the distributions are well comparable between the two experiments. Hence also some confidence can be derived for the modeling of these quantities for the prediction of backgrounds to \ta\ final states in the ATLAS experiment.

\begin{figure}
\begin{center}
\begin{tabular}{rl}
\includegraphics[width=7cm]{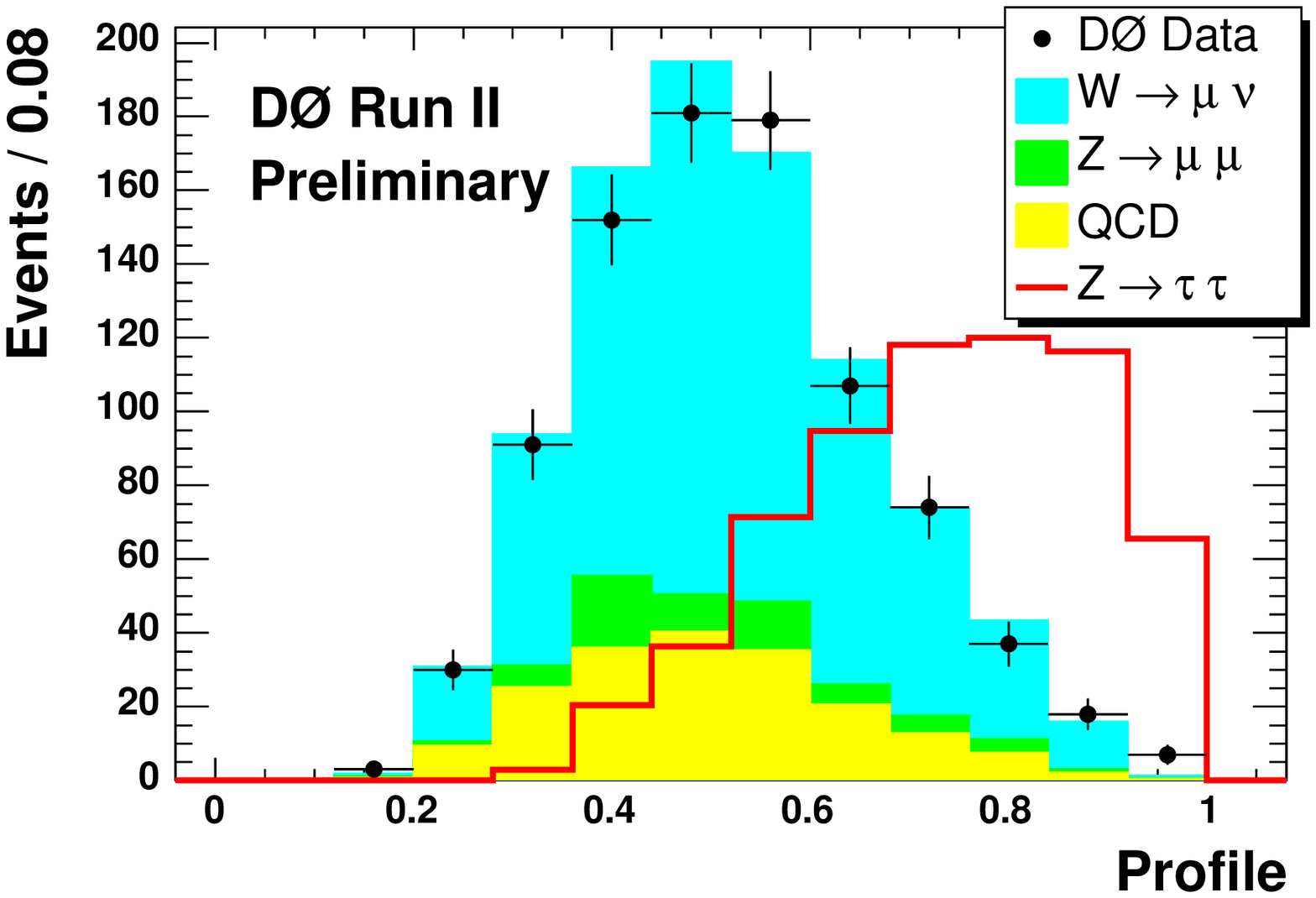} & 
\includegraphics[width=7cm]{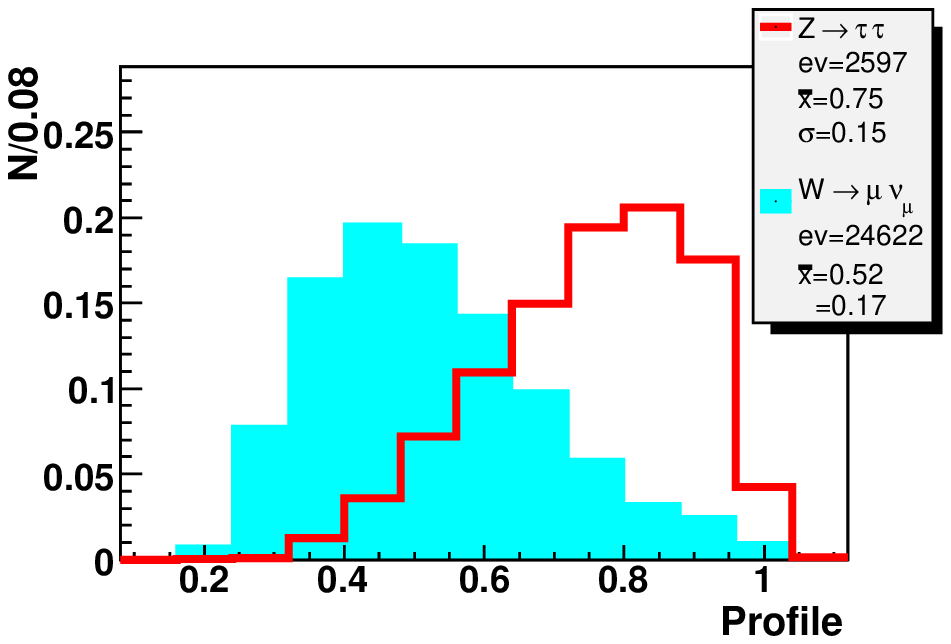} \\
\includegraphics[width=7cm]{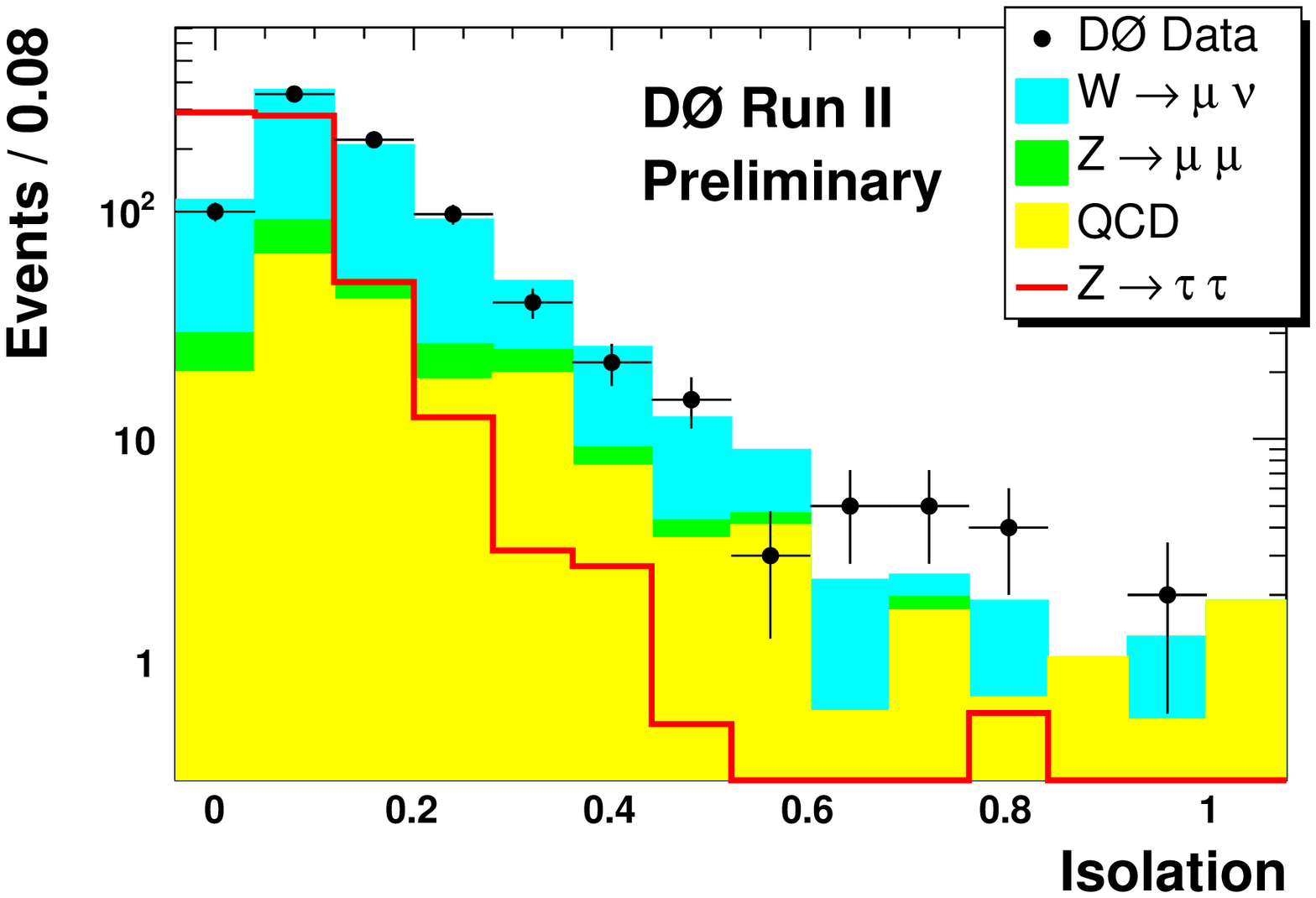} & 
\includegraphics[width=7cm]{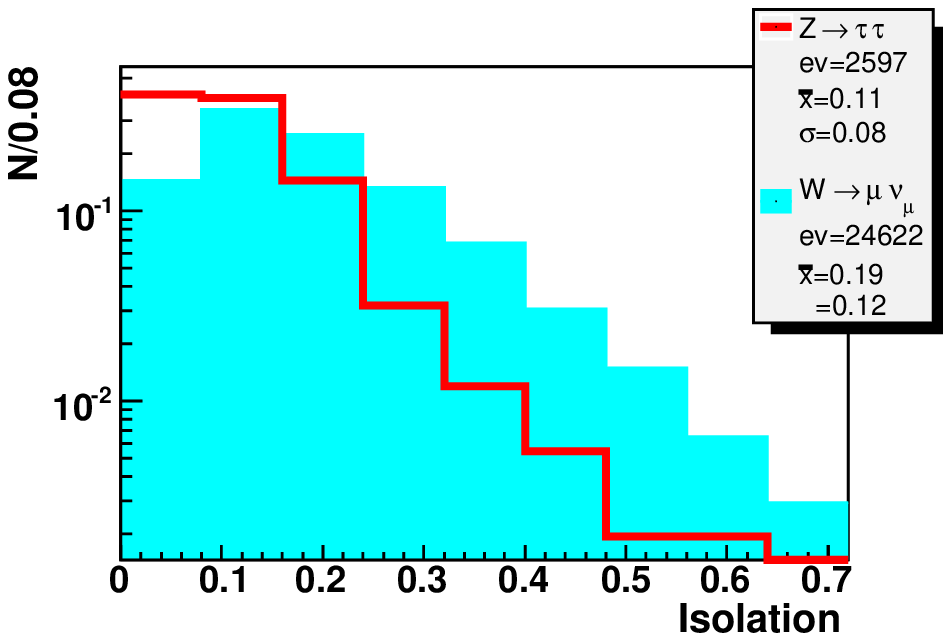} \\
\includegraphics[width=7cm]{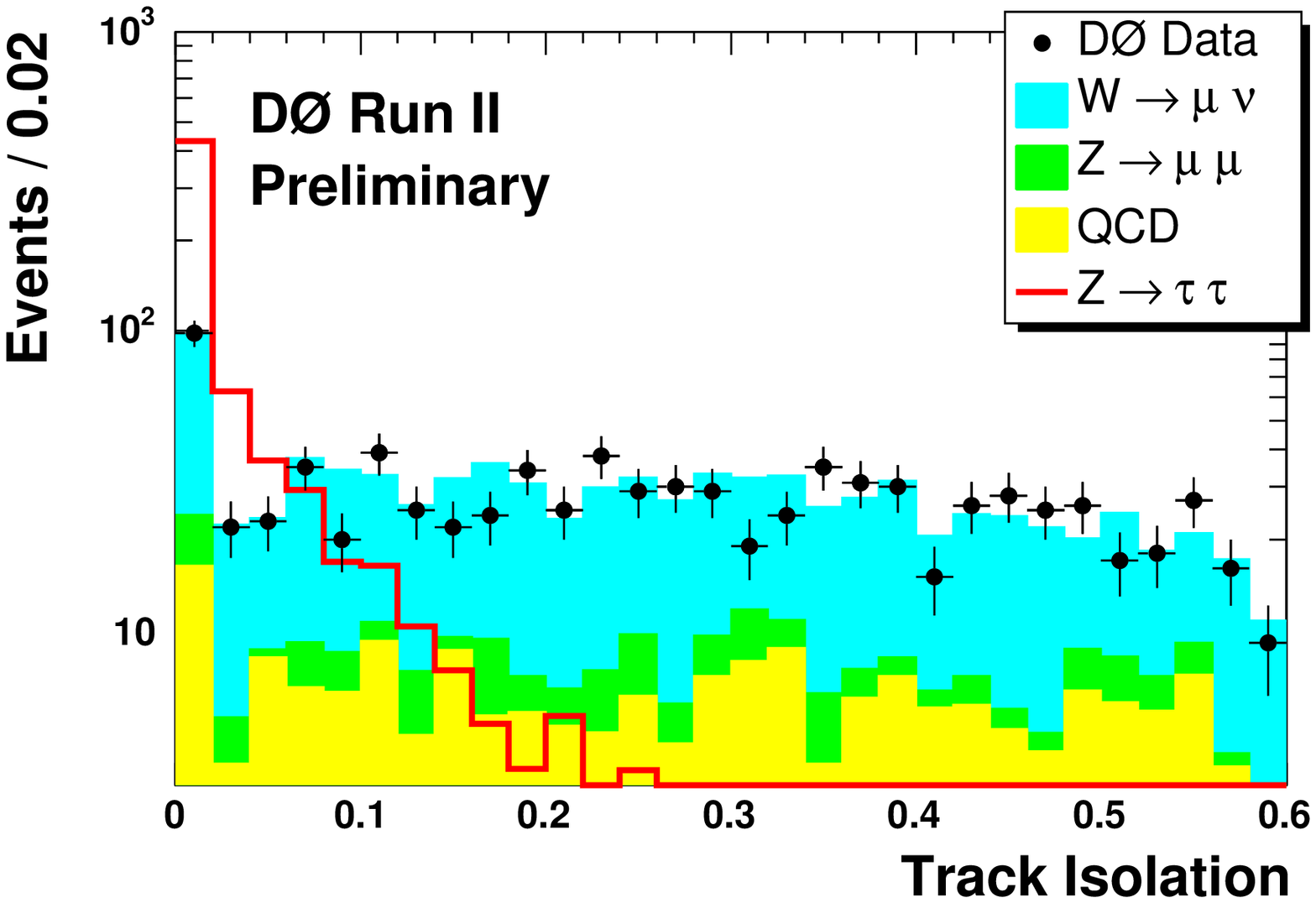} &
\includegraphics[width=7cm]{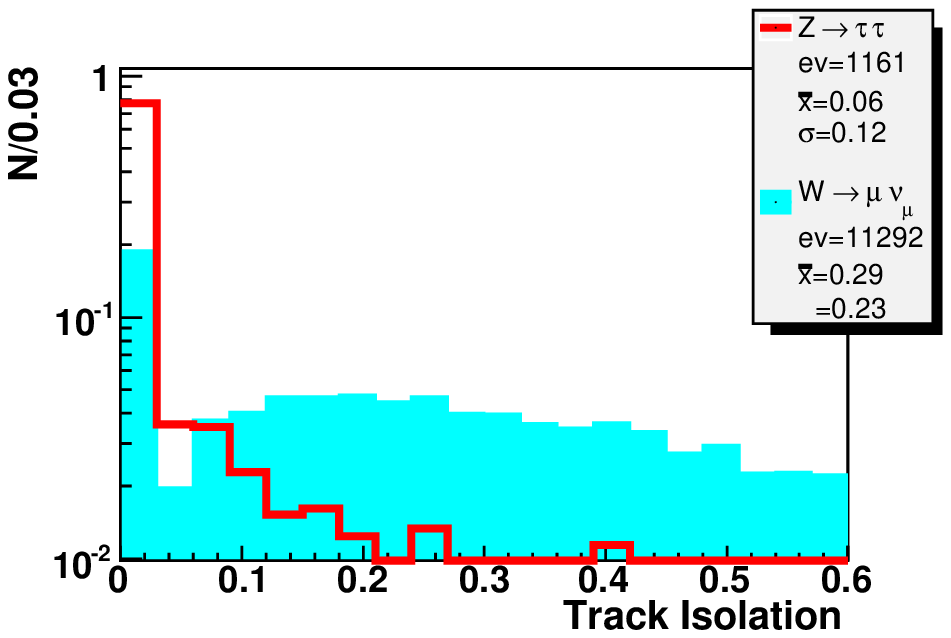}
\end{tabular}

\caption[]{{\small Distributions of three key variables, \ta-profile (upper row), \ta-calorimeter isolation (middle row) and \ta-track isolation (lower row). The left (right) column shows distributions for D\O\ (ATLAS). Distributions are given for D\O\ data (crosses), \Wmunu (cyan), \zintautau (red), $\rm{Z} \rightarrow \mu \mu$ (green) and QCD-jets (yellow).}}
\label{ATLASWMUNUJETSCALARSUM}
\end{center}
\end{figure}

\subsubsection{Quantities used in Tau identification}

Both \dzero and ATLAS make use of several quantities to separate tau leptons from jets. To keep the environment simple we concentrate on three principal variables, defined in Section~\ref{d0tauid}, which are important both for \dzero and ATLAS. For the following comparisons all \dzero discrimination variables were implemented in the ATLAS software framework, even though only three are used in this document. All are available in the official ATLAS software. In the implementation several details had to be faced that are consequences of the difference in detector layout. This included for example grouping ATLAS calorimeter cells to simulate a granularity and longitudinal segmentation similar to D\O\. Also differences in noise levels and general activity had to be taken into account. 
 
Figure~\ref{ATLASWMUNUJETSCALARSUM} shows ``Profile'', ``Calorimeter Isolation'' and ``Track Isolation'' for \dzero data, background and a possible signal in the left column. Modelling of these depends on complex details of jet fragmentation and detector simulation. Hence the agreement within statistical errors between data and Monte Carlo for \dzero should be considered as very good. This shows that a high precision in the prediction of these variables can be achieved with current Monte Carlo event and detector simulations. 


The right column of Figure~\ref{ATLASWMUNUJETSCALARSUM} shows the same quantities for ATLAS Monte Carlo simulation. All three variables show a similar behavior for \dzero and ATLAS. For ``Profile'' the agreement is very good and for ``Calorimeter Isolation'' reasonable. The ``Track Isolation'' shows the biggest discrepancy which we mainly attribute to the different \PT threshold on tracks which is 1 GeV for ATLAS and 0.4 GeV for \dzero \!\!. This could explain the rather deep dip for low values of track isolation for jets. \\

\begin{figure}
\begin{center}
\includegraphics[width=8.1cm]{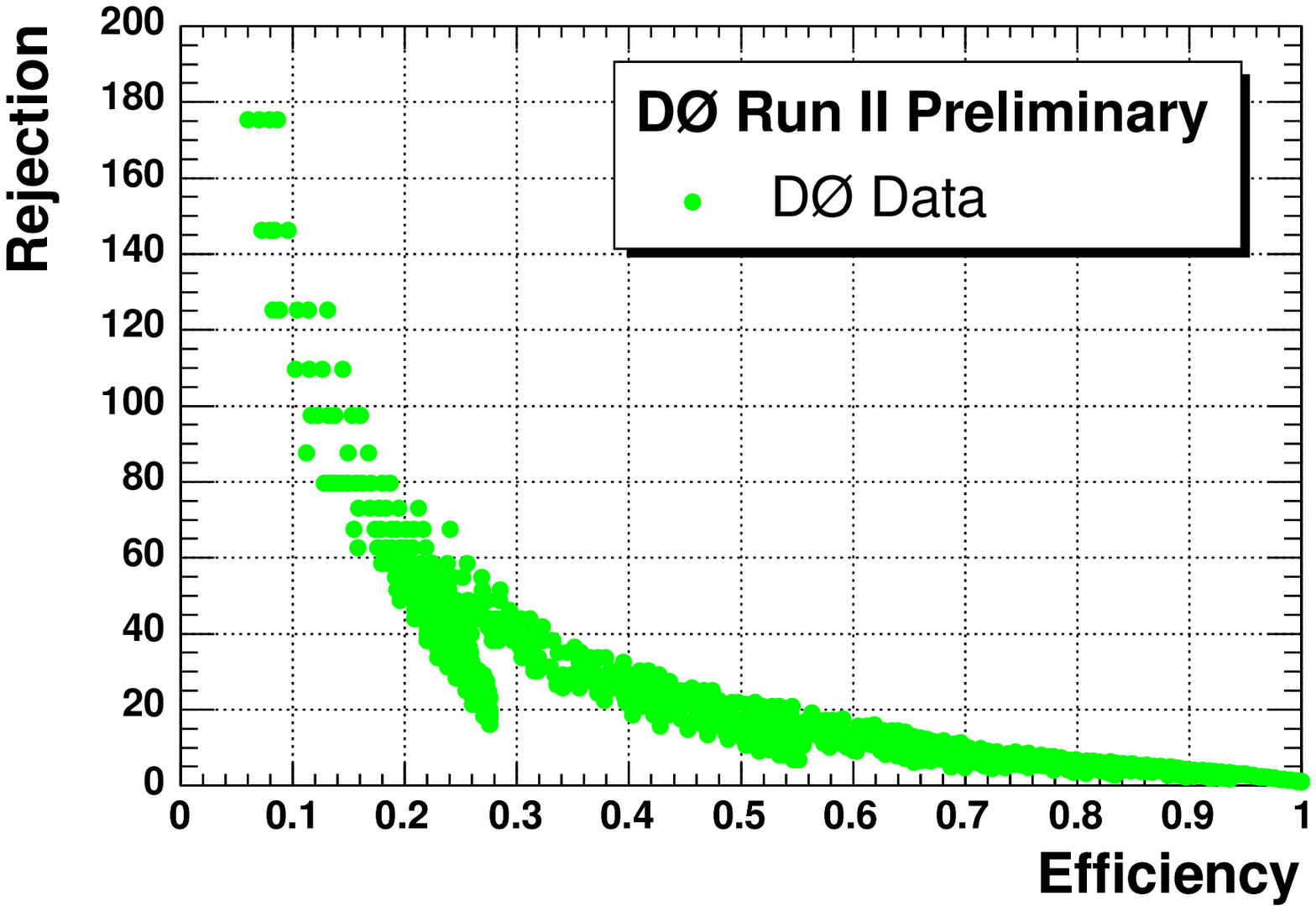}
\includegraphics[width=8.1cm]{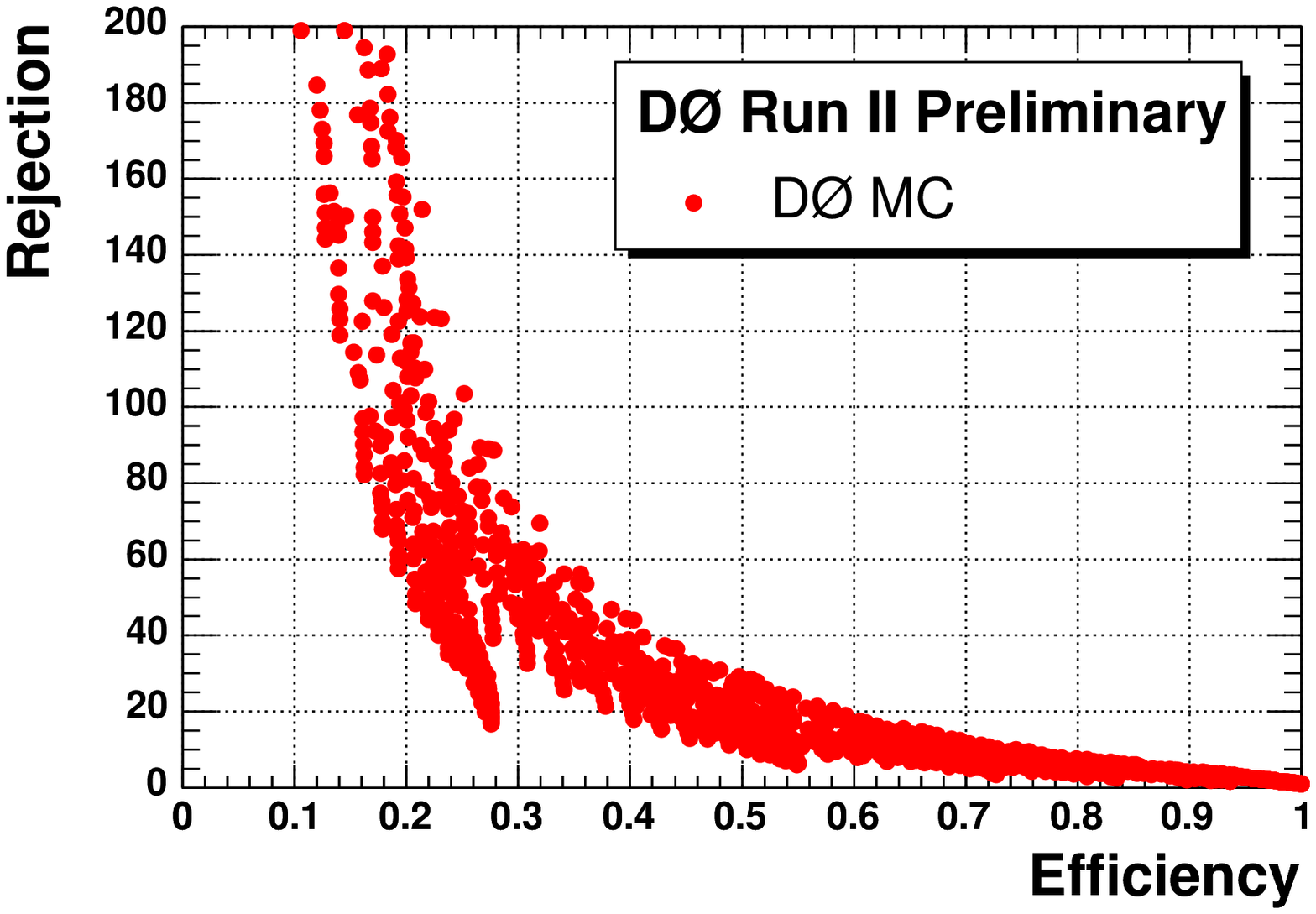}
\caption[]{{\small Rejection against jets vs. efficiency for \ta\ jets. Each point in the scatter plot stand for one possible combination of cuts of the three variables ``Profile'', ``Isolation'' and ``Track Isolation''. The efficiency is always determined on the $Z^0\rightarrow\tau\tau$ MC sample. The left shows the performance on \dzero \wmunu\ data while the right plot the result of the same procedure on \dzero \wmunu\ MC.}}
\label{DOEFF3CUTS}
\end{center}
\end{figure} 

\begin{figure}[t]
\begin{center}
\includegraphics[width=8.1cm]{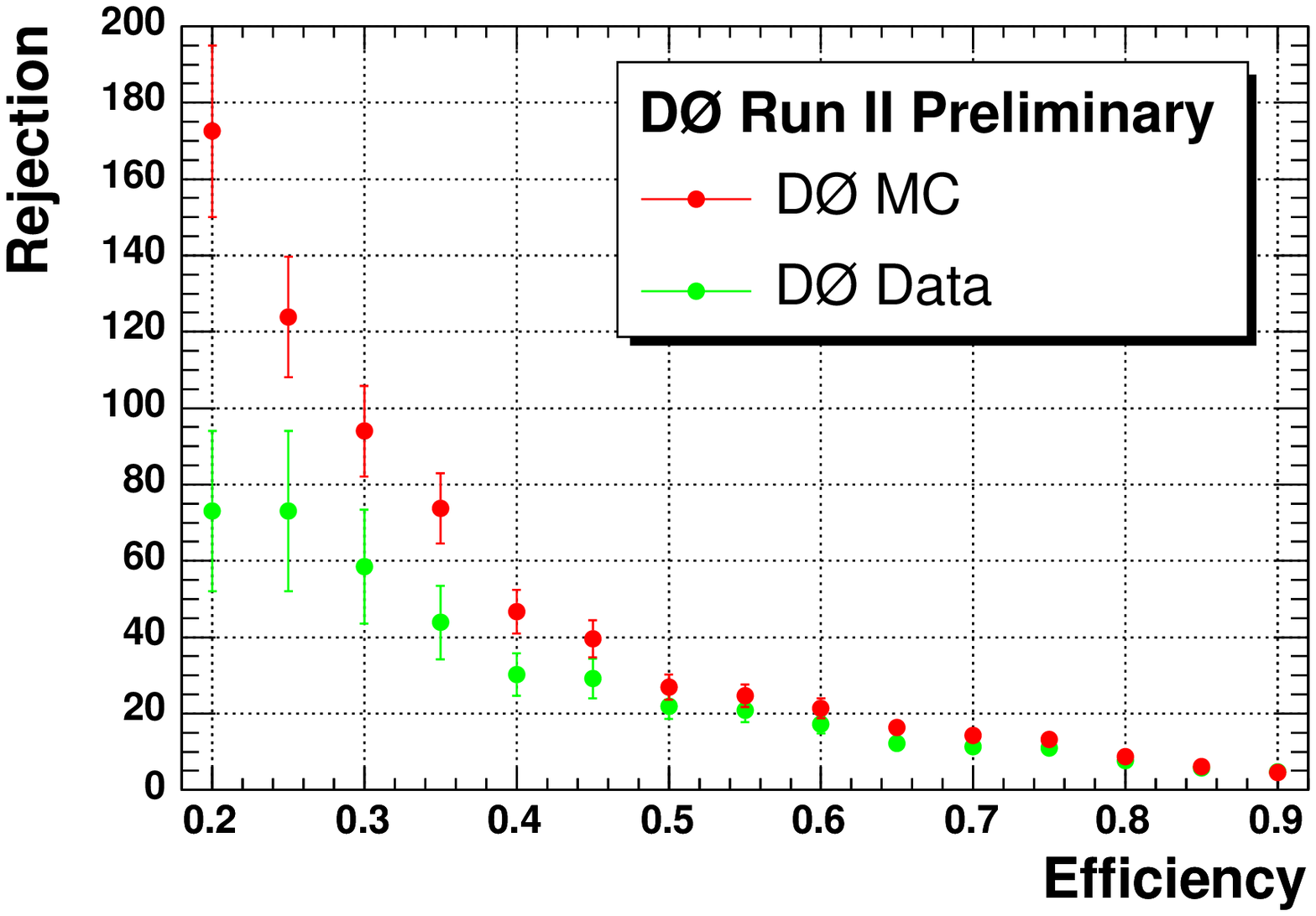}
\includegraphics[width=8.1cm]{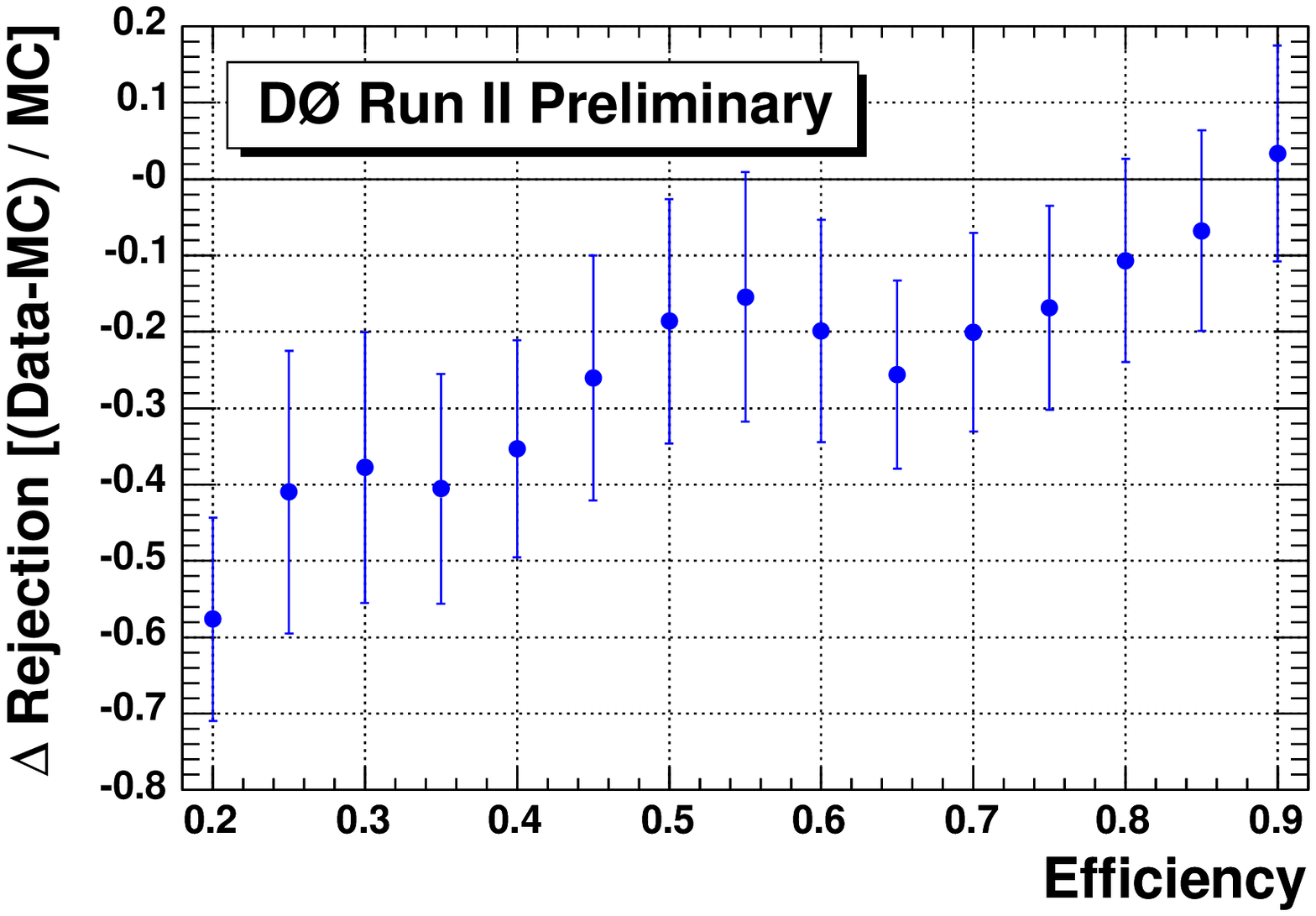}
\caption[]{{\small Left figure shows ``best'' points from Figure~\ref{DOEFF3CUTS} for MC and data. Right plot shows the relative difference.}}
\label{D0DATAMC}
\end{center}
\end{figure} 

\begin{figure}[h]
\begin{center}
\begin{tabular}{rl}
\includegraphics[width=8.1cm]{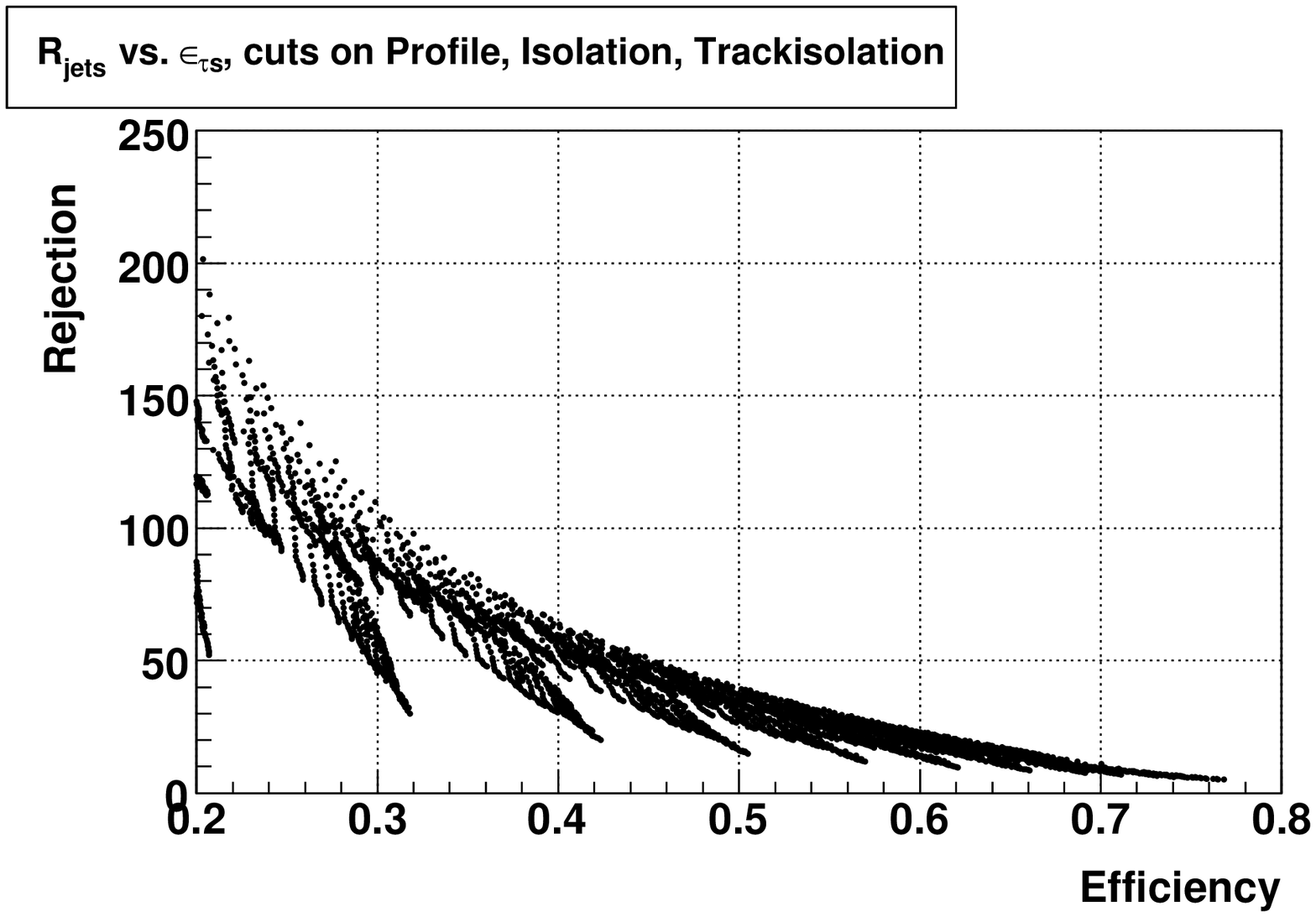} &
\includegraphics[width=8.1cm]{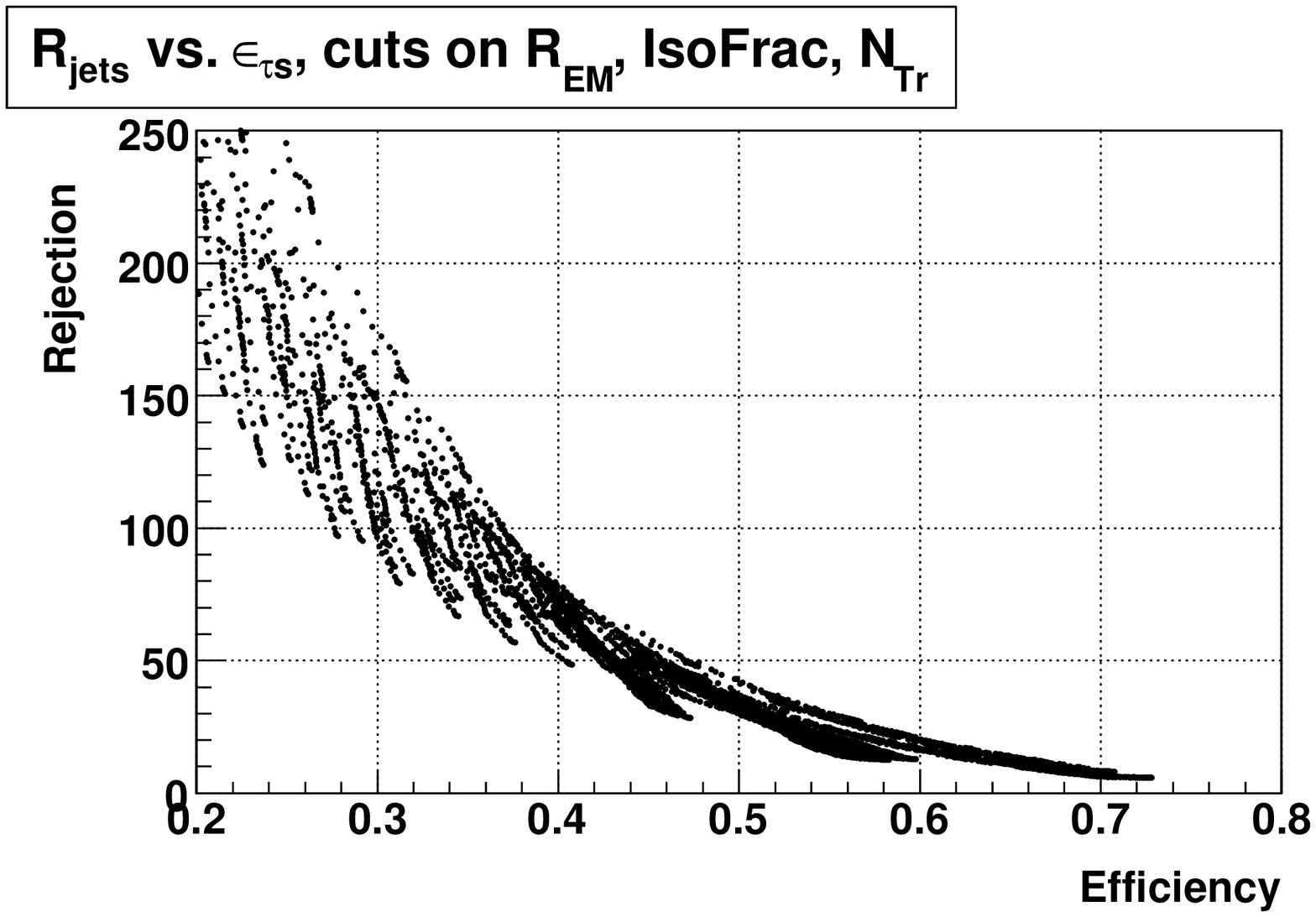} \\
\end{tabular}
\caption[]{{\small Rejection against jets vs. efficiency for \ta-jets on ATLAS samples. For the left (right) distribution D\O\ (ATLAS) variables have been used. Every point stands for a given combination of cuts on \Rem, \Isofrac, \Ntr. }}
\label{ATLASREFF3CUTS}
\end{center}
\end{figure}

\subsubsection{Simple ``benchmark tau identification'' based on three variables}

\begin{figure}[t]
\begin{center}
\includegraphics[width=9.8cm]{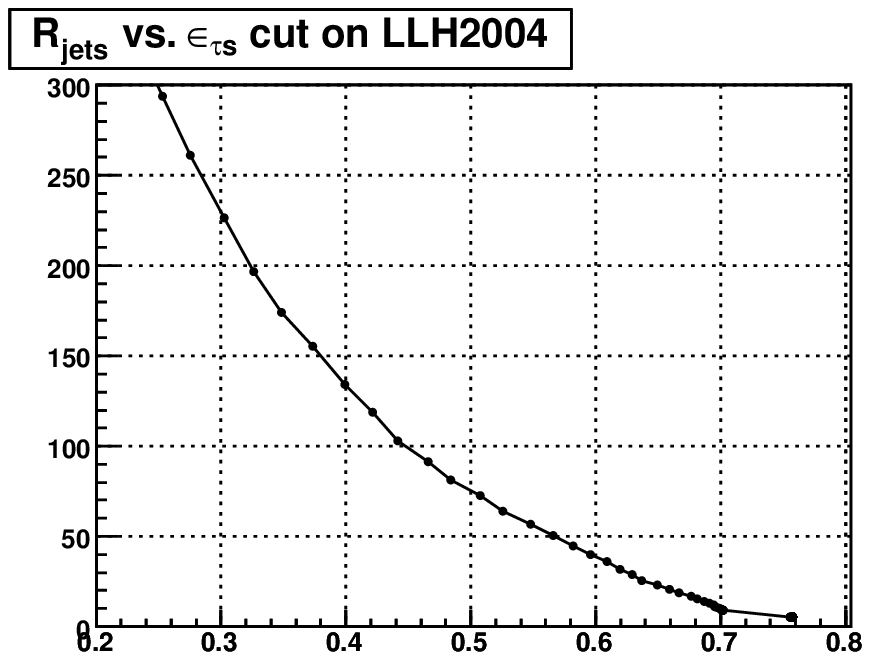} 
\caption[]{{\small Rejection against jets vs. efficiency for \ta-jets on ATLAS samples using the ATLAS LLH.}}
\label{ATLASREFFLLH}
\end{center}
\end{figure}
\begin{figure}[t]
\begin{center}
\includegraphics[width=15cm]{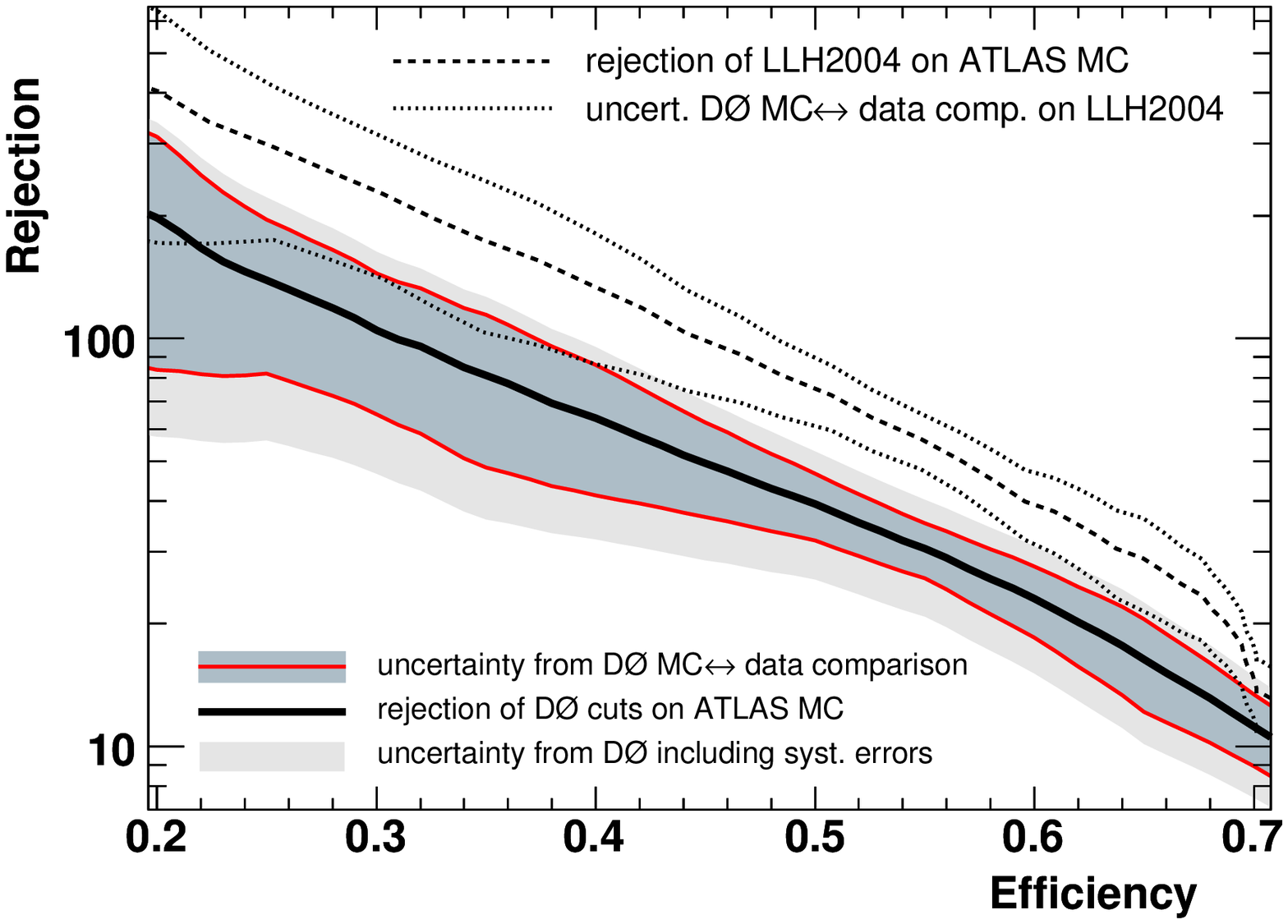} 
\caption[]{{\small Rejection against jets vs. efficiency for \ta-jets on ATLAS samples using the ATLAS LLH including the uncertainties from \dzero \!\!.}}
\label{ATLASFINAL}
\end{center}
\end{figure}

To be able to judge the similarities and differences of the \dzero and the ATLAS experiment with respect to $\tau$-identification we decided to implement a very basic tau selection for both \dzero and ATLAS. We choose a cut analysis on three variables, namely Profile, Isolation and Track Isolation. To obtain the optimal cut values a scan in these three variables has been performed, optimizing for the best rejection for all efficiencies.
The performance of such an analysis on \dzero data and MC is given in Figure~\ref{DOEFF3CUTS}, while Figure~\ref{D0DATAMC} gives a comparison between data and MC. Every point in the graph corresponds to a set of three cut values. For a given efficiency the highest point is the optimal set if cut values. For example, this simple method gives a rejection of $\approx 25$ at 50~\% efficiency. The right plot of Figure~\ref{D0DATAMC} shows the difference between the rejections obtained on data and MC for the same set of cuts on the before mentioned variables. Even though the input variables show a good agreement of MC and data, the analysis performs systematically worse on data than what is expected from MC. We assume this is due to significant correlations. It can be observed that the difference grows from ~0~\% at 80~\% efficiency up to 60~\% at 20~\% efficiency. It should be noted that the usual \dzero working point is at around 80~\% , where the difference is smaller than 10~\%. These differences are now translated into an uncertainty on the prediction of jet rejection in the ATLAS experiment.


\subsubsection{Transfer to ATLAS}

The left plot of Figure~\ref{ATLASREFF3CUTS} shows the performance for the same kind of simple cut analysis as was used to produce Figure~\ref{DOEFF3CUTS} but analyzing ATLAS Monte Carlo samples. It can be seen that the performance is roughly comparable to \dzero results.

The right plot of Figure~\ref{ATLASREFF3CUTS} shows the results of such a cut analysis using the ``standard'' ATLAS variables. The performance at higher efficiencies is very similar, while at lower efficiencies the ``standard'' ATLAS variables perform significantly better.

Since the cuts were optimized separatly for D\O\ and ATLAS the exact cut values used to produce Figures~\ref{DOEFF3CUTS} and ~\ref{ATLASREFF3CUTS} were not the same. Therefore it has been checked that the contribution of a certain variable to the rejection is very similar for ATLAS and D\O\ (not shown).
Therefore the assumption, that results obtained for \dzero using \dzero variables are portable to ATLAS performance expectations, using ATLAS variables, seems justified.

Figure~\ref{ATLASREFFLLH} shows the performance for ATLAS using the ``standard'' ATLAS method to identify tau leptons, which is a likelihood ratio method based on 8 variables (not all shown), including the three ``standard'' ATLAS variables we showed before. As expected, it can be seen that this method outperforms the three cut analysis significantly. The difference varies between 100~\% and 50~\% and again leaves the performance in the same order of magnitude. 

Even though this improvement might result from additional variables, the assumption that their uncertainty behaves in a similar way can be made. This would mean that the differences seen in Figure~\ref{D0DATAMC} can be directly translated into an uncertainty on the prediction of rejection for the ATLAS experiment.
Figure~\ref{ATLASFINAL} shows the expected rejection obtained for ATLAS using the three cuts analysis (thick black line) and the ``standard'' approach (slashed black line).  The uncertainty band given by the deviation at \dzero between MC and data is shown as a thick grey line. The dotted black line shows the same uncertainty for the ``standard'' ATLAS \ta-identification. In light gray the ``worst'' case is shown when including also the uncertainties on the deviation, assuming the worst performance within uncertainties. \\

\subsubsection{Conclusion}

Identification of hadronically decaying \ta\ leptons is an important and challenging issue at both Tevatron and LHC experiments. To conduct feasibility studies at the LHC it is important to be able to estimate the uncertainty on the prediction of \ta -identification performance. To obtain such an estimate, we selected $\Wmunu + \rm{jets}$ events in \dzero data, \dzero Monte Carlo and ATLAS Monte Carlo. We compared the \ta -identification related properties of these jets between \dzero MC and data. The comparison shows good agreement for all jet shape variables. 

On the same sample a simplified \ta -identification, based on three key variables, was studied. The cut values were optimized to yield highest rejection for a given efficiency. The efficiency was obtained from a $Z^0\rightarrow\tau\tau$ Monte Carlo sample. The rejection was compared between Monte Carlo and D\O\ data and no significant discrepancies were found. Taking into account correlations the agreement is within one sigma deviation for efficiencies above 50~\% and two sigma deviations for lower efficiencies.


To justify that these results are meaningful also for the ATLAS experiment, the \dzero variables have been implemented in the ATLAS software as far as possible. It was found that the \dzero variables perform in a very similar way on an ATLAS \Wmunu sample, after the same preselection. Also for ATLAS a simplified \ta -identification was optimized on ATLAS Monte Carlo. Regardless the obvious differences in detector and machine, it shows good agreement with the \dzero results, in terms of the overall performance as well as the relative dependency on the variables. 

This gives us confidence to quote an uncertainty between 0~\% (at 90~\% efficiency) and 50~\% (at 20~\% efficiency) for the prediction of rejection against jets in the ATLAS experiment. 


\clearpage\setcounter{equation}{0}\setcounter{figure}{0}\setcounter{table}{0}
\newcommand{\rs}        {\sqrt{s}}
\newcommand{\Ecm}       {E_{\mathrm{cm}}}
\newcommand{\Ebeam}     {E_{\mathrm{b}}}
\newcommand{\ee}        {{\mathrm e}^+ {\mathrm e}^-}
\newcommand{\pp}        {pp}

\newcommand{\Evis}      {E_{\mathrm{vis}}}
\newcommand{\Rvis}      {E_{\mathrm{vis}}\,/\roots}
\newcommand{\Mvis}      {M_{\mathrm{vis}}}
\newcommand{\Rbal}      {R_{\mathrm{bal}}}
\newcommand{\mjet}      {\bar{M}_{\mathrm{jet}}}
\newcommand{\mET}       {$\not \!\!\! {E_T}$\ }
\newcommand{\degree}    {^\circ}
%
%
\newcommand{\thrust}    {T}
\newcommand{\nthrust}   {\hat{n}_{\mathrm{thrust}}}
\newcommand{\thethr}    {\theta_{\,\mathrm{thrust}}}
\newcommand{\phithr}    {\phi_{\mathrm{thrust}}}
\newcommand{\acosthr}   {|\cos\thethr|}
\newcommand{\thejet}    {\theta_{\,\mathrm{jet}}}
\newcommand{\acosjet}   {|\cos\thejet|}
\newcommand{\thmiss}    { \theta_{\mathrm{miss}} }
\newcommand{\cosmiss}   {| \cos \thmiss |}
\newcommand{\pbinv}     {\mathrm{pb}^{-1}}

\newcommand{\phiacop}       {\phi _{\mathrm {acop}}}
\newcommand{\cosjet}        {\cos\thejet}
\newcommand{\costhr}        {\cos\thethr}

%
%
\newcommand{\ff}            {{\mathrm f} \bar{\mathrm f}}
\newcommand{\allqq}         {\sum_{q \neq t} q \bar{q}}
\newcommand{\qq}            {{\mathrm q}\bar{\mathrm q}}
\newcommand{\bb}            {{\mathrm b}\bar{\mathrm b}}
\newcommand{\toppair}        {{\mathrm t}\bar{\mathrm t}}

\newcommand{\tq}            {\mathrm t} 
\newcommand{\bq}            {\mathrm b} 
\newcommand{\cq}            {\mathrm c} 
\newcommand{\qu}            {\mathrm q} 
\newcommand{\ele}           {\mathrm e} 

\newcommand{\nunu}          {\nu \bar{\nu}}
\newcommand{\ellell}        {\ell^+ \ell^-}

\newcommand{\Zboson}        {{\mathrm Z}^{0}}
\newcommand{\Zv}            {{\mathrm Z}^{*}}
\newcommand{\Wp}            {{\mathrm W}^{+}}
\newcommand{\gv}            {\gamma^*}
\newcommand{\Zg}            {{\mathrm Z} \gamma}
\newcommand{\Zgv}           {{\mathrm Z} \gamma^*}
\newcommand{\Zvgv}          {{\mathrm Z^*} \gamma^*}

\newcommand{\HSM}           {\mathrm H^{0}_{SM}}
\newcommand{\MHSM}          {\mathrm M(H^{0}_{SM})}

\newcommand{\eetautau}      {\ee-\rightarrow {\tau^+}{\tau^-}}
\newcommand{\nulqq}         {\nu \ell {\mathrm q} \bar{\mathrm q}'}
\newcommand{\Wenu}          {{\mathrm{We}} \nu}

%
%
\newcommand{\hsusy}         {h^{0}}
\newcommand{\Hsusy}         {\mathrm H^{0}}
\newcommand{\Asusy}         {\mathrm A^{0}}
\newcommand{\CHsusy}         {\mathrm H^{\pm}}

\newcommand{\sfermion}      {\tilde{\mathrm f}}
\newcommand{\sfpair}        {\tilde{\mathrm{f}}^{+} \bar{\tilde{\mathrm{f}}}^{-}}
\newcommand{\sele}          {\tilde{\mathrm e}}
\newcommand{\smu}           {\tilde{\mu}}
\newcommand{\stau}          {\tilde{\tau}}
\newcommand{\staum}         {\tilde{\tau}_{1}}
\newcommand{\sell}          {\tilde{\ell}}
\newcommand{\slepton}       {\tilde{\ell}^{\pm}}
\newcommand{\sellsell}      {\sell^+ \sell^-}
\newcommand{\snu}           {\tilde{\nu}}

\newcommand{\nt}            {\tilde{\chi}^0}
\newcommand{\neutralino}    {\tilde{\chi }^{0}_{1}}
\newcommand{\neutrala}      {\tilde{\chi }^{0}_{2}}
\newcommand{\neutralb}      {\tilde{\chi }^{0}_{3}}
\newcommand{\neutralc}      {\tilde{\chi }^{0}_{4}}
\newcommand{\bino}          {\tilde{\mathrm B}^{0}}
\newcommand{\wino}          {\tilde{\mathrm W}^{0}}
\newcommand{\higgsino}      {\tilde{\mathrm H}^{0}}
\newcommand{\higginoa}      {\tilde{\mathrm H_{1}}^{0}}
\newcommand{\higginob}      {\tilde{\mathrm H_{1}}^{0}}

\newcommand{\ch}            {\tilde{\chi}^\pm}
\newcommand{\chp}           {\tilde{\chi}_{1}^+}
\newcommand{\chm}           {\tilde{\chi}_{1}^-}
\newcommand{\chpm}          {\tilde{\chi}_{1}^\pm}
\newcommand{\cwino}         {\tilde{\mathrm W}^{\pm}}
\newcommand{\chiggsino}     {\tilde{\mathrm H}^{\pm}}
\newcommand{\chargino}      {\tilde{\chi }^{\pm}_{1}}
\newcommand{\charginop}     {\tilde{\chi }^{+}_{1}}
\newcommand{\gra}           {\tilde{\mathrm G}}
\newcommand{\gluino}        {\tilde{\mathrm g}}
\newcommand{\supq}          {\tilde{\mathrm u}}
\newcommand{\sdownq}        {\tilde{\mathrm d}}
\newcommand{\stopm}         {\tilde{\mathrm{t}}_{1}}
\newcommand{\stops}         {\tilde{\mathrm{t}}_{2}}
\newcommand{\stopbar}       {\bar{\tilde{\mathrm{t}}}_{1}}
\newcommand{\stopx}         {\tilde{\mathrm{t}}}
\newcommand{\stopl}         {\tilde{\mathrm{t}}_{\mathrm L}}
\newcommand{\stopr}         {\tilde{\mathrm{t}}_{\mathrm R}}
\newcommand{\stoppair}      {\tilde{\mathrm{t}}_{1} \bar{\tilde{\mathrm{t}}}_{1}}
\newcommand{\sbotm}         {\tilde{\mathrm{b}}_{1}}
\newcommand{\sbots}         {\tilde{\mathrm{b}}_{2}}
\newcommand{\sbotbar}       {\bar{\tilde{\mathrm{b}}}_{1}}
\newcommand{\sbotx}         {\tilde{\mathrm{b}}}
\newcommand{\sbotl}         {\tilde{\mathrm{b}}_{\mathrm L}}
\newcommand{\sbotr}         {\tilde{\mathrm{b}}_{\mathrm R}}
\newcommand{\sbotpair}      {\tilde{\mathrm{b}}_{1} \bar{\tilde{\mathrm{b}}}_{1}}
%
%
\newcommand{\MGUT}          {M_{\mathrm{GUT}}}
\newcommand{\mscalar}       {m_{0}}
\newcommand{\Mgaugino}      {M_{1/2}}

\newcommand{\mixstop}       {\theta _{\stopx}}
\newcommand{\mixsbot}       {\theta _{\sbotx}}
\newcommand{\mchar}         {m_{\chpm}}
\newcommand{\mstop}         {m_{\stopm}}
\newcommand{\msbot}         {m_{\sbotm}}
\newcommand{\mchi}          {m_{\neutralino}}
%
%
\newcommand{\PhysLett}  {Phys.~Lett.}
\newcommand{\PRL}       {Phys.~Rev.\ Lett.}
\newcommand{\PhysRep}   {Phys.~Rep.}
\newcommand{\PhysRev}   {Phys.~Rev.}
\newcommand{\NPhys}     {Nucl.~Phys.}
\newcommand{\NIM}       {Nucl.~Instr.\ Meth.}
\newcommand{\CPC}       {Comp.~Phys.\ Comm.}
\newcommand{\ZPhys}     {Z.~Phys.}
\newcommand{\IEEENS}    {IEEE Trans.\ Nucl.~Sci.}
%
%
\newcommand{\OPALColl}  {OPAL Collab.}
\newcommand{\JADEColl}  {JADE Collab.}

\subsection{Jets and Missing \et: Standard Model Background for SUSY
searches at the LHC} 
\label{sec:jetmet}

S. Asai and T. Sasaki \\ [2mm]
{\em University of Tokyo, E-mail: Shoji.Asai@cern.ch}\\ 

{\em
The search for supersymmetry is one of the main purposes of the LHC. 
The Standard Model background processes are 
estimated with the Matrix Elements calculation, and
we find out that the background 
contributions become larger than what we have expected, 
and that the distributions are similar to the  SUSY signal.
The careful studies of the Standard Model processes 
are useful using Tevatron Run-II data, 
especially studies of the slope of the \pt  and 
missing $E_{T}$ distributions of $\Wpm$ + jets, $\Zboson$ + jets and
$\toppair$ + jets. 
}\\

\subsubsection{Introduction}

Supersymmetric (SUSY) Standard Models~\cite{Nilles:1983ge,Haber:1984rc}
are promising extensions of the Standard Model(SM), 
because the SUSY can naturally deal with the 
problem of the quadratic Higgs mass divergence.
Furthermore, the SUSY models provides a natural 
candidate for cold dark matter, 
and they have given a hint of the Grand Unification of gauge couplings
around $2 \times 10^{16}$~GeV\@.
In these theories, each elementary particle 
has a superpartner whose spin differs
by 1/2 from that of the particle. 
Discovery of these SUSY particles 
should open a window of new epoch, and 
is one of the important purposes of the LHC project~\cite{LHC,Armstrong:1994it}\@.

Dominant SUSY production processes at LHC are 
$\gluino\gluino$, $\gluino\squark$ and $\squark\squark$ through 
the strong interaction.
These production cross-sections, $\sigma$, do not strongly depend on 
the SUSY parameters except for masses of $\gluino$ and $\squark$~\cite{Eichten:1984eu,Spira:2000vf}\@.
When these masses are 500~GeV, $\gluino\gluino$ is main production process, 
and total $\sigma$($\gluino\gluino$, $\gluino\squark$
and $\squark\squark$) is 100~pb.
$\gluino\gluino$ production is dominate process for this case, since the population of
gluon in the proton is very huge. 
$\sigma$ becomes 3~pb for $m_{\squark}$=$m_{\gluino}$=1TeV.
Even when these masses are 2~TeV, sizable production cross-section
of about 20~fb is expected.
$\supq\supq$ and $\supq\sdownq$ are main production processes for such a heavy 
case, since u and d quarks are valence quarks.

\begin{figure}[t]
\vspace{7cm}
\includegraphics{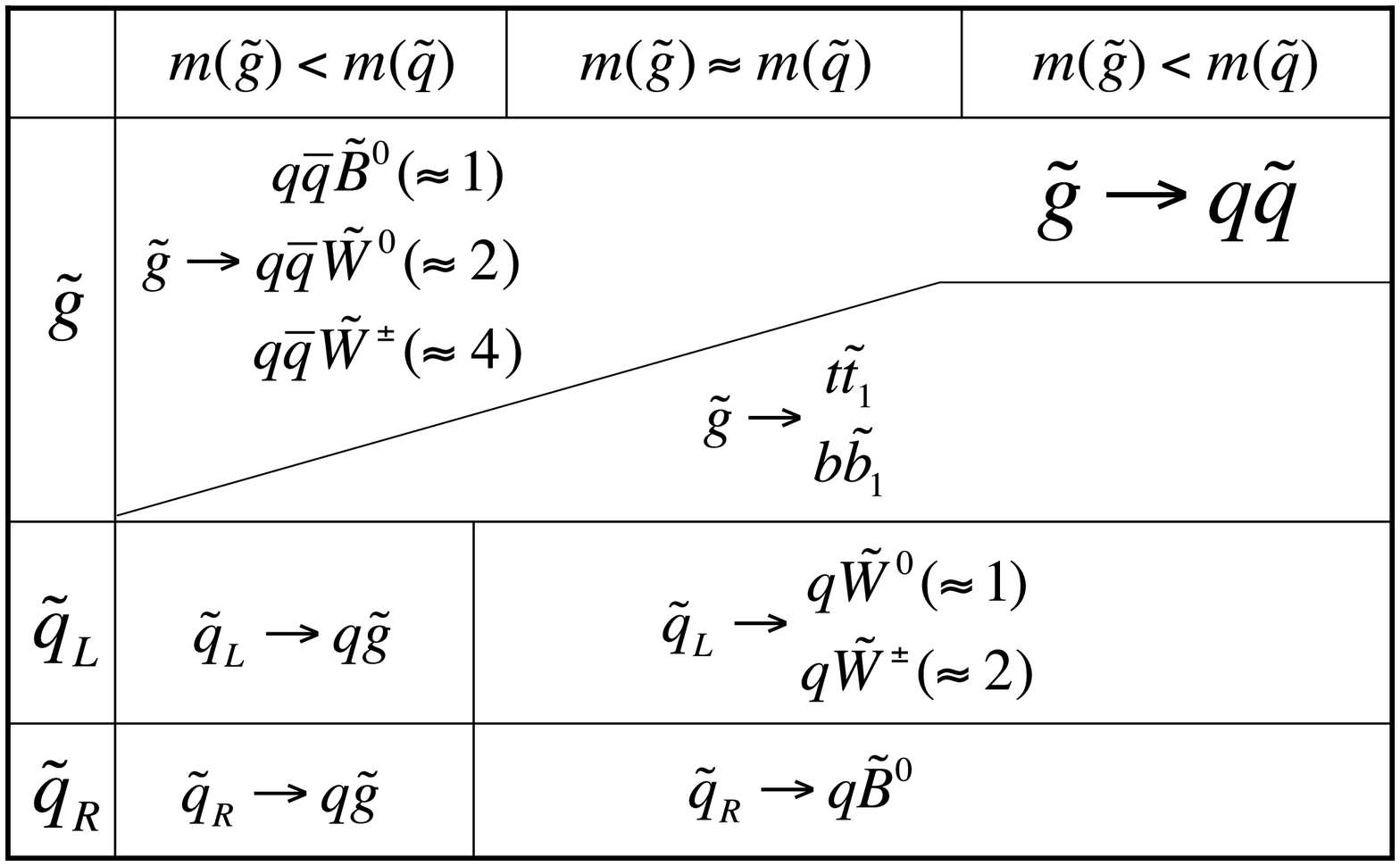}
\vspace{-1.8cm}
\caption{Decay table of squarks and gluino:
\label{fig:decay1}} 
\end{figure}

Decay modes of $\gluino$ and $\squark$ are controlled by the mass-relation 
between each other, and are summarized in the Fig.~\ref{fig:decay1}\@.
If kinematically possible, they decay into 2-body through the strong interaction.
Otherwise, they decay into a Electroweak gaugino plus quark(s)\@.
Bino/Wino-eigenstates presented in this table become
simply mass-eigenstate, ($\bino \sim \nt_{1}$, 
$\wino \sim \nt_{2}$, and $\cwino \sim \chpm$),
when $m_{0}$ is not too larger than  $m_{1/2}$. 
In this case, Higgsino mass ($|\mu|$) becomes larger than gaugino mass
at the EW scale, then Higgsino component decouples from 
lighter mass-eigenstates as already mentioned.
Decay modes of third generation squarks ($\stopm$ and $\sbotm$)
are more complicated, since they have enough coupling to Higgsino 
due to non-negligible Yukawa couplings.

There are four leading decay modes of $\nt_{2}$
depending on mass spectrum.
When the scalar lepton, $\slepton$, is lighter than $\nt_2$,
2-body decay chain, $\nt_2 \ra \ell \slepton (\ra \ell \nt_1)$ becomes dominant
decay mode. 
Branching fraction of $\nt_2 \ra \tau \staum$ is significantly large
in the case of $\tan\beta \gg$ 1\@.
$\nt_2 \ra {\mathrm h} \nt_1$ is dominant mode, if the mass difference
between $\nt_2$ and $\nt_1$ is larger than Higgs boson mass.
When the mass difference is smaller than $m_{\Zboson}$, 
three body decay is main decay process.   
$\chpm$ has three leading decay modes, $ \chpm \ra \slepton \nu$, $\Wpm\nt_1$ and
$\ff^{'} \nt_{1}$ as the similar manner to $\nt_2$\@.
 
$\gluino$ and/or $\squark$ are copiously produced at the LHC,
and the cascade decay follows after.
The colored SUSY particles decays into the EW gauginos and jets 
as shown in the figure\@.
Transverse momenta \pt  of these jets are expected to be large due to 
the large mass-difference between the colored sparticles 
and EW gauginos.
Cascade decay via scalar top and scalar bottom quarks also contributes, if they are
significantly lighter than the other scalar quarks.
Each event contains two $\nt_1$'s in the final state.
If R-parity~\cite{Fayet:1977yc} is conserved, $\nt_1$ is stable, and
it is neutral and weakly interacting and escape from the detection.
Then missing transverse energy, \, \mET, carried away by two $\nt_1$'s plus 
multiple high \pt \, jets is the leading experimental signature of 
SUSY at LHC.

Also the other activities of additional jets, leptons and $\bb$  
are possible, coming from the decays of $\nt_2$ and $\chpm$.
These additional informations are important 
to confirm SUSY signals, and to investigate its properties.

The following four SM processes can potentially 
have \mET event topology with jets.
\begin{itemize}
\item $\Wpm$ + jets, $\Wpm \ra \ell \nu$
\item $\Zboson$ + jets, $\Zboson \ra \nunu, \tautau $
\item $\toppair$ + jets 
\item Heavy flavor quarks (b,c) with semi-leptonic decay and the light flavor QCD jets with mis-measurement
\end{itemize}

Supersymmetry will be observed as an excess of these SM processes,
and it should be discovered at LHC within one or two years (L=1-10 fb$^{-1}$) 
after the LHC starts, 
if $\gluino$ and $\squark$ are lighter than about 2.0~TeV\@.
Quick but well understanding of these SM processes plays important role in the discovery of 
SUSY, especially 'high \pt  jet' and '\mET ' measurements.
These two subject are very important and we have good chance to study them with 
Tevatron Run-II data.

\begin{figure}[t]
\vspace{4.5cm}
\includegraphics{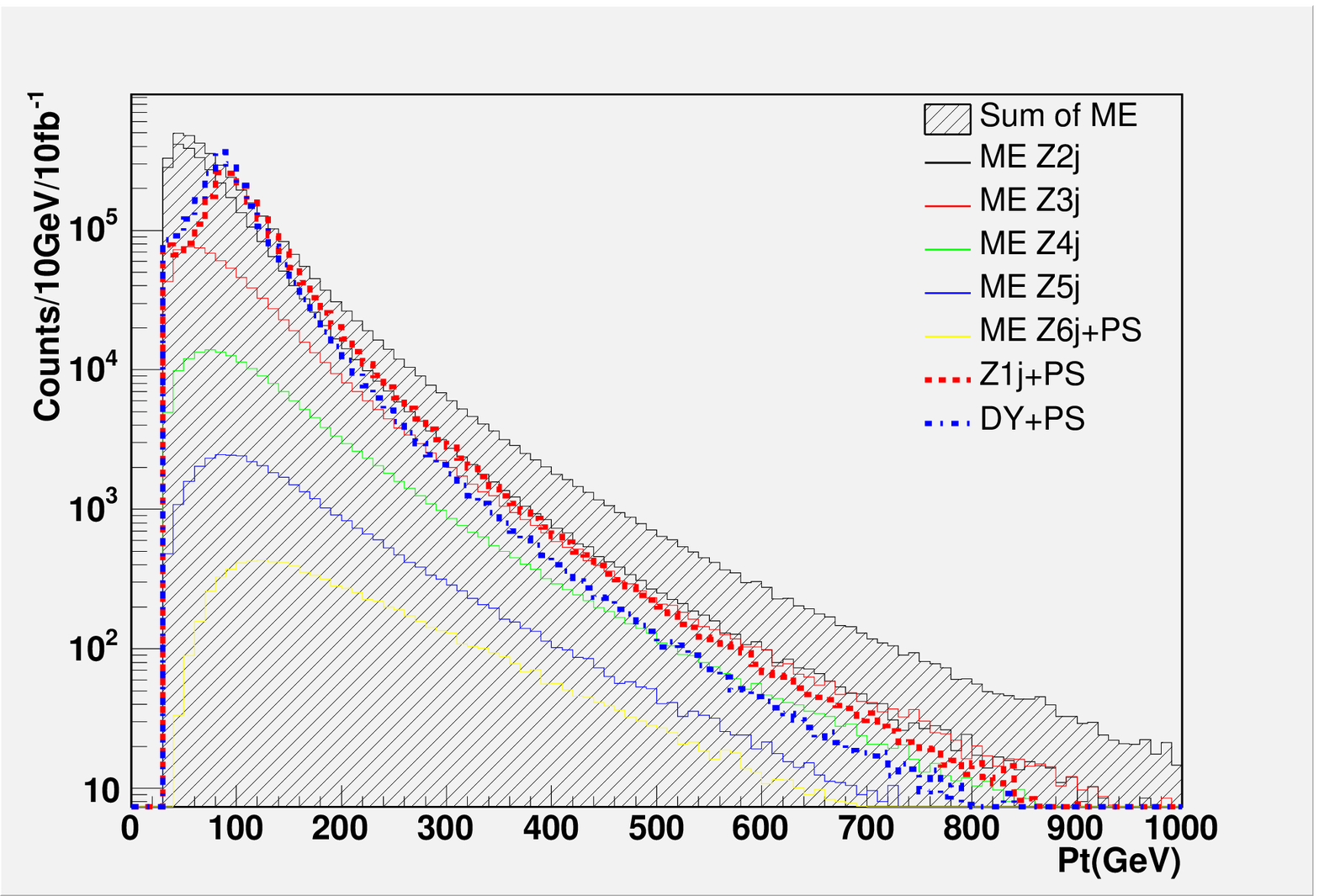}
\includegraphics{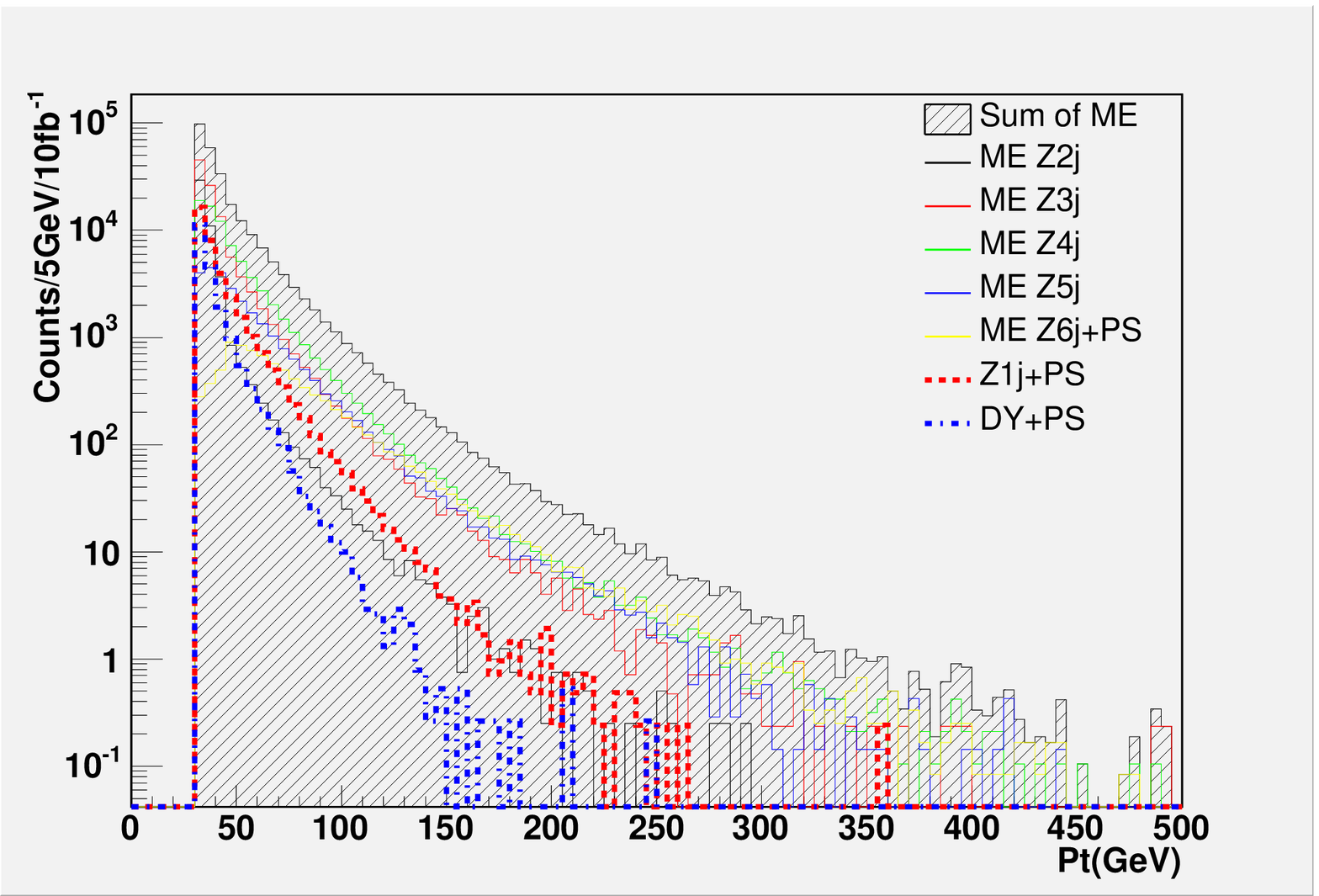}
\caption{$P_T$ distribution of (a) the leading jet and (b) 4th jet. Red dotted shows the PS calculation 
and hatched histogram show the ME predictions. It is obtained after sum up of all different jet multiplicity samples as shown in colored histogram in both figures.  
\label{fig:ptjet}} 
\end{figure}

\subsubsection{High \pt jets}

High \pt  jets and \mET are vital for the SUSY search.
The Parton Shower (PS) is the good model in the collinear and soft regions,
since all leading logs are summed up.
But the high \pt  jets are not emitted in the PS model, and 
the high \pt  jets should be estimated with the Matrix Element (ME) calculation.

\begin{figure}[t!]
\vspace{15.5cm}
\includegraphics{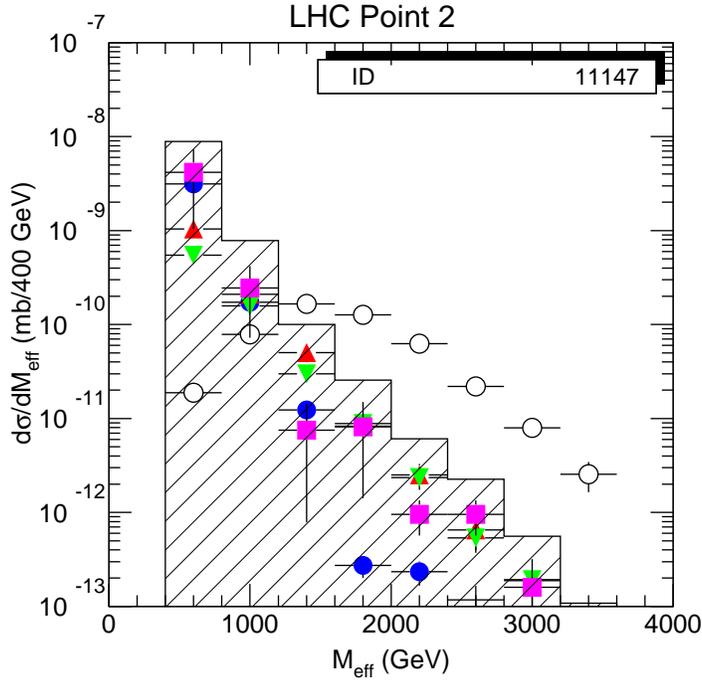} 
\includegraphics{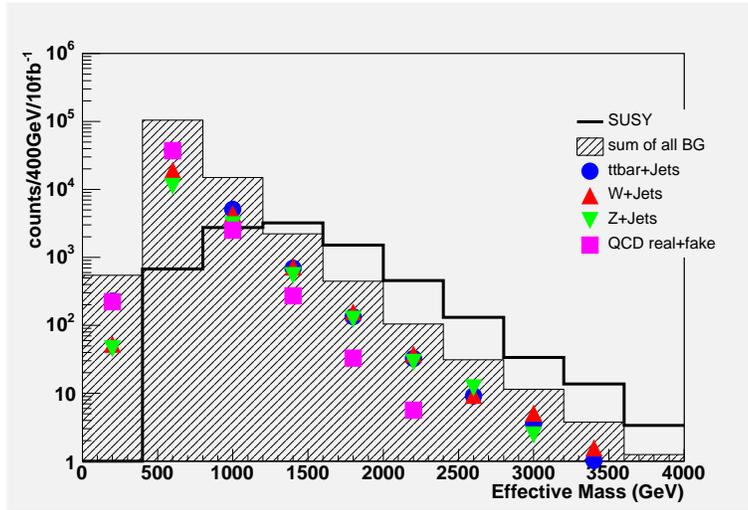}
\caption{Effective mass distributions of the SUSY signal and the SM background 
processes for no lepton mode: (a) Prediction with the PS model (b) Prediction with the ME\@.
Mass of the scalar quarks and gluino is 1TeV in both figures.Blue circle, red triangle,
green triangle and magenta box show the top,W,Z and QCD processes, respectively.  
\label{fig:meff1}} 
\end{figure}

260 million  events are generated with the ALPGEN~\cite{Mangano:2002ea} for 
the SM processes mentioned above.
Up to 6 quarks or gluons are emitted with the ME calculation in ALPGEN.
The region of \pt  $> 40$~GeV is covered with these partons.
The produced events are fed into the PS generator (PYTHIA6.2~\cite{Sjostrand:2000wi}) 
in order to evolute the QCD shower, which covers the soft and collinear regions.
Special treatments are necessary in order to remove the 
double count of jet produced with the ME calculation 
and the PS jets.
When the jet from the PS evolution  emits into
 the phase space which is covered with
the ME, the event is discarded~\cite{Mangano}\@.
This rejection factor is related to Sudakov factor.  
Figures~\ref{fig:ptjet} show the \pt  distribution of the leading and 4th jets of Drell-Yan processes.
In both figures the hatched histogram shows the ME predictions, 
and dotted line shows the 
predictions of the PS generator.
The \pt  distributions  calculated with the PS 
model is softer than the ME prediction,
and this difference becomes larger for higher  jet-multiplicity.
About 2nd order of magnitude is different for 4th leading jet as shown
in Fig.~\ref{fig:ptjet}. 

\begin{figure}[t]
\vspace{7cm}
\includegraphics{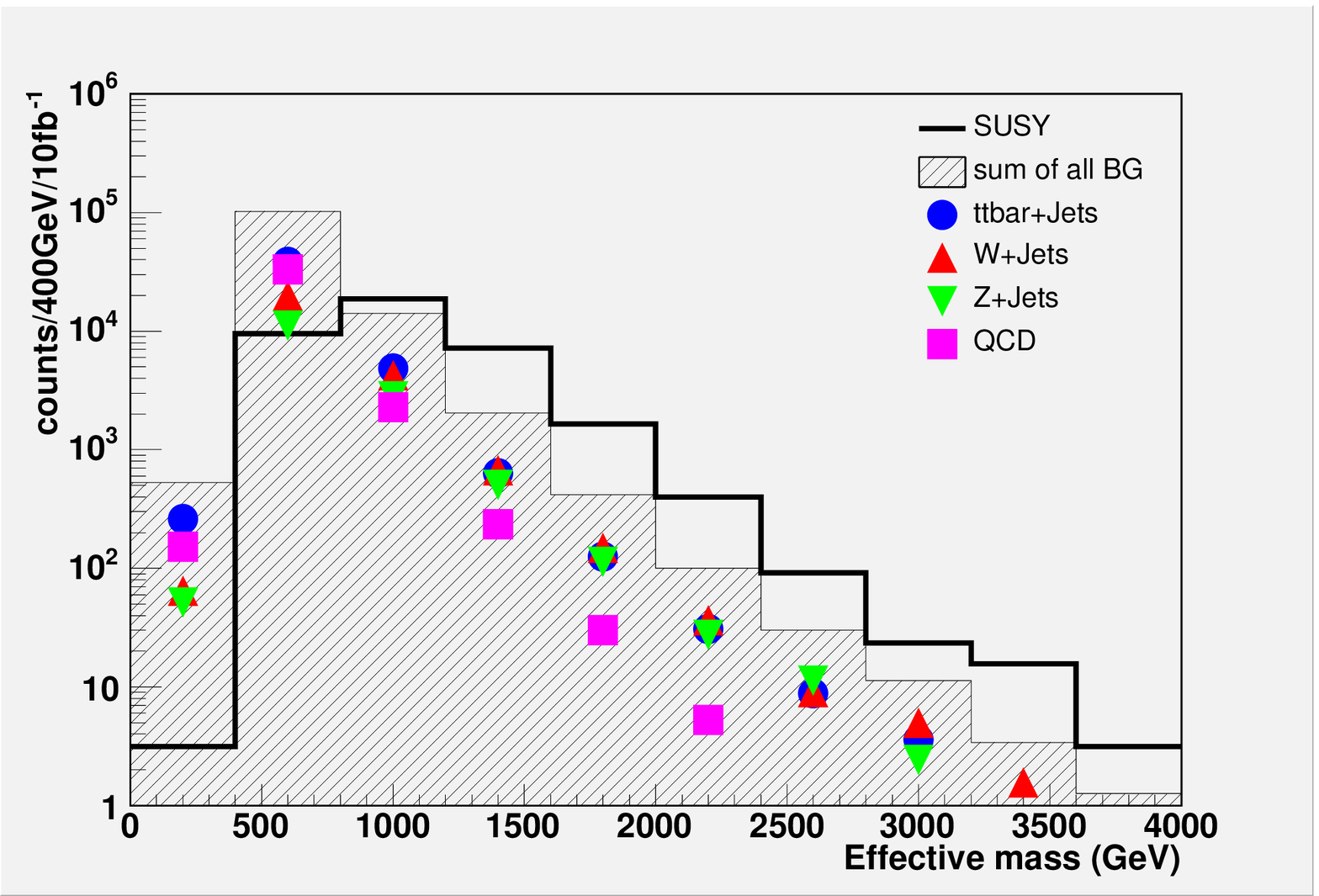}
\includegraphics{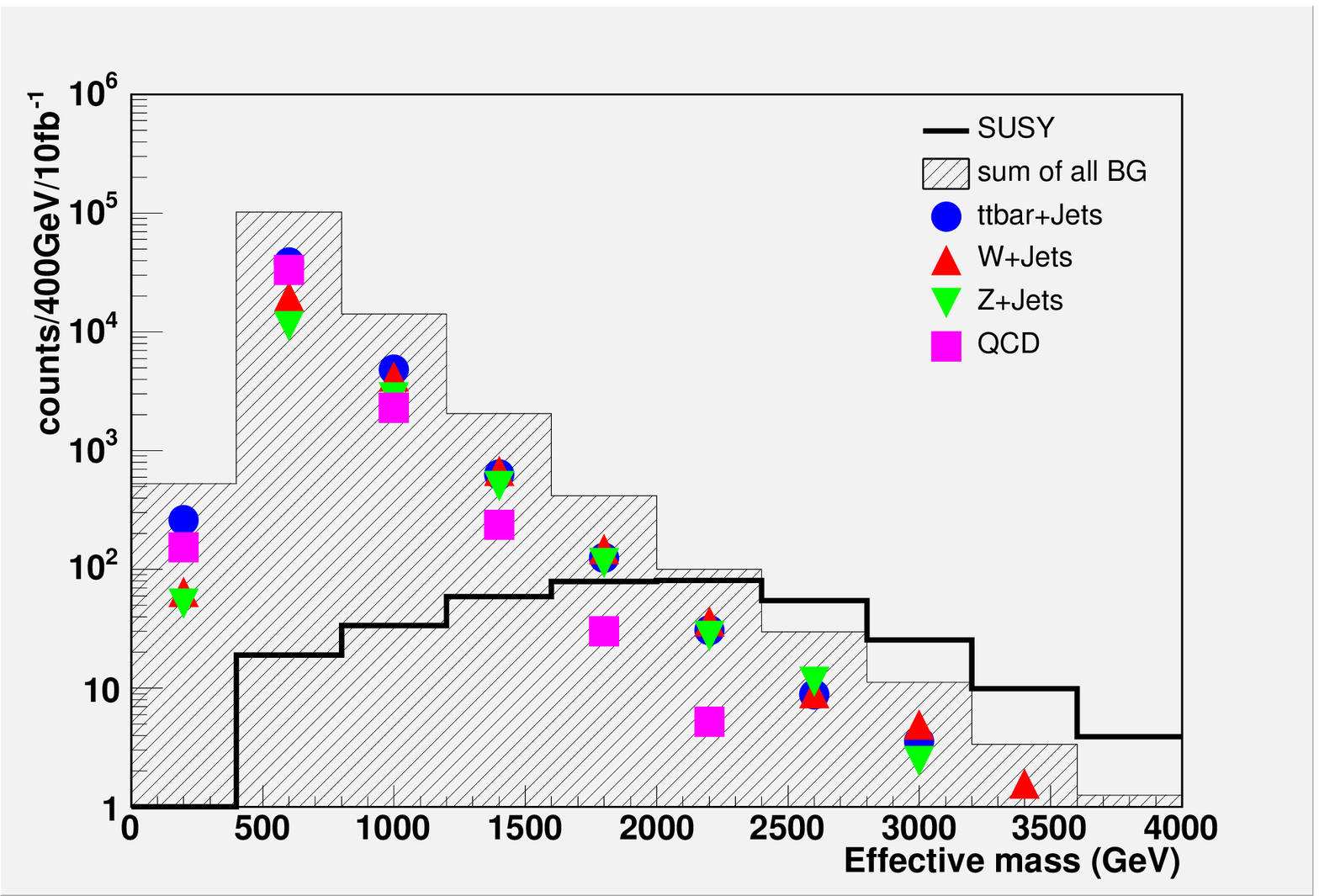}
\vspace{-1.5cm}
\caption{Effective mass distributions of the SUSY signal and the SM background.
The masses of both the squarks and gluino are 
(a) 700 GeV and (b) 1.5 TeV.
Notation is the same as in Fig.~\ref{fig:meff1}. 
\label{fig:meff2}} 
\end{figure}

The showered events are fragmented and decayed with the PYTHIA, 
and the detector effect is taken into account using the smearing 
Monte Carlo simulation of the ATLAS detector (ATLFAST~\cite{TDR})\@.
The following event selections are applied, 
which are the standard selections for SUSY searches and not yet optimized 
depending on the sparticles mass.
\begin{itemize}
\item \mET is larger than 100~GeV
\item \pt is larger than 100~GeV for at least one jet
\item Number of jets with \pt $>$ 50~GeV is larger than or equal to 4\@.
\item Transverse Sphericity is larger than 0.2, that means event are not back to back.
\item If the event contains one isolated lepton (e or $\mu$),
whose \pt is larger than 10~GeV (defined as `one lepton mode'), 
the transverse mass between the lepton and \mET is required 
to be larger than 100~GeV to reduce leptonic decay of the W boson.
Event without the isolated lepton is also accepted as `no lepton mode'\@.
\end{itemize}

The effective mass, which is define as \mET $+ \sum_{4jets}$ \pt ,
is a good variable, since the both \mET and \pt of 
jets would have discrimination power between the SUSY signal and 
the SM background processes.
Figures.~\ref{fig:meff1} show the effective mass distributions of the SUSY 
signal, in which mass of squarks and gluino is 1~TeV, 
and the SM background processes.
These are results of no lepton mode after the selection criteria are 
applied.
Left side figure shows the old results estimated with the PS model~\cite{TDR} and 
right shows the new results estimated with the ME + PS prediction.

(1) The SM background contributions become larger by factor 2-4 depending on
the effective mass. They become the same order of the SUSY signal.
There are many uncertainties in these estimations, 
for example effect of the higher order, 
choice of the various scales, and uncertainties of PDF\@.
Especially the region of the 
`high \pt ' and `high jet-multiplicity' are important
and they have to be understood directly using real data.
We will have enough luminosity to study these regions at Tevatron Run-II.
(2) The slope of the distribution of the background is
more gentle and the similar to that of the SUSY signal.
Figures.~\ref{fig:meff2} show the same distributions 
for the different SUSY mass scales.
The slope of the background processes is very similar for the various 
SUSY mass scales, and they show we have to understand these slope well.
Effect of the higher order, choice of the various scales are important
to be understood.
(3) Four types of background processes contribute at the same level.

\begin{figure}[t]
\vspace{5cm}
\includegraphics{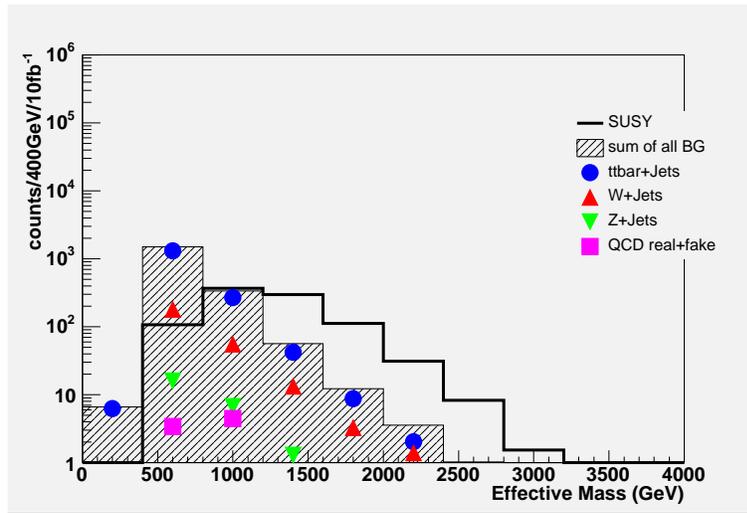}
\vspace*{1.8cm}
\caption{Effective mass distributions of the SUSY signal and the SM background 
processes for one lepton mode:
Notations are the same as in Fig.~\ref{fig:meff1}\@.
\label{fig:meff3}} 
\end{figure}

The Z+jets and the QCD background processes can be significantly suppressed
if one isolated lepton is required.
Furthermore the transverse mass cut removes effectively the background including the leptonic decay
of W boson (W+jets and tt+jets).
Total number of background can be suppressed by factor about 20.
On the other hand, the SUSY signal is reduced only by factor 2-5 depending the SUSY parameters. 
Fig.~\ref{fig:meff3} shows the effective mass distribution for the one lepton mode.
As you can see, clear excess will be observed.
The dominant background processes are tt + jets and W +jets, which are more
controllable than 
the QCD background processes.

\subsubsection{Fake \mET } 

\begin{figure}[t]
\vspace{5cm}
\includegraphics{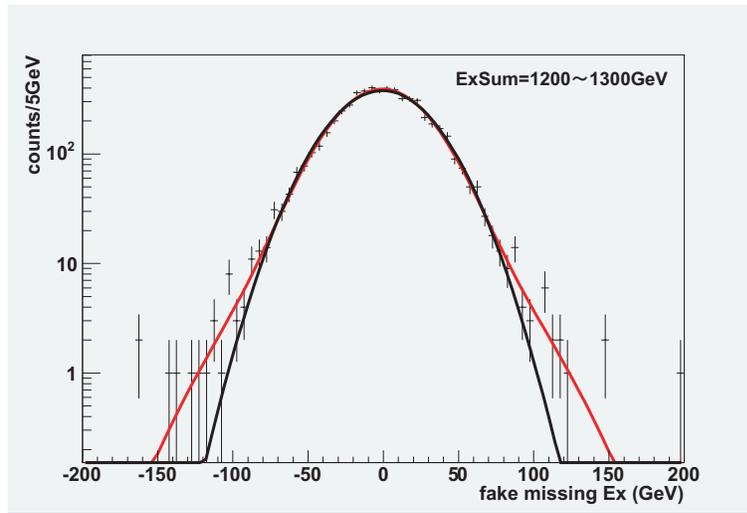}
\vspace*{1.8cm}
\caption{Distributions of the fake missing Ex: 
points with error show the simulated results.
Black and red lines show the fitted result with single and double  Gaussian,
respectively.
\label{fig:met2}} 
\end{figure}

The missing transverse energy, \mET, is vital variable for SUSY searches.
It is mainly produced with neutralino and neutrino, 
but also produced with the limited energy resolution of Hadronic jets.
This is called as ` fake \mET ' .
The non-Gaussian response of the jet energy 
measurements makes non-Gaussian tail in the fake \mET distribution.
Various detector components makes the different response(resolution) 
for the jet.
Then the fake \mET distribution becomes more complicated 
as shown in Fig.~\ref{fig:met2}(a)\@.

There are two important points:
\begin{itemize}
\item The resolution of the bulk part depends on the sum of energies deposited
on the calorimeters, and also the dead material in front 
of or the inside of the calorimeters.
If the event-topologies are different, the fraction of the energy in 
the various detector components becomes different, 
and the resolutions will be different.
\item Distribution has a tail apart from the Gaussian as shown in the figure.
This tail becomes dangerous background, 
since the QCD multi-jet processes have a huge cross-section.
\end{itemize}

Now the statistics of the simulated samples is very limited, 
and we can not conclude the shape/dependence of the tail.
More careful studies of the tail are necessary, 
i.e. the dependence of the tail  on the materials in front of 
the detectors, \pt of the jets, the contributions from the pileup and 
underlying.
The dependence of the event-topologies is also important.
The Fake \mET points simply the direction of the jet, 
if jet multiplicity is small.
\mET is required not to point the jet direction for SUSY searches at Tevatron,
and this selection removes the events including the large fake \mET \@.
But the SUSY signals at LHC are, I have already mentioned, 
multi-jets topologies.
Direction of the fake \mET will be smeared in the multi-jets events.
These studies using Tevatron run-II data are useful for the quick startup
of the LHC\@.
   
\subsubsection{Conclusions}

The Standard Model background processes are 
estimated with the Matrix Elements calculation, and
it is found out that the background 
contributions become larger than that we have expected. 
The distributions of the background processes are header and 
are similar to the  SUSY signal.
The careful studies of the Standard Model processes 
are useful using Tevatron Run-II data, 
especially studies of the slope of the \pt  and 
missing $E_{T}$ distributions of $\Wpm$ + jets, $\Zboson$ + jets and
$\toppair$ + jets. 

\mET is the vital variable for SUSY searches and will be complicated 
distribution depending on the sum of the energies, the dead materials, and 
event-topologies. 
Detail studies using Tevatron Run-II are useful for quick start up of LHC\@.

\clearpage\setcounter{equation}{0}\setcounter{figure}{0}\setcounter{table}{0}

\def\ra{\rightarrow}
\newcommand{\um}{\ensuremath{\mathrm{\mu m}}}
\newcommand{\bszero}{\ensuremath{B_s^0}}
\newcommand{\bd}{\ensuremath{B_d^0}}
\newcommand{\bsd}{\ensuremath{B_{s,d}^0}}
\newcommand{\bu}{\ensuremath{B^{+}}}
\newcommand{\mus}{\ensuremath{\mu^{+}\mu^{-}}}
\newcommand{\musss}{\ensuremath{\mu^{\pm}\mu^{\pm}}}
\newcommand{\hh}{\ensuremath{h^{+}h^{-}}}
\newcommand{\bsmm}{\ensuremath{\bszero\ra\mus}}
\newcommand{\bdmm}{\ensuremath{\bd\ra\mus}}
\newcommand{\bsdmm}{\ensuremath{\bsd\ra\mus}}
\newcommand{\bjk}{\ensuremath{\bu\ra J/\psi K^{+}}}
\newcommand{\bjp}{\ensuremath{\bu\ra J/\psi \pi^{+}}}
\newcommand{\bjf}{\ensuremath{\bszero\ra J/\psi \phi}}
\newcommand{\brbsmm}{\ensuremath{\mathcal{B}(\bsmm)}}
\newcommand{\brbdmm}{\ensuremath{\mathcal{B}(\bdmm)}}
\newcommand{\brbsdmm}{\ensuremath{\mathcal{B}({B_{s,d}^0}\ra\mus)}}
\newcommand{\brbjk}{\ensuremath{\mathcal{B}(\bjk)}}
\newcommand{\brbjf}{\ensuremath{\mathcal{B}(\bjf)}}
\newcommand{\musumu}{\ensuremath{M_{\mu\mu}}}
\newcommand{\Lxy}{\ensuremath{\vec{L}_{T}}}
\newcommand{\Lxyz}{\ensuremath{{L}_{\rm{3D}}}}
\newcommand{\ctau}{\ensuremath{\lambda}}
\newcommand{\pting}{\ensuremath{\Delta\Theta}}
\newcommand{\iso}{\ensuremath{\mathit{I}}}
\newcommand{\ptmm}{\ensuremath{\vec{p}^{\:\mu\mu}_{T}}}
\newcommand{\pmm}{\ensuremath{\vec{p}^{\:\mu\mu}}}
\newcommand{\cdf}{CDF~II}
\newcommand{\svx}{SVX~II}
\newcommand{\jmm}{\ensuremath{J/\psi\ra\mus}}

\subsection{Search for \bsmm\ and \bdmm\ Decays at the Tevatron}
\label{sec:BsMM}

D.~Glenzinski$^1$, C.-J.~Lin$^1$, R. Bernhard$^2$ and F. Lehner$^2$ \\ [2mm] 
{\em $^1$ Fermilab (CDF) \\ $^2$ University of Z\"urich (\dzero)} \\



In the Standard Model (SM), Flavor-Changing Neutral Current (FCNC) decays are
highly suppressed and can only occur through higher order diagrams.
The decay rate for the FCNC decay \bsmm~\footnote{Throughout this section inclusion of charge conjugate modes is implicit.} is proportional to the
CKM matrix element $\left|V_{ts}\right|^{2}$. The rate of \bdmm\ decays
is further suppressed by the ratio of $\left|V_{td}/V_{ts}\right|^2$.
The SM expectations for these branching fractions are
$\brbsmm=(3.42\pm0.54)\times10^{-9}$ and
$\brbdmm=(1.00\pm0.14)\times10^{-10}$~\cite{Buchalla:1993bv,Buras:2003td}, 
which are about two orders
of magnitude smaller than the current experimental sensitivity.
However, new physics contributions can significantly enhance these branching
fractions.
In the absence of an observation, any improvements 
to the limits can be used to set significant constraints on many SUSY 
models.


The best existing
experimental limits are $\mathcal{B}(\bsmm)<4.1\times10^{-7}$ using $240$~\pb\
of \dzero data~\cite{Abazov:2004dj,D0BsMMprl2} and $<1.5\times10^{-7}$ using $364$~\pb\ of
CDF data~\cite{Abulencia:2005pw} at the 90\% confidence level (C.L.), and
$\mathcal{B}(\bdmm)<8.3\times 10^{-8}$ using $111$~\fb\ of
Babar data~\cite{Aubert:2004gm} and $<3.9\times10^{-8}$ using $364$~\pb\ of
CDF data~\cite{Abulencia:2005pw} at the 90\% confidence level.

In the following sections we briefly describe the analysis strategies
employed by CDF and \dzero in these \bsmm\ and \bdmm\ rare decay searches.
Since the techniques employed are quite similar in the two experiments,
we begin with a discussion of the general analysis strategy in
section~\ref{sec:BsMMGeneral}.

\subsubsection{Analysis Strategy}
\label{sec:BsMMGeneral}

\paragraph{General}
The CDF and \dzero\ collaborations have published several papers describing
searches for the rare \bsmm\ and \bdmm\ decays at the
Tevatron~\cite{Abazov:2004dj,D0BsMMprl2}
\cite{Abulencia:2005pw,Acosta:2004xj,Abe:1998ah}.
Although they all use a very similar analysis strategy, we concentrate here on
describing the most recent CDF and \dzero\ analyses which yield the
most stringent limits to date.

In general, the strategy is to collect
the signal sample on a di-muon trigger that is as inclusive as possible. The
analysis is simplified and the systematic uncertainties reduced by
collecting both the signal and normalization ($B\ra J/\psi X$) decays on the
same trigger.  Similarly, the reconstruction requirements are also chosen to
be as loose possible while still maintaining a high purity so that the
same \mus\ reconstruction requirements are made for both the signal and
normalization modes. 
After a the pre-selection, four discriminating variables are used to further 
reduce the expected background while maintaining good signal efficiency. Both 
collaborations employ a ``blind'' analysis strategy when choosing the
final selection criteria. For these optimizations the signal mass region
in the data is kept hidden, the backgrounds are estimated from the data mass
sidebands, and the signal efficiency is estimated from Monte Carlo (MC) samples.
Only after the final selection criteria have been chosen, and after the
background estimates and MC modeling of the signal efficiency have both
been verified using data control samples, is the data signal region revealed.

\paragraph{Event Selection and Optimization}
\label{sec:BsMMEvtSelAndOpt}

For \bsmm\ decays the final state is fully reconstructed.  This yields
invariant mass and vertex resolutions that are much better than for the
partially reconstructed final states which dominate the background.
In addition to the pre-selections requirements, four different variables are 
used to suppress the background.   We exploit the long {\it B} lifetime and
use $|\vec{L}|$, the 3D decay length of the di-muon pair relative to the 
primary interaction vertex. Combinatoric background and partially 
reconstructed {\it B} hadrons are 
removed with requirements on the 3D opening angle between the di-muon momentum 
(\pmm) and decay length ($\vec{L}$) vectors, \pting.  
Since {\it b}-quark fragmentation is hard, 
{\it B} hadrons carry most of the transverse momentum of the {\it b} quark, and 
thus are isolated.  An Isolation variable~\footnote{The $B$-candidate isolation is defined as
 $\iso=| \ptmm | /(\sum_{i}p^{i}_{T} + | \ptmm | )$, where the
 sum is over all tracks with $\sqrt{\Delta\eta^2+\Delta\phi^2} \le 1$;
 $\Delta\phi$ and $\Delta\eta$ are the azimuthal angle and
 pseudorapidity of track $i$ with respect to \pmm.  Only tracks that
 extrapolate within 5~cm of the di-muon vertex along the $z$-axis
 are included in order to exclude tracks orginating from other
 \pp\ interactions which may have occured in the same bunch crossing.}, 
\iso,  is therefore 
used to enhance the heavy flavor content of the sample and also to reject 
partially reconstructed {\it B} hadrons, which are less isolated since some of
the daughter tracks are included in the isolation cone.   The last variable is
\musumu, the di-muon invariant mass.

To further enhance signal and background separation CDF constructs a multivariate
likelihood ratio based on the input variables: \iso, \pting, and
$P(\ctau)=e^{-\ctau/c\tau_{B_{s(d)}}}$, where $\ctau=c\musumu |\vec{L}|/|\pmm|$
and $\tau_{B_{s(d)}}$ is the world average $B_{s(d)}$ lifetime.  The
$P(\ctau)$ variable offers the same background discrimination as \ctau\ or
$\vec{L}$ but with reduced sensitivity to the modeling of the vertex resolution. 
The likelihood ratio is then defined to be
\begin{equation}\label{eq:BsMMlh}
  L_R = \frac{\prod_{i} \mathbf{P}_{s}(x_{i})}
       {\prod_{i} \mathbf{P}_{s}(x_{i})
      + \prod_{i} \mathbf{P}_{b}(x_{i})},
\end{equation}
where $x_1=\iso$, $x_2=\pting$, $x_3=P(\ctau)$, and $\mathbf{P}_{s(b)}(x_{i})$
is the probability that a signal (background) event has an observed $x_i$.
The probability distributions for the signal events are obtained from the signal
MC and the background distributions are taken from the data sidebands. 
Using the procedure detailed in reference~\cite{Abulencia:2005pw} CDF optimized
the $L_R$ and \ptmm\ requirements as well as the width of the signal search
window.  The optimization resulted in this choice of final selection
criteria: $L_{R}>0.99$, $\ptmm > 4$~GeV/c, and a search window of 
$\pm60$~GeV/c$^2$ centered on the world average \bsd\ mass.


The expected number of background events in the CDF analysis is estimated by 
extrapolating the number of sideband events passing the pre-selection 
requirements to the signal window and then scaling by the expected rejection, 
$\kappa$, for a given $L_R$ requirement.  The parameter $\kappa$ is determined 
from the background $L_R$ distribution, which is generated by randomly 
sampling the $P(\ctau)$, \pting, and \iso\ distributions from the data 
sidebands to improve statistical precision on $\kappa$. The procedure for 
estimating the background is cross-checked using control samples from the data: 
like sign $\mu^{\pm}\mu^{\pm}$ events, \mus\ events with $\ctau<0$, and a 
fake-muon enhanced \mus\ sample in which at least one muon candidate fails 
the muon quality requirements.   The background predictions are compared to 
the number of events observed in the search window for a wide range of $L_R$ 
requirements.   No statistically significant discrepancies are observed.

In the \dzero analysis background is suppressed by making stringent requirements
on the variables \pting, \iso, and the transverse decay length significance, 
$L_{T}/\sigma_{L_{T}}$, where $L_{T}$ is the projection of $\vec{L}$ onto the 
plane transverse to the beamline.  The $L_{T}$-significance was found to have 
a better discriminating power than $L_{T}$ alone. \dzero optimizes the 
requirements on these three variables using the procedure described in
reference~\cite{Abazov:2004dj,D0BsMMprl2}.  
The optimization resulted in this choice of
final selection criteria: $\pting<0.2$~rad, $\iso>0.56$, and 
$L_{T}/\sigma_{L_{T}}>18.5$.  The width of the search window is 
$\pm180$~MeV/c$^2$ around the mean reconstructed \bszero\ mass.  The background
is estimated by interpolation using an unbinned likelihood fit to the
sideband data.


\paragraph{Normalization}
\label{sec:BsMMNorm}

Both collaborations choose to normalize the \brbsmm\ to \bjk\ decays rather
than \bjf\ decays. The reasons for this are that the $B^{+}$
decay yields larger statistics and the lifetime and branching ratio are well
known from $e^{+}e^{-}$ experiments.  In contrast the \bszero\ branching ratio
has only been measured at the Tevatron~\cite{Abe:1993fa} with limited
precision\footnote{In addition, normalizing to this measured \bszero\ branching
ratio is a bit disingenuous since in reference~\cite{Abe:1993fa} \brbjf\ is
determined by normalizing to \brbjk.}.  One might alternatively consider
using the measured \brbjk\ and inferring \brbjf\ using $SU(3)$ symmetries.
However, at present the theoretical uncertainties associated with these
symmetry assumptions are about $20\%$, which is larger than the 
$13\%$ uncertainties associated with the $f_{u}/f_{s}$ needed
for the \bjk\ normalization.  Moreover, the \bjf\ final state presents
additional challenges.  For example, the reconstructed final state has four
tracks, two from the \jmm\ decay and two from the $\phi\ra K^{+}K^{-}$ decay,
which results in larger kinematic differences with respect to the two-body
\bsmm\ decays compared to the differences observed in the (effective)
three-body $\bjk\ra\mus K^{+}$ final state.
The understanding of the efficiency for the \bjf\ final state is further
complicated by the presence of CP-even and CP-odd decay components with
significantly different lifetimes.

\paragraph{Limit Setting}
Both CDF and \dzero\ normalize to the \brbjk\ and calculate the upper limit 
on \brbsmm\ using the following expression\footnote{The expression for 
\brbdmm\ is derived from equation~\ref{eq:brbsmm} by replacing \bszero\ with 
\bd\ and the fragmentation ratio $f_{u}/f_{s}$ with ${f_{u}}/{f_{d}}=1$.}:
\begin{eqnarray}\label{eq:brbsmm}
\brbsmm \leq \frac{N_{\bszero}^{ul}}{N_{\bu}}\cdot
     \frac{A_{\bu}}{A_{\bszero}}\cdot
     \frac{\mathcal{B}(B^{+}\ra J/\psi(\ra\mus)K^{+})}{(f_{s}/f_{u}) + R},
\end{eqnarray}
where $N_{\bszero}^{ul}(N_{o},N_{b})$ is the upper limit on the number of \bsmm\
decays observed at some confidence level given $N_{o}$ observed events
in the mass signal region when expecting $N_{b}$ background events;
$N_{\bu}$ is the number of observed \bjk\ candidates; $A_{\bszero}$ and $A_{\bu}$
are the total acceptance for the \bsmm\ and \bjk\ decays, respectively,
including trigger, reconstruction, and final selection efficiencies; the
$\mathcal{B}(\bjk\ra\mus K^{+})=(5.88\pm0.26)\times10^{-5}$ is taken from
reference~\cite{Eidelman:2004wy}; the fragmentation ratio $f_{s}/f_{u}$ is
taken from~\footnote{$f_x$ is the fraction of weakly decaying $B_x$ hadrons in $b$ quark
 fragmentation. We use values from~\cite{Eidelman:2004wy}.}; 
and $R=(\brbdmm / \brbsmm)\cdot(A_{\bd} / A_{\bszero})$.
Since the \bd\ and \bszero\ are only $90$~$\mathrm{MeV}/c^2$ apart, the
factor $R$ corrects for \bdmm\ ``contamination'' in the \bsmm\ signal
region.  In practice, the CDF \musumu\ resolution of about
$24$~$\mathrm{MeV}/c^{2}$ is good enough to resolve the two states so that
the \bsmm\ and \bdmm\ signal windows are chosen to have a small overlap.
This overlap is ignored (ie. $R$ is set to 0) when estimating the branching
ratio limits.  Similarly, \dzero\ also conservatively sets $R=0$. The
\musumu\ resolution of \dzero\ is about $88$~$\mathrm{MeV}/c^{2}$ so that the
\bsmm\ and \bdmm\ final states can not be separately resolved. \dzero
chooses to interpret their result as a limit on \brbsmm\ since, as
discussed in section~\ref{sec:BsMM}, \brbdmm\ is expected to be CKM
suppressed in most models. The experimental inputs and observed limits are 
summarized in Table~\ref{tab:BsMMCombo}.

\subsubsection{Tevatron Combination}
\label{sec:BsMMCombination}
The most recent CDF and \dzero\ results have been 
combined~\cite{Bernhard:2005yn}
properly accounting for the correlated uncertainties. The
two measurements are summarized in Table~\ref{tab:BsMMCombo}.  The
single-event-sensitivity, {\it ses}, is defined as the \brbsmm\ obtained
when setting $N_{\bszero}^{ul} =1$ in equation~\ref{eq:brbsmm}.  Despite using
approximately $20\%$ less luminosity, the \dzero\ analysis maintains a
slightly better {\it ses}, largely owing to their superior muon acceptance.
On the other hand, the CDF analysis has a much smaller background expectation,
even with $20\%$ more luminosity, largely owing to their superior mass resolution.
The expected limits are defined as the average limit obtained by summing over
all possible experimental outcomes, $N_{o}$, weighted by their Poisson
probability when expecting $N_b$ background events.  It is a measure of the
exclusion power of given method assuming no signal is observed and takes
into account the {\it ses} and background expectations.

\begin{table}[tbh]
\begin{center}
\begin{tabular}{c||ccc} \hline\hline
             & CDF U-U    & CDF U-X    & \dzero     \\ \hline
  Luminosity & $364\:\pb$ & $336\:\pb$ & $300\:\pb$ \\ \hline\hline
  $(A_{\bu}/A_{\bszero})$
             & $0.852\pm0.084$ & $0.485\pm0.048$ & $0.247\pm0.019$ \\
  $N_{\bu}$
             & $1785\pm60$     & $696\pm39$      & $906\pm41$      \\
  $N_b$
             & $0.81\pm0.12$   & $0.66\pm0.13$   & $4.3\pm1.2$     \\
  $N_o$
             & 0               & 0               & 4               \\
  {\it ses} $(\times10^7)$
             & $1.04\pm0.16$   & $1.52\pm0.25$   & $0.59\pm0.09$   \\
             & \multicolumn{2}{c}{(0.62 combined)} &               \\
  expect. limit 90\% C.L.
             & $3.5\times10^{-7}$ & $5.6\times10^{-7}$ & $3.5\times10^{-7}$ \\
             & \multicolumn{2}{c}{($2.0\times10^{-7}$ combined)} & \\ \hline
  obsvd. limit 90\% C.L.
             & \multicolumn{2}{c}{($1.5\times10^{-7}$ combined)}
             & $3.2\times10^{-7}$                                  \\ \hline\hline
\end{tabular}
\caption{\label{tab:BsMMCombo} Summary of CDF and \dzero\ results used to
  calculate the Tevatron combined limit on \brbsmm.  The CDF analysis is divided in two separated search channels (U-U and U-X) with different muon acceptance. The {\it ses}, and
  expected and observed limits for the combination of the two CDF channels
  are given parathetically.}
\end{center}
\end{table}

The combined limit was obtained using a Bayesian 
technique~\cite{Bayes,Bayes2,Eidelman:2004wy}
that parameterizes the uncertainties as gaussian distributions in the
integration.  A flat prior was used for the unknown \brbsmm\ and it
was verified that the resulting limit is insensitive to reasonable
variations of the cut-off assumed in the definition of the prior.
In the combination the uncertainties on the fragmentation ratio and
the normalization branching ratios were added in quadrature and treated
as fully correlated.  All the other uncertainties were assumed to be
uncorrelated.  The resulting Tevatron combined limit is
\begin{equation}
{\cal B}(\bszero\ra\mus)_{comb.} < 1.2\, (1.5)\times 10^{-7}\,\, {\rm at \,\,a\,\, 90\%\,(95\%)\, C.L.}
\end{equation}
assuming for the fragmentation ratio the standard PDG value~\cite{Eidelman:2004wy} of
$f_u/f_s=3.71\pm 0.41$. Using an evaluation of the fragmentation function based
on Tevatron data alone ($f_u/f_s=3.32\pm 0.59$) would improve the limit 
by $10\%$.

The excellent CDF mass resolution allows them to carry out an independent 
search for \bdmm\ decays.  As previously reported the
resulting limit is $\brbdmm\leq3.9\times10^{-8}$ at 90\% C.L.  An independent
search is not possible for the \dzero\ experiment.  However, the \dzero\ 
results can be interpreted as limits on \brbdmm\ by assuming there is 
\bsmm\ contribution to the signal region.  Interpreting the \dzero\ results 
in this way and combining with the CDF limit gives
\begin{equation}
  {\cal B}(\bd\ra\mus)_{comb.} < 3.2\, (4.0)\times 10^{-8}\,\, {\rm at\,\, a\,\, 90\%\,(95\%)\, C.L.}
\end{equation}
It should be stressed that the Tevatron combined limit on \brbdmm\ is not
independent of the \bszero\ limit, since the same \dzero\ information is used in
both.

\subsubsection{Tevatron Outlook}
\label{sec:BsMMOutlook}
\begin{figure}[h]
  \begin{center}
  \includegraphics[width=0.8\linewidth]{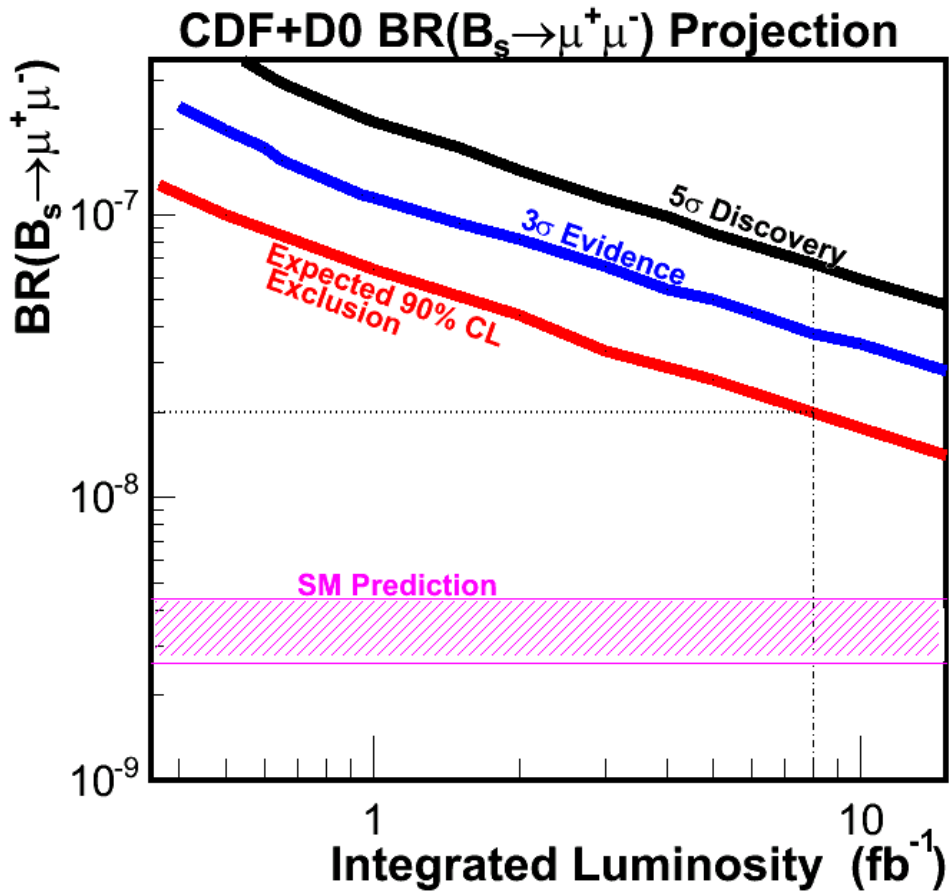}
  \caption{\label{fig:BsMMProj} The projected Tevatron combined reach for
    \brbsmm\ as a function of the integrated luminosity collected by
    each experiment.}
\end{center}
\end{figure}

The projected Tevatron reach for the \dzero\ and CDF combined search for \bsmm\
decays is shown in Figure~\ref{fig:BsMMProj} as a function of the luminosity 
collected per experiment.   The projection assumes the analysis techniques 
are unchanged and that the trigger and reconstruction efficiencies are 
unaffected with increasing luminosity.   If, then, each experiment collects 
$8\:\fb$ the Tevatron combination will allow for a ``$5\sigma$ discovery'' 
down to \bsmm\ branching ratios of about $7\times10^{-8}$ and for $90\%$ 
C.L. exclusions down to branching ratios of about $2\times10^{-8}$.  
Both experiments are pursuing further improvements to the analysis sensitivity,
which would push the Tevatron combined sensitivity to still lower branching 
ratios.  Even if no signal is observed, the resulting stringent limit would 
eliminate a very large part of the high $\tan\beta$ parameter space in many 
supersymmetric models.

\subsubsection{Relevance for LHC}

At the Tevatron, the search for the decay \bsmm\ is a part of the core
Run~II physics program.  An observation of \bsmm\ at the Tevatron would 
unambiguously signal the presence of new physics.  At the LHC Atlas, CMS, 
and LHCb are all expected to have sensitivity down to the SM branching ratio. 
In the following we discuss a few issues relevant for these future LHC 
analysis.  In particular we offer comments concerning the mass resolution, 
background composition, choice of normalization mode, and trigger definitions.

Obviously a crucial aspect in the \bsdmm\ analysis is the experimental di-muon
mass resolution.  It's important for two reasons.  First, an excellent mass 
resolution helps to reject background events in the signal region since high
efficiency is maintained for narrow mass windows.  The different background 
expectation numbers from \dzero\ and CDF roughly reflect their mass resolutions.
Second, it will be important to measure \brbsmm\ and \brbdmm\ separately, and 
this is most cleanly accomplished with mass resolutions that
are small compared to $(M_{\bszero} - M_{\bd})=90$~MeV/c$^2$.  The ratio of these
branching ratios can help in determining the flavor structure of any new
physics which might be present.  For minimal flavor violating (MFV) models, 
\brbdmm\ remains CKM suppressed relative to \brbsmm.  This is not necessarily 
true for non-MFV models such as R-parity violating SUSY, which can produce 
large enhancements, even for low values of $\tan\beta$, in either or both of 
the \bszero\ and \bd\ FCNC decay rates.

The exact background composition at the Tevatron is not precisely known and 
is modeled from sideband data.  It includes contributions from sequential
semileptonic decays ($b\to c\mu^{-} X\to\mus X$), gluon splitting 
($g\to b\overline{b}\to \mus X$), and fakes ($b\to\mu X$ + fake muon).  
Backgrounds from $\bsd\to h^{+}h^{-}$ ($h^{\pm}=\pi^{\pm},\: K^{\pm}$) are, 
at present, negligible.  This will not be the case at the LHC, where these 
decays may form an irreducible background.  Other exclusive decays from 
$B_c$ (e.g. $B_c^{\pm}\to J/\psi(\to \mu^+\mu^-) \mu^{\pm}\nu$) or {\it B} 
baryons might also become important at the LHC.  These types of backgrounds 
are difficult since they cannot be trivially estimated from the data 
sidebands, but instead require dedicated estimates that rely on a detailed 
understanding of the trigger performance and hadron-to-muon fake rates.  It 
will thus be important to have a set of triggers which allow for the 
necessary studies.

Systematic effects can be minimized with an intelligent choice of the 
normalization decay.  For the reasons discussed in 
Section~\ref{sec:BsMMNorm}, the Tevatron searches have normalized to the 
decay $B^{+}\to J/\psi(\mus)K^{+}$.  This choice incurs a $\pm13\%$ systematic
uncertainty associated with the fragmentation ratio, $f_{u}/f_{s}$, which
is correlated across all measurements.  And although it is plausible to assume 
that the ratios determined from $e^{+}e^{-}\to Z^{0}\to b\overline{b}$ 
experiments can be extrapolated to high energy $p\bar{p}\to b\bar{b}$ 
collisions, there is no strong theoretical argument in support of it. 
It is thus desirable to find a fragmentation independent normalization using 
a known \bszero\ decay - e.g. \bjf\ decays.  The present Tevatron searches 
suffer from a relatively low yield of \bjf\ decays.  However, as the Tevatron
dataset grows it is very likely that measurements on this important decay 
will considerably improve such that \bjf\ will become a well-known ``standard''
before LHC turns on.  Given the large LHC datasets eventually available, it may 
be possible to normalize to one of the $B\to\hh$ decays if an efficient 
trigger with a managable rate can be defined.  The consideration of which
normalization mode to choose should be coupled with the analysis trigger 
strategy.

The trigger plan for this analysis should be carefully considered.  Generally 
speaking three types of triggers will be needed: 
A) one that collects the signal sample used for the analysis, 
B) a second that is unbiased with respect to the first so that the efficiency 
of trigger A can be determined, and C) a third that collects samples of 
heavy flavor hadronic decays (e.g. $B\to \hh$).   We'll briefly discussed
issues relevant to each of these.  The 
analysis is greatly simplified if trigger A simultaneously collects 
both the \bsdmm\ decays and the normalization decays.  It is important to 
note that for the Tevatron results, the single largest correction in 
equation~\ref{eq:brbsmm} comes from the ratio of trigger acceptance between 
the \bsmm\ and normalization decay.  This
ratio is driven by the trigger requirements and reflects the fact that the
$p_{T}(\mu)$ spectrum is softer for the $B^{+}\to J/\psi(\to\mus) K^{+}$ decays
than for the two body \bsmm\ decays.  Thus the \bjk\ trigger acceptance
can have a much stronger dependence on the $p_{T}(B)$ than the \bsmm\ 
acceptance.  Depending on the trigger, this could be a source of significant
systematic uncertainty, especially once model uncertainties associated with the
$p_{T}(B)$ spectrum are folded-in.  These effects should be considered 
when choosing a normalization mode.  For example, normalizing to \bjf\ 
($B\to\hh$) decays would likely exacerbate (mitigate) these effects.
The other two trigger paths must be designed to avoid introducing any
kinematic bias relative to the requirements of trigger A.   The principal 
aim of trigger B is to determine whether or not trigger A has any strong 
kinematic dependence which might effect the signal and normalization 
decay differently (e.g. $p_{T}(\mu)$ or $p_{T}(B)$ dependencies).  Trigger 
C is used to collect clean $K^{\pm}$ and $\pi^{\pm}$ samples 
(e.g. from the decay $D^{+*}\to D^{0}\pi^{+}\to K^{-}\pi^{+}\pi^{+}$) from 
which to determine the kaon- and pion-to-muon fake rates needed to accurately
estimate the $B\to\hh$ backgrounds.  Ideally, large samples of $B\to\hh$ 
decays would also be available for detailed study.  Careful attention should 
be given to the expected rates of these triggers.   It may happen that some
or all of them may be rate limited at higher luminosities. If
pre-scales are employed, care should taken to ensure that 
the data collected with triggers B and C have a luminosity profile similar 
to that of trigger A so that trigger efficiencies and backgrounds can be
reliably estimated over the full range of relevant luminosities.

\clearpage


\section{Particle-based Phenomenology}
\label{sec:particle}

Let us assume that a signal for new physics will be observed at the
Tevatron or the LHC.  The question is how should one go about
explaining what that signal is in a well defined theory.  The simplest
first attempt would be to try to fit the signal by assuming the
existence of a single particle beyond the ones discovered already, in
a Lorentz invariant quantum field theory.  This is a good description
at the TeV scale for many models that include several additional
particles, with only one of them being relatively easy to discover.

If the signal cannot be convincingly explained by the existence of a
single new particle, then one should attempt to explain it using
several new particles.  This situation occurs in a wide class of
models where the signatures of different particles are correlated, for
instance through cascade decays.  We concentrate in this section on
signatures produced by individual particles, leaving the discussion of
models with multiple particles for Sec.~\ref{sec:models}.

The new particle is specified by its spin and its $SU(3)_c\times
SU(2)_w\times U(1)_y$ charges.  All other properties are described by
some continuous parameters: mass, mixings and couplings.  The number
of types of new particles that are likely to be discovered at the
Tevatron and the LHC is rather limited.  The majority of the theories
beyond the Standard Model discussed in the literature include only
particles of spin 0, 1/2, 1, or 2.  Higher-spin particles could exist,
but they would require complicated strongly-coupled theories, or
departures from quantum field theory (for a study of higher-spin
particles at hadron colliders, see Ref.~\cite{Burikham:2004su}).

Under $SU(3)_c$, new particles are most likely to transform as color
singlets, triplets, octets, or sextets.  Higher $SU(3)_c$
representations could exist, but would pose a variety of theoretical
challenges.  Under $SU(2)_w$, new particles may transform as singlets,
doublets or triplets, while higher representations are not usually
present in the models proposed so far.

Finally, the hypercharges of new particles are constrained by the
requirement that the electric charge of any color-singlet (or of the
ensuing hadrons in the case of colored particles) is an integer.
Otherwise, the lightest particle with non-integer electric charge
would be stable on cosmologial time scales, and would be ruled out for
most interesting regions of parameter space by a variety of searches
for stable charged particles.

We will not attempt to study here all these possible particles.  We
only display several representative examples, and urge the readers to
analyze as many of the other cases before the start of the LHC.
Sec.~\ref{sec:zprime} describes the case of a heavy spin-1 particle
that is a singlet under the Standard Model gauge group, usually
referred to as a $Z^\prime$ boson, emphasizing the case where the
$Z^\prime$ interacts with the quarks and leptons.
Sec.~\ref{sec:higgsless} also discusses the collider implications of a
$Z^\prime$ boson, but in the case where it couples exclusively to
gauge bosons.  Sec.~\ref{sec:wprime} analyzes a heavy spin-1 particle
that is color singlet and has electric charge $\pm 1$, usually
referred to as a $W^\prime$ boson

Sec.~\ref{sec:vectorlike} presents a study of a spin-1/2 particle
which is color-triplet, and has the same charges for the left- and
right-handed components.  This commonly referred to as a vectorlike
quark.  Sec.~\ref{sec:triplet} deals with spin-0 particles that are
$SU(2)_w$ triplets. Sec.~\ref{sec:stable} presents a study of the
collider signatures of a new electrically-charged particle which is
stable enough to escape the detector.


\clearpage\setcounter{equation}{0}\setcounter{figure}{0}\setcounter{table}{0}
\newcommand{\zp}{\ensuremath{Z^\prime} }
\subsection{\zp at the LHC}
\label{sec:zprime}

Fabienne~Ledroit, Julien~Morel, Benjamin~Trocm\'e\\ [3mm]
{\em Laboratoire de Physique Subatomique et de Cosmologie,\\ Grenoble, France} \\


{\em 
In this section, we develop a model independent determination of the ATLAS efficiency in detecting \zp bosons decaying to an electron-positron pair. This efficiency is then applied to the cross section predicted in the CDDT parameterization of \zp models, a model independent parameterization already used by CDF. We then derive the ATLAS \zp discovery potential in this framework. Finally we compare this potential to the results from 
LEP and the Tevatron.}\\

The existence of an additional spin 1 neutral boson -denoted as \zp in this section- is predicted in many extensions of the Standard Model (SM), such as $E_6$ or $SO(10)$ Grand Unified Theories, extra dimensions theories, little Higgs models...\\
At the LHC, the \zp production process mainly consists in a quark-antiquark annihilation, the $u\bar{u}$ and $d\bar{d}$ cases being largely dominant in most models\footnote{In the following, the $s\bar{s}$, $c\bar{c}$, $b\bar{b}$ processes are ignored in order to ease the reading; the treatment of $u\bar{u}$ and $d\bar{d}$ events can be generalised to these marginal cases.}. In all the following, only decays to known particles are considered; because of the high QCD backgound, there is very little hope to detect the hadronic decays of a \zp boson; with a large missing energy coming from two neutrinos, the $\tau^+\tau^-$ decay is also a very challenging channel. With the Drell Yan process as unique irreducible background and a very limited reducible background, the electron and muon channels are much more promising and can be considered as golden channels to discover a new neutral gauge boson. Up to \zp masses of about 5~TeV, the signature simply consists in a high invariant mass peak above the Drell Yan line shape.

In this section, the ATLAS potential in term of discovery of a \zp decaying to an electron-positron pair is studied. It is especially detailed in the CDDT parameterization\cite{Carena:2004xs} adopted by the CDF collaboration\cite{Abulencia:2006iv}; this parameterization takes into account both experimental limits and general theoretical assumptions to constrain the models with an additionnal neutral gauge boson. Given these, four classes of solutions are found, three parameters remaining totally free in the four classes; these parameters are the mass of the additionnal gauge boson, $M_{\zp}$, the global coupling strength, $g_{\zp}$, and a parameter $x$ describing the relative coupling strength to the different fermions. An original method to extract a realistic efficiency depending on the model is presented. Since the reducible background is expected to be small, only the irreducible background was considered. 

\subsubsection{Samples}
In order to study the reconstruction efficiency in a realistic context, several samples of $q\bar q\to \gamma / Z / \zp\rightarrow e^+e^-$ events were generated with PYTHIA\cite{Sjostrand:2000wi}, and simulated with GEANT 3 for the ATLAS detector response. The response of the particles with a pseudo-rapidity out of the range [-2.5,2.5] was not simulated. The events were then reconstructed in the official ATLAS reconstruction framework\cite{Athena}.\\
The samples were generated with two different \zp masses (1.5~TeV and 4~TeV) and for a variety models -- SM like, $E_6$ derived models, Left-Right model -- each model being fully determined by its coupling constants to the known fermions. The knowledge of the exact characteristics of these models, which can be found in \cite{nous}, is not useful here since a model independent approach was chosen. The CTEQ5L parton distribution functions were used and initial/final state radiations were switched on. A total of 150,000 events with di-electron masses above 500~GeV and 15,000 above 2000~GeV were simulated.

\subsubsection{Event selection}

\begin{figure}[ht]
  \begin{center}
    \psfrag{YTITLE}{}
    \subfigure[Energy resolution]{
      \psfrag{XTITLE}[t]{\small$\frac{E_{gen}-E}{E_{gen}}$}
      \epsfig{file=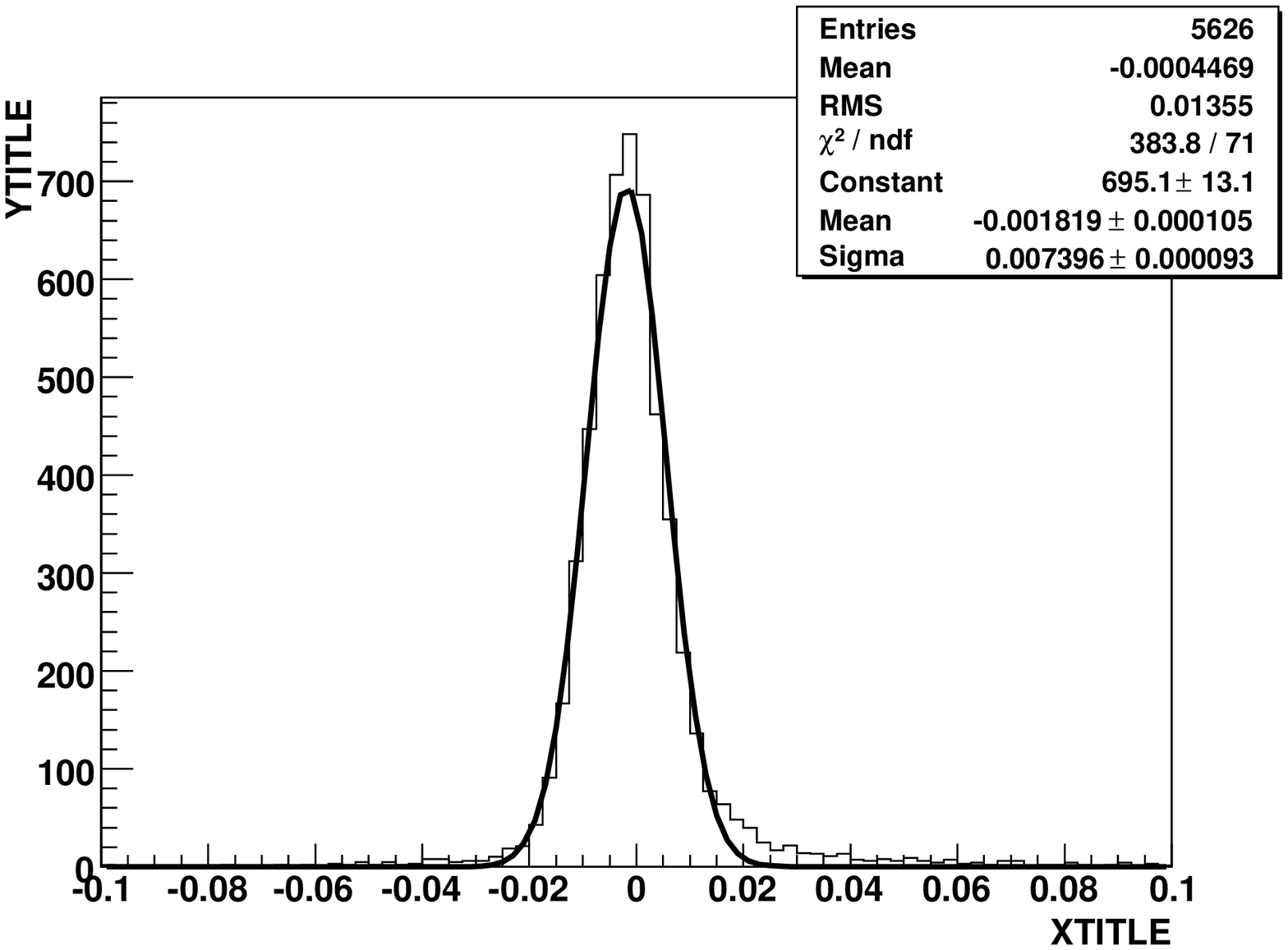,height=5.5cm}
    }
    \subfigure[Pseudo rapidity resolution.]{
      \psfrag{XTITLE}[t]{\small$\frac{\eta_{gen}-\eta}{\eta_{gen}}$}
      \epsfig{file=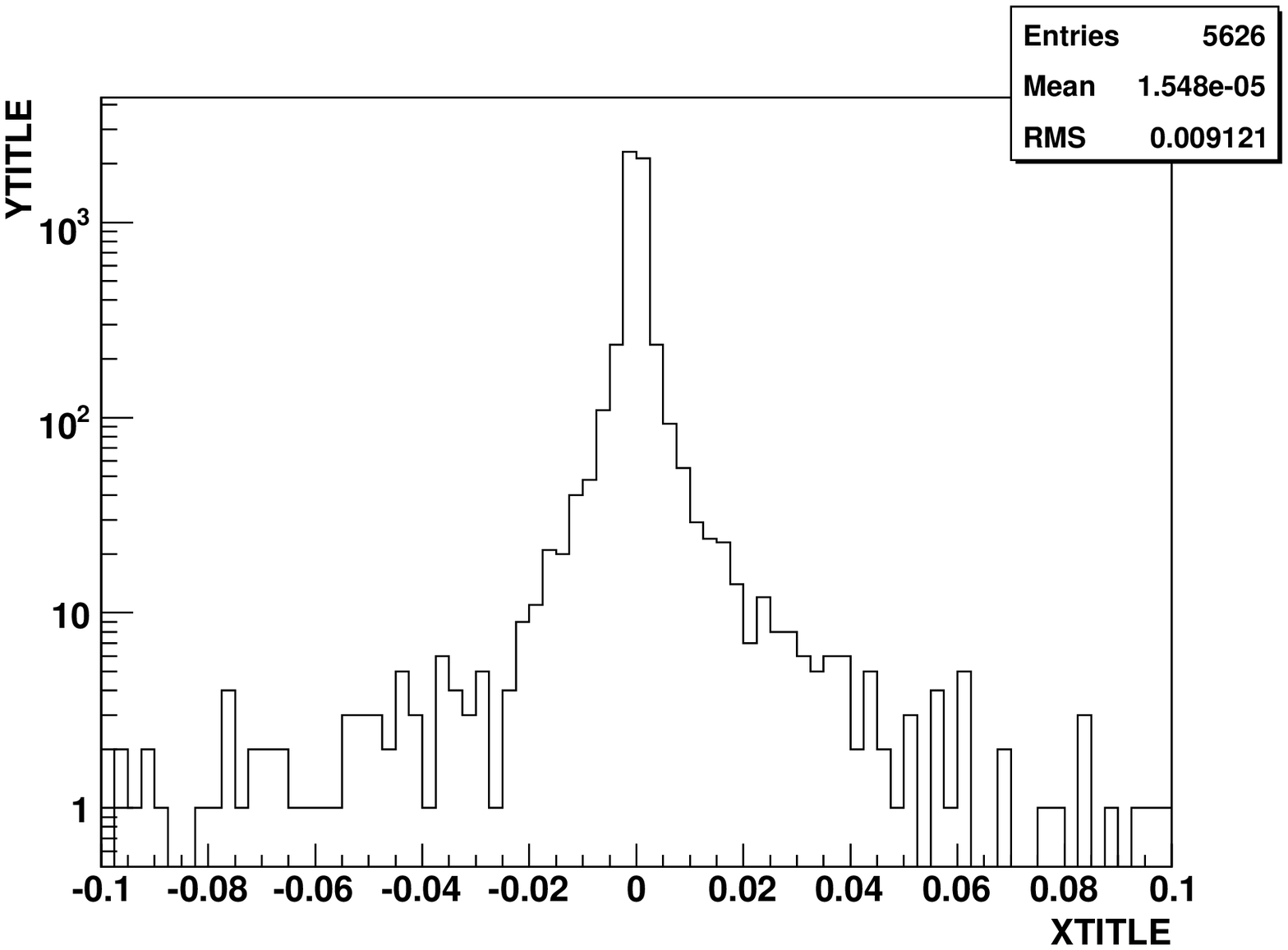,height=5.5cm}
    }
  \end{center}
  \caption{Reconstruction resolutions estimated by considering electrons/positrons decaying from a 1.5~TeV SM like \zp.}
\bigskip
  \label{f_ele1000}
\end{figure}

First the electron (positron) candidates are reconstructed using the standard ATLAS electron identification: additionally to criteria on shower shape and energy leakage, one requires to have a good track quality, with a total number of hits in the tracking detectors greater than~6. The absence of any additional track in a broad cone (0.05 in $\eta$ and 0.1 in $\phi$) around the matched track is also required in order to reduce the QCD and tau backgrounds.\\
Although being optimized on low energy electrons, these simple criteria lead to satisfactory results with reasonable angular and energy resolutions (see figure \ref{f_ele1000}) and an acceptable efficiency; this procedure will have to be optimized in a near future but is good enough for our present purpose.

Only events with exactly two electrons candidates are kept; these two candidates were also required to be isolated in the calorimeter, i.e. with no cluster of transverse energy greater than 40~GeV in a cone of radius $\sqrt{(\Delta\phi)^2+(\Delta\eta)^2}$ equal to 0.5.\\
Finally, the two reconstructed electrons must be of opposite charge and back to back in the transverse plane, the absolute difference of azimuthal angles having to be greater than 2.9 radians.\\
The typical efficiencies for this selection are detailed in table \ref{t_accept} for a sample of SM like \zp generated with a mass of 1.5~TeV.

\begin{table}[ht]
\centering
\begin{tabular}{|c|c|}\hline
At least two electrons with $|\eta|<2.5$  & (82.1$\pm 0.3$ )\% \\ \hline \hline
At least two identified and isolated  electrons  & (57.1$\pm 0.4$ )\% \\ \hline 
Electrons candidate of opposite charge &  (53.4$\pm 0.4$ )\% \\ \hline 
Electrons candidate back to back &  (45.9$\pm 0.4$ ) \% \\ \hline \hline
Accepted events& (45.9$\pm 0.4$ )\% \\ \hline
\end{tabular}
\caption{Step by step event selection efficiency - SM like \zp with a mass of 1.5~TeV}
\label{t_accept}
\end{table}

\subsubsection{Model dependence of the efficiency.}

With a full detector simulation and the use of the official ATLAS reconstruction framework, the efficiency estimate can be considered as realistic. Nevertheless, it strongly depends on the \zp mass and on the underlying model; therefore it cannot be used to derive any model independent limit.\\
The different leptons kinematic characteristics, directly inducing different detector efficiencies, can be controlled by considering two characteristics of the model : the forward backward asymmetry (related to the coupling constants to quark and leptons), and the \zp boost distribution (related to the \zp mass and the coupling constants to quarks).\\

Introducing the angle $cos\theta^\star$ -the angle between the negative lepton and the incoming quark in the \zp rest frame-, the \zp production cross section is:
\begin{equation}
\frac{d\sigma}{d\cos\theta^{\star}} \propto \frac{3}{8}(1+\cos^2\theta^{\star})+A_{FB}\cos\theta^{\star}
\end{equation}
The  $A_{FB}$ coefficient depends on the boson coupling constants to incoming quarks and decay products and therefore strongly depends on the underlying model; this coefficient however vanishes when integrating over two $cos\theta^\star$ intervals symmetric around 0. Keeping in mind that flipping the $cos\theta^\star$ sign corresponds to swapping the electron and the positron, one can deduce the following property: in any positive (or negative) interval of pseudo rapidity in the \zp rest frame, the total number of leptons, electrons plus positrons, is independent of the $A_{FB}$ coefficient. Therefore, if the detector is assumed to have  equivalent detection and reconstruction efficiencies for electron and positron, the event selection efficiency does not depend on the $A_{FB}$ coefficient; this however does not mean that it is independent on the model, since the \zp boost still has to be taken into account. There is however a way to control this effect, as explained below. Notice that a forward/backward asymmetric efficiency does not spoil this result since the LHC is a pp collider, and hence the probability for the incoming quark to be forward is the same as the probability for the incoming antiquark.\\

\begin{figure}[ht]
  \begin{center}
    \subfigure[Rapidity distributions (arbitrary scale).]{
      \psfrag{XTITLE}[r]{\small \zp rapidity}
      \psfrag{YTITLE}[r]{}
      \epsfig{file=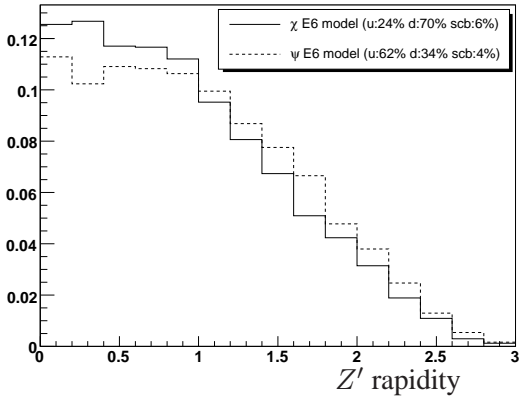,width=7.9cm}    
    }
    \subfigure[Rapidity ratios for the two models, considering  $u\bar u$ and $d\bar d$ production separately.]{
      \psfrag{xTitle1}[t]{\small $Y_{u\bar{u}}\rightarrow\zp$}
      \psfrag{xTitle2}[t]{\small $Y_{d\bar{d}}\rightarrow\zp$}
      \psfrag{yTitle1}{}
      \psfrag{yTitle2}[b]{\small \hspace{1cm} ratio($\psi$ $E_6$ model/$\chi$ $E_6$ model)}    
      \epsfig{file=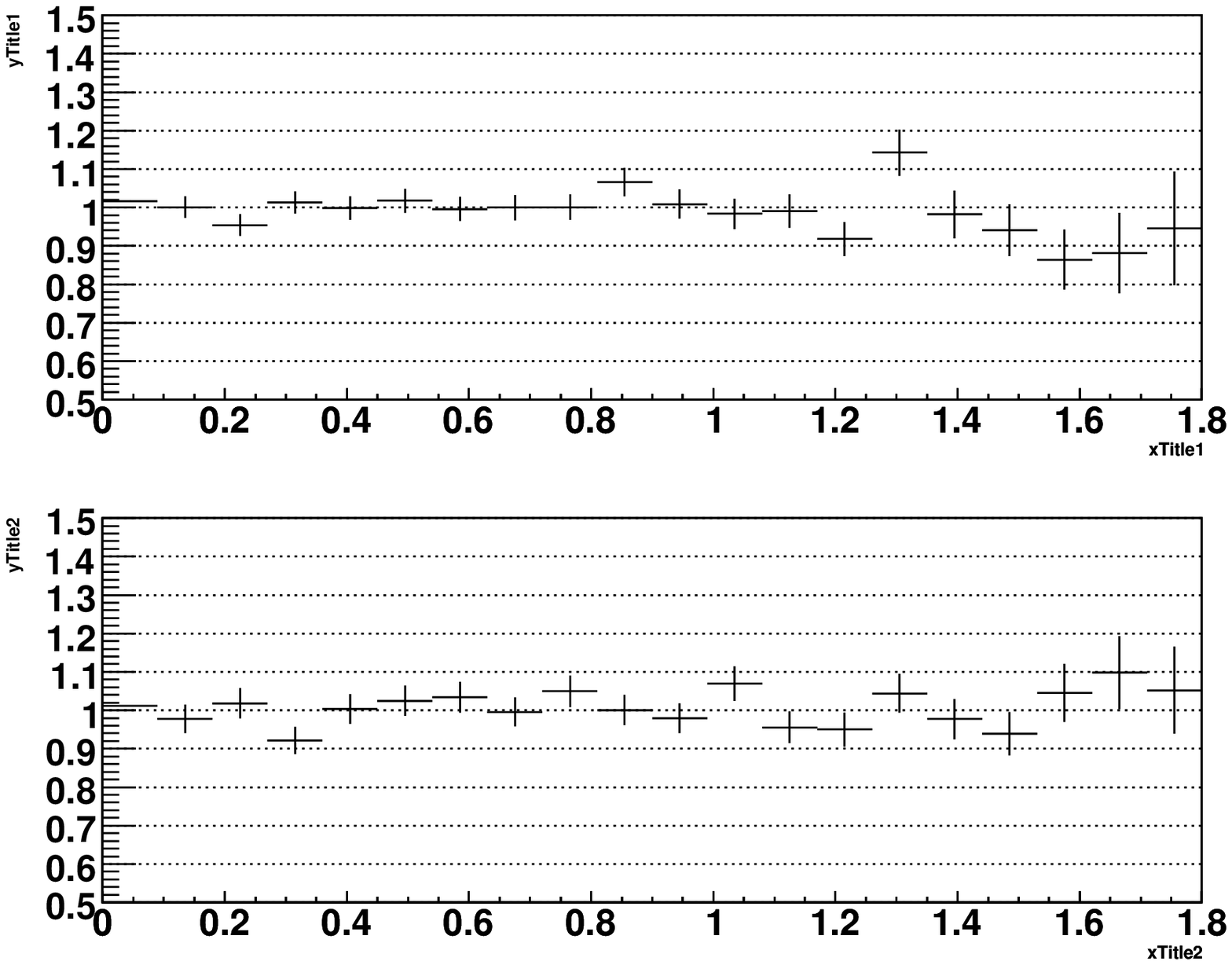,width=7.9cm}    
    }
  \end{center}
\caption{\zp rapidity for two different $E_6$ models in the mass range [1.48~TeV,1.52~TeV] ($M_{\zp} = 1.5$TeV)}
\label{f_rapid}
\end{figure}

The \zp boost can be deduced from its mass and its rapidity. The \zp  rapidity distribution is represented on figure~\ref{f_rapid} (left) for two different models. Their shapes are different only due to the different $u$/$d$ parton density functions in the proton and, because of different coupling constants of the \zp to $u$ and $d$, due to the different fractions of di-electron coming from $u\bar u$ and $d\bar d$. When splitting each sample in two subsamples according on the incoming quark flavour, all \zp rapidity distributions become similar, independently of the model as can be seen on figure~\ref{f_rapid}b.

\begin{figure}[t]
  \begin{center}
    \psfrag{YTITLE}[b]{\small Efficiency}
    \subfigure{
      \psfrag{XTITLE}[r]{\small Effective \zp mass}
      \epsfig{file=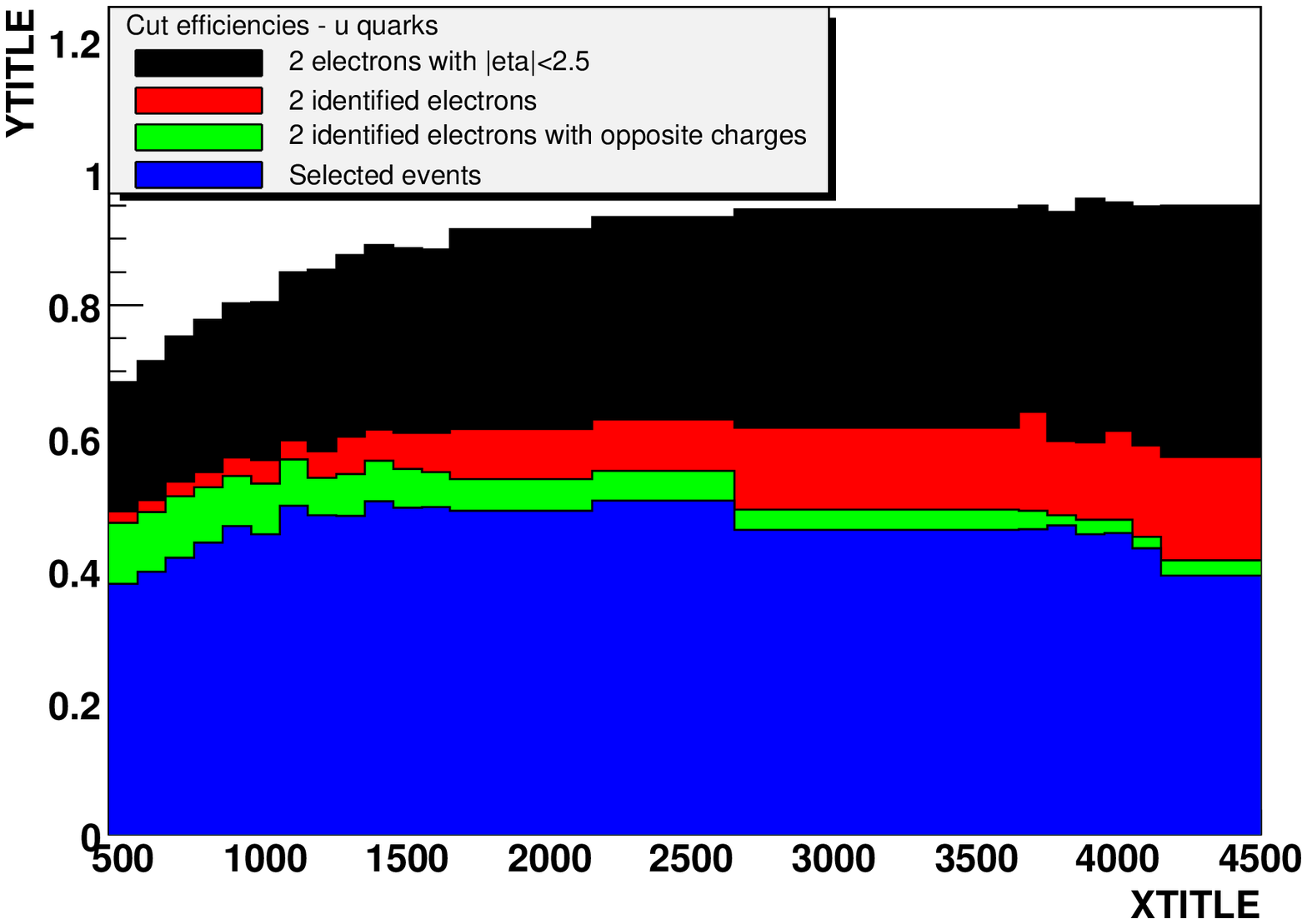,width=7.8cm}
    }
    \subfigure{
      \psfrag{XTITLE}[r]{\small Effective \zp mass}
      \epsfig{file=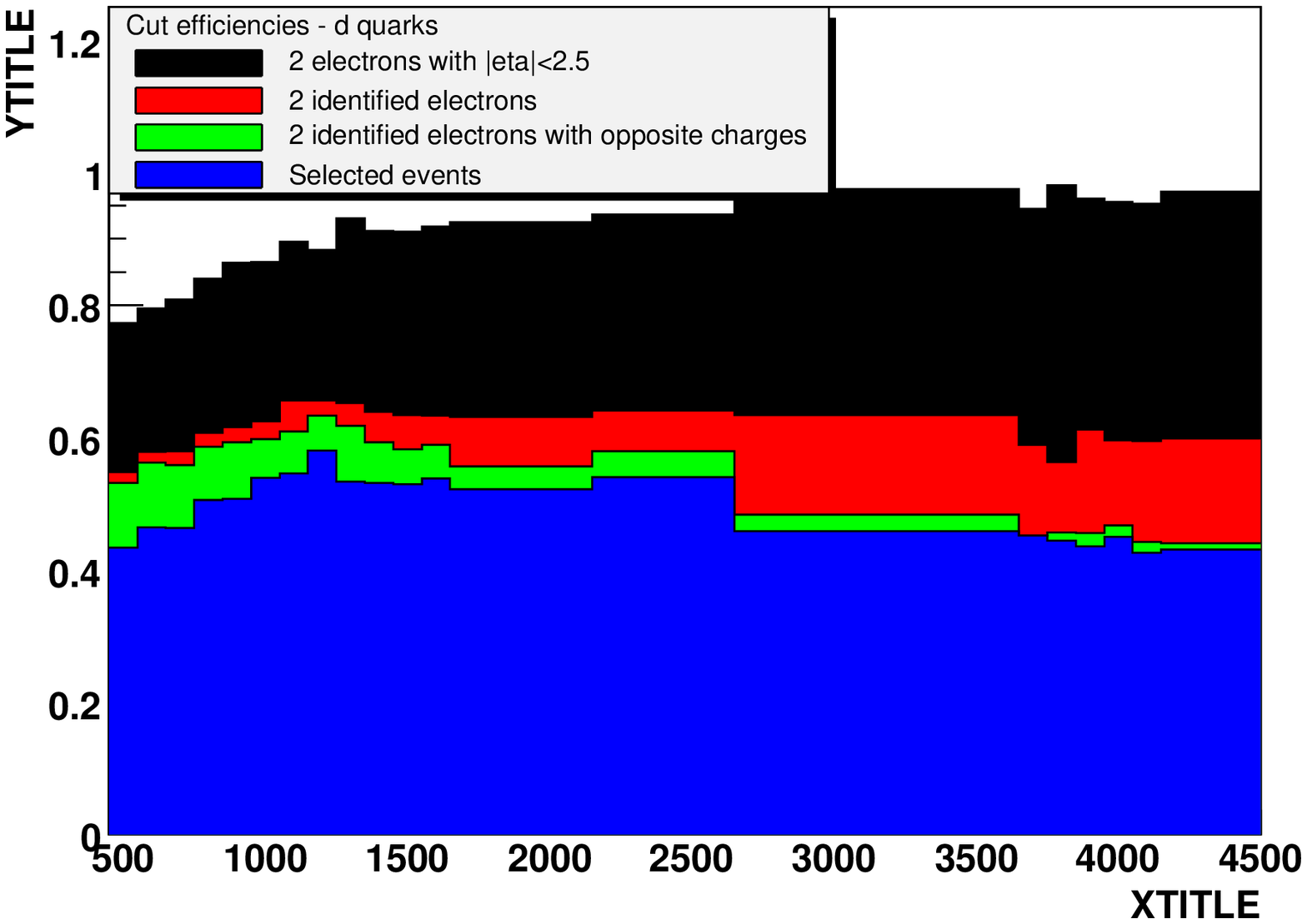,width=7.8cm}
    }
  \end{center}
  \caption{Efficiencies for $u$ and $d$ flavours.}
  \label{f_splitFlavEff}
\end{figure}

The same is true for the leptons decaying from the \zp: for a given \zp mass, their kinematic properties only depend on the flavour of the incoming quarks. Consequently, an average reconstruction efficiency can be extracted event by event in any \zp model, with the only knowledge of the incoming quark flavour and the effective \zp mass. These efficiencies are summarized on figure \ref{f_splitFlavEff} for the two main quark flavours, including the intermediate efficiencies for each selection step\footnote{The variable bin size was chosen in order to optimize the number of events by bins.}.

In both cases, one observes an increase of the number of events fully contained in the considered detector acceptance, when the di-electron mass becomes larger: this is a pure kinematic effect, the leptons being less boosted at large \zp mass. This effect is however counterbalanced by a degraded electron reconstruction efficiency; this can be explained by the fact that the electron identification algorithm is optimized on low energy electrons; there is some hope to recover from such effect by tuning the algorithm for higher energy objects. When the electron transverse momentum becomes larger, the detector charge identification is degraded, whereas the accuracy on the azymuthal angle measurement is improved. Consequently, at higher di-electron mass, the efficiency related to the charge criterion becomes lower and, on the opposite, the efficiency related to the acoplanarity becomes higher.\\
Finally, the reconstruction efficiency of the $d\bar d\rightarrow{\zp}$ is found to be always higher than the one of the $u\bar u\rightarrow{\zp}$. This can be explained by the particle density function differences, the $d$ quark one being less peaked at low x, therefore inducing less boosted events, and consequently events which are more contained in the detector.\\
These efficiencies can be exploited in two ways: 
\begin{itemize}
\item considering only a cross section production and ignoring the incoming quark flavour, the $u\bar{u}$ efficiency can be taken conservatively in order to derive a discovery reach or an exclusion limit.
\item in the context of a given model, where the relative fractions of incoming $u$ and $d$ quarks are known, the discovery reach and exclusion limit can be precisely extracted.
\end{itemize}
With a limited number of generated and simulated Monte Carlo events, it is therefore possible to derive realistic limits with reasonable reconstruction efficiency in any model. In any case, this reconstruction efficiency is more realistic than the ones estimated with a fast detector simulation.

\subsubsection{ATLAS discovery reach.}

\begin{figure}[t]
  \begin{center}
    \subfigure{
      \epsfig{file=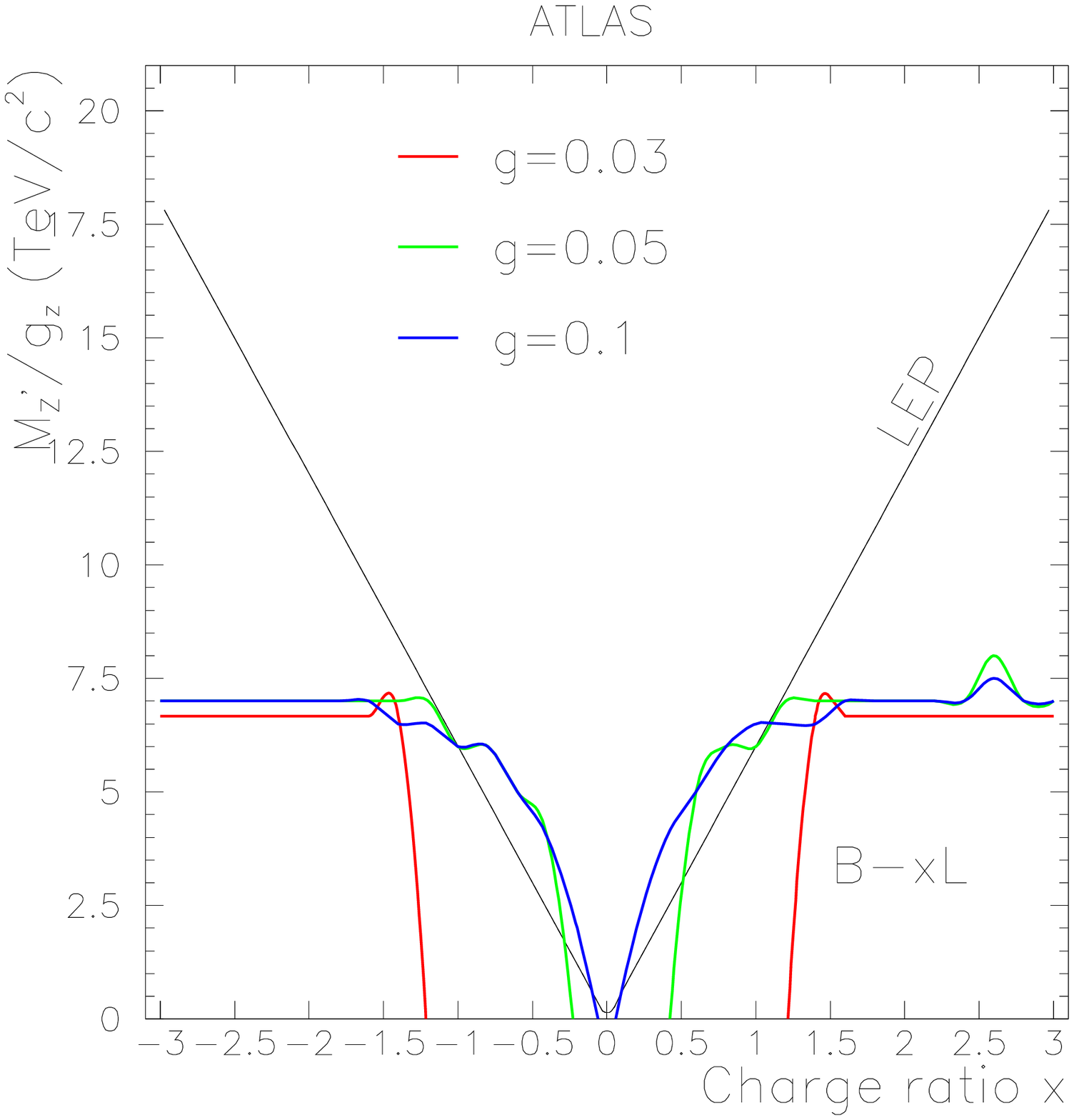,height=7.9cm}
    }
    \subfigure{
      \epsfig{file=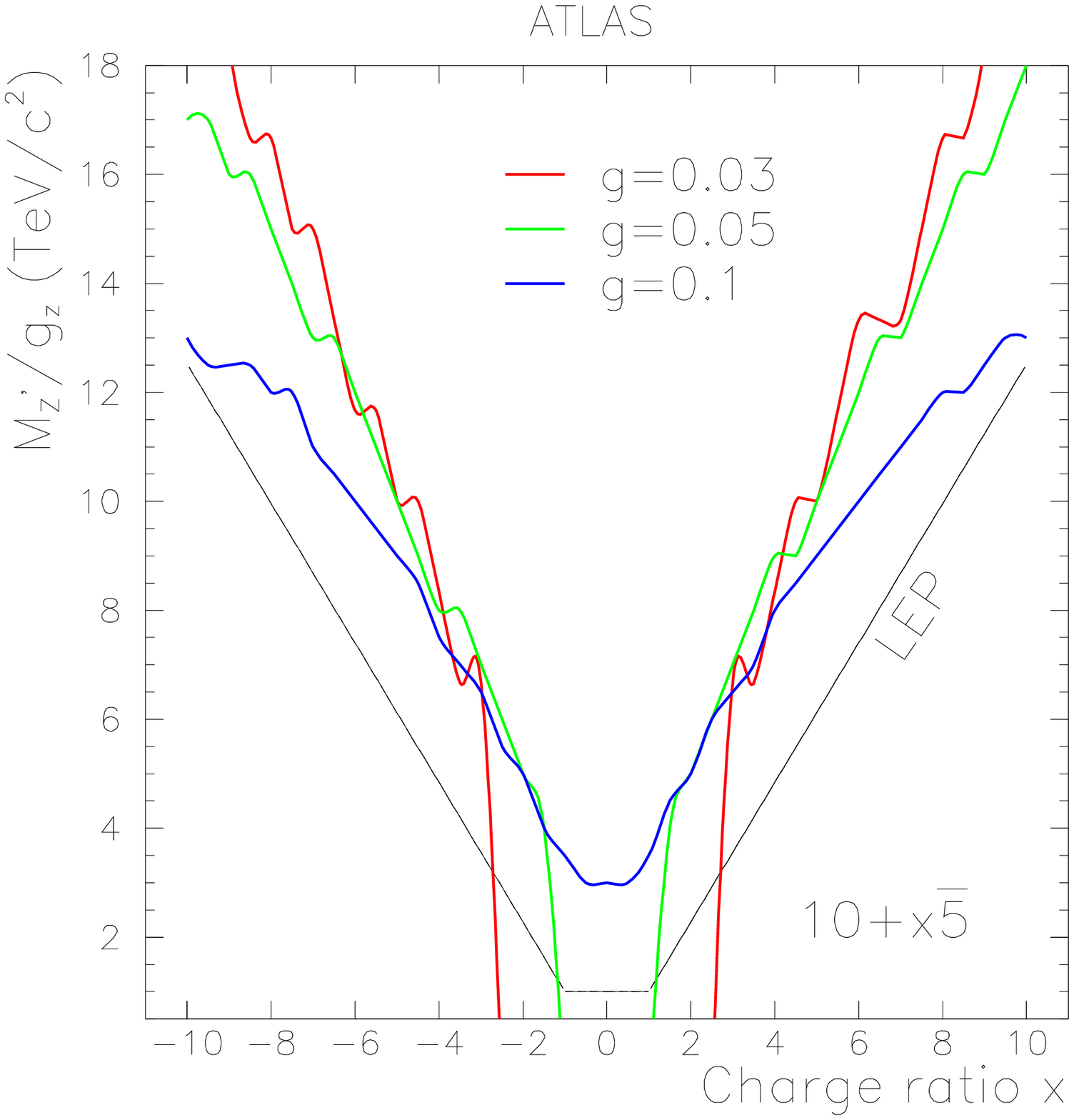,height=7.9cm}
    }
  \end{center}
  \begin{center}
    \subfigure{
      \epsfig{file=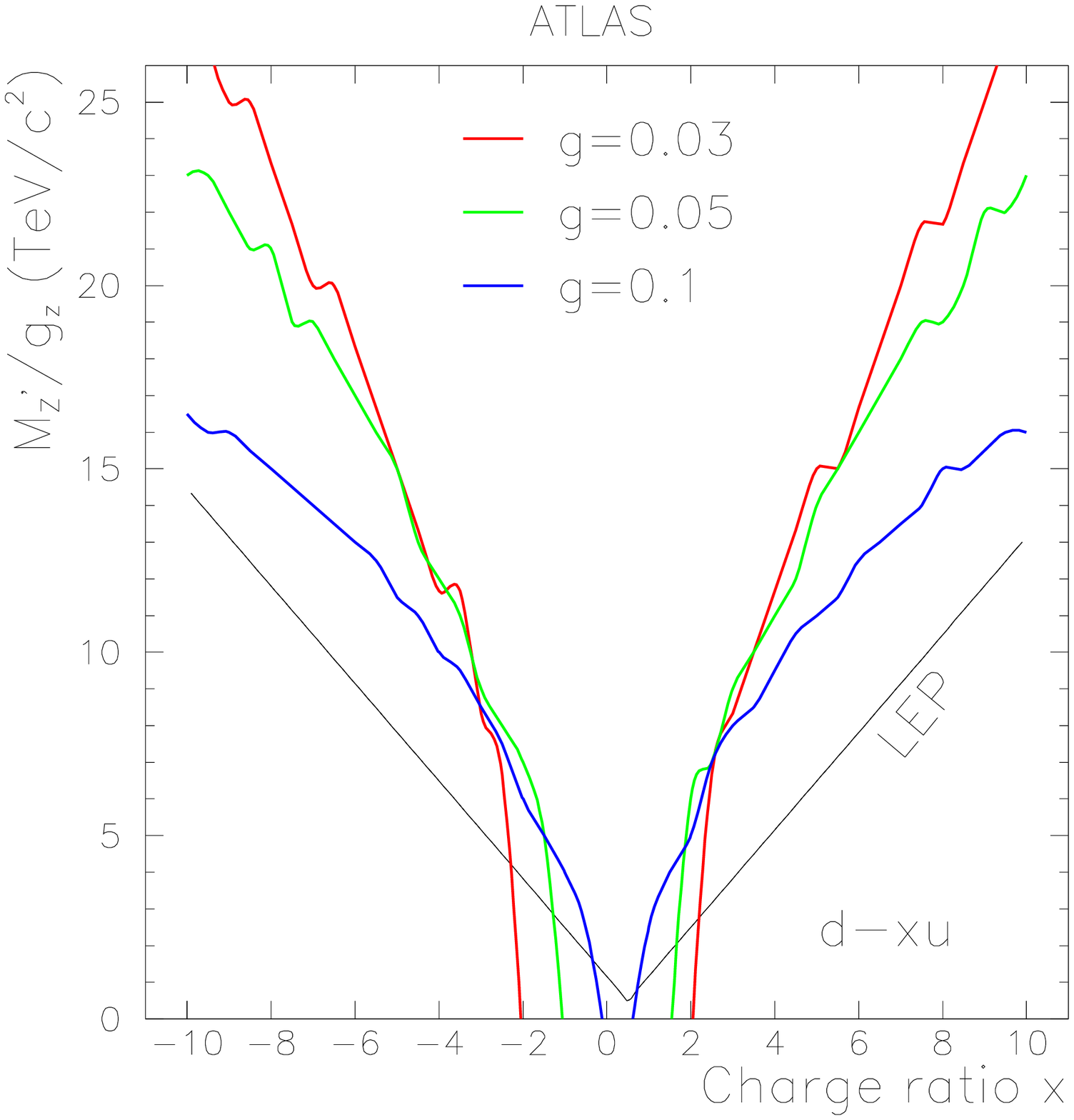,height=7.9cm}
    }
    \subfigure{
      \epsfig{file=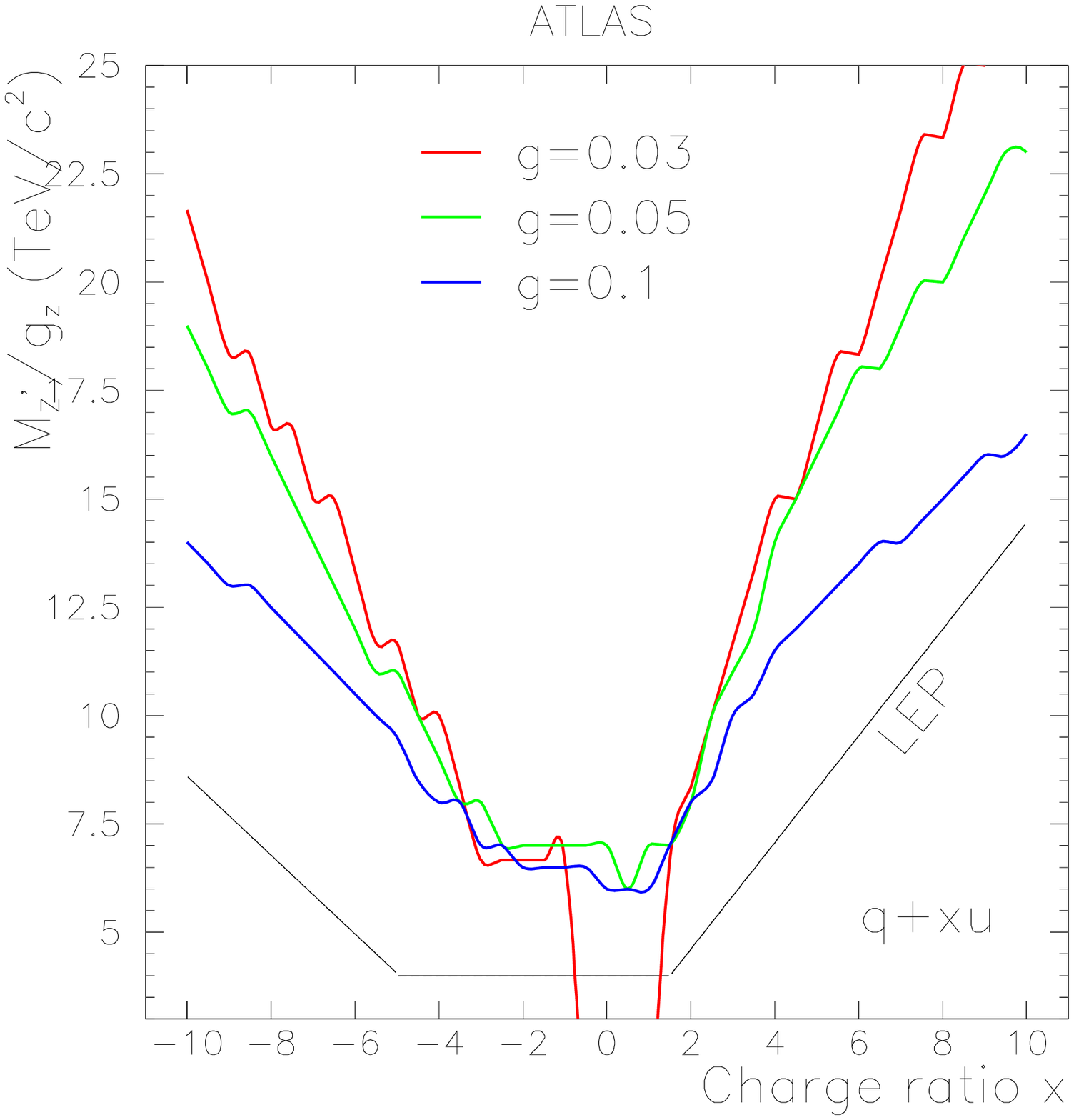,height=7.9cm}
    }
  \end{center}
  \caption{ATLAS discovery reach with an integrated luminosity of $400~pb^{-1}$ in the 4 classes of CDDT models.}
  \label{f_atlDisco100pb}
\end{figure}

\begin{figure}[t]
  \begin{center}
    \subfigure{
      \epsfig{file=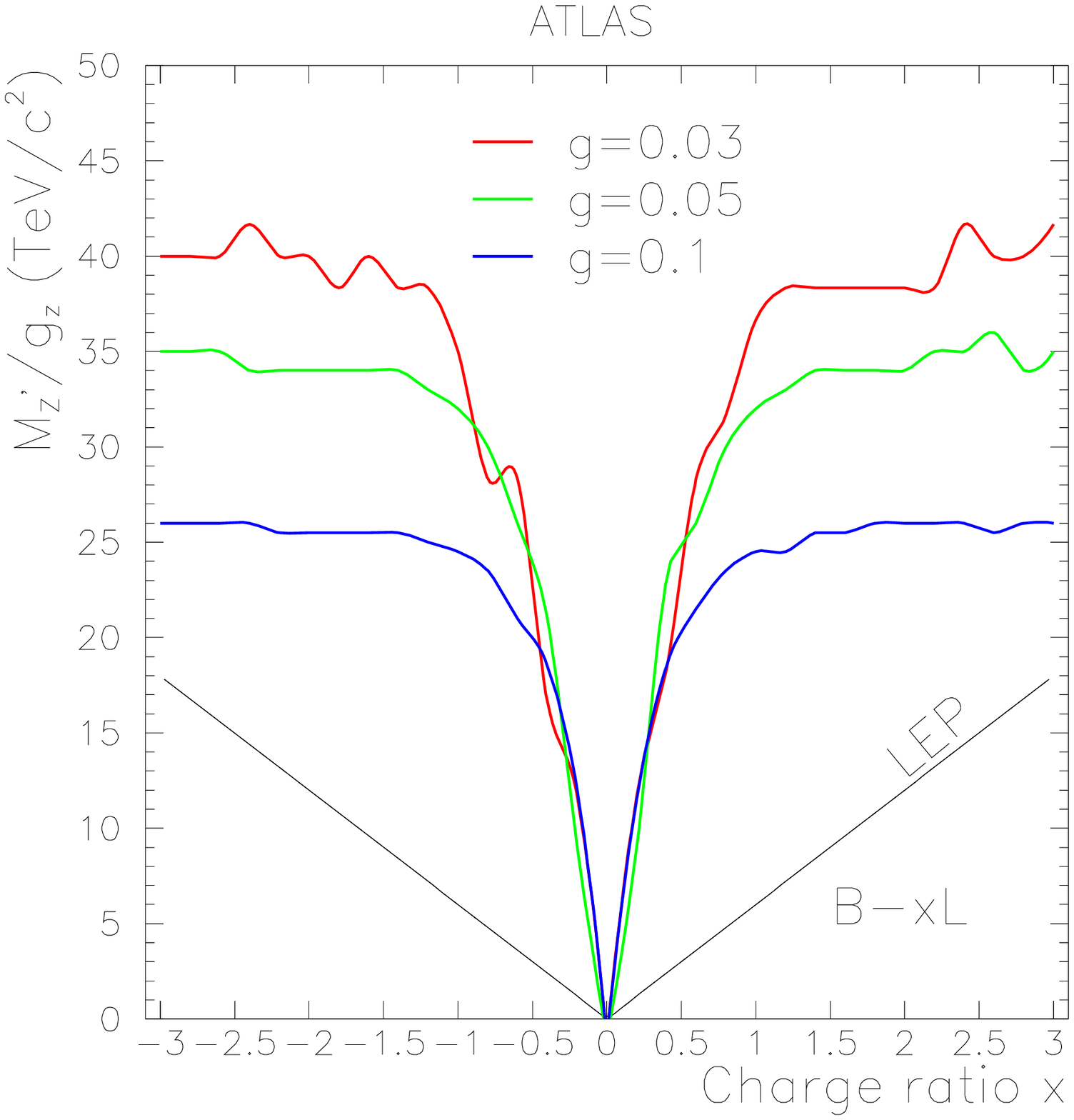,height=7.9cm}
    }
    \subfigure{
      \epsfig{file=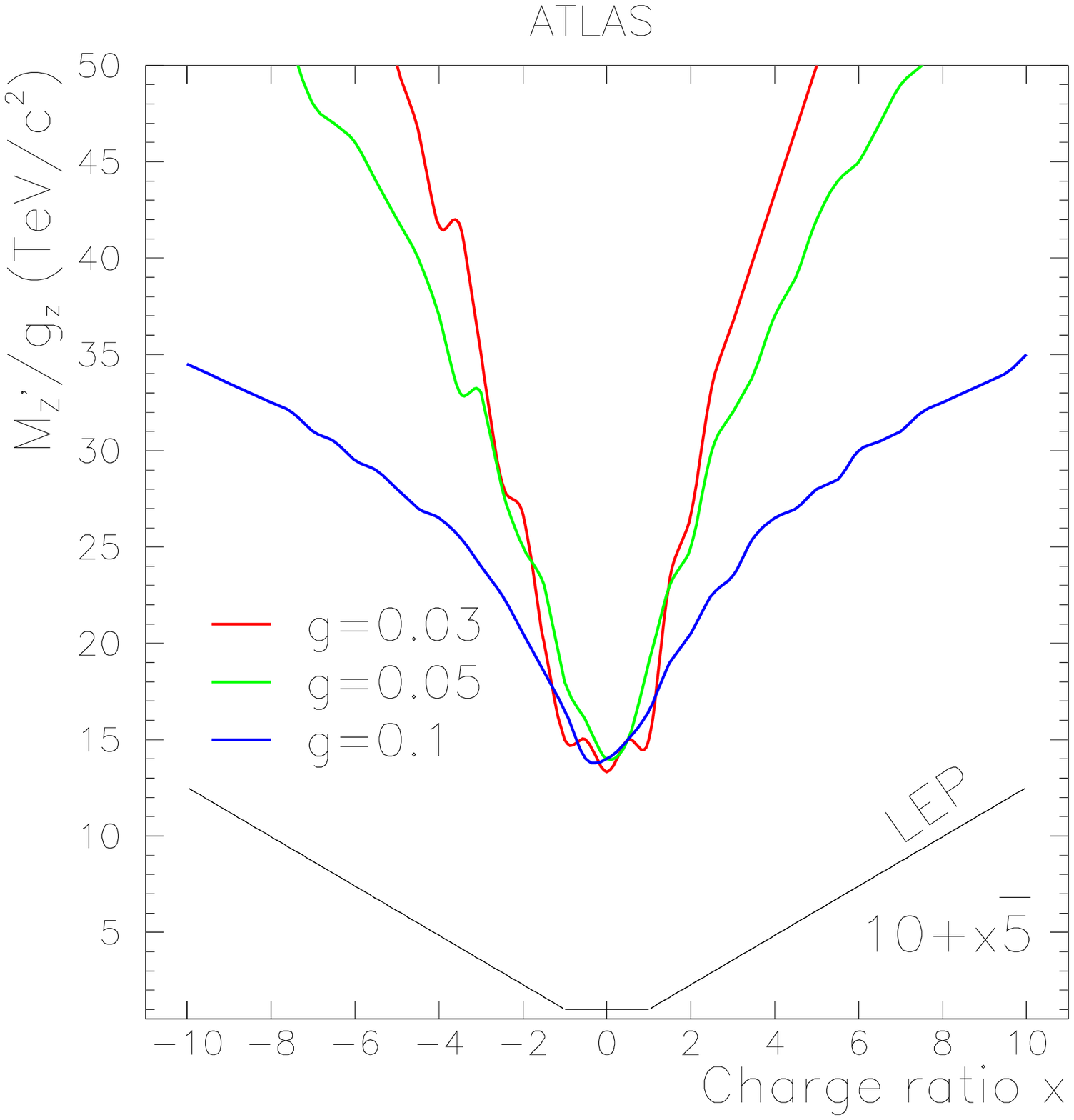,height=7.9cm}
    }
  \end{center}
  \begin{center}
    \subfigure{
      \epsfig{file=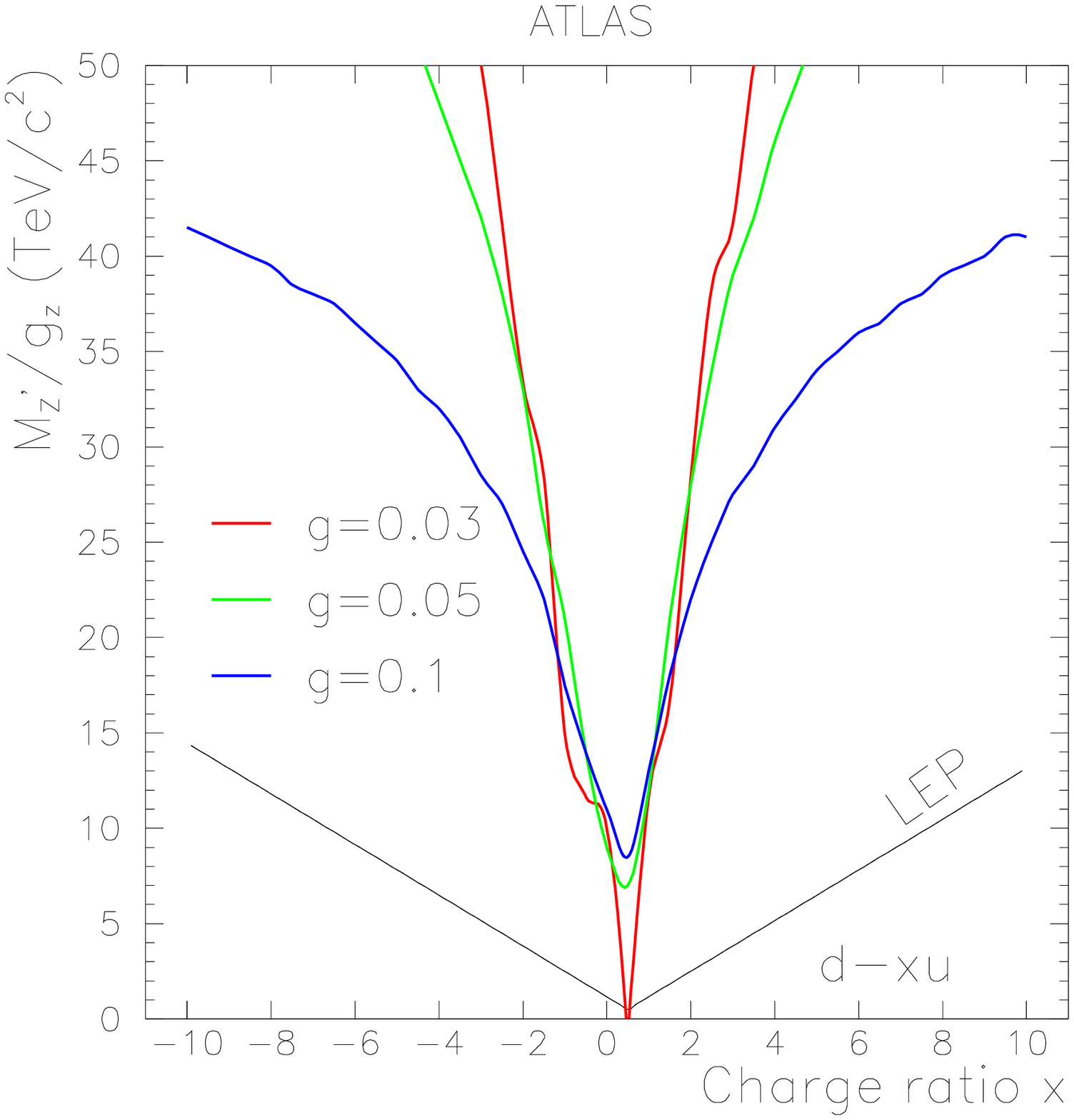,height=7.9cm}
    }
    \subfigure{
      \epsfig{file=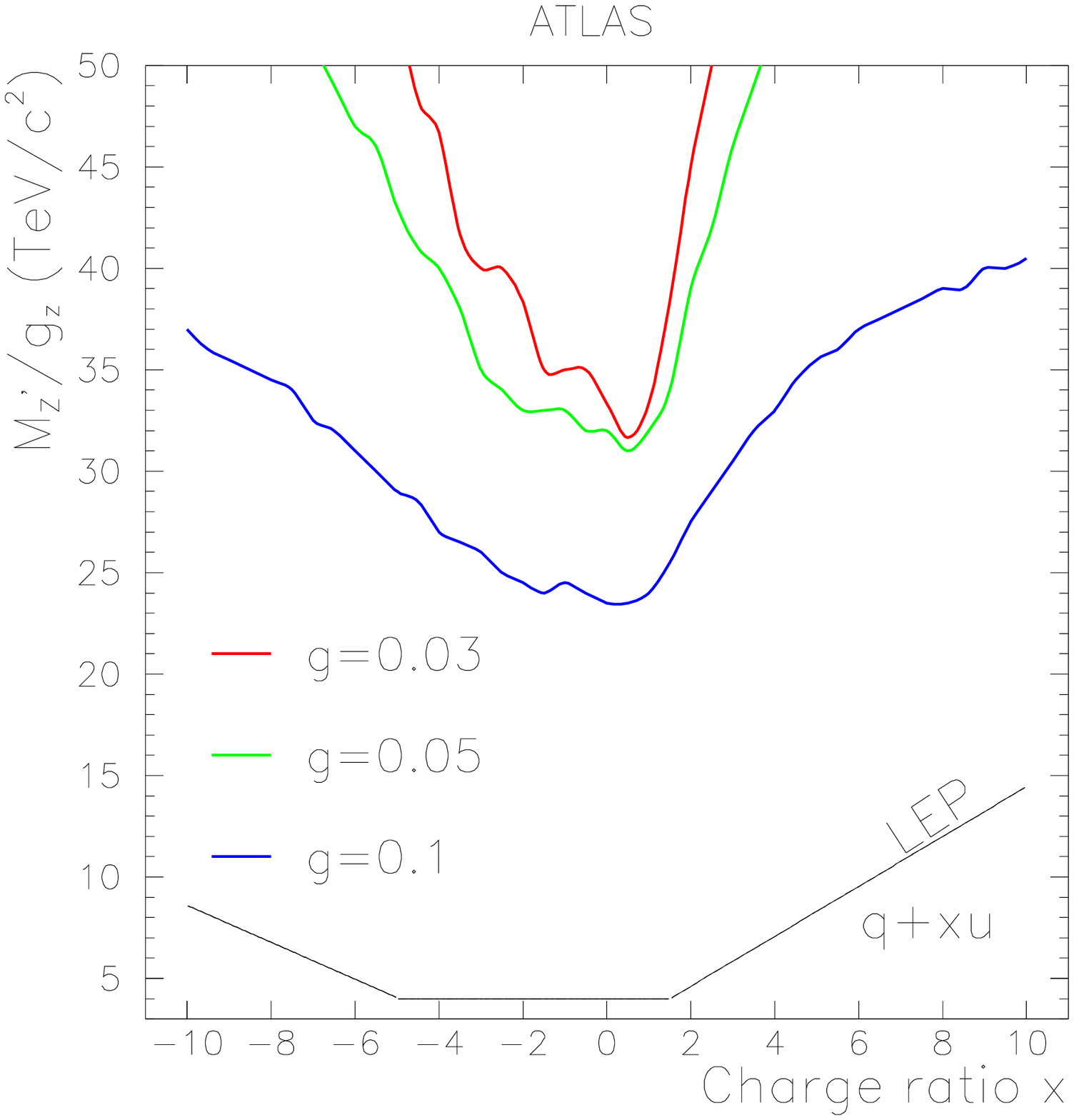,height=7.9cm}
    }
  \end{center}
  \caption{ATLAS discovery reach with an integrated luminosity of $100~fb^{-1}$ in the 4 classes of CDDT models.}
  \label{f_atlDisco100fb}
\end{figure}

The four classes of CDDT solutions were considered with three different values of the $g_{\zp}$ coupling strength, and a wide range of $x$ parameter values. The events generated by Pythia were efficiency weighted according to the incoming quark flavour and \zp mass, as explained in the previous section, in order to derive an effective production cross section. This procedure was also applied to the irreducible background.  Then a significance estimator, called $S_{12}$\cite{SignifSc12},  was used in order to extract the discovery reach in the $(x,g_{\zp}/M_{\zp})$ plane for several values of integrated luminosities. This estimator is defined by $S_{12} = \sqrt{S+B} - \sqrt{B}$ where $S$ (resp. $B$) is the expected number of signal (resp. background) events; this definition is supposed to be more realistic than the usual $S/\sqrt{B}$ or $S/\sqrt{S+B}$.\\
The results are presented on figures \ref{f_atlDisco100pb} and \ref{f_atlDisco100fb} for two different values of luminosities : $400~pb^{-1}$  and $100~fb^{-1}$. The ATLAS discovery reach goes beyond the LEP exclusion limits in most scenarios, already in the first months of LHC running ($400~pb^{-1}$ case); furthermore, with such a limited luminosity, it is also possible to probe regions of parameters space not yet excluded by CDF. The long term limits ($100~fb^{-1}$ case), as for them, illustrate the very promising LHC discovery potential which is, as expected, far beyond the ultimate TeVatron one. This would be even more striking when including expected analysis refinements, such as : optimization of the electron reconstruction, performing a bump hunt analysis instead of a basic counting method, including the forward backward asymmetry measurement as done by CDF, ...\\

\subsubsection{Conclusions.}

The ATLAS \zp discovery reach has been presented in the context of the CDDT parameterization, by taking into account the efficiency, derived from a detector full simulation, independently of the model. The potential was found to be very promising, even with a limited amount of data.


\clearpage\setcounter{equation}{0}\setcounter{figure}{0}\setcounter{table}{0}
\def\leqn#1{(\ref{#1})}
\def\({\left(}
\def\){\right)}
\def\stacksymbols #1#2#3#4{\def\theguybelow{#2}
    \def\vp{\lower#3pt}
    \def\sp{\baselineskip0pt\lineskip#4pt}
    \mathrel{\mathpalette\intermediary#1}}
\def\intermediary#1#2{\vp\vbox{\sp
     \everycr={}\tabskip0pt
     \halign{$\mathsurround0pt#1\hfil##\hfil$\crcr#2\crcr
              \theguybelow\crcr}}}
\def\gapproxeq{\stacksymbols{>}{\sim}{2.5}{.2}}
\def\lapproxeq{\stacksymbols{<}{\sim}{2.5}{.2}}

\def\Mz{M_{\rm Z}}
\def\Mw{M_{\rm W}}

\newcommand{\bspace}{\!\!\!\!}
\def\mini#1{\leavevmode\hbox{\tiny #1}}
\def\g#1{g_{\mini{#1}}}

\subsection{Phenomenology of Higgsless Models at the LHC and ILC}
\label{sec:higgsless}

A.~Birkedal$^1$, K.~Matchev$^2$ and M.~Perelstein$^3$ \\ [3mm]
{\em
$^1$ Santa Cruz Institute for Particle Physics\\ [2mm]
$^2$ Institute for Fundamental Theory, University of Florida\\ [2mm]
$^3$ Institute for High-Energy Phenomenology, Cornell University
}\\

{\em We investigate the signatures of the recently proposed Higgsless
models at future colliders.  We focus on tests of the mechanism of
partial unitarity restoration in the longitudinal vector boson
scattering, which do not depend on any Higgsless model-building
details.  We study the LHC discovery reach for charged massive vector
boson resonances and show that all of the preferred parameter space
will be probed with 100~fb$^{-1}$ of LHC data.  We also discuss the
prospects for experimental verification of the Higgsless nature of the
model at the LHC.  In addition, we present new results relevant for
the discovery potential of Higgsless models at the International
Linear Collider (ILC).}\\


One of the greatest unsolved mysteries of the Terascale is the origin
of electroweak symmetry breaking (EWSB).  Within the usual description
of the Standard Model (SM), a weakly coupled Higgs boson performs this
task.  However, it still has not been experimentally verified whether
the electroweak symmetry is broken by such a Higgs mechanism, strong
dynamics~\cite{NUPHA.B155.237,NUPHA.B155.237,PHRVA.D19.1277}, or something else.  This is one of the crucial
questions particle physicists hope to answer in the upcoming
experiments at the Large Hadron Collider (LHC) at CERN.

Experiments have already been able to put some constraints on
theoretical ideas about EWSB.  In theories involving EWSB by strong
dynamics, the scale $\Lambda$ at which new physics enters can be
guessed from the scale at which massive gauge boson scattering becomes
non-unitary.  A simple estimate gives a value of
\begin{equation}
\label{lambda}
\Lambda\sim4\pi M_W/g \sim 1.8~{\rm TeV},
\end{equation}
which is disfavored by precision electroweak constraints
(PEC)~\cite{PRLTA.65.964,PHRVA.D46.381}.  Thus, strong dynamics would seem to be largely ruled
out as the source of EWSB.  However, a new class of models, termed
``Higgsless''~\cite{KK1,KK2,Nomura,KK3}, have been able to raise the
scale of strong dynamics, allowing agreement with 
PEC~\cite{HEP-PH/0409126,HEP-PH/0409266,HEP-PH/0504240,HEP-PH/0505001}.

Realistic Higgsless models contain new TeV-scale weakly coupled states
accessible at the LHC.  Among those, new massive vector bosons (MVB),
heavy cousins of the $W$, $Z$ and $\gamma$ of the SM, which are of
primary interest.  It is those states which delay unitarity violation
and hence allow the scale $\Lambda$ to be
raised~\cite{Chivukula:2001hz}.  Unfortunately, the details of the
fermion sector of the theory are highly model-dependent.  For
instance, early Higgsless models did not allow sufficient change in
$\Lambda$ to agree with 
PEC~\cite{HEP-PH/0310285,HEP-PH/0312247,HEP-PH/0401160,DHLR,HEP-PH/0403300,HEP-PH/0405040,HEP-PH/0407059}, and
modifications of the fermion sector were necessary.  However, the
basic mechanism by which $\Lambda$ is raised is identical in all
``Higgsless'' models, even regardless of the number of underlying
dimensions~\cite{HEP-PH/0312324,HEP-PH/0405188,HEP-PH/0406077,HEP-PH/0408067,HEP-PH/0408072,HEP-PH/0508014}.  It is this mechanism which was studied
in~\cite{us}, focusing on its collider signatures.  We review the
analysis of Ref.~\cite{us} and present some new results relevant for
the International Linear Collider (ILC).  In Sec.~\ref{sec:sumrules}
we derive a set of sum rules which should be obeyed by the couplings
between the new MVBs and the SM $W/Z$ gauge bosons.  We identify
discovery signatures of the new MVBs at the LHC which rely only on the
couplings guaranteed by sum rules, and compare to the SM Higgs search
signals.  In Sec.~\ref{sec:lhc} we discuss the LHC reach for charged
MVBs and methods for testing the sum rules of Sec.~\ref{sec:sumrules}
in order to identify the ``Higgsless'' origin of the MVB resonances.
In Sec.~\ref{sec:ilc} we discuss the corresponding Higgsless
phenomenology at the ILC.


\subsubsection{Unitarity sum rules}
\label{sec:sumrules}

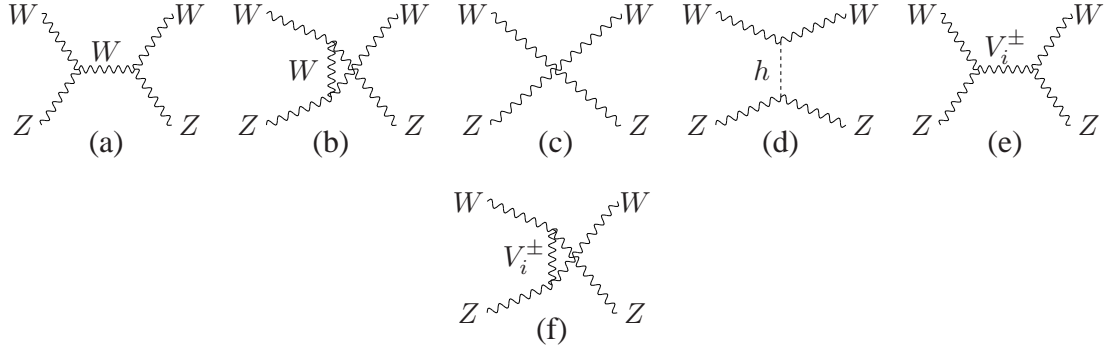
\begin{figure*}[t!]
\begin{center} 
{
\unitlength=0.7 pt
\SetScale{0.7}
\SetWidth{0.5}      
\normalsize    
{} \allowbreak
\begin{picture}(100,100)(0,0)
\Photon(15,80)(35,50){2}{6}
\Photon(15,20)(35,50){2}{6}
\Photon(35,50)(65,50){2}{6}
\Photon(65,50)(85,80){2}{6}
\Photon(65,50)(85,20){2}{6}
\Text(50,10)[c]{(a)}
\Text(50,60)[c]{\small $W$}
\Text( 5,80)[c]{\small $W$}
\Text( 5,20)[c]{\small $Z$}
\Text(95,80)[c]{\small $W$}
\Text(95,20)[c]{\small $Z$}
\end{picture}
\quad
\begin{picture}(100,100)(0,0)
\Photon(15,80)(50,65){2}{6}
\Photon(15,20)(50,35){2}{6}
\Photon(50,65)(50,35){2}{6}
\Photon(50,65)(85,20){2}{9}
\Photon(50,35)(85,80){2}{9}
\Text(50,10)[c]{(b)}
\Text(35,50)[c]{\small $W$}
\Text( 5,80)[c]{\small $W$}
\Text( 5,20)[c]{\small $Z$}
\Text(95,80)[c]{\small $W$}
\Text(95,20)[c]{\small $Z$}
\end{picture}\
\quad
\begin{picture}(100,100)(0,0)
\Photon(15,80)(85,20){2}{13}
\Photon(15,20)(85,80){2}{13}
\Text(50,10)[c]{(c)}
\Text( 5,80)[c]{\small $W$}
\Text( 5,20)[c]{\small $Z$}
\Text(95,80)[c]{\small $W$}
\Text(95,20)[c]{\small $Z$}
\end{picture}\
\quad
\begin{picture}(100,100)(0,0)
\Photon(15,80)(50,65){2}{6}
\Photon(15,20)(50,35){2}{6}
\DashLine(50,65)(50,35){2}
\Photon(50,65)(85,80){2}{6}
\Photon(50,35)(85,20){2}{6}
\Text(50,10)[c]{(d)}
\Text(40,50)[c]{\small $h$}
\Text( 5,80)[c]{\small $W$}
\Text( 5,20)[c]{\small $Z$}
\Text(95,80)[c]{\small $W$}
\Text(95,20)[c]{\small $Z$}
\end{picture}\
\quad
\begin{picture}(100,100)(0,0)
\Photon(15,80)(35,50){2}{6}
\Photon(15,20)(35,50){2}{6}
\Photon(35,50)(65,50){2}{6}
\Photon(65,50)(85,80){2}{6}
\Photon(65,50)(85,20){2}{6}
\Text(50,10)[c]{(e)}
\Text(50,62)[c]{\small $V_i^\pm$}
\Text( 5,80)[c]{\small $W$}
\Text( 5,20)[c]{\small $Z$}
\Text(95,80)[c]{\small $W$}
\Text(95,20)[c]{\small $Z$}
\end{picture}\
\quad
\begin{picture}(100,100)(0,0)
\Photon(15,80)(50,65){2}{6}
\Photon(15,20)(50,35){2}{6}
\Photon(50,65)(50,35){2}{6}
\Photon(50,65)(85,20){2}{9}
\Photon(50,35)(85,80){2}{9}
\Text(50,10)[c]{(f)}
\Text(35,50)[c]{\small $V_i^\pm$}
\Text( 5,80)[c]{\small $W$}
\Text( 5,20)[c]{\small $Z$}
\Text(95,80)[c]{\small $W$}
\Text(95,20)[c]{\small $Z$}
\end{picture}
}
\caption{Diagrams contributing to the $W^\pm Z\to W^\pm Z$ 
scattering process: (a), (b) and (c) appear both in the SM and in
Higgsless models, (d) appears only in the SM, while (e) and (f) appear
only in Higgsless models.}
\label{wzwz}
\end{center}
\end{figure*}

Consider the elastic scattering process $W^\pm_L Z_L\to W^\pm_L Z_L$.
In the absence of the Higgs boson, this process receives contributions
from the three Feynman diagrams shown in Figs.~\ref{wzwz}(a)--(c).
The resulting amplitude contains terms which grow with the energy $E$
of the incoming particle as $E^4$ and $E^2$, ultimately causing
unitarity violation at high energy.  In the SM, both of these terms
are precisely cancelled by the contribution of the Higgs exchange
diagram in Fig.~\ref{wzwz}(d).  In Higgsless theories, the diagram of
Fig.~\ref{wzwz}(d) is absent, and the process instead receives
additional contributions from the diagrams in Figs.~\ref{wzwz}(e) and
\ref{wzwz}(f), where $V_i^\pm$ denotes the charged MVB of mass
$M_i^\pm$.  The index $i$ corresponds to the KK level of the state in
the case of a 5D theory, or labels the mass eigenstates in the case of
a 4D deconstructed theory.  Remarkably, the $E^4$ and $E^2$ terms can
again be exactly cancelled by the contribution of the MVBs, provided
that the following sum rules are satisfied~\cite{us}:
\begin{eqnarray}
\label{sumW} 
\g{WWZZ} &=& \g{WWZ}^2 \,+\, \sum_i (\g{WZV}^{(i)})^2, \\
2(\g{WWZZ}-\g{WWZ}^2)(\Mw^2+\Mz^2) + \g{WWZ}^2\,\frac{\Mz^4}{\Mw^2} 
&=& \sum_i (\g{WZV}^{(i)})^2
\left[3(M^\pm_i)^2-\frac{(\Mz^2-\Mw^2)^2}{(M^\pm_i)^2}\right]. 
\nonumber
\end{eqnarray}
Here $M_W$($M_Z$) is the $W(Z)$-boson mass and the notation for triple
and quartic gauge boson couplings is self-explanatory.  In 5D
theories, these equations are satisfied exactly if all the KK states,
$i=1\ldots\infty$, are taken into account.  This is not an accident,
but a consequence of the gauge symmetry and locality of the underlying
theory.  While this is not sufficient to ensure unitarity at all
energies (the increasing number of inelastic channels ultimately
results in unitarity violation), the strong coupling scale can be
significantly higher than the naive estimate~\leqn{lambda}.  For
example, in warped-space Higgsless models~\cite{KK2,HEP-PH/0409126,HEP-PH/0409266,HEP-PH/0504240,HEP-PH/0505001} unitarity
is violated at the scale~\cite{Pap}
\begin{equation}
\label{NDA}
\Lambda_{\rm NDA} \sim \frac{3\pi^4}{g^2}\frac{\Mw^2}{M^\pm_1},
\end{equation}
which is typically of order 5--10 TeV.  In 4D models, the number of
MVBs is finite and the second of the sum rules~\leqn{sumW} is
satisfied only approximately; however, our numerical study of sample
models indicates that the sum rule violation has to be very small, at
the level of $1\%$, to achieve an adequate improvement in $\Lambda$.

\begin{figure*}[t]
\begin{center} 
{
\unitlength=0.7 pt
\SetScale{0.7}
\SetWidth{0.7}      
\footnotesize    
{} 
\allowbreak
%
\begin{picture}(100,100)(0,0)
\Photon(15,80)(50,65){2}{6}
\Photon(15,20)(50,35){2}{6}
\Photon(50,65)(50,35){2}{6}
\Photon(50,65)(85,80){2}{6}
\Photon(50,35)(85,20){2}{6}
\Text(50,10)[c]{(a)}
\Text(35,50)[c]{\small $W$}
\Text( 5,80)[c]{\small $W$}
\Text( 5,20)[c]{\small $W$}
\Text(95,80)[c]{\small $Z$}
\Text(95,20)[c]{\small $Z$}
\end{picture}\
\quad
\begin{picture}(100,100)(0,0)
\Photon(15,80)(50,65){2}{6}
\Photon(15,20)(50,35){2}{6}
\Photon(50,65)(50,35){2}{6}
\Photon(50,65)(85,20){2}{9}
\Photon(50,35)(85,80){2}{9}
\Text(50,10)[c]{(b)}
\Text(35,50)[c]{\small $W$}
\Text( 5,80)[c]{\small $W$}
\Text( 5,20)[c]{\small $W$}
\Text(95,80)[c]{\small $Z$}
\Text(95,20)[c]{\small $Z$}
\end{picture}\
\quad
\begin{picture}(100,100)(0,0)
\Photon(15,80)(85,20){2}{13}
\Photon(15,20)(85,80){2}{13}
\Text(50,10)[c]{(c)}
\Text( 5,80)[c]{\small $W$}
\Text( 5,20)[c]{\small $W$}
\Text(95,80)[c]{\small $Z$}
\Text(95,20)[c]{\small $Z$}
\end{picture}\
\quad
%
\begin{picture}(100,100)(0,0)
\Photon(15,80)(35,50){2}{6}
\Photon(15,20)(35,50){2}{6}
\DashLine(35,50)(65,50){2}
\Photon(65,50)(85,80){2}{6}
\Photon(65,50)(85,20){2}{6}
\Text(50,10)[c]{(d)}
\Text(50,62)[c]{\small $h$}
\Text( 5,80)[c]{\small $W$}
\Text( 5,20)[c]{\small $W$}
\Text(95,80)[c]{\small $Z$}
\Text(95,20)[c]{\small $Z$}
\end{picture}\
\quad
%
\begin{picture}(100,100)(0,0)
\Photon(15,80)(50,65){2}{6}
\Photon(15,20)(50,35){2}{6}
\Photon(50,65)(50,35){2}{6}
\Photon(50,65)(85,80){2}{6}
\Photon(50,35)(85,20){2}{6}
\Text(50,10)[c]{(e)}
\Text(35,50)[c]{\small $V^\pm$}
\Text( 5,80)[c]{\small $W$}
\Text( 5,20)[c]{\small $W$}
\Text(95,80)[c]{\small $Z$}
\Text(95,20)[c]{\small $Z$}
\end{picture}\
\quad
%
\begin{picture}(100,100)(0,0)
\Photon(15,80)(50,65){2}{6}
\Photon(15,20)(50,35){2}{6}
\Photon(50,65)(50,35){2}{6}
\Photon(50,35)(85,80){2}{9}
\Photon(50,65)(85,20){2}{9}
\Text(50,10)[c]{(f)}
\Text(35,50)[c]{\small $V^\pm$}
\Text( 5,80)[c]{\small $W$}
\Text( 5,20)[c]{\small $W$}
\Text(95,80)[c]{\small $Z$}
\Text(95,20)[c]{\small $Z$}
\end{picture}\
}
\caption{Diagrams contributing to the $W^\pm W^\mp\to ZZ$ scattering 
process: (a), (b) and (c) appear both in the SM and in Higgsless
models, (d) appears only in the SM, and (e) and (f) appear only in
Higgsless models.}
\label{wwzz}
\end{center}
\end{figure*}
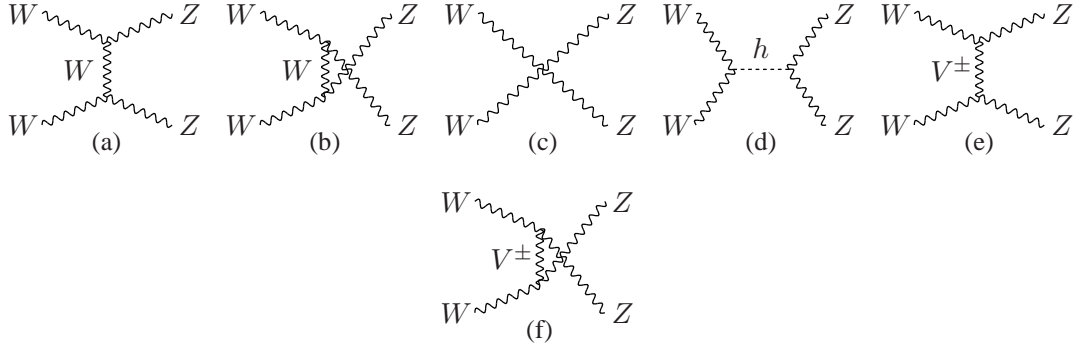

Considering the $W^+_LW^-_L\to W^+_LW^-_L$ scattering process yields
sum rules constraining the couplings of the neutral MVBs $V_i^0$ (with
masses denoted by $M^0_i$)~\cite{KK1}:
\begin{eqnarray}
\label{sumZ} 
\g{WWWW} &=& \g{WWZ}^2 + \g{WW$\gamma$}^2 \,+\, \sum_i (\g{WWV}^{(i)})^2,
\\ \nonumber
4\g{WWWW}\,\Mw^2 &=& 3\,\left[\g{WWZ}^2 \Mz^2 + \sum_i (\g{WWV}^{(i)})^2
\,(M^0_i)^2\right]\, .
\end{eqnarray}
Considering other channels such as $W_L^+W_L^-\to ZZ$ (see
Fig.~\ref{wwzz}) and $ZZ\to ZZ$ does not yield any new sum rules.  The
presence of multiple MVBs, whose couplings obey
Eqs.~\leqn{sumW},~\leqn{sumZ}, is a generic prediction of Higgsless
models.

Our study of collider phenomenology in Higgsless models will focus on
vector boson fusion processes.  These processes are attractive for two
reasons.  Firstly, the production of MVBs via vector boson fusion is
relatively model-independent, since the couplings are constrained by
the sum rules~\leqn{sumW},~\leqn{sumZ}.  This is in sharp contrast
with the Drell-Yan production mechanism~\cite{DHLR}, which dominates
for the conventional $W^\prime$ and $Z^\prime$ bosons but is likely to
be suppressed for the Higgsless MVBs due to their small couplings to
fermions, as needed to evade PEC~\cite{HEP-PH/0409126,HEP-PH/0409266,HEP-PH/0504240,HEP-PH/0505001}.  In the following,
unless specified otherwise, we shall assume that the MVBs have no
appreciable couplings to SM fermions.  Secondly, if enough couplings
and masses can be measured, these processes can provide a {\em test}
of the sum rules, probing the mechanism of partial unitarity
restoration.

Eq.~\leqn{NDA} indicates that the first MVB should appear below $\sim
1$ TeV, and thus be accessible at the LHC.  For $V^\pm_1$, the sum
rules~\leqn{sumW} imply an inequality
\begin{equation}
\label{bound}
\g{WZV}^{(1)}  
\lapproxeq \frac{\g{WWZ}\Mz^2}{\sqrt{3}M^\pm_1\Mw}.
\end{equation}
This bound is quite stringent: $\g{WZV}^{(1)}\lapproxeq0.04$ for
$M^\pm_1=700$ GeV.  Also, sum rule~\leqn{sumW} convergence requires
$\g{WZV}^{(k)}\,\propto\,k^{-1/2}\,(M^\pm_k)^{-1}$.  The combination
of heavier masses and lower couplings means that the heavier MVBs may
well be unobservable, so that only the $V_1$ states can be studied.
The "saturation limit", in which there is only a single set of MVBs
whose couplings saturate the sum rules, is likely to provide a good
approximation to the phenomenology of the realistic Higgsless models.
In this limit, the partial width of the $V^\pm_1$ is given by
\begin{equation}
\Gamma(V^\pm_1\to W^\pm Z)\approx \frac{\alpha\ (M^\pm_1)^3}{144\, 
\sin^2\theta_W\, M_W^2}\, .
\label{width}
\end{equation}

\begin{figure*}[t]
\begin{center}
\includegraphics[width=100mm]{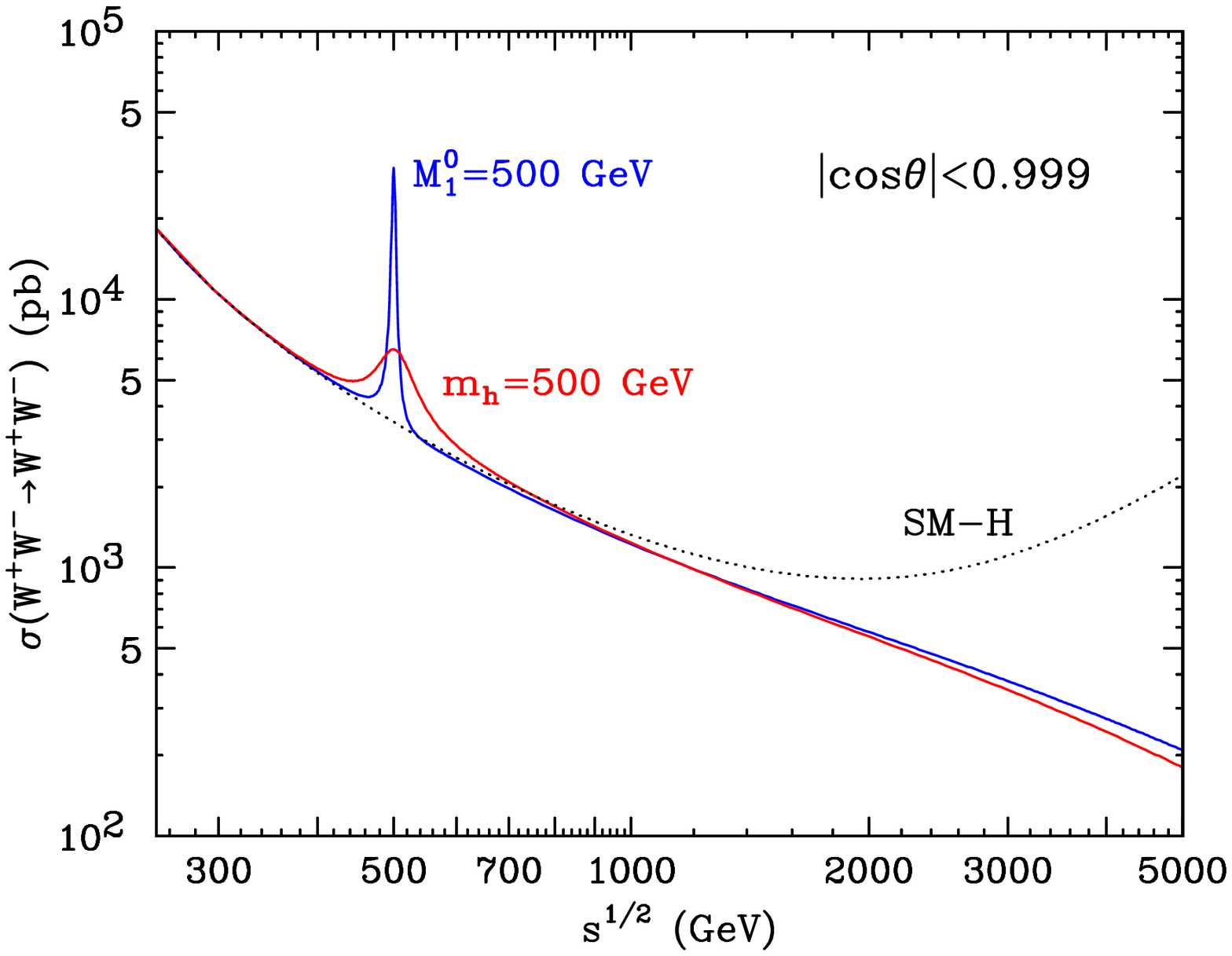}
\includegraphics[width=100mm]{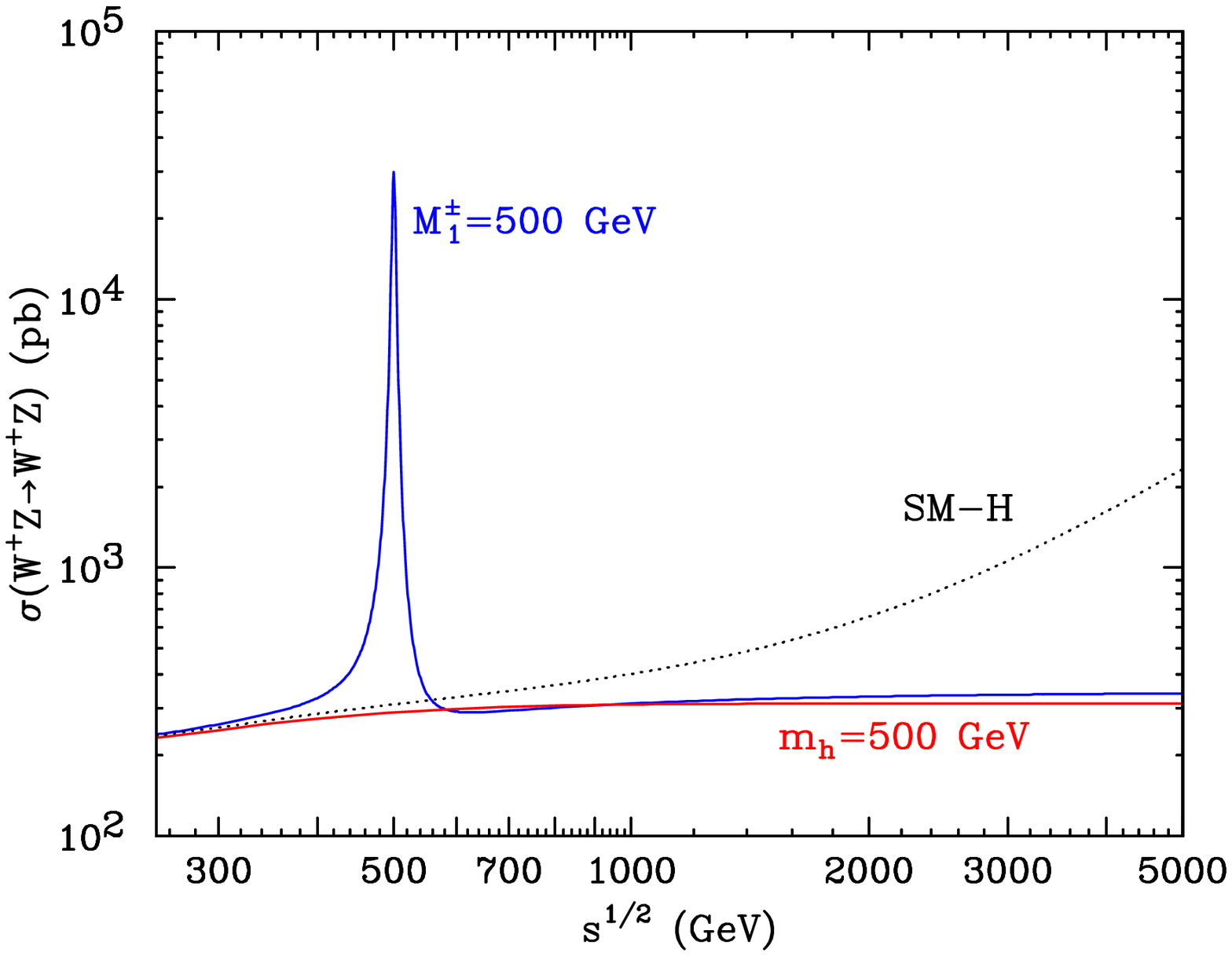}
\caption{Elastic scattering cross-sections for $WW\to WW$ (left) and
$WZ\to WZ$ (right) in the SM without a Higgs boson (SM-H) (dotted),
the SM with a 500~GeV Higgs boson (red) and the Higgsless model with a
500~GeV MVB (blue).}
\label{fig:parton}
\end{center}
\end{figure*}

Given the couplings of the MVBs to the SM $W$ and $Z$, we can now
predict (at the parton level) the size of the new physics signals in
the various channels of vector boson fusion.  Fig.~\ref{fig:parton}
provides an illustration for the case of $WW\to WW$ and $WZ\to WZ$.
We show the expected signal for either a SM Higgs boson of mass
$m_h=500$ GeV, or the corresponding MVB $V_1$ of mass $500$ GeV in the
saturation limit.  The sum rules (\ref{sumZ}) govern the signal in the
$WW\to WW$ channel shown in the left panel of Fig.~\ref{fig:parton}.
However, the $WW$ final state is difficult to observe over the SM
backgrounds at the LHC: in the dilepton channel there is no resonance
structure, while the jetty channels suffer from large QCD backgrounds.
It is therefore rather challenging to test the sum rules (\ref{sumZ}).
Notice that even if a $WW$ resonance is observed, without a test of
the sum rules (\ref{sumZ}), its interpretation is unclear, since the
SM Higgs boson is {\em also} expected to appear as a $WW$ resonance
(see the left panel in Fig.~\ref{fig:parton}).

We shall therefore concentrate on the $WZ\to WZ$ channel, in which the
Higgsless model predicts a series of resonances as in
Fig.~\ref{wzwz}(e), while in the SM the amplitude is unitarized by the
$t$-channel diagram of Fig.~\ref{wzwz}(d) and has no resonance (see
the right panel in Fig.~\ref{fig:parton}).  Conventional theories of
EWSB by strong dynamics may also contain a resonance in this channel,
but it is likely to be heavy ($\sim 2$ TeV for QCD-like theories) and
broad due to strong coupling.  In contrast, the MVB resonance is very
narrow, as can be seen from Fig.~\ref{fig:parton} and
Eq.~(\ref{width}).  In fact it is almost a factor of 20 narrower than
a SM Higgs boson of the same mass.  This is primarily due to the
vector nature of the MVB and our assumption that it only has a single
decay channel.  We therefore conclude that a resonance in the $WZ\to
WZ$ channel would be a smoking gun for the Higgsless model (for
alternative interpretations involving extended Higgs sectors,
see~\cite{Asakawa:2005gv} and references therein).  Finally, the
$WW\to ZZ$ channel is a good discriminator as well, since it will
exhibit a resonance for the case of the SM but not the Higgsless
models (see Fig.~\ref{wwzz}).  A comparison of the resonant structure
of the three vector boson fusion final states is shown in
Table~\ref{tab:res}.

\begin{table}[htb]
\begin{center}
\caption{Comparison of the resonance structure of the SM and Higgsless models 
in different vector boson fusion channels.}
\vspace{5mm}
\begin{tabular}{||c||c|c|c||}
\hline\hline
Model
& $WW\to WW$
& $WZ\to WZ$
& $WW\to ZZ$ \\ 
\hline\hline
SM
& Yes
& No
& Yes \\ 
\hline
Higgsless
& Yes
& Yes
& No  \\ 
\hline\hline
\end{tabular}
\label{tab:res}
\end{center}
\end{table}


\subsubsection{Collider phenomenology at the LHC}
\label{sec:lhc}

\begin{figure*}[t]
\centering
\includegraphics[width=100mm]{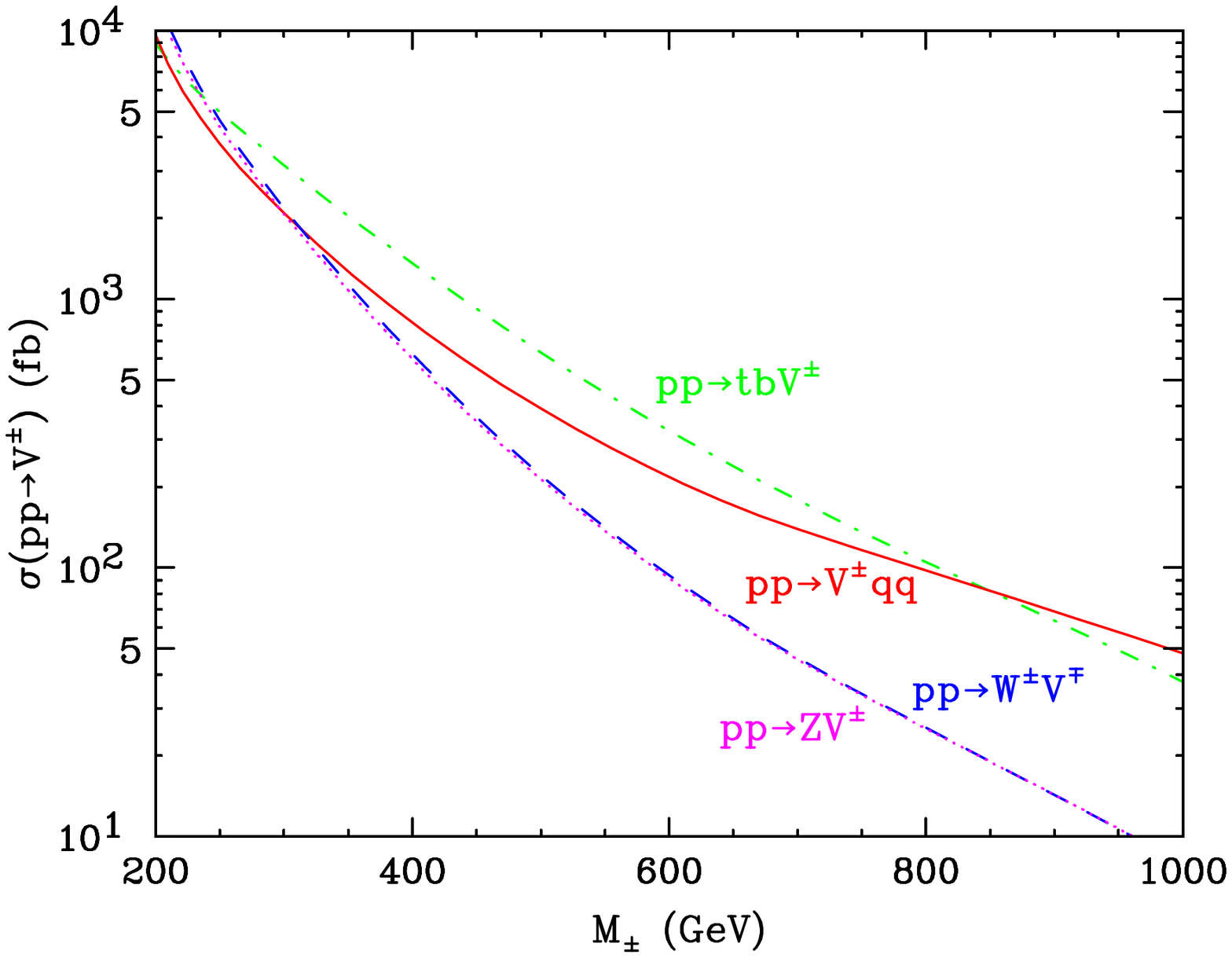}
\includegraphics[width=100mm]{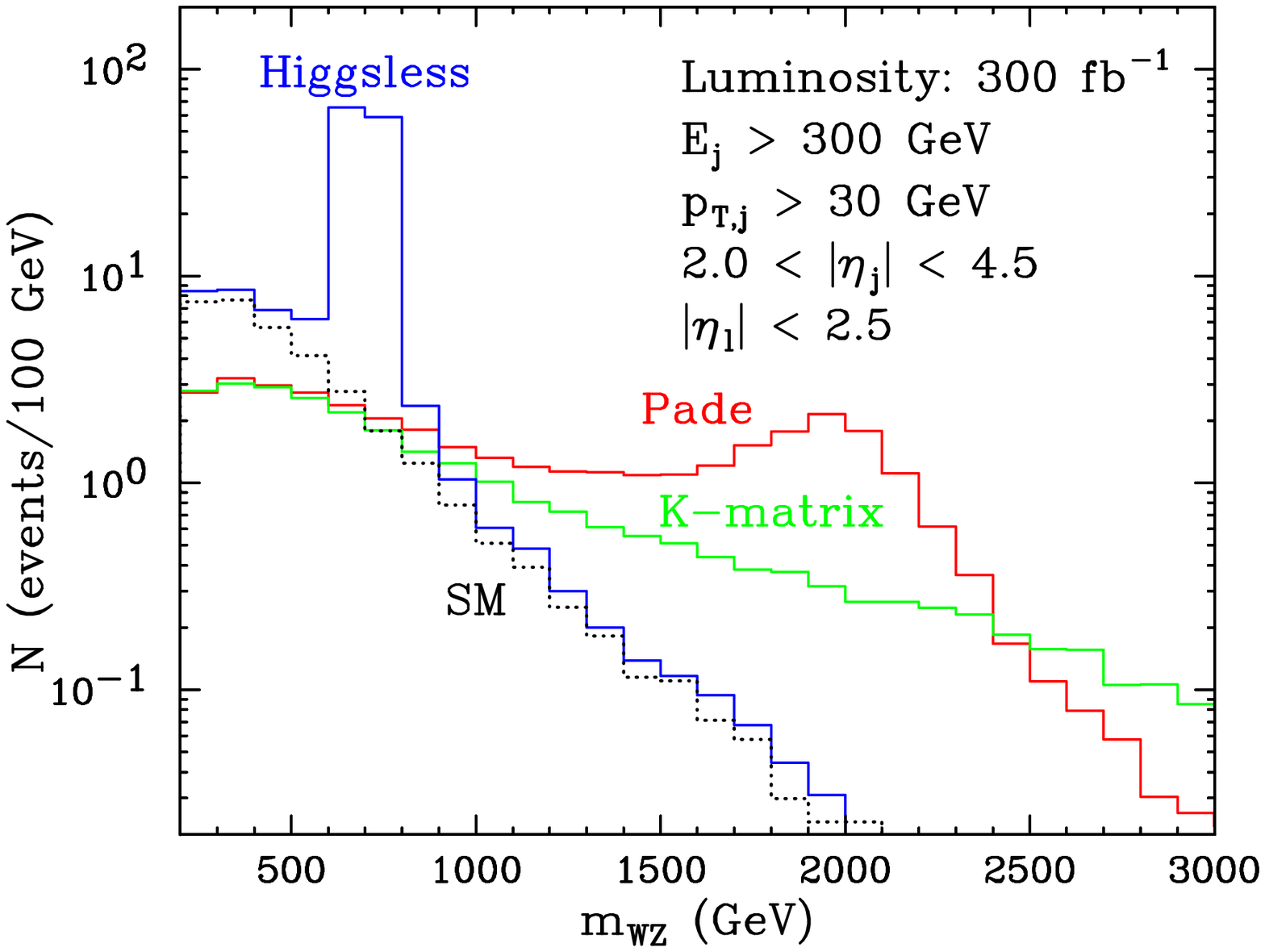}
\caption{Left: Production cross-sections for $V^\pm$ at the LHC. 
Here $tbV^\pm$ production assumes SM-like couplings to third
generation quarks.  Right: The number of events per 100~GeV bin in the
$2j+3\ell+\nu$ channel at the LHC with an integrated luminosity of
300~fb$^{-1}$ and cuts as indicated in the figure.  Results are shown
for the SM (dotted), the Higgsless model with $M^\pm_1=700$~GeV
(blue), and two "unitarization" models: Pad\'e (red) and K-matrix
(green)~\protect\cite{ZEPYA.C50.205,ZEPYA.C50.465}.} 
\label{fig:lhc}
\end{figure*}

At the LHC, vector boson fusion processes will occur as a result of
$W/Z$ bremsstrahlung off quarks.  The typical final state for such
events includes two forward jets in addition to a pair of gauge
bosons.  The production cross section of $V^\pm_1$ in association with
two jets is shown by the solid line in the left panel of
Fig.~\ref{fig:lhc}.  To estimate the prospects for the charged MVB
search at the LHC, we require that both jets be observable (we assume
jet rapidity coverage of $|\eta|\leq 4.5$), and impose the following
lower cuts on the jet rapidity, energy, and transverse momentum:
$|\eta|>2$, $E>300$~GeV, $p_T>30$~GeV.  These requirements enhance the
contribution of the vector boson fusion diagrams relative to the
irreducible background of the non-fusion $q\bar{q}'\to WZ$ SM process
as well as Drell-Yan $q\bar{q}'\to V_1^\pm$.  The ``gold-plated''
final state~\cite{HEP-PH/9306256,HEP-PH/9504426,HEP-PH/0412203} for this search is $2j+3\ell$+\met,
with the additional kinematic requirement that two of the leptons have
to be consistent with a $Z$ decay.  We assume lepton rapidity coverage
of $|\eta|<2.5$.  The $WZ$ invariant mass, $m_{WZ}$, can be
reconstructed using the missing transverse energy measurement and
requiring that the neutrino and the odd lepton form a $W$.  The number
of "gold-plated" events (including all lepton sign combinations) in a
300~fb$^{-1}$ LHC data sample, as a function of $m_{WZ}$, is shown in
Fig.~\ref{fig:lhc} for the SM (dotted), Higgsless model with
$M^\pm_1=700$ GeV (blue), and two "unitarization" models: Pad\'e (red)
and K-matrix (green)~\cite{ZEPYA.C50.205,ZEPYA.C50.465} (for details, see Ref.~\cite{us}).  A
Higgsless model can be easily identified by observing the MVB
resonance: for the chosen parameters, the dataset contains 130
$V_1^\pm\to W^\pm Z \to 3\ell+\nu$ events.  The irreducible non-fusion
SM background is effectively suppressed by the cuts: the entire
dataset shown in Fig.~\ref{fig:lhc} contains only $6$ such events.  We
therefore estimate the discovery reach for $V^\pm_1$ resonance by
requiring 10 signal events after cuts.  The efficiency of the cuts for
$500\le M^\pm_1\le 3$~TeV is in the range $20-25\%$.  We then find
that with 10~fb$^{-1}$ of data, corresponding to 1 year of running at
low luminosity, the LHC will probe the Higgsless models up to
$M^{\pm}_1\lapproxeq 550$~GeV, while covering the whole preferred
range up to $M^{\pm}_1=1$~TeV requires 60~fb$^{-1}$.  Note, however,
that one should expect a certain amount of reducible background with
fake and/or non-isolated leptons.

Once the $V^\pm_1$ resonance is discovered, identifying it as part of
a Higgsless model would require testing the sum rules (\ref{sumW}) by
measuring its mass $M^\pm_1$ and coupling $g^{(1)}_{WZV}$.  The
coupling can be determined from the total $V^\pm_1$ production cross
section $\sigma_{\rm tot}$.  However, we are observing the $V^\pm_1$
resonance in an exclusive channel, which only yields the product
$\sigma_{\rm tot}\,BR(V^\pm_1\to W^\pm Z)$.  A measurement of the
total resonance width $\Gamma(V^\pm_1\to{\rm anything})$ would remove
the dependence on the unknown branching fraction $BR$.  The accuracy
of this measurement is severely limited by the poor missing energy
resolution.  Even though a Higgsless origin of the resonance can be
ruled out if the value of $g^{(1)}_{WZV}$, inferred with the
assumption of $BR=1$, violates the bound (\ref{bound}), the LHC alone
will not be able to settle the issue and precise measurements at an
ILC appear to be necessary for the ultimate test of the theory.


\subsubsection{Collider phenomenology at an ILC}
\label{sec:ilc}

\begin{figure*}[t]
\centering
\includegraphics[width=100mm]{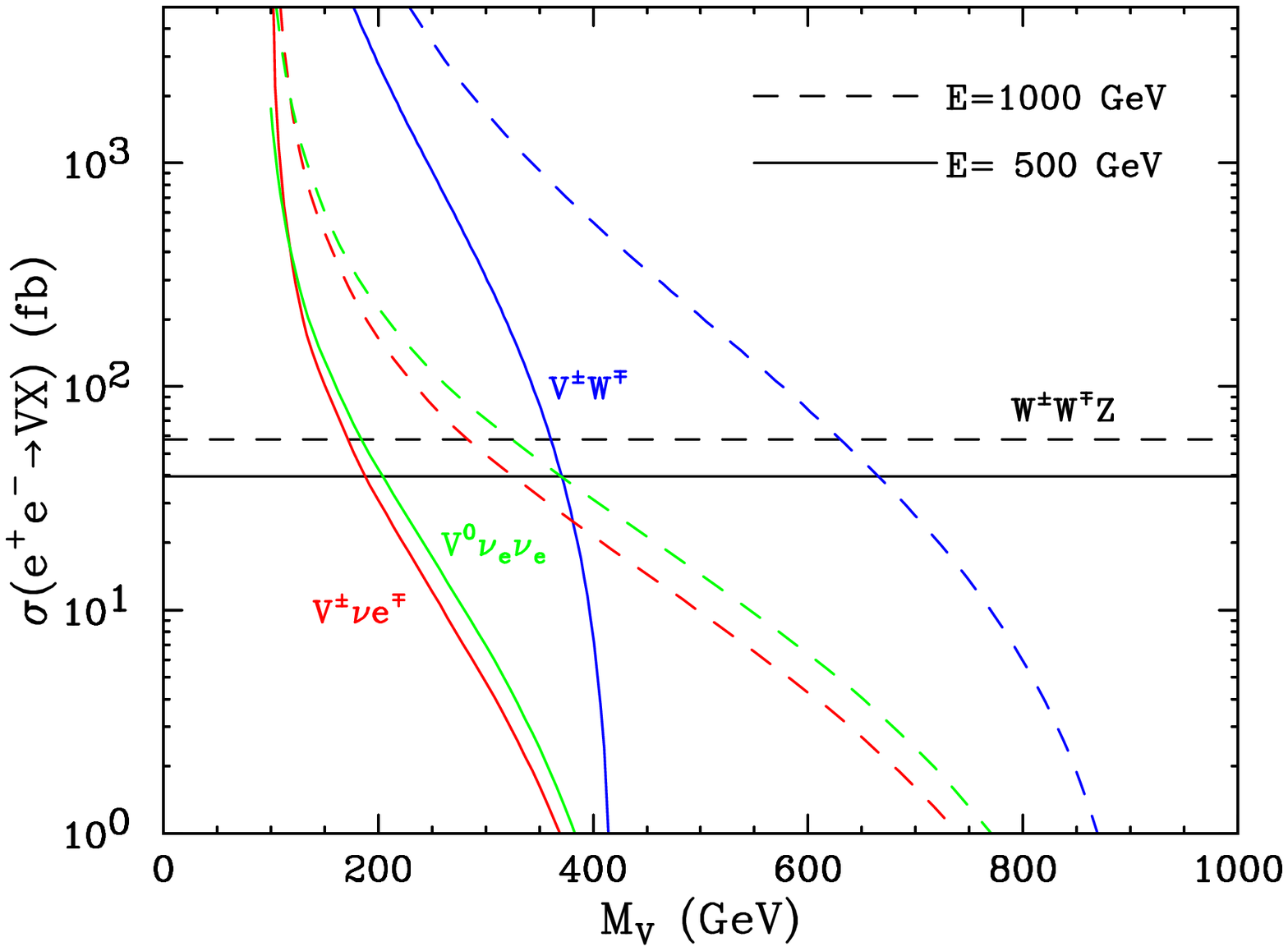}
\includegraphics[width=100mm]{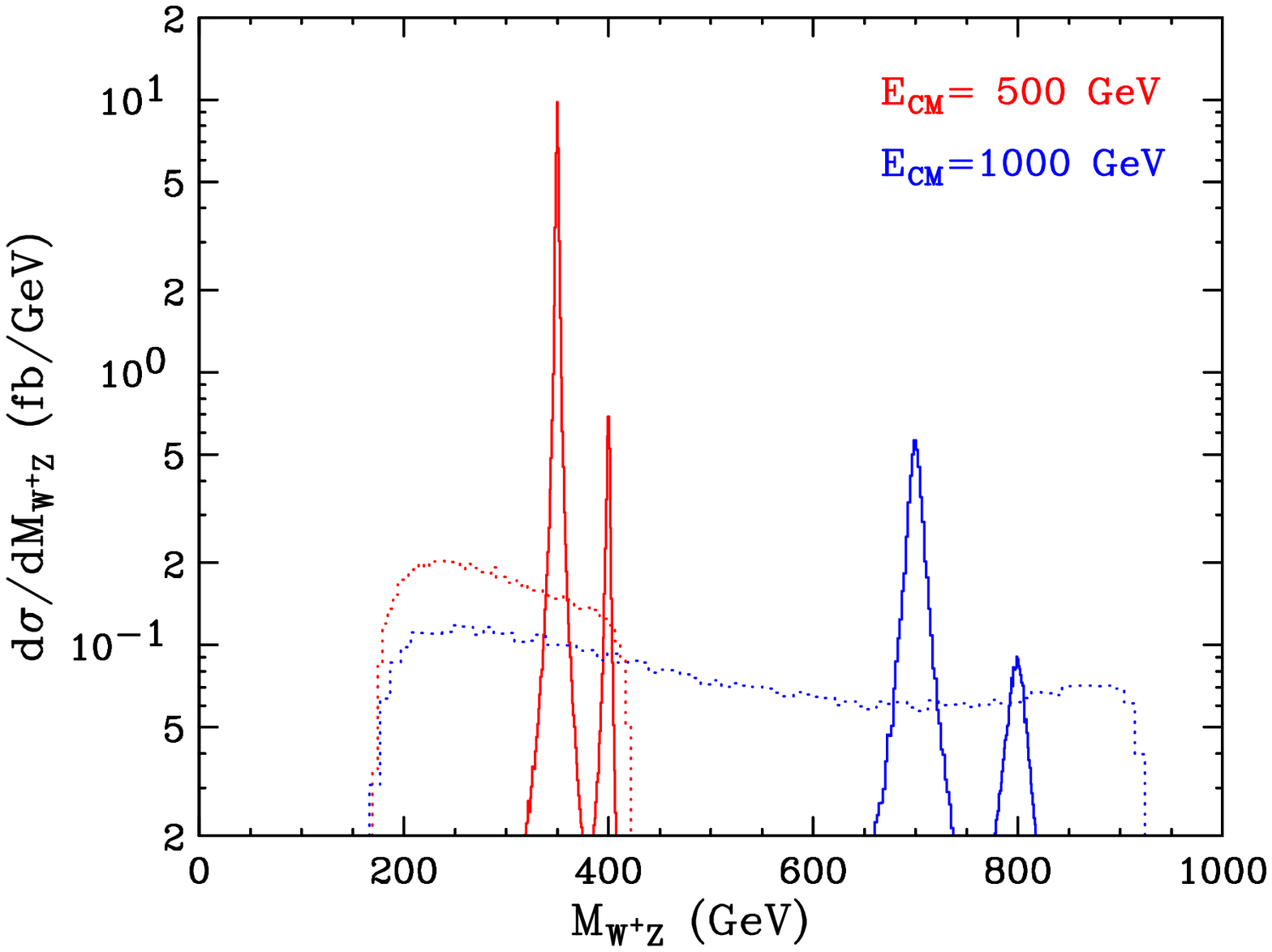}
\caption{Left: $V_1$ production cross-sections and the continuum 
SM background at an $e^+e^-$ lepton collider of center of mass energy
500~GeV (solid) or 1~TeV (dashed).  Right: $WZ$ invariant mass
distribution for Higgsless signals (solid) and SM background (dotted),
at $E_{CM}=500$~GeV (red, $M^\pm=350,400$~GeV) and $E_{CM}=1$~TeV
(blue, $M^\pm=700,800$~GeV).}
\label{fig:ilc}
\end{figure*}

Unlike traditional technicolor, Higgsless models offer new discovery
opportunities for a lepton collider with a center-of-mass energy in
the sub-TeV range.  From Eq.~\leqn{NDA} we have seen that the masses
of the new MVBs are expected to be below 1~TeV, and they can be
produced at an ILC through the analogous vector boson fusion process
by bremsstrahlung of $W$'s and $Z$'s off the initial state $e^+$ and
$e^-$.  The $V_1$ production cross sections for vector boson fusion
$e^+e^-\to V^\pm_1 e^\mp\nu_e$ and $e^+e^-\to V^0_1\nu_e\bar{\nu}_e$,
as well as associated production $e^+e^-\to V^\pm W^\mp$, are shown in
the left panel of Fig.~\ref{fig:ilc}.  The horizontal lines correspond
to the total cross sections of the continuum SM background.  We see
that for a large range of $V_1$ masses, ILC searches appear promising,
already at the level of total numbers of events, before cuts and
efficiencies.  Furthermore, because of the cleaner environment of a
linear lepton collider, one could use the dominant hadronic decay
modes of the $W$ and $Z$ and easily reconstruct the invariant mass of
the $V_1$ resonance, which provides an extra handle for background
suppression (see the right panel in Fig.~\ref{fig:ilc}).  Further
detailed studies are needed to better evaluate ILC potential for
testing the generic predictions (\ref{sumW}) and (\ref{sumZ}) of the
Higgsless models.

\clearpage\setcounter{equation}{0}\setcounter{figure}{0}\setcounter{table}{0}
\subsection{Model independent searches for $W^\prime$ bosons}
\label{sec:wprime}

Zack~Sullivan \\ [2mm]
{\em High Energy Physics Division,
Argonne National Laboratory \\
Argonne, Illinois 60439, USA}\\


A new charged current interaction mediated by a particle with vector and/or
axial-vector couplings to fermions is generically called a $W^\prime$ boson.  Many
classes of models of physics beyond the standard model predict the existence
of $W^\prime$ bosons with a wide range of masses and couplings to fermions.  From an
experimental point of view, it is desirable to perform a search for these
particles that is independent of any particular model.  Fortunately, a
completely model independent search for any finite-width $W^\prime$ boson exists
\cite{Sullivan:2002jt}.

The most general Lorentz invariant Lagrangian describing the coupling
of a $W^\prime$ to fermions may be written as \cite{Langacker:1989xa}
\begin{equation}
{\cal L} = \frac{1}{\sqrt{2}} \overline{f}_i\gamma_\mu \bigl( g_R
e^{i\omega} \cos\zeta\, V^R_{f_if_j} P_R + g_L \sin\zeta\, V^L_{f_if_j}
P_L\bigr) W^{\prime} f_j + \mathrm{H.c.} \,, \label{eq:lagrangian}
\end{equation}
where $\zeta$ is a left-right mixing angle, and $\omega$ is a CP-violating
phase that can be absorbed into $V^R$.  In this notation, $g_{R(L)}$ are the
right (left) gauge couplings, and $V^{R,L}_{f_if_j}$ are generalized
Cabibbo-Kobayashi-Maskawa (GCKM) matrices.  In models where the $W$ and $W^\prime$
mix, the mixing angle $\zeta$ is usually constrained to be small ($|\zeta|<$ a
few $\times 10^{-5}$--$10^{-2}$ \cite{Groom:2000in}).  Hence, searches are
usually performed for purely right- or left-handed states, but that is not
necessary in the analysis below.

\begin{figure}[tbh]
\centering
\includegraphics[width=1.5in]{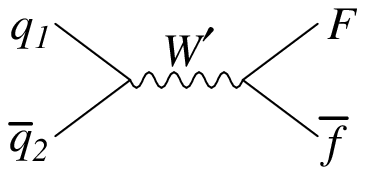}
\includegraphics[width=2.25in]{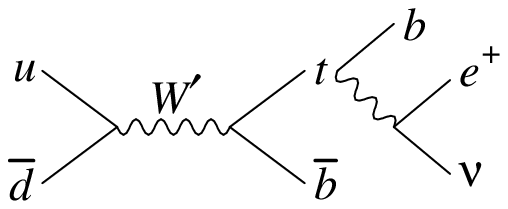}
\caption{Feynman diagram for $W^\prime$ production and decay into (left) any two
fermions, and (right) the single-top-quark final state ($Wbj$).
\label{fig:wpfeyn}}
\end{figure}

In Ref.\ \cite{Sullivan:2002jt} the fully differential next-to-leading order
(NLO) cross section for the production of a $W^\prime$ with arbitrary couplings, and
decay into any pair of fermions was published.  This paper proved that both
the width and differential cross section factorize completely through NLO.
Hence, a simple rescaling of naive right- or left-handed simulations can be
mapped onto any arbitrary model as a function of generic couplings (denoted
$g^\prime$), $W^\prime$ mass, and $W^\prime$ total width.  If the $W^\prime$ boson only decays
into fermions (as in Fig.\ \ref{fig:wpfeyn}), then the $W^\prime$ width dependence
is redundant.

It was demonstrated in Ref.\ \cite{Sullivan:2002jt} that the most effective
model independent search for $W^\prime$ bosons at either the Tevatron or LHC (see
also \cite{Sullivan:2003xy}) looks for the decay of the $W^\prime$ into the $Wbj$
final state (Fig.\ \ref{fig:wpfeyn}).  This final state allows a
straight-forward peak search for the $W^\prime$ invariant mass; and spin
correlations provide the promise of disentangling the exact Dirac structure if
a $W^\prime$ is found.

\begin{figure}[tbh]
\centering
\includegraphics[width=0.46\textwidth]{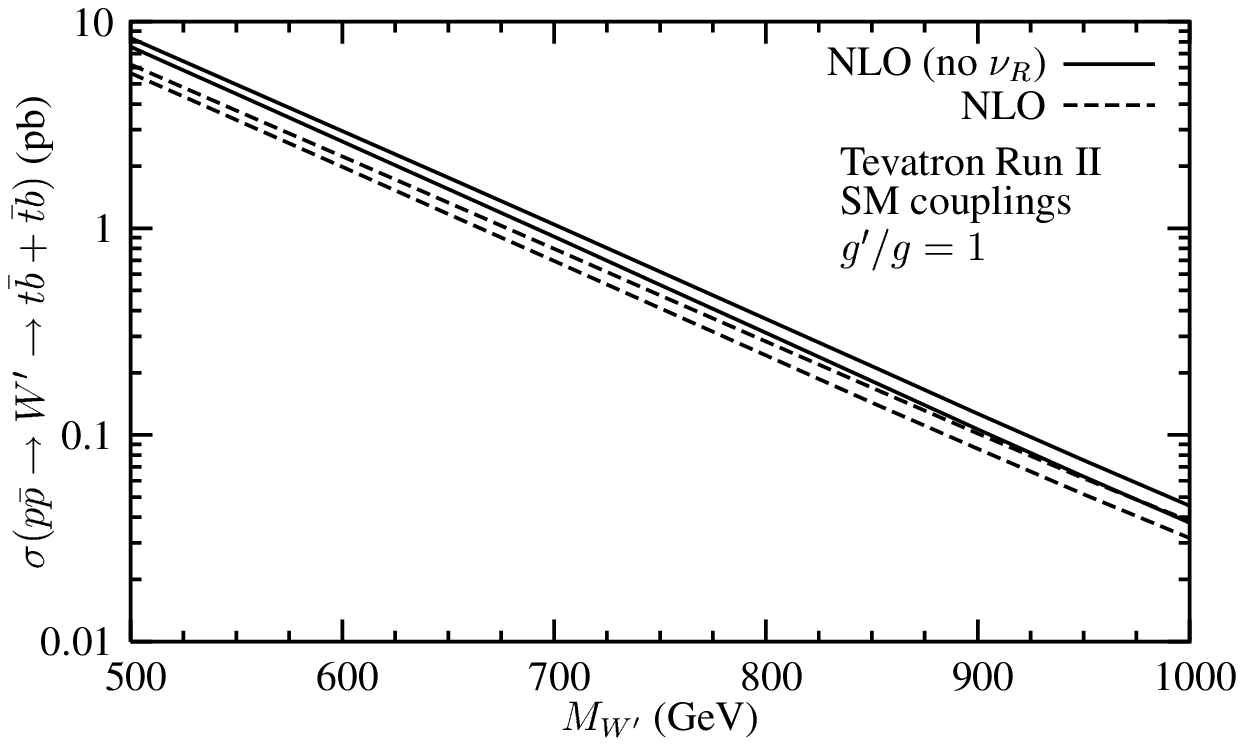}
\includegraphics[width=0.46\textwidth]{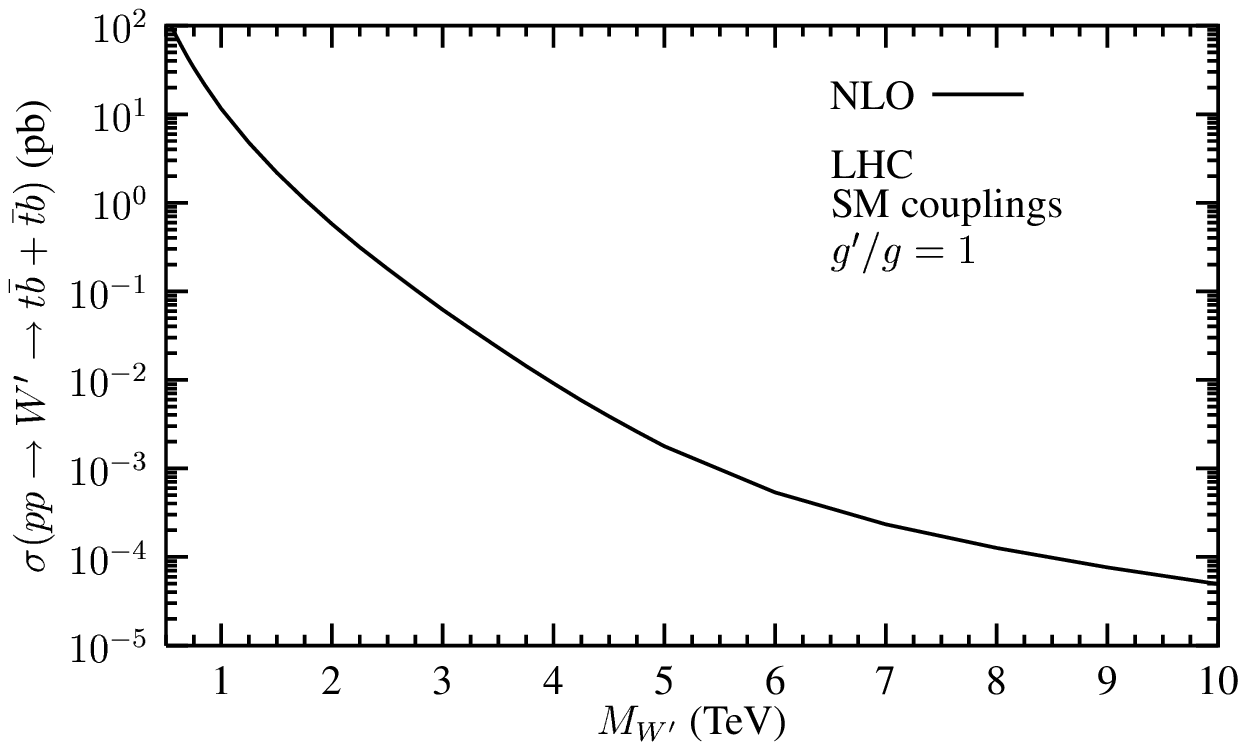}
\caption{Cross section at the (left) Tevatron and (right) LHC for $W^\prime$ boson
production plus decay into the $t\bar b$ final state.  Full theoretical error
bands are shown for the Tevatron.
\label{fig:wptbsig}}
\end{figure}

The cross section for a 700 GeV $W^\prime$ at the Tevatron is comparable to the
single-top-quark cross section (see Fig.\ \ref{fig:wptbsig}), but with a much
smaller background.  The CDF Collaboration looked for a mass peak in the run I
single-top-quark analysis \cite{Acosta:2002nu}, and set a lower mass bound on
$W^\prime$ bosons of 536(566) GeV assuming standard model-like couplings, where
decays to right-handed neutrinos are (not) allowed.  For pure left-handed $W^\prime$
bosons, the current best bound is 786 GeV \cite{Affolder:2001gr} based on the
decay into an electron and neutrino.  Using the current single-top-quark
bounds, an analysis of run II data could already surpass this limit for all
$W^\prime$ bosons.

The $s$-channel production of single top quarks via $W^\prime$ bosons can occur at
an extremely large rate at the LHC.  In Fig.\ \ref{fig:wptbsig}, the cross
section for this channel is shown for SM-like couplings as a function of $W^\prime$
mass up to 10~TeV.  In high-luminosity (100 fb$^{-1}$) years there could be 50
$W^\prime$ bosons produced with masses of 10~TeV that decay into this channel.  The
question is, can these be observed over the background?

\begin{figure}[tbh]
\centering
\includegraphics[width=0.46\textwidth]{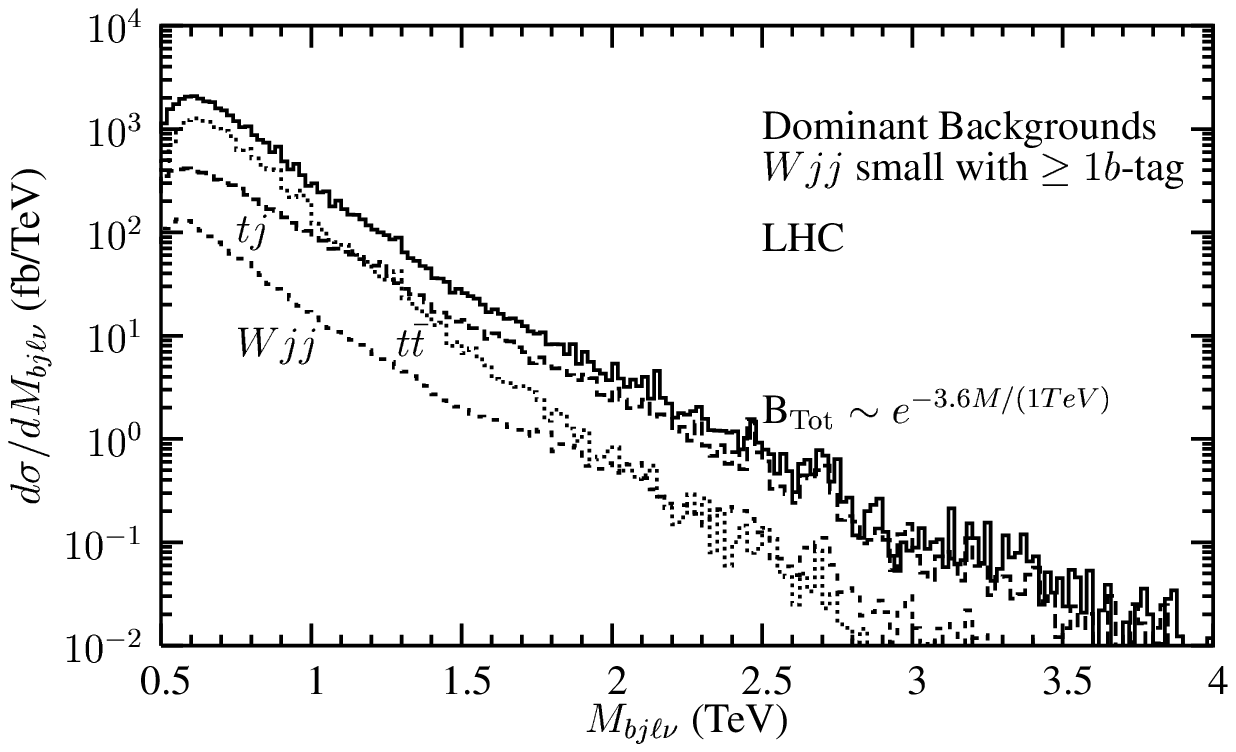}
\includegraphics[width=0.46\textwidth]{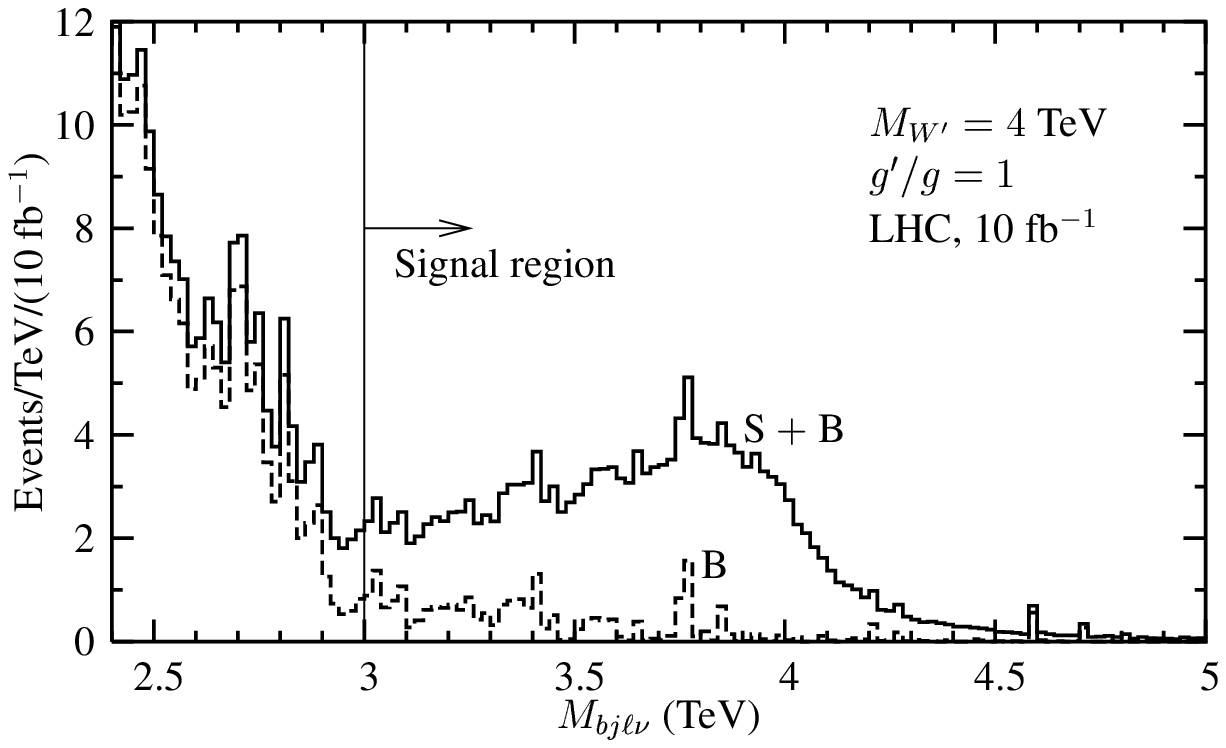}
\caption{(left) Dominant backgrounds for $Wbj$ production at the LHC.  After
$b$-tagging the second jet, $Wjj$ is roughly $1/5$ of the $t$-channel
single-top-quark background $tj$.  (right) Number of events expected per
low-luminosity year (10 fb$^{-1}$) at the LHC in the $Wbj$-invariant mass final
state vs.\ background.\label{fig:wpsandb}}
\end{figure}

In order to address this question, a fully simulated analysis of the signal
and background was performed \cite{Sullivan:2003xy,ZScoming}.  The signal was
evaluated using PYTHIA \cite{Sjostrand:2001yu} run through the SHW detector
simulation \cite{SHW} with parameters updated to match the ATLAS detector
\cite{TDR}. The final state of interest contains a lepton ($e$ or $\mu$),
2 $b$-jets, and missing energy.  The backgrounds come from $t\bar t$,
$t$-channel single-top-quark production (i.e.\ $tj$), $Wjj, Wcj, Wb\bar b,
Wc\bar c, WZ, Wt$, and $s$-channel single-top-quark production. As is apparent
from Fig.\ \ref{fig:wpsandb}, the most important of these are $t\bar t$, $tj$,
and $Wjj$.  The cross section for the backgrounds falls exponentially with
$M_{bj\ell\slash\!\!\!\!E_T}$ the reconstructed invariant mass, and drops to
less than one event above 3~TeV.

Unfortunately, the event generators do not currently model the $tj$ or $Wjj$
backgrounds correctly.  In Fig.\ \ref{fig:wpsandb} matrix-element calculations
are normalized to the correct fully-differential NLO calculations of the $tj$
\cite{Harris:2002md,Sullivan:2004ie} and $Wjj$ \cite{Campbell:2002tg} cross
sections.  The Tevatron will play an important role in validating these
matching schemes.

\begin{figure}[tbh]
\centering
\includegraphics[width=0.46\textwidth]{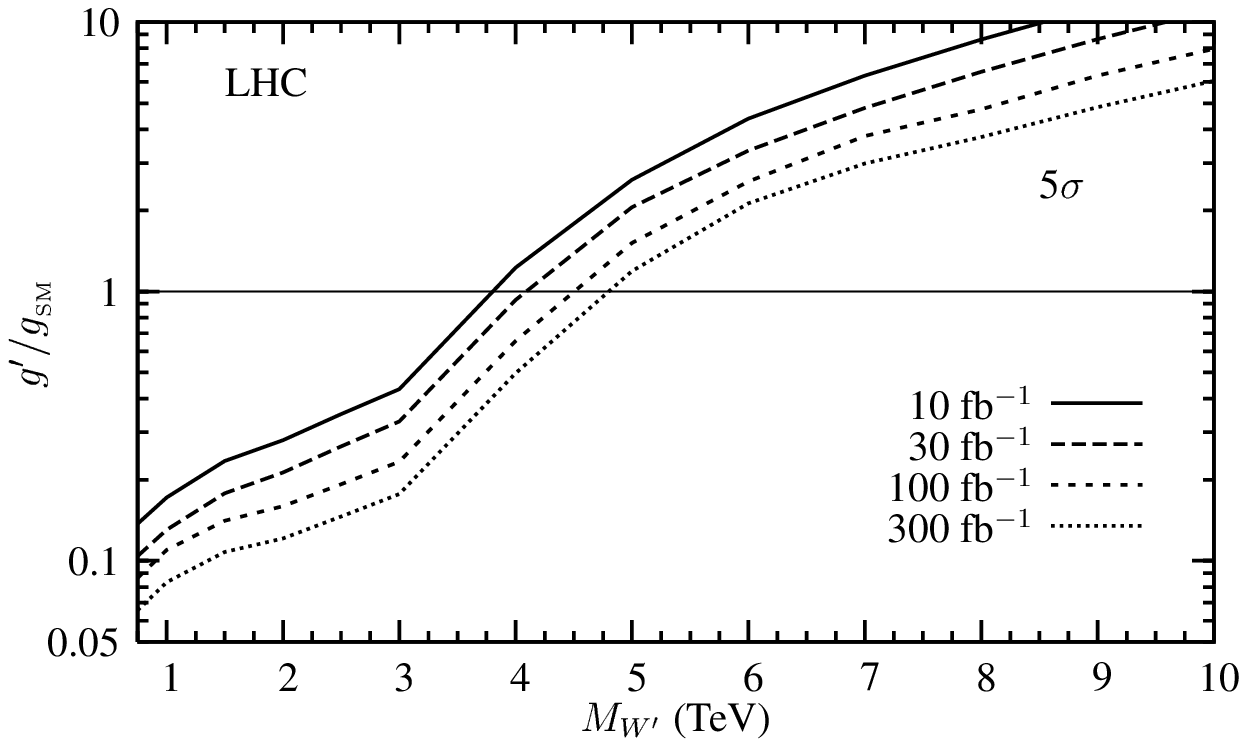}
\includegraphics[width=0.46\textwidth]{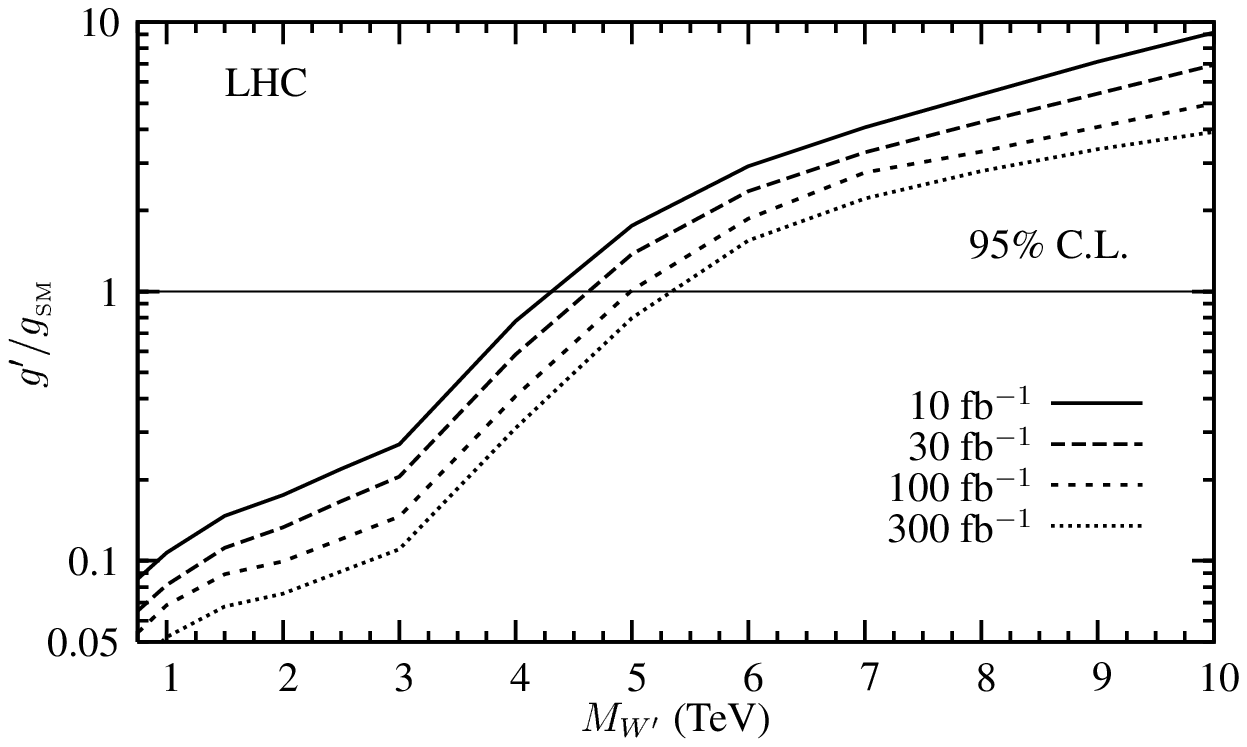}
\caption{Reach in relative coupling $g^\prime/g_{_\mathrm{SM}}$ at the LHC
for $5\sigma$ discovery and 95\% C.L.\ exclusion.
\label{fig:wplhcsig}}
\end{figure}

Figure \ref{fig:wplhcsig} shows that the LHC should have $5\sigma$ discovery
reach for standard model-like $W^\prime$ bosons of $3.7$ TeV in the first 10 fb$^{-1}$,
and $4.7$ TeV with 300 fb$^{-1}$.  There is an effective hard cutoff in mass reach
due to an almost exponentially falling parton luminosity above $5.5$ TeV.
While Fig.\ \ref{fig:wptbsig} shows a large cross section for 6--10 TeV, most
events are produced well below resonance, and just add to the single-top-quark
rate near the single-top threshold.  More remarkable than mass reach is that
couplings up to 20 times smaller than standard model-like couplings can be
probed in the 1 TeV range.  This allows complete coverage of Littlest Higgs
parameter space in 1 year \cite{Sullivan:2003xy,ZScoming}.  Perturbative
models based on ratios of couplings have effective couplings $g^\prime$ that
do not differ from the standard model by more than a factor of 5, and
typically average to $g^\prime\approx g_{\mathrm{SM}}$
\cite{Sullivan:2003xy,ZScoming}.  Hence, these models will be accessible over
the full mass reach.

The most sensitive analysis of left-handed $W^\prime$ bosons completed so far looked
for $W^\prime$ decay into an electron and neutrino \cite{Affolder:2001gr}.  While
the reach is impressive, the lepton final state is not model independent.  In
particular, many new physics models have $W^\prime$ bosons with right-handed
couplings, so this final state would never be produced.  Leptophobic models,
such as some versions of top color, also do not produce this final state.

\begin{figure}[tbh]
\centering
\includegraphics[width=0.46\textwidth]{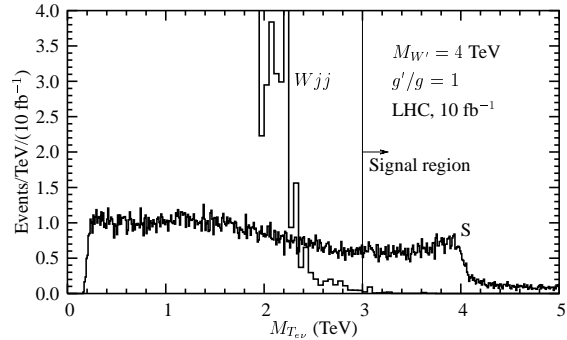}
\caption{Number of events expected per low-luminosity year at the LHC vs.\ the
reconstructed transverse mass of the $W^\prime$, and the previously missed $Wjj$
background.\label{fig:wplnu}}
\end{figure}

The lepton final state suffers from several challenges in going to high
energy.  First, the background studies for LHC physics missed the dominant
background of $Wjj$ production \cite{Sullivan:2003xy,ZScoming}.
Valence-valence scattering opens a new production channel at the LHC that
completely overwhelms the high-mass cross section, and is not produced by
showering evolution of $Wj$ production.  Second, no mass can be reconstructed
from fits to the transverse mass without huge data sets.  Both problems are
highlighted in Fig.\ \ref{fig:wplnu}, where a 4 TeV $W^\prime$ will produce at most
1 event above background per low-luminosity (10 fb$^{-1}$) year with a fairly flat
transverse mass distribution.  Finally, the prevalence of higher-order
radiation makes it unlikely that a $W^\prime$ will be produced without additional
jets, which will degrade reconstruction.

There are several key issues affecting $W^\prime$ production for which Tevatron
studies will provide vital assistance to the LHC effort.
\begin{itemize}
\item The real $W$ background to all final states is much larger than
estimated in the LHC TDRs.  Since this background must be modeled by a mixture
of $Wj$ and $Wjj$ events, it is vital to understand the performance and
limitations of the new NLO and NLO-matched Monte Carlos in the context of the
much simpler Tevatron environment.
\item Similarly, single-top-quark production survives to much larger invariant
masses than is predicted by the event generators used for the TDRs.  The
measurement of single-top-quark production at the Tevatron will provide a
vital test of the methods and theoretical tools for modeling that background
to new physics.
\item How well high-energy leptons and jets will be measured is not a settled
issue at the LHC.  For $W^\prime$ bosons it has been assumed that only the $\sim$200
GeV $b$ from top-quark decay can be tagged, but there is a TeV $b$ jet
recoiling against the top quark.  Very little is known about high-energy $b$
tagging.  The Tevatron could provide valuable insight into the causes of the
expected reduction of $b$-tagging efficiency at large $E_T$.
\end{itemize}

The Tevatron and LHC are complementary machines for $W^\prime$ searches.  While the
Tevatron can reach $\sim$900 GeV with 2 fb$^{-1}$ of data \cite{Sullivan:2002jt},
it will be a challenge for the LHC to go below 750 GeV in the $Wbj$ final
state because the background is rising exponentially.  However, a 5.5 TeV $W^\prime$
with standard-model couplings can be probed.  For an indication of where
several classes of models fall in the $g^\prime$-$M_{W^\prime}$ plane, including
additional width effects, see Refs.\ \cite{Sullivan:2003xy,ZScoming}.


\clearpage\setcounter{equation}{0}\setcounter{figure}{0}\setcounter{table}{0}
\subsection{Vectorlike Quarks}
\label{sec:vectorlike}

Bogdan A. Dobrescu$^1$ and  Tim M.P. Tait$^2$ \\ [3mm]
{\em
\noindent
{$^1$ Fermilab, Batavia, IL 60510, USA} \\ [2mm]
{$^2$ Argonne National Laboratory, Argonne, IL 60439, USA} \\
}

\subsubsection{Motivation}

All observed elementary fermions are chiral: their left- and right-handed
components have different charges under the $SU(2)_w\times U(1)_y$
gauge group. Additional chiral fermions could exist, but they would induce
rather large one-loop contributions to electroweak observables, so that
their number and properties are tightly restricted by the electroweak data.
By contrast, if non-chiral fermions carrying Standard Model gauge charges
exist, they would decouple from the observed particles in the limit where 
their masses are large compared to the electroweak scale.
Non-chiral fermions are commonly called ``vectorlike'' because 
the $Z$ and $W$ bosons have vector couplings to them.

A particularly interesting type of vectorlike fermion is one that has the 
same gauge charges as the right-handed top quark.
Such a vectorlike quark, usually labelled by $\chi$, 
plays an essential role in the top-quark seesaw 
model \cite{Dobrescu:1997nm,Chivukula:1998wd,Dobrescu:1999gv,He:2001fz}, 
where a Higgs doublet arises as a bound state of $\chi$ and the 
top-bottom doublet.
The same vectorlike quark is used in 
little Higgs models~\cite{
Arkani-Hamed:2001nc,Perelstein:2003wd,Han:2003wu},
where it serves to cancel the top quark contribution to quadratic divergences
of the Higgs mass. 

From a phenomenological point of view, $\chi$ is interesting 
because of its potentially large mixing with the top quark.
Here, we discuss the potential for
discovery of $\chi$ at the LHC and the Tevatron. 
It should be mentioned, that although we limit our discussion to
only up-type vectorlike quarks,
down-type vectorlike quarks also appear in a variety of models,
and may be phenomenologically interesting
\cite{Rizzo:1986kq,He:1999vp,Choudhury:2001hs,Morrissey:2003sc}.
The potential for their discovery at the LHC is evaluated
in~\cite{Andre:2003wc,Mehd_iso}.  

\subsubsection{Couplings of the $\chi$ quark}

Let us concentrate on the top and $\chi$ quarks, ignoring  
the mixing with the first two generations of quarks (this can also be
included, but the effects are expected to be small).
We write down a
low-energy effective theory whose parameters are sufficiently general to 
include
the models mentioned above as particular cases.

Let us denote the gauge eigenstate quarks by a subscript 0.
The gauge interactions of $\chi_{0_L}, \chi_{0_R}$ and $t_{0_R}$ are identical:
they are all color triplets, $SU(2)_w$ singlets, and have hypercharge 4/3.
The left-handed top quark, $t_{0_L}$, is part of an $SU(2)_w$ doublet
of hypercharge 1/3.  
The effective Lagrangian includes two gauge invariant 
mass terms and two Yukawa interactions of $\chi_0$ and $t_0$ to the Higgs doublet.
Given that $SO(2)$ transformations that mix $\chi_R$  and $t_R$ are not physically 
observable, one can arrange
that either one of the mass parameters or one of the Yukawa couplings vanishes.
Therefore, after electroweak symmetry breaking, the quark mass matrix and Higgs boson 
interactions are given by 
\begin{equation}
{\cal L} = - \left( \overline{t}_{0_L} \ , \ \overline{\chi}_{0_L} \right)
\left( \ba{cc} 0  & \lambda_\chi \left(v_h + h /\sqrt{2}\right) \\
    m_{\chi t}  & m_{\chi \chi} \ea \right)
\left( \ba{c} t_{0_R} \\ \chi_{0_R} \ea \right) + {\rm h.c.} ~,
\label{massterm}
\end{equation}
where $h$ is the Higgs boson and $v_h \simeq 174$ GeV is the 
Higgs VEV.
The two mass parameters, $m_{\chi t}$ and $m_{\chi \chi}$, and 
the $\lambda_\chi$ Yukawa coupling are taken to be real parameters, 
as their complex phases can be absorbed 
by U(1) transformations of the quark fields.
Hence, there are only three real parameters 
that describe the mass sector of heavy $t$ and $\chi$ quarks. 

To relate the parameters in the Lagrangian to physical observables, we 
transform the gauge eigenstates $t_{0_L}, t_{0_R}, \chi_{0_L}, \chi_{0_R}$ to the mass 
eigenstates $t_L, t_R, \chi_L, \chi_R$, where
$t$ is the top quark observed at the Tevatron, of mass $m_t\approx 175$ GeV,
and $\chi$ is a new quark of mass $m_\chi$, which remains to be measured.
The relation between the two bases depends on two angles, $\theta_L$ and $\theta_R$,
but $\theta_R$ is not related to any physical observable, as explained above.
The mixing angle $\theta_{L}$ affects the electroweak interactions of the 
top quark as 
well as the Yukawa couplings of the Higgs boson.
We use $s_L$ and $c_L$ as short-hand notation for $\sin\theta_L$
and $\cos\theta_L$, respectively. 

The relations between the physical parameters $m_t, m_\chi$ and $\theta_{L}$ 
and the initial parameters $\lambda_\chi, m_{\chi t}, m_{\chi \chi}$ are given by
\begin{equation}
m_{t,\chi}^2 = \frac{1}{2} \left[ m_{\chi\chi}^2 + m_{\chi t}^2 +  \lambda_{\chi}^2v_h^2
 \mp \sqrt{ \left(m_{\chi\chi}^2 + m_{\chi t}^2 + \lambda_{\chi}^2v_h^2 \right)^2
- 4 \left( \lambda_{\chi} v_h  m_{\chi t}\right)^2 } \, \right]
\label{mtmc}
\end{equation}
for the masses, and by
\begin{equation}
s_L = \frac{1}{\sqrt{2}} \left( 1 -
\frac{ m_{\chi\chi}^2  + m_{\chi t}^2 - \lambda_{\chi}^2v_h^2 }
{m_{\chi}^2 - m_{t}^2} \right)^{\!\! 1/2} ~.
\label{mix}
\end{equation}
for the mixing angle.  Note that in the limit of 
$m_{\chi} \rightarrow \infty$, the mixing vanishes ($s_L \rightarrow 0$)
so that the new physics decouples from the Standard Model.

The interactions of $t$ and $\chi$ 
with the electroweak gauge bosons, which depend on $\theta_{L}$, can be 
computed straightforwardly.
There are charged current interactions,
\begin{eqnarray}
t - \overline{b} - W_\mu^+ & : & 
-i \frac{g}{\sqrt{2}} c_L \gamma_\mu P_L ~,
\nonumber \\ [3mm]
\chi - \overline{b} - W_\mu^+ & : & 
-i \frac{g}{\sqrt{2}} s_L \gamma_\mu P_L~,
\end{eqnarray}
where $P_L = (1-\gamma_5)/2$ is the left-handed projector and 
$g \equiv e / \sin \theta_W$ is the $SU(2)_w$ gauge coupling.  
The charge-conjugate interactions have the same vertex factors.
The neutral current interactions contain the photon 
interactions, which are
standard for both $t$ and $\chi$, as demanded by gauge
invariance under $U(1)_{EM}$.
The $Z$ boson interactions with the left-handed quarks are modified,
and include $t$-$\chi$ flavor-changing neutral currents:
\begin{eqnarray}
t_L - \overline{t}_L - Z_\mu & : &
-i \frac{e}{\sin \theta_W \cos \theta_W} \left( \frac{1}{2} c_L^2 
- \frac{2}{3} \sin^2 \theta_W \right) \gamma_\mu P_L ~,
\nonumber \\ [3mm]
t_L - \overline{\chi}_L - Z_\mu & : &
-i \frac{e}{\sin \theta_W \cos \theta_W} \frac{1}{2} s_L c_L \gamma_\mu P_L ~,
\nonumber \\ [3mm]
\chi_L - \overline{\chi}_L - Z_\mu & : &
-i \frac{e}{\sin \theta_W \cos \theta_W} \left( \frac{1}{2} s_L^2 
- \frac{2}{3} \sin^2 \theta_W \right) \gamma_\mu P_L ~,
\end{eqnarray}
and the $\overline{t}_L-\chi_L-Z_\mu$  
interaction is the same as the $t_L - \overline{\chi}_L - Z_\mu$
interaction given above.  The interactions of $t_R$ and $\chi_R$ are with 
the $Z$ boson are
identical with those of the right-handed top quark in the Standard Model.

The Higgs interactions with $t$ and $\chi$ can be expressed
in terms of $\theta_L$, $m_t/v$ and $m_\chi/v$:
\begin{eqnarray}
h^0 - \overline{t}_L - {t}_R & : & -i c_L^2 \frac{m_t}{v\sqrt{2}}
\nonumber \\ [3mm]
h^0 - \overline{t}_L - {\chi}_R & : & -i c_L s_L \frac{m_\chi}{v\sqrt{2}} 
\nonumber \\ [3mm]
h^0 - \overline{\chi}_L - {t}_R & : & -i c_L s_L \frac{m_t}{v\sqrt{2}} 
\nonumber \\ [3mm]
h^0 - \overline{\chi}_L - {\chi}_R & : & -i s_L^2 \frac{m\chi}{v\sqrt{2}}
\end{eqnarray}
The charge-conjugate vertex factors are the same as their 
counter-parts given above.

\subsubsection{Decays of the $\chi$ quark}

The charge-current interactions allow for the $\chi \rightarrow W^+ b$ decay,
while the flavor-changing neutral-current interactions  
allow for the $\chi \rightarrow Z t$ decay, assuming that $m_\chi$ is above 
$\sim$ 250 GeV.
These Higgs interactions allow decays of $\chi$ into a top and a Higgs boson,
which competes with the $\chi \rightarrow Z t$ and $\chi \rightarrow W b$ decays
for some regions of parameter space.
In the heavy $\chi$ limit, the decay widths are given by
\begin{eqnarray} \label{eq:widths}
&& \Gamma(\chi \rightarrow W^+ b) \simeq \frac{s_L^2 m_\chi^3}{32 \pi v^2} ~,
\nonumber \\ [2mm]
&& \Gamma(\chi \rightarrow Z t ) \simeq \Gamma(\chi \rightarrow h t ) 
\simeq \frac{c_L^2}{2} \, \Gamma(\chi \rightarrow W^+ b) ~. 
\end{eqnarray}

Clearly the $\chi \rightarrow W^+ b$ decay is dominant, but if $c_L$
is not much smaller than unity, then 
the decay $\chi \rightarrow h t$ could be very interesting. 
In the presence of the $\chi$ quark the bounds from precision electroweak
measurements on the mass of the Higgs boson are considerably
loosened \cite{Collins:1999rz}. 
It is likely that the Higgs boson is heavier than 
about 180 GeV, and it decays most of the time to $W^+W^-$ and $ZZ$.

For $m_\chi > 350$ GeV, 
we will consider the LHC signal induced by the decay $\chi \rightarrow h t$ followed by 
$h \to Z Z$ with one of the $Z$ bosons decaying leptonically. 
Previous studies in the context of the little Higgs model 
\cite{Azuelos:2004dm}, have assumed a light Higgs boson ($m_H$= 120 GeV), which leads 
to a signal harder to see at the LHC.

For $m_\chi < 350$ GeV, which is the region of interest at the Tevatron, 
$\chi$ decays predominantly to $Wb$, and would look like a heavier top quark.

\subsubsection{Single-$\chi$ production at the LHC}

Both single-$\chi$ and $\chi$-pair productions are possible at the 
Tevatron and the LHC.
We first discuss single-$\chi$ production, which has the advantage for
large $\chi$ masses that only one massive quark must be produced, and thus
parton luminosity and phase space are favorable.  However, it has the feature
that the process is only possible because of the mixing between the $\chi$
and the top quarks, and thus the cross section depends strongly on the mixing
angle, $\sigma \propto s_L^2$.  The difference in the mass of the final state
objects implies that single-$\chi$ production is usually the dominant
production for large $\chi$ masses, reachable by the LHC, whereas 
$\chi$-pair production is usually dominant for lower $\chi$ masses, testable
at the Tevatron.

The process proceeds through the $t$-channel exchange of a $W$ boson 
from any light quark in one of the colliding hadrons to a bottom quark in
the other one.  The final state thus consists of a single $\chi$ quark and
a jet which tends to be in the forward region of the detector.  This process
has been considered in the past, particularly in the little Higgs context
\cite{Han:2003wu, Perelstein:2003wd, Burdman:2002ns}.  
We point out here that there is another process of single-$\chi$
production (though one that does not interfere with the $W$-exchange process)
not usually considered in the literature in which a $t$-channel
$Z$ boson is exchanged between a light quark and a top quark in the initial 
state, resulting from gluon splitting.  This process is important \cite{DHT},
and should be included in phenomenological studies of single-$\chi$
production.

The LHC discovery potential with 300 fb$^{-1}$, estimated in 
Ref.~\cite{Azuelos:2004dm} 
(without the inclusion of the $Z$-exchange contribution)
and assuming that the three branching fractions shown in Eq~(\ref{eq:widths}) 
are in the proportions 2:1:1 (corresponding to $s_L \ll 1$), 
reaches about 1 TeV for the $\chi \to Zt$ 
decay, and 2 TeV for $\chi \to Wb$, with some
dependence on the model parameters as expected for single-$\chi$ production.

Another interesting signature at the LHC is provided by 
single-$\chi$ production followed by the decay 
$\chi \to ht$. Given that the Higgs boson 
mass in this model is not tightly constrained by electroweak fits,
and the leptonic decays of gauge bosons allow for
clean event reconstruction, 
a clear signature of the process would be obtained from the case $h \to Z Z$.

The event topologies are complex and their reconstruction from jets and 
leptons leads to some combinatorial background.  
The topologies with two, three or four leptons
in the final state, originate from either a $Z$ or $t$ decay, are quite promising.
The case with one lepton has overwhelming background, especially if there 
is some misidentification of jets as leptons.  The case with five leptons 
has a cross section which is too low to lead to an observable signal. 
The events can be selected to contain 
at least two leptons of the same flavor and opposite
charge, and at least 2 non $b$-tagged jets.  Because of the massive
parent particles, all of these objects are expected to be central and at
relatively large $p_T$.  The $W$, $Z$, and top resonances provide an excellent
means to unravel the structure of the events, and should prove efficient
to reduce the dominant backgrounds, $t\bar{t}$, $Wjt\bar t$ and $Zt\bar t$.

\subsubsection{Pair production}

Recently, it has been
shown~\cite{Aguilar-Saavedra:2005pv} that with 100 fb$^{-1}$, a 1 TeV
top-like quark can be discovered at the LHC in the channel 
$gg,qq \to T\bar T \to W^+ b W^- \bar b$.  
Pair production of $\chi$ quarks with their subsequent decays leads
to very complex events containing many jets and leptons. Since more
center of mass energy is required to produce a pair of heavy quarks,
the mass reach will be in general lower than for single
production. However, the production cross section is through the strong
force, and does not depend on any of the other parameters than the $\chi$
mass.

Furthermore, detector resolution, efficiencies, and
combinatorial effects make it difficult to reconstruct these
events. The case with one lepton in the final state, dominated by
$\chi\chi \to Wb~Wb \to \ell\nu~jj$, has been studied
by~\cite{Aguilar-Saavedra:2005pv}.
 Here, we consider only the case with two leptons in the final
state, $\chi\chi \to Wb~Zt \to jjb~\ell\ell~jjb$, as other cases are
limited by statistics or by combinatorial backgrounds.  

The principal backgrounds for the $\chi\chi \to Wb~Zt \to jjb~\ell\ell~jjb$ 
channel are $t\bar{t}$, $WWj t\bar{t}$ and $Zt\bar{t}$. 
The main components are $t\bar{t}$ production and $Zt\bar{t}$, as they
have respectively a high cross section and similar event topology.
The $WWt\bar{t}$ background has a low cross section and can be strongly
suppressed by the requirement of recontruction of the intermediate resonance.
The events will contain 2 leptons of the same flavor and opposite charge,
whose invariant mass should reconstruct close to the $Z$ boson mass.
In addition, we expect 2 $b$-tagged jets and at least 4 untagged jets,
two of which will reconstruct a $W$.

\bigskip

{\it Acknowledgments:} \ We would like to thank Georges Azuelos for 
many helpful discussions, and for describing to us
the plans of the ATLAS collaboration to search for vectorlike quarks.

\clearpage\setcounter{equation}{0}\setcounter{figure}{0}\setcounter{table}{0}
\subsection{
Triplet Higgs Boson}
\label{sec:triplet}

Mu-Chun~Chen \\ [3mm]
{\em Fermi National Accelerator Laboratory, Batavia, IL, USA} \\

The Standard Model requires a Higgs boson to explain the generation
of fermion and gauge boson masses.  Precision electroweak measurements suggest
that the Higgs boson must be relatively light, $m_{H} <219~GeV$. 
Currently, experimental data overwhelmingly support the SM with 
a light Higgs boson. The simplest version of the Standard Model with a 
single Higgs boson, however, has the theoretical problem that 
the Higgs boson mass is quadratically sensitive 
to any new physics which may arise at high energy scales.
Little Higgs models are a new approach to understanding
the hierarchy between the $TeV$ scale of possible
new physics and the electroweak scale. These models have an expanded
gauge structure at the TeV scale which contains the Standard Model
$SU(2)\times U(1)$ electroweak gauge groups.  
The LH models are constructed
such that an approximate global symmetry prohibits the Higgs boson from 
obtaining a quadratically divergent mass until at least two loop order.
The Higgs boson is a pseudo-Goldstone 
boson resulting from the spontaneous breaking of the
approximate global symmetry and so
is naturally light. We present in this talk, which 
is based on the work done in Ref.~\cite{Chen:2003fm,Chen:2004ig,Chen:2005jx}, 
the one-loop electroweak precision constraints 
in the Littlest Higgs model (LLH)~\cite{Arkani-Hamed:2002qy}, 
which contains a gauged 
$[SU(2) \otimes U(1)]_{1} \otimes [SU(2) \otimes U(1)]_{2}$ 
symmetry as its subgroup. We  
include the logarithmically enhanced 
contributions from both fermion and scalar loops, and  emphasize 
the role of triplet scalars in constructing a consistent 
renormalization scheme. \\

Precision electroweak measurements give stringent bounds on the scale of
little Higgs type models. One of the strongest bounds comes from fits
to the $\rho$ parameter, since in the LLH model the relation $\rho=1$ 
is modified  at the tree level. A special feature of the SM with the assumption of one Higgs doublet 
is the validity of the tree level relation, $\rho = 1 = 
\frac{M_{W}^{2}}{M_{Z}^{2} c_{\theta}^{2}}$ due to the tree level 
custodial symmetry. There is thus a definite relation between the 
W-boson mass and the Z-boson mass. 
Of course, one can equivalently choose any three physical observables 
as the input parameters in the gauge sector. If we choose $G_{\mu}, \, 
M_{Z}$ and $\alpha$ as the three input parameters in the gauge sector,
 the W-boson mass, $M_{W}$, then is predicted in the usual way via muon-decay,
\begin{equation}
M_{W}^{2} = \frac{\pi \alpha}{\sqrt{2} G_{\mu} s_{\theta}^{2}} \biggl[ 1 
+ \Delta r \biggr] \; ,
\end{equation}
where $\Delta r$ summarizes the one-loop radiative corrections, 
and it is given in terms of the gauge boson self-energy two point functions as,
\begin{eqnarray}
\Delta r & = & -\frac{\delta G_{\mu}}{G_{\mu}} - 
\frac{\delta M_{W}^{2}}{M_{W}^{2}} + \frac{\delta \alpha}{\alpha}
 - \frac{\delta s_{\theta}^{2}}{s_{\theta}^{2}}
\\
& = & \frac{\Pi^{WW}(0)-\Pi^{WW}(M_{W})}{M_{W}^{2}} + \Pi^{\gamma\gamma
 \, \prime}(0)  + 2 \frac{s_{\theta}}{c_{\theta}} 
\frac{\Pi^{\gamma Z}(0)}{M_{Z}^{2}} -  \frac{\delta s_{\theta}^{2}}{s_{\theta}^{2}}
\nonumber \; .
\end{eqnarray}
The counter term for the weak mixing angle $s_{\theta}$ which is defined 
through the W- and Z-boson mass ratio, $s_{\theta}^{2} = 1 
- \frac{M_{W}^{2}}{M_{Z}^{2}}$, is then given by,
\begin{equation}
\frac{\delta s_{\theta}^{2}}{s_{\theta}^{2}} = \frac{c_{\theta}^{2}}{s_{\theta}^{2}} 
\biggl[ 
\frac{\Pi^{ZZ}(M_{Z})}{M_{Z}^{2}}-\frac{\Pi^{WW}(M_{W})}{M_{W}^{2}}\biggr]
\; .
\end{equation}
Both of the two point functions, $\Pi^{WW}(M_W)$ and $\Pi^{WW}(0)$,
 have identical leading quadratic $m_{t}$ dependence, 
$\frac{\sqrt{2}G_{\mu}}{16\pi^{2}}3m_{t}^{2}\bigl( 1 
+ 2 \ln\frac{Q^{2}}{m_{t}^{2}}\bigr)$, and thus their difference
 is only logarithmic.The two-point function, $\Pi^{\gamma\gamma \prime}(0)$, 
is also logarithmic in $m_{t}$. However, in the counter term $\delta s_{\theta}^{2}/s_{\theta}^{2}$, 
the difference between $\Pi^{WW}(M_W)$ and $\Pi^{ZZ}(M_Z)$ is quadratic in $m_{t}$. 
The prediction for $M_{W}^{2}$ thus depends on $m_{t}$ quadratically in this case.\\

While the Standard Model requires 
three input parameters in the weak sector, a model with $\rho\ne 1$ 
at tree level, such as the LLH model, requires an additional input 
parameter in the gauge-fermion sector, 
which can be taken to be the 
VEV of the Higgs triplet, $v^\prime$. Many of the familiar predictions 
of the Standard Model are drastically changed by the need for an extra
input parameter~\cite{Blank:1997qa,Czakon:1999ha}. We choose as our input parameters
the muon decay constant $G_{\mu}$, the physical Z-boson mass $M_{Z}^{2}$, 
the effective lepton mixing angle $s_{\theta}^{2}$ and the fine-structure 
constant $\alpha(M_{Z}^{2})$ as the four independent input parameters in 
the renormalization procedure. The $\rho$ parameter, defined as,
$\rho \equiv M_{W_{L}}^{2}/(M_{Z}^{2}c_{\theta}^{2})$, where $s_{\theta}^{2}$ 
is the effective leptonic mixing angle at the Z-resonance, 
and the $W$-boson mass, which is defined through muon decay, 
are then derived quantities.
Since the loop factor occurring in radiative corrections,  
$1/16\pi^2$, is similar in magnitude to the expansion parameter, 
$v^{2}/f^{2}$, of chiral perturbation 
theory, the one-loop radiative corrections can
be comparable in size to  the next-to-leading order contributions at tree 
level. We compute the loop corrections to the $\rho$ parameter which are
enhanced by large logarithms; we focus on terms of order 
$1/(16\pi^2) \ln (M^2/Q^2)$, where 
$Q\sim M_Z$ and $M\sim f \sim {\cal O}(TeV)$.
At the one-loop level, we have to take into account the radiative correction 
to the muon decay constant $G_{\mu}$, the counterterm for the electric 
charge $e$, the mass counterterm of the Z-boson, and the counterterm 
for the leptonic mixing angle $s_{\theta}^{2}$. \\

The effective leptonic mixing angle is defined through the ratio of the 
vector to axial vector parts 
of the $Zee$ coupling, 
\begin{equation}
4s_{\theta}^{2} -1 = \frac{\mbox{Re}(g_{V}^{e})}{\mbox{Re}(g_{A}^{e})},
\end{equation}
which differs from the naive definition of the Weinberg angle in the
 littlest Higgs model, 
$s_{W}^{2} = g^{\prime 2}/(g^{\prime 2} + g^{2})$, by,
\begin{equation}
\Delta s_{\theta}^{2} \equiv s_{W}^{2} - s_{\theta}^{2} 
 = -\frac{1}{2\sqrt{2} G_{\mu} f^{2}}
\left[ s_{\theta}^{2} c^{2} (c^{2} - s^{2}) 
- c_{\theta}^{2} (c^{\prime 2} - s^{\prime 2}) (-2 + 5 c^{\prime 2})
\right] \; .
\end{equation}
The W-boson mass is defined through muon decay,
\begin{equation}
M_{W}^{2} = \frac{\pi \alpha}{\sqrt{2} G_{\mu} s_{\theta}^{2}}
\left[ 1 + \Delta r_{\mbox{tree}} + \Delta r^{\prime} \right] \; ,
\end{equation}
where $\Delta r_{\mbox{tree}}$ summarize the tree level corrections due to the 
change in definition in the weak mixing angle as well as the 
contributions from exchange 
of the heavy gauge bosons,
\begin{equation}
\Delta r_{\mbox{tree}} = -\frac{\Delta s_{\theta}^{2}}{s_{\theta}^{2}} + 
\frac{c^{2} s^{2}}{\sqrt{2} G_{\mu} f^{2}} \; ,
\end{equation}  
and the one-loop radiative corrections are collected in $\Delta r^{\prime}$,
\begin{eqnarray}
\Delta r^{\prime} & = & -\frac{\delta G_{\mu}}{G_{\mu}} - 
\frac{\delta M_{W}^{2}}{M_{W}^{2}}
+ \frac{\delta \alpha}{\alpha} - \frac{\delta s_{\theta}^{2}}{s_{\theta}^{2}}
\\
& = & \frac{1}{M_{W}^{2}} \left[ \Pi^{WW}(M_{W}) - \Pi^{WW}(0) \right]
+ \Pi^{\gamma\gamma}(0)^{\prime} - 
\frac{c_{\theta}}{s_{\theta}} \frac{\Pi^{\gamma Z}(M_{Z})}{M_{Z}^{2}}
\nonumber\; .
\end{eqnarray} 
When deriving this equation, we have used 
\begin{equation}
\frac{\delta s_{\theta}^{2}}{s_{\theta}^{2}}  = \mbox{Re}
\bigg\{ \frac{c_{\theta}}{s_{\theta}} \biggl[
\frac{\Pi^{\gamma Z}(M_{Z})}{M_{Z}^{2}} - 
\frac{v_{e}}{2s_{\theta}c_{\theta}}
\biggl(\frac{a_{e}^{2}-v_{e}^{2}}{a_{e}v_{e}} \Sigma_{A}^{e}(m_{e}^{2}) +
\frac{\Lambda_{V}^{Zee}(M_{Z})}{v_{e}}
-\frac{\Lambda_{A}^{Zee}(M_{Z})}{a_{e}}\biggr)\biggr]\bigg\}
\; ,
\end{equation}
where $\Sigma_{A}^{e}$ is the axial part of the electron 
self-energy and $\Lambda_{V,A}^{Zee}$ are the vector and 
axial vector parts of the $Ze\overline{e}$ vertex corrections. 
This follows from the fact that the counter term for $s_{\theta}^{2}$ 
is formally related to the wave function renormalizations for 
$\gamma$ and $Z$. 
The dominant contribution, $\Pi^{\gamma Z}(M_Z)$, depends on 
$m_{t}$ only logarithmically. Due to this logarithmic dependence,
 the constraint on the model is weakened. On the other hand,
 the scalar contributions become important as they are quadratic 
due to the lack of the tree level custodial symmetry.\\

We find that the one-loop contribution to $\Delta r_{\prime}$ due to 
the SU(2) triplet scalar field, $\Phi$, scales as 
\begin{equation}
\frac{1}{16\pi^{2}} \frac{v^{\prime 2}}{v^4} M_{\Phi}^{2} ~.
\end{equation} 
In the limit $v^{'} = 0$ while keeping $f$ fixed, which is equivalent to turning 
off the coupling $\lambda_{h \Phi h}$ in the Coleman-Weinberg potential, 
the one loop contribution due to the SU(2) triplet, 
$\Delta r_{Z}^{s}$, vanishes. 
The large $f$ limit of the scalar one-loop contribution, 
$\Delta r_{Z}^{s}$, vanishes depending upon how the limit 
$f \rightarrow \infty$ is taken~\cite{Chen:2003fm,Chen:2004ig,Chen:2005jx}.
As $f$ approaches infinity, the parameter $\mu^{2}$ (thus $v^{2}$) 
can be kept to be of the weak scale by fine-tuning the unknown coefficient 
in the mass term $\mu^{2}$  
in the Coleman-Weinberg potential  
while all dimensionless parameters remain of order one. 
The scalar one-loop contribution in this limit does {\it not} de-couple because 
$M_{\Phi}^{2}$ 
increases as $f^{2}$ which compensates the $1/f^{2}$ suppression from 
$v^{\prime 2} /v^{2}$. In this case, the SM Higgs mass $m_{H}$ is of the weak 
scale $v$. 
On the other hand, without the fine-tuning mentioned above, 
$v$ can be held constant while varying $f$, 
if the quartic coupling $\lambda_{h^{4}}$ (thus $\lambda_{\Phi^{2}}$) 
approaches infinity as $f^{2}/v^{2}$. 
This can be done by taking $a \sim f^{2}/v^{2}$ while keeping 
$a^{\prime}$ finite and $s$ and $s^{\prime}$ having specific values. 
The scalar one-loop contribution then scales as
\begin{equation}
\Delta r_{Z}^{s} 
\sim 
\frac{1}{v^{2}} (\frac{v^{\prime}}{v})^{2} M_{\Phi}^{2}
\sim 
(\frac{1}{v^{2}}) 
(\frac{\lambda_{h\Phi h}}{\lambda_{\Phi^{2}}})^{2} \frac{v^{2}}{f^{2}} 
\lambda_{\Phi^{2}} f^{2}
\rightarrow \frac{\lambda_{h\Phi h}^{2}}{\lambda_{\Phi^{2}}} \quad .
\end{equation}
Since the coupling constant $\lambda_{\Phi^{2}}$ must approach 
infinity in order to keep $v$ constant as we argue above, 
the scalar one-loop contribution 
$\Delta r_{Z}^{s}$ 
thus vanishes in the limit $f \rightarrow \infty$ with $v$ held fixed and
no fine tuning. 
In this case, $m_{H} \sim \mu$ scales with $f$.  \\

\begin{figure}[tb]
\begin{center}
\psfrag{M(theory)-M(exp) (GeV)}[][]{$M_{\mbox{theory}}
-M_{\mbox{exp}}$ (GeV)}
\psfrag{x_L}[][]{$x_{L}$}
\psfrag{Mw  }[][]{$\delta M_{W_{L}}$(total) $\qquad $}
\psfrag{Mz  }[][]{$\delta M_{Z}$(total) $\qquad $}
\psfrag{Mwtree  }[][]{$\delta M_{W_{L}}$(tree) $\quad $}
\psfrag{deltaMw}[][]{1 $\sigma$ limit on 
$\delta M_{W_{L}}$ (exp) $\qquad \qquad \qquad$ }
\psfrag{deltaMz}[][]{1 $\sigma$ limit on 
$\delta M_{Z}$ (exp) $\quad \qquad \qquad \qquad $}
\includegraphics*[angle=270,width=11cm]{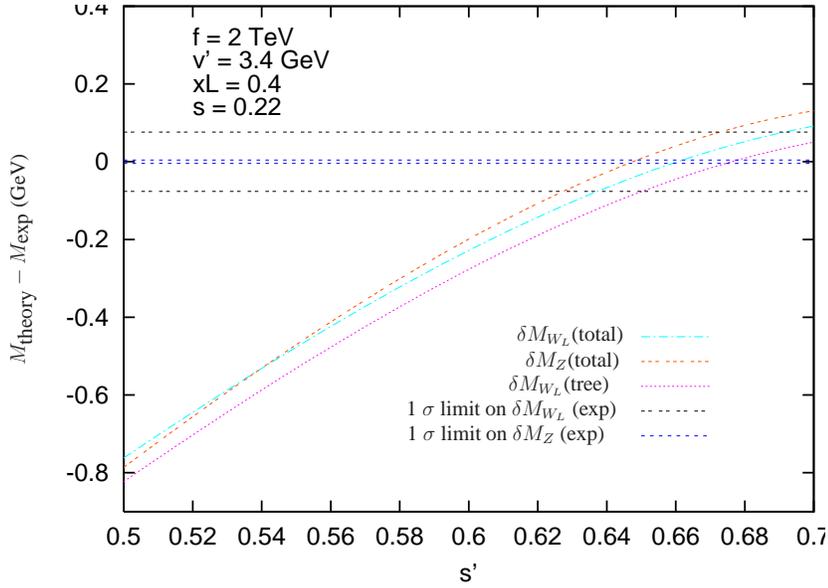}
\caption{%
Prediction for $M_{W_{L}}$ as a function of the mixing angle 
$s^\prime$ at the tree level and the one-loop level. 
Also plotted is the correlation between $M_{Z}$ and 
$s^\prime$ for fixed $s$, $v^{\prime}$ and $f$. 
The cutoff scale $f$ in this plot is $2$ $TeV$, the $SU(2)$ triplet VEV 
$v^\prime = 3.4 \; GeV$, the mixing angle $s=0.22$, and $x_{L}=0.4$.
}
\label{fig1}
\end{center}
\end{figure}

We analyze the dependence of the W-boson mass, $M_{W_{L}}$, on the 
mixing between 
$SU(2)_{1}$ and $SU(2)_{2}$, described by $s^{\prime}$, 
the mixing between $U(1)_{1}$ 
and $U(1)_{2}$, described by $s$, the mixing parameter in $t-T$ 
sector, $x_{L}$, 
and the VEV of the $SU(2)$, 
$v^{\prime}$. The predictions for $M_{W_{L}}$ with and without the 
one-loop contributions for $f=2$ TeV is given in Fig.~\ref{fig1}, 
which demonstrates that a low value of $f$ ($f\sim 2~TeV$) is allowed by 
the experimental restrictions from the $W$ and $Z$ boson masses, 
provided the VEV of the $SU(2)$ triplet scalar field is non-zero. 
This shows the importance of the $SU(2)$ triplet in placing the electroweak 
precision constraints. In order to have experimentally acceptable 
gauge boson masses, however, 
the parameters of the model must be quite finely tuned, regardless 
of the value of the scale $f$. 
On the other hand, the prediction for $M_{W_{L}}$ is very 
sensitive to the values of $s^{\prime}$ as well as $v^{\prime}$.

\begin{figure}
\begin{center}
\psfrag{xL = 0.4}[][]{\large $\large x_{L} = 0.4$}
\psfrag{s ' = 0.5}[][]{\large $\large s^{\prime} = 0.5$}
\psfrag{s = 0.2}[][]{\large $\large s = 0.2$}
\psfrag{v ' = 1 GeV}[][]{\large $\large v^{\prime} = 1 GeV$}
\psfrag{Dtree  }[][]{\large $\large \Delta_{\mbox{tree}}\qquad \qquad $  }
\psfrag{Df  }[][]{\large $\large \Delta r_{Z}^{f} \qquad \qquad \qquad \; $}
\psfrag{Ds  }[][]{\large $\large \Delta r_{Z}^{S} \qquad \qquad \qquad \; $}
\psfrag{D1loop  }[][]{\large $\large \Delta \hat{r}_{Z} -\Delta_{\mbox{tree}} \qquad   $}
\psfrag{piww  }[][]{\large $\large \Pi^{WW}(0)/M_{Z}^{2}  \qquad $}
\includegraphics*[angle=270,width=11cm]{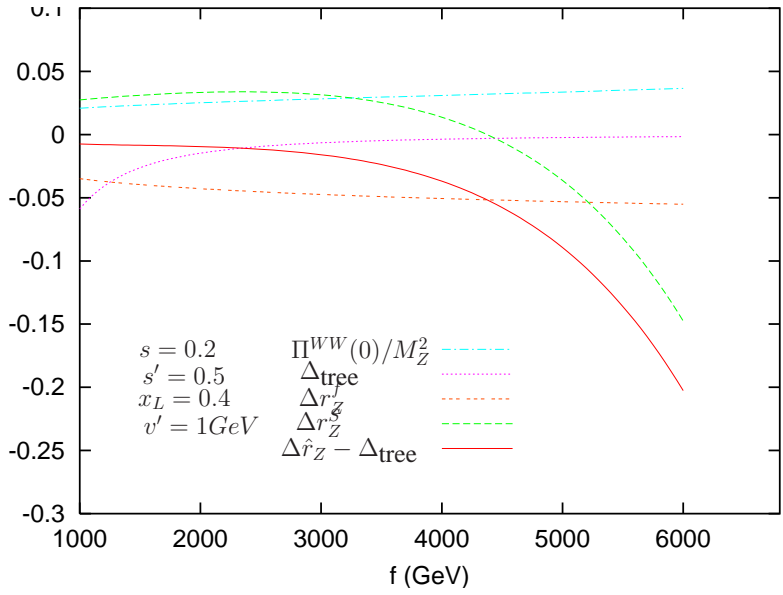}
\caption{%
The tree level correction, $\Delta_{\mbox{\tiny tree}}$, the 
fermionic and scalar contributions to the one loop correction, 
$\Delta r_{Z}^{f} $  
and $\Delta r_{Z}^{S}$,
the total one loop correction, $\Delta \hat{r}-\Delta_{\mbox{tree}}$, 
and $\Pi^{WW}(0)/M_{Z}^2$ 
as functions of the cutoff scale $f$ at fixed $s$, $s^{\prime}$, 
$x_{L}$ and $v^{\prime}$.
}
\label{fig2}
\end{center}
\end{figure}

The non-decoupling of the SU(2) triplet scalar field shown in 
Fig.~\ref{fig2} implies the 
importance of the inclusion of the scalar one-loop 
contributions in the analyses.
In the region below $f=4~TeV$, where the tree level corrections are large, 
the vector boson self-energy is about half of the size of the tree 
level contributions, but 
with an opposite sign. (Other one-loop contributions roughly 
cancel among themselves in this region). Due to this cancellation 
between the tree level correction and the one-loop
correction, there is an allowed region of  parameter space
with low cutoff scale $f$.
Fig.~\ref{fig2} also shows that the tree level contribution of the LH
model get smaller as $f$ increases, as is expected.
In order to be consistent with experimental data, the triplet 
VEV $v^{\prime}$ 
must approach zero as $f$ goes to infinity.
Our results emphasize the need for a full one loop calculation.

The forth input parameter in the gauge sector is needed in any new models where a $SU(2)_{L}$ triplet with a non-vanishing VEV is present. In addition to the littlest Higgs model, models of this kind include the SM with a  triplet Higgs and the left-right symmetric model based on $SU(2)_{L} \times SU(2)_{R} \times U(1)_{B-L}$. A unique collider signature of models with a triplet Higgs is the decay of the doubly charged component of the triplet into same sign di-leptons, $\phi^{--} \rightarrow \ell^{-} \ell^{-}$. The Tevatron and the LHC thus have the capability to discover a triplet Higgs if its mass is of the order of a TeV. This decay mode  is unique in the sense that it does not exist in MSSM or  other extensions of the SM having only Higgs doublets or singlets. It is interesting to note that the operator which leads to the decay $\phi^{--} \rightarrow \ell^{-}\ell^{-}$ also contributes to the LH Majorana neutrino masses. This thus provides an interesting possibility of probing the neutrino mixing parameters at collider  experiments.


\clearpage\setcounter{equation}{0}\setcounter{figure}{0}\setcounter{table}{0}
\subsection{Expected Signatures of Charged Massive Stable Particles at the 
Tevatron}
\label{sec:stable}

C.~Cl\'{e}ment$^a$, Y.~de Boer$^{b,c}$, M.~Karlsson$^d$, D.~Milstead$^a$\\ [2mm]
{\em
$^a$ Stockholm University, Stockholm, Sweden \\  [2mm]
$^b$ Institute for Theoretical and Experimental Physics, Moscow, Russia \\ [2mm]
$^c$ University of Twente, Enschede, the Netherlands \\ [2mm]
$^d$ Lund University, Lund, Sweden \\
}


{\it The expected signatures of a range of charged massive stable
particles produced in proton-antiproton collisions at $2$ TeV
centre-of-mass energy were investigated using QCD-based models. The
fragmentation properties of jets containing $R$-hadrons formed from
stable stops and gluinos allow discrimination between $R$-hadron and
Standard Model jets. Interactions of stable massive particles in material 
were shown to give sensitivity to the species of scattering particle.
}

\subsubsection{Introduction} 

Many candidate theories beyond the SM
predict the presence of charged massive stable\footnote{The term
stable is taken to refer to particles which do not decay over a time
scale corresponding to their passage through a typical detector.}
particles (CMSPs). Different types of CMSPs arise in a number of
scenarios of SUSY, Universal Extra Dimensions, leptoquarks, and
various unification models.


One of the primary tasks of the Tevatron and LHC programs is
therefore to demonstrate or disprove the existence of CMSPs. A
number of experimental searches have been performed at the Tevatron
and other 
colliders\cite{Perl:2001xi,Aktas:2004pq,Abdallah:2002qi,Abe:1992vr,Acosta:2002ju,Pagliarone:2003ya,Eads:2005bz} 
and much preparatory work is underway at the
LHC\cite{Nisati:1997gb,zal:1999,giac:2001,hellman:2005,Kraan:2005ji}. The extracted
exclusion limits are dependent on the models used for the
cross-sections and span up to approximately $200$ GeV in mass in the
most optimistic scenarios. One typical search strategy which has
been employed is to use time-of-flight information to isolate slow
moving muon-like tracks. However, hadronic CMSPs (usually referred
to as $R$-hadrons in the context of SUSY\footnote{In this paper, the
term $R$-hadron refers to an exotic hadron containing an unspecified
sparticle, while the terms R$_{\tilde {g}}$-hadron
 and R$_{\tilde q}$-hadron refer to particles containing a gluino and squark,
 respectively.}) can evade detection in
this way through charge exchange
interactions\cite{Baer:1998pg,Raby:1998xr,Mafi:1999dg,Kraan:2004tz} with passive detector
material. For example, a charged $R$-hadron can convert to a neutral
state through nuclear interactions in the calorimetry and thus not
be recorded in the muon chambers. However, since the available
charge exchange reactions depend on the species of $R$-hadron
undergoing scattering, charge exchange interactions also offer a
possibility to differentiate between CMSP scenarios.

As part of this workshop the detector signatures of stable staus,
gluinos and stops were studied as a means to develop analysis tools
for the possible discovery and quantification of these sparticles.
The selected sparticles are predicted to be stable in a number of
scenarios. The observables studied here are relevant for any generic
search for a colourless particle with charge $\pm e$, a charge 
$\pm\frac{2}{3}e$ colour triplet, or
an electrically uncharged colour octet state. QCD-based models were used 
to study the production mechanisms, fragmentation properties and 
subsequent scattering in matter of the sparticles. In performing this work
several potential experimental challenges were highlighted which may
need to be addressed in order to discover CMSPs or definitively
exclude their presence at either the Tevatron or the LHC.

\subsubsection{Theoretical Background} The topic of CMSPs was recently
brought to the fore by the emergence of the theory of
Split-SUSY\cite{Arkani-Hamed:2004fb,Giudice:2004tc} in which the
gluino can be stable. Within this approach, the hierarchy problem
and the fine-tuning of the Higgs mass are accepted. SUSY is still 
necessary to unify the gauge couplings, but by accepting the fine-tuning 
of the Higgs mass, Split-SUSY proposes a way to break the symmetry at
scale above $1000$ TeV. The scalar particles, except for a single
neutral Higgs boson acquire masses at this high scale. Chiral
symmetries assure that the fermions possess masses around the TeV
scale. Split-SUSY still provides a dark matter candidate and
furthermore, possesses none of the difficulties in describing
electric dipole moments\cite{Polchinski:1983zd,Giudice:2005rz} or flavour changing neutral
currents\cite{Ellis:1981ts} which challenge the Minimal Supersymmetric
Standard Model. A further consequence  of Split-SUSY is that the
gluino can then become meta-stable since it decays through a squark and 
the decay is therefore suppressed. For values of squark
masses above around $10^6$ GeV, a produced gluino can form a
$R_{\tilde{g}}$-hadron which is sufficiently stable so as to
propagate through a Tevatron detector. The potential of the Tevatron
and the LHC to discover stable gluons has been investigated in a
number of works\cite{Hewett:2004nw,Kilian:2004uj,Kraan:2005ji,hellman:2005}. In
addition to Split-SUSY, stable gluinos also arise in other SUSY
scenarios\cite{Baer:1998pg,Kaplunovsky:1993rd,Chen:1996ap,Raby:1997pb,Starkman:1990nj,Mohapatra:1997sc} 
including GMSB.

In the context of GMSB, it is, however, more common that searches
are performed for meta-stable staus\cite{Raby:1997pb,Drees:1996ca}. 
A very light
gravitino as a LSP which couples very weakly to the other particles
is a characteristic of GMSB models. The NLSP is usually a neutralino
or one of the sleptons. If the mixing of the stau states $\tilde{\tau_L}$
and $\tilde{\tau_R}$ is non-negligible then the lightest stau
$\tilde{\tau_1}$ can also become lighter than the other sleptons and
the neutralino and therefore be the only NLSP. The lifetime of the
NLSP depends on the gravitino mass (or equivalently the SUSY
breaking scale) and meta-stable staus can be expected over a
sizable part of the parameter space open to the Tevatron and LHC.

Long lived charged particles are also predicted in five-dimensional
SUSY\cite{Barbieri:2000vh}. In this model the Standard Model is embedded in a
supersymmetric theory with a compactified extra dimension. In this 
scenario, a stable stop
with mass around 200 GeV is predicted.

CMSPs are also predicted in a number of alternative exotic
scenarios. Theories of leptoquarks\cite{Friberg:1997nn}, Universal Extra
Dimensions\cite{Appelquist:2000nn}, certain unification models\cite{Ingelman:1986dp}, and
theories which postulate new SM fermions\cite{Frampton:1997up,Fritzsch:1978xk,Fishbane:1983hf}. Magnetic monopoles are a further type of CMSP
which have been sought. The existence of Dirac Monopoles addresses
the question of electric charge quantisation\cite{Dirac:1931kp,Dirac:1948um} and Dirac
Monopoles are themselves predicted within unification
models\cite{'tHooft:1974qc,Polyakov:1974ek}.

\subsubsection{$R$-hadrons in Jets}
\section*{\normalsize Fragmentation of $R$-hadrons}

Hard scattering events for $p\bar{p}$ interactions
 are simulated using the leading order generator 
{\sc Pythia/Jetset} 6.3\cite{Sjostrand:2006za}. The effects of initial and
final state QCD radiation are described in {\sc Pythia} by leading
logarithm parton showers. In {\sc Pythia} the fragmentation of
partons into hadrons follows the Lund string model\cite{Andersson:1983ia}.
The hadronisation of gluinos and stops was performed within {\sc
Jetset} using special routines for this purpose\cite{rout}. A
Peterson fragmentation function parameter\cite{Peterson:1982ak},
extrapolated to the $R$-hadron mass region under study, was used. The
fragmentation parameter is computed according to
\begin{equation}
\frac{\epsilon_{ {\tilde{q}\tilde{g}} }}{\epsilon_b} = \frac{m_b^2}
{m_{{\tilde{q}}{\tilde{g}}}^2}
\end{equation}

Since a gluino is a colour octet, two colour strings are attached to
it and a gluino-induced jet is thus expected to possess a larger particle 
multiplicity than a jet initiated by a squark.

Mesonic states
 and baryonic states are produced in the hadronisation model.
 Furthermore, neutral $R_{\tilde{g}}$-hadrons can be formed as gluino ball
($\tilde{g}g$) states. The probability $P_{{\tilde{g}}g}$ of forming
a gluino ball in the hadronisation step is set by default to $0.1$
within {\sc Pythia}. However, the fraction of gluino balls is {\it a 
priori}
unknown and any comprehensive search strategy should therefore also
consider scenarios in which CMSPs are produced dominantly as neutral
states.

Each state is set stable and it has been predicted
\cite{Kraan:2004tz} that the mass splitting is sufficiently small so
as to exclude the decay into a low lying neutral mass state.


\section*{\normalsize Jet Properties of $R$-hadrons}

Searches for stable staus and sleptons often impose isolation
criteria either in calorimeter or tracking systems to reject
background. Since a $R$-hadron is produced within a jet, searches
cannot rely on isolation. However, measurements of $R$-hadrons in a
jet could then be used to distinguish between $R$-hadrons and staus.
Furthermore, the jet structure can be used to characterise
$R$-hadrons and as a search tool. This can be particularly useful in
scenarios in which the $R$-hadron is not recorded in a muon system
owing to the effects of nuclear interactions.

\begin{figure}
\center \epsfxsize = 16.0cm \epsfbox{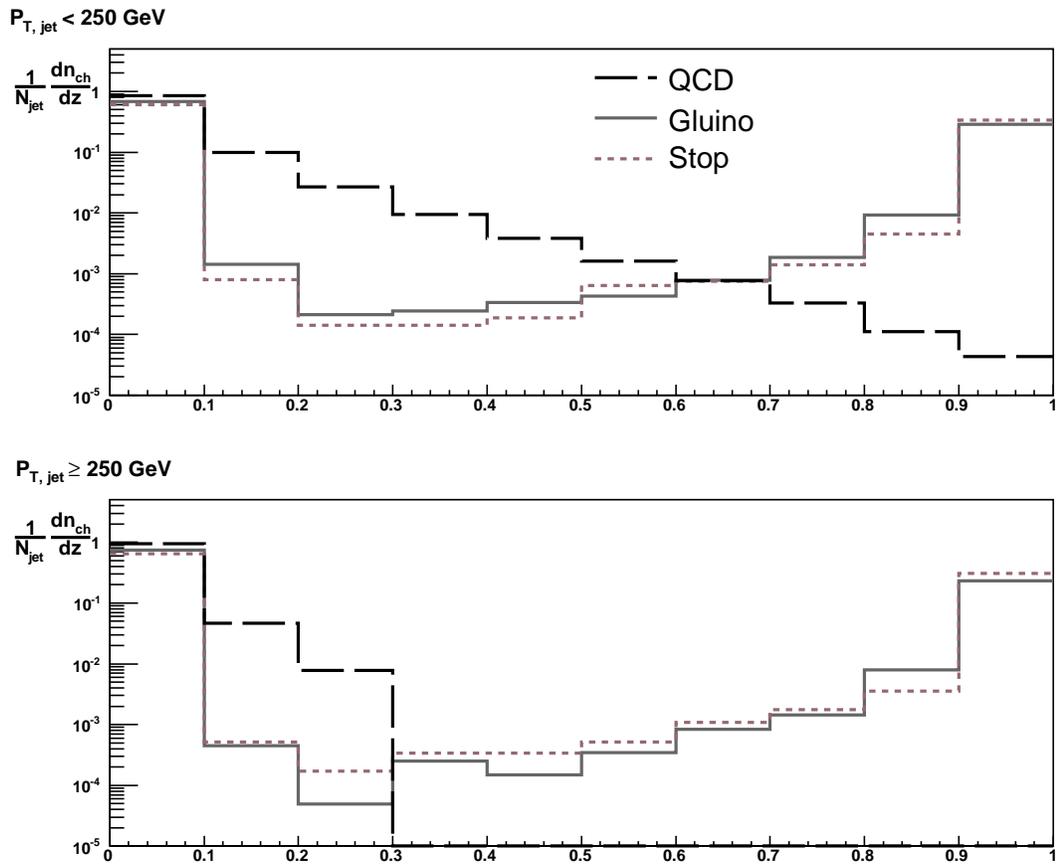} \caption{The
distribution of $(1/N_{jet})(dn_{ch}/dz)$ for charged
particles in jets in $R$-hadron and QCD events. The distributions
are shown for two intervals of jet transverse momentum.}
\label{fig:jet2}
\end{figure}

Samples of pair produced gluinos and stop-antistops with masses of
$300$ GeV/c$^2$ were studied and compared with QCD dijet events. The
distribution $\frac{1}{N_{jet}}\frac{dn_{ch}}{dz}$ is shown in
Fig.~\ref{fig:jet2} for different intervals of jet transverse momentum. 
Here, $z$ is a fragmentation variable
$z=p_{ch}/p_{jet}$ defined for all charged particles which are
reconstructed within a jet. The variable $z$ is formed from momenta
of the charged particle ($p_{ch}$) and the jet ($p_{jet}$).
$N_{jet}$ and $n_{ch}$ are the number of jets and charged particles,
respectively. A minimum cut on the jet transverse energy of $20$
GeV has also been applied.

Jets were found using a cone algorithm\cite{Abe:1991ui} in the
pseudorapidity region $|\eta|<2$. The jets were formed from stable
particles produced following hadronisation and a cone radius of
$R$=0.7 was used.

The $z$ distribution for the QCD sample shows an expected behaviour
with a large rate of low momentum particles produced from QCD
radiation and leading particles. The $R$-hadron samples show a
large rate of low momentum tracks but in addition a leading
$R$-hadron populates the high $z$ region.

The dependence of $\frac{1}{N_{jet}}\frac{dn_{ch}}{dz}$ on the jet
transverse momentum can be used to discriminate signal from background. 
Fig.~\ref{fig:jet2} shows evidence of 
the classic scaling
violations for the QCD dijet case with the depopulation
of the high $z$ region, as would be expected from perturbative
QCD\cite{Altarelli:1979kv}. However, for the $R$-hadron samples, the peak at
high $z$ remains relatively constant.

\begin{figure}
\center \epsfxsize = 11.0cm \epsfbox{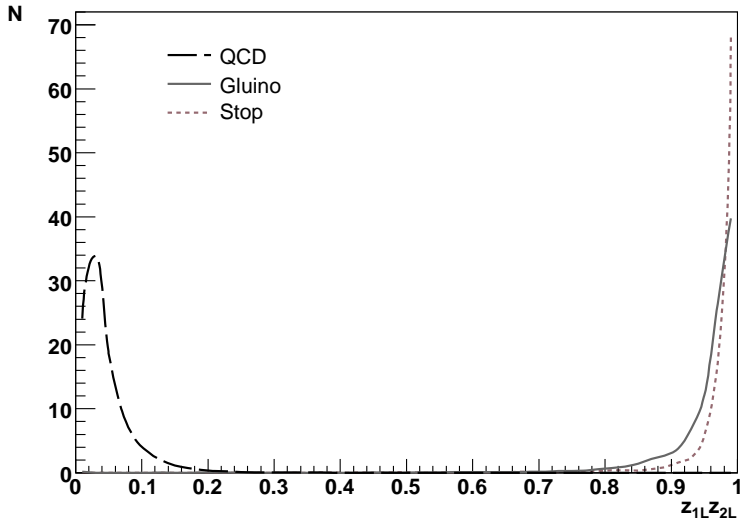} \caption{The
distribution of $z_{1L}z_{2L}$ for leading charged particles in
events containing two jets. Distributions are shown for $R$-hadron
and QCD events. The distributions are normalised to contain the same
yield of two-jet events.} \label{fig:dijet}
\end{figure}

Since $R$-hadrons can be produced in pairs, the correlation between
leading $R$-hadrons can offer further discrimination. The
distribution $z_{1L}z_{2L}$ is shown in Fig.~\ref{fig:dijet}. Here
$z_{1L}$ and $z_{2L}$ are the values of $z$ for the leading
particles in the first and second jets, respectively. The 
normalisation is arbitrarily chosen to
provide same-sized samples of QCD and $R$-hadron 2-jet events. The
$R$-hadron distributions remain peaked above $0.9$ while the QCD
spectra now peaks at around $0.05$. The stop sample 
peaks at a higher value of $z$ than the gluino sample, as would be expected from the 
different colour string topologies associated with the different types of 
sparticle.

\begin{figure}
\center \epsfxsize = 11.0cm \epsfbox{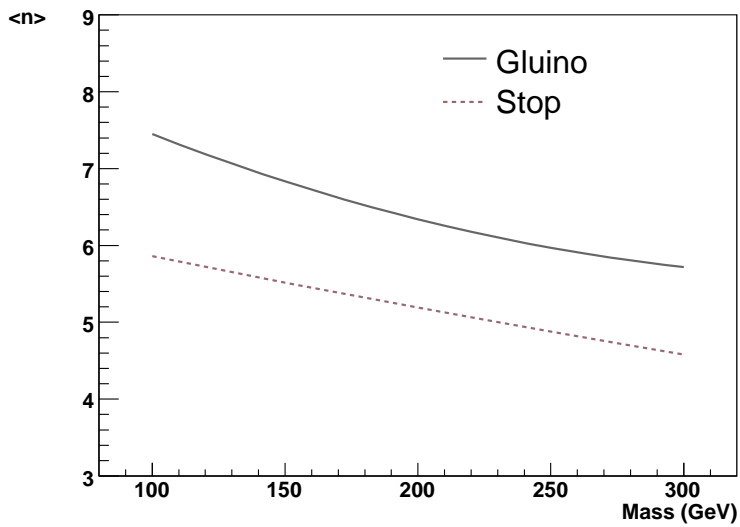} \caption{The average
multiplicity of gluino and stop jets as a function of sparticle
mass.} \label{fig:mul}
\end{figure}

The multiplicity of gluino and stop jets is shown in
Fig.~\ref{fig:mul}. The multiplicity falls with mass as less phase
space is available for QCD emissions. The gluino multiplicity
exceeds that from stop jets. An enhanced multiplicity at low
momentum would be expected owing the different QCD colour
factors involved in hadronising squark and 
gluinos\cite{Brodsky:1976mg}.

The fine details of the fragmentation spectra presented here are
subject to a number of theoretical uncertainties such as those due
to the treatment of the gluino constituent mass and the choice of
the gluino fragmentation function. However, the gross features of
the distributions presented would not be expected to be sensitive to
these effects.

While the golden channel for observing a CMSP would be a slow
penetrating particle, the study of jet properties would provide
supplementary information regarding the fragmentation of a heavy
coloured object in the case of a discovery. Furthermore, it could
also act as a component of a search strategy in its own right for
events not containing a slow muon-like candidate. Such events could
arise due to charge exchange interactions as described in
section~\ref{sec:scatt}. However, there would be important
experimental issues to address. An inefficiency in a muon detector
could lead to the measurement of a $R$-hadron like $z$ spectra for
heavy quark production. Of greater importance would be the
triggering of $R$-hadrons which were not recorded as muons. Since
the energy deposition due to a R-hadron is likely to be maximally
around $15$ GeV this may not be enough to trigger on an event. Thus,
an event with two R-hadron jets and no muon signature would not be
recorded unless there was evidence of a third jet arising from a
higher order process. Three-jet processes and their
implication for Tevatron limits in the case $R_{\tilde{g}}$-hadron
production have already been considered in\cite{Hewett:2004nw}.

\subsubsection{Scattering of CMSPs in Matter}\label{sec:scatt}

\section*{\normalsize Modelling CMSP scattering}

Non-coloured CMSPs such as staus  are usually
treated as heavy muons when modelling their propagation through a
detector. Continuous ionisation and repeated elastic Coulomb
scatterings with nuclei are handled by modified {\sc Geant}
routines\cite{geant3}.

However, the situation is more complicated for $R$-hadrons for which
electromagnetic and hadronic energy loss are important. Calculations
of the fine details of $R$-hadron nuclear scattering in matter are
uncertain, and a number of models have been proposed in the
literature\cite{Baer:1998pg,Raby:1998xr,Mafi:1999dg,Kraan:2004tz}. 
Although the
phenomenology and predictions differ between the various approaches,
some generic, well-motivated features of $R$-hadron scattering
exist.

The probability of an interaction between a heavy coloured parton in
the $R$-hadron and a quark in the target nucleon is low since the
cross-section varies with the inverse square of the parton mass
according to perturbative QCD. When modelling the scattering of a
$R$-hadron in material, one can thus use the central picture of a
stable non-interacting gluino accompanied by a coloured hadronic
cloud of light constituents, which are responsible for the
interactions. The effective interaction energy is therefore small
and equivalent to the interactions of a pion of energy of around
several GeV with a stationary nucleus.

A further feature of $R$-hadron scattering is that, following
multiple scatterings, a mesonic $R$-hadron will almost always have
converted into a baryonic state\cite{Kraan:2004tz}. This is due to
phase space suppression for baryon-to-meson conversion and the lack
of available pions, with which the R-baryon must interact in order
to give up its baryon number. The available baryon states are $S^0$,
$S^+$,$S^-$, and $S^{++}$. The state $S^{--}$, if formed from either
a squark or gluino $R$-hadron, would possess negative baryon number
and would immediately interact to become a meson\footnote{In this
paper $R$, $M$ and $S$ are generic labels used to denote a hadron,
meson, and a baryon, respectively. When appropriate, a superscript
denotes the charge and a subscript denotes the species of heavy
sparticle.}.

An important consequence of the nuclear reactions is that they allow
a $R_{\tilde{g}}$-hadron to reverse the sign of its charge in
nuclear interactions. However, a hadron containing a stop-like
squark can not reverse the sign of its charge through hadronic
interactions alone.  Charge reversal could nevertheless take place
for $R_{\tilde{q}}$-hadrons via the oscillation of intermediate
$M_{\tilde{q}}^0$ and $M_{\tilde{\bar{q}}}^0$-states which could be
formed in the calorimeter. Since $u$-type sflavour violation
involving the third generation is largely unconstrained it has been
shown that extremely rapid oscillations over the scale of a detector
are conceivable\cite{Sarid:1999zx}, as are minimal oscillations.

Thus, an observation of $R$-hadrons which reverse the sign of their
charge and which form a doubly charged state could, on the basis of
fundamentally allowed and forbidden reactions indicate the existence
of a $R_{\tilde{g}}$-hadron or a $R_{\tilde{t}}$-like hadron in
which oscillations have occured, thereby providing information
concerning the squark couplings. The observation of a doubly charged
state and no charge reversal processes could be used to identify a
$R_{\tilde{t}}$-like hadron if a lower limit on the rate of charge
reversal processes for gluino $R$-hadrons can be calculated. A
further way of discriminating between stop and gluino $R$-hadron
hypotheses would be to use information on the charges of the tracks
before scattering. A pair of charged
$R_{\tilde{t}}$,$R_{\tilde{\bar{t}}}$ would always have unlike signs
unlike an equivalent pair of charged $R_{\tilde{g}}$-hadrons which
can either have like or unlike charges.


The model\cite{Kraan:2004tz} for scattering which is used in this
study is implemented in {\sc Geant-3}\cite{geant3}. The model
provides a simple and general framework of simulating nuclear
interactions of heavy hadrons. The total cross-section is set
constant for gluino R-meson and R-baryon interactions to 24mb and
36mb, respectively, based on arguments from quark counting and the
values obtained from low energy hadron-hadron scattering. The
relative fraction of 2-2 and 2-3 processes is determined by a phase
space factor. Since it would be impossible to calculate individual
Clebsch-Gordon coefficients for each matrix element for each
reaction, the matrix elements are assumed to be the same. A
$R_{\tilde{g}}$-hadron will typically interact around $10$ occasions
as it propagates through the calorimeter systems of the Tevatron and
LHC experiments. For this work, {\sc Geant-3} scattering routines
for $R_{\tilde{g}}$-hadrons were adapted to simulate the scattering
of $R_{\tilde{t}}$-hadrons within the framework of the existing
model.


\section*{\normalsize Signatures of CMSPs after Scattering }

Both CDF\cite{Abe:1992vr} and D0\cite{Eads:2005bz} have used time-of-flight
information to search for CMSPs. One advantage of this technique is
that it is highly effective in suppressing background and a search
becomes largely a counting exercise when systematic uncertainties
are under control. Furthermore, it is possible to reconstruct the
mass of a CMSP from timing information alone. This has been studied
at the Tevatron\cite{Eads:2005bz} and in preparatory physics studies for
the LHC\cite{zal:1999,hellman:2006}.

As previously mentioned in section~\ref{sec:scatt}, the propagation of
a stau presents fewest experimental difficulties. However,
$R$-hadron scattering is subject to a number of experimental
challenges. Charge exchange processes can lead to tracks which
possess oppositely signed electric charge in the inner and muon
tracking chambers. Similarly, the production of doubly charged
states following nuclear scattering gives rise to tracks in which
the inner track appears to have twice the transverse momentum of the
track reconstructed in the muon system. The response of track
reconstruction software to such tracks would be a critical
experimental issue in any search. Should these effects be prominent,
they may well already have impeded searches for $R$-hadrons at many
colliders. A study of the discovery potential of ATLAS to
$R_{\tilde{g}}$-hadrons using tracks which have reversed the sign of
their electric charge has already been performed\cite{hellman:2005}. The
Tevatron offers the possibility to develop such searches using
collision data.

For this workshop, a toy MC implementing the resolution of D0
tracking systems was used in order to gain an estimate of the expected
visibility of CMSP tracks and in particular of $R$-hadrons which
have undergone charge exchange. The resolutions of the muon and
inner tracking chambers were parameterized according to ref.
\cite{d0mutdr}. An additional 25\% smearing was applied to the muon
resolution terms and a charge misidentification rate of around
$20\%$ was assumed.

Nuclear interactions involving stop and gluino $R$-hadrons were
calculated using the model\cite{Kraan:2004tz} described in
section~\ref{sec:scatt}. A thickness of $11\lambda_T(\pi)$ was
assumed to model the D0 calorimetry.

\begin{figure}
\center \hspace*{-1.2cm} \epsfxsize = 18.0cm \epsfbox{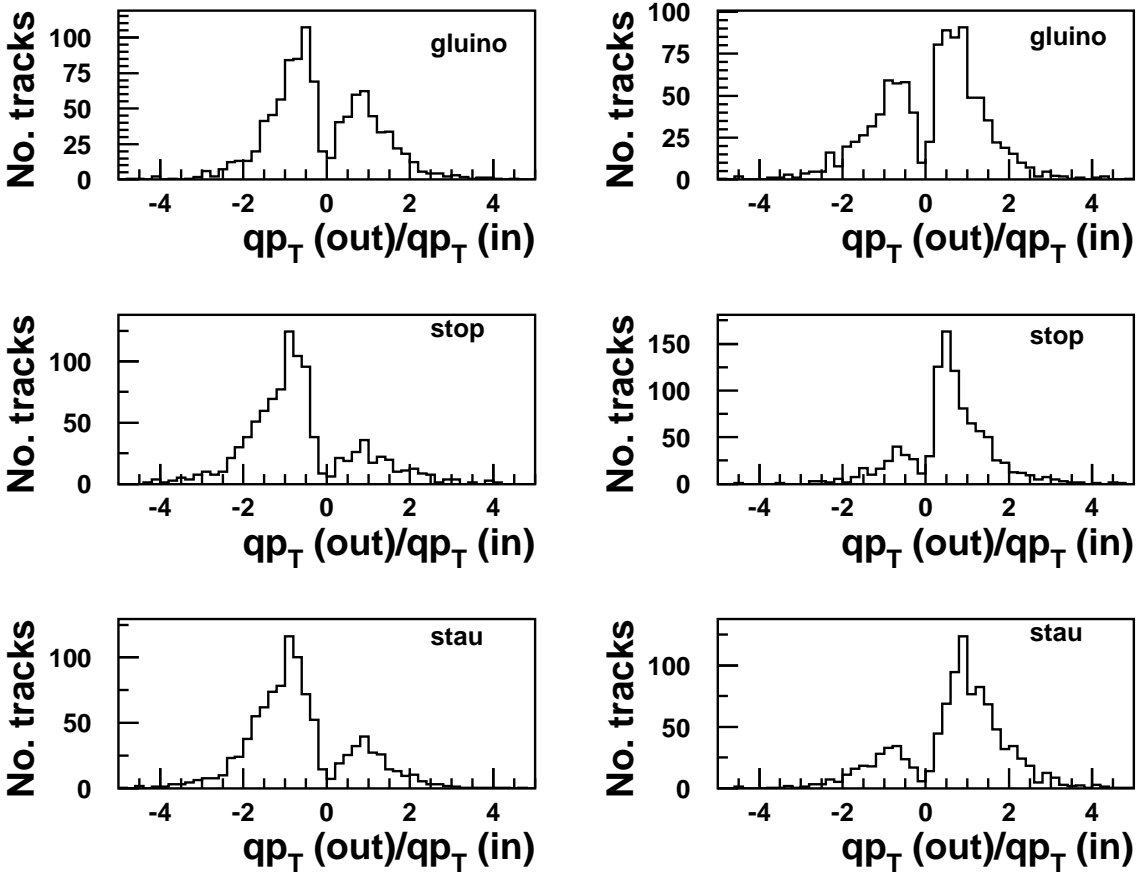} \caption{The ratio
$r=\frac{qp_T (in)}{qp_T (out)}$ for gluino and stop $R$-hadrons and
staus. Distributions are shown for tracks with positively (right)
and negatively (left) charged inner tracks. } \label{fig:flip1}
\end{figure}

Pairs of staus, stops and gluinos were generated using {\sc Pythia}
and subjected to acceptance cuts necessary for them to be identified
as slow moving particles\cite{Eads:2005bz}.

\begin{itemize}
\item The scaled speed of the CMSP $\beta$ was required to be less
than $0.65$.
\item The CMSPs were restricted to the central pseudorapidity region
($|\eta|<1.5$).
\item The CMSP transverse momentum was required to be greater than
$15$ GeV.
\end{itemize}

Fig.~\ref{fig:flip1} shows the ratio $r=\frac{qp_T (out)}{qp_T
(in)}$, where $q$ is the charge of the CMSP and $p_T (in)$ ($p_T
(out)$) is the momentum in the inner (muon) tracking chamber.
Negative values of $r$ would denote evidence of charge 'flipping'.
Spectra are shown for stop and gluino $R$-hadrons and staus, and are
shown for the cases in which the inner track has negative and positive
charge.

Interesting features emerge following scattering. The gluino spectra
shows a substantial (approximately $50\%$) rate of charge
'flipping'. A lower rate of charge reversal is expected for
positively charged inner tracks than negatively charged tracks since
there are two possible positive charge states for the emerging
$R$-hadron: $S_{\tilde{g}}^+$ and $S_{\tilde{g}}^{++}$. In this
model, the stop $R$-hadrons undergo minimal mixing of neutral mesino
states, and the only 'flipping' which arises comes from charge
misidentification. Maximal mixing would bring the stop spectra close
to the gluino one. A further feature of the stop and gluino spectra is 
that
the average momentum of the positive tracks in the muon system is
lower than the negative tracks. This arises due to the presence of
the doubly charged state which would be reconstructed with half of
the transverse momentum of the singly charged state. The staus flip
purely from charge misidentification.

\begin{figure}
\center \epsfxsize = 11.0cm \epsfbox{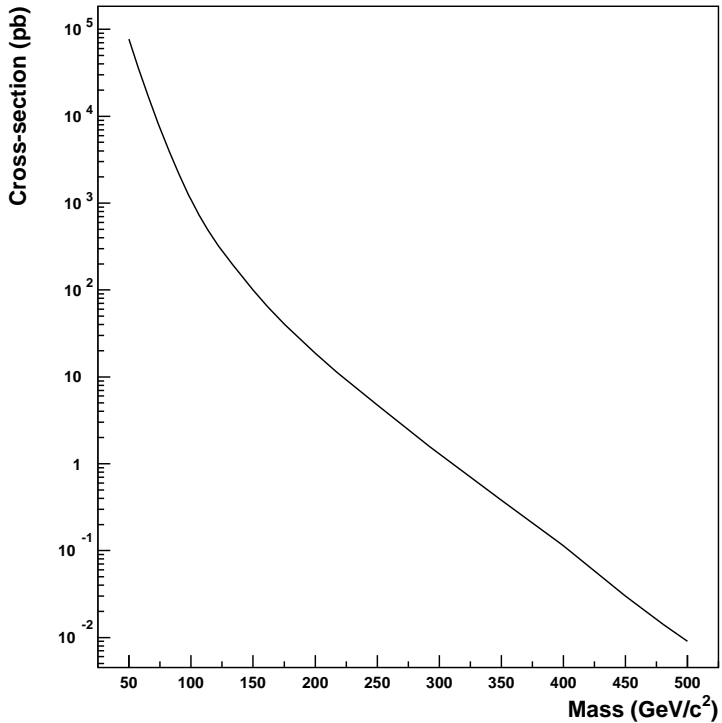} \caption{The NLO
cross-section for gluino pair production as a function of gluino
mass.} \label{fig:xsec}
\end{figure}

It is interesting to study the expected rate of tracks arising in a
Split-SUSY scenario in which the gluino is stable. Using
next-to-leading QCD 
calculations\cite{Beenakker:1995fp,Beenakker:1996ch} in a Split-SUSY
scenario\cite{hellman:2005}, the cross-section for the pair production
of gluinos is shown in Fig.~\ref{fig:xsec}. For an accumulated
luminosity of $2$ fb$^{-1}$ several thousand gluinos could be
expected for $300$ GeV mass.

\begin{figure}
\center \hspace*{-0.8cm} \epsfxsize = 18.0cm \epsfbox{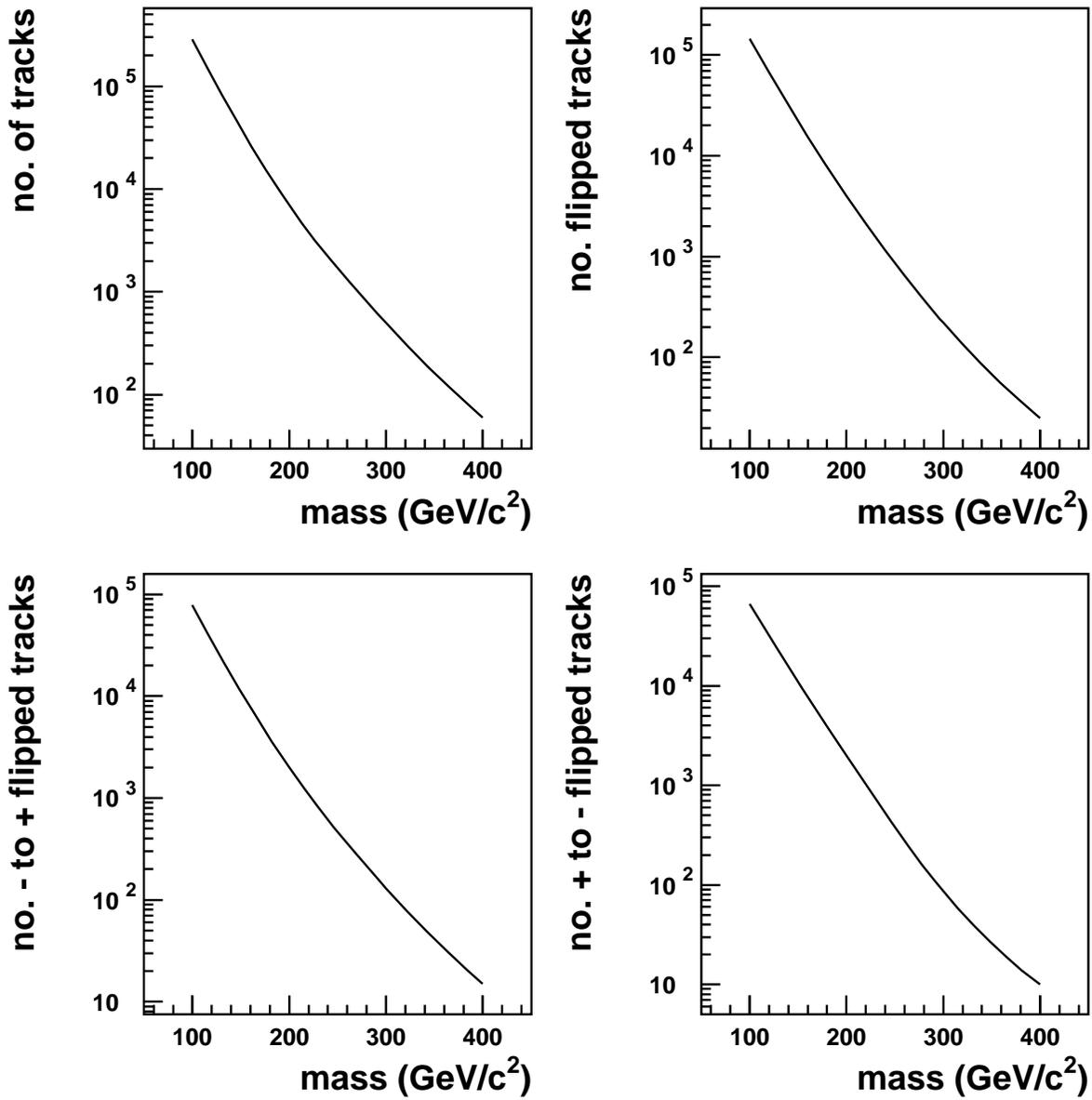} 
\caption{The yield of
gluino $R$-hadron tracks expected for $2$fb$^{-1}$ and the yields of
tracks which change the sign of their charge.} \label{fig:flip2}
\end{figure}

For a luminosity of $2$ fb$^{-1}$, the expected yield of gluino
$R$-hadron tracks passing the acceptance cuts described above is
shown in Fig.~\ref{fig:flip2}. The expected yields of tracks which
reverse the sign of their charge are also shown. The total yield of
charge reversing tracks and the amount of tracks undergoing
positive-to-negative and negative-to-positive changes are shown. In
an optimistic scenario, several hundred tracks would be accumulated
at $300$ GeV mass. However, it should be again be pointed out that
this represents a best-case scenario and detector effects will
undoubtedly degrade any signal. Nevertheless, if an excess could be
observed and charge reversal could be established, it would be
striking evidence for the existence of $R$-hadrons.

\subsubsection{Summary}

The existence of stable, heavy, charged particles has been predicted
within a number of different scenarios of physics beyond the
Standard Model. One of the tasks of the Tevatron and LHC programs 
will be to discover and characterise these particles or to exclude their 
production.

In this work, experimental signatures of long-lived staus, stops
and gluinos were studied. The fragmentation of stops and gluinos in
jets allows discrimination from QCD jets. Hadronic interactions of
$R$-hadrons with matter were considered. These give rise to
challenging experimental effects which can assist and impede any
search. Charge exchange processes provide striking signatures of
tracks which change the sign of their charges. Rates of such
processes were estimated and expected track yields were estimated for the 
D0 detector.

\clearpage\setcounter{equation}{0}\setcounter{figure}{0}\setcounter{table}{0}

\section{Model-based Phenomenology}
\label{sec:models}

In addition to the particle-based approach to phenomenology, there is
considerable focus on specific models of new physics, or classes of
models, and their phenomenological implications, both at colliders and
in cosmology and astrophysics, and indeed where these fields
intersect.  This approach is not exactly tuned to how experimental
searches at colliders tend to be conducted, but is still of great
value as many of the mainstream models are highly motivated by a
variety of theoretical and phenomenological arguments.  The recent
proliferation of new classes of models has, however, muddied the
waters somewhat.  Unless one closely follows these trends, it is easy
to become lost in the landscape.

A complicating factor is the realization that models of new physics
which incorporate dark matter candidates typically yield collider
signatures of cascade decays of heavy particles through lighter ones
down to SM particles plus the dark matter.  That is, many very
different types of new physics can appear in experiments with the
classical SUSY signature of high-multiplicity leptons, jets and
missing energy.  This realization presented a phenomenological crisis,
and drives much of the current effort to glean more information from
cascade decays and other data to disentangle a signature from a
largish set of possible explanations.  The first contribution of this
chapter, Sec.~\ref{sec:spin}, addresses this problem.

Much effort is still directed, however, at the initial fleshing-out of
production and decay channels, and in creating generators for these
processes which can be used for practical phenomenology.  Sections
\ref{sec:UED},\ref{sec:ED},\ref{sec:LHT} and~\ref{sec:TC}
do this for extra-dimensional, Little Higgs T-parity, and technicolor
models, the last via a ``straw-man'' framework meant to establish a
working language for generic signatures anticipated by technicolor
models.  It is especially notable because Tevatron has significant
potential to observe these signatures, and Run~I data contains a few
interesting hints which have not yet been followed up on in Run~II.

Sec.~\ref{sec:slep} on slepton mass measurement improves existing
techniques for determing supersymmetric lepton partner masses at the
LHC, in cases where established techniques would have considerable
difficulty.  The technique should be extendible to other types of
models as well, and would be important for formulating statements
about dark matter from collider data on new observed states.

Moving even further in this direction,
Sections~\ref{sec:lstop1},\ref{sec:lstop2} and~\ref{sec:msugra}
address specifically the issue of combining data from Tevatron, LHC
and a future international linear collider (ILC) to connect
supersymmetry, dark matter and cosmology in both general mSUGRA and
electroweak baryogensis scenarios.

The final two contributions, Sections~\ref{susytools}
and~\ref{sec:sfitter}, deal exclusively with SUSY scenarios.  The
first is a guide to SUSY tools publicly available for phenomenology,
while the second is a brief description of the leading scheme for how
collider data from various machines but especially the Tevatron can be
used to determine where in SUSY parameter space nature lies if SUSY is
indeed discovered by current experiments.


\clearpage\setcounter{equation}{0}\setcounter{figure}{0}\setcounter{table}{0}
\subsection{Spin Determination at the LHC}
\label{sec:spin}

{\em AseshKrishna Datta$^1$, 
Kyoungchul Kong$^2$ and Konstantin T.~Matchev$^2$ \\
$^1$ MCTP, University of Michigan, Ann Arbor, USA
\footnote{Current address: Harish-Chandra Research Institute, Allahabad, India} \\
$^2$ Institute for Fundamental Theory, Physics Dept.,
University of Florida, USA}\\


{\em We contrast the experimental signatures of low energy
supersymmetry and the model of Universal Extra Dimensions and discuss
methods for their discrimination at LHC.  We consider on-shell cascade
decay from squarks or KK quarks for two different types of mass
spectrum: a degenerate case (UED) and SPS1a.  For the dilepton
invariant mass, we find that it is difficult to discriminate two
models for both mass spectra, although for some parameter space in
MSSM, it can be used.  We also investigate the possibility of
differentiating the spins of the superpartners and KK modes by means
of the asymmetry method of Barr. In the case of the SPS1a mass
spectrum, we conclude that the UED model can not fake the SUSY
asymmetry through the entire parameter space.}


\subsubsection{Introduction}
\label{sec:intro}

With the highly anticipated run of the Large Hadron Collider (LHC) at
CERN we will begin to explore the Terascale in earnest.  There are
very sound reasons to expect momentous discoveries at the LHC.  Among
the greatest mysteries in particle physics today is the origin of
electroweak symmetry breaking, which, according to the Standard Model,
is accomplished through the Higgs mechanism.  The Higgs particle is
the primary target of the LHC experiments and, barring some unexpected
behavior, the Higgs boson will be firmly discovered after only a few
years of running of the LHC.  With some luck, a Higgs signal might
start appearing already at Tevatron Run~II.

The discovery of a Higgs boson, however, will open a host of new
questions.  As the first fundamental scalar to be seen, it will bring
about a worrisome fine tuning problem: why is the Higgs particle so
light, compared to, say, the Planck scale?  Various solutions to this
hierarchy problem have been proposed, and the most aesthetically
pleasing one at this point appears to be low energy supersymmetry
(SUSY).  In SUSY, the problematic quadratic divergences in the
radiative corrections to the Higgs mass are absent, being cancelled by
loops with superpartners.  The cancellations are enforced by the
symmetry, and the Higgs mass is therefore naturally related to the
mass scale of the superpartners.

While the solution of the hierarchy problem is perhaps the most
celebrated virtue of SUSY, supersymmetric models have other side
benefits.  For one, if the superpartners are indeed within the TeV
range, they would modify the running of the gauge couplings at higher
scales, and gauge coupling unification takes place with astonishing
precision.  Secondly, a large class of SUSY models, which have a
conserved discrete symmetry ($R$-parity), contain an excellent dark
matter candidate: the lightest neutralino $\tilde\chi^0_1$.  One
should keep in mind that the dark matter problem is by far the most
compelling {\em experimental} evidence for particles and interactions
outside the Standard Model (SM), and provides a completely independent
motivation for entertaining supersymmetry at the TeV scale.  Finally,
$R$-parity implies that superpartners interact only pairwise with SM
particles, which guarantees that the supersymmetric contributions to
low energy precision data only appear at the loop level and are small.
In summary, supersymmetric extensions of the SM are the primary
candidates for new physics at the TeV scale.  Not surprisingly,
therefore, signatures of supersymmetry at the Tevatron and LHC have
been extensively discussed in the literature.  In typical scenarios
with superpartners in the range of a few TeV or less, already within
the first few years of running the LHC would discover a signal of new
physics in several channels.  Once such a signal of physics beyond the
Standard Model is seen, it will immediately bring up the question: is
it supersymmetry or not?

The answer to this question can be approached in two different ways.
On the theoretical side, one may ask whether there are well-motivated
alternatives to low-energy supersymmetry, which would give similar
signatures at hadron colliders.  In other words, if the new physics is
not supersymmetry, what else can it be?  Until recently, there were no
known examples of other types of new physics which could ``fake''
supersymmetry sufficiently well.  The signatures of supersymmetry and
its competitors (Technicolor, new gauge bosons, large extra
dimensions, etc.) were sufficiently distinctive, and there was little
room for confusion.  However, it was recently realized that the
framework of Universal Extra Dimensions (UED), originally proposed
in~\cite{Appelquist:2000nn}, can very effectively masquerade as
low-energy SUSY at a hadron collider such as the LHC or the
Tevatron~\cite{Cheng:2002ab}.  It therefore became of sufficient
interest to try to prove SUSY at the LHC from first principles,
without resorting to model-dependent assumptions and without
theoretical bias.  The experimental program for proving SUSY at a {\em
lepton} collider was outlined a long time ago~\cite{Feng:1995zd} and
can be readily followed to make the discrimination between SUSY and
UED~\cite{Battaglia:2005zf,Bhattacharyya:2005vm,Bhattacherjee:2005qe,Riemann:2005es}.
Recently there has been a lot of interest regarding the ``inversion''
problem and spin measurements at LHC
~\cite{Bechtle:2004pc,Lafaye:2004cn,Meade:2006dw,Datta:2005vx,Arkani-Hamed:2005px,Barr:2004ze,Barr:2005dz,Datta:2005zs,Battaglia:2005ma,Smillie:2005ar,KK,Alves:2006df}
(see also Sec.~\ref{sec:sfitter}).  However, as we shall see below,
the case of hadron colliders is much more challenging.


\subsubsection{\label{sec:comp}UED versus SUSY}

The couplings of the SM particles and their superpartners are equal,
being related by supersymmetry and the generic collider signatures of
supersymmetric models with weakly-interacting massive particle (WIMP)
lightest SUSY particles (LSPs) is missing energy.  In UED, KK-parity
guarantees the lightest KK particle (LKP) is stable and UED can
explain dark matter
problem~\cite{Servant:2002aq,Kakizaki:2005uy,Burnell:2005hm,Kong:2005hn,Cheng:2002ej}.
The new couplings are also the same as SM couplings.  Therefore the
above two features are common to both SUSY and UED and cannot be used
to distinguish the two cases.  We see that while $R$-parity-conserving
SUSY implies a missing energy signal, the reverse is not true: a
missing energy signal would appear in any model with a dark matter
candidate, and even in models which have nothing to do with the dark
matter issue, but simply contain new neutral quasi-stable particles,
e.g.~gravitons~\cite{Arkani-Hamed:1998rs,Giudice:1998ck,Mirabelli:1998rt}. 
Similarly, the equality of couplings is a celebrated test of SUSY, but
we see that it is only a necessary, not sufficient, condition to prove
supersymmetry.  In addition, the measurement of superpartner couplings
in order to test the SUSY relations is a very challenging task at a
hadron collider.  For one, the observed production rate in any given
channel is sensitive only to the product of the cross-section times
the branching fractions, so any attempt to measure the couplings from
a cross section would have to make certain assumptions about the
branching fractions.  An additional complication arises from the fact
that at hadron colliders all kinematically available states can be
produced simultaneously, and the production of a particular species in
an exclusive channel is rather difficult to isolate.  The couplings
could also in principle be measured from the branching fractions, but
that also requires a measurement of the total width, which is
impossible in our case, since the Breit-Wigner resonance cannot be
reconstructed, due to the unknown momentum of the missing LSP (LKP).

The fundamental difference between SUSY and UED is first the number of
new particles, and second their spins.  The KK particles at $n=1$ are
analogous to superpartners in SUSY.  However, the particles at the
higher KK levels have no analogues in $N=1$ supersymmetric models.
Discovering the $n\ge2$ levels of the KK tower would therefore
indicate the presence of extra dimensions rather than SUSY.  However,
these KK particles can be too heavy to be observed.  Even if they can
be observed at LHC, they can be confused with other new
particles~\cite{Datta:2005zs,Battaglia:2005ma} such as $Z'$ or
different types of resonances from extra
dimensions~\cite{Burdman:2006gy}.

The second feature -- the spins of the new particles -- also provides
a tool for discrimination between SUSY and UED: the KK partners have
identical spin quantum numbers as their SM counterparts, while the
spins of the superpartners differ by $1/2$ unit.  However, spin
determination may in some cases be difficult at the LHC (or at hadron
colliders in general), where the parton-level center of mass energy
$E_{CM}$ in each event is unknown.  In addition, the momenta of the
two dark matter candidates in the event are also unknown.  This
prevents the reconstruction of any rest frame angular decay
distributions, or the directions of the two particles at the top of
the decay chains.  The variable $E_{CM}$ also rules out the
possibility of a threshold scan, which is one of the main tools for
determining particle spins at lepton colliders.  We are therefore
forced to look for new methods for spin determinations, or at least
for finding spin correlations\footnote{Notice that in simple processes
with two-body decays like slepton production
$e^+e^-\to\tilde\mu^+\tilde\mu^-\to\mu^+\mu^-\tilde\chi^0_1\tilde\chi^0_1$
the flat energy distribution of the observable final state particles
(muons in this case) is often regarded as a smoking gun for the scalar
nature of the intermediate particles (the smuons).  Indeed, the smuons
are spin zero particles and decay isotropically in their rest frame,
which results in a flat distribution in the lab frame.  However, the
flat distribution is a necessary but not sufficient condition for a
scalar particle, and UED provides a counterexample with the analogous
process of KK muon production~\cite{Battaglia:2005zf}, where a flat
distribution also appears, but as a result of equal contributions from
left-handed and right-handed KK fermions.}.  The purpose of this paper
is to investigate the prospects for establishing SUSY at the LHC by
discriminating it from its look-alike scenario of Universal Extra
Dimensions by measuring spins~\footnote{Another recent
work~\cite{Alves:2006df} showed how one can clearly distinguish a SUSY
gluino from a UED heavy gluon partner at the LHC.} of new particles in
two models~\footnote{The same idea can apply in the case of Little
Higgs models since the first level of the UED model looks like the new
particles in the Little
Higgs~\cite{Cheng:2003ju,Cheng:2004yc,Hubisz:2004ft,Cheng:2005as}.}.


\subsubsection{Spin Determination in Squark/KK Quark Cascade Decay}
\label{sec:spin-decay}

\begin{center}
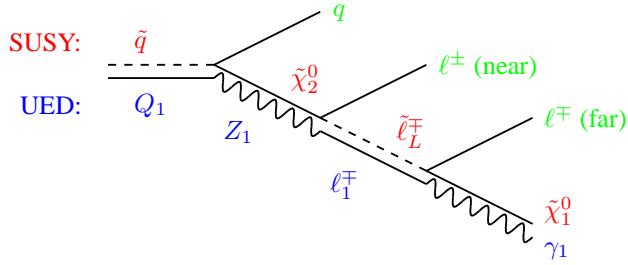
\begin{figure*}[t]
\unitlength=1.0 pt
\SetScale{1.0}
\SetWidth{0.7}      
\footnotesize    
%
\begin{picture}(200,100)(-130,0)
\Text(  0.0,79.0)[r]{\color{red} SUSY:}
\Text( 20.0,79.0)[l]{\color{red} $\tilde q$}
\Text( 90.0,65.0)[r]{\color{red} $\tilde \chi^0_2$}
\Text(130.0,45.0)[r]{\color{red} $\tilde \ell^\mp_L$}
\Text(175.0,15.0)[l]{\color{red} $\tilde \chi^0_1$}
\DashLine(10.0,70.0)(50.0,70.0){3}
\Line(50.0,70.0)(90.0,50.0)
\DashLine(90.0,50.0)(130.0,30.0){3}
\Line(130.0,30.0)(170.0,10.0)
\Text(  0.0,55.0)[r]{\color{blue} UED:}
\Text( 20.0,55.0)[l]{\color{blue} $Q_1$}
\Text( 65.0,45.0)[r]{\color{blue} $Z_1$}
\Text(105.0,25.0)[r]{\color{blue} $\ell^\mp_1$}
\Text(175.0, 0.0)[l]{\color{blue} $\gamma_1$}
\Line(10.0,65.0)(50.0,65.0)
\Photon(50.0,65.0)(90.0,45.0){3}{6}
\Line(90.0,45.0)(130.0,25.0)
\Photon(130.0,25.0)(170.0,5.0){3}{6}
\Text( 95.0,90.0)[l]{\color{green} $q$}
\Text(135.0,70.0)[l]{\color{green} $\ell^\pm$ (near)}
\Text(175.0,50.0)[l]{\color{green} $\ell^\mp$ (far)}
\Line( 50.0,70.0)( 90.0,90.0)
\Line( 90.0,50.0)(130.0,70.0)
\Line(130.0,30.0)(170.0,50.0)
\end{picture}
%
\caption{Twin diagrams in SUSY and UED.  The upper (red) line
corresponds to the cascade decay $\tilde{q}\to q\tilde\chi^0_2 \to
q\ell^\pm\tilde\ell^\mp_L\to q\ell^+\ell^-\tilde\chi^0_1$ in SUSY.
The lower (blue) line corresponds to the cascade decay $Q_1\to q
Z_1\to q\ell^\pm\ell^\mp_1\to q\ell^+\ell^-\gamma_1$ in UED.  In
both cases the observable final state is the same:
$q\ell^+\ell^-\rlap{\,/}E_T$.}
\label{fig:diagrams}
\end{figure*}
\end{center}
As discussed in the previous section, the second fundamental
distinction between UED and supersymmetry is reflected in the
properties of the individual particles.  Recently it was suggested
that a charge asymmetry in the lepton-jet invariant mass distributions
from a particular cascade (see fig.~\ref{fig:diagrams}) can be used to
discriminate SUSY from the case of pure phase space
decays~\cite{Barr:2004ze} and is an indirect indication of the
superparticle spins.  (A study of measuring sleptons spins at the LHC
can be found in~\cite{Barr:2005dz}).  It is therefore natural to ask
whether this method can be extended to the case of SUSY versus UED
discrimination. Following~\cite{Barr:2004ze}, we concentrate on the
cascade decay $\tilde{q}\to q\tilde\chi^0_2 \to
q\ell^\pm\tilde\ell^\mp_L
\to q\ell^+\ell^-\tilde\chi^0_1$ in SUSY and the analogous decay chain
$Q_1 \to q Z_1\to q\ell^\pm\ell^\mp_1\to q\ell^+\ell^-\gamma_1$ in
UED.  Both of these processes are illustrated in
Fig.~\ref{fig:diagrams}.  Blue lines represent the decay chain in UED
and red lines the decay chain in SUSY. Green lines are SM particles.


\section*{\normalsize Dilepton Invariant Mass}

\begin{figure}[t]
\includegraphics[width=7.5cm]{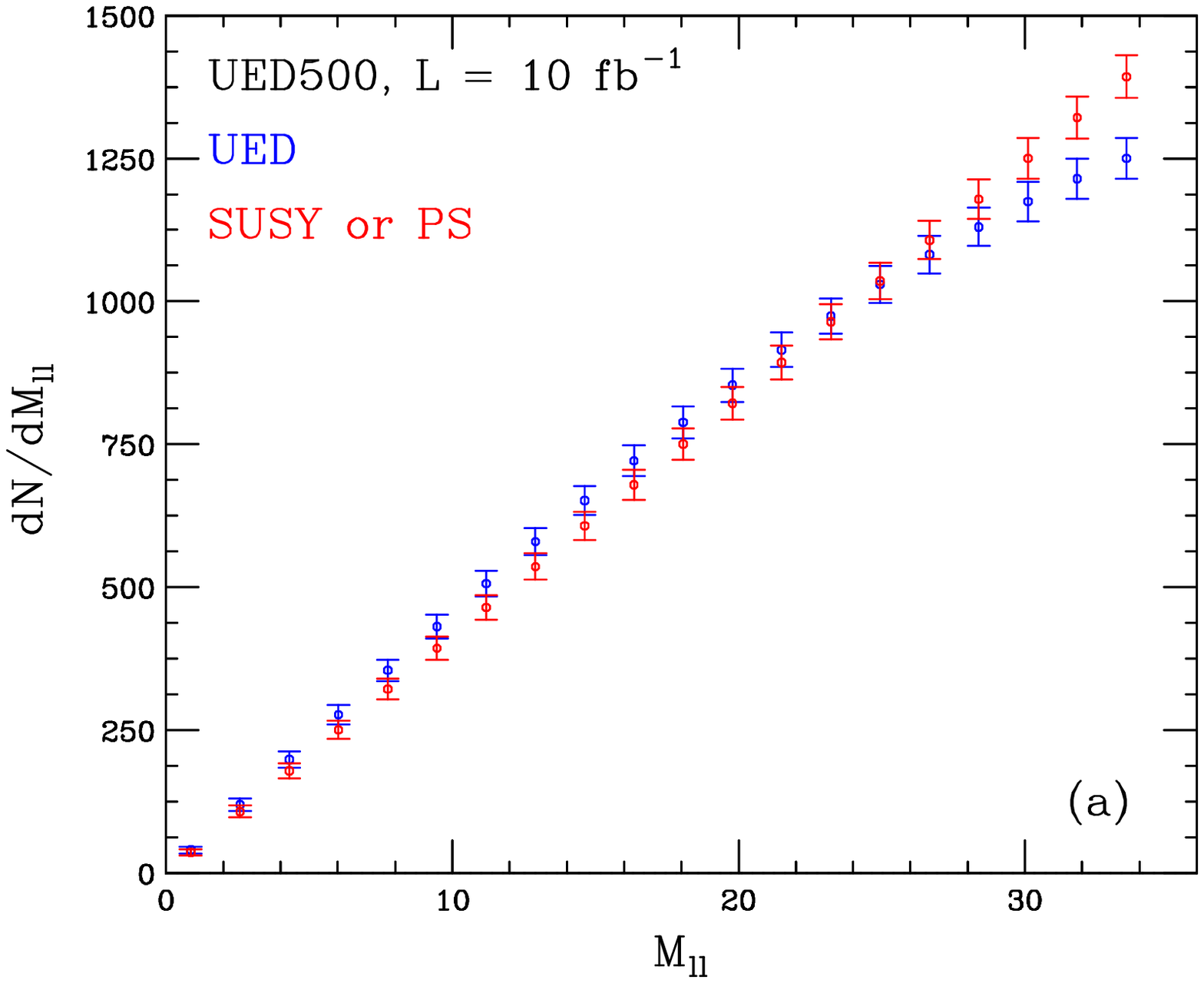}
\includegraphics[width=7.5cm]{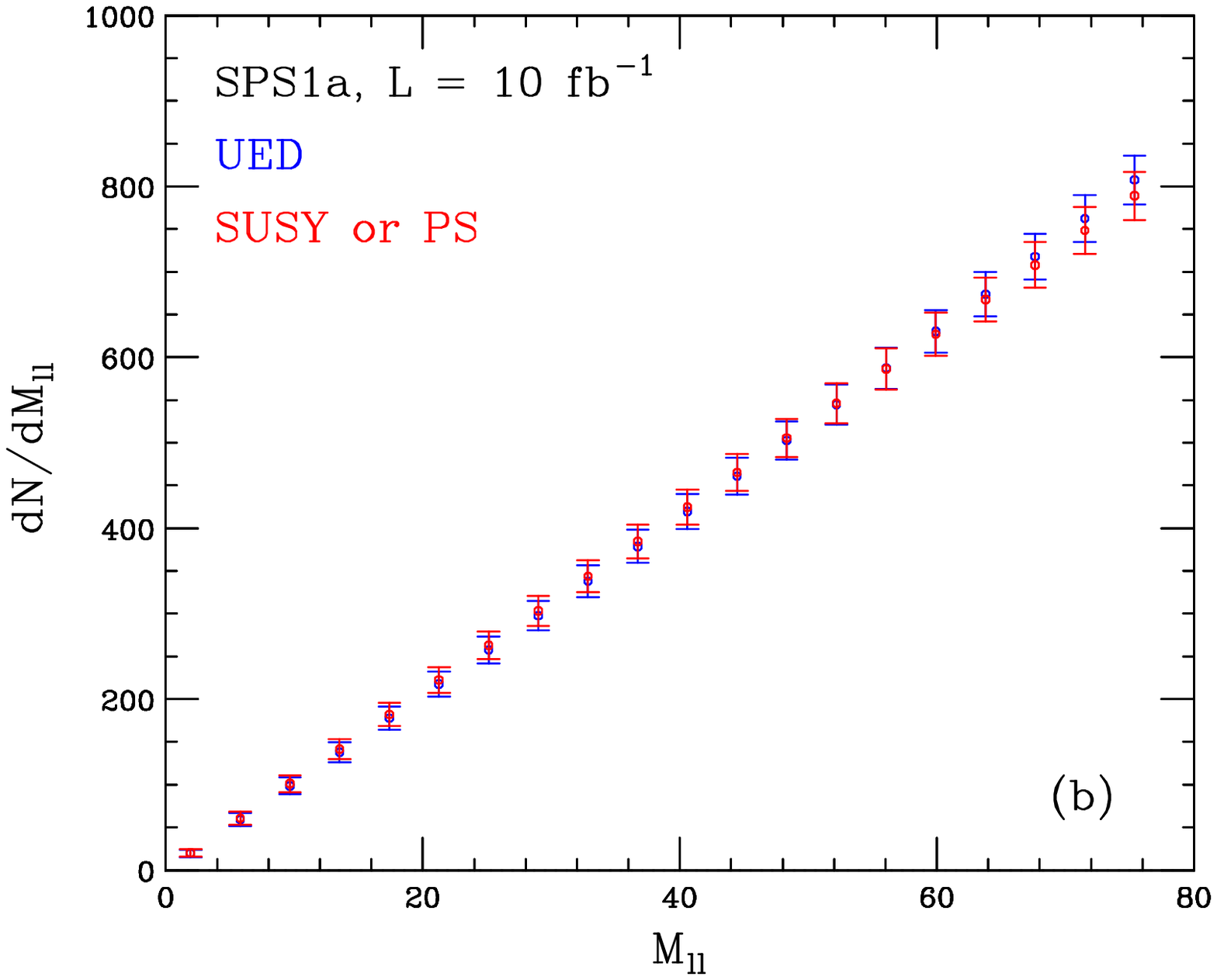}
\caption{Comparison of dilepton invariant mass distributions in the case
of (a) UED mass spectrum with $R^{-1}=500$ GeV (b) mass spectrum from
SPS1a.  In both cases, UED (SUSY) distributions are shown in blue
(red).  All distributions are normalized to ${\cal L}=10$ fb$^{-1}$
and the error bars represent statistical uncertainty.}
\label{fig:dilepton}
\end{figure}
First we will look for spin correlations between the two SM leptons in
the final state.  In SUSY, a slepton is a scalar particle and
therefore there is no spin correlation between the two SM leptons.
However in UED, a slepton is replaced by a KK lepton and is a fermion.
We might therefore expect a different shape in the dilepton invariant
mass distribution.  To investigate this question, we first choose a
study point in UED (SPS1a in mSUGRA) with $R^{-1}=500$ GeV taken
from~\cite{Cheng:2002ab,Cheng:2002iz} and then adjust the relevant
MSSM parameters (UED parameters) until we get a matching spectrum.  So
the masses are exactly same and cannot be used for discrimination.

In Fig.~\ref{fig:dilepton} we show invariant mass distributions in UED
and SUSY for two different types of mass spectra.  In
Fig.~\ref{fig:dilepton}(a), all UED masses are adjusted to be the same
as the SUSY masses in SPS1a ($m_0=100$ GeV, $m_{1/2}=250$ GeV,
$A_0=-100$, $\tan\beta=10$ and $\mu>0$) while in
Fig.~\ref{fig:dilepton}(b) the SUSY masses are replaced by KK masses
for $R^{-1}=500$.  In both cases, UED (SUSY) distributions are shown
in blue (red).  Squark/KK quark pair-production cross sections are
taken from Ref.~\cite{Smillie:2005ar} and the relevant branching
fractions are obtained from Ref.~\cite{Cheng:2002ab} for UED and
Ref.~\cite{Paige:2003mg} for SUSY.  All distributions are normalized
to ${\cal L}=10$~fb$^{-1}$ and the error bars represent statistical
uncertainty.  For SUSY, the distribution is the same as that in the
case of pure phase space decay since the slepton has no spin.  As we
see, the two distributions are identical for both UED and SUSY mass
spectrum even if the intermediate particles in UED and SUSY have
different spins.  Small differences in the distributions will
completely disappear once the background, radiative corrections and
detector simulation are included.

\begin{figure}[t]
\includegraphics[width=7cm]{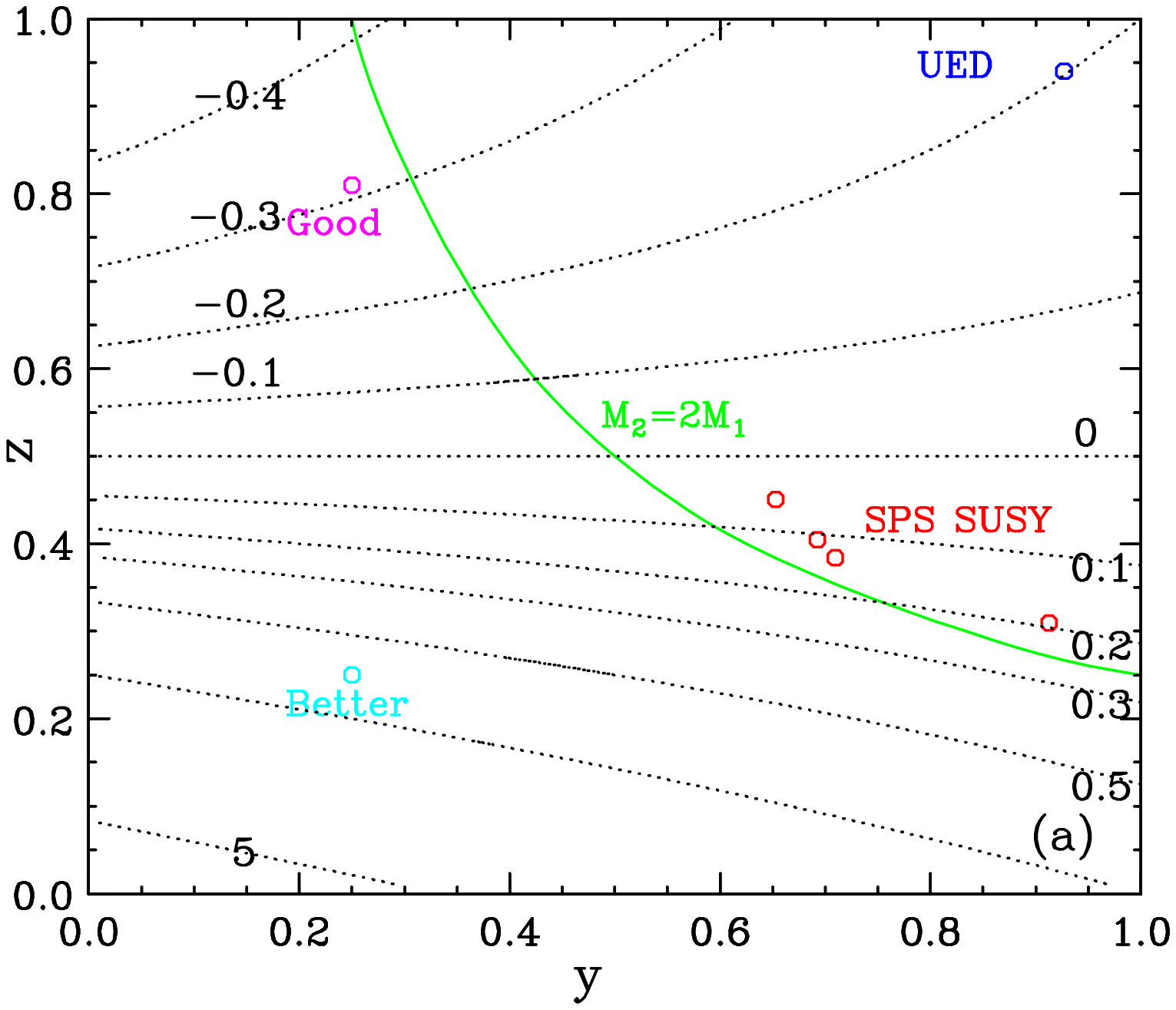}
\includegraphics[width=7.6cm]{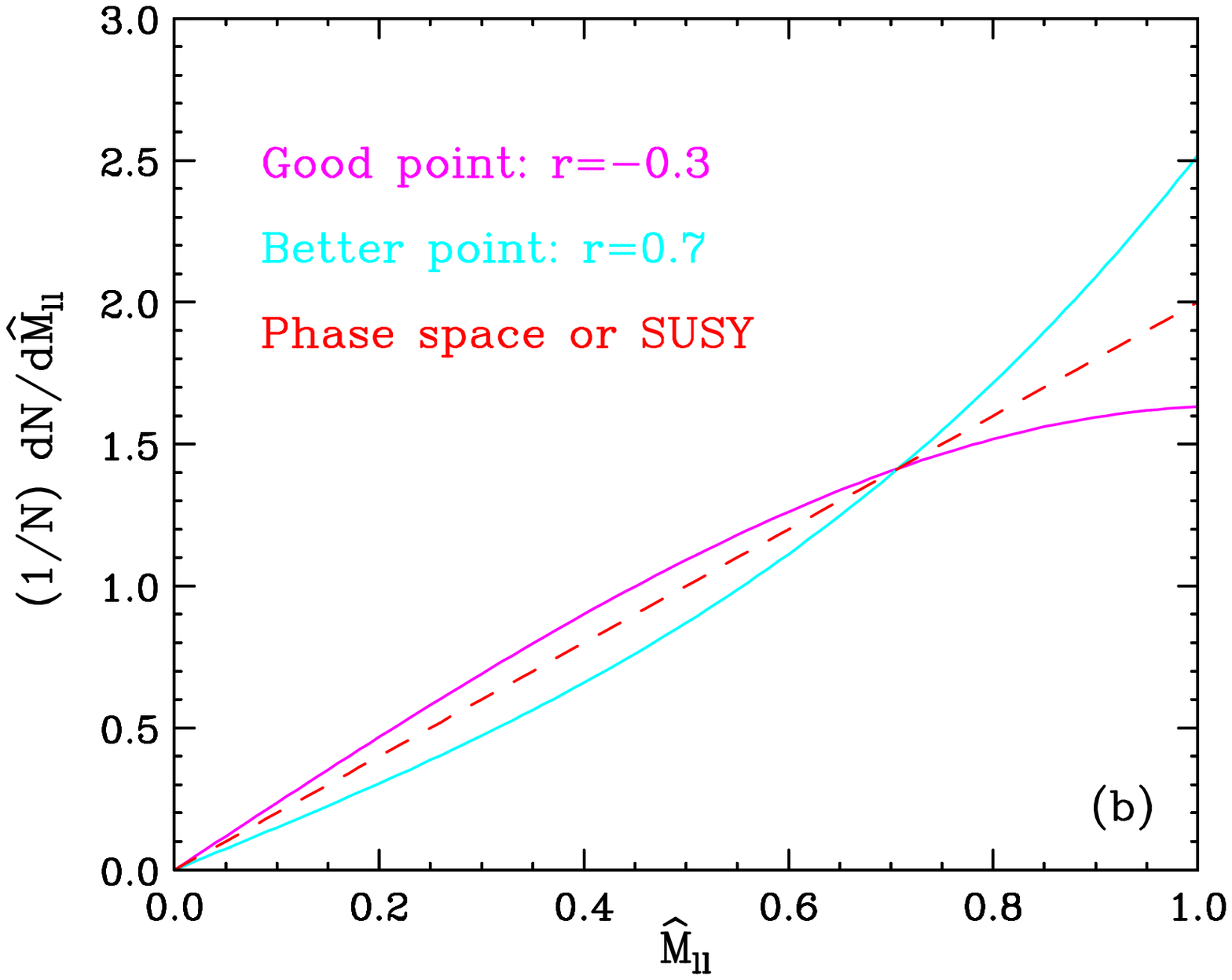}
\caption{(a) Contour dotted lines represent the size of the coefficient
$r$ in Eq.~\ref{r}.  UED is the blue dot in the upper-right corner
since $y$ and $z$ are almost 1 due to the mass degeneracy.  The red
dots represent several snowmass points: SPS1a, SPS1b, SPS5 and SPS3,
from left to right.  The green line represents gaugino unification, so
all SUSY benchmark points lie close to this line.  $r$ is small for
both UED and snowmass points.  (b) The dashed line represents the
dilepton distribution in SUSY or pure phase space.  Solid cyan
(magenta) line represents dilepton distribution in UED for $r=-0.3$
($r=0.7$).}
\label{fig:dilepton2}
\end{figure}

The invariant mass distributions for UED and SUSY/Phase space can be
written as~\cite{Smillie:2005ar,KK}
\begin{eqnarray}
{\rm Phase \,\, Space\,:\,\,} \frac{dN}{d  \hat{m} } &=& 2 \hat{m} \nonumber \\
{\rm SUSY  \,:\,\,}\frac{dN}{d  \hat{m} } &=& 2 \hat{m} \label{eqndilepton} \\
{\rm UED \,:\,\,}\frac{dN}{d  \hat{m} } &=& \frac{4(y+4z)}{(1+2z)(2+y)} \left ( \hat{m} + r \, \hat{m}^3\right ) \nonumber
\end{eqnarray}
where the coefficient $r$ in the second term of the UED distribution is defined as
\begin{equation}
        r = \frac{(2-y)(1-2z)}{y+4z} \, , \label{r}
\end{equation}
$\hat{m}=\frac{m_{\ell\ell}}{m_{\ell\ell}^{max}}$ is the rescaled
invariant mass,
$y=\left(\frac{m_{\tilde\ell}}{m_{\tilde{\chi}_2^0}}\right)^2$ and
$z=\left(\frac{m_{\tilde{\chi}_1^0}}{m_{\tilde\ell}}\right)^2$ are the
ratios of masses involved in the decay; $y$ and $z$ are less than 1 in
the case of on-shell decay.  From Eq.~\ref{eqndilepton}, there are two
terms in UED.  The first term is phase-space-like and the second term
is proportional to $\hat{m}^3$.  So we see that whether or not the UED
distribution is the same as the SUSY distribution depends on the size
of the coefficient $r$ in the second term of the UED distribution.
Note that the UED distribution becomes exactly the same as the SUSY
distribution if $r=0.5$.  Therefore we scan $(y,z)$ parameter space,
calculate the coefficient $r$ and show our result in
Fig.~\ref{fig:dilepton2}(a).
In Fig.~\ref{fig:dilepton2}(a), contour dotted lines represent the
size of the coefficient $r$ in Eq.~\ref{r}.  UED is the blue dot in
the upper-right corner since $y$ and $z$ are almost 1 due to the mass
degeneracy, while red dots represent several snowmass
points~\cite{Allanach:2002nj}: SPS1a, SPS1b, SPS5 and SPS3, from left
to right.  The green line represents gaugino unification so all SUSY
benchmark points are close to this green line.  As we can see, $r$ is
small for both UED and snowmass points.  This is why we did not see
any difference in the distributions from Fig.~\ref{fig:dilepton}.  If
the mass spectrum is either narrow (UED mass spectrum) or generic
(mSUGRA type), the dilepton distributions are very similar and we
cannot extract any spin information from this distribution.  However,
away from the mSUGRA model or UED, we can easily find regions where
the coefficient $r$ is large and the spin correlation is big enough so
that we can see a shape difference.  We show two points (denoted by
`Good' and `Better') in Fig.~\ref{fig:dilepton2}(a) and show the
corresponding dilepton distributions in Fig.~\ref{fig:dilepton2}(b).
For the `Good' point, the mass ratio is $m_{\tilde{\chi}_1^0} :
m_{\tilde\ell} : m_{\tilde{\chi}_2^0} = 9:10:20$ and for the `Better'
point, $m_{\tilde{\chi}_1^0} : m_{\tilde\ell} : m_{\tilde{\chi}_2^0} =
1:2:4$.  In Fig.~\ref{fig:dilepton2}(b), the dashed line represents
dilepton distribution in SUSY or pure phase space and the solid cyan
(magenta) line represents the dilepton distribution in UED for
$r=-0.3$ ($r=0.7$).  Indeed, for larger $r$, the distributions look
different, but background and detector simulation need to be included.
Notice that in the mSUGRA model, the maximum of the coefficient $r$ is
0.4.


\section*{\normalsize Lepton-Jet Invariant Mass - Charge Asymmetry}

Now we look at spin correlations between $q$ and $\ell$ in
Fig.~\ref{fig:diagrams}.  In this case, there are several
complications.  First of all, we don't know which lepton we need to
choose.  There are two leptons in the final state.  The lepton we call
`near' comes from the decay of $\tilde{\chi}_2^0$ in SUSY or $Z_1$ in
UED, while the other lepton we call `far' comes from the decay of
$\tilde{\ell}$ in SUSY or $\ell_1$ in UED.  The lepton-quark invariant
mass distributions $m_{\ell q}$ turns out to be useful.  The spin of
the intermediate particle ($Z_1$ in UED or $\tilde\chi^0_2$ in SUSY)
governs the shape of the distributions for the near lepton.  However,
in practice we cannot distinguish the near and far lepton, and one has
to include the invariant mass combinations with both leptons (it is
impossible to tell near and far leptons event-by-event, but there can
be an improvement for the selection~\cite{KK}.).  Second, we do not
measure jet (quark) charge.  Therefore we do not know whether a
particular jet (quark) came from the decay of a squark or an
anti-squark.  This doubles the number of diagrams that we need to
consider.  These complications tend to wash out the spin correlations,
but a residual effect remains, which is due to the different number of
quarks and anti-quarks in the proton, which in turn leads to a
difference in the production cross-sections for squarks and
anti-squarks~\cite{Barr:2004ze}.  Most importantly, we do not know
which jet is actually the correct jet in this cascade decay chain.  We
pair produce two squarks (or KK quarks) and each of them decays to one
jet.  Once initial-state radiation (ISR) is included, there are many
jets in the final state.  For now, as in~\cite{Smillie:2005ar}, we
assume that we know which jet is the correct one and choose it.  One
never knows for sure which is the correct jet, although there can be
clever cuts to increase the probability that we picked the right
one~\cite{KK}.  There are two possible invariant distributions in this
case: $\left(\frac{d\sigma}{dm}\right)_{q\ell^+}$ with positively
charged leptons and $\left(\frac{d\sigma}{dm}\right)_{q\ell^+}$ with
negatively charged leptons.  In principle, there are 8 diagrams that
need to be included (a factor of 2 from quark/anti-quark combination,
another factor of 2 from sleptons with different chiralities, and a
factor 2 from the ambiguity between near and far leptons).

\begin{figure}[t]
\includegraphics[width=7.5cm]{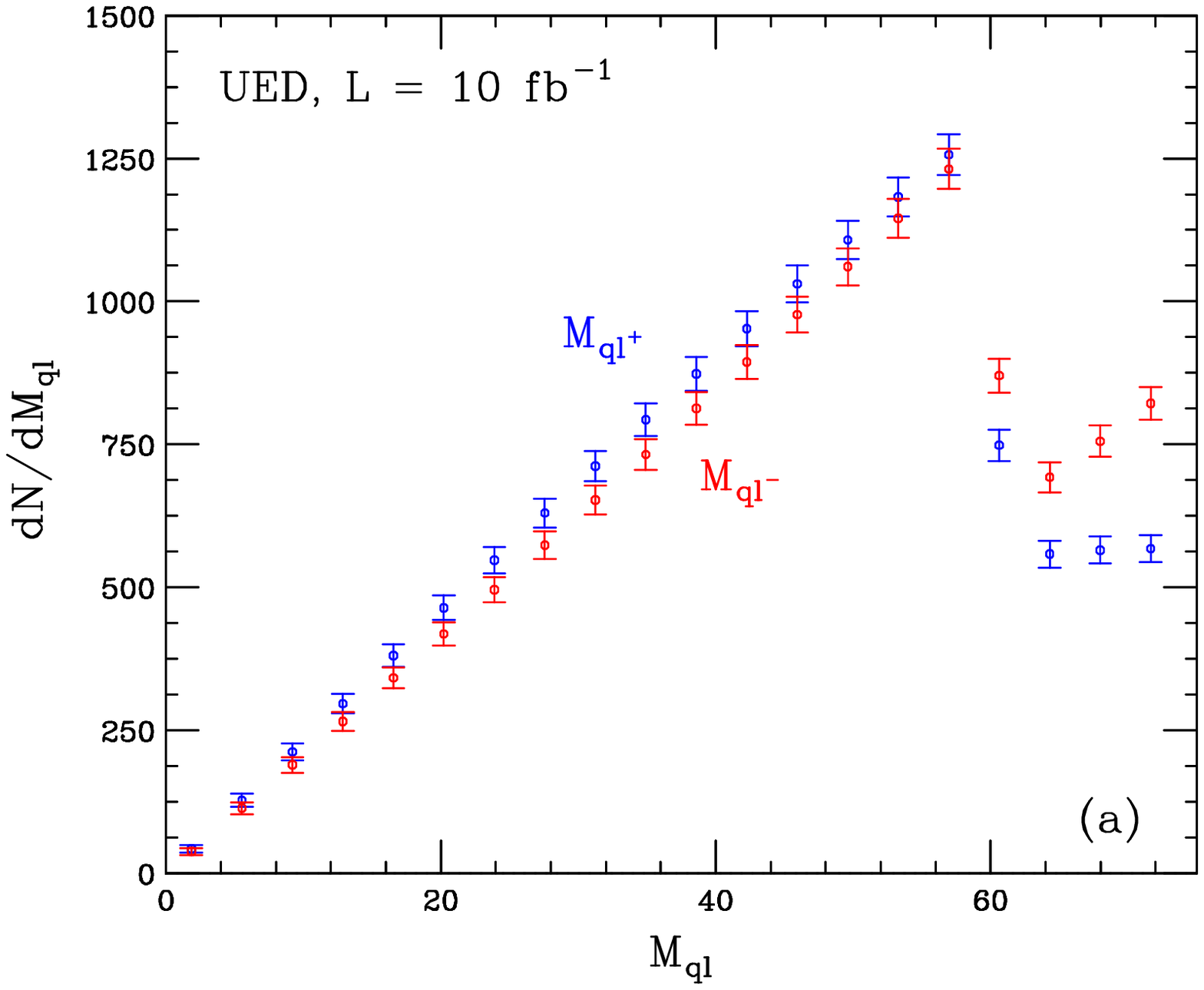}
\includegraphics[width=7.5cm]{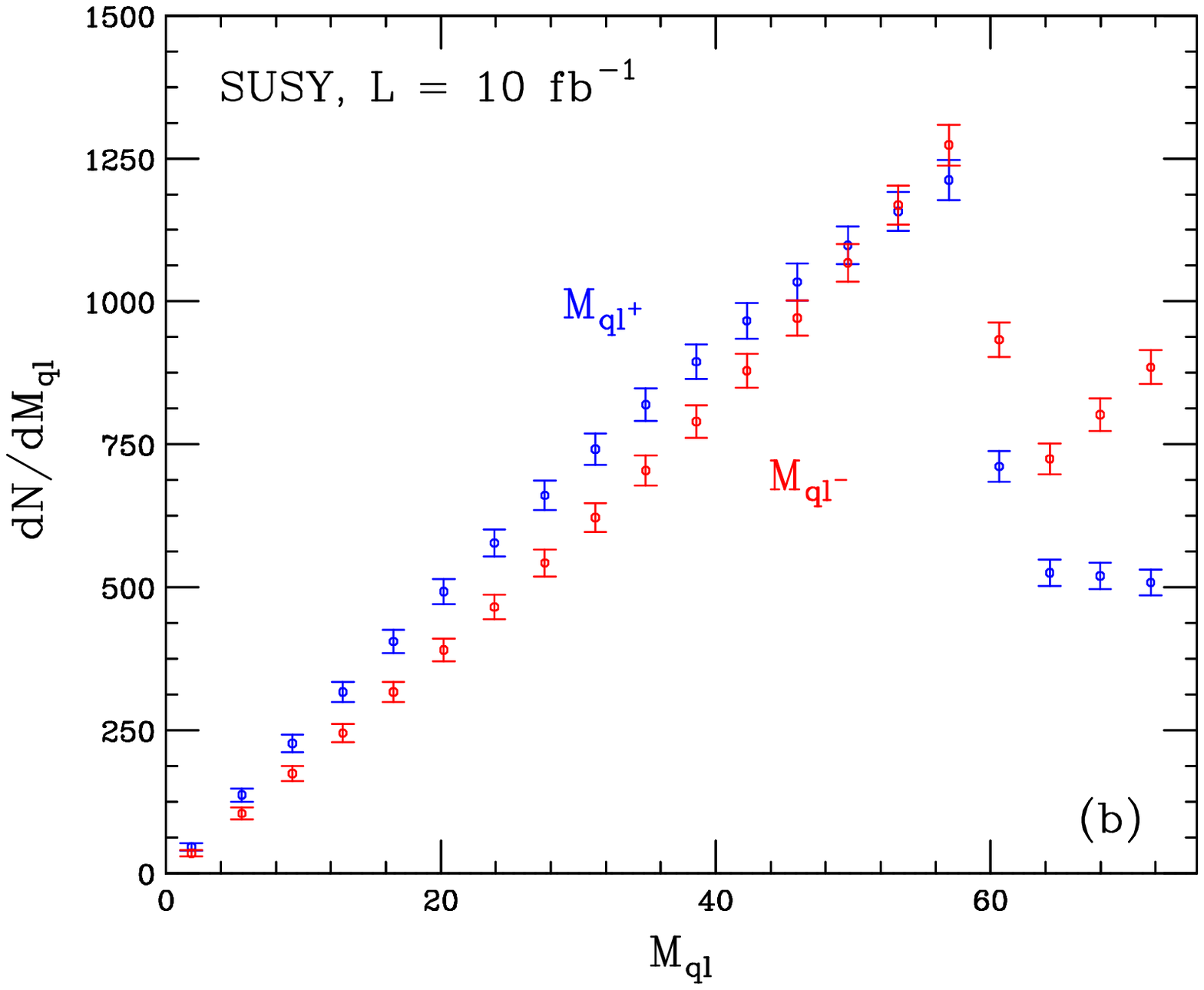}
\caption{$\left(\frac{dN}{dm}\right)_{q\ell^+}$ (blue) and 
$\left(\frac{dN}{dm}\right)_{q\ell^+}$ (red) in the case of (a) UED
and (b) SUSY for UED mass spectrum with $R^{-1}=500$ GeV.  $q$ stands
for both a quark and an anti-quark, and $N(q\ell^+)$ ($N(q\ell^-)$) is
the number of entries with a positively (negatively) charged lepton.
The distributions are normalized to ${\cal L}=10fb^{-1}$.  A very
sharp edge near $m_{q\ell}\sim 60$~GeV ($m_{q\ell}\sim 75$~GeV) is due
to the near (far) lepton.  Once background and detector resolutions
are included, these clear edges are smoothed out.}
\label{fig:mql}
\end{figure}

For this study, as in the dilepton case, we first start from a UED
mass spectrum and adjust the MSSM parameters until we get perfect
spectrum match.  In this case, $Z_1$ does not decay into a
right-handed lepton.  There are 4 contributions and they all
contribute to both $\left(\frac{d\sigma}{dm}\right)_{q\ell^+}$ and
$\left(\frac{d\sigma}{dm}\right)_{q\ell^+}$ distributions which are in
fig.~\ref{fig:mql},
\begin{eqnarray}
\left ( \frac{d \sigma}{d m} \right )_{q\ell^+} \hspace{-0.4cm}&=&\hspace{-0.3cm} f_q \left ( {\color[named]{Red}{\frac{dP_2}{d m_{n}}}}
                                                              + {\color[named]{Green}{\frac{d P_1}{dm_{f}}}} \right ) 
                                         +  f_{\bar{q}} \left ( {\color[named]{Blue}{\frac{dP_1}{d m_{n}}} }
                                                              + {\color[named]{Cyan}{\frac{d P_2}{dm_{f}}}} \right ) \nonumber \\
\left ( \frac{d \sigma}{d m} \right )_{q\ell^-} \hspace{-0.4cm}&=&\hspace{-0.3cm} f_q \left ( {\color[named]{Blue}{\frac{dP_1}{d m_{n}}}} 
                                                              + {\color[named]{Cyan}{\frac{d P_2}{dm_{f}}}} \right ) 
                                         +  f_{\bar{q}} \left ( {\color[named]{Red}{\frac{dP_2}{d m_{n}}} }
                                                              + {\color[named]{Green}{\frac{d P_1}{dm_{f}}}} \right ) \, ,
\end{eqnarray}
where $P_1$ ($P_2$) represents the distribution for a decay from a
squark or KK quark (anti-squark or anti-KK quark) and $f_q$
($f_{\bar{q}}$) is the fraction of squarks or KK quarks (anti-squarks
or anti-KK quarks) and by definition, $f_q+f_{\bar{q}}=1$.  The
quantity $f_q$ tells us how much squarks or KK quarks are produced
compared to the anti-particles.  For the UED mass spectrum and SPS1a,
$f_q\sim 0.7$~\cite{Smillie:2005ar}.  These two distributions in UED
(SUSY) are shown in Fig.~\ref{fig:mql}(a) (Fig.~\ref{fig:mql}(b)) in
different colors.  The distributions are normalized to ${\cal
L}=10fb^{-1}$ and the very sharp edge near $m_{q\ell}\sim 60$~GeV
($m_{q\ell}\sim 75$~GeV) is due to the near (far) lepton.  However,
once the background and detector resolutions are included, these clear
edges are smoothed out.

Now with these two distributions, a convenient quantity,
`asymmetry'~\cite{Barr:2004ze} is defined below
\begin{eqnarray}
A^{+-} = \frac{\left ( \frac{d \sigma}{d m} \right )_{q\ell^+} - \left( \frac{d \sigma}{d m}\right)_{q\ell^-}}
              {\left ( \frac{d \sigma}{d m} \right )_{q\ell^+} + \left( \frac{d \sigma}{d m}\right)_{q\ell^-}} \label{asymmetry} \ .
\end{eqnarray}
Notice that if $f_q=f_{\bar{q}}=0.5$,
$\left(\frac{d\sigma}{dm}\right)_{q\ell^+} =
\left(\frac{d\sigma}{dm}\right)_{q\ell^+}$ and $A^{+-}$ becomes
zero.  This is the case for pure phase space decay.  Zero asymmetry
means we don't obtain any spin information from this decay chain,
i.e., if we measure a non-zero asymmetry, it means that the
intermediate particle ($\tilde{\chi}_2^0$ or $Z_1$) has non-zero spin.
So for this method to work, $f_q$ must be different from
$f_{\bar{q}}$.  This method does not apply at a $p\bar{p}$ collider
such as the Tevatron since a $p\bar{p}$ collider produces the same
amount of quarks as anti-quarks.  The spin correlations are encoded in
the charge asymmetry~\cite{Barr:2004ze}.  However, even at a $pp$
collider such as the LHC, whether or not we measure a non-zero
asymmetry depends on parameter space.  E.g., in the focus point
region, gluino production dominates and the gluino produces equal
amounts of squarks and anti-squarks.  Therefore we expect $f_q\sim
f_{\bar q}\sim 0.5$ and any asymmetry will be washed out.

\begin{figure}[t]
\includegraphics[width=7.5cm]{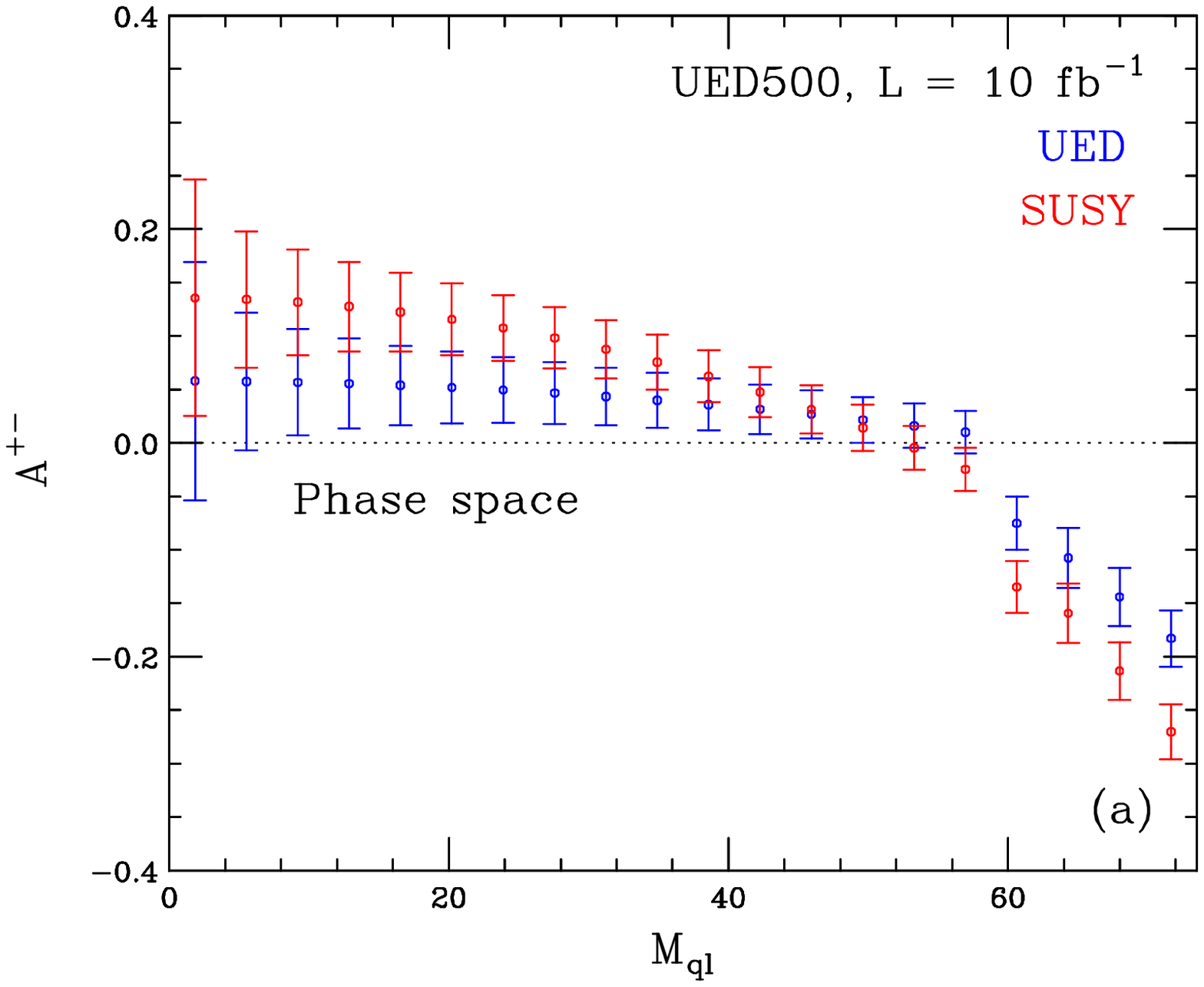}
\includegraphics[width=7.5cm]{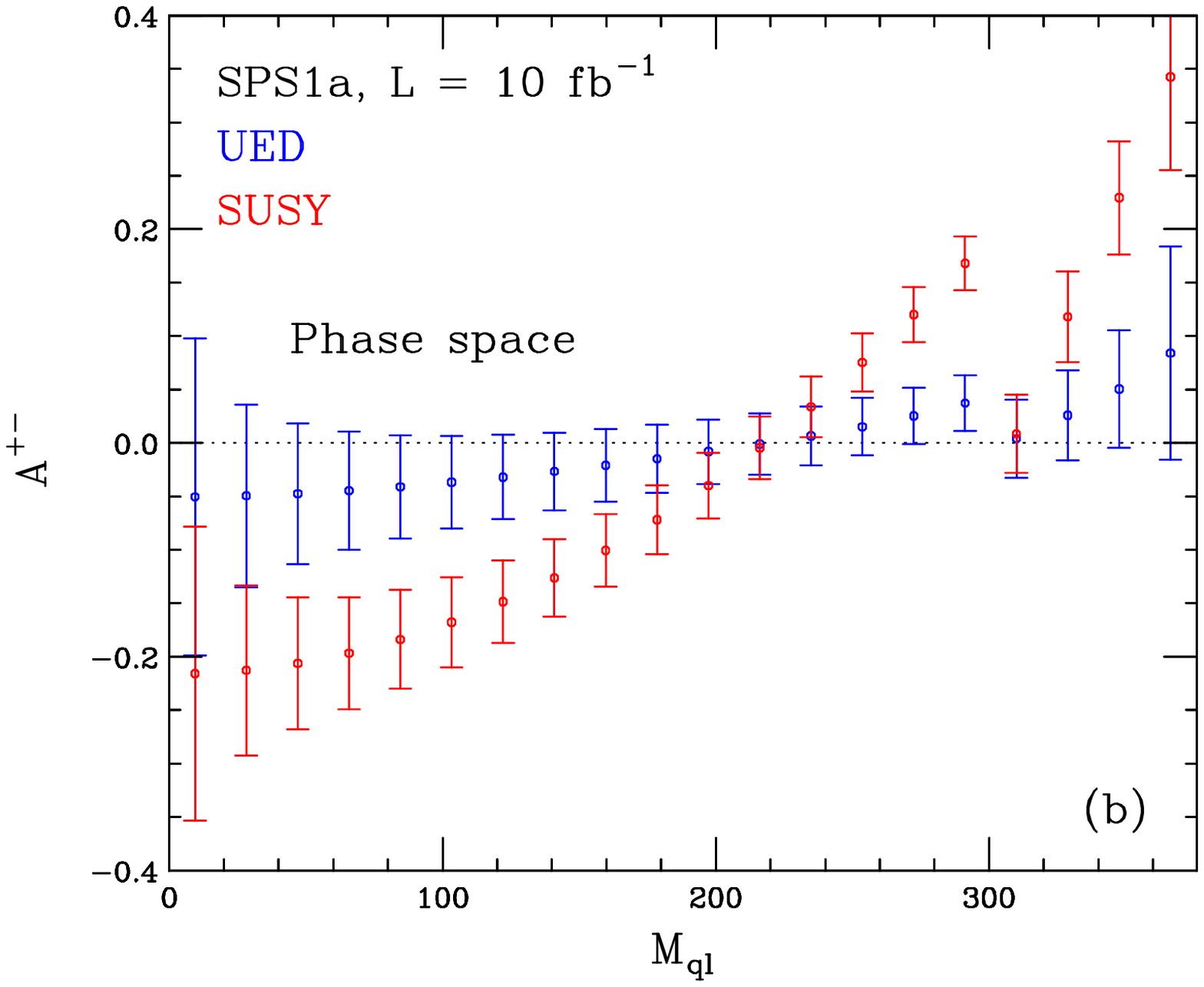}
\caption{Asymmetries for UED (SUSY) are shown in blue (red) in the
case of (a) a UED mass spectrum with $R^{-1}=500$ GeV and (b) the
SPS1a mass spectrum.  The horizontal dotted line represents pure phase
space.  The error bars represent the statistical uncertainty for
${\cal L}=10$ fb$^{-1}$.}
\label{fig:asymmetry}
\end{figure}

Our comparison between $A^{+-}$ in the case of UED and SUSY for the
UED mass spectrum is shown in Fig.~\ref{fig:asymmetry}(a).
We see that although there is some minor difference in the shape of
the asymmetry curves, overall the two cases appear to be very
difficult to discriminate unambiguously, especially since the regions
near the two ends of the plot, where the deviation is the largest,
also happen to suffer from poor statistics.  Notice that we have not
included detector effects or backgrounds.  Finally, and perhaps most
importantly, this analysis ignores the combinatorial background from
other jets in the event, which could be misinterpreted as the starting
point of the cascade depicted in Fig.~\ref{fig:diagrams}.  Overall,
Fig.~\ref{fig:asymmetry} shows that although the asymmetry
(Eq.~\ref{asymmetry}) does encode some spin correlations,
distinguishing between the specific cases of UED and SUSY appears to
be challenging.

Similarly in Fig.~\ref{fig:asymmetry}(b), we show the asymmetry for
UED and SUSY for a mass spectrum of mSUGRA point SPS1a.  In this case,
the mass spectrum is broad compared to the UED spectrum and
$\tilde{\chi}_2^0$ in SUSY ($Z_1$ in UED) does not decay into
left-handed sleptons ($SU(2)_W$ KK leptons).  Unlike the narrow mass
spectrum, here we experience larger mass splittings, as expected in
typical SUSY models, and the asymmetry distributions appear to be more
distinct than the case shown in Fig.~\ref{fig:asymmetry}(a), which is
a source of optimism.  These results have been recently confirmed in
Ref.~\cite{Smillie:2005ar}.  It remains to be seen whether this
conclusion persists in a more general setting, and once the
combinatorial backgrounds are included~\cite{KK}.  Notice that
comparing (a) and (b) in Fig.~\ref{asymmetry}, the signs of the two
asymmetries have changed.  The difference is the chirality of sleptons
or KK leptons.  In Fig.~\ref{asymmetry}(a) (Fig.~\ref{asymmetry}(a)),
left-handed sleptons or $SU(2)_W$-doublet KK leptons (right-handed
sleptons or $SU(2)_W$-singlet KK leptons) are on shell and the
asymmetry starts positive (negative) and ends negative (positive).  By
looking at the sign of the asymmetry, we can determine which chirality
was on-shell.

What we did so far was, first choose a study point in one model and
adjust parameters in other models until we match the mass spectra.
However, not all masses are observable and sometimes we get fewer
constraints than the number of masses involved in the decay.  So what
we need to do is to match endpoints in the distributions instead of
matching mass spectra, and ask whether there is any point in parameter
space which is consistent with the experimental data.  In other words,
we have to ask which model fits the data better.
We consider three kinematic endpoints: $m_{q\ell\ell}$, $m_{q\ell}$
and $m_{\ell\ell}$ (see Fig.~\ref{fig:diagrams}).  In principle, we
can find more kinematic endpoints such as the lower edge.  Here we are
conservative and take upper edges
only~\cite{Bachacou:1999zb,Allanach:2000kt,Gjelsten:2004ki}.  In the
case of an on-shell decay of $\chi^0_2$ and $\tilde\ell$, these three
kinematic endpoints are written in terms of the invariant masses
%
\begin{eqnarray}
m_{q\ell\ell} &=& m_{\tilde{q}} \sqrt{ (1-x) (1-y z)  } \nonumber \\
m_{q\ell}     &=& m_{\tilde{q}} \sqrt{ (1-x) (1- z)   } \\
m_{\ell\ell}  &=& m_{\tilde{q}} \sqrt{ x (1-y) (1- z) } \nonumber
\end{eqnarray}
where $m_{\tilde{q}}$ is the squark or KK quark mass and
$x=\left(\frac{m_{\tilde{\chi}_2^0}}{m_{\tilde{q}}}\right)^2$,
$y=\left(\frac{m_{\tilde\ell}}{m_{\tilde{\chi}_2^0}}\right)^2$ and
$z=\left(\frac{m_{\tilde{\chi}_1^0}}{m_{\tilde\ell}}\right)^2$ are the
ratios of masses in the cascade decay chain.  By definition, $x$, $y$
and $z$ are each less than 1.

We are now left with 2 free parameters, $f_q$ and $x$, from which we
solve for $y$, $z$ and $m_{\tilde{q}}$.  We minimize $\chi^2$,
\begin{equation}
\chi^2 = \sum_{i=1}^{n} \frac{(x_i - \mu_i)^2}{\sigma_i^2} \, ,
\end{equation}
between the two asymmetries in the ($x$, $f_q$) parameter space to see
whether we can fake a SUSY asymmetry in a UED model.  $x_i$ is the
theory prediction and $\mu_i$ is the experimental value with
uncertainty $\sigma_i$.  $\chi^2_{dof}=\chi^2/n$ is the `reduced'
$\chi^2$ or $\chi^2$ for $n$ degrees of freedom.

\begin{figure}[t]
\includegraphics[width=7.5cm]{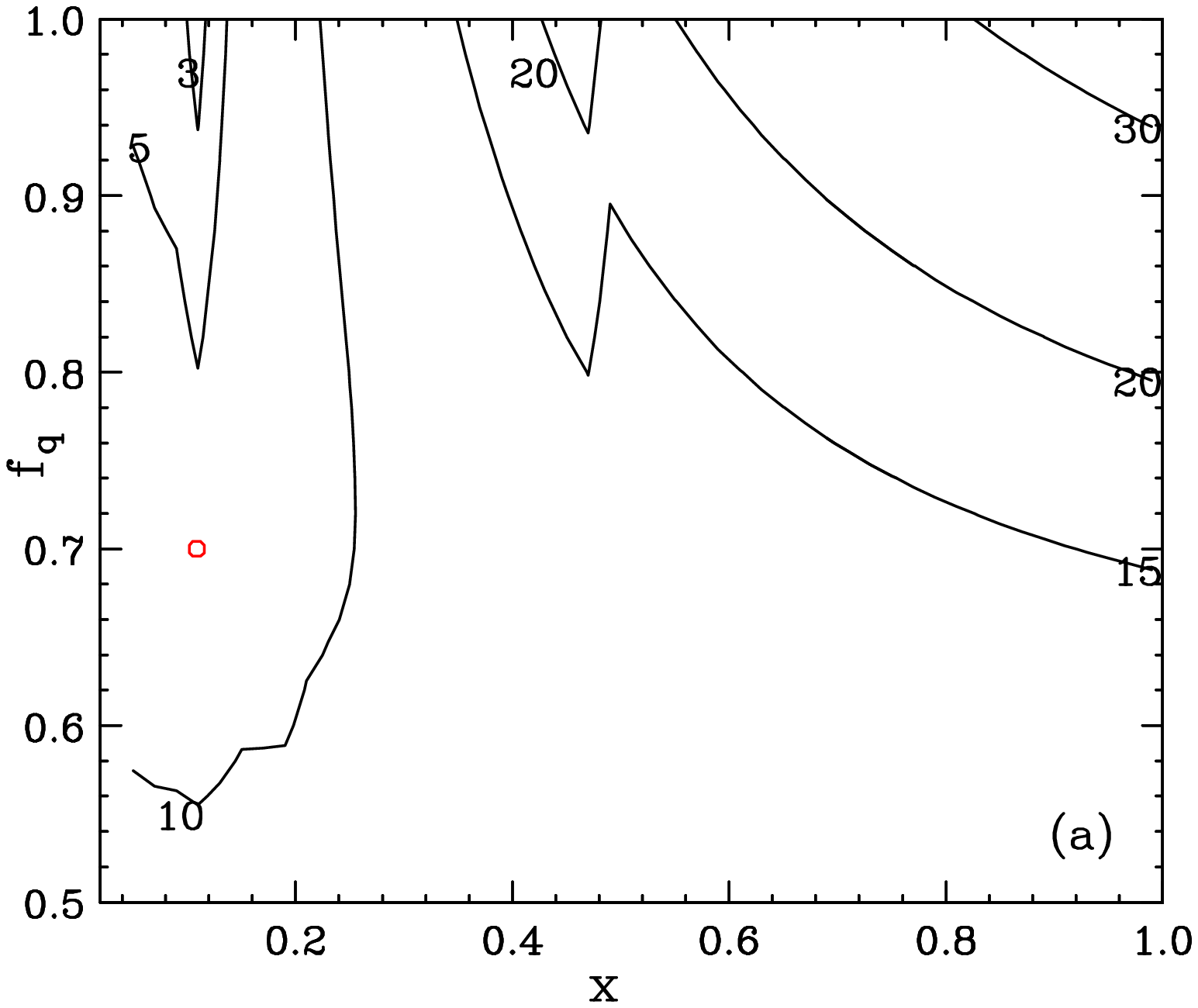}
\includegraphics[width=7.6cm]{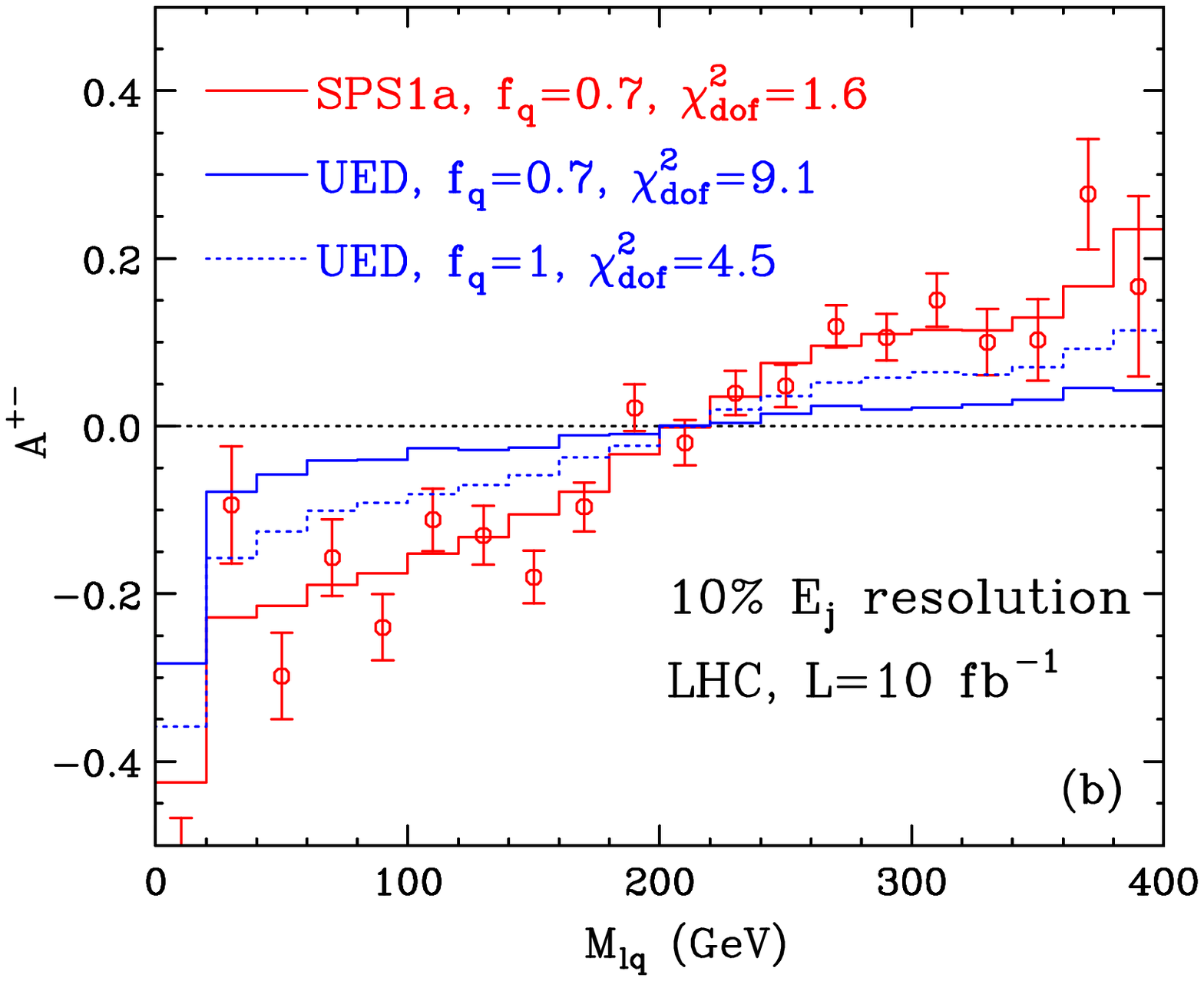}
\caption{(a) The contour lines show $\chi^2$ in ($x$, $f_q$) parameter
space and the red dot represents the SPS1a point.  $\chi^2$ is
minimized when $f_q \sim 1$ and $x$ are the same as for SPS1a.  (b)
Red dots represent the data points generated from SPS1a with ${\cal
L}=10$ fb$^{-1}$ including a $10\%$ jet energy resolution.
$\chi^2$-minimized UED (SUSY) fits to the Monte Carlo are shown in
blue (red).  Since the Monte Carlo was generated from SUSY, a small
$\chi^2$ for the SUSY fit is expected.  $\chi^2$ for UED fits is 9.1
(blue sold) and 4.5 (blue dotted) for $f_q=0.7$ and $f_q$=1,
respectively.}
\label{fig:smeared}
\end{figure}

Our result is shown in Fig.~\ref{fig:smeared}(a).  We found a minimum
$\chi^2$ of around 3 in the region where all KK masses are the same as
the SUSY masses in the decay and $f_q$ is large.  This means that
$\chi^2$ is minimized when we have a perfect match in mass spectrum.
The red circle is the point SPS1a.

Now since we don't yet have experimental data, we generated data
samples from SPS1a assuming $10fb^{-1}$ and construct the asymmetries
in SUSY and UED in Fig.~\ref{fig:smeared}(b).  We included a $10\%$
jet energy resolution.  Red dots represent data points and the red
line is the SUSY fit to the data points.  The blue lines are the UED
fits to data points for two different values of $f_q$.  For SUSY,
$\chi^2$ is around 1 as we expect.  We can get better $\chi^2$ for
UED, from 9.1 to 4.5, by increasing $f_q$.  It is still too large to
fit to the Monte Carlo.  So our conclusion for this study is that a
particular point like SPS1a can not be faked throughout the entire
parameter space of UED.  However, we need to check whether this
conclusion will remain the same when we include the wrong jet
assignment, i.e. jets which have nothing to do with this decay
chain~\cite{KK}.  Notice that the clear edge at $m_{q\ell}\sim
300$~GeV in Fig.~\ref{fig:asymmetry}(b) disappeared in
Fig.~\ref{fig:smeared}(b) after including jet energy resolution.  From
Fig.~\ref{fig:asymmetry}, we see that SUSY has a larger asymmetry.


\subsubsection{Conclusions}
\label{sec:conclusions}

The fundamental difference between UED and SUSY is 1) the number of
partners of SM particles and 2) the spins of new particles.  The
second level of KK particles can be confused with e.g. a $Z'$ and its
existence is not a direct proof of UED, although the smoking gun for
UED is degeneracy in resonance masses.  Therefore to discriminate
these two models, we need to measure the spins of new particles.  Two
methods are discussed in this paper and the key was a mass spectrum.
First in the dilepton mass, with a narrow mass spectrum (UED type) and
a mass spectrum from mSUGRA model, UED and SUSY predict very similar
distributions.  In some regions of MSSM parameter space away from the
mSUGRA, the spin correlation becomes more important and the
differences in distributions start to appear.  Second, if we measure a
non-zero asymmetry, this means that the new particle
($\tilde{\chi}^0_2$ or $Z_1$) in the cascade decay has non-zero spin.
An asymmetry study also tells about the relative chirality of sleptons
and KK leptons.  However, whether one can measure $A^{+-}\neq 0$ or
not depends on the particular point in parameter space.  For instance,
in the focus point region, $\tilde{g}$ production dominates and equal
numbers of quarks and anti-quarks are produced, which wash out
$A^{+-}$.  In the case of off-shell sleptons decays, the spin
correlation is small, and it is not clear whether an asymmetry would
be observable in this case.  Also, we can easily find parameter space
where two contributions from sleptons with different chiralities can
cancel each other.  Even if we measure a non-zero asymmetry, it is not
clear whether the new particle has spin 1 or 1/2 for the degenerate
mass spectrum.  For a particular point like SPS1a in mSUGRA, we can
tell that the new particle is indeed a SUSY partner.  However, even in
this case, we don't know the effect of wrong choice of jets, and
further study is needed.


\subsubsection*{Acknowledgements}

The work of KK and KM is supported in part by 
a US Department of Energy Outstanding Junior Investigator 
award under grant DE-FG02-97ER41209.

\clearpage\setcounter{equation}{0}\setcounter{figure}{0}\setcounter{table}{0}
\def\red#1{{\color{red}#1}}
\def\green#1{{\color{green}#1}}
\def\blue#1{{\color{blue}#1}}
\subsection{Collider Search for Level 2 Kaluza-Klein Gauge Bosons at Hadron Colliders}
\label{sec:Level-2}

{\em AseshKrishna Datta$^1$, 
Kyoungchul Kong$^2$ and Konstantin T.~Matchev$^2$ \\
$^1$ MCTP, University of Michigan, Ann Arbor, USA
\footnote{Current address: Harish-Chandra Research Institute, Allahabad, India} \\
$^2$ Institute for Fundamental Theory, Physics Dept.,
University of Florida, USA}\\


{\em We contrast the experimental signatures of low-energy
supersymmetry and the model of Universal Extra Dimensions, and discuss
their discrimination at hadron colliders.  We study the discovery
reach of the Tevatron and LHC for level 2 Kaluza-Klein modes, which
would indicate the presence of extra dimensions.  We find that with
100~fb$^{-1}$ of data the LHC will be able to discover the $\gamma_2$
and $Z_2$ KK modes as separate resonances if their masses are below 2
TeV.}


\subsubsection{Introduction}

Supersymmetry (SUSY) and Extra Dimensions (ED) offer two different
paths to a theory of new physics beyond the Standard Model (SM).  They
both address the hierarchy problem, play a role in a more fundamental
theory aimed at unifying the SM with gravity, and offer a candidate
particle for dark matter, compatible with present cosmology data.  If
either SUSY or ED exist at the TeV scale, signals of new physics
should be found by the ATLAS and CMS experiments at the Large Hadron
Collider (LHC) at CERN.  However, as we discuss below, the proper
interpretation of such discoveries may not be straightforward.

A particularly interesting scenario of TeV-size extra dimensions is
offered by the so called Universal Extra Dimensions (UED) model,
originally proposed in~\cite{Appelquist:2000nn}, where all SM
particles are allowed to freely propagate into the bulk.  The case of
UED bears interesting analogies to SUSY and sometimes has been
referred to as ``bosonic supersymmetry''~\cite{Cheng:2002ab}.  In
principle, disentangling UED and supersymmetry appears highly
non-trivial at hadron
colliders~\cite{Cheng:2002ab,Datta:2005zs,Battaglia:2005ma}.  For each
SM particle, both models predict the existence of a partner (or
partners) with identical interactions.  Unfortunately, the masses of
these new particles are model-dependent and cannot be used to
unambiguously discriminate between the two theories\footnote{Notice
that the recently proposed little Higgs models with
$T$-parity~\cite{Cheng:2003ju,Cheng:2004yc,Hubisz:2004ft,Cheng:2005as}
are reminiscent of UED, and may also be confused with SUSY.}.  Both
theories have a good dark matter
candidate~\cite{Servant:2002aq,Kakizaki:2005uy,Burnell:2005hm,Kong:2005hn,Cheng:2002ej}
and the typical collider signatures contain missing energy.  One would
therefore like to have experimental discriminators which rely on the
fundamental distinctions between the two models.  In what follows we
shall discuss methods for experimental discrimination between SUSY and
UED and study the discovery reach for level 2 Kaluza-Klein (KK) gauge
boson particles and the resolving power of the LHC to see them as
separate resonances.


\subsubsection{Phenomenology of Universal Extra Dimensions}
\label{sec:ued}


\subsubsection*{The Minimal UED Model}
\label{sec:MUED}
 
Models of UED place all SM particles in the bulk of one or more
compactified ED.  In the simplest, most popular version, there is a
single extra dimension of size $R$, compactified on an $S_1/Z_2$
orbifold~\cite{Appelquist:2000nn}.

A peculiar feature of UED is the conservation of KK number at tree
level, which is a simple consequence of momentum conservation along
the extra dimension.  However, bulk and brane radiative
effects~\cite{Georgi:2000ks,vonGersdorff:2002as,Cheng:2002iz} break KK
number down to a discrete conserved quantity, the so-called KK parity,
$(-1)^n$, where $n$ is the KK level.  KK parity ensures that the
lightest KK partners (level one) are always pair-produced in collider
experiments, just like in the $R$-parity conserving supersymmetry
models discussed in Section~\ref{sec:intro}.  KK parity conservation
also implies that the contributions to various low-energy
observables~\cite{Agashe:2001ra,Agashe:2001xt,Appelquist:2001jz,Petriello:2002uu,Appelquist:2002wb,Chakraverty:2002qk,Buras:2002ej,Oliver:2002up,Buras:2003mk,Iltan:2003tn,Khalil:2004qk}
arise only at loop level and are small.  As a result, limits on the
scale $R^{-1}$ of the extra dimension from precision electroweak data
are rather weak, constraining $R^{-1}$ to be larger than approximately
250~GeV~\cite{Appelquist:2002wb}.  An attractive feature of UED models
with KK parity is the presence of a stable massive particle which can
be a cold dark matter
candidate~\cite{Servant:2002aq,Kakizaki:2005uy,Burnell:2005hm,Kong:2005hn,Cheng:2002ej}.

\begin{figure}[t]
\includegraphics[width=7.5cm]{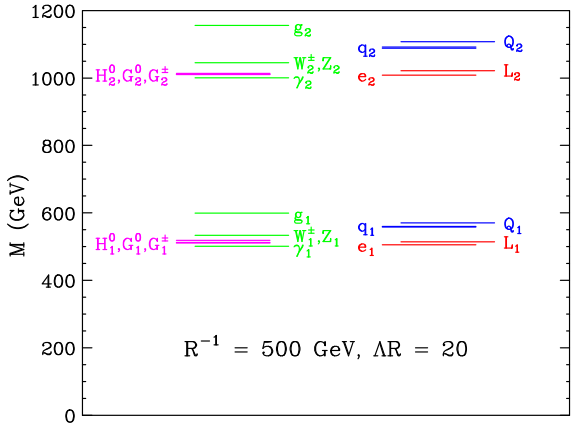}
\caption{One-loop-corrected mass spectrum of the $n=1$ and $n=2$ KK 
levels in Minimal UED, for $R^{-1}=500$ GeV, $\Lambda R=20$ and
$m_h=120$ GeV.  We show the KK modes of gauge bosons, Higgs and
Goldstone bosons and first generation fermions.}
\label{fig:spectrum}
\end{figure}

In Fig.~\ref{fig:spectrum} we show the mass spectrum of the $n=1$ and
$n=2$ KK levels in Minimal UED (MUED), for $R^{-1}=500$ GeV, $\Lambda
R=20$ and SM Higgs boson mass $m_h=120$ GeV.  We include the full
one-loop corrections from Ref.~\cite{Cheng:2002iz}.  We have used
RGE-improved couplings to compute the radiative corrections to the KK
masses.  It is well known that in UED the KK modes modify the running
of the coupling constants at higher scales.  We extrapolate the gauge
coupling constants to the scale of the $n=1$ and $n=2$ KK modes, using
the appropriate $\beta$ functions dictated by the particle
spectrum~\cite{Dienes:1998vg,Perez-Lorenzana:1999qb,Cheng:1999fu}.  As
a result, the spectrum shown in Fig.~\ref{fig:spectrum} differs
slightly from the one in Ref.~\cite{Cheng:2002iz}.  Most notably, the
colored KK particles are somewhat lighter, due to a reduced value of
the strong coupling constant, and overall the KK spectrum at each
level is more degenerate.

\begin{figure}[t]
\includegraphics[width=8cm]{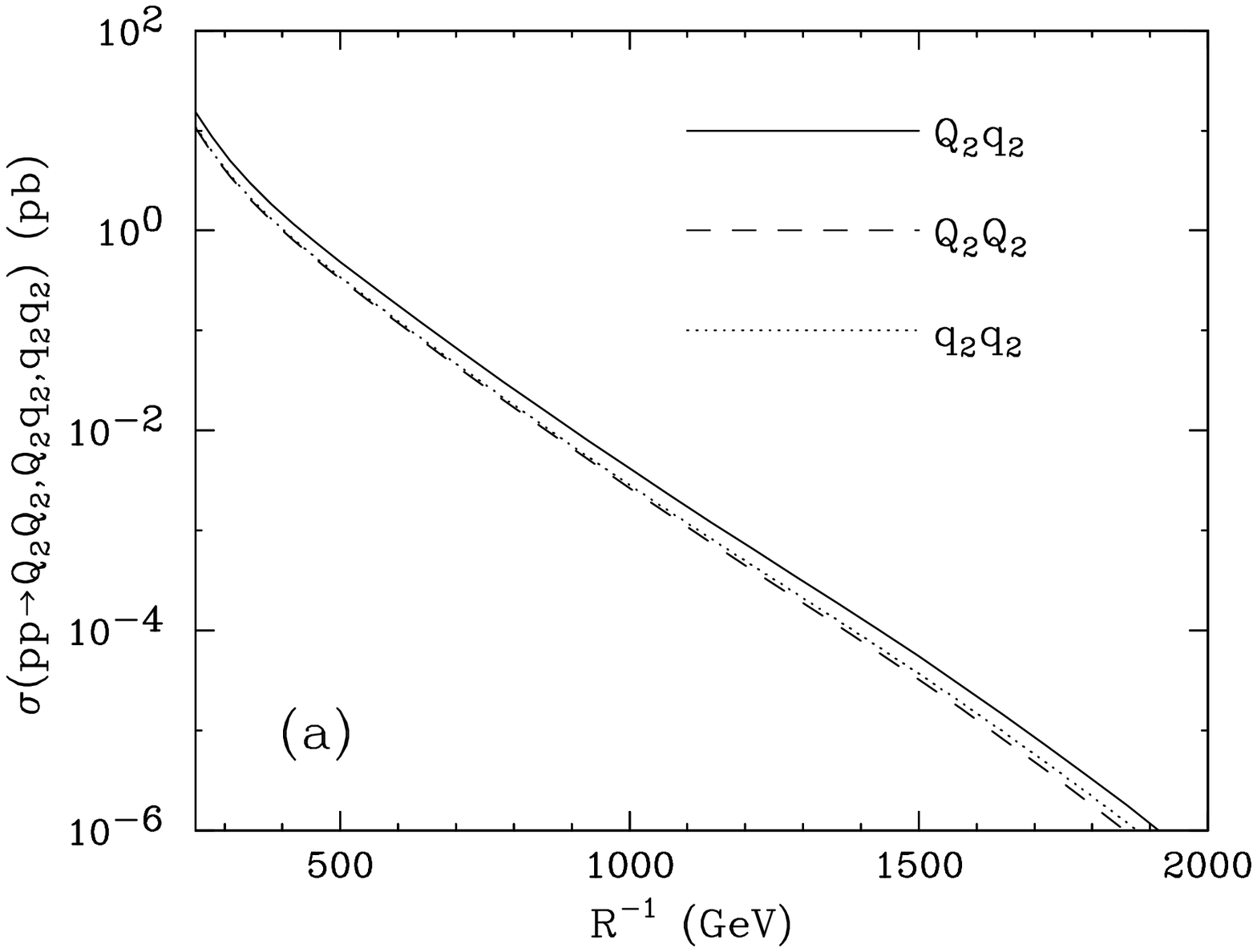}
\includegraphics[width=8cm]{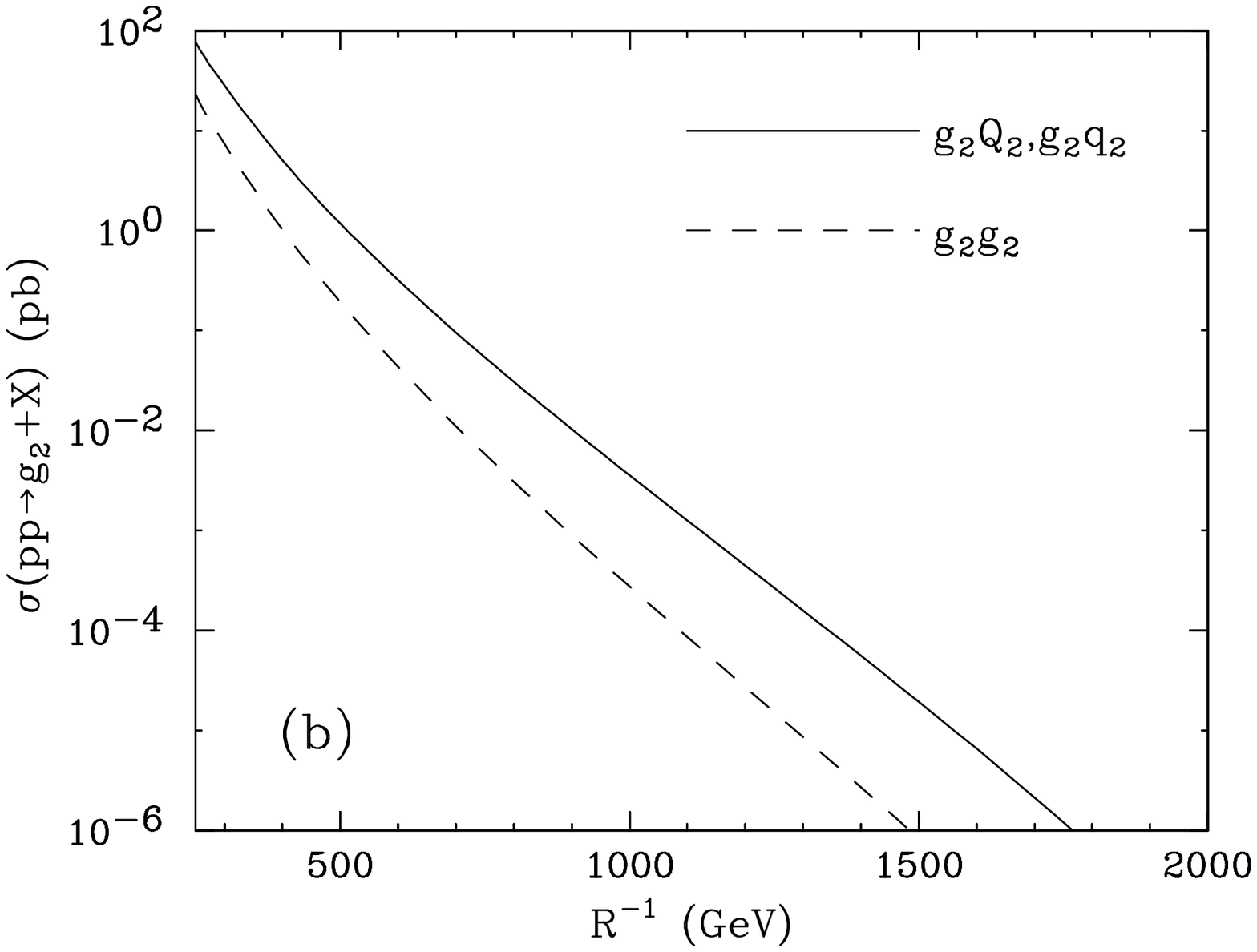}
\caption{Strong production of $n=2$ KK particles at the LHC:
(a) KK-quark pair production; (b) KK-quark/KK-gluon associated
production and KK-gluon pair production.  The cross-sections have been
summed over all quark flavors and also include charge-conjugated
contributions such as $Q_2\bar{q}_2$, $\bar{Q}_2q_2$, $g_2\bar{Q}_2$,
etc.}
\label{fig:sigma_q2g2}
\end{figure}
%


\subsubsection*{Comparison of UED and Supersymmetry}

There is a wide variety of SUSY models, with very diverse
phenomenology.  Nevertheless, they all share the following common
features which define a supersymmetric framework:
\begin{enumerate}
\item For each particle of the Standard Model, supersymmetry predicts
a new particle (superpartner).
\item The spins of the superpartners differ by $1/2$ unit.
\item The couplings of the particles and their superpartners are
equal, being related by supersymmetry
\item The generic collider signature of supersymmetric models with 
WIMP LSPs is missing energy.
\end{enumerate}

The last property makes exact reconstruction of the event kinematics
practically impossible.  At a hadron collider, the center of mass
energy is not known on an event-per-event basis.  In addition, the
momenta of {\em both} $\tilde\chi^0_1$ particles are unknown, and what
is measured is only the transverse component of the sum of their
momenta, provided there are no other sources of missing energy in the
event (such as neutrinos, $b$-jets, $\tau$-jets, etc.).  This
incomplete information is the main stumbling block in proving the
basic properties of SUSY at the LHC.

In complete analogy, the discussion of Minimal UED model leads to the
following generic features of UED:
\begin{enumerate}
\item For each particle of the Standard Model, UED models predict
an infinite\footnote{Strictly speaking, the number of KK modes is
$\Lambda R$.} tower of new particles (Kaluza-Klein partners).
\item The spins of the SM particles and their KK partners are the same.
\item The couplings of the SM particles and their KK partners are equal.
\item The generic collider signature of UED models with WIMP LKPs is 
missing energy.
\end{enumerate}
Notice that defining features 3 and 4 are common to both SUSY and UED
and cannot be used to distinguish the two cases.  We see that while
$R$-parity conserving SUSY implies a missing energy signal, the
reverse is not true: a missing energy signal would appear in any model
with a dark matter candidate, and even in models which have nothing to
do with the dark matter issue, but simply contain new neutral
quasi-stable particles.  Similarly, the equality of the couplings
(feature No.~3) is a celebrated test of SUSY, but from the above
comparison we see that it is only a necessary, not sufficient
condition in proving SUSY.

We are therefore forced to concentrate on the first two identifying
features as the only promising discriminating criteria.  Let us begin
with feature 1: the number of new particles.  The KK particles at
$n=1$ are analogous to superpartners in SUSY.  However, the particles
at the higher KK levels have no analogues in $N=1$ supersymmetric
models.  Discovering the $n\ge2$ levels of the KK tower would
therefore indicate the presence of extra dimensions rather than SUSY.
We shall concentrate on the $n=2$ level and investigate the discovery
opportunities at the LHC and the Tevatron (for linear collider studies
of $n=2$ KK gauge bosons, see
Ref.~\cite{Battaglia:2005ma,Battaglia:2005zf,Bhattacherjee:2005qe,Riemann:2005es}).
Notice that the masses of the KK modes are given roughly by $m_n\sim
n/R$, where $n$ is the KK level number, so that the particles at
levels 3 and higher are rather heavy and their production is severely
suppressed.

The second identifying feature -- the spins of the new particles --
also provides a tool for discrimination between SUSY and UED.
Recently it was suggested that a charge asymmetry in the lepton-jet
invariant mass distributions from a particular cascade can be used to
discriminate SUSY from the case of pure phase space
decays~\cite{Datta:2005zs,Battaglia:2005ma,Barr:2004ze,Smillie:2005ar}.
The possibility of discriminating SUSY and UED by this method was the
subject of Sec.~\ref{sec:spin}.  For the purposes of our study we
implemented the relevant features of MUED in the {\tt CompHEP} event
generator \cite{Pukhov:1999gg}.  The Minimal Supersymmetric Standard
Model (MSSM) is already available in {\tt CompHEP} (since
version~41.10).


\subsubsection{Collider Search for Level 2 KK Gauge Bosons}
\label{sec:level2}


\subsubsection*{Phenomenology of Level 2 Fermions}
\label{sec:f2}

In principle, there are two mechanisms for producing $n=2$ KK quarks
at the LHC: through KK-number conserving interactions, or through
KK-number violating (but KK-parity conserving) interactions.  The KK
number conserving QCD interactions allow production of KK quarks
either in pairs or singly (in association with the $n=2$ KK mode of a
gauge boson).  The corresponding production cross sections are shown
in Fig.~\ref{fig:sigma_q2g2} (the cross sections for producing $n=1$
KK quarks were calculated
in~\cite{Rizzo:2001sd,Macesanu:2002db,Smillie:2005ar}).  In
Fig.~\ref{fig:sigma_q2g2}a we show the cross sections (pb) for $n=2$
KK-quark pair production, while in Fig.~\ref{fig:sigma_q2g2}b we show
the results for $n=2$ KK-quark/KK-gluon associated production and for
$n=2$ KK-gluon pair production.  We plot the results versus $R^{-1}$,
and one should keep in mind that the masses of the $n=2$ particles are
roughly $2/R$.  In calculating the cross sections of
Fig.~\ref{fig:sigma_q2g2}, we consider 5 partonic quark flavors in the
proton along with the gluon.  We sum over the final state quark
flavors and include charge-conjugated contributions.  We used CTEQ5L
parton distributions~\cite{Lai:1999wy} and choose the scale of the
strong coupling constant $\alpha_s$ to be equal to the parton level
center of mass energy.  All calculations are done with {\tt
CompHEP}~\cite{Pukhov:1999gg} with our implementation of MUED.  One
could consider single production of $n=2$ KK quarks through KK-number
violation but the lowest-order coupling of an $n=2$ KK quark to two SM
particles is suppressed by the cutoff scale, which is unknown.

Having determined the production rates of level 2 KK quarks, we now
turn to the discussion of their experimental signatures.  To this end
we need to determine the possible decay modes of $Q_2$ and $q_2$.  At
each level $n$, the KK quarks are among the heaviest states in the KK
spectrum and can decay promptly to lighter KK modes.  As can be seen
from Fig.~\ref{fig:spectrum}, the KK gluon is always heavier than the
KK quarks, so the two body decays of KK quarks to KK gluons are
closed.  Instead, $n=2$ KK quarks will decay to the KK modes of the
electroweak gauge bosons which are lighter.  The branching fractions
for $n=2$ KK quarks are almost independent of $R^{-1}$, unless one is
close to threshold.  This feature persists for all branching ratios of
KK particles.

\begin{figure}[t]
\includegraphics[width=8cm]{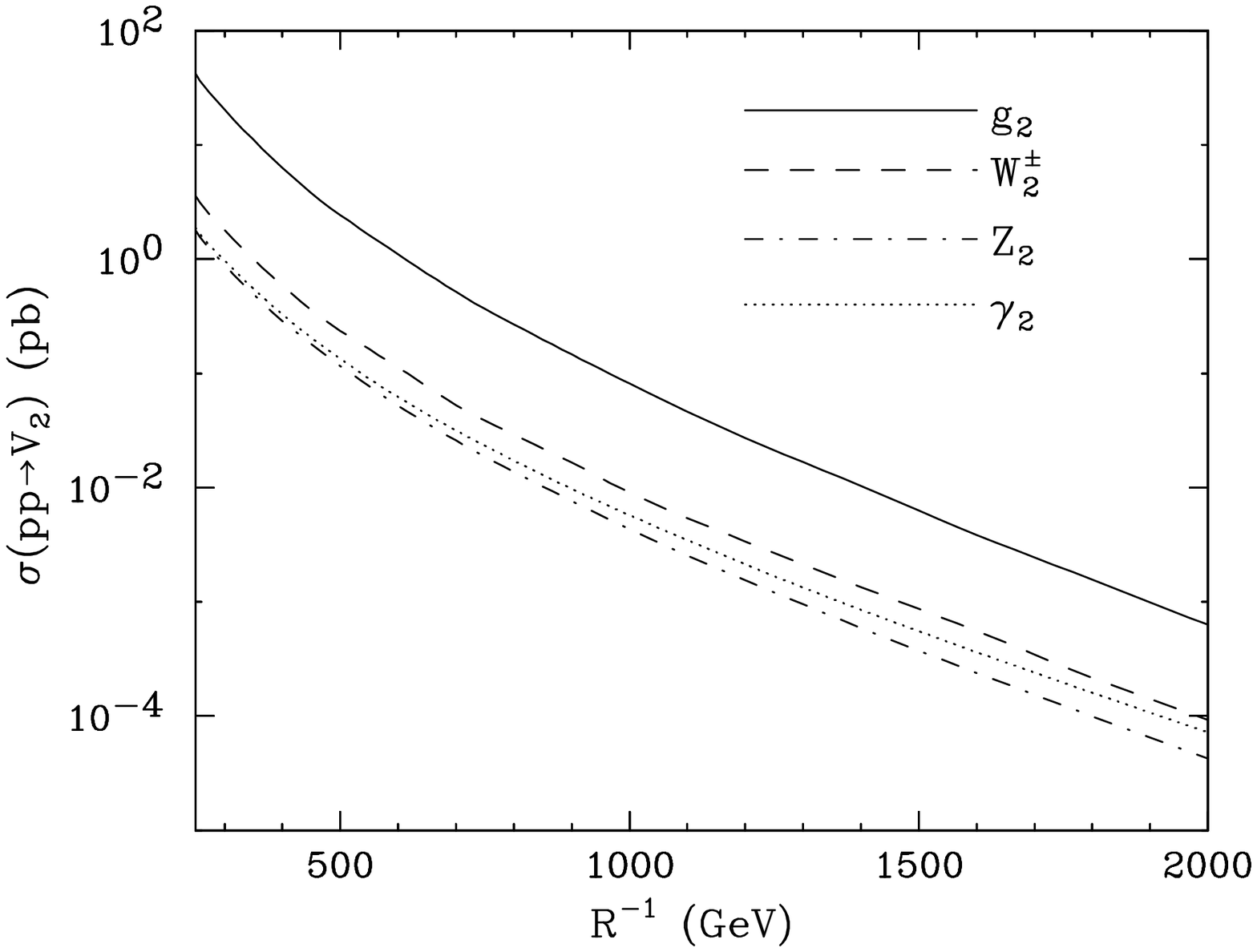}
\includegraphics[width=8cm]{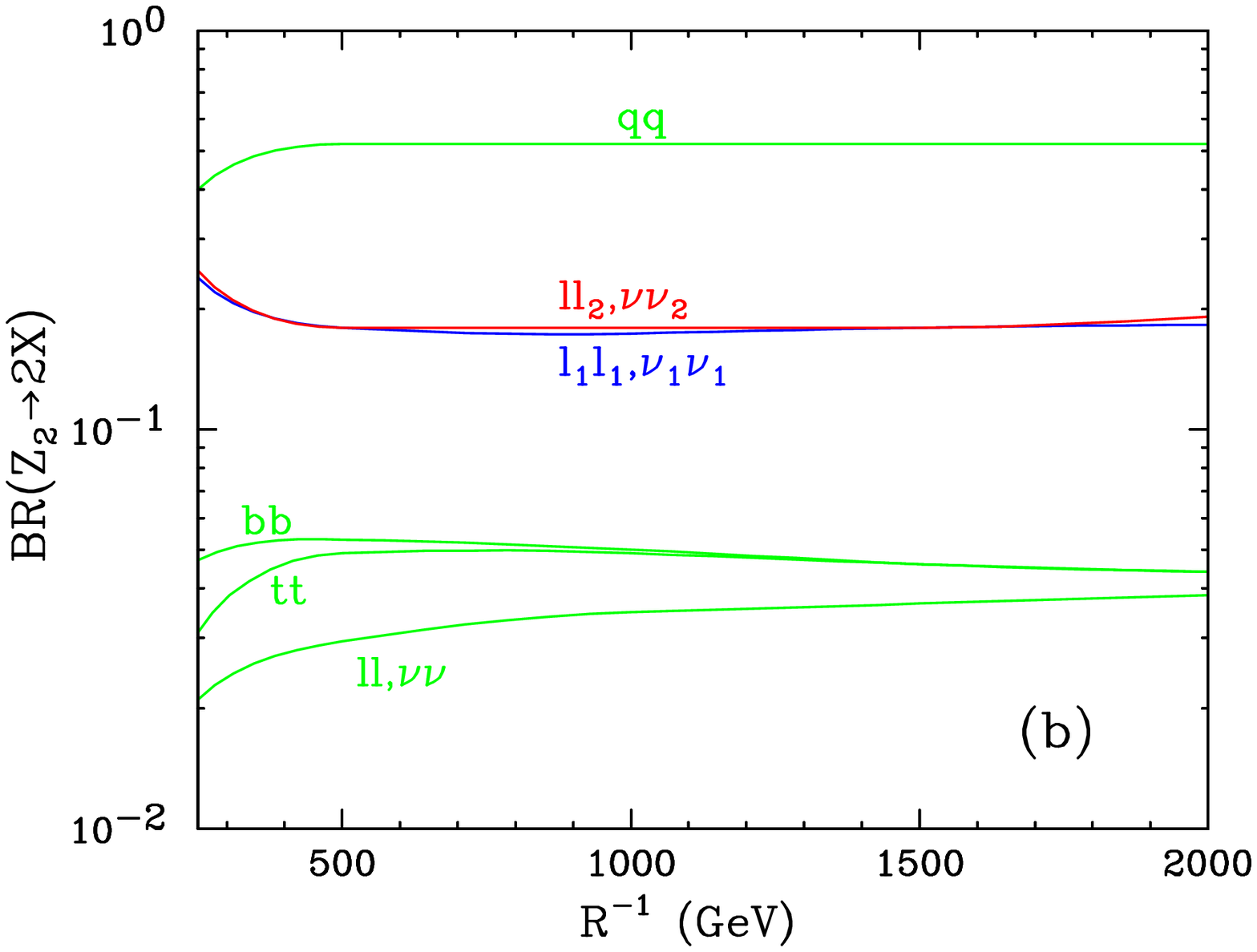}
\caption{(a) Cross sections for single production of level 2 KK gauge 
bosons through the KK number violating couplings.  (b) Branching
fraction of the $Z_2$ KK gauge boson and branching fractions of other
$n_2$ KK gauge bosons are very similar.}
\label{fig:sigma_V2}
\end{figure}

The case of the $SU(2)_W$-singlet quarks such as $q_2$ is simple,
since they only couple to the hypercharge gauge bosons.  At $n\ge1$
the hypercharge component is almost entirely contained in the $\gamma$
KK mode~\cite{Cheng:2002iz}.  We therefore expect a singlet KK quark
$q_2$ to decay to either $q_1\gamma_1$ or $q_0\gamma_2$, and in fact
they have roughly same branching fractions ($50\%$).  The case of an
$SU(2)_W$-doublet quark $Q_2$ is much more complicated, since $Q_2$
couples to the (KK modes of the) weak gauge bosons as well, and many
more two-body final states are possible.  Since the weak coupling is
larger than the hypercharge coupling, the decays to $W$ and $Z$ KK
modes dominate, with $BR(Q_2\to Q'_0W_2)/BR(Q_2\to Q_0Z_2)=2$ and
$BR(Q_2\to Q'_1W_1)/BR(Q_2\to Q_1Z_1)=2$.  The branching fractions to
the $\gamma$ KK modes are only on the order of a few percent.


\subsubsection*{Level 2 Gauge Bosons}
\label{sec:V2}

There are four $n=2$ KK gauge bosons: the KK ``photon'' $\gamma_2$,
the KK ``$Z$-boson'' $Z_2$, the KK ``$W$-boson'' $W^\pm_2$, and the KK
gluon $g_2$.  Recall that the Weinberg angle at $n=2$ is very small,
so that $\gamma_2$ is mostly the KK mode of the hypercharge gauge
boson and $Z_2$ is mostly the KK mode of the neutral $W$-boson of the
SM.  An important consequence of the extra-dimensional nature of the
model is that all four of the $n=2$ KK gauge bosons are relatively
degenerate; the masses are all roughly equal to $2/R$.  Mass
splittings are almost entirely due to radiative corrections, which in
MUED yield the hierarchy $m_{g_2}>m_{W_2}\sim m_{Z_2}>m_{\gamma_2}$.
The KK gluon receives the largest corrections and is the heaviest
particle in the KK spectrum at each level $n$.  The $W^\pm_2$ and
$Z_2$ particles are degenerate to a very high degree.

In Fig.~\ref{fig:sigma_V2}(a), we show single production cross
sections for level 2 KK gauge bosons.  Notice the roughly similar size
of the four cross sections.  This is somewhat surprising, since the
cross sections scale as the corresponding gauge coupling squared, and
one would have expected a wider spread in the values of the four cross
sections.  This is due to a couple of things.  First, for a given
$R^{-1}$, the masses of the four $n=2$ KK gauge bosons are different,
with $m_{g_2}>m_{W_2}\sim m_{Z_2}>m_{\gamma_2}$.  Therefore, for a
given $R^{-1}$, the heavier particles suffer a suppression.  This
explains to an extent why the cross section for $\gamma_2$ is not the
smallest of the four, and why the cross section for $g_2$ is not as
large as one would expect.  There is, however, a second effect, which
goes in the same direction.  The coupling is also proportional to the
mass corrections of the corresponding particles:
\begin{equation}
\frac{\bar\delta m_{V_2}}{m_{V_2}} - \frac{\bar\delta m_{f_2}}{m_{f_2}}\ .
\label{gQ0Q0A2}
\end{equation}
Since the QCD corrections are the largest, for
$V_2=\{\gamma_2,Z_2,W^\pm_2\}$, the second term dominates.  However,
for $V_2=g_2$, the first term is actually larger, and there is a
cancellation, which further reduces the direct KK gluon couplings to
quarks.

In Fig.~\ref{fig:sigma_V2}(b), we show branching fraction of $Z_2$
only as an example.  Again we observe that the branching fractions are
very weakly sensitive to $R^{-1}$, just as the case of KK quarks.
This can be understood as follows.  The partial for the KK number
conserving decays are proportional to the available phase space, while
the partial width for the KK number violating decay is proportional to
the mass corrections.  Both the phase space and mass corrections are
proportional to $R^{-1}$, which then cancels out in the branching
fraction.

\begin{figure}[t]
\includegraphics[width=8cm]{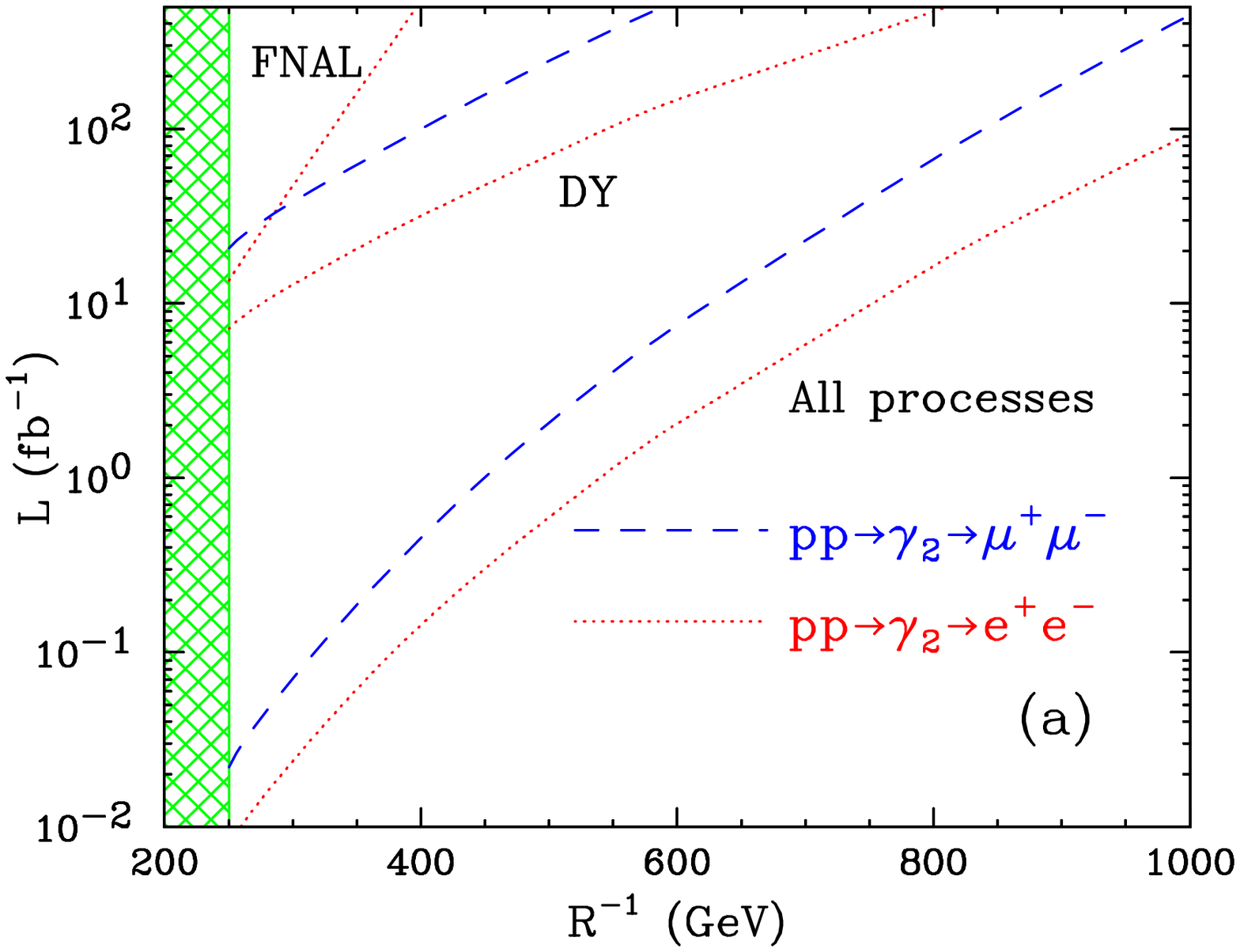}
\includegraphics[width=8cm]{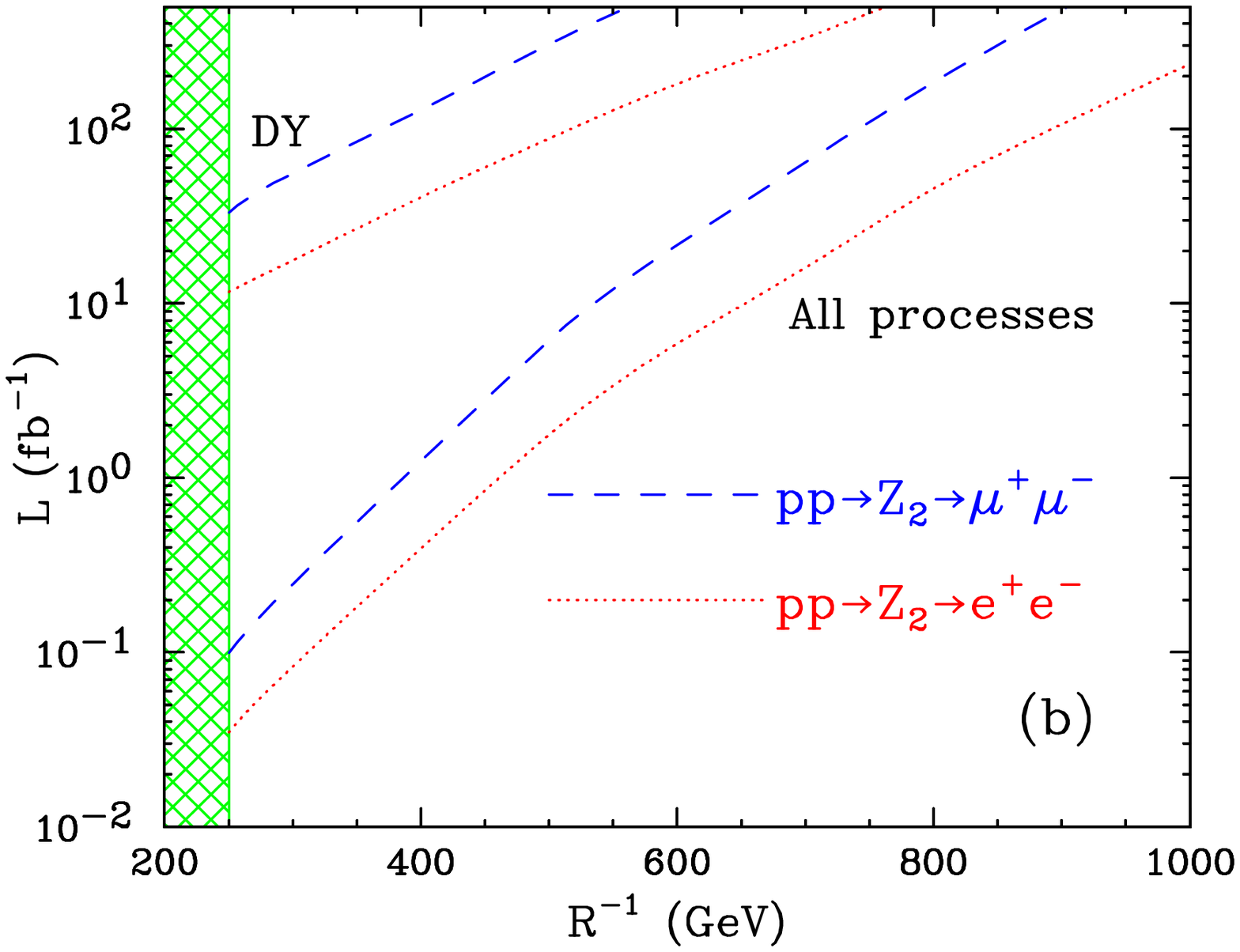}
\caption{$5\sigma$ discovery reach for (a) $\gamma_2$ and (b) $Z_2$.
We plot the total integrated luminosity ${\rm L}$ (fb$^{-1}$) required
for a $5\sigma$ excess of signal over background in the dielectron
(red, dotted) and dimuon (blue, dashed) channel, as a function of
$R^{-1}$.  In each plot, the upper set of lines labelled ``DY'' makes
use of the single $V_2$ production of Fig.~\ref{fig:sigma_V2} only,
while the lower set of lines (labelled ``All processes'') includes
indirect $\gamma_2$ and $Z_2$ production from $n=2$ KK quark decays.
The red dotted line marked ``FNAL'' in the upper left corner of (a)
reflects the expectations for a $\gamma_2\to e^+e^-$ discovery at
Tevatron Run~II.  The shaded area below $R^{-1}=250$~GeV indicates the
region disfavored by precision electroweak
data~\protect\cite{Appelquist:2002wb}.}
\label{fig:reach}
\end{figure}

The electroweak KK modes $\gamma_2$, $Z_2$ and $W^\pm_2$ can be
produced in the decays of heavier $n=2$ particles such as the KK
quarks and/or KK gluon.  This is well-known from the case of SUSY,
where the dominant production of electroweak superpartners is often
indirect -- from squark and gluino decay chains.  The indirect
production rates of $\gamma_2$, $Z_2$ and $W^\pm_2$ due to QCD
processes can be readily estimated from Figs.~\ref{fig:sigma_q2g2} and
branching fractions.  $BR(Q_2\to W^\pm_2)$, $BR(Q_2\to Z_2)$ and
$BR(q_2\to \gamma_2)$ are among the largest branching fractions of the
$n=2$ KK quarks, and we expect indirect production from QCD to be a
significant source of electroweak $n=2$ KK modes.

The $n=2$ KK modes can also be produced directly in pairs, through KK
number conserving interactions.  These processes, however, are
kinematically suppressed, since we have to make {\em two} heavy
particles in the final state.  One would therefore expect that they
will be the least relevant source of $n=2$ KK gauge bosons.  The only
exception is KK gluon pair production which is important and is shown
in Fig.~\ref{fig:sigma_q2g2}b.  We see that it is comparable in size
to KK quark pair production and $q_2g_2$/$Q_2g_2$ associated
production.  We have also calculated the pair production cross
sections for the electroweak $n=2$ KK gauge bosons and confirmed that
they are very small, hence we shall neglect them in our analysis
below.


\subsubsection*{Analysis of the LHC reach for $Z_2$ and $\gamma_2$}
\label{sec:analysis}

\begin{figure}[t]
\includegraphics[width=8cm]{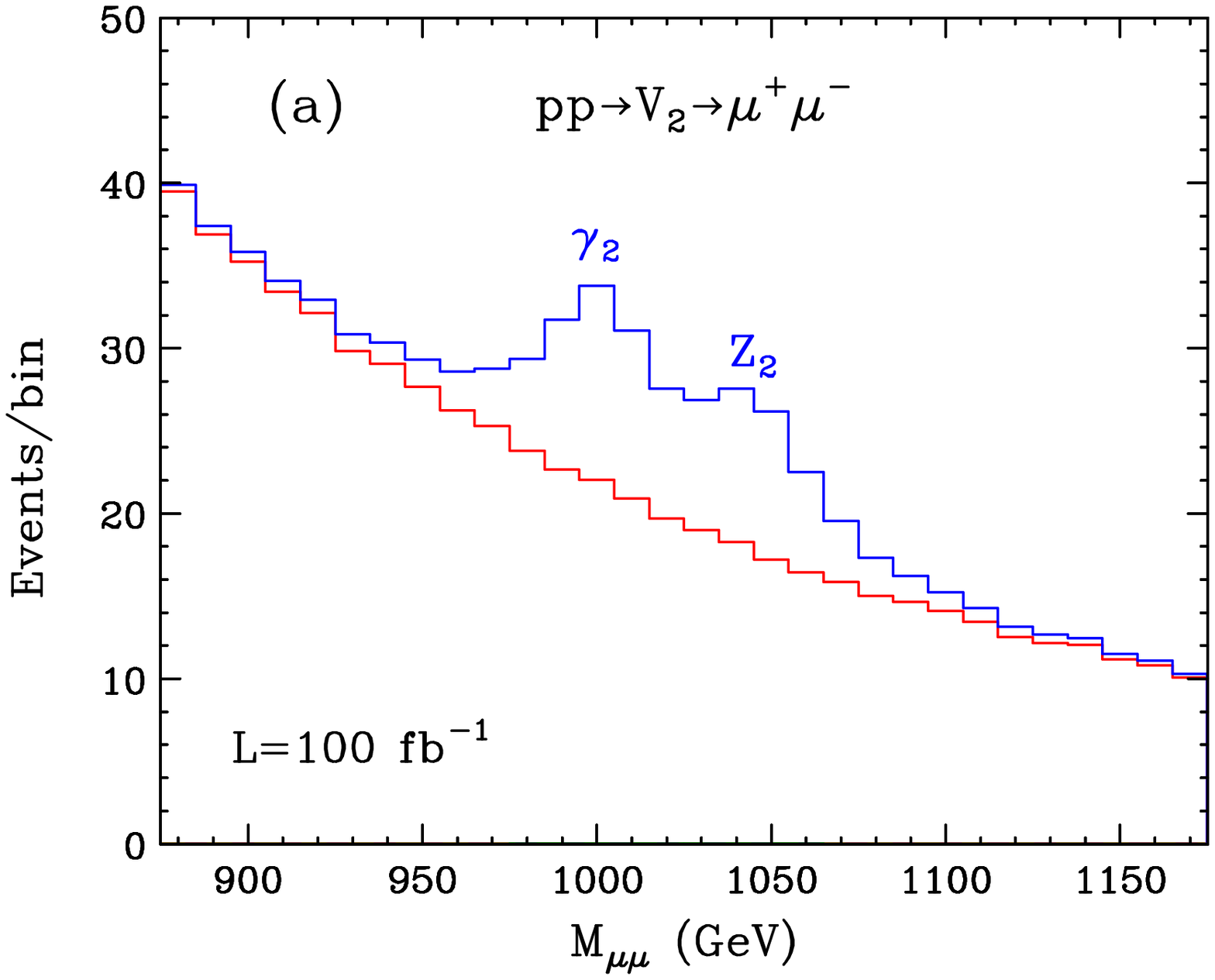}
\includegraphics[width=8cm]{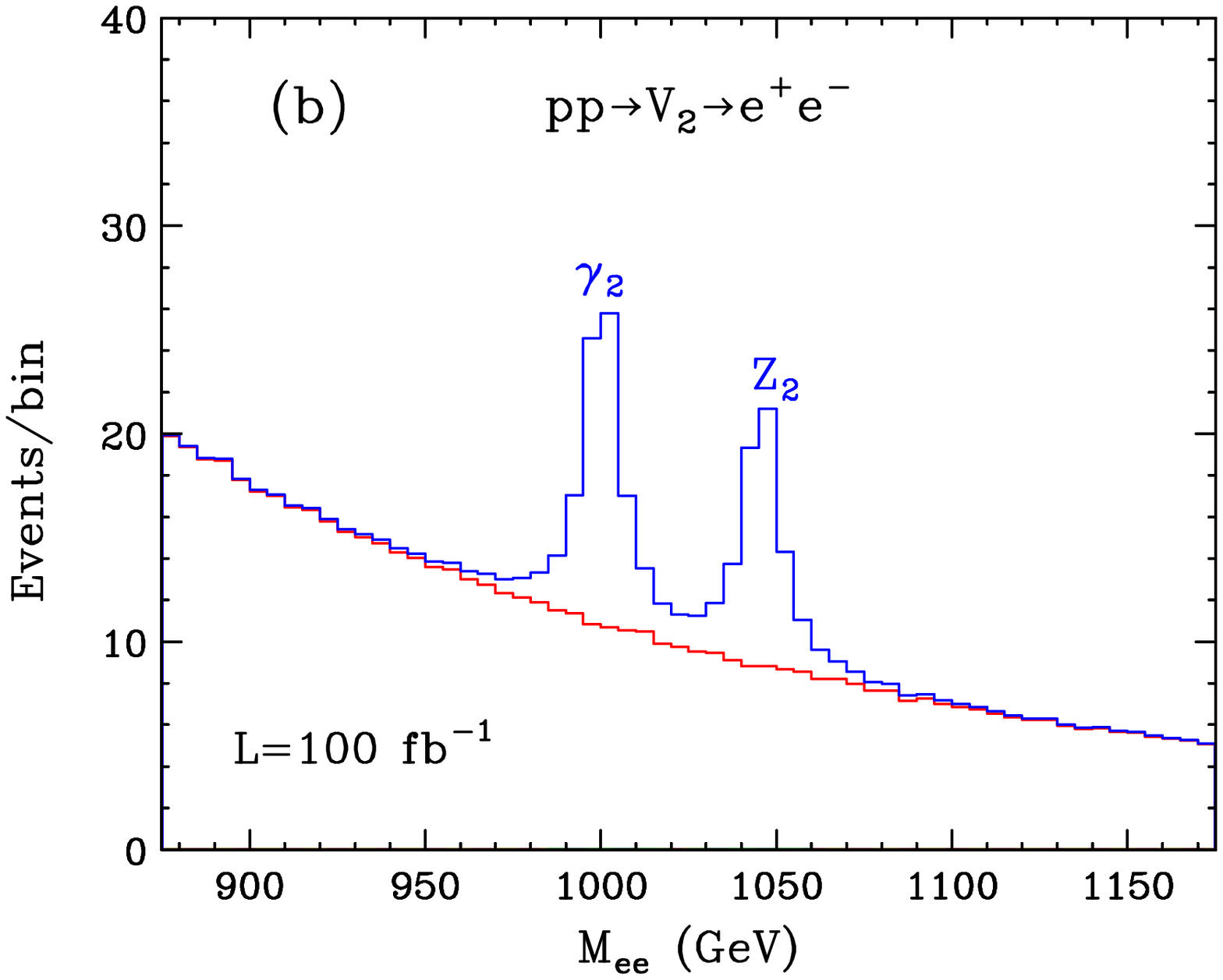}
\caption{The $\gamma_2-Z_2$ diresonance structure in UED with 
$R^{-1}=500$~GeV, for the (a) dimuon and (b) dielectron channels at
the LHC with ${\rm L}=100\ {\rm fb}^{-1}$. The SM background is shown
with the (red) continuous underlying histogram.}
\label{fig:mll}
\end{figure}

We now consider the inclusive production of $Z_2$ and $\gamma_2$ and
look for a dilepton resonance in both the $e^+e^-$ and $\mu^+\mu^-$
channels.  An important search parameter is the width of the
reconstructed resonance, which in turn determines the size of the
invariant mass window selected by the cuts.  Since the intrinsic width
of the $Z_2$ and $\gamma_2$ resonances is so small, the mass window is
entirely determined by the mass resolution in the dimuon and
dielectron channels.  For electrons, the resolution in CMS is
approximately constant, on the order of $\Delta m_{ee}/m_{ee}\approx
1\%$ in the region of interest~\cite{Darin}.  On the other hand, the
dimuon mass resolution is energy dependent, and in preliminary studies
based on a full simulation of the CMS detector has been parametrized
as~\cite{muonres} $\frac{\Delta m_{\mu\mu}}{m_{\mu\mu}}=
0.0215+0.0128\left(\frac{m_{\mu\mu}}{1\ {\rm TeV}}\right)$.  Therefore
in our analysis we impose the following cuts
\begin{enumerate}
\item Lower cuts on the lepton transverse momenta $p_T(\ell)>20$ GeV.
\item Central rapidity cut on the leptons $|\eta(\ell)|<2.4$.
\item Dilepton invariant mass cut for electrons
$m_{V_2}-2\Delta m_{ee}<m_{ee}<m_{V_2}+2\Delta m_{ee}$ and muons
$m_{V_2}-2\Delta m_{\mu\mu}<m_{\mu\mu}<m_{V_2}+2\Delta m_{\mu\mu}$.
\end{enumerate}
With these cuts the signal efficiency varies from $65\%$ at
$R^{-1}=250$ GeV to $91\%$ at $R^{-1}=1$ TeV.  The main SM background
to our signal is Drell-Yan, which we have calculated with the {\tt
PYTHIA} event generator~\cite{Sjostrand:2003wg}.

With the cuts listed above, we compute the discovery reach of the LHC
and the Tevatron for the $\gamma_2$ and $Z_2$ resonances.  Our results
are shown in Fig.~\ref{fig:reach}.  We plot the total integrated
luminosity ${\rm L}$ (fb$^{-1}$) required for a $5\sigma$ excess of
signal over background in the dielectron (red, dotted) and dimuon
(blue, dashed) channels, as a function of $R^{-1}$.  In each panel of
Fig.~\ref{fig:reach}, the upper set of lines labelled ``DY'' utilize
only the single $V_2$ production cross sections from
Fig.~\ref{fig:sigma_V2}.  The lower set of lines (labelled ``All
processes'') include in addition indirect $\gamma_2$ and $Z_2$
production from the decays of $n=2$ KK quarks to $\gamma_2$ and $Z_2$
(we ignore secondary $\gamma_2$ production from $Q_2\to Z_2\to \ell_2
\to \gamma_2$).  The shaded area below $R^{-1}=250$ GeV indicates the
region disfavored by precision electroweak
data~\cite{Appelquist:2002wb}.  Using the same cuts also for the case
of the Tevatron, we find the Tevatron reach in $\gamma_2\to e^+e^-$
shown in Fig.~\ref{fig:reach}a and labelled ``FNAL''.  For the
Tevatron we use electron energy resolution $\Delta
E/E=0.01\oplus0.16/\sqrt{E}$~\cite{Blair:1996kx}.  The Tevatron reach
in dimuons is worse due to the poorer resolution, while the reach for
$Z_2$ is also worse since $m_{Z_2}>m_{\gamma_2}$ for a fixed $R^{-1}$.

Fig.~\ref{fig:reach} reveals that there are good prospects for
discovering level 2 gauge boson resonances at the LHC.  Within one
year of running at low luminosity (${\rm L}=10\ {\rm fb}^{-1}$), the
LHC will have sufficient statistics in order to probe the region up to
$R^{-1}\sim 750$~GeV.  Notice that in MUED, the ``good dark matter''
region, where the LKP relic density accounts for all of the dark
matter component of the Universe, is at $R^{-1}\sim
500-600$~GeV~\cite{Servant:2002aq,Burnell:2005hm,Kong:2005hn}.  This
region is well within the discovery reach of the LHC for both $n=1$ KK
modes~\cite{Cheng:2002ab} and $n=2$ KK gauge bosons
(Fig.~\ref{fig:reach}).  If the LKP accounts for only {\em a fraction
of} the dark matter, the preferred range of $R^{-1}$ is even lower and
the discovery at the LHC is easier.

From Fig.~\ref{fig:reach} we also see that the ultimate reach of the
LHC for both $\gamma_2$ and $Z_2$, after several years of running at
high luminosity (${\rm L}\sim300\ {\rm fb}^{-1}$), extends up to just
beyond $R^{-1}=1$~TeV.  One should keep in mind that the actual KK
masses are at least twice as large: $m_{V_2}\sim m_2=2/R$, so that the
KK resonances can be discovered for masses up to $2$~TeV.

While the $n=2$ KK gauge bosons are a salient feature of the UED
scenario, any such resonance by itself is not a sufficient
discriminator, since it resembles an ordinary $Z'$ gauge boson.  If
UED is discovered, one could then still make the argument that it is
in fact some sort of non-minimal SUSY model with additional gauge
structure containing neutral gauge bosons.  Important corroborating
evidence in favor of UED would be the simultaneous discovery of
several, rather degenerate, KK gauge boson resonances.  While SUSY
also can accommodate multiple $Z'$ gauge bosons, there would be no
good motivation behind their mass degeneracy.  A crucial question
therefore arises: can we separately discover the $n=2$ KK gauge bosons
as individual resonances?  For this purpose, one would need to see a
double peak structure in the invariant mass distributions. Clearly,
this is rather challenging in the dijet channel, due to relatively
poor jet energy resolution.  We shall therefore consider only the
dilepton channels, and investigate how well we can separate $\gamma_2$
from $Z_2$.

Our results are shown in Fig.~\ref{fig:mll}, where we show the
invariant mass distribution in UED with $R^{-1}=500$ GeV, for the (a)
dimuon and (b) dielectron channels at the LHC with ${\rm L}=100\ {\rm
fb}^{-1}$.  We see that the diresonance structure is easier to detect
in the dielectron channel, due to the better mass resolution.  In
dimuons, with ${\rm L}=100\ {\rm fb}^{-1}$ the structure also begins
to emerge.  We should note that initially the two resonances will not
be separately distinguishable, and each will in principle contribute
to the discovery of a bump, although with a larger mass window.  In
our reach plots in Fig.~\ref{fig:reach} we have conservatively chosen
not to combine the two signals from $Z_2$ and $\gamma_2$, and instead
show the reach for each one separately.


\subsubsection{Conclusions}

We studied the discovery reach for level 2 KK modes in UED at hadron
colliders.  We showed that the $n=2$ KK gauge bosons offer the best
prospects for detection, in particular the $\gamma_2$ and $Z_2$
resonances can be {\em separately} discovered at the LHC.  However,
this is not a proof of UED.  These resonances could still be
interpreted as $Z'$ gauge bosons, but their close degeneracy is a
smoking gun for UED.  Furthermore, although we did not show any
results to this effect in this paper, it is clear that the $W_2^\pm$
KK mode can also be looked for and discovered in its decay to SM
leptons.  One can then measure $m_{W_2}$ and show that it is very
close to $m_{Z_2}$ and $m_{\gamma_2}$, which would further strengthen
the case for UED.  The spin discrimination is not so straightforward,
and requires further study.  The asymmetry method of Barr is discussed
in Sec.~\ref{sec:spin}.

While in this paper we concentrated on the Minimal UED model, it
should be kept in mind that there are many interesting possibilities
for extending the analysis to a more general setup.  For example,
non-vanishing boundary terms at the scale $\Lambda$ can distort the
MUED spectrum beyond recognition.  The UED collider phenomenology is
also very different in the case of a ``fat''
brane~\cite{Macesanu:2002ew,Macesanu:2004nb}, charged
LKPs~\cite{Byrne:2003sa}, KK graviton
superwimps~\cite{Feng:2003xh,Feng:2005gj} or resonances in two
universal extra dimensions~\cite{Burdman:2006gy}. Notice that Little
Higgs models with
$T$-parity~\cite{Cheng:2003ju,Cheng:2004yc,Hubisz:2004ft} are very
similar to UED, and can also be confused with SUSY.


\subsubsection*{Acknowledegments}

We thank H.-C.~Cheng and B.~Dobrescu for stimulating discussions.  AD
is supported by the US Department of Energy and the Michigan Center
for Theoretical Physics.  The work of KK and KM is supported in part
by a US Department of Energy Outstanding Junior Investigator award
under grant DE-FG02-97ER41209.

\clearpage\setcounter{equation}{0}\setcounter{figure}{0}\setcounter{table}{0}
\subsection{Universal extra dimensions with KK number violation}
\label{sec:UED}

{\em Cosmin Macesanu, Dept. of Physics, Syracuse University}\\


We discuss in this section the phenomenological signals associated
with a Universal Extra Dimensions (UED) model~\cite{Appelquist:2000nn}
with KK number violation.  The breaking of KK number may arise in
different ways, but here we take it to be a consequence of
gravitational interactions.  Such interactions are natural in a model
in which matter and gravity both propagate in the bulk.  If one adopts
the framework advocated in the Arkani-Hamed, Dimopulos and Dvali (ADD)
model~\cite{Arkani-Hamed:1998rs,Arkani-Hamed:1998nn,Antoniadis:1998ig},
the size of the compactified extra dimensions where gravity propagates
is of order inverse eV.  Naturally, matter cannot propagate all the
way into these extra dimensions (or one would observe eV spaced KK
excitations of the SM particles).  However, instead of being stuck on
the 4D brane, the matter fields might be able to propagate a limited
length of order $1/M$ into the bulk.  One can moreover conjecture that
the matter fields are confined close to the 4D brane by some
interactions generated by physics at the string scale, $M_D$.  Then
one may expect that the two scales $M$ and $M_D$ are related, and
roughly of the same order of magnitude.

The phenomenology of such a model differs significantly from that of
the standard UED models~\cite{Cheng:2002iz,Cheng:2002ab}.  In the
later case, there typically exists a lightest KK particle (LKP), which
is stable due to KK number conservation.  KK excitations of quarks and
gluons, which dominate KK production at a hadron colider, will decay
to the LKP, radiating semisoft quarks and leptons in the process.
Being neutral and weakly interacting, the LKP would not leave energy
in the detector, thus the experimental signal would consist of
relatively soft jets and leptons plus missing energy (note also that
while the absolute missing energy would be quite large, the observable
transverse missing energy would be rather small).  In the case of the
UED model with KK number violation, the LKP will also decay by
radiating a graviton and the SM partner of the LKP in our case.
Although the coupling of an individual graviton to matter is extremely
weak (of order $1/M_{Pl}^2$) the large number of gravitons available
to contribute to the decay will give a sizeable total gravitational
decay width $\Gamma_h\sim M^{N+3}/M_D^{N+2}$ (with $N$ the number of
extra dimensions), which for values of $M_D$ not much larger than $M$
is of the same order of magnitude as electroweak or strong decay
widths.  Moreover, since the masses of gravitons contributing to this
decay may be significantly smaller that the mass of the KK matter
particle, the momentum of the visible SM particle can be quite large.

One can in fact obtain different signals, depending of the parameters
of the model.  If the gravitational decay widths of the quark/gluon
excitations are larger than the decays widths to the LKP, the KK
particles produced at a hadron collider will decay gravitationally,
leaving behind two high $p_T$ jets~\cite{Macesanu:2002db}.  If the
opposite is true, then the KK quarks or gluons will first decay to the
LKP, radiating soft jets and leptons.  The LKP will in turn decay
gravitationally, leaving behind a high $p_T$ photon (assuming that the
LKP is the photon excitation) and missing energy, taken away by the
graviton.  The phenomenology of this scenario was discussed
in~\cite{Macesanu:2002ew}.

As a consequence of having high energy jets (or photons) in the final
state, it will be much easier to discover extra dimensions in a
scenario with KK number violation.  For example, for pair production
processes the Tevatron Run~II will be able to probe values of the
inverse compactification scale $M=1/R$ up to 400~GeV (in the case of
the dijet signal) or 500~GeV (for the diphoton signal).  At the LHC,
one can probe values of $1/R$ close to 3~TeV, which is almost double
the discovery reach compared to the case when the LKP is stable.
Moreover, one may have a better chance of differentiating the UED
model from competing models like SUSY by analyzing kinematical
observables which are not accessible in the stable LKP
scenario~\cite{Smillie:2005ar,Datta:2005zs}.

Another particular behavior of models where the gravitational
interaction violates KK number is the possibility of producing a
single first level KK excitation of matter.  Since these processes are
gravity-mediated, the production cross section will depend on the
strength of the gravitational interaction (the parameter $M_D$).  Thus
one can expect that the effective interaction strength is smaller than
in the case of KK pair production (where the interaction strength is
given by the strong coupling constant).  However, this might be
mitigated by the fact that one needs produce a single massive particle
in the final state, rather that two.  It is then possible that in the
case when the fundamental gravity scale is not much larger than the
collider energy, it would be easier to observe production of a single
KK matter excitation rather than the usual pair production process.

There are two types of processes which lead to a single KK matter
excitation in the final state: one in which gravitons play the role of
virtual particles mediating the process, and one in which gravitons
appear as final state real particles.  We start the discussion with
the first case~\cite{Macesanu:2004nb}.  The phenomenological signal
for such a process will be either two jets plus missing transverse
energy (associated with the graviton appearing in the decay of the KK
quark/gluon), or a single jet plus a photon and missing energy (if the
KK quark or gluon decays first to the LKP, and this particle decays
gravitationaly).  It turns out that for the most of the parameter
space where the cross section is observable, the gravitational decay
of KK quarks/gluons will take place, and the signal will be two jets
plus missing energy.  This can be understood by noting that an
observable cross section requires a relatively strong gravitational
interaction, which ensures in turn that the gravitational decay width
of the produced KK excitation will be larger that the decay width to
the LKP.

\begin{figure}[t!] 
\centerline{
   \includegraphics[height=3.5in]{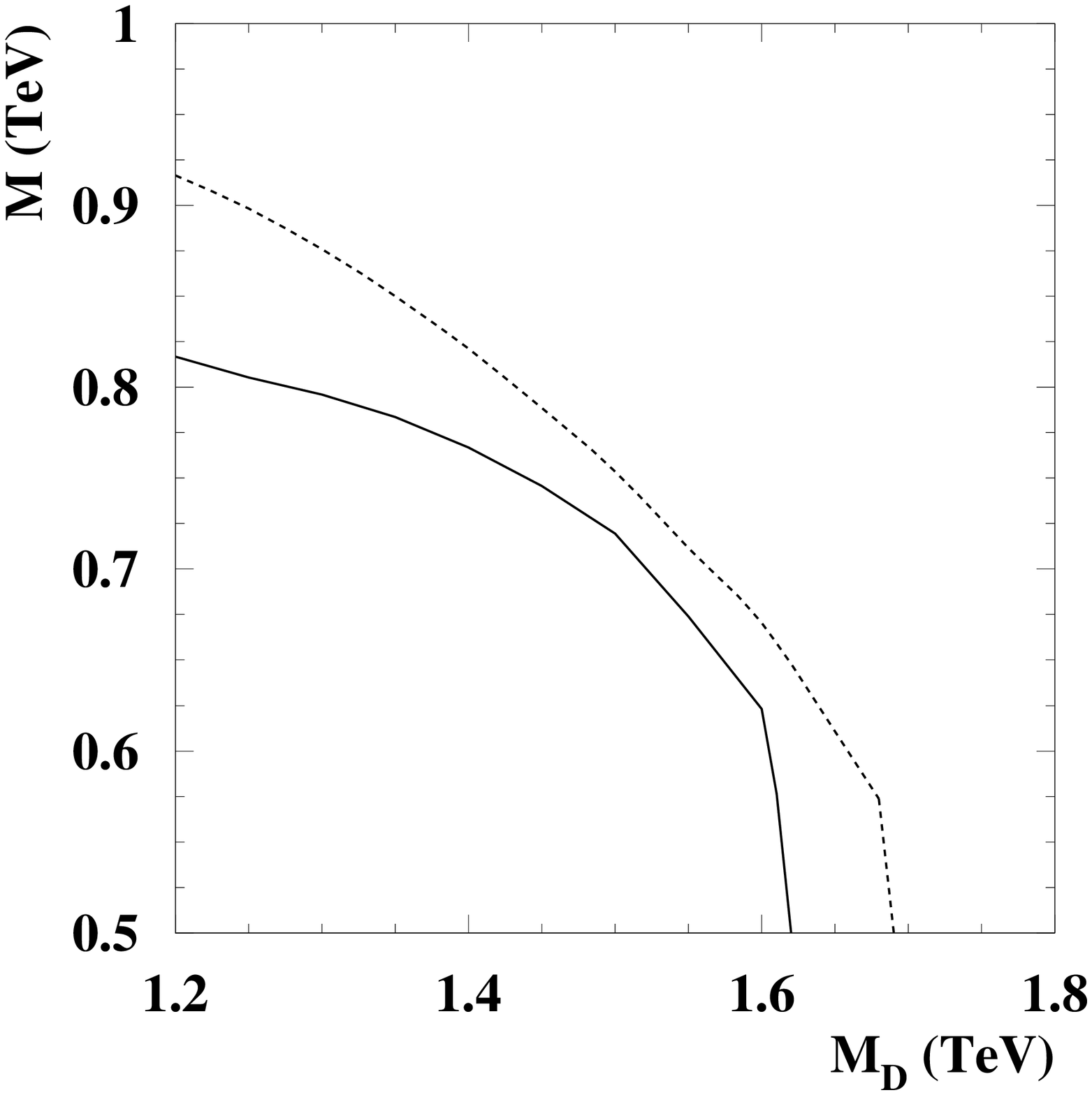}
   \includegraphics[height=3.5in]{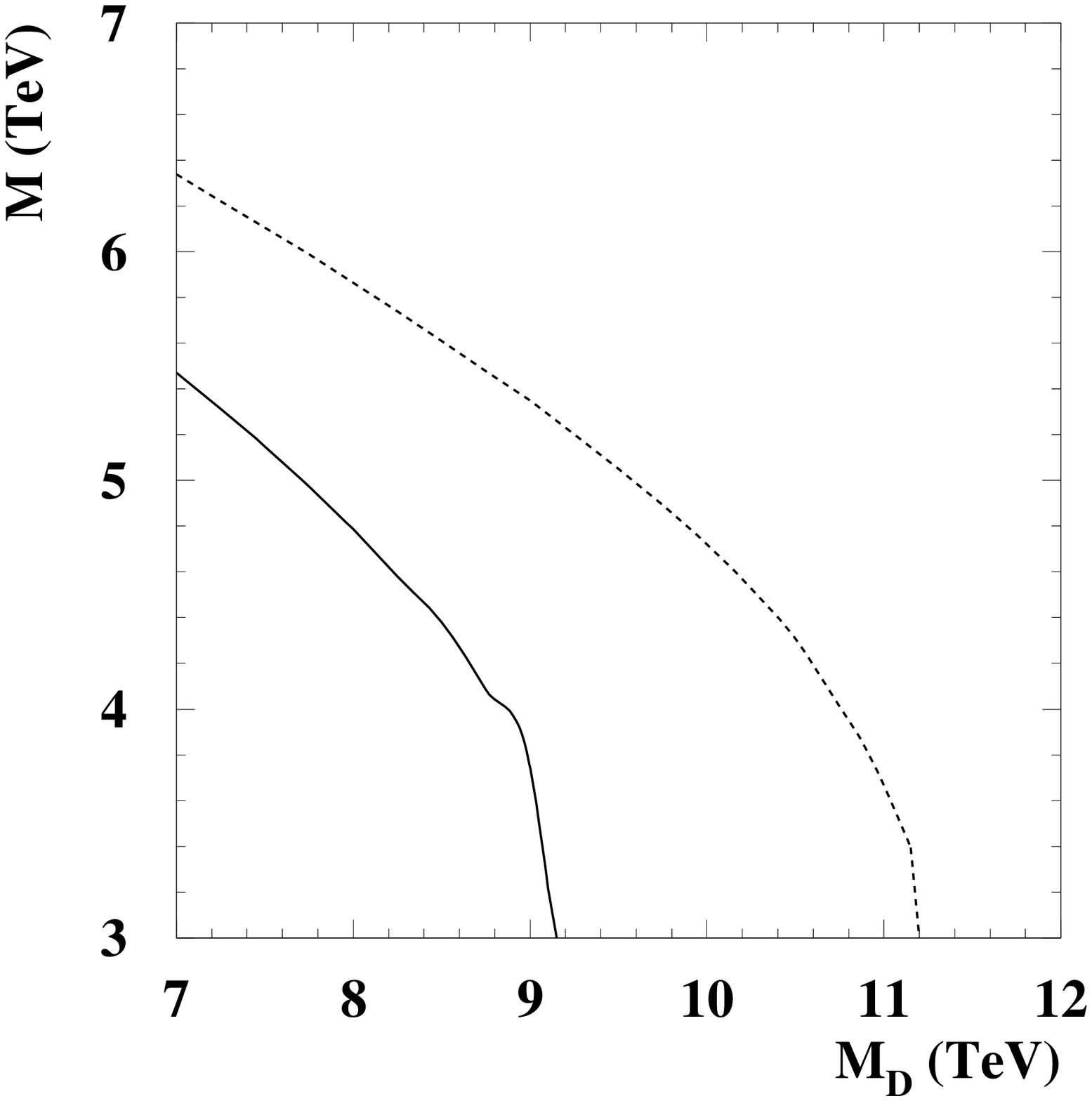}
   }
\caption{Tevatron Run~II (left) and LHC (right) $5\sigma$ discovery
reach contours for single KK and SM quark or gluon production, for
$N=2$ (dashed lines) and $N=6$ (solid lines) extra dimensions.}
\label{coll_reach}
\end{figure}

Some illustrative results for this type of process are presented in
Fig.~\ref{coll_reach}.  The dashed and solid lines are contours in the
($M_D,M$) plane for a $5\sigma$ discovery at the Tevatron Run~II with
2~fb$^{-1}$ integrated luminosity (left panel), and at LHC with
100~fb$^{-1}$ (right panel).  The dashed lines correspond to $N=2$
extra dimensions, while the solid lines correspond to $N=6$.  The
contour lines correspond to an observable cross section of 25~fb at
the Tevatron, with cuts of $p_T>150$~GeV and $\not{E_T}>300$~GeV on
the transverse momentum of the observable jets and on the missing
energy.  At the LHC, the cross-section is 1~fb, and the corresponding
cuts are $p_T>800$~GeV and $\not{E_T}>1.6$~TeV.  The large cuts
imposed on the missing energy and transverse momentum eliminate most
of the Standard Model background, which in this kinematic range is due
mostly to $Z$ plus two jet production.

We observe that for small values of $M_D$, one can indeed probe a
large range of values for the compactification scale associated with
matter, $1/M$, reaching to almost double the maximum accessible in
pair production processes.  However, this result is strongly dependent
on the fundamental gravity scale $M_D$: the production cross-section
behaves as $\sim 1/M_D^{10}$, and for larger values of this parameter,
this process would become unobservable, as one can note from the
abrupt drop of the contour lines in Fig.~\ref{coll_reach} once $M_D$
has increased past a certain value.  Note also that for these values
of $M_D$, one would be able to observe KK gravitons through processes
associated with typical ADD phenomenology.  That is, SM particle
production mediated by gravity, or processes with a graviton and a SM
particle in the final state.
 
Additionally, one could observe the direct production of a photon and
a photon KK excitation, or a lepton and KK lepton, through s-channel
processes mediated by gravitons.  The production cross sections for
these processes will be of the same order of magnitude as for
processes involving quarks or gluons and their KK excitations (since
the strength of the effective interaction is the same in both cases).
However, since the hadron collider background will be smaller for
final state photons or leptons, one can expect that the reach in this
channel will be somewhat larger than for processes involving quarks
and gluons.

The second type of processes mediated by gravity are processes with a
graviton and a KK matter excitation in the final
state~\cite{Macesanu:2005wj}.  The KK excitation can decay either
directly to another graviton and a gluon or quark, or first to the
LKP, which in turn decays to a photon and a graviton.  The signal will
then be a single jet or photon, with missing energy due to the two
gravitons in the final state.  Then one has to take into account
contributions to this signal coming from the standard ADD-type
processes with a KK graviton and a SM particle in the final state.

\begin{figure}[b!] 
\centerline{
   \includegraphics[height=3in]{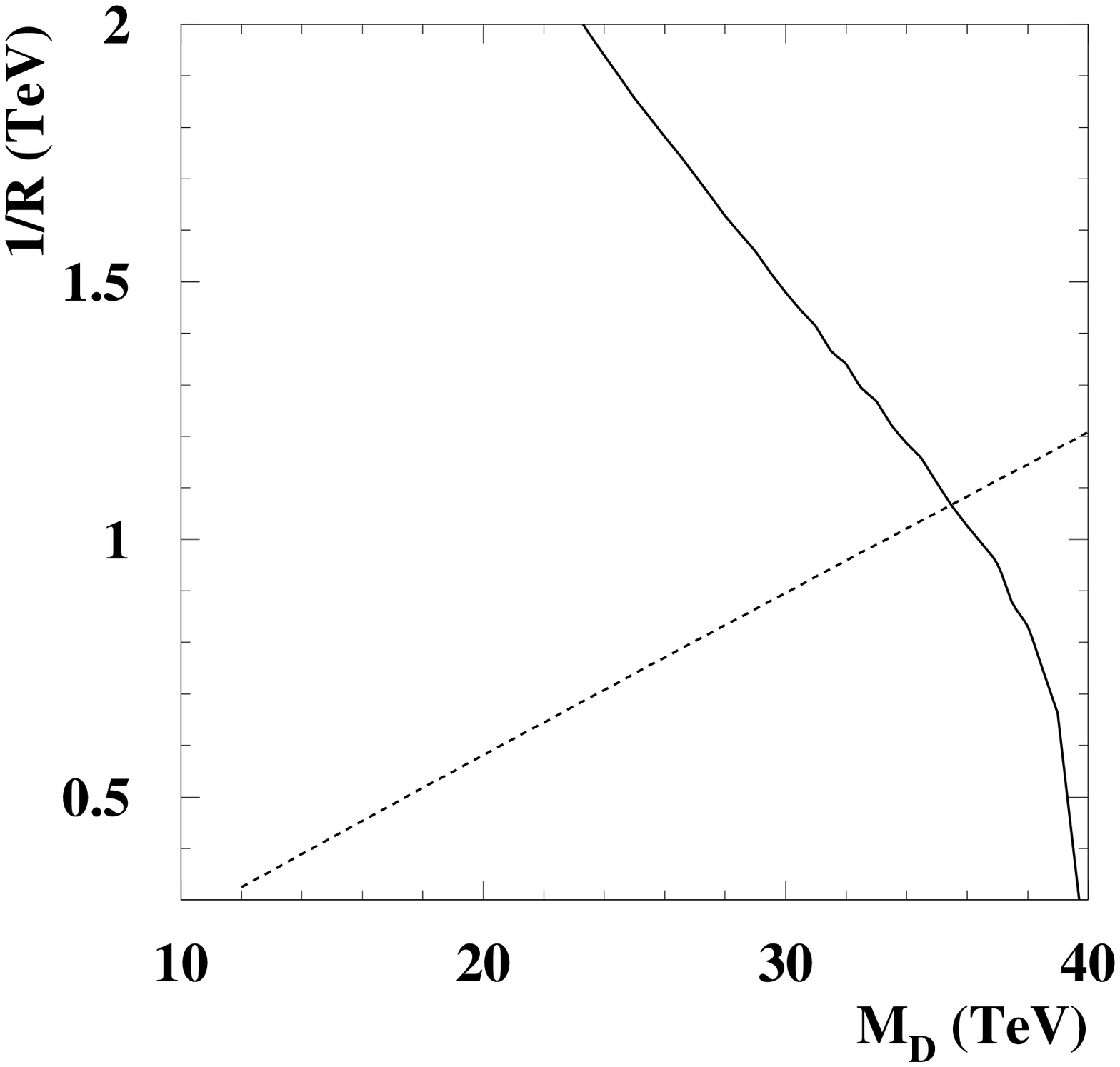}
   \includegraphics[height=3in]{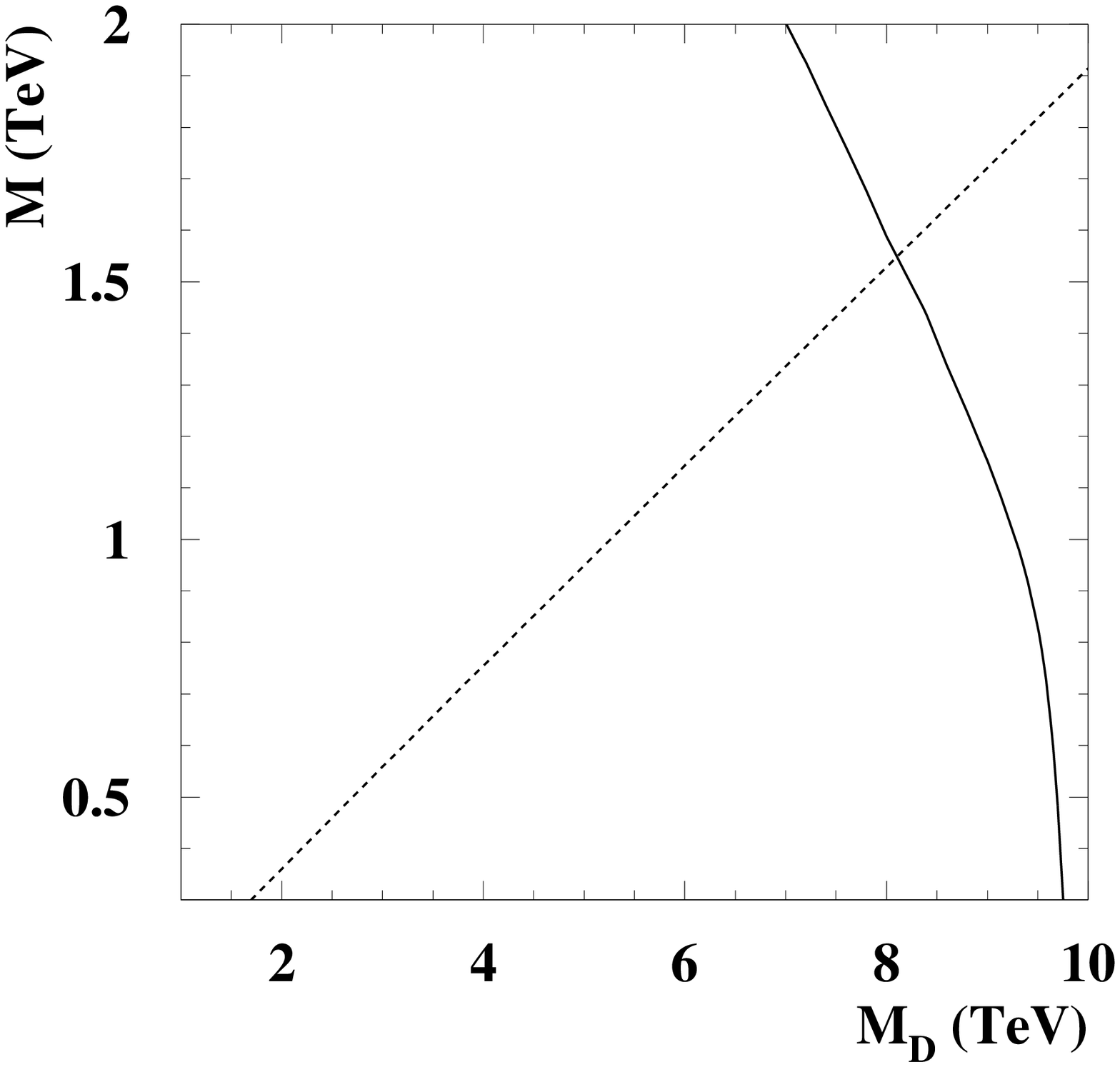}
}
\caption{Solid lines: the $5\sigma$ discovery reach at the LHC in the 
photon + $\not{E_T}$ channel for $N=2$ (left panel) and $N=4$ (right
panel).  For values of $M_D,1/R$ below the dashed lines, the KK quarks
and gluons decay first to the LKP.}
\label{reach_n24}
\end{figure}

In the first case, with a jet in the final state, the cross section
for production of a KK quark and gluon is smaller than the cross
section for associated production of a SM quark or gluon and a KK
graviton.  This is due partly the fact that one has an extra massive
particle in the final state, and partly to a suppression effect of the
form factor associated with the graviton--KK matter interaction
vertex~\cite{Macesanu:2003jx}.  As a consequence, one cannot directly
observe the production of a single KK excitation plus a graviton in
the jet plus missing energy channel.  However, if we are in a region
of parameter space where the KK quark or gluon decays first to the
LKP, one may observe such production in the photon plus missing energy
channel.  For such a signal, the cross section for direct production
of a SM photon and a graviton will be suppressed by the electroweak
coupling constant as well as, in the LHC case, by the small $q\bar{q}$
content in the initial state.

\begin{figure}[t!] 
\centerline{
   \includegraphics[height=3in]{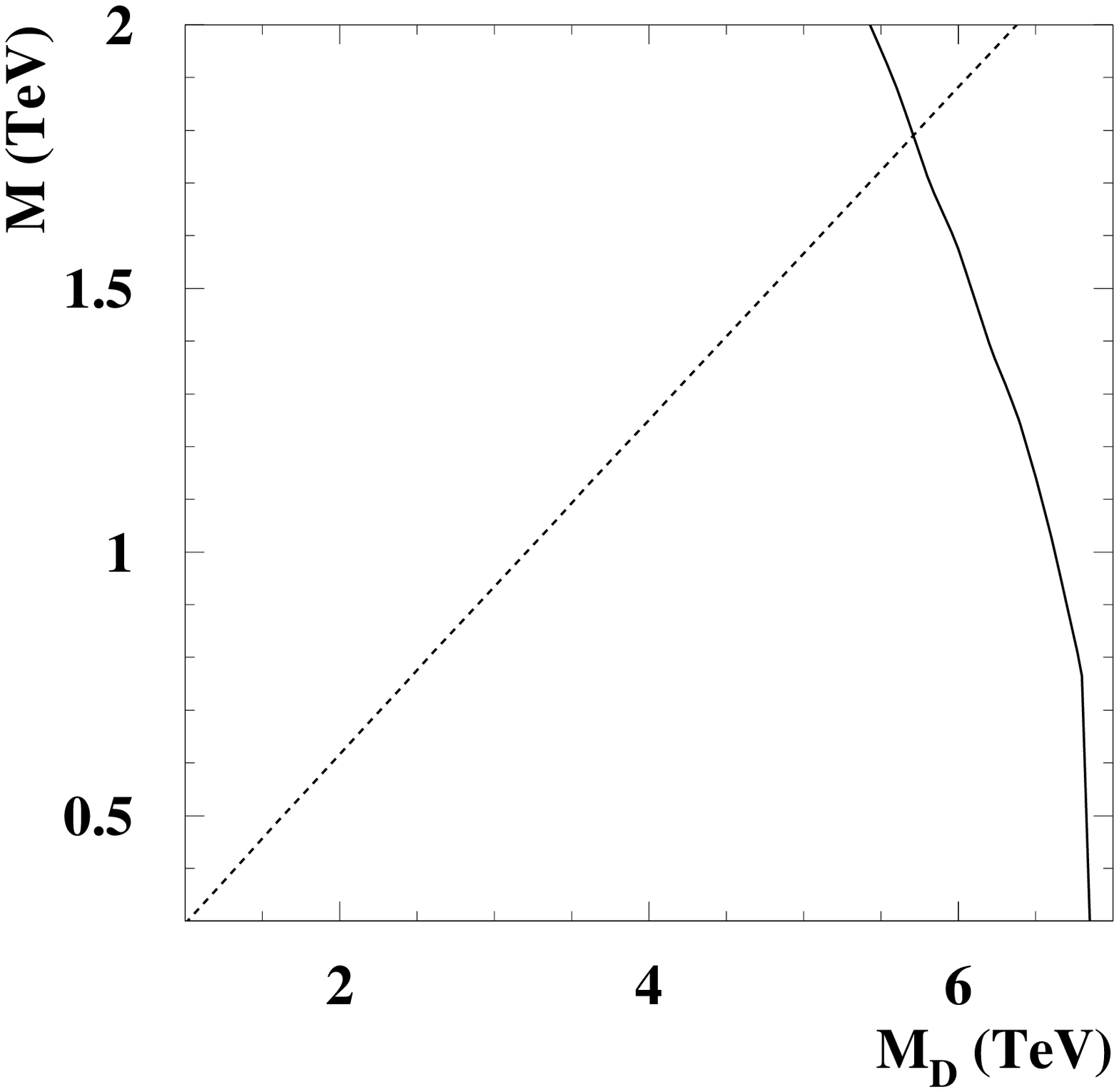}
   \includegraphics[height=3in]{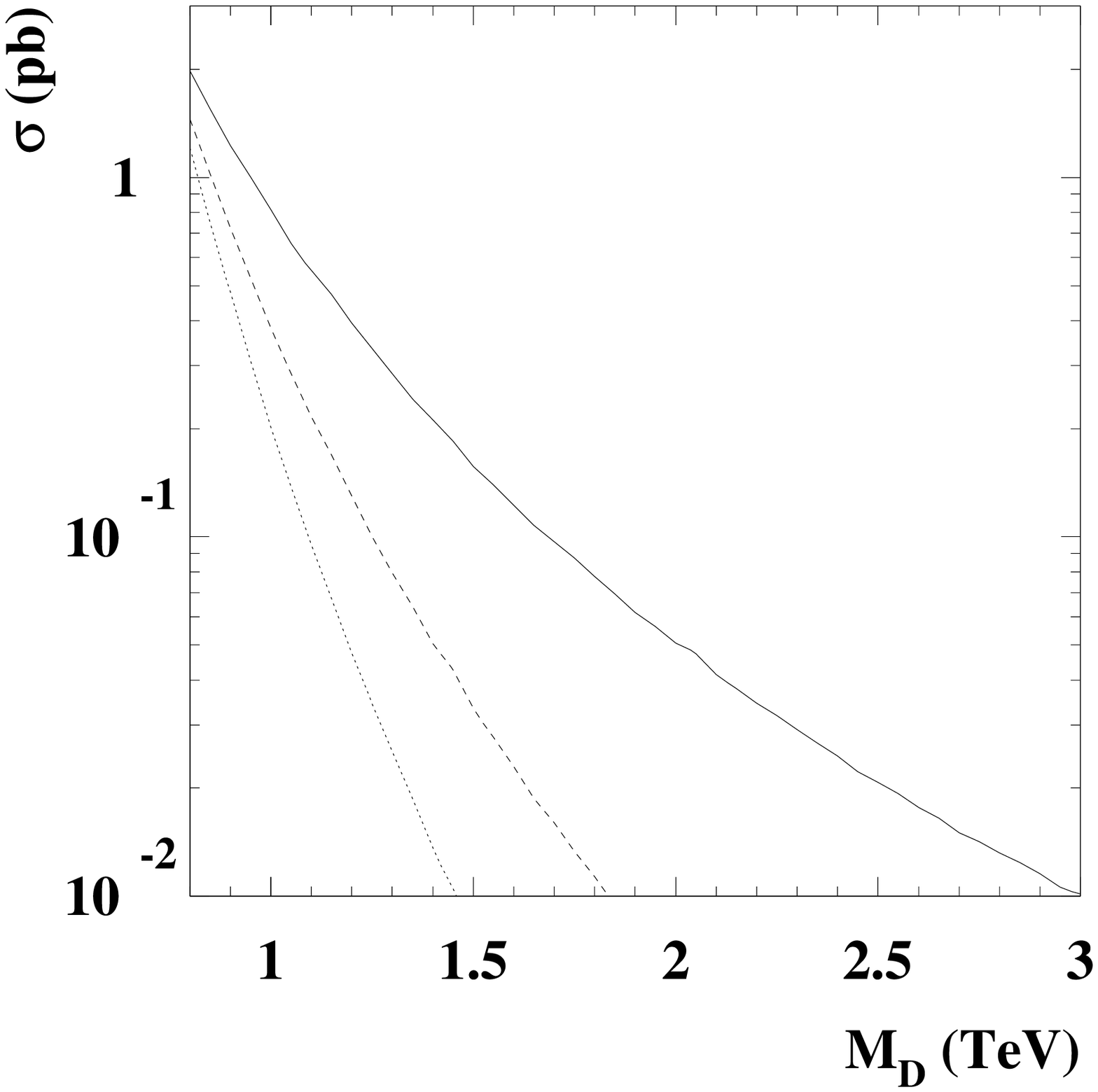}
}
\caption {Left panel: the $5 \sigma$ discovery reach at the LHC in the 
photon + $\not{E_T}$ channel for $N=6$.  Right panel: the SM photon +
$\not{E_T}$ cross-section at the Tevatron Run II, with $p_T>100$~GeV.
Solid, dashed and dotted lines correspond to $n=2$, 4 and 6 extra
dimensions respectively.}
\label{reach_n6}
\end{figure}

We show in the left panels of Figs.~\ref{reach_n24},\ref{reach_n6}
with solid curves the discovery reach for KK quark/gluon plus graviton
at LHC in the photon + missing energy channel.  The dashed curves show
the upper limits in parameter space where the decay of the KK
quark/gluon takes place through the LKP first, such that above those
lines, the signal would be jet + $\not{E_T}$.  We note that for small
values of the matter compactification scale $M$, one can probe quite
large values for the fundamental gravitational scale $M_D$, especially
for the case of $N=2$.  This is due to an enhancement of the
production cross section for very light
gravitons~\cite{Macesanu:2005wj}, which is the dominant contribution
for $N=2$.  Also, the cross section has a weaker dependence on $M_D$,
$\sigma\sim 1/M_D^{N+2}$, compared to the $1/M_D^{10}$ dependence
valid for the processes with KK production mediated by virtual
gravitons.

The Tevatron case is somewhat different.  Here, the cross section for
direct production of a SM photon and graviton plays a more important
role than in the case of the LHC, due in part to the initial state
containing a large $q\bar{q}$ fraction, which is required for the
process with a photon in the final state.  Conversely, the cross
section for production of a quark/gluon excitation which decays to the
LKP is highly supressed by large values of $M_D$.  Then the cross
section production for a single photon plus a graviton depends mostly
on $M_D$, and is almost the same as the one evaluated for the simple
ADD scenario (with no matter in extra dimensions).  This cross section
is shown in the right panel of Fig.~\ref{reach_n6} with a cut of
$p_T>100$~GeV on the photon transverse momentum (which corresponds to
a SM background of $\sim 80$ fb).  From such a process one can then
set the same order of magnitude limits on the fundamental gravity
scale as the limits obtained from the jet + $\not{E_T}$ signal in the
ADD scenario~\cite{Giudice:1998ck,Mirabelli:1998rt}.


\clearpage\setcounter{equation}{0}\setcounter{figure}{0}\setcounter{table}{0}
\subsection{The Need for Beyond-SUSY Tools for the LHC}
\label{sec:ED}

{\em Albert De Roeck, CERN, Geneva, Switzerland} \\


\subsubsection{Introduction}

One of the main missions of the LHC will be to test and discover new
physics beyond the Standard Model (BSM) in the data.  There is an
impressive variety of proposals for BSM physics at the LHC, the most
popular scenarios presently being supersymmetry and Extra Dimensions,
but others include Technicolor, Little Higgs models, new gauge bosons,
compositness, etc.

The experiments explore their sensitivity to these models mostly by
means detector simulations using event generators.  This is necessary
since the main challenges at the LHC to discover new physics will be
to find a signal on top of an often large background, and to have an
efficient trigger path in place for the signature of this new physics
channel.  Hence, to prepare for LHC data taking, Monte Carlo event
generators for these new processes are extremely useful tools.  Also,
once data is collected, event generators and programs to calculate
cross sections will be indispensable tools for the analysis and
interpretation of the data, whether a signal is observed or not.

The rapid increase of new models in the last few years has not been
followed by an organized effort on specialized Monte Carlo generators.
Often the experiments have made private implementations of (part of)
the physics of such new scenarios or models.  This contribution is a
plea to have a more systematic approach to event generators for new
physics for the LHC.  It results from collecting information on the
needs of the CMS and ATLAS experiments during their studies to 
prepare  for
physics.  A number of suggestions are made on possible ways
forward, taking the physics of Extra Dimensions (EDs) as an explicit
example.

Since the first presentation of these ideas at the TeV4LHC workshop,
this has been further developed at other workshops such as Les Houches
05, MC4BSM at FNAL in March 06, and the Tools for SUSY and BSM
workshop at LAPP, Annecy in June 06.  In particular after the Les
Houches workshop, some action was taken as reported below.


\subsubsection{Extra Dimensions as an example}

Issues for the ED analyses for experimentalists at the LHC are as
follows.
\begin{itemize}
\item
Include new processes in Monte Carlo generators, usable for LHC
(Tevatron) analyses.  Hence they become a "standard" which can be
used for comparison between experiments.
\item 
Include complete information into Monte Carlo generators, such as spin
correlations.  This will allow e.g. the study of measurability of spin
effects (e.g. SUSY versus UED, Z' versus KK states and so on).
\item
Cross checks between different codes/Monte Carlo results are important
(problems have been found in the past).
\item
Agree on parameter space available for EDs at LHC (\& ILC, CLIC,
Tevatron).  Under which conditions do e.g. astrophysical and other
limits apply.  How can certain scenarios (e.g MSLED) escape these
limits?  Presently the information is too scattered.
\item
Agree on a number of benchmark scenarios, like we have for SUSY and
successfully used e.g. in LHC/ILC studies.
\item
Time to think about K-factors?  These K-factors can be large and
affect the search reach.
\item
The accuracy of the SM process background understanding is important
(e.g. Drell-Yan, Z+jet, effects from PDFs..).  This is particularly
important for tails of distributions where the statistics and
experimental checks will be limited.
\item
Use the same formalism in the models, e.g. for the definition of the
effective Planck Scale.
\end{itemize}

So far the following generators are in use in the experiments
for ED studies

\begin{itemize}
\item
RS gravitons are included in the standard workhorses HERWIG and PYTHIA.
\item
ADD scenarios: several private codes for both the graviton radiation
and graviton exchange processes, for PYTHIA or HERWIG, are circulating
in the experiments.  Recently the situation was improved by SHERPA,
which contains complete ADD FeynmanRules and is now used in ATLAS/CMS.
\item
UEDs existed mainly in a private code for COMHEP (Matchev et al.)  Now
PYTHIA\_UED~\cite{Skands:2005vi} is available including UED without KK 
conservation.
\item
Plenty of other specific channels in private codes (e.g G. Azuelos 
et al.).
\item
Thanks to the Les Houches accord of 2001, an agreed exchange format
exists such that one can think of a tool kit for ED processes.
\end{itemize}

Typically the specialized generators deal with the hard process only
and workhorse generators such as PYTHIA and HERWIG, and now also
Sherpa, deal with the hadronization and fragmentation aspects of event
generation.  The Les Houches accord~\cite{Boos:2001cv} offers a common
interface for communication between the generators.  Hence one can
construct a toolbox for generators for ED or more general BSM
processes.  Note that the SUSY generators are already organized in a
way such that these can be used interchangeably to a large extent,
albeit not quite in a toolbox format.  Particularly useful has been
the SPA (SUSY Parameter Analysis)~\cite{Aguilar-Saavedra:2005pw}
project.

The future of HEP software architecture will be based on frameworks.
For the generators there is a proposal for such a framework called
ThePEG~\cite{Lonnblad:2006pt}.  Such a framework could well become in
future the host of a possible toolkit for generators for BSM
processes.


\subsubsection{Wishlist for ED process generators}

The variety of ED process is very large, with largely different
signatures for different processes.  A compiled wishlist (in 2005) of
processes to become available or to be implemented is as follows:

\begin{itemize}
\item Universal Extra Dimensions with KK number conservation.
\item Universal Extra Dimensions without KK number conservation.
\item Bulk scalars with Higgs interference.
\item Radions and interference with the Higgs.
\item RS generator with SM fields in the bulk .
\item Implementation of different running couplings.
\item More flexible/complete generation of KK resonances in TeV$^{-1}$ and 
RS models, which include many resonances, effects of brane kinematic
terms.
\item Branons.
\item More sophisticated black hole generators? (remnant treatement, 
radiation phases, spin)
\item String ball effects (black hole-like but different radiation/lower 
mass).
\item Trans-Planckian effects, especially high-$E_T$ dijet production.
\item SUSY + ED scenarios.
\item Thick branes, brain tension, rigid and soft branes.
\item Even more recent scenarios (such as intersecting branes, Higgsless 
EWSB).
 
\end{itemize}

At this stage, for LHC studies, MCs of new scenarios are important if
these imply new signatures and require new sorts of experimental
checks or the need for new triggers.


\subsubsection{K-factors, PDF and scale uncertainties}

\begin{figure}[htbp]
\centering 
\includegraphics[scale=0.7]{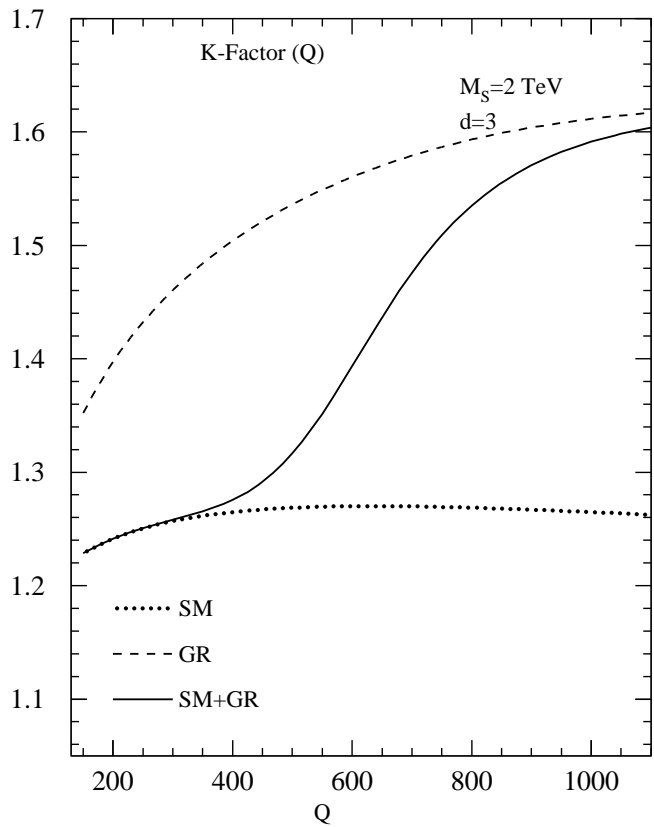}
\caption{
The $K$-factor for the cross section $d\sigma/dQ$ at $M_s=2~{\rm TeV}$
and $d=3$.  The plot is made for the LHC ($\sqrt{S}=14~{\rm TeV}$).
Standard Model (dotted line), gravity (long dashed line), Standard
Model plus gravity (solid line).}
\label{tev_mc:fig1}
\end{figure}

NLO estimates exist for Higgs production and SUSY particle production
 for a
number of processes and variables.  Recently K-factors have been
determined also for a few ED processes.  For example, it was shown
that for ADD/RS dilepton production the K-factor can be large.  At the
LHC the factor is typical of order 1.6, shown in
Fig.~\ref{tev_mc:fig1}, larger than for the Drell-Yan background
process\cite{Mathews:2004xp}.  K-factors can make a difference in both
discovery and extracted limits and are already useful to have now for
Tevatron analyses.  Only a few processes have been calculated so far.
E.g., even though we have the K-factor for a RS graviton decaying into
$G\to$ dileptons, it can't be transported to the $G\to\gamma\gamma$ 
process.

Another important effect is the effect of the parton density functions
(PDFs) and scale uncertainties. A recent analysis of PDF and scale
uncertainties for EDs was reported in Ref.~\cite{Kumar:2006id}.
Earlier studies\cite{Ferrag:2004ca} showed that these uncertainties
may reduce the search reach by up to a factor two, given the present
PDF uncertainties.  The HERA/LHC
workshop~\cite{Alekhin:2005dy,Alekhin:2005dx} offers a forum for the
study of PDF uncertainties and strategies to reduce them.

Clearly understanding the SM background processes is imperative for
BSM searches.  Other contributions in this and the HERA/LHC workshop
deal with these questions in detail.


\subsubsection{Further items}

Finally, for the physics TDR studies of CMS and ATLAS, it would have
been useful to have agreed-upon benchmark points.  Such points have
been defined for SUSY studies and have been instrumental in common
studies of ATLAS/CMS and the ILC study groups, as is demonstrated in
the huge report of Ref.~\cite{Weiglein:2004hn}.  While it is getting
late for benchmarks for the LHC now since data taking is approaching
in 2008, one aspect of the benchmark selection studies remains useful,
namely to get a complete review of the constraints of existing data
(HEP or other) on the available parameter space.

Finally, a unification of different formalisms would be useful,
e.g. for virtual graviton effects in ADD models for which there are at
least three distinct formalisms.


\subsubsection{Recent activities}

As mentioned at the start: since the Les Houches 05 workshop some
coherent activity has started.  As a first step, the existing Monte
Carlo generators and tools are cataloged and collected in one
repository.  In a next step one can try to unify them more and perhaps
create a toolkit.

A BSM tool repository, which now contains a collection of 25 programs,
is described in Ref.~\cite{Skands:2005vi} and available under: {\tt
http://www.ippp.dur.ac.uk/montecarlo/BSM/}

Other useful information is the summary paper of the recent MC4BSM
workshop~\cite{Hubisz:2006sn}, and a discussion forum for LHC tools is
available under {\tt http://www-theory.lbl.gov/tools/}.


\subsubsection{Summary}

The wishlist for BSM tools for the LHC is as follows:
\begin{itemize}
\item an ED or BSM Monte Carlo process tool box
\begin{itemize}
\item include the (still many) missing processes into generators. 
\item Keep track of details in the MC, such as spin correlations. These 
are likely to become very important when LHC will discover a new object.
\item Les Houches accords and frameworks such as ThePEG should facilitate
this task
\end{itemize}
\item SM background processes: These need to be known with high
precision.
\item Higher order QCD (EW) corrections to the processes
\item ED constraints from existing data
\item SUSY has the SPA project; do we need something similar for the EDs
or other BSM processes?
\end{itemize}
On some of these items on the list activity is already ongoing, but some 
more central coordination and intiative would be extremely useful.

\clearpage\setcounter{equation}{0}\setcounter{figure}{0}\setcounter{table}{0}
\subsection{Collider Phenomenology of a Little Higgs Model with T-Parity}
\label{sec:LHT}
\newcommand{\arline}{\nonumber \\}

{\em Jay Hubisz,
Fermi National Accelerator Laboratory, Batavia, IL, USA}\\

{\em Little Higgs models are an interesting approach to solving the
LEP paradox of precision data expectations to find a light Higgs but
not finding such a state.  The original littlest Higgs model was
plagued by strong electroweak precision constraints, requiring a
reintroduction of the fine tuning problem.  An economical solution is
to introduce a discrete symmetry called T-parity.  T-parity solves the
electroweak precision constraint issues, while also providing a dark
matter candidate.  We give the relic abundance of this dark matter
candidate as a function of the parameters in the model.  In addition,
we discuss the LHC phenomenology, presenting the production cross
sections and decay channels for the new particles in the model.}\\


The turn-on date for the LHC is fast approaching.  The electroweak
hierarchy problem has many physicists convinced that the data which
will pour out of this experiment will give many hints towards the way
in which this problem is solved by nature.  The most well studied
extension of the Standard Model (SM) which stabilizes the electroweak
hierarchy is supersymmetry.

Little Higgs models are a more recent attempt to solve the hierarchy
problem by the introduction of additional global symmetries which are
spontaneously broken at the TeV scale.  In these theories, the Higgs
is an approximate Goldstone boson of the global symmetry breaking
pattern~\cite{Georgi:1975tz,Georgi:1974yw,Dugan:1984hq,Georgi:1984af,Georgi:1984ef,Kaplan:1983sm,Kaplan:1983fs}.
The interactions which explicitly violate these symmetries generate
non-derivative interactions for these fields.  In Little Higgs models,
these interactions are introduced in a way such that any single
interaction preserves enough of the global symmetries to keep the
Higgs precisely massless~\cite{Arkani-Hamed:2001nc}.  However, all the
interactions together break all the global symmetries.  In this way,
the quadratically divergent contributions to the Higgs mass are
suppressed by additional loop factors.

We focus on the littlest Higgs model, based on an $SU(5)/SO(5)$
non-linear sigma model~\cite{Arkani-Hamed:2002qy}.  The earliest
implementations of this structure were not reconcileable with
electroweak precision
constraints~\cite{Csaki:2002qg,Csaki:2003si,Hewett:2002px}, but
recently the structure has been extended to include a discrete
symmetry which eliminates the tree-level contributions to SM
electroweak observables~\cite{Cheng:2003ju,Cheng:2004yc,Low:2004xc}.
This discrete symmetry is called T-parity.  Most new particles
introduced beyond the SM spectrum are odd under this parity, leading
to drastically modified collider
phenomenology~\cite{Burdman:2002ns,Han:2003wu,Hubisz:2004ft}.


\subsubsection{Model description}

The new interactions can be described by an $SU(5)/SO(5)$ non-linear
sigma model, as we describe below.  The breaking pattern $SU(5)\to
SO(5)$ is accomplished by a symmetric tensor of $SU(5)$, $\Sigma$.
The vacuum expectation value of this field is assumed to be near
1~TeV, so that fine tuning in the Higgs mass squared parameter is
minimized.  Embedded within the $SU(5)$ global symmetries is a
$\left[SU(2)\times U(1)\right]^2$ gauge symmetry.  The VEV $\Sigma$
breaks this gauge symmetry down to the diagonal subgroup which is then
associated with the SM $SU(2)_L\times U(1)_Y$.  The other gauge bosons
become massive at the TeV scale.

In the non-linear sigma model parametrization, the field $\Sigma$ can
be written as
\begin{equation}
\Sigma = e^{2 i \pi/f} \Sigma_0 \, ,
\end{equation}
where $\pi$ is a matrix containing all of the Goldstone degrees of
freedom associated with the breaking of the $SU(5)$ global symmetry to
the $SO(5)$ subgroup.

The Goldstone bosons associated with this breaking decompose under
$SU(2)_L\times U(1)_Y$ as
\begin{equation}
3_0 \oplus 3_{1/2} \oplus 2_{1/2} \oplus 1_0 \, .
\end{equation}
The $3_0$ and $1_0$ are eaten by the gauge bosons that become massive
at the scale $f$.  We associate the $2_{1/2}$ with the Higgs
multiplet.  The $3_{1/2}$ is a triplet of $SU(2)_L$.

In addition to the new gauge and scalar degrees of freedom, a new
vectorlike $SU(2)_L$ singlet quark is required in the theory to cancel
the quadratic divergence due to the top quark Yukawa interaction.  The
top quark is a mixture of this singlet and the $T_3=1/2$ component of
the $SU(2)_L$ third generation doublet.  We label the orthogonal mass
eigenstate $T_+$.  This new colored state obtains a mass slightly
larger than $f$.

The earliest implementations of this structure suffered from
electroweak precision constraints.  After electroweak symmetry
breaking, mixing would be induced between the standard model gauge
bosons and their TeV-scale partners.  This mixing leads to violations
of custodial $SU(2)$~\cite{Csaki:2002qg,Csaki:2003si,Hewett:2002px}
causing, for example, a tree-level shift in the $\rho$ parameter, a
tightly-constrained relation between the $W$ and $Z$ boson masses.
T-parity is a postulated discrete symmetry which forbids mixing
between the standard model fields, and their heavier counterparts.

T-parity exchanges the two copies of $SU(2)\times U(1)$.  In this way,
the diagonal subgroup (the SM gauge group) is T-even, while the other
combinations, which receive $f$ scale masses, are T-odd.  In addition,
if one wishes to implement this symmetry consistently throughout the
entire model, the matter sector of the model must also be symmetric
under this interchange.  For every multiplet that transforms under
$[SU(2)\times U(1)]_1$, there must be a partner multiplet that
transforms under $[SU(2)\times U(1)]_2$.  This discrete symmetry,
while it eliminates the tree-level shifts in SM observables,
drastically changes the phenomenology of little Higgs models.

Enforcing T-parity requires that the gauge couplings for the two
$SU(2)\times U(1)$ gauge groups be equal.  This fixes the mass
spectrum of the new gauge bosons with respect to the overall breaking
scale $f$:
\begin{equation}
M_{W_H^\pm} = M_{Z_H} =  g f \ \ \ M_{A_H} = \frac{g' f}{\sqrt{5}} \, ,
\end{equation}
where $g$ and $g'$ are the weak and hypercharge gauge couplings,
respectively.

If the discrete symmetry is made exact, the lightest T-odd particle is
stable and a potential dark matter candidate.  In collider
phenomenology, this lightest particle becomes a missing energy signal,
making observation of this new physics more complicated.  In
particular, it is likely that this type of model will look very much
like supersymmetry in certain regions of parameter space.  This is
similar to studies of universal extra dimensions, where the signals
are also similar to supersymmetry~\cite{Cheng:2002ab}.  In the
littlest Higgs model with T-parity, the heavy partner of the
hypercharge gauge boson, the $A_H$, is the dark matter candidate, and
can account for the WMAP observed relic density~\cite{Hubisz:2004ft}.

A consequence of implementing this discrete symmetry in the
$SU(5)/SO(5)$ Littlest Higgs model is that the fermion spectrum must
be substantially enlarged.  This is due to constraints on four-fermion
operators involving the standard model left handed quark and lepton
doublets~\cite{Cheng:2004yc}.  The additional T-odd fermions serve to
cut these contributions off.  In this analysis, we take these fermions
to be rather heavy (out of reach for the LHC).

In addition, a T-odd partner of the $T_+$ is necessary, which we label
$T_-$.  This T-odd singlet changes the collider phenomenology, as the
$T_+$ generically has a sizable branching fraction to missing energy:
$T_+\to T_- A_H$.  In particular, earlier studies of the phenomenology
of this new state~\cite{Perelstein:2003wd,Azuelos:2004dm} are
modified.

A more complete model description containing all details and
interactions may be found in Ref.~\cite{Hubisz:2004ft}.


\subsubsection{The dark matter Candidate}

\begin{figure}[h!]
\centerline{\includegraphics[width=0.7\hsize]{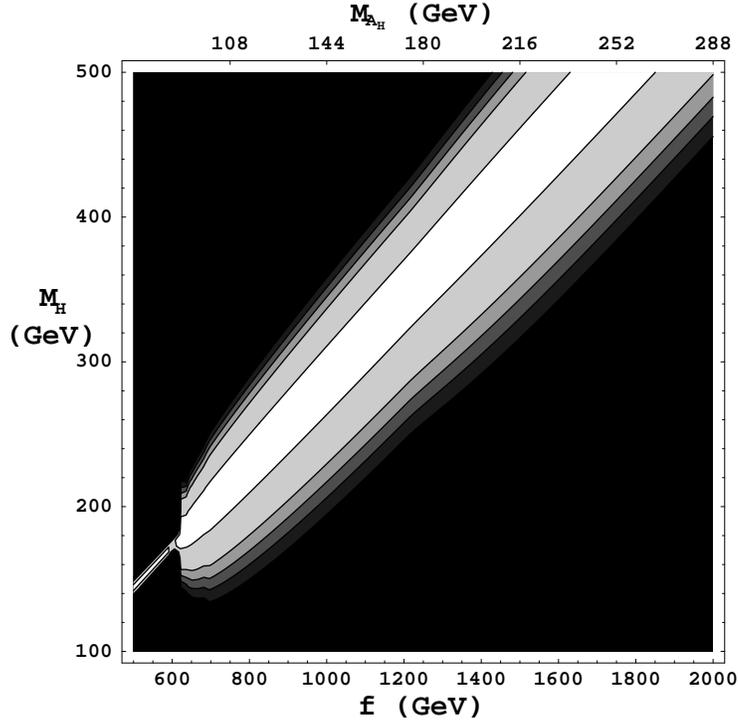}}
\caption{Variation of the dark matter relic density with respect to 
the Higgs mass and the symmetry breaking scale, $f$.  In order from
lightest to darkest regions, the $A_H$ makes up ($0-10\%$, $10-50\%$,
$50-70\%$, $70-100\%$, $100\%$, $>100\%$) of the observed relic
abundance.}
\label{limit-fig}
\end{figure}

We calculate the relic density of the lightest T-odd particle assuming
that T-parity is an exact symmetry, and that the T-odd fermions are
heavy.  The mass spectrum is sufficiently non-degenerate that
coannihilation effects are unimportant, and only direct annihilation
channels need be considered.  The dominant channels are those
involving s-channel Higgs exchange.  As a result, the annihilation
cross section is primarily a function of the Higgs and dark matter
candidate masses.  Imposing the constraints from
WMAP~\cite{Spergel:2003cb} leads to Fig.~\ref{limit-fig}.  We see that
there is a strong correlation between the scale $f$ and the Higgs mass
if the dark matter is to come purely from Little Higgs physics.  This
is due to the s-channel pole present when $m_{A_H}=m_H/2$.  Noteably,
larger values of $f$ prefer larger Higgs masses than the SM best fit
value.

We consider regions as ruled out where the relic density exceeds the
$95\%$ confidence limits imposed by the WMAP bound.  In regions where
the relic density of $A_H$ is below the WMAP $95\%$ confidence band,
there is the possibility that there is another form of dark matter,
such as an axion, which could make up the difference.

A study of the one-loop electroweak precision corrections in this
model reveals that certain contributions to $\Delta\rho$ from one-loop
diagrams arise with opposite sign as the terms which are logarithmic
in the Higgs mass~\cite{Hubisz:2005tx}.  This effect is due to the
contributions from singlet-doublet quark mass mixing in the third
generation Yukawa couplings.  Consequentially, the Higgs mass can be
raised far above its standard electroweak precision bound while
remaining consistent with LEP.  Thus, for certain ranges of the
parameters in the top-quark Yukawa sector, both dark matter and EWP
bounds may be satisfied simultaneously.


\subsubsection{Collider Phenomenology}

After entering all new interactions into COMPHEP~\cite{Pukhov:1999gg},
we calculate the production cross sections for the new particles in
this model, comparing with the original little Higgs model, where the
phenomenology was considered orginally
in~\cite{Burdman:2002ns,Han:2003wu}.  We briefly discuss some of the
signals, and summarize the primary decay modes of these particles.

Because SM particles are T-even while most new particles are T-odd,
the energy cost of creating these particles is doubled due to the need
to pair produce them.  In addition, most new states are not charged
under QCD, meaning that their production cross sections are somewhat
small.

\begin{figure}[h]
\centerline{\includegraphics[width=0.7\hsize]{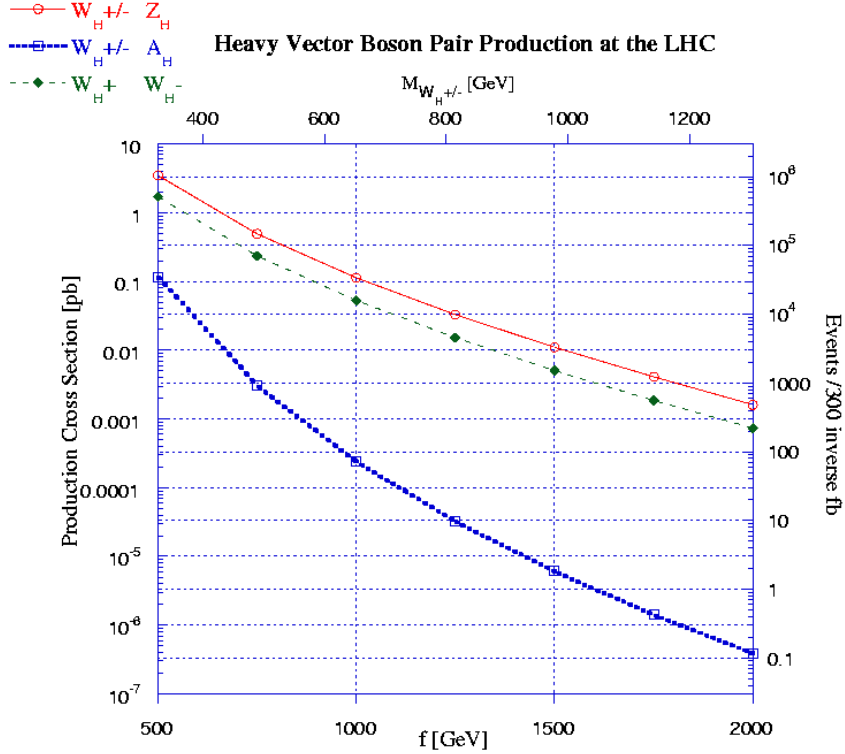}}
\caption{Cross section for production of a pair of T-odd heavy vector 
bosons at the LHC as a function of the symmetry breaking scale $f$.
The number of events for 300~fb$^{-1}$ is plotted on the second
y-axis.  $M_{W_H^{\pm}}$ is plotted on the second x-axis. $M_{Z_H}$ is
degenerate in mass with $M_{W_H^{\pm}}$, and $M_{A_H}\sim .16 f$.}
\label{vectorproduction}
\end{figure}

In the T-odd gauge boson sector, the only free parameter relevant for
production cross section calculations is the global symmetry breaking
scale, $f$.  The possible pairings are
\begin{eqnarray}
&&p p \rightarrow W_H^+ W_H^- \arline
&&p p \rightarrow W_H^\pm A_H \arline
&&p p \rightarrow W_H^\pm Z_H \, .
\end{eqnarray}
The cross sections can be seen in Fig.~\ref{vectorproduction}.  The
dominant diagrams involve s-channel exchange of SM gauge bosons.
Production of $W_H^\pm A_H$ is suppressed by the analog of the
Weinberg angle in the T-odd gauge boson sector, which is of order
$v/f$, which explains why the cross section for $pp\to W_H^\pm A_H$ is
smaller than the others.  The decays of the new gauge bosons are as
follows:
\begin{eqnarray}
&&W_H^\pm \rightarrow W^\pm A_H \arline
&&Z_H \rightarrow A_H h \, .
\end{eqnarray}

In our analysis, we assume that the additional T-odd fermion doublets
are significantly heavier than the new gauge bosons.  This way they do
not contribute significantly to the tree-level production cross
sections, and are not themselves produced in large number.

\begin{figure}[h]
\centerline{\includegraphics[width=0.7\hsize]{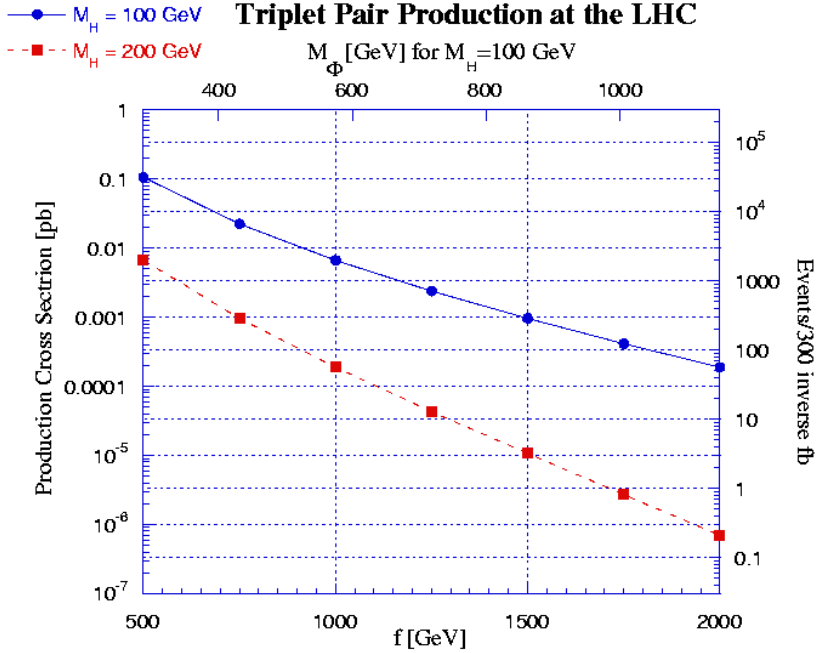}}
\caption{Cross sections for the production of a pair of T-odd triplets
at the LHC is plotted as a function of the symmetry breaking scale
$f$, plotted for $m_H=100,200\;\mathrm{GeV}$ since the triplet mass,
$M_\Phi$, is determined by $f$ and $m_H$.  The number of events for
300~fb$^{-1}$ is plotted on the second y-axis.  $M_\Phi$ for a Higgs
mass of 100~GeV is plotted on the second x-axis, for a Higgs mass of
200~GeV simply scale the second x-axis by a factor of 2.}
\label{tripprodfig}
\end{figure}

It is also possible to pair produce the components of the scalar
triplet.  The triplet mass is proportional to the Higgs mass:
\begin{equation}
m_{\Phi}^2 \approx \frac{2 m_H^2 f^2}{v^2} \, .
\end{equation}
The degeneracy of the electromagnetic-charged eigenstates of the
triplet is slightly split by electroweak symmetry breaking effects.
The production cross sections are plotted in Fig.~\ref{tripprodfig}.
When the Higgs is heavy enough, the relevant decay modes for the
triplet are given by
\begin{eqnarray}
&&\phi^{++} \rightarrow W^+ W_H^+ \rightarrow W^+ W^+ A_H \arline
&&\phi^+\ \ \rightarrow W^+ A_H \arline
&&\phi^P\ \ \rightarrow H A_H \arline
&&\phi^0\ \ \rightarrow Z A_H \, .
\end{eqnarray}
In the case that the triplet mass is below the threshold for the
doubly charged components to decay through the above channel, they
must decay directly to the three body final state.  We note that in
these regions, the Higgs is below 130~GeV.  This region is excluded by
WMAP, since such light values of the Higgs mass imply an excessive
relic abundance, as shown in Fig.~\ref{limit-fig}.

\begin{figure}[h]
\centerline{\includegraphics[width=0.6\hsize]{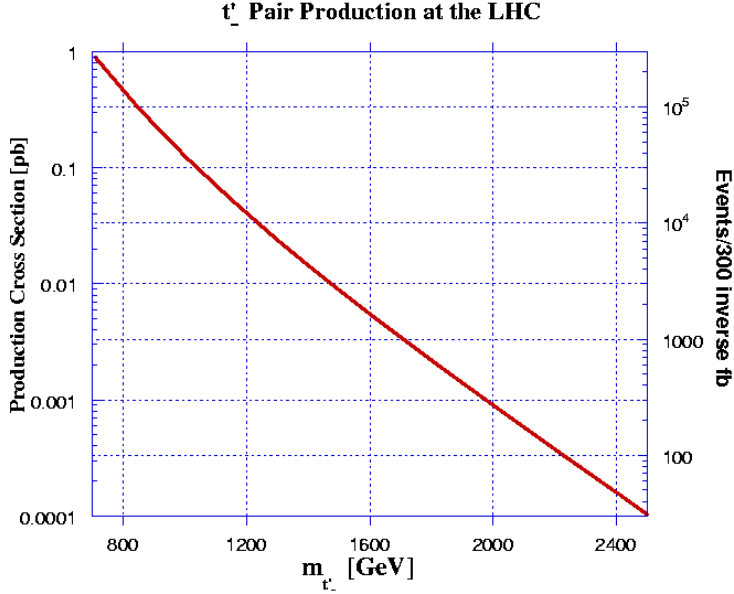}}
\caption{Cross section for LHC production of a pair of T-odd heavy
quarks $T_-$, plotted as a function of $m_{T_-}$.  The number of
events for 300~fb$^{-1}$ is plotted on the second y-axis.}
\label{tminus}
\end{figure}

We also consider pair production of the T-odd colored fermion, $T_-$.
In contrast to the gauge bosons and scalars, the $T_-$ is produced
primarily through gluon exchange.  Therefore its production cross
section, shown in Fig.~\ref{tminus}, is comparatively large.  The
$T_-$ decays through the channel $T_-\to t A_H$.

The backgrounds to these signals are likely to be considerable.  A
rough estimate is given in~\cite{Hubisz:2004ft}, but more thorough
study is necessary.

\begin{figure}[h]
\centerline{\includegraphics[width=0.7\hsize]{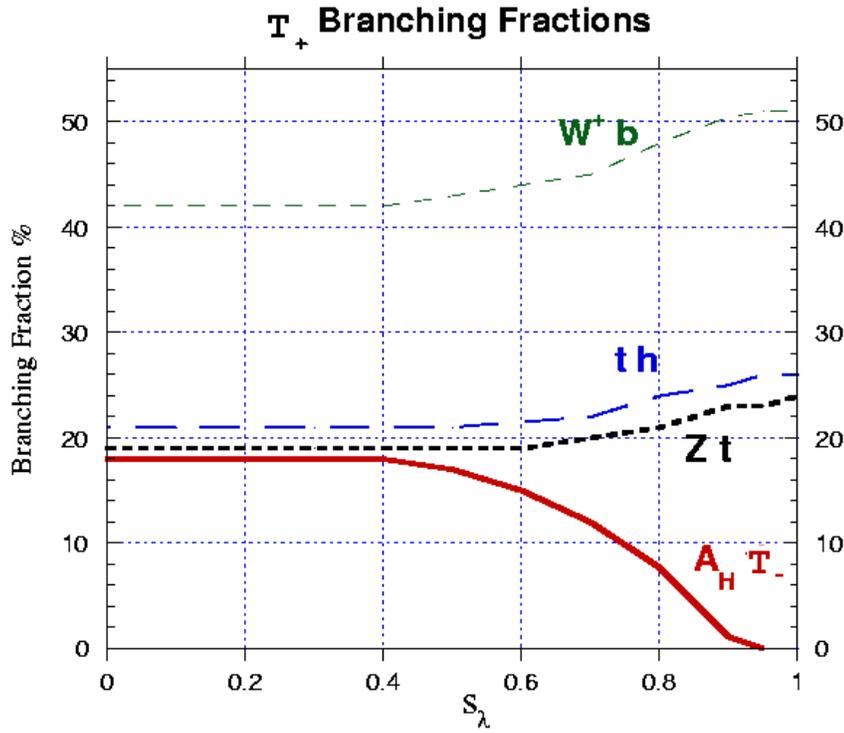}}
\caption{Branching fractions for $t'_{+}$ decay as a function of 
$s_{\lambda}$, which parameterizes the ratio of masses of the $T_{+}$
and $T_{-}$; for $f=1$~TeV.}
\label{tbranch}
\end{figure}

The phenomenology of the T-even partner of the top quark is similar to
the original Littlest Higgs implementation.  The $T_+$ production
cross sections are identical, as the top sector is not drastically
modified.  The primary difference is in the $T_+$ decay modes.  A
consistent implementation of T-parity in this model requires the
introduction of a T-odd partner of the $T_+$, which we call the $T_-$.
The $T_-$ is generically lighter than the $T_+$.  The mass difference
between these two states is large enough in most of the available
parameter space to allow the decay mode $T_+\to T_- A_H\to t A_H A_H$.
Thus, a sizeable fraction of the decay channels will have missing
energy in the final state, which will complicate reconstruction of the
$T_+$ width.  The branching fractions of the $T_+$ are given in
Fig.~\ref{tbranch}.

%


\subsubsection{Conclusions}

We have reviewed the phenomenology of the Littlest Higgs with T-parity
in the limit that the partners of the standard model $SU(2)_L$ fermion
doublets are taken to be above the reach of the LHC.  We have shown
that the T-parity symmetry provides a dark matter candidate that can
account for the WMAP best fit value for the relic abundance.  Due to
the requirement of decaying to a missing energy signal, LHC
phenomenology is more difficult.  In particular, T-parity models may
resemble supersymmetry in certain regions of parameter space.

\clearpage\setcounter{equation}{0}\setcounter{figure}{0}\setcounter{table}{0}
\subsection{Search for Low-Scale Technicolor at the Tevatron}
\label{sec:TC}


{\em Kenneth Lane, Dept. of Physics, Boston University}\\

{\em CDF and D\O\ each have more than $1\,\ifb$ of data on tape, and
their stores are increasing.  This should be sufficient to carry out
significant searches for low-scale technicolor in $\tro\ra W\tpi$ and
$\tom\,,\tro\ra\gamma\tpi$, processes whose cross sections may be as
large as several picobarn.  We motivate and describe the Technicolor
Straw Man framework for these processes and urge that they be sought
soon in the Run~2 data.}


\subsubsection*{1. Preamble}

\begin{figure}[t]
\vspace{9.0cm}
\includegraphics{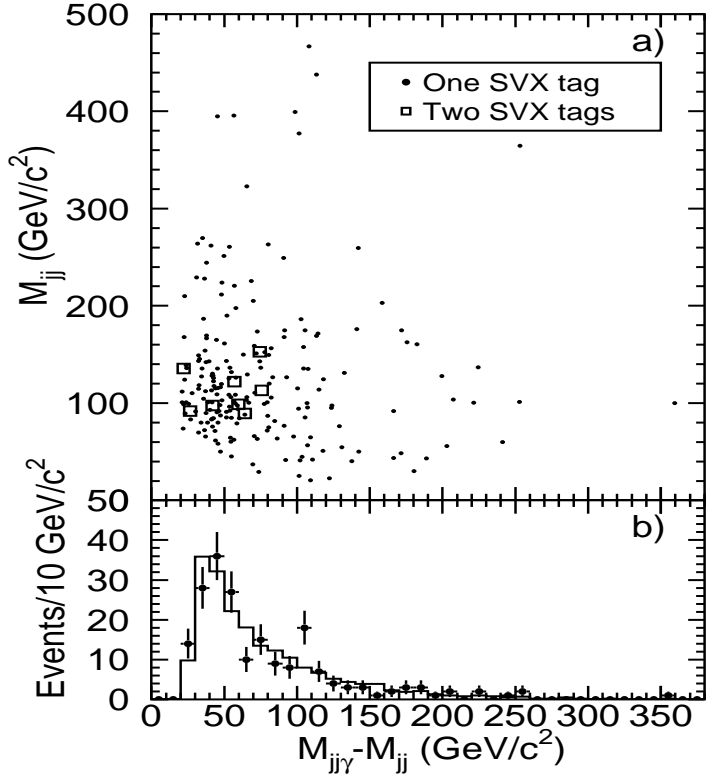}
\vskip2.0truecm
\caption{(a) The distribution of $M_{jj}$ vs. $M_{jj\gamma}-M_{jj}$ for
events with a photon, $b$--tagged jet and a second jet, and (b)
projection of this data in $M_{jj\gamma}-M_{jj}$; from
Ref.~\protect\cite{Abe:1998jc}.}
\label{TeV4LHC_fig_1}
\end{figure}

Fig.~\ref{TeV4LHC_fig_1} is from CDF in Run~I.  It shows a search for
$\tom\ra\gamma\tpi$, with $\tpi\ra b+$jet, based on about
100~pb$^{-1}$ of data, published in 1999~\cite{Abe:1998jc}.  Note the
$\sim 2\sigma$ excess near $M_{jj\gamma}-M_{jj}=100$~GeV.  This search
has not been repeated in Run~2.\footnote{Both detectors induce jet
backgrounds to photons that require much effort to suppress; the
effort should be made.}  Fig.~\ref{fig:TeV4LHC_fig_2} is from CDF in
Run~II.  It shows results of an unpublished CDF study looking for
$\tro\ra W^\pm\tpi$.\footnote{CDF's Run~I version of this search is
published in Ref.~\cite{Affolder:1999du}.}  The data were posted in
July 2004 and are based on 162~pb$^{-1}$ of data.  There are small
excesses in the dijet and $Wjj$ masses near 110~GeV and 210~GeV,
respectively.  Assuming $M_{\tom}=M_{\tro}\simeq 230$~GeV, and taking
into account losses from semileptonic $b$-decays, the excesses in
Figs.~1 and~2 are in about the right place for $M_{\tpi}\simeq
120$~GeV.

In December 2005, CDF reported on a search for $WH$-production with
$W\ra\ell\nu$ and $H\ra b\bar{b}$ and a single $b$-tag, based on
320~pb$^{-1}$ of data~\cite{Abulencia:2005ep}.  The dijet mass
spectrum appears in Fig.~\ref{fig:TeV4LHC_fig_3}.\footnote{I am
grateful to Y.-K.~Kim and her CDF collaborators for providing this
figure.}  There is a $2\sigma$ excess at $M_{jj}\simeq 110$~GeV.
The $Wjj$ spectrum was not reported and is not available.  The
expected rate for a $\sim 100$~GeV Higgs decaying to $b\bar{b}$ and
produced in association with a $W$ is about 0.1~pb.  If the excess
were real, it would correspond to a total $WH$ cross section of about
5~pb, about 50~times the expected cross section.
 
A $2\sigma$ excess does not constitute convincing evidence of a
signal, but does warrant follow-up investigation.  Both experiments
have now collected almost $1.5\,\ifb$.  This summer, CDF and D\O\ will
present new results for SUSY, large extra dimensions, Randall-Sundrum
gravitons, Little Higgs, and other new physics searches.  We hope they
present the searches for technicolor as well.  The most likely
processes and search modes are:
\bea\label{eq:TCsearch_modes}
&& \tro^\pm \ra W^\pm \tpi^0   \ra \ell^\pm \nu_\ell + b\bar{b} \\
&& \tro^0   \ra W^\pm \tpi^\mp \ra \ell^\pm \nu_\ell + b\bar{c} \,,\,b\bar{u} \\ 
&& \tom\,,\tro^0 \ra \gamma \tpiz\,,\gamma\tpipr \ra \gamma b \bar{b} \\
&& \tom\,,\tro^0 \ra e^+ e^-\,,\,\mu^+\mu^-\,.
\eea
These processes (and more) are available in
{\textsc{Pythia}}~\cite{Sjostrand:2000wi,Sjostrand:2003wg}.

In the rest of this section, we motivate low-scale technicolor ---
that technihadrons may be much lighter than $\sim 1\,\tev$ and, in
fact, may be readily accessible at the Tevatron.  Then we describe the
Technicolor Straw Man Model (TCSM) and present some rate estimates for
the most important color-singlet processes.  The TCSM is
\begin{figure}[ht!]
\begin{center}
\includegraphics[width=3.5in]{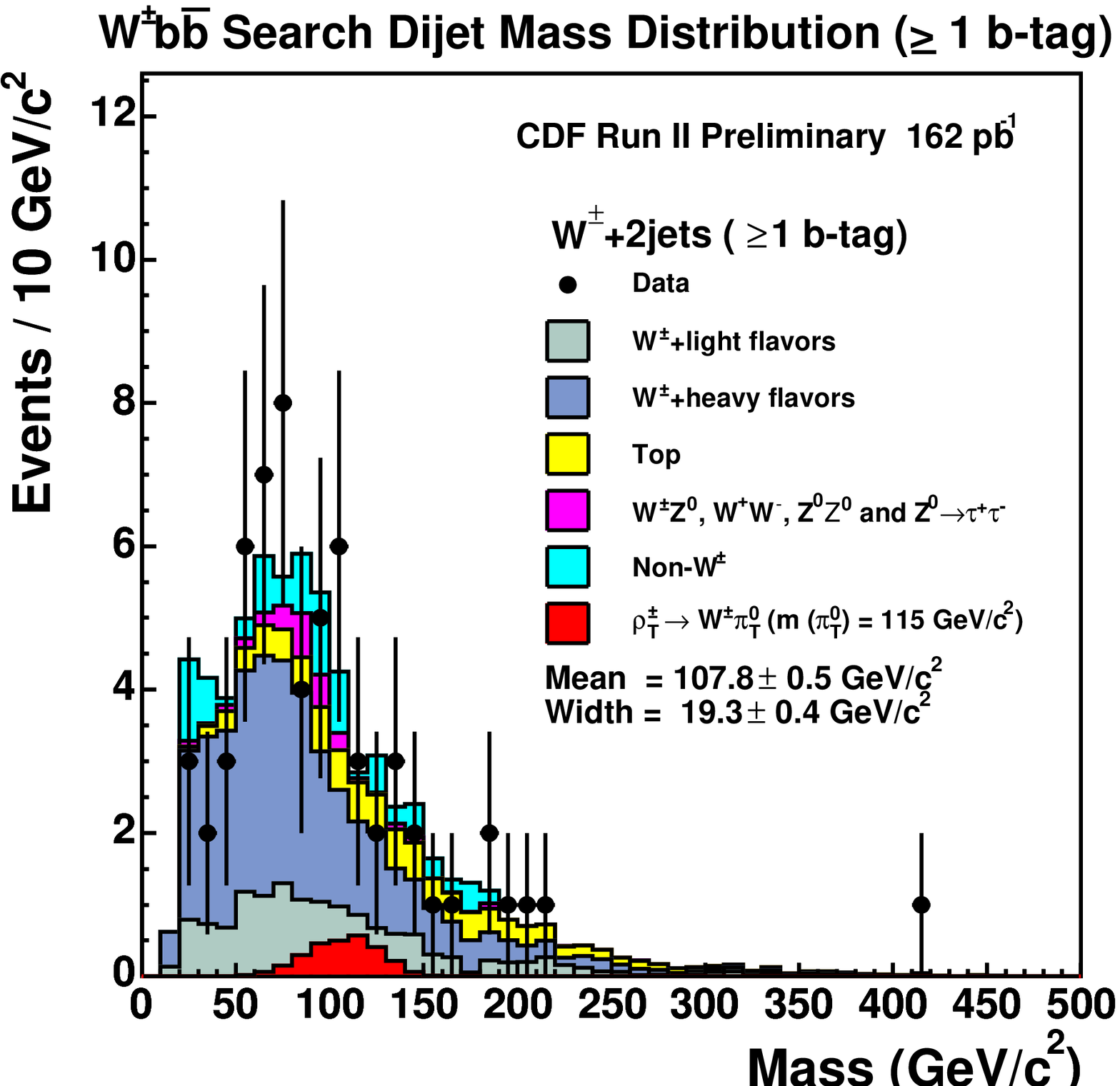}
\includegraphics[width=3.5in]{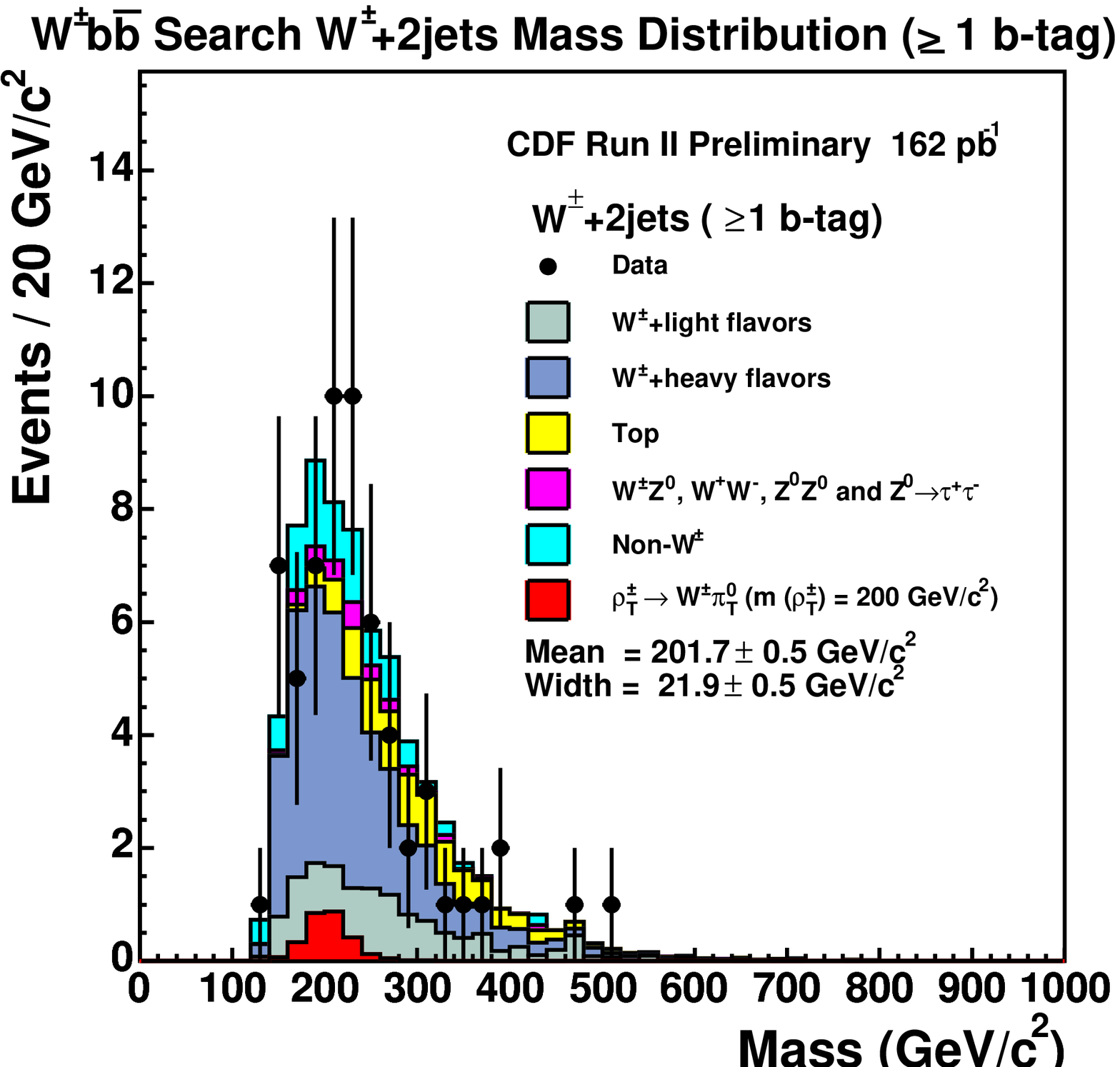}
\caption{(a) Invariant mass of the dijet system with $\ge 1$ $b$-tagged
jets, and (b) invariant mass of the $W+2\ts\jet$ system for the
$\ell+2\ts\jet$ mode in $\ge 1$ $b$-tagged jets; from Run~2 with
$162\,\ipb$ (see {\tt
http://www-cdf.fnal.gov/physics/exotic/r2a/20040722.lmetbj-wh-tc/}).}
\label{fig:TeV4LHC_fig_2}
\end{center}
\end{figure}
described in more detail in Refs.~\cite{Lane:1999uh,Lane:2002sm}, and
much of the last two subsections is taken from the second of these.

\begin{figure}[t]
\begin{center}
\includegraphics[width=3.5in]{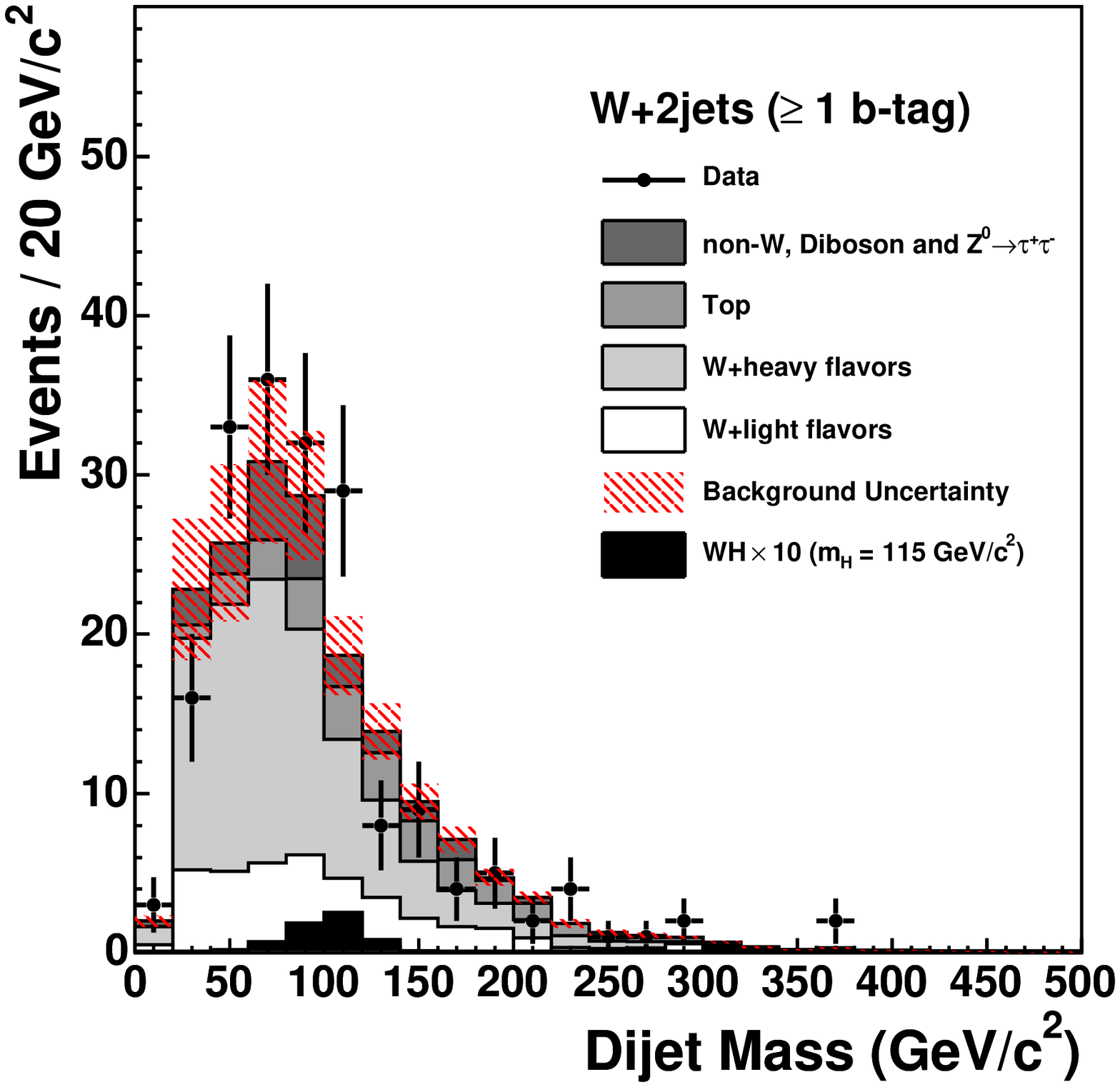}
\caption{Invariant mass of the $W+2\ts\jet$ system for the 
$\ell+2\ts\jet$ mode with $\ge 1$ $b$-tagged jets; from Run~2 with
$320\,\ipb$; see Ref.~\protect\cite{Abulencia:2005ep}.}
\label{fig:TeV4LHC_fig_3}
\end{center}
\end{figure}
%


\subsubsection*{2. Low-Scale Technicolor}

Technicolor (TC) is the only theory of electroweak symmetry breaking
(EWSB) by new strong dynamics whose characteristic energy scale is at
or below 1~TeV~\cite{Weinberg:1979bn,Susskind:1978ms}.  It is the most
natural scenario (not to mention the only precedent) for dealing with
the Standard Model's naturalness problem: it banishes {\em elementary}
scalar particles altogether.  TC by itself, however, cannot explain
--- or even describe in a phenomenological way, as the standard model
does --- the origin of quark and lepton masses and mixings.  The only
known way to do that in the dynamical context of TC is {\it extended
technicolor} (ETC)~\cite{Eichten:1979ah}.

Two elements of the modern formulation of TC (see the reviews and
references in Refs.~\cite{Lane:2002wv, Hill:2002ap}) strongly suggest
that its energy scale $\Ltc\simeq 4\pi F_T$, where $F_T$ is the
technipion decay constant --- and therefore the masses of
technihadrons ($\tro$ and $\tom$ as well as $\tpi$) --- are {\em much}
less than several TeV.  They are the notions of {\em walking
technicolor} (WTC)~\cite{Holdom:1981rm,Appelquist:1986an,
Yamawaki:1985zg,Akiba:1985rr} and {\em topcolor-assisted technicolor}
(TC2)~\cite{Hill:1994hp}.  Assuming for simplicity that the
technifermions form $N_D$ electroweak doublets, then $F_T\simeq
F_\pi/\sqrt{N_D}$, where $F_\pi=246$~GeV.  The EWSB condensate is
$\langle\bar{T}T\rangle_{TC}\simeq 4\pi F_T^3$.

Extended technicolor inevitably induces flavor-changing neutral
current interactions of quarks and leptons.  The most problematic of
these are the $|\Delta S|=2$ operators,
\begin{equation}\label{eq:dStwo}
\CH_{|\Delta S|=2} = \frac{g^2_{ETC}}{\Metc^2} \sum_{ij}
K_{ij}\, \ol{s} \Gamma_i d \, \ol{s} \Gamma_j d + {\rm h.c.} \,,
\end{equation}
which require effective ETC gauge boson masses
$\Metc/g_{ETC}\sqrt{K_{ij}}\gtrsim 1000\,\tev$.  If TC were a QCD-like
gauge theory, one in which asymptotic freedom sets in quickly near
$\Ltc$, the quark and lepton masses $m_{q,l}\simeq g^2_{ETC}\langle
\bar{T}T\rangle_{ETC}/M^2_{ETC}$ generated by such high-scale ETC
interactions would be unacceptably small because
$\langle\bar{T}T\rangle_{ETC}\simeq\langle\bar{T}T\rangle_{TC}$.  This
difficulty is cured by WTC, in which the technicolor gauge coupling
$\atc$ runs very slowly, i.e., the interaction is close to conformally
invariant, and the technifermion condensates
$\langle\bar{T}T\rangle_{ETC}$ renormalized at the ETC scale are
enhanced relative to $\langle\bar{T}T\rangle_{TC}$ by a factor not
much less than $\Metc/\Ltc$.  The small $\beta_{TC}$-function required
for WTC is readily achieved by having many technidoublets transforming
as the fundamental representation of the TC gauge
group.\footnote{Walking could in principle be achieved by having a few
technidoublets in higher-dimensional TC representations; see
Refs.~\cite{Lane:1989ej} and~\cite{Dietrich:2005jn,Evans:2005pu}.  It
is difficult to see how this could be done without some number of
doublets in the fundamental representation; see
Ref.~\cite{Lane:1991qh}.}  Thus, $N_D$ is large and $F_T$ is small.

Even with the enhancements of walking technicolor, there is no
satisfactory way in the context of ETC alone to understand the large
mass of the top quark.  Either the ETC mass scale generating $m_t$
must be too close to $\Ltc$ or the ETC coupling must be
fine-tuned.\footnote{A possible exception to this was proposed in
Ref.~\cite{Appelquist:2003hn}.  In this model, $N_D=4$ and $F_T$ is
not particularly small.  The model is genuinely baroque, but that is
probably true of any quasi-realistic ETC model.}  So far, the most
attractive scheme for $m_t$ is that it is produced by the condensation
of top quarks, induced at a scale near $1\,\tev$ by new strong
topcolor gauge interactions ($SU(3)\otimes U(1)$ in the simplest
scheme).  This top condensation scheme, topcolor-assisted technicolor,
accounts for almost all the top mass, but for only a few percent of
EWSB.  Realistic models that provide for the TC2 gauge symmetry
breaking and for the mixing of the heavy third generation with the two
light generations typically require many ($N_D\simeq 10$ (!)) 
technifermion doublets.  Therefore, in the following, we shall assume
$F_T\simle 100$~GeV.\footnote{The question of the effect of
technicolor on precisely measured electroweak quantities such as $S$,
$T$, and $U$ naturally arises because of the appearance of many
technifermion doublets in low-scale technicolor.  Calculations which
show TC to be in conflict with precision measurements have been based
on the assumption that TC dynamics are just a scaled-up version of
QCD.  However, because of its walking gauge coupling, this cannot be.
In WTC there must be something like a tower of spin-one technihadrons
reaching almost to the ETC scale, and these states must contribute
significantly to the integrals over spectral functions involved in
calculating $S$, $T$, and $U$.  Therefore, in the absence of detailed
experimental knowledge of this spectrum, including the spacing between
states and their coupling to the electroweak currents, it is not
possible to estimate these quantities reliably.}


\subsubsection*{3. The Technicolor Straw Man Model}

The TCSM provides a simple framework for light technihadron searches.
Its first and probably most important assumption is that the
lowest-lying bound states of the lightest technifermions can be
considered {\em in isolation}.  The lightest technifermions are
expected to be an isodoublet of color singlets, $(T_U,T_D)$.  Color
triplets, not considered here, will be heavier because of $\suc$
contributions to their hard (chiral symmetry breaking) masses.  We
assume that all technifermions transform under technicolor $SU(\Ntc)$
as fundamentals.  This leads us to make --- with no little trepidation
in a walking gauge theory --- large-$\Ntc$ estimates of certain
parameters.  The electric charges of $(T_U,T_D)$ are $Q_U$ and
$Q_D=Q_U-1$; they are important parameters of the TCSM.  The
color-singlet bound states we consider are vector and pseudoscalar
mesons.  The vectors include a spin-one isotriplet $\tro^{\pm,0}$ and
an isosinglet $\tom$.  Techni-isospin can be a good approximate
symmetry in TC2, so that $\tro$ and $\tom$ are nearly degenerate.
Their mixing with each other and the photon and $Z^0$ is described by
a neutral-sector propagator matrix.

The lightest pseudoscalar bound states of $(T_U,T_D)$ are the color-singlet
technipions. They also form an isotriplet $\Pi_T^{\pm,0}$ and an isosinglet
$\Pi_T^{0 \prime}$. However, these are not mass eigenstates. Our second
important assumption for the TCSM is that the isovectors may be described as
simple {\em two-state mixtures} of the longitudinal weak bosons $W_L^\pm$,
$Z_L^0$ --- the true Goldstone bosons of dynamical electroweak symmetry
breaking --- and mass-eigenstate pseudo-Goldstone technipions $\tpi^\pm, \tpiz$:
\begin{equation}\label{eq:pistates}
\vert\Pi_T\rangle = \sin\chi \ts \vert
W_L\rangle + \cos\chi \ts \vert\tpi\rangle\ts.
\end{equation}
Assuming that $\sutc$ gauge interactions dominate the binding of all
technifermions into technihadrons, the decay constants of
color-singlet and nonsinglet $\tpi$ are approximately equal,
$F_T\simeq F_\pi/\sqrt{N_D}$, and the mixing factor $\sin\chi$ ---
another important TCSM parameter --- is given by
\begin{equation}\label{eq:sinchi}
\sin\chi \simeq F_T/F_\pi \simeq 1/\sqrt{N_D}\ts,
\end{equation}
so that $\sin^2\chi\ll 1$.

Similarly,
$\vert\Pi_T^{0\prime}\rangle=\cos\chipr\ts\vert\tpipr\rangle+\cdots$,
where $\chipr$ is another mixing angle and the ellipsis refer to other
technipions needed to eliminate the two-technigluon anomaly from the
$\Pi_T^{0\prime}$ chiral current.  It is unclear whether, like $\tro$
and $\tom$, these neutral technipions will be degenerate.  If $\tpiz$
and $\tpipr$ are nearly degenerate {\em and} if their widths are
roughly equal, there may be appreciable $\tpiz$--$\tpipr$ mixing; then
the lightest neutral technipions will be ideally-mixed $\bar{T}_U T_U$
and $\bar{T}_D T_D$ bound states.  {\em Searches for these technipions
ought to consider both possibilities: they are nearly degenerate or
such that $M_{\pi_T^\pm}=M_{\pi_T^0}\ll M_{\tpipr}$}.

Color-singlet technipion decays are mediated by ETC and (in the case of
$\tpipr$) $\suc$ interactions.  In the TCSM they are taken to be:
\bea\label{eq:tpiwidths}
\Gamma(\tpi\ra\bar{f}'f) &=& \frac{1}{16\pi F^2_T}
\ts N_f \ts p_f \ts C^2_{1f} \, (m_f + m_{f'})^2 \nn
\\ \nn \\
\Gamma(\tpipr \ra gg) &=& \frac{1}{128 \pi^3 F^2_T}
\ts \alpha_C^2 \ts C^2_{1g} \ts \Ntc^2 \ts M_{\tpipr}^{\frac{3}{2}} \ts .
\eea
The number of colors of fermion~$f$ is $N_f$ and the fermion momentum
is $p_f$.  The QCD coupling $\alpha_C$ is evaluated at $M_{\tpi}$;
$C^2_{1g}$ is a Clebsch-Jordan coefficient of order one.  The default
values of these and other parameters are tabulated in
Ref.~\cite{Lane:2002sm}.  Like elementary Higgs bosons, technipions
are {\em expected} to couple to fermions proportional to the fermion
mass.  Thus, $C_{1f}$ is an ETC-model dependent factor of order one
{\em except} that TC2 implies a weak coupling to top quarks, $\vert
C_{1t}\vert\simle m_b/m_t$.  Thus there is no strong preference for
technipions to decay to (or radiate from) top quarks.  For
$M_{\tpi}<m_t+m_b$, these technipions are expected to decay mainly as
follows: $\tpip\ra c\bar{b}$, $u\bar{b}$, $c\bar{s}$ and possibly
$\tau^+\nu_\tau$; $\tpiz\ra b\bar{b}$ and, perhaps $c\bar{c}$,
$\tau^+\tau^-$; and $\tpipr\ra gg$, $b\bar{b}$, $c\bar{c}$,
$\tau^+\tau^-$.

In the limit that the electroweak couplings $g,g'=0$, the $\tro$ and
$\tom$ decay as
\bea\label{eq:vt_decays}
\tro &\ra& \Pi_T \Pi_T = \cos^2 \chi\ts (\tpi\tpi) + 2\sin\chi\ts\cos\chi
\ts (W_L\tpi) + \sin^2 \chi \ts (W_L W_L) \ts; \nn 
\\ \nn \\
\tom &\ra& \Pi_T \Pi_T \Pi_T = \cos^3 \chi \ts (\tpi\tpi\tpi) + \cdots \ts.
\eea
The $\tro$ decay amplitude is
\begin{equation}\label{eq:rhopipi}
\CM(\tro(q) \ra \pi_A(p_1) \pi_B(p_2)) = g_{\tro} \ts \CC_{AB}
\ts \epsilon(q)\cdot(p_1 - p_2) \ts,
\end{equation}
where $\epsilon(q)$ is the $\tro$ polarization vector, $\atro\equiv
g_{\tro}^2/4\pi=2.91(3/\Ntc)$ is scaled naively from QCD (and the
parameter $\Ntc=4$ is used in calculations), and
\begin{equation}\label{eq:ccab}
\ba{ll}
\CC_{AB} &= \left\{\ba{ll} \sin^2\chi  & {\rm for} \ts\ts\ts\ts W_L^+ W_L^-
\ts\ts\ts\ts {\rm or} \ts\ts\ts\ts  W_L^\pm Z_L^0 \\
\sin\chi \cos\chi & {\rm for} \ts\ts\ts\ts W_L^\pm \tpimp\,,
\ts\ts\ts\ts  {\rm or} \ts\ts\ts\ts W_L^\pm \tpiz\,, Z_L^0 \tpipm \\
\cos^2\chi & {\rm for} \ts\ts\ts\ts \tpip\tpim  \ts\ts\ts\ts {\rm or}
\ts\ts \ts\ts \tpipm\tpiz \ts.
\ea \right.
\ea
\end{equation}
The $\tro$ decay rate to two technipions is then (for use in cross
sections, we quote the energy-dependent width for a $\tro$ mass of
$\sqrt{\shat}$):
\begin{equation}\label{eq:trhopipi}
\Gamma(\troz \ra \pi_A^+ \pi_B^-) = \Gamma(\tropm \ra \pi_A^\pm \pi_B^0)
= \frac{2 \atro \CC^2_{AB}}{3} \ts {\ts\ts \frac{p^3}{\shat}} \ts,
\end{equation}
where $p=[(\shat-(M_A+M_B)^2)(\shat-(M_A-M_B)^2)]^\half/2\rshat$ is
the $\tpi$ momentum in the $\tro$ rest frame.


\subsubsection*{4. Sample TCSM Production Rates at the Tevatron}

The $\tro\ra\Pi_T\Pi_T$ decays are strong transitions, therefore we
might expect the $\tro$ to be quite wide.  Almost certainly this is
not so.  The enhanced technifermion condensate in WTC magnifies
technipion masses much more than it does technivector, so the channels
$\tro\ra\tpi\tpi$, $\tom\ra\tpi\tpi\tpi$ and even the
isospin-violating decay $\tom\ra\tpi\tpi$ are likely to be
closed~\cite{Lane:1989ej}.  A $\troz$ of mass $200$~GeV may then
decay mainly to $W_L^\pm\tpi^\mp$ or $W_L^+W_L^-$.  These channels are
also isospin-forbidden for the $\tom$, so all its important decays are
electroweak: $\tom\ra\gamma\tpiz$, $Z^0\tpiz$, $W^\pm\tpimp$, and
$\bar{f}f$ --- especially $e^+e^-$ and $\mu^+\mu^-$.  Here, the $Z$
and $W$ are transversely polarized.\footnote{Strictly speaking, the
identification of $W$ and $Z$ decay products as longitudinal or
transverse is approximate, becoming exact in the limit of very large
$M_{\tro,\tom}$.}  Furthermore, since $\sin^2\chi\ll 1$, the
electroweak decays of $\tro$ to the transverse gauge bosons
$\gamma,W,Z$ plus a technipion may be competitive with the
open-channel strong decays.  Thus, we expect $\tro$ and $\tom$ to be
{\underbar{\em very narrow}}.  For masses accessible at the Tevatron,
it turns out that $\Gamma(\tro)\sim 1$~GeV and $\Gamma(\tom)\simle
0.5$~GeV.

Within the context of the TCSM (and with plausible assumptions for its
parameters), we expect that $\tro^{\pm,0}$ and $\tom$ with masses
below about $250$~GeV should be accessible in Tevatron Run~II in one
channel or another.  Assuming $M_{\rho_T}<2M_{\pi_T}$, the $\tro\ra
W\tpi$ cross sections have rates of a few picobarn.  An example is
shown in Fig.~\ref{TeV4LHC_fig_4} for $M_{\rho_T}= 210$~GeV and
$M_{\pi_T} = 110$~GeV.\footnote{This figure does not include
contributions from transverse weak bosons, which are small for this
choice of parameters.}  The parameter $M_V$ against which these rates
are plotted is described below; it hardly affects them.  These cross
sections were computed with EHLQ structure
functions~\cite{Eichten:1984eu}, and should be multiplied by a
K-factor of about~1.4, typical of Drell-Yan processes such as these.
Searches for these modes at the Tevatron require a leptonic decay of
the $W$ plus two jets with at least one $b$-tag.

\begin{figure}[t]
\vspace{9.0cm}
\includegraphics{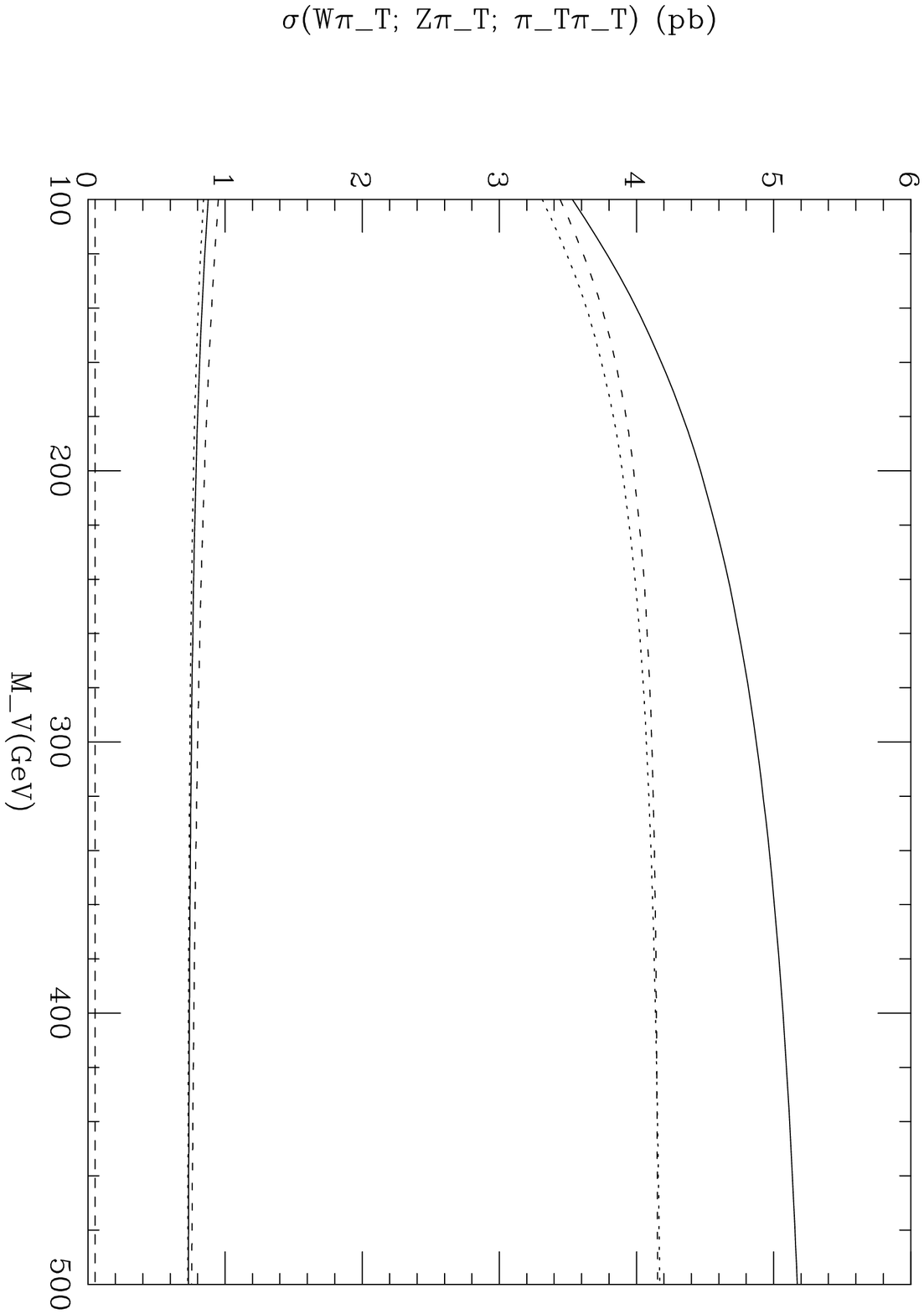}
\caption{Production rates in $p\bar{p}$ collisions at $\ecm=2\,\tev$ 
for $\tom$, $\troz$, $\tropm \ra W \tpi$ (upper curves) and $Z\tpi$
(lower curves) versus $M_V$, for $M_{\tro}=210$~GeV and
$M_{\tom}=200$ (dotted curve), 210 (solid), and $220$~GeV
(short-dashed); $Q_U+Q_D=\frac{5}{3}$ and $M_{\tpi}=110$~GeV.  Also
shown is $\sigma(\tro \ra \tpi\tpi)$ (lowest dashed curve).  From
Ref.~\protect\cite{Lane:1999uh}.}
\label{TeV4LHC_fig_4}
\end{figure}

\begin{figure}[ht!]
\vspace{9.0cm}
\includegraphics{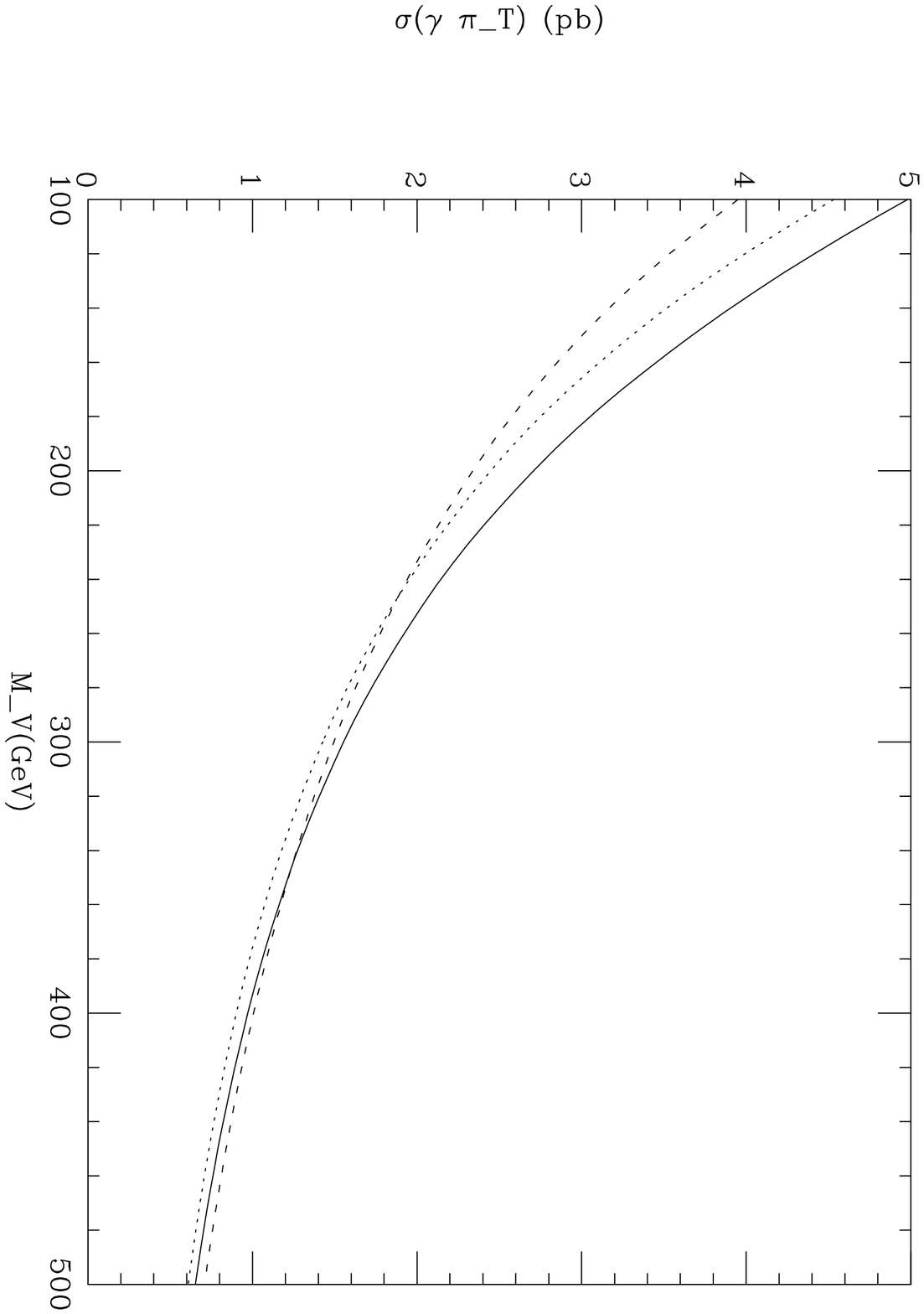}
\caption{Production rates in $p\bar{p}$ collisions at $\ecm=2\,\tev$
for the sum of $\tom$, $\troz$, $\tropm \ra \gamma \tpi$ and
$\gamma\tpipr$ versus $M_V$, for $M_{\tro}=210$~GeV and
$M_{\tom}=200$ (dotted curve), 210 (solid), and $220$~GeV
(short-dashed); $Q_U+Q_D=\frac{5}{3}$, and
$M_{\tpi}=M_{\tpipr}=110$~GeV.  From
Ref.~\protect\cite{Lane:1999uh}.}
\label{TeV4LHC_fig_5}
\end{figure}

The parameter $M_V$ appears inversely in the amplitude for $\tro,\tom
\ra\gamma\tpi$.  It is a typical TC mass-scale and, for low-scale TC,
should lie in the range 100--$500$~GeV. As long as the $\tro\ra
W\tpi$ channels are open, $\gamma\tpiz$ and $\gamma\tpipr$ production
proceeds mainly through the $\tom$ resonance.  Then $M_V$ and the sum
of the technifermion charges, $Q_U + Q_D$, control their rates, which
are approximately proportional to $(Q_U+Q_D)^2/M_V^2$.
Fig.~\ref{TeV4LHC_fig_5} shows the $\gamma\tpi$ cross sections
v.~$M_V$ for the favorable case $Q_U+Q_D=\frac{5}{3}$.  Again, a
K-factor of about 1.4 should be applied.  Here, $M_{\tpipr}=M_{\tpiz}$
and about half the rate is $\gamma\tpipr$.  Note that the $gg$ decays
of the $\tpipr$ will dilute the usefulness of the $b$-tag for these
processes.  On the other hand, decays involving $b$'s have two
$b$-jets.

Finally, for large $M_V$, $\tom$ decays mainly to $\bar{f}f$ pairs.
The most promising modes at the Tevatron (and the LHC) then are
$e^+e^-$ and $\mu^+\mu^-$.  Figs.~\ref{TeV4LHC_fig_6}
and~\ref{TeV4LHC_fig_7} show the effect of changing $M_V$ from~100 to
$500$~GeV on the $e^+e^-$ invariant mass distributions.  Note also
the $\tom$--$\tro$ interference effect when their masses are close.
This would be lovely to observe!  The cross section for
$M_{\tom}=M_{\tro}=210$~GeV, integrated from 200 to $220$~GeV and
including the Drell-Yan background, increases from 0.12 to $0.25\,\pb$
when $M_V$ is increased from 100 to $500$~GeV.  A first search
for $\tom,\tro\ra e^+e^-$ was carried out by D\O\ in Run~I and
published in Ref.~\cite{Abazov:2001qd}.  We look forward to a search
based on Run~II data soon; it shouldn't be difficult to carry out.

\begin{figure}[!ht]
\vspace{9.0cm}
\includegraphics{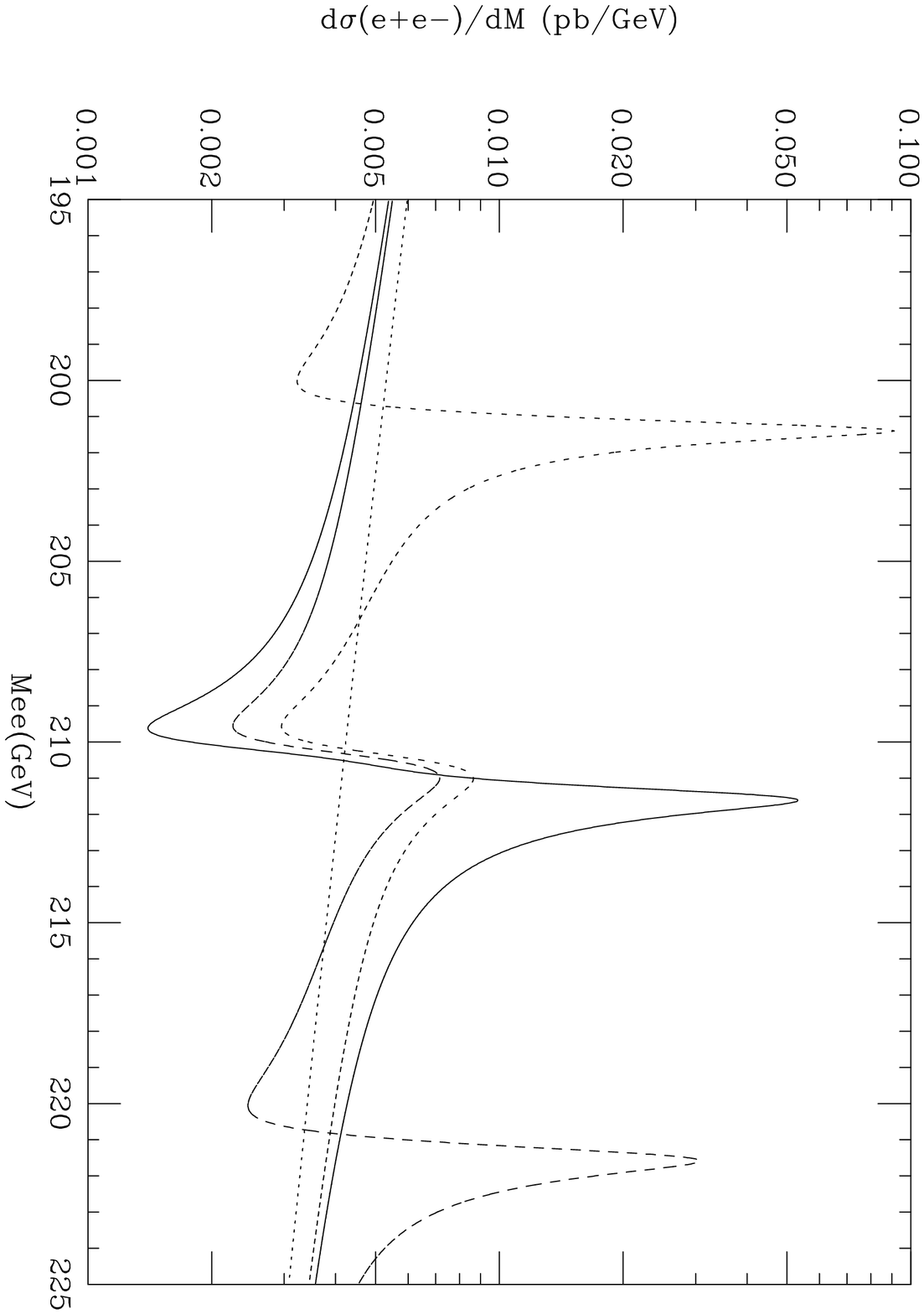}
\caption{Invariant mass distributions in $p\bar{p}$ collisions at 
$\ecm=2\,\tev$ for $\tom$, $\troz\ra e^+e^-$ for $M_{\tro}=210$~GeV
and $M_{\tom}=200$ (short-dashed curve), 210 (solid), and $220$~GeV
(long-dashed); $M_V=100$~GeV.  The Standard Model background is the
sloping dotted line.  $Q_U+Q_D=\frac{5}{3}$ and $M_{\tpi}=110$~GeV.
From Ref.~\protect\cite{Lane:1999uh}.}
\label{TeV4LHC_fig_6}
\end{figure}

\begin{figure}[!ht]
\vspace{9.0cm}
\includegraphics{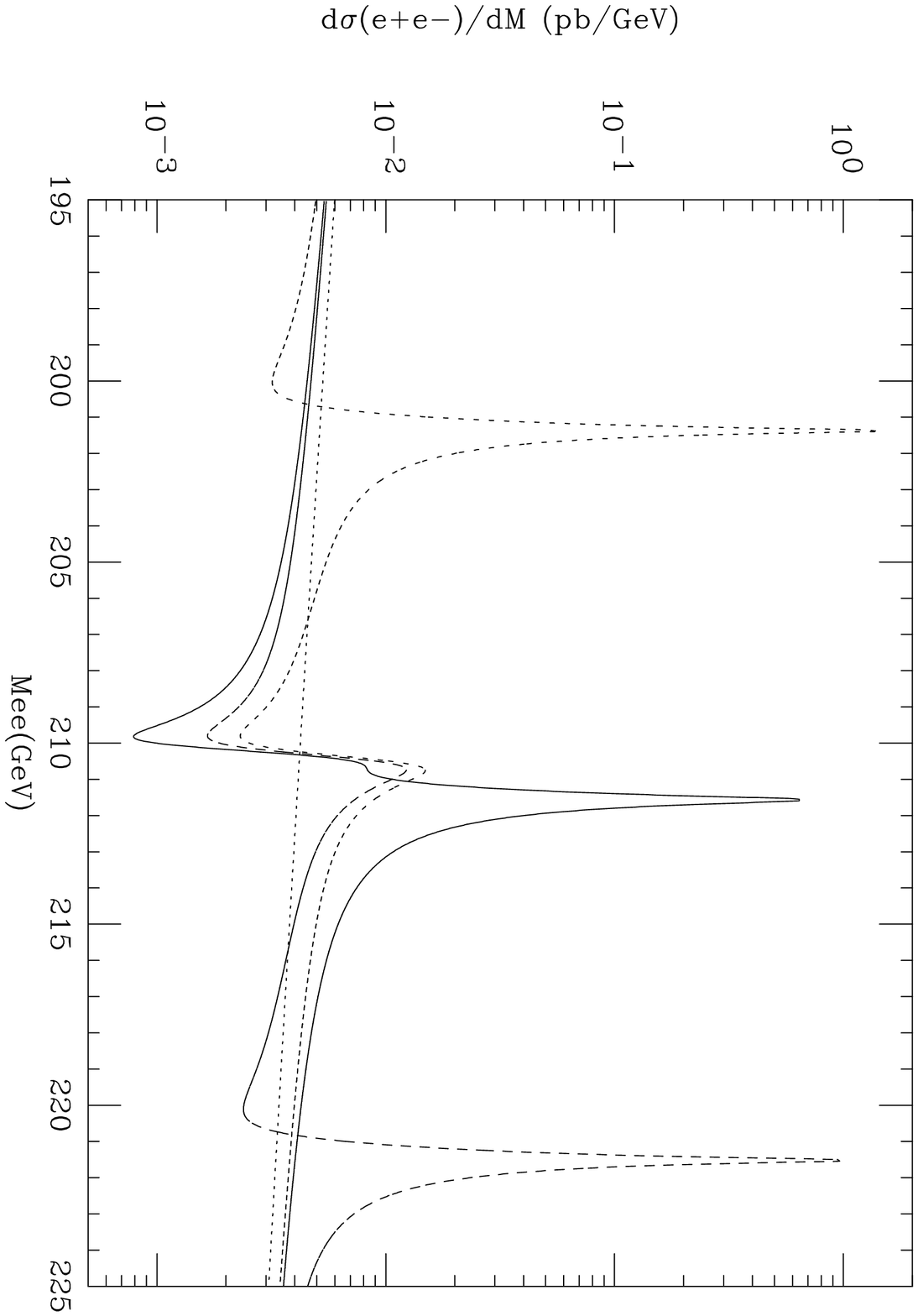}
\caption{Invariant mass distributions in $p\bar{p}$ collisions at 
$\ecm=2\,\tev$ for $\tom$, $\troz\ra e^+e^-$ for $M_{\tro}=210$~GeV
and $M_{\tom}=200$ (short-dashed curve), 210 (solid), and $220$~GeV
(long-dashed); $M_V = 500$~GeV.  The Standard Model background is
the sloping dotted line.  $Q_U+Q_D=\frac{5}{3}$ and $M_{\tpi} =
110$~GeV.  From Ref.~\protect\cite{Lane:1999uh}.}
\label{TeV4LHC_fig_7}
\end{figure}

To sum up, there are nagging little hints of something at $\sim
110$~GeV in dijets with a $b$-tag coming from some parent at $\sim
210$~GeV.  These have been around since Run~I and deserve a closer
look in Run~II.  We urge the Tevatron experimental collaborations to
settle this soon.


\subsubsection*{Acknowledgements}
I am grateful for many conversations with colleagues friends in CDF
and D\O.  I again thank Steve Mrenna for putting the TCSM into {\sc
pythia}.  Thanks also to David Rainwater and Bogdan Dobrescu of the
TeV4LHC Landscapes project for their patience with me. This research
was supported in part by the Department of Energy under
Grant~No.~DE--FG02--91ER40676.

\clearpage\setcounter{equation}{0}\setcounter{figure}{0}\setcounter{table}{0}
\subsection{Slepton Mass Measurements at the LHC}
\label{sec:slep}
\newcommand{\notE}{\not{\hspace{-.05in}E}}

Andreas Birkedal$^1$, Craig Group$^2$ and Konstantin Matchev$^2$ \\

{\noindent\em
$^1$Santa Cruz Institute for Particle Physics\\
$^2$Institute for Fundamental Theory, University of Florida
}\\

{\em The necessity of measuring slepton masses at the LHC is
discussed, emphasizing their importance for cosmology. The possibility
of making slepton mass determinations at the LHC in neutralino decays
is investigated. It is demonstrated that by studying the shape of the
dilepton invariant mass distribution in the decay
$\tilde\chi^0_2\to\tilde\chi^0_1\ell^+\ell^-$, one can determine
whether the slepton is real or virtual.  Furthermore, in case of
virtual sleptons, it is possible to bound the slepton mass within a
limited range.  In this note this method is applied to the special
case of mSUGRA via an approximate LHC detector simulation as a proof
of technique study.}\\


Low-energy supersymmetry remains the best-motivated extension of the
standard model (SM).  The search for superpartners is a prime
objective of the LHC.  Strong production of colored superpartners
(squarks and the gluino) would dominate, and there is an extensive
body of literature devoted to signatures.  In contrast, direct
production of non-colored superpartners (e.g. sleptons) is smaller,
posing a challenge for their
discovery~\cite{delAguila:1990yw,Baer:1993ew,Baer:1995va,Krasnikov:1996np,Bityukov:1997ck,Bityukov:1998va,Andreev:2004qq}.
A recent analysis~\cite{Andreev:2004qq} finds that CMS could discover
right-handed sleptons with mass up to 200~GeV and left-handed sleptons
up to 300~GeV with only 30~fb$^{-1}$ of data.

Supersymmetric theories conserving R-parity also generically contain
viable weakly-interacting massive particle (WIMP) dark matter
candidates.  This is typically is the lightest neutralino,
$\tilde\chi^0_1$, for which the discovery signatures contain missing
transverse energy due to two stable $\tilde\chi^0_1$'s in each event
escaping the detector.  A missing energy signal at the LHC would fuel
the WIMP hypothesis.  However, a missing energy signal at a collider
only implies that particles have been created which are stable on a
timescale characteristic of the detector size.  To prove that the
missing energy particle is indeed viable dark matter, one needs to
calculate its relic abundance.  To this end, one needs to measure all
parameters which enter this calculation.

The relic abundance of a dark matter particle is determined in large
part by its annihilation cross section
$\sigma\left(\chi\chi\to\sum_{i}X_{i}\right)$, where $\chi$ is used to
represent a generic dark matter particle, and $X_{i}$ is any allowed
final state.  The post-WMAP determination of the dark matter abundance
is accurate to about $10\%$~\cite{Spergel:2003cb}.  Assuming a
standard cosmology, one can then deduce a value for the cross section
$\sigma\left(\chi\chi\to\sum_{i}X_{i}\right)$.  This can in turn be
translated into a model-independent prediction for the rates of
$e^+e^-\to\chi\chi\gamma$, $q\bar{q}\to\chi\chi\gamma$, and
$q\bar{q}\to\chi\chi\tilde{g}$ at colliders~\cite{Birkedal:2004xn}.
However, these searches are challenging at both the
ILC~\cite{Birkedal:2004xn} and LHC~\cite{Feng:2005gj}.

In typical models, slepton masses are among the key parameters in
determining whether $\tilde\chi^0_1$ is a good dark matter
candidate~\cite{Drees:1992am,Nihei:2002ij,Birkedal-Hansen:2002sx}.
For example, if the slepton is light, then slepton-mediated
annihilation diagrams are important.  In this case the slepton mass is
required to determine the relic abundance.  Conversely, if the slepton
is heavy, its mass is unimportant for the relic abundance
calculation~\cite{Birkedal-Hansen:2001is,Birkedal-Hansen:2002wd,Binetruy:2003ad,Mizuta:1992ja,Corsetti:2000yq,Bertin:2002sq,Birkedal-Hansen:2002am,Birkedal-Hansen:2003gy,Feng:1999mn,Feng:1999zg,Feng:2000gh,Feng:2000bp}.
But, without a collider measurement of the slepton mass, there may be
significant uncertainty in a relic abundance calculation.

To summarize, the importance of slepton discovery is two-fold.  First,
supersymmetry predicts a superpartner for every standard model
particle.  Therefore, the discovery of the superpartners of the
leptons is an important step in verifying supersymmetry.  Second,
knowledge of slepton masses is {\it always} important for an accurate
determination of the relic abundance of $\tilde\chi_1^0$.

Here we show that the LHC will indeed have sensitivity to slepton
masses, even in the case of heavy sleptons, and describe the details
of how slepton masses can be determined from neutralino decays.  In a
previous note~\cite{Birkedal:2005cm} it was illustrated how this
analysis can be done for the example of minimal supergravity (mSUGRA).
It was shown that the difference between real and virtual sleptons can
clearly be seen.  Establishing the presence of a real slepton in a
cascade decay by the method described is equivalent to a slepton
discovery.  In the case of virtual sleptons, it is possible to limit
the allowed range of their masses with this method, which is
equivalent to a rough indirect measurement of the slepton mass.

The previous analysis did not include backgrounds, detector effects,
or realistic LHC event rates.  The goal of this work is to confirm
that dominant backgrounds can be reduced, find characteristics of the
neutralino decays survive detector effects, and obtain realistic
estimates of the LHC luminosity and event rate will provide suitable
statistics for this study.


\subsubsection{Slepton Phenomenology}

\underline{Sleptons at the LHC}

Direct slepton production suffers from large backgrounds, mostly due
to $W^+W^-$ and $t\bar{t}$ production~\cite{Andreev:2004qq}.  Direct
methods for slepton mass determination available at a linear
colliders, such as threshold scans, are not applicable here.
Fortunately, sleptons would be produced in sizable quantities at the
LHC through cascade decays.  These events can be easily triggered on
and separated from the SM backgrounds.  In principle, these slepton
events present an opportunity for a slepton mass measurement.  A
common situation in supersymmetric models is the hierarchy
$|M_1|<|M_2|<|\mu|$.  In that case, sleptons affect the decay
$\tilde\chi_2^0\to\ell^\pm\tilde{\ell}^\mp\to\ell^\pm\ell^\mp\tilde\chi_1^0$.
The resulting dilepton distribution, in principle, contains
information about the slepton mass $m_{\tilde{\ell}}$.  This situation
is complicated by the fact that $\tilde\chi_2^0$ can also decay
through a real or virtual $Z$: $\tilde\chi_2^0\to
Z\tilde\chi_1^0\to\ell^\pm\ell^\mp\tilde\chi_1^0$.  The Feynman
diagrams for both decay channels are shown in Figure~\ref{Diags}.  In
the next subsection we investigate the process
$\tilde\chi_2^0\to\ell^\pm\ell^\mp\tilde\chi_1^0$ in detail.

\underline{Slepton Masses through Neutralino Decays}

\begin{figure*}[t]
\centering
\includegraphics[width=80mm]{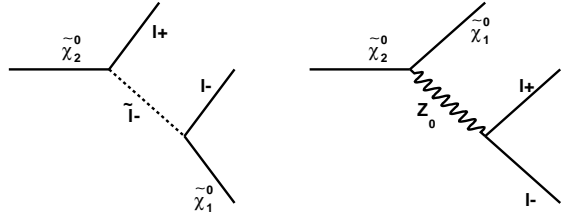}
\caption{Feynman diagrams for
$\tilde\chi_2^0\to\tilde{\ell}^{\pm}\ell^{\mp}\to\ell^\pm\ell^\mp\tilde\chi_1^0$
and $\tilde\chi_2^0\to
Z\tilde\chi_1^0\to\ell^\pm\ell^\mp\tilde\chi_1^0$.}
\label{Diags}
\end{figure*}

What is the observable in these events that is sensitive to the
slepton mass?  We consider the dilepton invariant mass distribution,
$m_{\ell\ell}$, in this analysis.  It is already known that the
endpoint of the $m_{\ell\ell}$ spectrum contains information about the
masses of the real particles involved in the
decay~\cite{Hinchliffe:1996iu}.

\begin{itemize}
\item If the decay occurs through a real $Z$, $\tilde\chi_2^0\to Z\tilde\chi_1^0\to\ell^\pm\ell^\mp\tilde\chi_1^0$, then almost all such events will occur in the $Z$ mass peak, and the endpoint information will be lost.
\item In the case of a virtual intermediate particle ($\tilde\chi_2^0\to Z^*\tilde\chi_1^0\to\ell^\pm\ell^\mp\tilde\chi_1^0$ or $\tilde\chi_2^0\to\tilde{\ell}^{\pm *}\ell^{\mp}\to\ell^\pm\ell^\mp\tilde\chi_1^0$), this process is a three-body decay and the endpoint value is:

\begin{equation}
m_{\ell\ell,max} = m_{\tilde\chi_{2}^{0}} - m_{\tilde\chi_{1}^{0}} \, .
\label{virtual}
\end{equation}

\item Finally, if the decay is through a real slepton ($\tilde\chi_2^0\to\ell^\pm\tilde{\ell}^\mp\to\ell^\pm\ell^\mp\tilde\chi_1^0$), the endpoint is at:

\begin{equation}
m_{\ell\ell,max} =
\sqrt{\frac{\left(m_{\tilde\chi_{2}^{0}}^2-m_{\tilde{\ell}}^2\right)\left(m_{\tilde{\ell}}^2
- m_{\tilde\chi_{1}^{0}}^2\right)}{m_{\tilde{\ell}}^2}}.
\label{real}
\end{equation}

\end{itemize}

The endpoint can be measured; however, its interpretation is ambiguous
since it is not known {\it a priori} which formula is applicable
(Eqn.~\ref{virtual} or Eqn.~\ref{real}).  More information is
contained in the $m_{\ell\ell}$ distribution than just the value of
the endpoint.  One would expect the shapes of the $Z$ and
$\tilde{\ell}$ mediated distributions to be different.  Furthermore,
the shape of the total decay distribution (including both $Z$ and
$\tilde{\ell}$ contributions) changes as a function of the slepton
mass.  The slepton mass dependence is illustrated in
Fig.~\ref{InvMassDists}, where the dilepton invariant mass
distribution resulting from the interference of the $Z$ and
$\tilde{e}_R$-mediated diagrams is shown.  Since the kinematic
endpoint is kept fixed, this illustrates that the endpoint analysis is
largely insensitive to the slepton mass.  In all four cases, the
slepton is virtual, but there exists a clear difference in the shape
of the distribution.  This implies that virtual slepton masses can be
determined by studying the shape of the decay distributions.  In the
case of two-body decay through a real slepton, the $m_{\ell\ell}$
distribution will be triangular~\cite{Miller:2005zp,Kong:2006}.

\begin{figure*}[t]
\centering
\includegraphics[width=80mm]{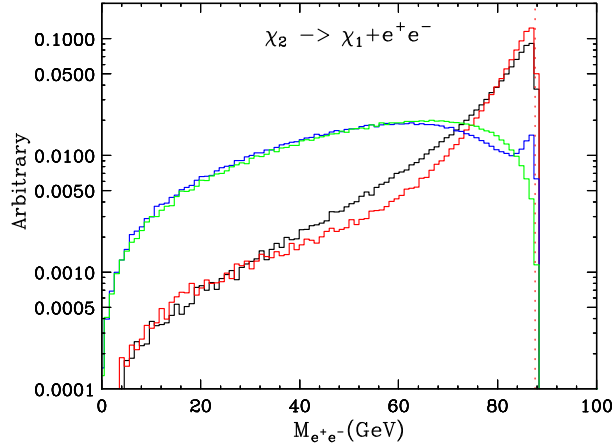}
\caption{$M_{e^+ e^-}$ distributions for different selectron masses.
We consider only the $Z$- and $\tilde{e}_R$-mediated diagrams.  All
parameters are held fixed except for $m_{\tilde{e}_R}$.  The (green,
blue, red, black) line is for a (300, 500, 1000~GeV, and $\infty$)
mass selectron.  The neutralino masses, $m_{\tilde\chi_{1}^{0}}$ and
$m_{\tilde\chi_{2}^{0}}$, are kept constant, and their difference is
88~GeV.}
\label{InvMassDists}
\end{figure*}
%


\subsubsection{Neutralino decay distributions at the LHC}

We asssume in this analysis that LHC experiments have observed the
dilepton mass distribution and have measured a kinematic endpoint at
59~GeV.  What are the implications of this measurement for the SUSY
mass spectrum?  Generally speaking, this reduces the parameter space
by one degree of freedom.  This is illustrated in Fig.~\ref{Points},
where a two-dimensional slice of the mSUGRA parameter space is defined
by fixing $A_0=0$ and $\tan\beta=10$.  The measurement of the
kinematic endpoint reduces the two-dimensional parameter space to
\begin{figure*}[ht!]
\centering
\includegraphics[width=80mm]{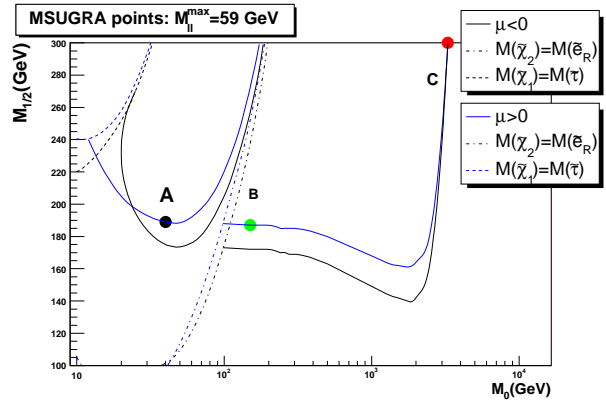}
\caption{Slepton mass determination in a slice of mSUGRA parameter
space with $A_0=0$ and $\tan\beta=10$.  Here $M_0$ ($M_{1/2}$) is the
universal scalar (gaugino) mass parameter.  The effect on mSUGRA
parameter space of fixing the dilepton kinematic endpoint of the
$\chi_{2}^{0}\to e^+e^-\chi_{1}^{0}$ decay to be
$m_{\ell\ell,max}=59$~GeV.}
\label{Points}
\end{figure*}
\begin{figure*}[h!]
\centering
\includegraphics[width=80mm]{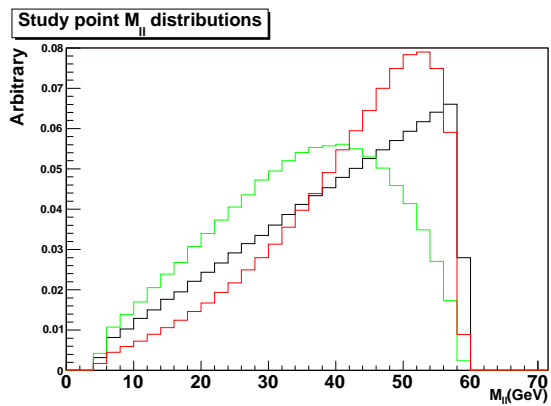}
\caption{Color-coded distributions for the 3 study points from 
Fig.~\ref{Points}.  Points A, B, and C are shown in black, green, and
red respectively. The distributions are normalized to one.  Details
are explained in the text.}
\label{Lines}
\end{figure*}
one-dimensional line segments.  These are the solid lines in
Fig.~\ref{Points}.  In mSUGRA, there is also the binary choice of
$\mu>0$ or $\mu<0$, and their respective results are shown in blue and
black.  The dashed lines in the upper left corner indicate where
$m_{\tilde\chi_{1}^{0}}=m_{\tilde{\tau}_1}$.  Any points to the left
of these lines are ruled out by constraints on charged dark matter.
The dashed-dotted lines running through the middle of the plot
indicate where $m_{\tilde{e}_R}=m_{\tilde\chi_{2}^{0}}$.  This is
where the slepton-mediated neutralino decays change from being
three-body ($\tilde{e}_R$ is virtual to the right of these lines) to
two-body ($\tilde{e}_R$ is real to the left of these lines).  The
three points labeled in Fig.~\ref{Points} are the study points
analyzed for this analysis.  Point A is a low-mass SUSY point in which
$m_{\tilde{l}_R}<m_{\tilde\chi_{2}^{0}}$, so that the decay channel
through a real slepton is open.  Point B is also a low mass point,
however $m_{\tilde{l}_R}>m_{\tilde\chi_{2}^{0}}$ and therefore the
real decay channel is closed.  Finally, point C is a high-mass SUSY
point which can decay only through virtual slepton or Z boson
channels.  This difference in the SUSY scale of point C is obvious in
the rate of production shown in the Table~\ref{Sigma}.

The plot in Fig.~\ref{Lines} shows the invariant mass distribution
expected from $\tilde\chi_{2}^{0}$ decays for the the three study
points described above.  These distributions are normalized to one so
that their shapes may be compared.  The black line displays the
triangular shape of $\tilde\chi_{2}^{0}$ decay through a real slepton.
Also, note the extreme difference in shape between the two virtual
decays (red and green lines).  ISAJET 7.69~\cite{Paige:2003mg} is
employed for this and all other Monte Carlo-generated results in the
analysis.

\underline{Event Rate}

Depending on the SUSY model, and the particular point in its parameter
space, SUSY events expected during the life of the LHC can vary by
many orders of magnitude.  The mSUGRA production cross sections, as
well as the number of events expected for 10~fb$^{-1}$ for the three
study points for this analysis and dominant SM background ($t\bar{t}$
production) can be found in Table~\ref{Sigma}.

\begin{table}[htb]
\begin{center}
\begin{tabular}{|l|l|l|l|l|l|}
\hline
Point                     &  $M_{0}$   &  $M_{\frac{1}{2}}$ & $M_{\tilde{\ell}}$& $\sigma$& N(10fb$^{-1})$ \\ \hline
A                         & 40GeV      &189 GeV             & 92 GeV            &  170 pb  & $1.7*10^{6}$  \\ \hline
B                         & 150GeV     & 187 GeV            & 96 GeV            & 150 pb   & $1.5*10^{6}$   \\ \hline
C                         & 3280GeV    & 300 GeV            & 3277 GeV          & 4.4 pb   & 44,000         \\ \hline\hline
$t\bar{t}$ (SM background)&  NA        &  NA                &  NA               & 425 pb    & $4.25*10^{6}$ \\ \hline
\end{tabular}
\caption{Event rates for mSUGRA study points at the LHC.  All SUSY 
points have $\mu>0$, $A_{0}=0$ and $tan(\beta)=10$.  In the second and
third columns we show the point in the $(M_{0},M_{\frac{1}{2}})$
plane.  The forth column contains the mass of the lightest slepton.
The total inclusive SUSY cross section as well as number of events
produced for 10~fb$^{-1}$ of integrated luminosity is also included.}
\label{Sigma}
\end{center}
\end{table}

\underline{ID efficiencies, jet clustering, and energy smearing}

As mentioned above, it was shown in a previous work that with
reasonable statistics and a perfect detector it is possible to
discriminate quite well between the regions of parameter space based
on the shape of $m_{\ell\ell}$ distributions~\cite{Birkedal:2005cm}.
Here, major detector effects are included to study their affect on
these distributions.

We include particle-level jet clustering, lepton ID efficiencies, and
smearing parameterizations for the muon, electron, photon, and jets.
A basic cone algorithm ($R_{cone}=0.7$) was used to combine all stable
hadronically-interacting particles into jets.  The jet energy was then
conservatively smeared according to $\frac{\Delta
E}{E}=\frac{120\%}{\sqrt{E}}+7\%$, where E is in units of GeV.  It
should be noted that jets are used in this anlysis only as a tool to
cut out the SM background.  Their definitions affect only the signal
multiplicity and the signal to background ratio.  Jet characteristics
do not affect the shapes of the distributions studied.  A $90\%$ ID
efficiency was used for leptons.  Photons and electrons were smeared
according to $\frac{\Delta E}{E}=\frac{5\%}{\sqrt{E}}+0.5\%$, with E
given in GeV.  Muons with $|\eta|<1$ were smeared according to
$\frac{\Delta P}{P}=.01\%P+1\%$ while muons with $|\eta|>1$ were
smeared by $\frac{\Delta P}{P}=.04\%P+2\%$, where P is in units of
GeV.  Missing transverse energy is calculated by taking the magnitude
of the vector $E_T$ sum of all leptons, photons, and jets
reconstructed in the event.

\underline{Backgrounds and kinematic cuts}

\begin{table}[htb]
\begin{center}
\begin{tabular}{|l|l|}
\hline
Variable                     &  Cuts   \\ \hline
$N_{jets}$(P$_T>$50GeV, $|\eta|<$2)      &  4       \\ \hline
$N_{jets}$(P$_T>$100GeV, $|\eta|<$2)      & 1        \\ \hline
$\notE _T$                & max(100GeV,.2$M_{eff}$)  \\ \hline
$N_{leps}$(P$_T>$20GeV, $|\eta|<$2.5)      & 2 (opposite sign)             \\ \hline
\end{tabular}
\caption{Base Cuts}
\label{Cuts}
\end{center}
\end{table}

In this analysis, there are two main types of backgrounds to be
considered: those which come from the Standard Model, and those which
arise from SUSY processes.  In general, any SM or SUSY process which
contains opposite-sign lepton pairs in the final state must be
considered.  After applying the standard set of cuts shown in
Table~\ref{Cuts}, $t\bar{t}\to W^{+}W^{-}$ is the dominant SM
background.  The events surviving these base cuts as a function of
$M_{\rm{eff}}\equiv\notE_{T}+\sum_{i=1}^{4}P^{Jet}_{T_{i}}$ with
10~fb$^{-1}$ of data are shown in Fig.~\ref{MEFF}.  The total number
of events surviving the base cuts are included in the legend for each
sample.  The low-mass slepton points (A and B) statistically dominate
the background, while the the high-mass slepton point (point C)
clearly needs more than 10~fb$^{-1}$ for any statistical study.  A cut
on $M_{\rm{eff}}$ has not been included in the analysis but could be
used to increase the significance.  More luminosity as well as an
optimized cut on $M_{\rm{eff}}$ will be necessary to apply this
analysis to high slepton mass points (such as point C).

\begin{figure*}[t]
\centering
\includegraphics[width=80mm]{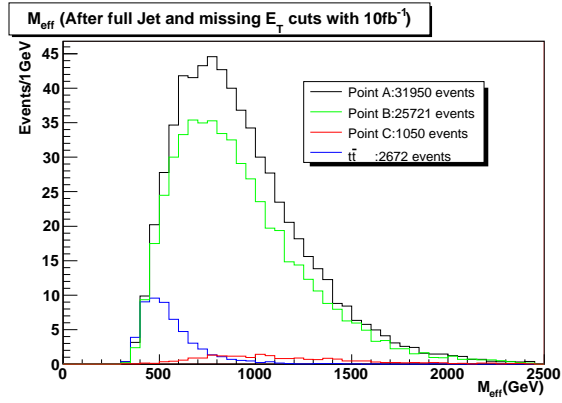}
\caption{The number of events per GeV surviving the base cuts of
Table~\ref{Cuts} versus $M_{eff}$ for the low-mass virtual decay, real
decay, and the high mass-points, as well as the t$\bar{t}$ background.
The total number of events surviving the cuts appears in the legend
for each sample.}
\label{MEFF}
\end{figure*}

The cuts in Table~\ref{Cuts} are designed to cut out SM events so
exotic physics can be studied.  However, lepton pairs that come from
unrelated SUSY decays in the cascade are indistinguishable from the
pairs which originate from a single neutralino decay.  These SUSY
background events will pass the base cuts with high efficiency,
therefore they must be dealt with in a different manner than the
Standard Model background.

\underline{Background subtraction}

Both SUSY and SM backgrounds are uncorrelated in the sense that the
opposite sign leptons do not originate from the decay of a single
parent particle (except for the $Z\to\ell ^{+}\ell ^{-}$ decays).  A
\emph{subtracted distribution} will be used to reduce both of these
uncorrelated backgrounds.  The idea is that $\mu^{\pm}e^{\mp}$
distributions will have the same rate and distribution as the
uncorrelated e$^\pm$e$^\mp$ and $\mu^{\pm}\mu^{\mp}$
distributions~\cite{Bartl:1996dr}.  Thus, distributions such as:
\begin{equation}
 \left.{\frac{d\sigma}{dM}}\right\vert_{\rm sub} =
 \left.{\frac{d\sigma}{dM}}\right\vert_{e^+e^-}
+\left.{\frac{d\sigma}{dM}}\right\vert_{\mu^+\mu^-}
-\left.{\frac{d\sigma}{dM}}\right\vert_{e^+\mu^-}
-\left.{\frac{d\sigma}{dM}}\right\vert_{e^-\mu^+}\
\label{sub}
\end{equation}
will be independent of these uncorrelated backgrounds. 

\underline{Results}

The goal of this work is to extract information from the shape of the
invariant mass distributions.  An immediate concern is whether or not
this subtraction method preserves the shape of the distribution.

Fig.~\ref{Sub_dist} puts this concern to rest.  The reconstructed
invariant mass distributions for all opposite-sign electron pairs
added to the same distribution for opposite-sign muon pairs is shown
in solid black.  This distribution contains signal events as well as
SUSY background.  In solid blue, the opposite-sign invariant mass
distributions of opposite+sign electron--muon mixed pairs are shown.
This distribution should be similar to the opposite-sign same-flavor
SUSY background as long as $M_{\tilde{\mu}}\simeq M_{\tilde{e}}$.  The
dotted black line is the subtracted distribution defined in
Eq.~\ref{sub} (solid black minus solid blue).  This distribution
should be independent of the SUSY background and represents
experimental results after background subtraction with 10~fb$^{-1}$ of
LHC data.  The actual decay distribution (template distribution) for
the SUSY point is shown in red.  This template represents the
theoretical distribution expected without any cuts, smearing, or
misidentification.  The template distribution is normalized to the
subtracted distribution over the range 0--60~GeV.  The subtracted
distribution of Fig.~\ref{Sub_dist} matches the shape of the template
quite well for the two low-mass points: A (left) and B (center).
Finally, shown on the right of Fig.~\ref{Sub_dist} is the invariant
mass distribution for the $t\bar{t}$ background.  The subtraction
method effectively reduces this background to zero (within statistical
fluctuations).

\begin{figure*}[t]
\centering
\includegraphics[width=160mm]{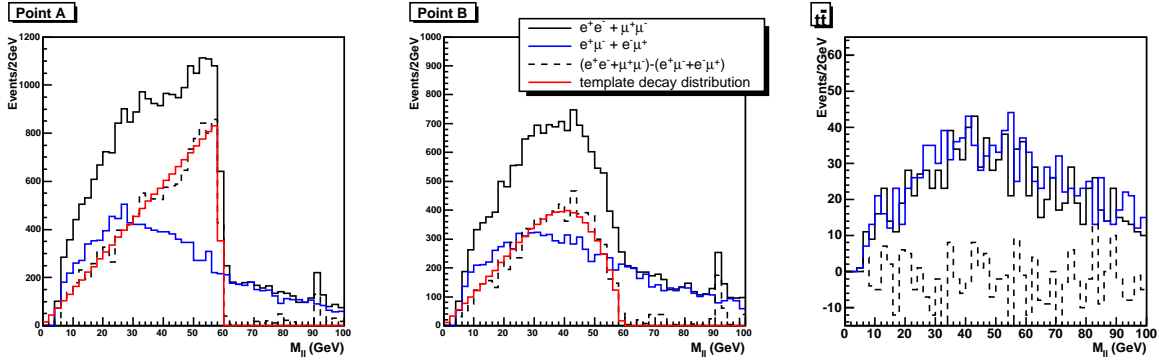}
\caption{The plots show the various lepton pair distributions, as 
well as, the subtracted and template distributions for points A and B
and the $t\bar{t}$ background.  For points A and B, the template
distributions(red line) match the subtracted distributions (black
dashed line) quite well.  See the text for more information.}
\label{Sub_dist}
\end{figure*}

Several conclusions may be drawn from these results:         
\begin{itemize}
\item The subtraction method does not distort the shape of the invariant mass distribution of the decay products of the $\tilde\chi_2^0$, or any other correlated lepton pair contribution (notice that the $Z$ peak survives).   
\item Smearing effects from the detector do not distort the shape of the invariant mass distribution of the decay products of the $\tilde\chi_2^0$.
\item With as little at 10~fb$^{-1}$ of integrated luminosity, an analysis based on shapes of lepton pairs' invariant mass distributions will be possible in the case of a light slepton at the LHC (points A and B). 
\item The subtraction method described above is relevant for both standard model and SUSY background subtractions.
\end{itemize} 

As stated above, if the slepton is heavy (point C), it is clear from
Fig.~\ref{MEFF} that it is not possible to do a shape analysis without
more luminosity and a further optimization of cuts.  Therefore, this
point is not included in Fig.~\ref{Sub_dist}.


\subsubsection{Conclusions}

In this report we addressed the importance of measuring slepton masses
at the LHC.  An inability to significantly bound the slepton masses
would introduce large uncertainties into any subsequent calculation of
the neutralino dark matter abundance.  Dilepton invariant mass
distributions from neutralino decays were identified as one avenue for
determining slepton masses at the LHC.  We investigated the decay
$\tilde\chi_{2}^{0}\to e^+e^-\tilde\chi_{1}^{0}$ for a specific value
of the dilepton kinematic endpoin, performing an analysis in the
mSUGRA paradigm with $A_0=0$ and $\tan\beta=10$ -- though this
analysis is clearly extendible to more general theories.  In a
previous note it was shown that whether the intermediate slepton is
real or virtual can be determined based on the shape of the lepton
pair invariant mass distribution.  This provides one clean bound on
the slepton mass.  In the case of light virtual sleptons, one can
place significant lower and upper bounds on the slepton mass.  For
very heavy virtual sleptons, only a lower bound can be placed.
However, this bound is generally above 1~TeV, except for the case of
cancellation between $Z$ and slepton diagrams with $\mu<0$.  This work
extends the previous analysis to include detector affects, dominant
backgrounds, and realistic event rates for the LHC.  We conclude that
statistics will be reasonable for studies if the slepton is light
(points A and B) with as little at 10~fb$^{-1}$ of integrated
luminosity.  Furthermore, the shape of the lepton invariant mass
distribution is not distorted due to the background subtraction
methods described or major detector effects included in the
simulation.  This result adds reassurance that constraints on the
slepton mass based on the shape of invariant mass distributions from
neutralino decays will be a useful technique at the LHC.

Further studies are needed to determine mass constraints based on the
results including detector simulation and background subtraction.  In
addition, more effort will extend this method to the general MSSM.
Furthermore, the exact extent to which this measurement assists the
determination of the neutralino dark matter density needs to be
quantified.


\clearpage\setcounter{equation}{0}\setcounter{figure}{0}\setcounter{table}{0}
\subsection{Light stop searches}
\label{sec:lstop1}

{\em Sabine Kraml$^1$ and Are R. Raklev$^{1,2}$ \\
$^1$ CERN, Geneva, Switzerland\\
$^2$ University of Bergen, Norway}\\

{\em If searches at the Tevatron find an excess in $c\bar
c\!\not{\!\!E}_T$ events, this will hint at a light stop with
$m_{\tilde t_1}\lsim m_t$, decaying into $c\tilde\chi^0_1$.  The
nature of this excess may be confirmed at the LHC using the signature
of 2$b$-jets + 2 same-sign letpons + jets + $\not{\!\!E}_T$, stemming
from gluino-pair production followed by gluino decays into top and
stop.}


\subsubsection{Introduction}

Within the MSSM, electroweak baryogenesis motivates a very light
$\tilde{t}_1$ with $m_{\tilde{t}_1}\lsim m_t$
\cite{Delepine:1996vn,Carena:1997ki,Cline:1998hy,Balazs:2004bu}; see
also Sec.~\ref{sec:lstop2}.  The Tevatron reach for such a light stop
was studied in Ref.~\cite{Demina:1999ty}.  It was found that if the
$\tilde{t}_1$ decays into $c\tilde\chi^0_1$, giving a signature of
$c\bar{c}+{\not\!\!E_T}$, the Tevatron can cover the
baryogenesis-motivated region with 4~fb$^{-1}$ of integrated
luminosity provided the $\tilde{t}_1$--$\tilde\chi^0_1$ mass
difference is large enough, that is $\gsim 30$~GeV, see
Fig.~\ref{fig:tev_reach}.  For smaller mass differences, especially in
the stop coannihilation region where
$m_{\tilde{t}_1}-m_{\tilde\chi^0_1}\lsim 25$~GeV, the $c$-jets are too
soft and hence the number of events not significant enough for a
discovery.

Should an excess of $c$-jets plus missing energy events be observed at
the Tevatron, this will trigger dedicated searches for light stops at
the LHC.  Although stop pair production has a large cross section in
the interesting mass range (see Table~\ref{tab:stopxs}) the signal
will be buried in the background at the LHC.  Moreover, an
interpretation of $c\bar{c}\!\not{\!\!E}_T$ as a signal of light stops
is equivocal, and gives only weak bounds on the stop mass.  The
alternative at the LHC is to exploit gluino pair production with the
gluinos decaying into top and stop as proposed in
Ref.~\cite{Kraml:2005kb,Allanach:2006fy}: since gluinos are Majorana
particles, they can decay either into $t\tilde{t}_1^*$ or
$\bar{t}\tilde{t}_1$.  Pair-produced gluinos therefore give same-sign
top quarks in half of the gluino-to-stop decays.  If the stops decay
into $c\tilde\chi^0_1$ and the $W$'s from the $t\to bW$ decay
leptonically, we obtain
\begin{equation}
    pp \to \tilde g\tilde g\to 
    bb\,l^+l^+\: ({\rm or}\: \bar b\bar b\, l^-l^-) 
    + {\rm jets\:} + \not\!\!E_T\,.
\label{eq:bbllsignature}
\end{equation}
This peculiar signature has little background and could be used to
prove that the Tevatron excess of $c\bar{c}\!\not{\!\!E}_T$ indeed
originated from $\tilde{t}_1^{}\tilde{t}_1^*$ production.

\begin{table}
\begin{center}
\begin{tabular}{l|rrrrrrr}
\hline
$m_{\tilde{t}_1}$ [GeV] & \quad 120 & \quad 130 & \quad 140 & \quad 150 
                        & \quad 160 & \quad 170 & \quad 180 \\
\hline
$\sigma(\tilde{t}_1^{}\tilde{t}_1^*)$, Tevatron 
                        & 5.43 & 3.44 & 2.25 & 1.50 & 1.02 & 0.71 & 0.50 \\
$\sigma(\tilde{t}_1^{}\tilde{t}_1^*)$, LHC      
                        & 757 & 532 & 382 & 280 & 209 & 158 & 121 \\
\hline
\end{tabular}
\end{center}
\caption{NLO cross sections in pb for $\tilde{t}_1$ pair-production 
at the Tevatron and the LHC, computed with {\sc
Prospino2}~\protect\cite{Beenakker:1996ed}.
\label{tab:stopxs}}
\end{table}

%
\begin{figure}[t!]
\centerline{\psfig{file=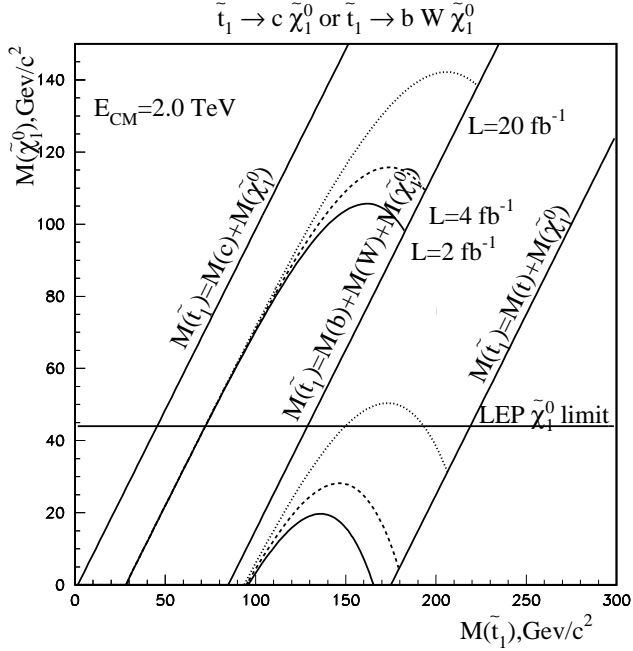,height=9cm}}
\caption{Tevatron reach for a light stop, from 
Ref.~\protect\cite{Demina:1999ty}.}
\label{fig:tev_reach}
\end{figure}


\subsubsection{Simulation and Results}

To demonstrate the use of the signature in
Eq.~(\ref{eq:bbllsignature}), we performed a case study for the LST1
benchmark point with $m_{\tilde{g}}=660$~GeV,
$m_{\tilde{t}_1}=150$~GeV, and $m_{\tilde\chi^0_1}=105$~GeV.  All
other squarks (in particular the sbottoms) are taken to be heavier
than the gluinos.  This suppresses the SUSY background, and gluinos
decay to $100\%$ into $t\tilde{t}_1$.  Sleptons are also assumed to be
fairly heavy, $m_{\tilde l}\sim 250$~GeV.  A neutralino relic density
within the WMAP bound is achieved by annihilation through a Higgs for
$m_A=250$~GeV.  Assuming BR$(\tilde{t}_1\to c\tilde\chi^0_1)\simeq 1$,
experiments at the Tevatron should see a clear excess in the
$c\bar{c}+{\not\!\!E_T}$ channel for this
scenario~\cite{Demina:1999ty}.

At the LHC, gluino pair production has a cross section of 5.4~pb at
LST1.  We generated events equivalent to 30~fb$^{-1}$ of integrated
luminosity with PYTHIA~6.321\cite{Sjostrand:2000wi}.  The events were
run through the detector simulation program
AcerDET~1.0~\cite{Richter-Was:2002ch} to simulate a generic LHC
detector.  In the SM background we included $t\bar{t}$, $W$+jet,
$Z$+jet, $WW/WZ/ZZ$ and QCD $2\rightarrow 2$ events, assuming that
FCNCs are too small to lead to significant same-sign top production.
Other sources of SM background were found to be negligible; see
Ref.~\cite{Kraml:2005kb} for details on the simulation and the LST1
benchmark point.  We applied the following cuts to isolate the signal:
\begin{itemize}
\item
Two same-sign leptons ($e$ or $\mu$) with $p^{\mathrm{lep}}_T>20$~GeV.
\item
At least four jets with $p^{\mathrm{jet}}_T>50$~GeV, at least two of
which are $b$-tagged.\footnote{We assume a $b$-tagging efficiency of
$43\%$.  Light-jet rejection is set according to the $p_T$
parametrization for a low luminosity environment, given in
Ref.~\cite{unknown:1997fs}.}
\item
Missing transverse energy $\not{\!\!E}_T > 100$~GeV.
\item
Two combinations of the two hardest leptons and $b$-jets with
$m_{bl}<160$~GeV.
\end{itemize}
The effects of these cuts are summarized in Table~\ref{tab:cut_eff}:
``2lep 4jet'' is the cut on two leptons and four jets; ``2b'' the
requirement of two $b$-jets; ``$\not{\!\!\!E}_T$'' the cut on missing
transverse energy and ``SS'' the requirement of two same-sign leptons.
Note the central importance of the same-sign cut in removing the SM
background, which at that point consists only of $t\bar{t}$ events.
The cuts on transverse momentum and invariant mass ``2$t$'' were used
to further reduce the background.  We find that the signature of
Eq.~(\ref{eq:bbllsignature}) is easily separated from both SM and SUSY
backgrounds.

\begin{table}\begin{center}
\begin{tabular}{l|rrrrrrr}
\hline
Cut & 2lep 4jet & $p_T^{\mathrm{lep}}$ & $p_T^{\mathrm{jet}}$ 
    & 2$b$ & $\not{\!\!E}_T$  & 2$t$ & SS \\
\hline
$\tilde{g}\tilde{g}$ signal & 10839 & 6317 &  4158 &  960 &  806 &  628 & 330 \\
SUSY bkgd                   &  1406 &  778 &   236 &   40 &   33 &   16 &   5 \\
SM   bkgd                   & 25.3M & 1.3M & 35977 & 4809 & 1787 & 1653 &  12 \\
\hline
\end{tabular}\end{center}
\caption{Number of events left for 30~fb$^{-1}$ of data after each
stage of cuts.
\label{tab:cut_eff}}
\end{table}

Isolating this same-sign top signature at the LHC would strengthen the
interpretation of the signal observed at the Tevatron.  The next aim
would be to measure the masses of the newly discovered particles.
With the missing energy and momentum of the neutralino, reconstruction
of a mass peak would be impossible. The well-studied alternative to
this~\cite{Hinchliffe:1996iu,Bachacou:1999zb,Allanach:2000kt,Lester:2001zx,Gjelsten:2004ki},
is to use the SM decay products' invariant-mass distributions.  Their
endpoints can be given in terms of the SUSY masses, and these
equations can then in principle be solved to give the masses.
However, among the four possible endpoints, one is simply a
relationship between SM masses, and two are linearly dependent, so
that we are left with three unknown masses and only two equations.
Also, because of the information lost with the escaping neutrino, the
distributions of interest all fall very gradually to zero.
Determining exact endpoints in the presence of background, taking into
account smearing from the detector, etc., would be be very difficult.

%
\begin{figure}[t!]
\centerline{\epsfig{file=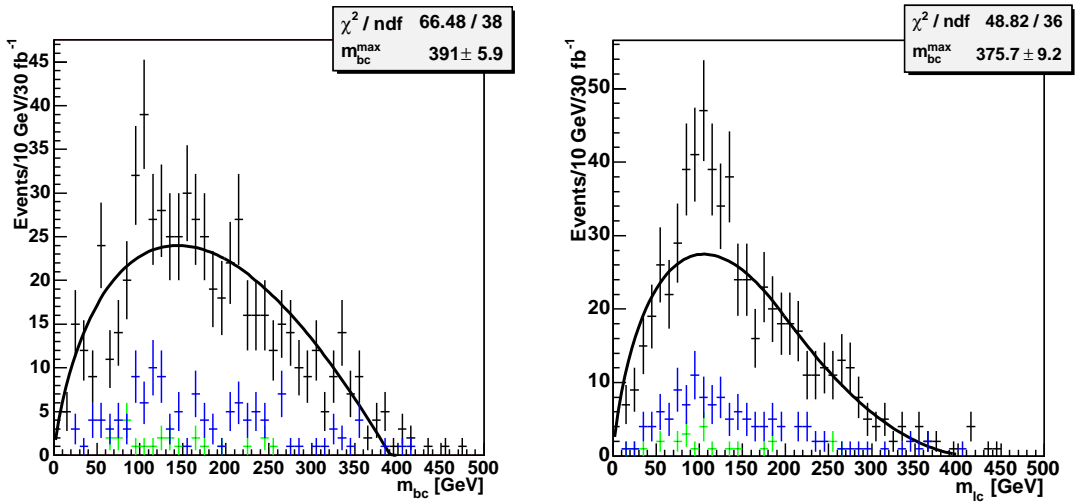,height=7cm}}
\caption{Invariant-mass distributions $m_{bc}$ (left) and $m_{lc}$
(right) for LST1, together with fits of the calculated
distributions. Also shown are the contributions from the SM background
(green) and the SUSY background (blue).}
\label{fig:im_ctag}
\end{figure}

We attacked this problem with an extension of the endpoint method,
deriving the complete shapes of the invariant-mass distributions for
$m_{bc}$ and $m_{lc}$; for details see Ref.~\cite{Kraml:2005kb}.
Fitting to the whole invariant mass distribution greatly reduces the
uncertainty involved in endpoint determination, and may give
additional information on the masses.  Extending this method to
include spin effects propounds the possibility of comfirming the
scalar nature of the stop.  Fitting to the $m_{bc}$ and $m_{lc}$
distributions can in principle be used to determine both of the two
linearly independent parameters
\begin{equation}
(m_{bc}^{\max})^2=\frac{(m_t^2-m_W^2)(m_{\tilde{t}_1}^2-m_{\tilde{\chi}_1^0}^2)
(m_1^2+m_2^2)}{2m_t^2m_{\tilde{t}_1}^2} \quad \textrm{and} 
\quad a=\frac{m_2^2}{m_1^2},
\end{equation}
where
\begin{equation}
m_1^2=m_{\tilde{g}}^2-m_t^2-m_{\tilde{t}_1}^2 \quad \textrm{and} 
\quad m_2^4=m_1^4-4m_t^2m_{\tilde{t}_1}^2.
\end{equation}
For light stops, models typically have $m_tm_{\tilde{t}_1}\ll
m_{\tilde{g}}^2$ and hence $a\approx 1$.  The distributions are
sensitive to such values only at very low invariant masses, so that
$a$ cannot be determined in our case.  We shoe the $m_{bc}$ and
$m_{lc}$ distributions for LST1, and the fits to them, in
Fig.~\ref{fig:im_ctag}.

In fitting the $m_{bc}$ and $m_{lc}$ distributions, the $b$-jets and
leptons are paired through the cut on invariant mass.  In some events
the $W$ decays to a tau, which in turn decays leptonically; these
events are an additional, irreducible background to our distributions.
The likelihoods in the $b$-tagging routine should help to discriminate
$c$-jets from other jets.  We assume a $20\%$ probability of
identifying a $c$-jet directly from the $b$-tagging likelihood.  For
events where one or both $c$-jets are missed, they are chosen as the
two hardest remaining jets with $p_T^{\mathrm{jet}}<100$~GeV.  This
upper bound is applied to avoid picking jets from the decay of heavy
squarks.  Note that the $c$-jets are expected to be relatively soft if
our signal exists and, depending on the
$\tilde{t}_1$--$\tilde\chi^0_1$ mass difference, the final results are
somewhat sensitive to the exact value of this cut.  Information from
the Tevatron on the kinematic distribution of the excess $c$-jets can
hence be helpful for determining the appropriate value.  Finally, our
$c$-jet candidates are paired to the top candidates by their angular
separation, and by requiring consistency with the endpoints of the
invariant-mass distributions we are not looking at.  The precision of
our mass determination is limited by systematics from these issues and
others that affect the distributions, such as final-state radiation,
finite-width effects and cuts.\footnote{For a more thorough discussion
of these issues, details on deriving invariant-mass distributions in
cascade decays, and the possible inclusion of spin effects, see
Ref.~\cite{Miller:2005zp}.}

The combined result of the two distributions, with statistical error,
is $m_{bc}^{\max}=389.8\pm5.3$~GeV, which compares well to the nominal
value of $391.1$~GeV.  However, the quality of the fits are rather
low, with large $\chi^2$ values, and the two separate results lie on
either side of the nominal value, fortuitously cancelling, indicating
that the systematic errors can be significant.  All in all, additional
information, e.g.\ a measurement of the effective-mass scale of
events, would be necessary to determine the masses of the SUSY
particles involved, in particular the mass of the light stop.

Finally, we want to comment on the robustnest of our method.  We
checked that the signal of Eq.~(\ref{eq:bbllsignature}) remains
significant enough for a $5\sigma$ discovery for gluino masses up to
$m_{\tilde g}\sim 900$~GeV and for sbottom masses lighter than the
gluino.  We also checked that lowering the stop mass to
$m_{\tilde{t}_1}=120$~GeV does not considerably reduce the
significance of the signal.  This implies that the same-sign signature
can be used to search for a light stop even in the stop-coannihilation
region.


\subsubsection{Conclusions}

If experiments at the Tevatron discover a light stop in the channel
$p\bar{p}\to\tilde{t}_1\tilde{t}_1^*\to c\bar{c}+{\not\!\!E_T}$ (or
see a significant excess of $c\bar{c}+{\not\!\!E_T}$ events), this may
be confirmed at the LHC using gluino pair production followed by
gluino decays into top and stop.  The signature of 2 $b$-jets + 2
same-sign leptons + jets + $\not\!\!E_T$ discussed in this
contribution has little background and will help determine whether
what has been discovered is indeed a light scalar top.  The kinematic
distribution of the $c$-jets in the Tevatron signal may be useful for
reducing systematic errors from mistagging at the LHC.


\subsubsection*{Acknowledgements}

S.K.\ is supported by an APART (Austrian Programme of Advanced
Research and Technology) grant of the Austrian Academy of
Sciences.  A.R.R.\ acknowledges support from the Norwegian Research
Council and the European Community through a Marie Curie Fellowship
for Early Stage Researchers Training.

\clearpage\setcounter{equation}{0}\setcounter{figure}{0}\setcounter{table}{0}
\subsection{Tevatron-LHC-ILC synergy: light stops, baryogenesis and dark matter}
\label{sec:lstop2}

Csaba Bal\'azs$^{1}$, Marcela Carena$^{2}$, Arjun Menon$^{1,3}$, 
David E.~Morrissey$^{4}$ and Carlos Wagner$^{1,3}$\\

{\noindent\em
$^1$Argonne National Laboratory\\[-1mm]
$^2$Fermi National Accelerator Laboratory\\[-1mm]
$^3$University of Chicago\\[-1mm]
$^4$University of Michigan\\[-1mm]
}\\

{\em After highlighting the basics and the consequences of electroweak
baryogenesis in the Minimal Supersymmetric Standard Model (MSSM), the
viability that the MSSM simultaneously provides the measured baryon
asymmetry and dark matter abundance is summarized. Examining a few
benchmark points within this scenario, we demonstrate a synergy
between the Tevatron, the LHC and the ILC.}


\subsubsection{Electroweak baryogenesis and neutralino dark matter}

The cosmological energy density of both main components of matter,
baryons and dark matter, is known with a remarkable
precision~\cite{Spergel:2003cb}.  In units of the critical density
$\rho_c=3H_0^2/(8\pi G_N)$\footnote{$H_0=h\times 100~km/s/Mpc$ is the
present value of the Hubble constant, $h=0.71^{+0.04}_{-0.03}$, and
$G_N$ is Newton's constant.}, they are:
\begin{equation}\label{odm}
\Omega_B h^2 = 0.0224\pm{0.0009} {\rm ~~~ and ~~~}
\Omega_{DM}h^2 = 0.1126^{+0.0161}_{-0.0181} \, .
\end{equation}
at $95\%$~CL.  According to the observations, the baryon density is
dominated by baryons while anti-baryons are only secondary products in
high energy processes.  The source of this baryon--anti-baryon
asymmetry is one of the major puzzles of particle physics and
cosmology.

Assuming that inflation washes out any initial baryon asymmetry after
the Big Bang, there should be a dynamic post-inflation mechanism to
regenerate the asymmetry.  Any microscopic mechanism for baryogenesis
must fulfill the three Sakharov requirements~\cite{Sakharov:1967dj}:
\begin{equation}
\bullet~{\rm baryon~number~(B)~violation} ~~~~~
\bullet~{\rm CP~violation} ~~~~~
\bullet~{\rm departure~from~equilibrium}.
\nonumber
\end{equation}

These requirements are satisfied in the MSSM during the electroweak
phase transition.  This is the basis for electroweak baryogenesis
(EWBG)~\cite{Cohen:1993nk,Quiros:1994dr,Rubakov:1996vz,Riotto:1999yt,Quiros:1999tx}.
Baryon number violation occurs in the MSSM due to quantum transitions
between inequivalent SU(2) vacua that violate
$(B\!+\!L)$~\cite{'tHooft:1976fv}.  These transitions are
exponentially suppressed at low temperatures in the electroweak broken
phase~\cite{Manton:1983nd,Klinkhamer:1984di}, but become active at
high temperatures when the electroweak symmetry is
restored~\cite{Bodeker:1998hm,Arnold:1999uy,Arnold:1999uz,Moore:1999fs,Moore:2000mx}.
If the electroweak phase transition is first order, bubbles of broken
phase nucleate within the symmetric phase as the universe cools below
the critical temperature.  These provide the necessary departure from
equilibrium.

To generate the observed baryon asymmetry the electroweak phase
transition has to be \emph{strongly} first
order~\cite{Bochkarev:1987wf},
\begin{equation}
v(T_c)/T_c \gtrsim 1 \, ,
\end{equation}
where $v(T_c)$ denotes the Higgs vacuum expectation value at the
critical temperature $T_c$.

For sufficiently light Higgs bosons, a first-order phase transition
can be induced by the loop effects of light bosonic particles, with
masses of the order of the weak scale and large couplings to the Higgs
fields.  Within the MSSM the most important contribution comes from a
light stop.  Detailed calculations show that for the mechanism of
electroweak baryogenesis to work, the lightest stop mass must be less
than the top mass but greater than about 120 GeV to avoid color
breaking minima.  Simultaneously, the Higgs boson involved in breaking
the electroweak symmetry must be lighter than 120
GeV~\cite{Carena:1996wj,Laine:1996ms,Losada:1996ju,Farrar:1996cp,deCarlos:1997ru,Bodeker:1996pc,Carena:1997ki,Laine:1998qk,Losada:1998at,Losada:1999tf,Laine:2000kv,Laine:2000rm},
and only slightly above the present experimental
bound~\cite{Barate:2003sz},
\begin{equation}\label{mhiggs} 
m_h \gtrsim 114~\mbox{GeV} \, , 
\end{equation} 
which is valid for a Standard Model (SM) Higgs boson.

To avoid generating too large a contribution to $\Delta\rho$, the
light stop must be mostly right-handed.  Since the stops generate the
most important radiative contribution to the Higgs boson mass in the
MSSM~\cite{Ellis:1990nz,Okada:1990vk,Haber:1990aw}, the other stop
must be considerably heavier in order to raise the Higgs boson mass
above the experimental bound, Eq.~(\ref{mhiggs}).  For the stop soft
supersymmetry breaking masses, this implies~\cite{Carena:1997ki}
\begin{eqnarray}\label{Eq:U3Q3}
m_{U_3}^2 \lesssim 0 {\rm ~~~ and ~~~ } 
m_{Q_3}^2 \gtrsim (1~{\rm TeV})^2 \, .
\end{eqnarray}
where $U_3$ ($Q_3$) is the soft mass of the third-generation
electroweak singlet up-type (doublet) scalar quarks at the electroweak
scale.  A similar balance is required for the combination of soft SUSY
breaking parameters defining the stop mixing~\cite{Carena:1997ki}:
\begin{eqnarray}\label{Eq:tanBXt}
5 \lesssim \tan\beta \lesssim 10 {\rm ~~~ and ~~~ } 
0.3 \lesssim |A_t-\mu^*/\tan\beta|/m_{Q_3} \lesssim 0.5 \, .
\end{eqnarray}

In addition to a strong electroweak phase transition, a CP-violating
source is needed to generate a chiral charge asymmetry.  This
translates into the following bounds on the chargino sector:
\begin{eqnarray}\label{Eq:phz}
|\arg(\mu\,M_2)| \gtrsim  0.1 {\rm ~~~ and ~~~ } 
\mu, M_2 \lesssim 500~{\rm GeV} \, .
\end{eqnarray} 
These conditions are relevant to the abundance of neutralino dark
matter, since the masses and mixing in the neutralino (and chargino)
sector are directly affected by the value of the soft gaugino masses
($M_i$) and the higgsino mass parameter ($\mu$) at the weak scale.

Low energy supersymmetry also provides a natural solution to the
origin of dark matter in the form of the lightest supersymmetric
particle (LSP).  In this summary, we consider only the case where the
LSP is the lightest neutralino.  To assess the viability of
simultaneous generation of the observed baryon--anti-baryon asymmetry
and neutralino dark matter, we focus on the narrow parameter region of
the MSSM defined by equations (\ref{mhiggs})-(\ref{Eq:phz}).  As
established earlier, in this parameter region EWBG is expected to
yield the observed amount of baryon density of the Universe.  To
further simplify the analysis, we assume that the gaugino mass
parameters $M_1$ and $M_2$ are related by the standard unification
relation, $M_2=(g_2^2/g_1^2)\,M_1\simeq 2\,M_1$.  The first and second
generation sfermion soft masses are taken to be very large,
$m_{\tilde{f}}\gtrsim 10$ TeV, to comply with the electron electric
dipole moment (EDM) constraints in the presence of sizable
phases.\footnote{As was shown in Ref.~\cite{Balazs:2004ae}, EDM limits
strongly constrain the EWBG mechanism in the MSSM.}  Only a phase
directly related to EWBG is introduced, namely $\arg(\mu)$, and for
convenience we set the phases of $A_f$ equal and opposite to it.  For
simplicity, we neglect the mixing between CP-even and CP-odd Higgs
bosons due to these phases.

\begin{figure}
\begin{center}
\includegraphics[height=10cm]{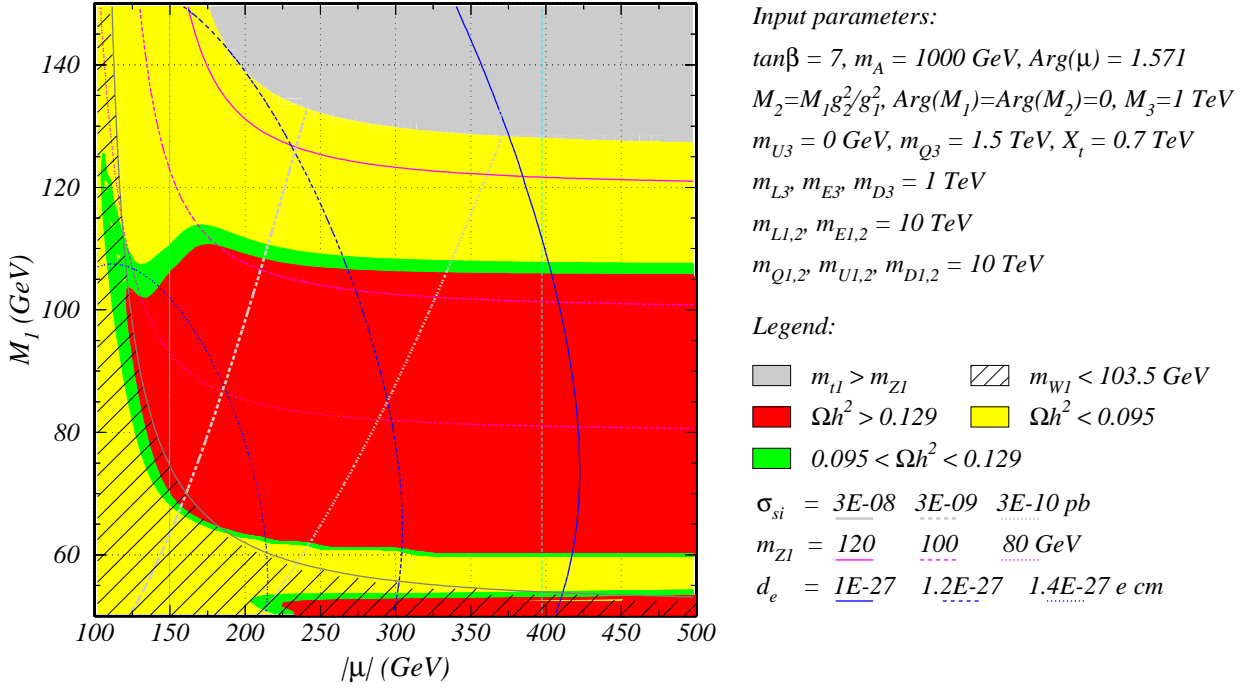}
\caption{Neutralino relic density as a function of $M_1$ vs. $|\mu|$
for $m_A = 1000$ GeV and $\arg(\mu)=\pi/2$.}
\label{fig:dm1}
\end{center}
\end{figure}

We compute the relic neutralino abundance as described in
Ref.~\cite{Balazs:2004ae}.  Fig.~\ref{fig:dm1} shows the typical
neutralino relic density dependence on $|\mu|$ and $M_1$ for typical
parameters inspired by EWBG: $\tan\beta=7$, $m_A=1000$ GeV, and
$\arg({\mu})=\pi/2$.  The green (medium gray) bands show the region of
parameter space where the neutralino relic density is consistent with
WMAP at $95\%$~CL.  The regions in which the relic density is above or
below this experimental bound are indicated by the red (dark gray) and
yellow (light gray) areas, respectively.  Finally, in the
(medium-light) gray region at the upper right corner, the lightest
stop becomes the LSP, while in the hatched area at the lower left
corner the mass of the lightest chargino is lower than is allowed by
LEP data\footnote{See 
$http://lepsusy.web.cern.ch/lepsusy/www/inos\_moriond01/charginos\_pub.html$
}.

In the upper allowed band the mass difference between the neutralino
LSP and the light stop is less than about 20-25~GeV, and
stop-neutralino coannihilation as well as stop-stop annihilation are
very efficient in reducing the neutralino abundance.  There is an area
below the disallowed band in which the neutralino mass lies in the
range 40-60~GeV and the neutralino annihilation cross section is
enhanced by s-channel resonant $h$ exchange.  The relic density is
also quite low for smaller values of $|\mu|$.  In these regions, the
neutralino LSP acquires a significant Higgsino component allowing it
to couple more strongly to the Higgs bosons and the $Z$.  For higher
$M_1$ values, the lightest neutralino and chargino masses are also
close enough that chargino-neutralino coannihilation and
chargino-chargino annihilation substantially increase the effective
cross section.

In summary, the requirement of a consistent generation of baryonic and
dark matter in the MSSM leads to a well-defined scenario with a light
stop and a light Higgs boson, light neutralinos and charginos,
sizeable CP-violating phases, and moderate values of
$5\lesssim\tan\beta\lesssim 10$.  These properties will be tested in a
complementary way by the Tevatron, the LHC and a prospective ILC, as
well as through direct dark matter detection experiments in the near
future.  The first tests of this scenario will probably come from
electron EDM measurements, stop searches at the Tevatron and Higgs
searches at the LHC within the next few years.


\subsubsection{Tevatron-LHC-ILC synergy}

A stop lighter than the top quark was and is being searched for at LEP
and the Tevatron, respectively, in various decay modes.  The Tevatron
reach depends on the decay properties of the lightest stop, and also
on the specific values of the light chargino and neutralino
masses~\cite{Affolder:1999wp,Abachi:1995jp,Carena:2002wz,Berger:1999zt,Chou:1999zb}.
Here we focus on the case in which the neuralino is the lightest (LSP)
and the lighter stop is the next-to-lightest supersymmetric partner
(NLSP).  In such a case, the Tevatron can find a light stop provided
its mass is smaller than about 200~GeV~\cite{Demina:1999ty}, a region
that overlaps substantially with the interesting one for EWBG.

To assess the light stop collider reach in the EWBG scenario, we
conducted a random scan over the following range of MSSM
parameters:\footnote{Parameters which are not scanned over are fixed
as in the right side of Fig.~\ref{fig:dm1}.}
\begin{eqnarray}\label{eq:RandomScanPars}
&& -(80 ~{\rm GeV})^2 < m_{\tilde U_3}^2 < 0 , ~~~ 
    100 < |\mu| < 500 ~{\rm GeV}, ~~~
     50 < M_1 < 150 ~{\rm GeV}, \nonumber \\
&&  200 < m_A < 1000 ~{\rm GeV}, ~~~
      5 < \tan\beta < 10 \, .
\end{eqnarray}
The result of the scan, projected to the stop mass versus neutralino
mass plane, is shown by Figure \ref{fig:mz1vsmt1}.  The region where
$m_{{\tilde{Z}}_1}>m_{{\tilde{t}}_1}$ is inconsistent with a
neutralino LSP, while for
$m_{{\tilde{t}}_1}>m_{W}+m_{b}+m_{{\tilde{Z}}_1}$ the three-body decay
$\tilde{t}_1\to W^+\bar{b}{\tilde{Z}}_1$ becomes accessible and
dominant.  For models marked by dots the neutralino relic density is
below or within the $2\sigma$ WMAP bound.  The lower left corner is
excluded by the LEP chargino mass limit of 103.5~GeV.

\begin{figure}[htb]
\begin{center}
\includegraphics[height=10cm]{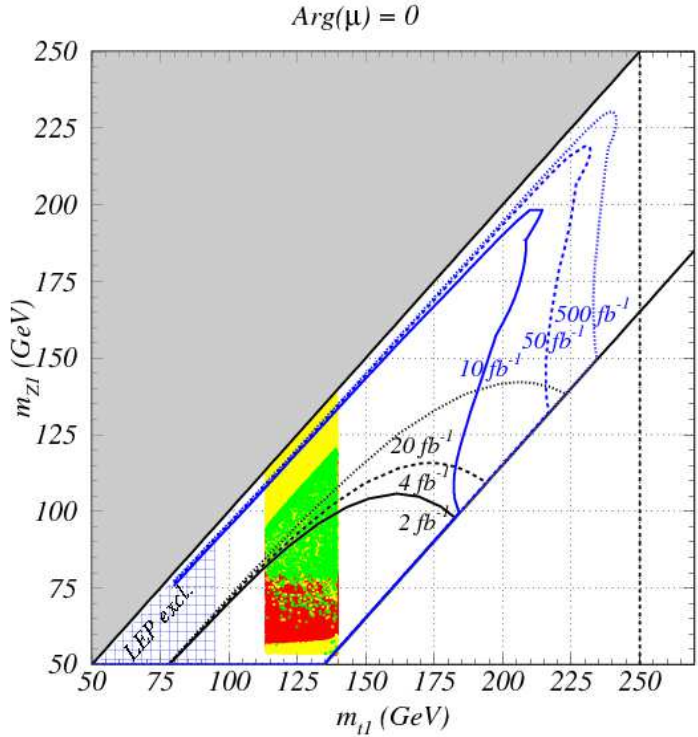}
\caption{Discovery reach of the Tevatron (black contours) and an ILC 
(blue contours) for production of light stop quarks in the decay
channel $\tilde{t}_1\to c\chi_1^0$.}
\label{fig:mz1vsmt1}
\end{center}
\end{figure}

Overlayed on Fig.\ref{fig:mz1vsmt1} is the Tevatron light stop search
sensitivity in the $cc\!\!\not\!\!{E_T}$ channel for 2, 4 and
20~pb$^{-1}$ pf integrated luminosity~\cite{Demina:1999ty}.  The
Tevatron can cover a considerable region of parameter space of the
EWBG-motivated scenario.  In the Tevatron covered region, resonant
annihilation via the light Higgs produces acceptable amount of dark
matter.  Coannihilation with the lightest stop is dominant where the
stop-neutralino mass gap is small.  As is apparent from the figure,
under the present missing-energy triggering requirements, the Tevatron
will not be able to detect a light stop in this region of parameter
space.  This region, on the other hand, is easily covered by even a
500~GeV ILC.

For the detailed exploration of the collider phenomenology in this
scenario, the common strategy of selecting and analysing individual
parameter space points, or benchmark points, was used at Les Houches
in 2005.  The benchmark points were defined taking into account the
discussion of the parameter values presented in the previous section.
All benchmark points were selected such that the baryon asymmetry of
the universe and the relic neutralino density is predicted to be close
to those measured by WMAP, and pass all known low energy, collider and
astronomy constraints.  The most important of these are the SUSY
particle masses, the electron EDM, $B(b\to s\gamma)$, and direct WIMP
detection.  A crucial constraint is the LEP~II Higgs boson direct
exclusion mass limit of $m_h>114.4$~GeV.  In the calculations of the
supersymmetric spectrum and the baryon asymmetry, we used tree-level
relations except for the Higgs mass, which was calculated at one loop.
In the parameter region of interest, the one-loop calculation results
in about a 6-8~GeV lower lightest Higgs mass than the two-loop
result~\cite{Heinemeyer:1998np,Degrassi:2002fi}.  Thus, if the soft
supersymmetric parameters defining the benchmark points are used in a
two-loop calculation, the resulting lightest Higgs mass is found to be
inconsistent with LEP~II.  A two-loop-level consistency with the
LEP~II limit can be achieved only when a baryon asymmetry calculation
becomes available using two-loop Higgs boson masses.

The main difference between the Les Houches benchmark points lies in
the mechanism that ensures the neutralino relic density also complies
with WMAP.  Keeping the unification-motivated ratio of the gaugino
mass parameters $M_2/M_1$ close to 2 (together with the baryogenesis
required $100\lesssim |\mu|\lesssim 500$~GeV) induces a lightest
neutralino with mostly bino admixture.  A bino typically overcloses
the universe, unless there is a special situation that circumvents
this.  For example, as in the supergravity-motivated minimal SUSY
scenario mSUGRA, neutralinos can coannihilate with sfermions,
resonantly annihilate via Higgs bosons, or acquire a sizable Higgsino
admixture in special regions of the parameter space.  This lowers the
neutralino density to a level that is consistent with observation.

Benchmark point LHS-1 features strong stop-neutralino coannihilation
which lowers the relic density of neutralinos close to the WMAP
central value.  Sizable coannihilation occurs only when the
neutralino--stop mass difference is small, less than about $30-40\%$.
A small neutralino-stop mass gap poses a challenge for the Tevatron
and the LHC while an ILC could cover this region efficiently.
 
Point LHS-2, resonant annihilation of neutralinos via s-channel Higgs
resonances lowers the neutralino abundance to the measured level.  In
this case, the neutralino mass must be very close to half of the
lightest Higss boson mass.  This point features a stop that, given
enough luminosity, can be discovered at the Tevatron due to the large
difference between the stop and the neutralino masses.  Even the
heavier stop can possibly be produced at the LHC together with the
third generation sleptons.  On the other hand, the resonance feature
implies that the lightest Higgs boson can decay into neutralinos,
which would reduce its visible width and therefore could make its
discovery more challenging.

Point LHS-3 satisfies the WMAP relic density constraint partly because
the lightest neutralino acquires some wino admixture and because it
coannihilates with the lightest stop and chargino.  The multiple
effects lowering the relic density allow for a little larger
neutralino-stop mass gap than in LHS-1.  This point has a
neutralino-stop mass gap that makes it detectable at the Tevatron and
the LHC.

LHS-4, a variation of LHS-1, is defined in detail in
Ref.~\cite{Carena:2005gc}.  Here the small neutralino-stop mass
difference makes the light stop inaccessible at the Tevatron and the
LHC.  On the other hand, an ILC could measure the parameters with
precision.  The discovery potential of this point is discussed in
detail in Ref.~\cite{Carena:2005gc}.

In summary, the four benchmark points offer various challenges for the
three colliders.  The Tevatron could resolve the stop quark in points
LHS-2 and LHS-3, where the $\tilde{t}_1$ decays into $\tilde\chi_1^\pm
b$, but not in LHS-1 and LHS-4, where it decays into $\tilde\chi_1^0
c$ with a small phase space.  The LHC on the other hand may explore
LHS-1 and LHS-2 as described in the Les Houches 2005 proceedings.  In
principle these methods are also applicable for LHS-4 and LHS-3; the
small mass differences at these points, however, make the analysis
much more difficult.  In LHS-1, LHS-2 and LHS-3 the LHC can pair
produce the heavier stop, which is needed to pin down the stop sector
so crucial for baryogenesis.  At an ILC, one can perform precision
measurements of the light stop.  Moreover, the weak ino sector
including the important phase(s) can be measured precisely (see
Ref.~\cite{Moortgat-Pick:2005cw} and references therein).


\subsubsection*{Acknowledgements}

Research at the HEP Division of ANL is supported in part by the US
DOE, Division of HEP, Contract W-31-109-ENG-38.  Fermilab is operated
by Universities Research Association Inc. under contract
no. DE-AC02-76CH02000 with the DOE.

\clearpage\setcounter{equation}{0}\setcounter{figure}{0}\setcounter{table}{0}
\subsection{Discovering SUSY at the LHC with Same-Sign Di-Muons}
\label{sec:SS-dimuon}

{\em Salavat Abdoulline$^1$, Darin Acosta$^2$, Paolo Bartalini$^2$,
Richard Cavanaugh$^2$, Alexey Drozdetskiy$^2$, Sven Heinemeyer$^3$,
Andrey Korytov$^2$, Guenakh Mitselmakher$^2$, Yuriy Pakhotin$^2$,
Bobby Scurlock$^2$, Georg Weiglein$^4$\\
$^1$ Fermi National Laboratory, Chicago, IL, USA\\
$^2$ University of Florida, Gainesville, FL, USA\\
$^3$ Dept. de F\'isica Te\'orica, Universidad de Zaragoza, Spain\\
$^4$ IPPP, University of Durham, UK}\\

{\em Within the framework of the Constrained Minimal Supersymmetric
Standard Model (CMSSM) we analyze the discovery potential of the LHC
for the same-sign di-muon signature.  The analysis focuses on
parameter space that will not be probed by the Tevatron, and that is
favored by current electroweak precision data and cosmological
observations.  With an integrated luminosity of 10~fb$^{-1}$,
corresponding to the first phase of LHC operations, fermionic mass
parameters $m_{1/2}$ can be probed up to $m_{1/2}<650$~GeV.  For
$\tan\beta=10$ this covers the full range favored by current
low-energy precision data.  For larger $\tan\beta$ values, the highest
favored $m_{1/2}$ values require a larger integrated luminosity.}


\subsubsection{Introduction}

Theories based on Supersymmetry
(SUSY)~\cite{Nilles:1983ge,Haber:1985rc,Barbieri:1987xf} are widely
considered as the theoretically most appealing extension of the
Standard Model
(SM)~\cite{Glashow:1961tr,Weinberg:1967tq,Salam:1968rm}.  They are
consistent with the approximate unification of the gauge coupling
constants at the GUT scale and provide a way to cancel the quadratic
divergences in the Higgs sector, stabilizing the huge hierarchy
between the GUT and Fermi scales.  Furthermore, in SUSY theories,
breaking of the electroweak symmetry is naturally induced at the Fermi
scale, and the lightest supersymmetric particle can be neutral, weakly
interacting and stable, providing therefore a natural solution for the
dark matter problem.

SUSY predicts the existence of scalar partners $\tilde{f}_L$,
$\tilde{f}_R$ to each SM chiral fermion, and spin--1/2 partners to the
gauge and scalar Higgs bosons.  So far, the direct search for SUSY
particles has not been successful.  One can only set lower bounds of
${\cal O}(100)$~GeV on their masses~\cite{Eidelman:2004wy}.  The
search reach is currently extended in various ways in the ongoing
Run~II at the upgraded Fermilab Tevatron~\cite{Abel:2000vs}.  The
LHC~\cite{TDR,CMS-PTDR-Vol1} and the proposed $e^+e^-$ International
Linear Collider
(ILC)~\cite{Aguilar-Saavedra:2001rg,Abe:2001nn,Abe:2001np,Abe:2001nq,Abe:2001gc}
have very good prospects for exploring SUSY at the TeV scale, which is
favored from naturalness arguments.  From the interplay of both
machines detailed information on the SUSY spectrum can be expected in
this case~\cite{Weiglein:2004hn}.

In the Minimal Supersymmetric Standard Model (MSSM), no further
assumptions are made on the structure of the soft SUSY-breaking
parameters, and a parameterization of all possible SUSY-breaking terms
is used.  The high dimensionality makes analyses in the MSSM without
any further constraints quite cumbersome.  For this reason,
simplifying assumptions that may be more or less well-motivated are
often made, so as to reduce the parameter space to a manageable
dimensionality.  Following many previous studies, we work here in the
framework of the constrained MSSM (CMSSM), in which the soft
supersymmetry-breaking scalar and gaugino masses are each assumed to
be equal at some GUT input scale.  In this case, the new independent
MSSM parameters are just four in number: the universal gaugino mass
$m_{1/2}$, the scalar mass $m_0$, the trilinear soft
supersymmetry-breaking parameter $A_0$, and the ratio $\tan\beta$ of
Higgs vacuum expectation values~\footnote{An economical way to ensure
the universality assumed in the CMSSM is by gravity-mediated SUSY
breaking in a minimal supergravity (mSUGRA) scenario.  The mSUGRA
scenario predicts in particular a relation between the gravitino mass
and $m_0$, which is not necessarily fulfilled in the CMSSM.  For
simplicity, we do not make the distinction between the CMSSM and the
mSUGRA scenario, and use the phrase ``CMSSM'' for both.}.

The non-discoveries of supersymmetric particles and the Higgs boson at
LEP and the Tevatron (so far) impose significant lower bounds on
$m_{1/2}$ and $m_0$.  An important further restriction is provided by
the density of dark matter in the universe, which is tightly
constrained by WMAP and other astrophysical and cosmological
data~\cite{Bennett:2003bz,Spergel:2003cb}.  These have the effect
within the CMSSM, assuming that the dark matter consists largely of
neutralinos~\cite{Goldberg:1983nd,Ellis:1983ew}, of reducing the
available parameter space and allowing only thin strips in ($m_{1/2}$,
$m_0$)-planes (for fixed $A_0$ and
$\tan\beta$)~\cite{Ellis:2003cw,Ellis:2004tc}.

\medskip

An important discovery signal for SUSY at the LHC is the same-sign
di-muon signature. A typical decay chain for a signal event is similar
to the following one: $gq \rightarrow \tilde{g}\tilde{q}_L$, where
$\tilde{g}\to\tilde{t}t\to\tilde{\chi}_1^+ +X\to W^+ +\tilde{\chi}_1^0
+X\to\tilde{\chi}_1^0 +X+\mu^+ +\nu$ and
$\tilde{q}_L\to\tilde{\chi}_1^+ +Y\to W^+ +\tilde{\chi}_1^0
+Y\to\tilde{\chi}_1^0 +Y+\mu^+ +\nu$.  Here we see two
$\tilde{\chi}_1^0$ stable neutral heavy SUSY particles providing MET,
two high $P_T$ $\mu^+$ and high $E_T$ jets included into $X$, $Y$
decay chain parts.  Two same-sign muons with relatively high $p_T$ (at
least $p_T>10$~GeV) significantly reduce background contamination with
respect to the ``multi-jets only'' signature, as well as provide high
trigger efficiency.  Additional cuts on missing transverse energy as
well as jet $E_T$ increase signal to background significance even
further.  A recent published theoretical study of that signature for
the Tevatron can be found in Ref.~\cite{Matchev:1999nb}.

The outline of the contribution is as follows.  In
Sec.~\ref{sec:CMSSM} we review the CMSSM landscape.  We review the
lower bounds on SUSY masses from the LEP and Tevatron searches and
describe briefly the effects of cold dark matter (CDM) density.  We
also outline the regions of CMSSM parameter space that are favored by
electroweak precision observables (where also Tevatron data plays an
important role).  Within the region allowed by direct searches and
favored by precision data we perform a simulated experimental analysis
for the same-sign di-muon signature at the LHC.  The reach within the
CMSSM parameter space for various SUSY mass scales is then explored in
Sec.~\ref{sec:expreach}.  We find that for $\tan\beta=10$ the full
range of $m_{1/2}$ values, favored at the $90\%$~C.L.\ by current
electroweak precision data, can be probed with the same-sign di-muon
signature in the first phase of LHC operations (10~fb$^{-1}$).  For
higher $\tan\beta$ values, the largest $m_{1/2}$ values favored by
precision data will require a higher integrated luminosity.


\subsubsection{The CMSSM landscape}
\label{sec:CMSSM}

Here we briefly review the parameters of the CMSSM, their experimental
bounds and the regions of parameter space favored by current
electroweak precision data.


\medskip
\noindent\underline{\it The CMSSM}

The study presented below has been performed in the framework of the
CMSSM, in which the soft SUSY-breaking scalar and gaugino masses are
each assumed to be equal at some GUT input scale.  The new independent
MSSM parameters are: the universal gaugino mass $m_{1/2}$, the scalar
mass $m_0$, the trilinear soft supersymmetry-breaking parameter $A_0$,
and the ratio $\tan\beta$ of Higgs vacuum expectation values.  Also,
the sign of the Higgs mixing parameter $\mu$ is in principle still
undetermined.  The anomalous magnetic moment of the muon, $(g-2)_\mu$,
shows a 2-3$~\sigma$ discrepancy from the SM
expectation~\cite{Czarnecki:2001pv,Knecht:2003kc,Bennett:2004pv}.
SUSY effects can easily account for this deviation if $\mu$ is
positive~\cite{Moroi:1995yh}.  For this reason, in the rest of this
contribution, we restrict our attention to $\mu>0$.  Even in view of
the possible size of experimental and theoretical uncertainties, it is
very difficult to reconcile $\mu<0$ with the present data on
$(g-2)_\mu$.

We furthermore assume that the CMSSM gives the right amount of CDM
density (with the lightest neutralino being the dark matter particle)
to be in the range $0.094<\Omega_{\rm CDM}h^2<0.129$ favored by a
joint analysis of WMAP and other astrophysical and cosmological
data~\cite{Bennett:2003bz,Spergel:2003cb}.  This strongly reduced the
available CMSSM parameter space and allows only thin strips in
($m_{1/2}$,$m_0$)-planes (for fixed $A_0$ and
$\tan\beta$)~\cite{Ellis:2003cw,Ellis:2004tc}.  For simplicity the
experimental analysis was performed for $A_0=0$ and certain fixed
values of $\tan\beta$: 10, 20, and 35.


\medskip
\noindent\underline{\it Limits from LEP and the Tevatron}

The four LEP experiments actively searched for SUSY particles without
seeing any significant excess of signal over
background~\cite{deschtalk,vivietalk,Kraan:2005vy,Pasztor:2005es}.
Limits of $\sim 100$~GeV could be set on the masses of all
electrically charged SUSY particles.  Within the CMSSM, the mass of
the lightest neutralino was limited to be above $\sim 50$~GeV.

The search for SUSY particles has been continued at the Tevatron in
Run II~\cite{Abel:2000vs}.  Currently, the limits for scalar quarks
(except scalar tops and bottoms) and gluinos have been extended to
350~GeV~\cite{heinemanntalk}.  Assuming that the Tevatron will not
discover SUSY even with 8~fb$^{-1}$ per detector, these bounds will be
extended to about 450~GeV~\cite{heinemanntalk}.

Within the CMSSM these SUSY particle limits can be translated into
limits on $m_{1/2}$, $m_0$, $A_0$ and $\tan\beta$.  Assuming the CDM
density constraints (see above), values of ${\cal O}(50)$~GeV can be
excluded for $m_{1/2}$ and $m_0$ (depending on the choice of $A_0$ and
$\tan\beta$).  In our analysis below we discard all CMSSM parameter
combinations that are not in agreement with the anticipated future
exclusion bounds for SUSY particles or Higgs
bosons~\cite{Barate:2003sz,:2006cr,Abazov:2005yr,Abulencia:2005kq,Tevcharged}
from the Tevatron and LEP searches.

 
\medskip
\noindent\underline{\it Indications from electroweak precision observables}

Measurements at low energies may provide interesting indirect
information about the masses of particles that are too heavy to be
produced directly.  A prime example is the use of precision
electroweak data from LEP, the SLC, the Tevatron and elsewhere to
predict (successfully) the mass of the top quark and to provide an
indication of the possible mass of the hypothetical Higgs
boson~\cite{Grunewald:2003ij,diaconutalk}.  Predicting the masses of
supersymmetric particles is much more difficult than for the top quark
or even the Higgs boson, because the renormalizability of the Standard
Model and the decoupling theorem imply that many low-energy
observables are insensitive to heavy sparticles.  Nevertheless,
present data on electroweak precision observables can already provide
interesting information on the scale of SUSY.

We consider the following observables: the $W$~boson mass, $M_W$, the
effective weak mixing angle at the $Z$~boson resonance,
$\sin^2\theta_{\rm eff}$, the anomalous magnetic moment of the muon,
$(g-2)_\mu$, the rare decay ${\rm BR}(b\to s\gamma)$ and the mass of
the lightest CP-even Higgs boson, $m_h$.  Within the CMSSM, a $\chi^2$
analyses for fixed values of $A_0/m_{1/2}$ for $\tan\beta=10$ and~50
with $\mu>0$ was performed~\cite{Ellis:2004tc,Ellis:2006ix}.  A
remarkably high sensitivity of the current data for the electroweak
precision observables to the scale of SUSY was observed.  For
$\tan\beta=10$, a preference for low values of $m_{1/2}\sim 300$~GeV
was found.  This increases to $m_{1/2}\sim 450,550,600$~GeV for
$\tan\beta=20,35,50$.  As an example, showing also the impact of
future Tevatron measurements, we reproduce here the result for the
{\em upper} limit on $m_{1/2}$ at the $90\%$~C.L.
Fig.~\ref{fig:varymt10}~\cite{Ellis:2006ix} shows this upper limit for
various top-quark mass measurements ($m_t$ enters the prediction of
the precision observables) and possible future uncertainties that
could be realized during RunII at the Tevatron~\footnote{Also, the
$M_W$ measurement will be improved during RunII, possibly down to an
uncertainty of $\delta M_W^{\rm RunII} \approx 20$~MeV, compared to
the current uncertainty of $\delta M_W=32$~MeV today.  This will
slightly tighten the constraints from precision observables.  However,
the impact is expected to be substantially smaller than from the
improved $m_t$ measurements.}.  We fixed $m_0$ via the CDM constraint,
varied $A_0$ from -2~$m_{1/2}$ to +2~$m_{1/2}$, and set
$\tan\beta=10$.  The upper limit on $m_{1/2}$ does not exceed 650~GeV
for 169~$<m_t<$~178~GeV (depending only slightly on the experimental
precision of $m_t$).  Only for $m_t$ values outside this interval much
are larger upper bounds obtained.  The bounds increase to about
850~GeV for $\tan\beta=20$, 1000~GeV for $\tan\beta=35$, and about
1200~GeV for $\tan\beta=50$.  Consequently we will focus on
$m_{1/2}<900$~GeV in our analysis.

\begin{figure}[htb!]
\begin{center}
\includegraphics[width=8cm,height=7.5cm]{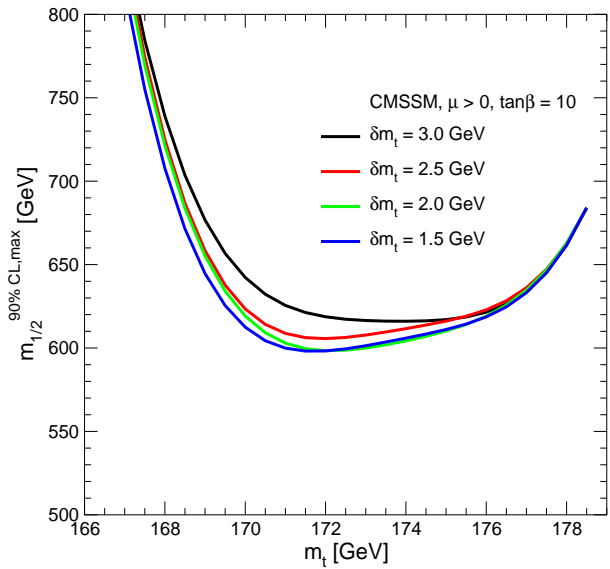}
\caption{The dependence of the {\em upper} limit on $m_{1/2}$ at 
the $90\%$~C.L.\ on $m_t$ and its uncertainty $\delta
m_t$~\protect\cite{Ellis:2006ix}.}
\label{fig:varymt10} 
\end{center}
\end{figure}
%


\medskip
\noindent\underline{Experimental analysis}

In this study we used the following values of the parameters: ${\rm
sign}(\mu)>0$, $A_0=0$, $\tan\beta=10,20,35$ and 20
$(m_{1/2},m_0)$-points.  All points were chosen so as to satisfy
recent theoretical and experimental constraints~\cite{Ellis:2003cw}.
The top quark mass, $m_t$, was fixed to $m_t=175$~GeV.

The study is based on the assumption of 10~fb$^{-1}$ integrated
luminosity collected at CMS.  The points are analyzed in view of the
same-sign di-muon signature.  More recent results on the two same-sign
di-muon signature analysis with CMS are now available in the CMS
Physics TDR, vol.2~\cite{CMS-PTDR-Vol2}.


\medskip
\noindent\underline{\it Simulation and Reconstruction}

The points chosen for this analysis are shown in
Fig.~\ref{fig:points}.  Points 2, 11, 16, 19 are CMSSM benchmark
points taken from Ref.~\cite{Battaglia:2003ab} (two of them, points 2
and 16, were modified for a top-quark mass of 175~GeV as used for all
other points).  All points shown on the plot by empty markers have a
cross section too small for a target integrated luminosity of
10~fb$^{-1}$ and are not considered in the study.  On the other hand,
these are exactly the points that are disfavored by current
electroweak precision data~\footnote{For larger $\tan\beta$ values it
is possible to find non-disfavored points that have a too small cross
section.}.

\begin{figure}[htb!]
  \begin{center}
    \resizebox{12cm}{!}{\includegraphics{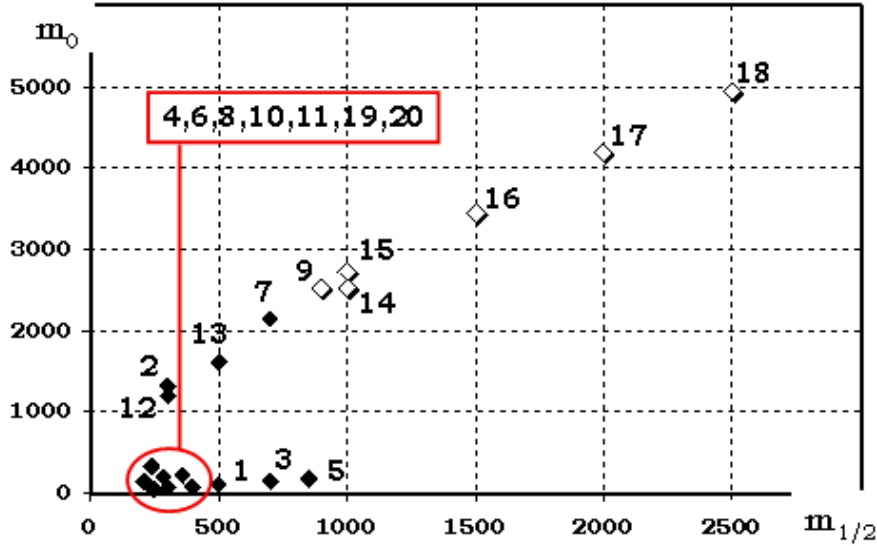}}
    \caption{The CMSSM benchmark points investigated in this analysis.}
    \label{fig:points}
  \end{center}
\end{figure}

Coupling constants and cross sections at leading order (LO) for SUSY
processes were calculated with ISASUGRA 7.69~\cite{Baer:1999sp}.
Next-to-Leading Order (NLO) corrections were calculated with the
PROSPINO program~\cite{Beenakker:1996ed} and used in the analysis.
The cross sections for the SM processes were calculated using PYTHIA
6.220~\cite{Sjostrand:2001yu} and CompHEP 4.2p1~\cite{comphep}.  For
several SM processes (${t\bar{t},ZZ,Zb\bar{b}}$), the NLO correction
are known and were used~\cite{thesisbartsch}.  All events were
generated with PYTHIA 6.220.  Some preselection cuts were applied at
the generator level: events were kept if at least two same-sign muons
with $p_T>10$~GeV and $|\eta|<2.5$.  After the event generation, a
GEANT-based simulation CMSIM~\cite{CMSSOFT} was performed.

Data digitization and reconstruction were done with the
ORCA~\cite{CMSSOFT} reconstruction package.  Pile-up events were not
taken into account in this study (and muon isolation cuts were not
used).  Muon reconstruction was performed using an algorithm
implemented for the CMS High-Level Triggers (HLT)~\cite{JETCORMET}
based on the muon and tracker sub-detector information.  Jets were
reconstructed with the Iterative Cone Algorithm with a cone size of
0.5\cite{JET}.  Jet $E_T$ corrections were applied and missing $E_T$
calculated as described in Ref.~\cite{JETCORMET}.

We checked that all events satisfying the selection criteria passed
both L1 and HLT muon triggers (single and di-muon triggers).  Details
on the trigger threshold and efficiencies can be found in
Ref.~\cite{JETCORMET}.


\medskip
\noindent\underline{\it Background processes}

Several of the important backgrounds were studied in detail, while for
several of the rare processes only an estimate was made based on the
cross section and branching ratio to the final state signature.

The cross sections and numbers of generated and selected events for
the important sources of backgrounds are listed in
Tab.~\ref{tab:bckg1}.  In all ${tb}$, ${tqb}$, ${\bar{t}b}$,
${\bar{t}qb}$ processes, the top quark was forced to decay to ${Wb}$
and the ${W}$ was forced to decay into a muon and neutrino.  The muon
from that decay chain was required to have $p_T>10$~GeV and
$|\eta|<2.5$ at the CompHEP generation level.  In the ${Zb\bar{b}}$
process, the ${Z}$ was allowed to be off mass shell (${Z/\gamma^*}$)
and was forced to decay to $\mu^+\mu^-$.  The invariant mass of the
two muons from the initial ${Z/\gamma^*}$ was required to be larger
than 5~GeV at the CompHEP generation level.

\begin{table}[htb!]
\small
\begin{center}
\begin{tabular}{|c||c|c|c|c|c|} \hline\hline
Process  & $\sigma$, pb & $N_{\rm generated}$ & 
$N_{\rm selected}$ & N1 & N2 \\ \hline\hline 
${tb}$ & 0.212 & 18,999 & 1,000 & 2,120 & 112  \\ \hline
${tqb}$ & 5.17 & 28,730 & 1,000 & 51,700 & 1,798  \\ \hline
${\bar{t}b}$ & 0.129 & 13,588 & 745 & 1,290 & 71  \\ \hline
${\bar{t}qb}$ & 3.03 & 28,359 & 1,000 & 30,300 & 1,067  \\ \hline
${ZZ}$ & 18(NLO) & 433,489 & 1,000 & 180,000 & 256  \\ \hline
${ZW}$ & 26.2 & 368,477 & 1,000 & 262,000 & 727  \\ \hline
${WW}$ & 26.2 & 894,923 & 41 & 702,000 & 39.7  \\ \hline
${t\bar{t}}$ & 886(NLO) & 931,380 & 15,000 & 8,860,000 & 142,691  \\ \hline
${Zb\bar{b}}$ & 232(NLO) & 359,352 & 2,000 & 2,320,000 & 12,924  \\ \hline
${\rm All}$ &  &  &  &  & 160,000  \\ \hline\hline
\end{tabular}
\end{center}
\vspace{-1em}
\caption{Cross sections and numbers of events for the SM processes
for which a detailed simulation was performed for an integrated
luminosity of 10~fb$^{-1}$.  $N_{\rm generated}$ is the unweighted
number of generated events, $N_{\rm selected}$ is the unweighted
number of pre-selected events, $N_1$ is the number of events for an
integrated luminosity of 10~fb$^{-1}$, $N_2$ is the number of events
after pre-selection cuts (at least two same-sign muons with
$p_T>10$~GeV.)}
\label{tab:bckg1}
\end{table}

In Tab.~\ref{tab:bckg2} the estimates of other potential background
processes are shown.  No detailed simulation was performed for these
processes.  An estimate was obtained from the process cross section
(calculated with CompHEP) and the branching fraction into muons.

\begin{table}[htb!]
\small
\begin{center}
\begin{tabular}{|c||c|c|c|} \hline\hline
Process  & $\sigma$, pb & N1 & N2 \\ \hline\hline
${\rm WWW}$ & 0.129 & 1,290 & $< 15$  \\ \hline
${\rm ZWW}$ & 0.0979 & 979 & $< 10$  \\ \hline
${\rm ZZW}$ & 0.0305 & 305 & $< 3$  \\ \hline
${\rm ZZZ}$ & 0.00994 & 99.4 & $< 1$  \\ \hline
${\rm WWWW}$ & 0.000574 & negl. & negl.  \\ \hline
${\rm ZWWW}$ & 0.000706 & negl. & negl.  \\ \hline
${\rm ZZWW}$ & 0.000442 & negl. & negl.  \\ \hline
${\rm ZZZW}$ & 0.000572 & negl. & negl.  \\ \hline
${\rm ZZZZ}$ & 0.0000161 & negl. & negl.  \\ \hline
${\rm t\bar{t}W}$ & 0.556 & 5,560 & $< 200$  \\ \hline
${\rm t\bar{t}Z}$ & 0.65 & 6,500 & $< 200$  \\ \hline
${\rm t\bar{t}WW}$ & negl. & negl. & negl.  \\ \hline
${\rm t\bar{t}ZW}$ & negl. & negl. & negl.  \\ \hline
${\rm t\bar{t}ZZ}$ & negl. & negl. & negl.  \\ \hline\hline
\end{tabular}
\end{center}
\vspace{-1em}
\caption{Cross sections and event numbers for the SM processes for
which only an estimate was obtained for an integrated luminosity of
10~fb$^{-1}$.  $N_1$ is the estimated number of events for integrated
luminosity of 10~fb$^{-1}$, $N_2$ is the estimated number of events.}
\label{tab:bckg2}
\end{table}

The main conclusion here is that all but the $t\bar{t}W$, $t\bar{t}Z$
processes are negligible and they are neglected in this analysis.
Both $t\bar{t}W$ and $t\bar{t}Z$ backgrounds need, however, further
future investigation.


\medskip
\noindent\underline{\it Selection and cut optimization}

The distributions of kinematic variables such as missing $E_T$, jet
$E_T$, muon $p_t$ (as chosen for this study), are very different for
SUSY and SM processes as shown in Fig.~\ref{fig:jetmet}.  Suitable
cuts on these variables help reducing the SM background.

\begin{figure}[htb!]
\vspace{4em}
\begin{center}
\resizebox{10cm}{!}{\includegraphics{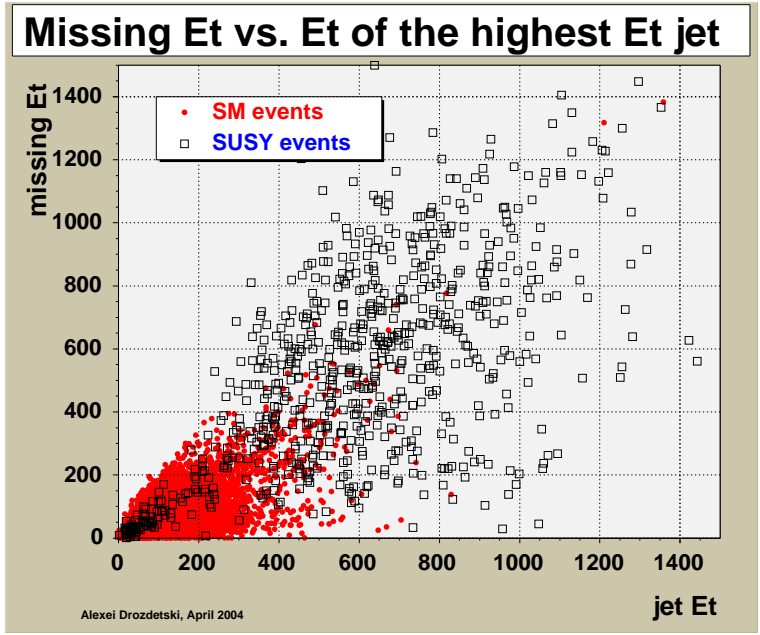}}
\caption{Missing transverse energy vs. jet $E_T$ for SUSY (open
squares) and SM events (points) after full simulation and
reconstruction.}
\label{fig:jetmet}
\end{center}
\end{figure}

A set of five variables (missing $E_T$, $E_{T,{\rm jet}_1}$,
$E_{T,{\rm jet}_3}$, $p_{T,\mu_1}$, $p_{T,\mu_2}$) was chosen.  For
each variable a set of possible selection cuts was defined.  A
combination of selection variables and cuts was performed, leading to
different sets of cuts.  The significance, signal-over-background
ratio and number of expected events for 10~fb$^{-1}$ were calculated
for each set of cuts.  Only the two sets of cuts shown in the
Tab.~\ref{tab:cutsfin} were finally chosen to be applied in the
analysis, since they cover all CMSSM points with significance greater
than $5\sigma$.

\begin{table}[htb!]
\small
\begin{center}
\begin{tabular}{|c||c|c|c|c|c|} \hline\hline
set & miss. $E_T$, GeV & $E_{T, {\rm jet}_1}$, GeV & 
$E_{T, {\rm jet}_3}$, GeV & $P_{T, \mu_1}$, GeV & $P_{T, \mu_2}$, GeV \\
\hline\hline 
1 & $>$ 200  & $>$   0 & $>$ 170 & $>$ 20 & $>$ 10 \\ \hline
2 & $>$ 100  & $>$ 300 & $>$ 100 & $>$ 10 & $>$ 10 \\ \hline\hline
\end{tabular}
\end{center}
\vspace{-1em}
\caption{Chosen cut sets.  $E_{T,{\rm jet}_1}$ is the $E_T$ of
the leading jet (maximum $E_T$) (GeV), $E_{T,{\rm jet}_3}$ is the
$E_T$ of the third highest-$E_T$ jet (GeV), $P_{T,\mu_1}$ and
$P_{T,\mu_2}$ are the two highest-$p_T$ values of the same-sign muons
(GeV).}
\label{tab:cutsfin}
\end{table}
%


\medskip
\noindent\underline{\it SUSY points characteristics}

For each CMSSM point, the LO cross section was calculated as well as
the number of events, for an integrated luminosity of 10~fb$^{-1}$.
NLO corrections were applied for all CMSSM points and corrected values
were used for this analysis.

Two other parameters were calculated for each CMSSM point: the
statistical significance and the signal-to-background ratio.  The
significance was calculated using the following
expression~\cite{Bityukov:2002ih}:
$S_{12}=2(\sqrt{N_S+N_B}-\sqrt{N_B})$, where $N_B$ is the total number
of background events and $N_S$ is the number of signal events for each
point in the $(m_{1/2},m_0)$-plane.

Six points (numbers 9 and 14-18) were excluded from the analysis
because of their small cross section.  These are exactly the points
that are disfavored by current electroweak precision data.  This
selection is affected by our initial choice of benchmark points, where
most have $\tan\beta=10$, only one has $\tan\beta=20$, and three have
$\tan\beta=35$.  For $\tan\beta>10$ it is in principle possible to
find non-disfavored points that have a too-small cross section.


\medskip
\noindent\underline{\it Results}

Details concerning the expected number of signal and background events
as well as the significance at each point studied are listed in the
Tab.~\ref{tab:results1}.

\begin{table}[htb!]
\small
\begin{center}
\begin{tabular}{|c||c|c|c||c|c|c|} \hline\hline
   & $N_{{\rm set}1}$ & $S_{12,{\rm set}1}$ & 
     $S/B_{{\rm set}1}$ & $N_{{\rm set}2}$ & 
     $S_{12,{\rm set}2}$ & $S/B_{{\rm set}2}$  \\ \hline\hline
SM & 69.5$\pm$6.0 &  &  & 432$\pm$8.8 &  &   \\ \hline
1  & 95.9$\pm$6.7 & 9.05 & 1.38 & 184$\pm$9.3 & 8.06 & 0.43 \\ \hline
2  & 282$\pm$20 & 20.8 & 4.06 & 560$\pm$29 & 21.4 & 1.3 \\ \hline
3  & 17.7$\pm$1.1 & 2 & 0.25 & 30.4$\pm$1.4 & 1.44 & 0.07 \\ \hline
4  & 365$\pm$73 & 25 & 5.26 & 1590$\pm$152 & 48.4 & 3.7 \\ \hline
5  & 6.54$\pm$0.37 & 0.77 & 0.094 & 9.6$\pm$0.45 & 0.46 & 0.002 \\ \hline
6  & 277$\pm$35 & 20.6 & 4.0 & 1030$\pm$67 & 35 & 2.4 \\ \hline
7  & 6.7$\pm$0.35 & 0.78 & 0.096 & 8.31$\pm$0.39 & 0.4 & 0.019 \\ \hline
8  & 188$\pm$17 & 15.5 & 2.71 & 530$\pm$28 & 20.5 & 1.2 \\ \hline
10 & 515$\pm$78 & 31.7 & 7.41 & 1950$\pm$151 & 56.1 & 4.5 \\ \hline
11 & 137$\pm$11 & 12.1 & 1.98 & 322$\pm$18 & 13.4 & 0.75 \\ \hline
12 & 409$\pm$30 & 27.1 & 5.89 & 781$\pm$42 & 28.1 & 1.8 \\ \hline
13 & 58.8$\pm$3.3 & 6 & 0.85 & 86.9$\pm$4 & 4 & 0.2 \\ \hline
19 & 377$\pm$59 & 26.5 & 5.43 & 1220$\pm$106 & 39.8 & 2.8 \\ \hline
20 & 279$\pm$36 & 20.6 & 4.01 & 996$\pm$67 & 34 & 2.3 \\ \hline\hline
\end{tabular}
\end{center}
\vspace{-1em}
\caption{The CMSSM benchmark points: the expected ``final'' number
of events after all cuts, $N$, the significance values, $S_{12}$, and
the signal-to-background ratio, $S/B$.  The errors quoted on $N_{\rm
set1,set2}$ account for Monte Carlo statistics only.  The indices,
set1 and set2, are for cut sets \# 1 and \#2 respectively, and the
``SM'' row gives the expected number of the SM background events after
all cuts for all considered processes.}
\label{tab:results1}
\end{table}

The number of points out of reach (significance less than $5\sigma$)
for 10~fb$^{-1}$ varies from nine to ten.  Results are also
illustrated at Fig.~\ref{fig:reach_}.  For the benchmark CMSSM points
with significance greater than $5\sigma$ the signal to background
ratios are greater than 0.4 (the excess of SUSY events over the SM is
greater than $40\%$).  The interpretation of these results in terms of
SUSY masses etc.\ can be found in Sec.~\ref{sec:expreach}.

\begin{figure}[htb!]
\vspace{2em}
\begin{center}
\resizebox{12cm}{!}{\includegraphics{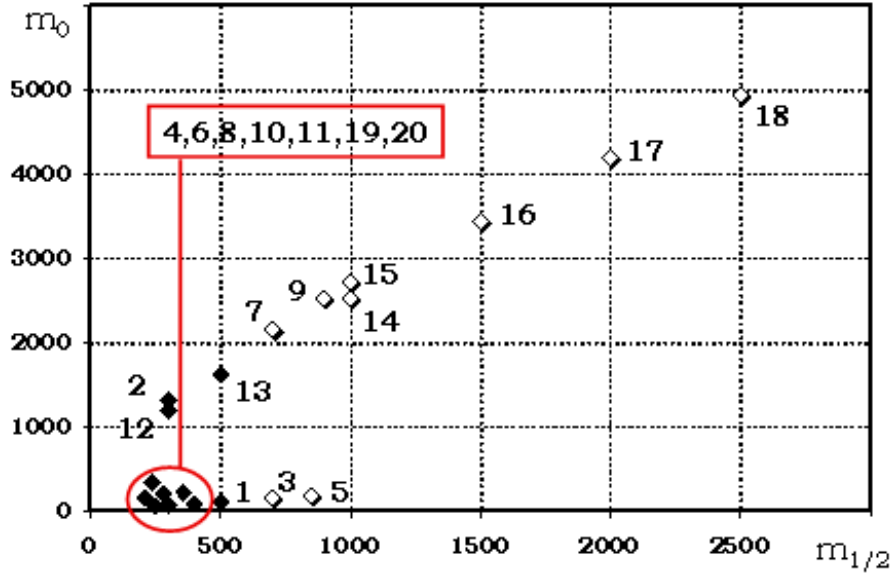}}
\caption{The black points in the $(m_{1/2},m_0)$-plane have
significance greater than five.  The white points are not reachable.}
\label{fig:reach_}
\end{center}
\end{figure}
%


\medskip
\noindent\underline{\it First estimate of systematic effects}

The stability of the significance as a function of the uncertainty on
the signal acceptance and the background normalization was verified.
A correlated variation of the SM event number ($+30\%$) and expected
number of SUSY events ($-30\%$) was applied.  As a result,
(Tab.~\ref{tab:res1}) the significance of only one CMSSM point (\# 13)
drops below discovery level.

\begin{table}[htb!]
\small
\begin{center}
\begin{tabular}{|c||c|c||c|c|} \hline\hline
   & $S_{12}$, set\#1 & $S/B$, set\#1 & 
     $S_{12}$, set\#2 & $S/B$, set\#2 \\ \hline\hline
1  & 6.09 & 0.743 & 5.15 & 0.23 \\ \hline
2  & 14.9 & 2.19 & 14.4 & 0.7 \\ \hline
3  & 1.26 & 0.137 & 0.889 & 0.038 \\ \hline
4  & 18.2 & 2.83 & 34.5 & 2 \\ \hline
5  & 0.476 & 0.0507 & 0.283 & 0.012 \\ \hline
6  & 14.7 & 2.15 & 24.3 & 1.3 \\ \hline
7  & 0.425 & 0.0516 & 0.245 & 0.01 \\ \hline
8  & 10.8 & 1.46 & 13.7 & 0.66 \\ \hline
10  & 23.5 & 3.99 & 40.5 & 2.4 \\ \hline
11  & 8.31 & 1.07 & 8.71 & 0.4 \\ \hline
12  & 19.8 & 3.17 & 19.2 & 0.97 \\ \hline
13  & 3.93 & 0.456 & 2.5 & 0.11 \\ \hline
19  & 18.6 & 2.93 & 27.9 & 1.5 \\ \hline
20  & 14.8 & 2.16 & 23.6 & 1.2 \\ \hline\hline
\end{tabular}
\end{center}
\vspace{-1em}
\caption{CMSSM benchmark points: results after the variation for
cuts set \#1 and \#2.}
\label{tab:res1}
\end{table}
%


\subsubsection{Experimental reach in the CMSSM}
\label{sec:expreach}

Figure~\ref{fig:reach_} shows which CMSSM points have a significance
greater than $5\sigma$ when plotted in ($m_{1/2}$,$m_0$)-plane.  An
approximate sensitive area for 10~fb$^{-1}$ is well defined on the
$m_{1/2}$ parameter axis.  The largest benchmark value of $m_{1/2}$
giving a significance higher than $5\sigma$ has $m_{1/2}=500$~GeV.  It
can be expected that also slightly larger values,
$m_{1/2}\;\raisebox{-.3em}{$\stackrel{\displaystyle
<}{\sim}$}\;650$~GeV can be tested with 10~fb$^{-1}$ at CMS.  For
$\tan\beta=10$ this covers the whole range preferred by electroweak
precision data.  For $\tan\beta=35$, however, a large interval of
$m_{1/2}$ values, 650~GeV$\;\raisebox{-.3em}{$\stackrel{\displaystyle
<}{\sim}$}\;m_{1/2}\;\raisebox{-.3em}{$\stackrel{\displaystyle
<}{\sim}$}\;1000$~GeV remains unexplored.  Higher luminosities are
required to cover this part of the CMSSM parameter space.  This gap
becomes even larger for largerer $\tan\beta$ values.  It should also
be noticed that at large $\tan\beta$ the production of $\tau$'s
increases compared to $\mu$~production.  This might lead to a weaker
signal and correspondingly weaker coverage.

\begin{table}[htb!]
\renewcommand{\arraystretch}{1.3}
\begin{center}
\begin{small}
\begin{tabular}{|c||r|r|r|r|}
\hline\hline
$\tan\beta$ & 10 & 10 & 20 & 35 \\ \hline
$m_{1/2}$ & 500 & 300  & 240 & 500  \\ \hline
$m_0$     & 107 & 1330 & 330 & 1620 \\ \hline
sign($\mu$) & + & + & + & + \\ \hline\hline
$m_{\tilde t_1}$  & 813.02 & 914.14 & 448.95 & 1257.40 \\ \hline
$m_{\tilde t_2}$  & 1015.81 & 1230.46 & 606.35 & 1557.96 \\ \hline
$m_{\tilde b_1}$  & 972.28 & 1216.66 & 552.76 & 1540.94 \\ \hline
$m_{\tilde b_2}$  & 1000.59 & 1422.56 & 593.76 & 1718.06 \\ \hline
$m_{\tilde u_L}$  & 1047.11 & 1451.59 & 619.98 & 1891.29 \\ \hline
$m_{\tilde u_R}$  & 1011.73 & 1447.76 & 607.91 & 1877.60 \\ \hline
$m_{\tilde d_L}$  & 1050.16 & 1453.76 & 625.09 & 1892.97 \\ \hline
$m_{\tilde d_R}$  & 1008.09 & 1448.01 & 607.96 & 1876.51 \\ \hline
$m_{\tilde g}$    & 1154.04 & 791.75 & 601.89 & 1237.18 \\ \hline
$m_{\tilde \tau_1}$  & 213.62 & 1321.88 & 323.45 & 1441.49 \\ \hline
$m_{\tilde \tau_2}$  & 358.84 & 1337.29 & 370.18 & 1562.59 \\ \hline
$m_{\tilde \nu_\tau}$  & 347.87 & 1333.95 & 354.88 & 1559.54 \\ \hline
$m_{\tilde e_L}$  & 358.00 & 1342.14 & 369.93 & 1650.46 \\ \hline
$m_{\tilde e_R}$  & 221.03 & 1333.49 & 344.24 & 1629.05 \\ \hline
$m_{\tilde \nu_e}$  & 348.89 & 1339.45 & 361.04 & 1648.21 \\ \hline
$m_{\tilde \chi^0_1}$  & 200.38 & 119.67 & 92.28 & 206.06 \\ \hline
$m_{\tilde \chi^0_2}$  & 386.88 & 233.35 & 174.04 & 400.77 \\ \hline
$m_{\tilde \chi^0_3}$  & 631.76 & 434.46 & 327.81 & 582.35 \\ \hline
$m_{\tilde \chi^0_4}$  & 646.55 & 452.22 & 348.19 & 599.03 \\ \hline
$m_{\tilde \chi^\pm_1}$  & 387.55 & 234.03 & 174.19 & 401.85 \\ \hline
$m_{\tilde \chi^\pm_2}$  & 646.00 & 451.72 & 348.18 & 598.98 \\ \hline
$m_h$  & 118.19 & 115.41 & 113.29 & 118.87 \\ \hline
$m_H$  & 717.66 & 1399.04 & 456.33 & 1400.62 \\ \hline
$m_A$  & 712.79 & 1389.77 & 453.32 & 1391.45 \\ \hline
$m_{H^\pm}$  & 721.93 & 1401.22 & 463.65 & 1403.49 \\ \hline\hline
\end{tabular}
\end{small}
\end{center}
\renewcommand{\arraystretch}{1}
\vspace{-1em}
\caption{Three points, one for each $\tan\beta$ value with the highest
$m_{1/2}$, and one for $\tan\beta=10$ with the highest $m_0$ value
(with $A_0=0$ and $m_t=175$~GeV) are shown together with some examples
of the corresponding physical spectrum.  All masses are in~GeV.}
\label{tab:spectra}
\end{table}

We now turn to the discovery reach in the physical spectrum.  For each
$\tan\beta$ value we show in Tab.~\ref{tab:spectra} the point with the
highest $m_{1/2}$ value.  For $\tan\beta=10$ we show a second point
with the highest $m_0$.  The CMSSM parameters are given together with
some examples of the corresponding physical spectrum~\footnote{It
should be noted that the $m_h$ evaluation has a large theoretical
error.  While with the most advanced
code~\cite{Heinemeyer:1998yj,Heinemeyer:1998np,Degrassi:2002fi} a
theory error of up to 3~GeV was estimated~\cite{mhiggsAEC}, this error
is even larger in the calculation as implemented in ISASUGRA~7.69.
Therefore the low value for $\tan\beta=20$ does not rule out this
parameter point.}.  If a signal at the LHC is seen in the initial
phase or running with 10~fb$^{-1}$, within the CMSSM framework this
corresponds to scalar quark masses below $\sim 1800$~GeV and gluino
masses below $\sim 1200$~GeV.  Scalar lepton masses are strongly
correlated with $m_0$.  In the chargino/neutralino sector the light
masses are below $\sim 400$~GeV, while the heavy masses can go up to
$\sim 650$~GeV.  The heavy Higgs masses can be as high as $\sim
1400$~GeV.  For the four cases shown in Tab.~\ref{tab:spectra}, the
heavy Higgs bosons can certainly not observed at the LHC directly.  On
the other hand, observation of SUSY particles at the LHC with
10~fb$^{-1}$ would in the framework of the CMSSM guarantee that
several charginos and neutralinos could be detected at the ILC.

\bigskip

The bounds on $m_0$ are obtained only indirectly via the requirement
of the correct CDM density and the choice $A_0=0$.  In order to get a
full overview of the CMS capabilities, more extended studies will be
necessary.  These should involve more CMSSM benchmark points with
$m_{1/2}<650$~GeV and $m_0>1500$~GeV and various $A_0$ values.  Also,
for larger $\tan\beta$, more points have to be studied in order to
correctly assess the LHC potential.


\subsubsection{Conclusions}

Within the framework of the Constrained Minimal Supersymmetric
Standard Model (CMSSM), we have analyzed the discovery potential of
the LHC for the same-sign di-muon signature.  The analysis was
performed in the parameter space that is bounded from below by the
non-observation of SUSY particles at the Tevatron.  Indications for
upper bounds are derived with the help of electroweak precision data,
where the top-quark mass measurement at the Tevatron plays a crucial
role.  At the $90\%$~C.L.\ the fermionic mass parameter $m_{1/2}$ is
bounded from above by $m_{1/2}\le 650,1000$~GeV for $\tan\beta=10,35$.
The analysis concentrated on 200~GeV~$\le m_{1/2}\le$~1000~GeV.  We
furthermore used $A_0=0$ and $m_t=175$~GeV.  All benchmark points that
entered the analysis also fulfilled the requirement that the lightest
neutralino gives the right amount of CDM density.  This fixes to a
large extent the scalar mass parameter~$m_0$.

The analysis was performed for the first phase of LHC operations,
corresponding to an integrated luminosity of 10~fb$^{-1}$.  We found
that $m_{1/2}$ can be tested up to $m_{1/2}<650$~GeV.  This covers the
full range favored by current low-energy precision data for
$\tan\beta=10$, but leaves unexplored regions for larger $\tan\beta$
values -- then a higher integrated luminosity would be necessary.
Thus, if SUSY in its simple version of the CMSSM with $A_0=0$ and
$\tan\beta=10$ is realized in nature, it will be discovered already in
the first phase of LHC running.  For larger $\tan\beta$ values, the
preferred range of $m_{1/2}$ is larger, and more luminosity has to be
collected to cover the correspondingly higher $m_{1/2}$ values.  In
order to get a full overview of the LHC potential more extended
studies will be necessary.  These should involve more CMSSM benchmark
points with larger $m_0$ values, $A_0\neq 0$ and larger $\tan\beta$.


\subsubsection*{Acknowledgements}

We would like to thank A.~Birkedal, N.~Marinelli, K.~Matchev, L.~Pape,
A.~De~Roeck, M.~Spira, M.~Spiropulu, G.~Wrochna, for their special
help and contributions to the analysis.  We thank B.~Heinemann for
helpful discussions on the Tevatron results.  S.H.\ was partially
supported by CICYT (grant FPA2004-02948) and DGIID-DGA (grant
2005-E24/2).

\clearpage\setcounter{equation}{0}\setcounter{figure}{0}\setcounter{table}{0}
\subsection{Collider Searches and Dark Matter Detection Prospects in mSUGRA}
\label{sec:msugra}
\def\eslt{\not\!\!{E_T}}
\def\mslash{\not\!\!{m}}
\def\to{\rightarrow}
\def\Phat{\hat{\Phi}}
\def\bi{\begin{itemize}}
\def\ei{\end{itemize}}
\def\be{\begin{equation}}
\def\ee{\end{equation}}
\def\te{\tilde e}
\def\tl{\tilde l}
\def\tu{\tilde u}
\def\ts{\tilde s}
\def\tb{\tilde b}
\def\tf{\tilde f}
\def\td{\tilde d}
\def\tQ{\tilde Q}
\def\tL{\tilde L}
\def\tH{\tilde H}
\def\tst{\tilde t}
\def\ttau{\tilde \tau}
\def\tmu{\tilde \mu}
\def\tg{\tilde g}
\def\tnu{\tilde\nu}
\def\tell{\tilde\ell}
\def\tq{\tilde q}
\def\tw{\widetilde W}
\def\tz{\widetilde Z}
\def\alt{\stackrel{<}{\sim}}
\def\agt{\stackrel{>}{\sim}}

Howard~Baer$^1$, Alexander~Belyaev$^2$, Tadas~Krupovnickas$^{1,3}$, 
Jorge~O'Farrill$^1$ and Xerxes~Tata$^4$ \\

{\noindent\em
$^1$ Florida State University at Tallahassee, USA \\
$^2$ Michigan State University at East Lansing, USA \\ [1mm]
$^3$ Brookhaven National Laboratory, USA \\
$^4$ University of Hawai'i, USA \\
}


In recent years, supersymmetric models have become increasingly
constrained by a variety of
measurements~\cite{Baer:2002gm,Baer:2002ay}.  These include
determination of the branching fraction $BF(b\to
s\gamma)$~\cite{Cronin-Hennessy:2001fk,Abe:2001hk,Baer:1996kv,Baer:1997jq},
the muon anomalous magnetic moment
$a_\mu=(g-2)_\mu/2$~\cite{Bennett:2002jb,Baer:2001kn} and most
recently, the tight restriction on the relic dark matter density from
the Big Bang, as determined by the WMAP
experiment~\cite{Spergel:2003cb}.  Analyses of WMAP and other data
sets have determined a preferred range for the abundance of cold dark
matter~\cite{Spergel:2003cb}:
\be
\Omega_{CDM}h^2=0.1126^{+0.0161}_{-0.0181} \, ,
\ \ \ \ 2\sigma\ {\rm level} \, .
\label{wmaprange}
\ee  

For phenomenologically viable ranges of parameters, the lightest
neutralino in the minimal supergravity (mSUGRA) framework is usually
the lightest SUSY
particle~\cite{Chamseddine:1982jx,Barbieri:1982eh,Ohta:1982wn,Hall:1983iz}.
Since $R$-parity is assumed to be conserved, this neutralino is stable
and provides a good candidate for cold dark matter.  The possibility
that dark matter, like visible matter, is made up of several
components cannot be excluded at this point.  In our analysis we
therefore interpret the WMAP measurement (\ref{wmaprange}) as an {\it
upper} bound, $\Omega_{\tz_1}h^2<0.129,$ on the neutralino relic
density, unless stated otherwise.  mSUGRA is characterized by four
SUSY parameters together with a sign choice,
\be
m_0,\ m_{1/2},\ A_0,\ \tan\beta\ \ {\rm and}\ sign(\mu ) \, .
\ee
Here $m_0$ is the common mass of all scalar particles at $M_{GUT}$,
$m_{1/2}$ is the common gaugino mass at $M_{GUT}$, $A_0$ is the common
trilinear soft term at $M_{GUT}$, $\tan\beta$ is the ratio of Higgs
field vacuum expectation values at the scale $M_Z$, and finally the
magnitude -- but not the sign -- of the superpotential $\mu$ term is
determined by the requirement of radiative electroweak symmetry
breaking (REWSB).

\begin{figure*}[ht!]
\centering
\includegraphics[width=10cm]{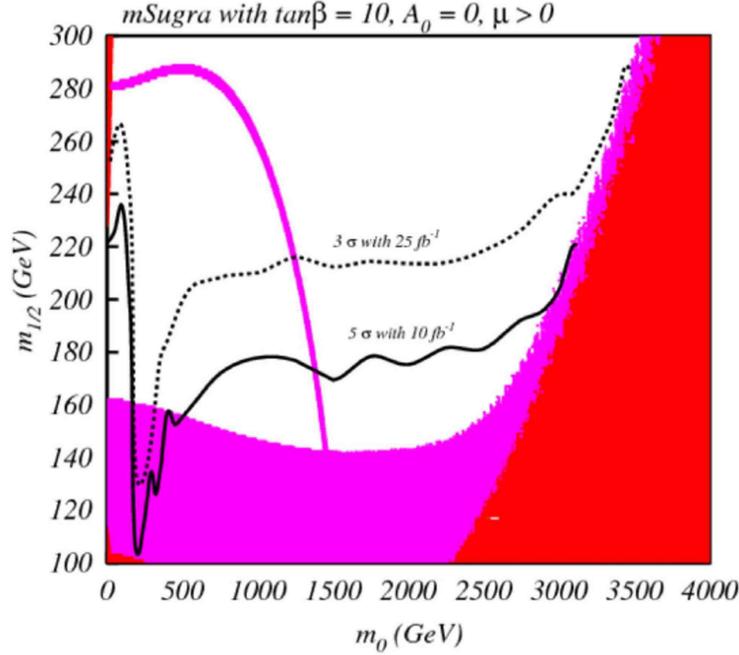}
\caption{The reach of Fermilab Tevatron in the $m_0\ vs.\ m_{1/2}$ 
mSUGRA parameter plane, with $\tan\beta =10$, $A_0=0$, $\mu>0$ and
$m_t=175$~GeV assuming a $5\sigma$ signal with 10 fb$^{-1}$ (solid)
and a $3\sigma$ signal with 25 fb$^{-1}$ (dashed) of integrated
luminosity.  The red (magenta) region is excluded by theoretical
(experimental) constraints.  The region below the magenta contour has
$m_h<114.1$ GeV, in violation of the Higgs mass limit from LEP~II.}
\label{fig:10p}
\end{figure*}

Evaluations of the neutralino relic
density~\cite{Ellis:1998kh,Ellis:1999mm,Drees:1992am,Baer:1995nc,Baer:1997ai,Baer:2000jj,Ellis:2001ms,Lahanas:2001yr,Feng:2000gh,Boehm:1999bj,Ellis:2001nx}
show four qualitatively different viable regions of mSUGRA parameter
space consistent with recent WMAP and other data
sets~\cite{Spergel:2003cb}.  These include 1.) the bulk region at low
$m_0$ and $m_{1/2}$ where neutralinos may annihilate in the early
universe via $t$-channel slepton exchange, 2.) the stau
co-annihilation region where $m_{\tz_1}\simeq
m_{\ttau_1}$~\cite{Ellis:1998kh,Ellis:1999mm}, 3.) the axial Higgs $A$
annihilation corridor at large
$\tan\beta$~\cite{Drees:1992am,Baer:1995nc,Baer:1997ai,Baer:2000jj,Ellis:2001ms,Lahanas:2001yr},
and 4.) the hyperbolic branch/focus poin (HB/FP) region where the
neutralino has a significant higgsino component and can readily
annihilate to $WW$ and $ZZ$ pairs in the early
universe~\cite{Feng:2000gh,Baer:1997ai}.  Somewhat less popular but
still viable scenarios in the literature include a region of
squark--neutralino co-annihilation which can exist for particular
values of the $A_0$ parameter that give rise, for instance, to
$m_{\tst_1}\simeq m_{\tz_1}$~\cite{Boehm:1999bj,Ellis:2001nx}.  Also,
Ref.~\cite{Baer:2004zk} showed that for a large value of the top quark
mass, $m_t=180$~GeV, there exists a narrow band just above the LEP~II
exclusion contour where neutralinos can annihilate through a light
Higgs resonance ($m_h\simeq 2m_{\tz_1}$).  The latter scenario seems
to be currently disfavored due to the new top quark mass measurement,
which pushes the light Higgs annihilation corridor into the region
already excluded by LEP~II searches for the Higgs boson.

\begin{figure*}[ht!]
\centering
\includegraphics[width=10cm]{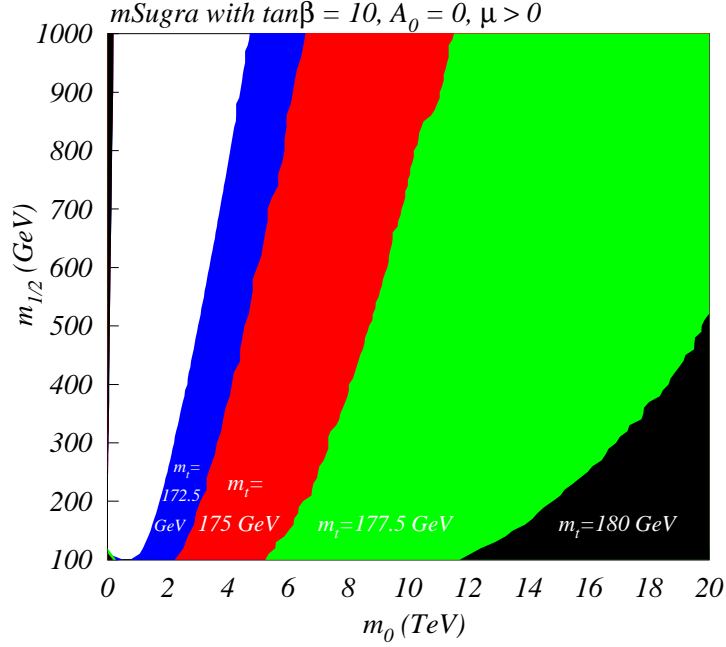}
\caption{Boundary of the mSUGRA $m_0\ vs.\ m_{1/2}$ parameter plane, 
with $\tan\beta =10$, $A_0=0$ and $\mu >0$, for $m_t=172.5$, 175,
177.5 and 180~GeV.}
\label{fig:10ptop}
\end{figure*}

We first turn our attention to mSUGRA's prospects at the Tevatron.
There, $\tw_1\tz_2$ production can lead to trilepton plus $\eslt$
final states which can be above SM background levels for significant
regions of parameter space.  This channel was found to be the most
promising at the Tevatron.  We extend the trilepton search results
presented in Ref.~\cite{Baer:1999bq} to large values of $m_0>1$ TeV,
including the HB/FP region.  Here, we adopt the set of cuts labelled
SC2 in Ref.~\cite{Baer:1999bq}, which generally give the best reach.
From Fig.~\ref{fig:10p}~\cite{Baer:2003dj}, we see that the $5\sigma$
reach for 10~fb$^{-1}$ approaches $m_{1/2}\sim 175$~GeV for $m_0\sim
1000-2000$~GeV, corresponding to a reach in $m_{\tw_1}$($m_{\tg}$) of
125(525)~GeV.

Tevatron also provides us with a very important measurement of the top
quark mass.  The impact of variation of $m_t$ on the allowed mSUGRA
parameter space is shown in Fig.~\ref{fig:10ptop}~\cite{Baer:2003dj}.
The boundary of the allowed parameter space exhibits very strong
sensitivity to the precise value of $m_t$.

\begin{figure*}[ht!]
\centering
\includegraphics[width=10cm]{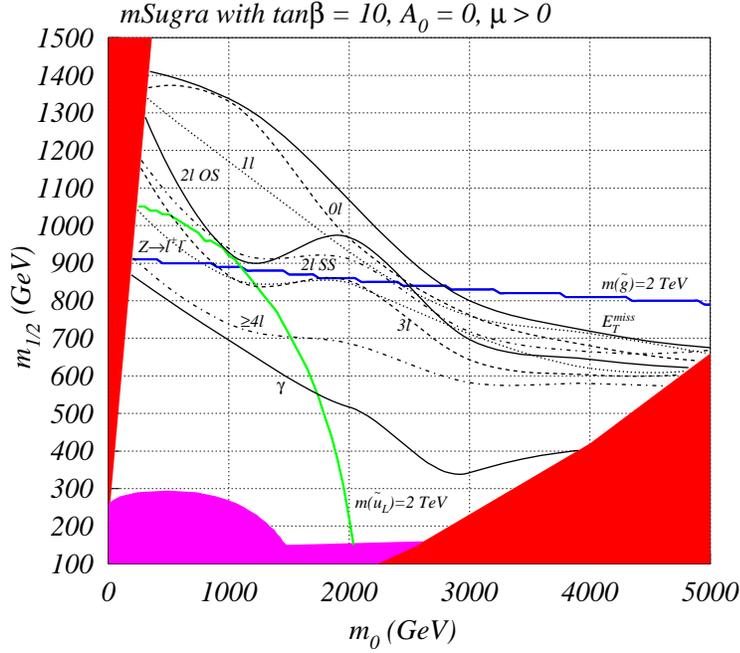}
\caption{The reach of the CERN LHC in the mSUGRA $m_0\ vs.\ m_{1/2}$ 
parameter plane, with $\tan\beta =10$, $A_0=0$, $\mu >0$ and
$m_t=175$~GeV, assuming 100 fb$^{-1}$ of integrated luminosity.  The
red (magenta) regions are excluded by theoretical (experimental)
constraints discussed in the text.  We show the reach in the $0\ell$,
$1\ell$, $OS$, $SS$, $3\ell$, $\ge 4\ell$, $\gamma$ and $Z$ channels,
as well as in the ``inclusive'' $\eslt$ channel.}
\label{fig:10plhc}
\end{figure*}

The CERN LHC is expected to accumulate a significant data sample in
2008 with $pp$ collisions at $\sqrt{s}=14$ TeV.  While the initial
luminosity is expected to be $\sim 10$~fb$^{-1}$ per year, an
integrated luminosity of several hundred fb$^{-1}$ is ultimately
anticipated.  When analyzing the prospects of mSUGRA at the LHC, we
adopt the approach of optimizing kinematic cuts for different
scenarios using computer code, rather than a detailed case-by-case
study.  In our study all events have to pass the following pre-cuts:
\begin{itemize}
\item $E_T^{miss} > 200$ GeV;
\item Number of jets, $N_j \ge 2$,
\end{itemize}
and then we try a large set of combinations of cuts on the most
important variables on signal and background~\cite{Baer:2003wx}.  We
divide the events into signal types according to the number of
isolated leptons (or photons for the isolated $\gamma$ signal).  In
the case of an $E_T^{miss}$ signal there can be any number of leptons:
0-lepton signal has no leptons, 1-lepton signal has 1 lepton,
2-OS-lepton signal has 2 opposite sign leptons, 2-SS-lepton signal has
2 same sign leptons, 3-lepton signal has 3 leptons, $\ge 4$-lepton has
more than 3 leptons, and $Z\to\ell^+ \ell^-$ has at least 2 OS,
same-flavor leptons with invariant mass in the interval $(M_Z-\Delta
M_Z,M_Z+\Delta M_Z)$ ($\Delta M_Z$ is varied during the optimization
procedure).  Finally, the isolated $\gamma$ signal has any number of
leptons plus at least one photon (the cut on the number of photons is
varied during the optimization procedure).  The resulting reaches for
different channels are shown in Fig.~\ref{fig:10plhc}.

Next we turn our attention to a future Linear Collider (LC).  We
explore two possibilities for center-of-mass energy: $500$~GeV and
$1$~TeV.  It is difficult to create a cut optimization algorithm for
the LC which would work well throughout all mSUGRA parameter space,
primarily because different sparticle processes (or at least different
sparticle kinematics) are accessed in different regions.  Therefore we
restrict ourselves to creating the best sets of cuts for
phenomenologically different regions of parameter space.  We find
several such regions~\cite{Baer:2003ru,Baer:2004zk}:
\bi
\item At low $m_0$ with $m_{1/2}\sim 300-500$~GeV, slepton pair
production occurs at large rates.  For low $\tan\beta\sim 10$, the
reach due to selectron, smuon or stau pair production is roughly the
same.  However, stau pair production extends the reach of LC in the
case of larger $\tan\beta$ values.
\item 
There exists a small region around $m_0\sim 200-500$~GeV and
$m_{1/2}\sim 300-350$~GeV where neither slepton nor chargino pairs are
kinematically accessible, but where $e^+e^-\to\tz_1\tz_2$ is.  In this
case, the decay $\tz_2\to\tz_1 h$ was usually found to be dominant.
\item 
For larger $m_0$ values, chargino pair production occurs at a large
rate.  We found that this region cannot be treated by applying the
same set of cuts throughout.  This is due to the fact that in the
lower range of $m_0$ the chargino $\tw_1$ and neutralino $\tz_1$ mass
gap is large, and consequently the visible decay products are hard,
while for larger $m_0$ the opposite is the case.
\ei
In Fig.~\ref{fig:10plc}, we show LC reach contours for $\tan\beta =10$.

\begin{figure*}[ht!]
\centering
\includegraphics[width=10cm]{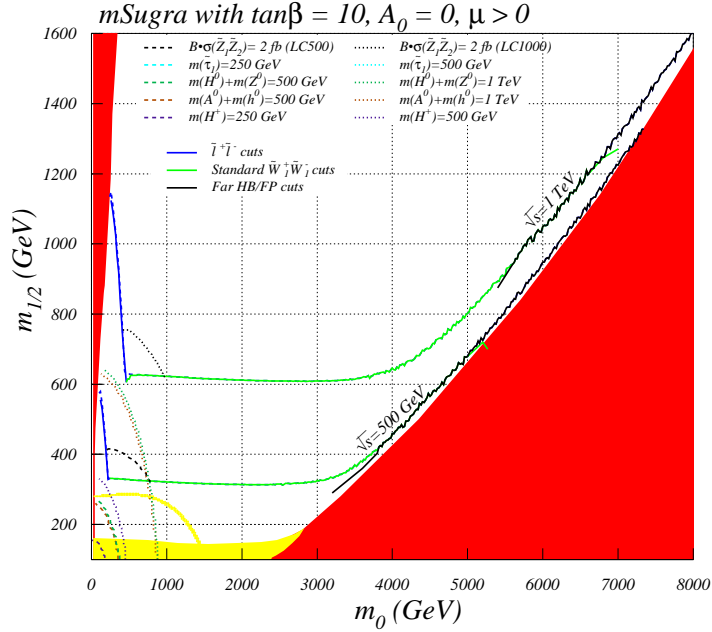}
\caption{Reach of a linear collider for supersymmetry in the mSUGRA
model for $\sqrt{s}=500$ and $1000$~GeV, for $\tan\beta=10$, $A_0=0$,
$\mu>0$ and $m_t=175$~GeV.  The slepton pair production reach is
denoted by the blue contour, while lower-$m_0$ cuts for chargino pair
production yield the green contour.  Larger-$m_0$ chargino pair cuts
yield the black contour in the HB/FP region.  The red region is
theoretically excluded, while the yellow region is excluded by LEP~II
measurements.  Below the yellow contour, $m_h\leq 114.4$~GeV.}
\label{fig:10plc}
\end{figure*}

One can summarize all the collider reaches and compare them with the
constraints from WMAP measurements.  The resulting contours are shown
in Fig.~\ref{fig:10plcall}.  The striking feature of
Fig.~\ref{fig:10plcall} is that the reach of the 1~TeV LC bypasses the
reach of LHC in the far HB/FP region, which is favored by dark matter
(DM) constraints.
\begin{figure*}[t]
\centering
\includegraphics[width=10cm]{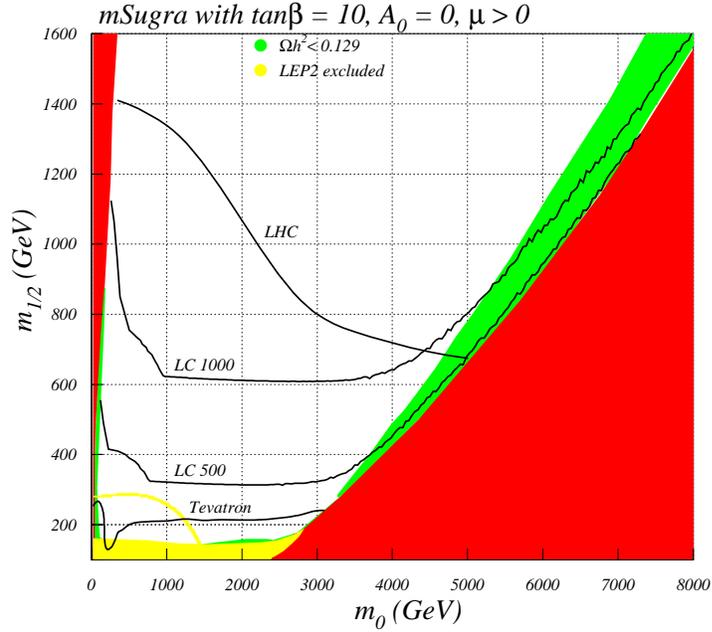}
\caption{Reach of a $\sqrt{s}=0.5$ and 1~TeV LC for sparticles in
mSUGRA for $\tan\beta =10$, $A_0=0$, $\mu>0$ and $m_t=175$~GeV.  We
also show the Fermilab Tevatron reach assuming 10~fb$^{-1}$ of
integrated luminosity (for isolated trileptons), and the CERN LHC
reach assuming 100~fb$^{-1}$ of data.  Finally, the green shaded
region shows points where the relic density $\Omega_{\tz_1}h^2<0.129$
as dictated by WMAP.}
\label{fig:10plcall}
\end{figure*}

One can also include the direct and indirect searches of relic
neutralinos in the analysis.  The bounds from future experiments were
summarized in Ref.~\cite{Baer:2004qq}.  We considered neutrino signals
from neutralino annihilation in the core of the Earth or the Sun,
$\gamma$'s from neutralino annihilation in the core of the galaxy,
positrons and antiprotons from neutralino annihilation in the galactic
halo, and direct searches for neutralino DM via neutralino scattering
off nuclei.  The projected reaches from all these experiments, along
with the favored DM density regions and the collider reaches are
presented in Fig.~\ref{fig10p}.  The intriguing point is that almost
the entire HB/FP region (up to $m_{1/2}\sim 1400$~GeV) can be explored
by the cubic-km-scale IceCube $\nu$
telescope~\cite{Ahrens:2002dv,Halzen:2003fi}.  It can also be explored
(apparently at later times) by the Stage 3 direct DM detectors such as
ZEPLIN4~\cite{Cline:2003pi}, XENON~\cite{Suzuki:2000ch} and
WARP~\cite{Brunetti:2004cf}.
\begin{figure*}[t]
\centering
\includegraphics[width=10cm]{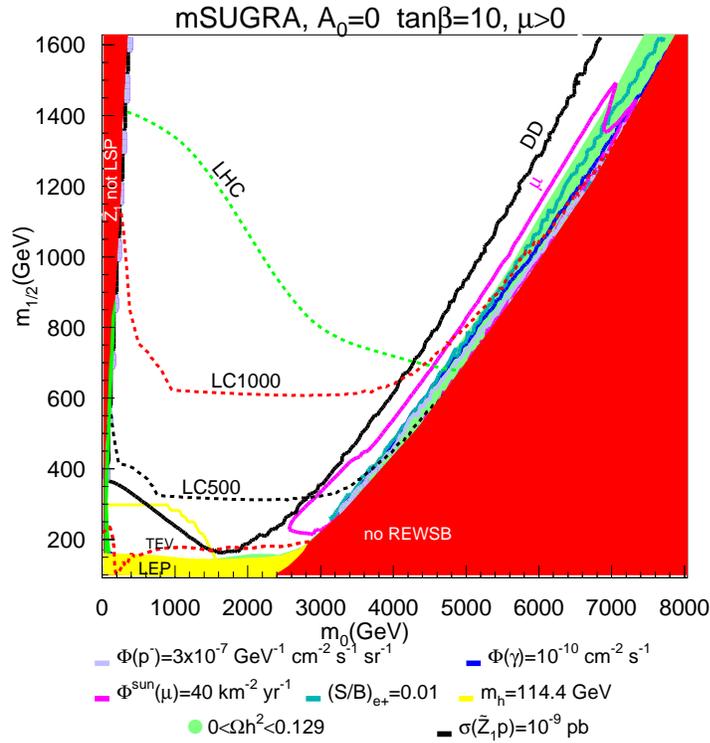}
\caption{A plot of the reach of direct, indirect and collider searches 
for neutralino dark matter in the $m_0\ vs.\ m_{1/2}$ plane, for
$\tan\beta=10$, $A_0=0$, $\mu>0$ and $m_t=175$~GeV.  We include the
reaches of Tevatron, LHC, and linear colliders of $E_{cm}=500$~GeV and
$E_{cm}=1$~TeV (dashed lines).  We also show the reaches of Stage 3
direct dark matter detection experiments (DD) and the IceCube $\nu$
telescope ($\mu$), $\Phi(\gamma)=10^{-10}$ $\gamma$s/cm$^2$/s contour,
$S/B>0.01$ contour for halo produced positrons and the antiproton flux
rate $\Phi(\bar p)=3\times 10^{-7}$ $\bar{p}$s/cm$^2$/s/sr (solid
lines).}
\label{fig10p}
\end{figure*}

In conclusion, Tevatron still has a chance of discovering mSUGRA,
although the region of the parameter space which will be probed is
relatively small.  Nevertheless, the more accurately measured top
quark mass will give us the information about which part of mSUGRA
parameter space is still theoretically allowed.  After accumulating
100~fb$^{-1}$ of integrated luminosity, LHC will have probed gluino
masses as large as $\sim 3$~TeV.  However, the reach of the LHC peters
out in the far HB/FP region.  It does not seem possible to extend the
LHC reach using $\tw_1\tz_2\to 3\ell$ production~\cite{Baer:2005ky},
and $b$-jet tagging extends the gluino reach by just
$10-15\%$~\cite{Mercadante:2005vx}; thus, accessing the far-HB/FP
region seems to be a real challenge for the LHC.  That provides even
more support for the case of a future LC, preferably with a large
center-of-mass energy (of order 1~TeV).  It is very encouraging that
direct and indirect DM search experiments will be able to probe the
far-HB/FP before the LC can be expected to start operating, even if
these experiments alone do not shed light on the physics origin of DM.
If we interpret this DM as the neutralino of mSUGRA, these
experiments, together with absence of signals at the LHC, will point
to the HB/FP region, and make a strong case for the construction of a
TeV LC.

\clearpage\setcounter{equation}{0}\setcounter{figure}{0}\setcounter{table}{0}
\newcommand{\tsc}[1]{\textsc{#1}}
\newcommand{\marm}[1]{\mathrm{#1}}

\newcommand{\SKB}{\ensuremath{\marm{B}}}
\newcommand{\SKZ}{\ensuremath{\marm{Z}}}
\newcommand{\brm}{\ensuremath{\marm{b}}}
\newcommand{\s}{\ensuremath{\marm{s}}}

\def\lsim{\mathrel{\rlap{\lower2pt\hbox{\hskip1pt$\sim$}}
\raise2pt\hbox{$<$}}}
\def\gsim{\mathrel{\rlap{\lower2pt\hbox{\hskip1pt$\sim$}}
\raise2pt\hbox{$>$}}}
\subsection{A Quick Guide to SUSY Tools}
\label{susytools}

{\em Peter Z.\ Skands\footnote{skands@fnal.gov},
Fermi National Accelerator Laboratory, Batavia, IL, USA}\\

{\em The last decade has seen the emergence of a wide range of
automated calculations for supersymmetric extensions of the Standard
Model. This guide contains a brief summary of these, with the main
focus on hadron collider phenomenology, as well as a brief
introduction to the so-called SUSY Les Houches Accord. See also the
Les Houches Web Repository for BSM Tools:
\texttt{http://www.ippp.dur.ac.uk/montecarlo/BSM/} }


\subsubsection{Introduction}

Among the most enticing possibilities for observable new physics both
at the Tevatron and at the LHC is supersymmetry (SUSY); for reviews,
see e.g.\ Refs.~\cite{Martin:1997ns,Tata:1997uf,Chung:2003fi}.  At the
most fundamental level, imposing SUSY on a quantum field theory
represents the most general (and only) possible way of extending the
Poincar\'e group of spacetime symmetries~\cite{coleman67,haag75}.  At
the same time it furnishes a desirable relation between the bosonic
and fermionic degrees of freedom.  Empirically, however, SUSY can at
most be a broken symmetry if it exists in nature, due to the
non-observation of mass-degenerate (or lighter) spin-partners for each
of the Standard Model (SM) particles.

However, even a softly-broken supersymmetry can have quite amazing
properties, as long as the mass splittings introduced by the breaking
are smaller than a TeV or so.  Among the most well-known consequences
of such SUSY are radiative breaking of electroweak symmetry, an
elegant solution to the so-called hierarchy problem, a natural
weakly-interacting dark matter candidate (in theories with conserved
$R$-parity), and unification of the strong, weak, and electromagnetic
gauge couplings at a (very) high energy scale.

For collider phenomenology, the most immediately relevant consequences
are 1) a minimal extension of the SM Higgs sector to two doublets, 2)
promotion of each of the Standard Model fields (plus the extra Higgs
content) to superfields, resulting in a spin-partner for each SM
particle, with mass splittings inside each boson--fermion doublet
$\lsim$~1 TeV, and 3) the special properties which accompany a
conserved $R$-parity, namely production of the new states only in
pairs, followed by individual cascade decays down to the Lightest
Supersymmetric Particle (LSP) which is stable and (usually) escapes
detection.

The large interest in ($N=1$) supersymmetric extensions of the SM and
their phenomenological consequences has carried with it the need for
automated tools to calculate SUSY mass spectra and couplings, cross
sections, decay rates, dark matter relic densities, event rates,
precision observables, etc.  To handle communication between the many
tools, the SUSY Les Houches
Accord~\cite{Skands:2003cj,Allanach:2004ub,slha2} (SLHA) is now in
widespread use.  Section~\ref{sec:slha} contains a brief introduction
to this accord.  Next, in Section~\ref{sec:tools}, an overview of the
presently available state-of-the-art tools is given, divided into four
main categories.  A more extensive collection of tools for BSM physics
as well as an online repository can be found in~\cite{bsmrepository}.
Another recent and comprehensive tools review is the Les Houches
Guidebook to MC Generators~\cite{Dobbs:2004qw}.


\subsubsection{The SUSY Les Houches Accord}
\label{sec:slha}

Given the long history of the subject, it is not surprising that
several different conventions for defining SUSY theories have been
proposed over the years, many of which are in active use by different
groups.  While this is not a problem per se (unique translations can
always be constructed), it does entail a few practical problems, in
particular when the results of one group are compared to or used in
the calculations of a different group.

In addition, even when the theoretical conventions are identical,
there remains the purely technical issue that each program has its own
native way of inputting and outputting parameters, each of which is
unintelligible to most other programs.

The SLHA was proposed to solve both these problems.  Due to the large
parameter space of unconstrained supersymmetric models, the SLHA in
its present form~\cite{Skands:2003cj} (SLHA1) is limited to the
Minimal Supersymmetric Standard Model (MSSM), with conservation of
$R$-parity, CP, and flavour.  Extensions to more general models are
underway~\cite{slha2} (SLHA2).

Technically, the accord is structured into 3 ASCII files (or strings):
1) model definition and measured SM parameters, 2) SUSY mass and
coupling spectrum, and 3) decay tables.  Though admittedly not
elegant, the ASCII format was chosen for its robustness across
platforms and compilers.  In general, all input parameters used for a
calculation are copied to the output, so that any subsequent
calculation also has access to the exact input parameters used for the
previous one.

\paragraph{The SLHA Conventions}

The backbone of the Accord is a unique set of conventions for defining
the SUSY parameters, fields, and couplings.  These conventions, which
have also been adapted for the Supersymmetry Parameter Analysis (SPA)
project~\cite{Aguilar-Saavedra:2005pw}, largely resemble the widely
used Gunion--Haber conventions~\cite{Gunion:1986yn}, with a few
differences as noted explicitly in Ref.~\cite{Skands:2003cj}.  Simply
stated, to define a SUSY model, one needs the field content, the
superpotential, the SUSY breaking terms, and the gauge couplings.  For
the field content, the SLHA assumes that of the MSSM, while SLHA2 will
include extensions for the NMSSM.

The MSSM superpotential is specified by the measured SM particle
masses (giving the Yukawa couplings) and by the $\mu$ term.  In SLHA1,
only the third-generation Yukawas are included.  The gauge couplings
are specified in terms of $M_\SKZ$, $G_F$,
$\alpha_s(M_\SKZ)^{\overline{\marm{MS}}}$, and the fine structure
constant at zero momentum transfer.  All of these are the usual SM
ones that one can get from a review text, i.e.\ no SUSY corrections
should be included here.  SLHA2 will include masses for all 3
generations, as well as the CKM matrix.

The SUSY breaking terms can be specified either by giving the
parameters for a minimal version of a particular SUSY breaking model
(SUGRA, GMSB, or AMSB), or individually, either by starting from a
minimal model and successively adding non-universal terms, or simply
by giving all terms explicitly.  For higher-order calculations, these
parameters are interpreted in the modified dimensional reduction
($\overline{\marm{DR}}$)
scheme~\cite{Siegel:1979wq,Capper:1980ns,Jack:1994rk}, either at the
(derived) unification scale or at a user-specifiable scale.  As
mentioned, CP, $R$-parity, and flavour are assumed conserved in SLHA1.

In the spectrum output, three kinds of parameters are given: 1) pole
masses of all (s)particles, 2) mixing matrices, and 3) Lagrangian
parameters.  While the precise definition of the mixing matrix
elements are left up to each spectrum calculator, the Lagrangian
parameters are defined as $\overline{\marm{DR}}$ ones at one or
several user-specifiable scales $Q$.

\paragraph{The SLHA Decay Tables}

A somewhat separate and self-contained aspect of the SLHA is the
possibility to pass total widths and partial branching ratios via a
file structure similar to that of the rest of the Accord.  A common
use for this is to improve or extend the width calculations of an
event generator by the numbers calculated with a specialised package.

{\bf Note!} An important potential pitfall when using these files is
on-shell intermediate resonances in final states with more than 2
particles.  If not treated properly, large problems both with
double-counting and with incorrect population of phase space can
occur.  Please see \cite{Skands:2003cj} for an explicit description of
the correct procedure to adopt in these cases.


\subsubsection{Computing SUSY}
\label{sec:tools}

This Section contains an overview of SUSY calculational tools, divided
into 1) spectrum calculators, 2) observables calculators, 3) matrix
element and event generators, and 4) data fitting programs.  For links
and references, the reader should consult the recently constructed
online repository for BSM tools~\cite{bsmrepository}.

\paragraph{Spectra}

Given assumptions about the underlying supersymmetric theory (field
content, superpotential, supersymmetry breaking terms) and a set of
measured parameters (SM particle masses, gauge couplings, electroweak
symmetry breaking), the masses and couplings of all particles in the
spectrum can be computed.  This is the task of spectrum calculators,
also called RGE packages.

The most commonly used all-purpose spectrum calculators are
\tsc{Isajet}~\cite{Baer:1993ae},
\tsc{SoftSusy}~\cite{Allanach:2001kg}, \tsc{SPheno}~\cite{Porod:2003um},
and \tsc{SuSpect}~\cite{Djouadi:2002ze}, all compatible with SLHA.  In
general, the codes agree with each other to within a percent or so,
though larger discrepancies can occur, in particular at large
$\tan\beta$.  For mSUGRA, a useful online tool for comparison between
them (and different versions of them) exists~\cite{Belanger:2005jk}.
Other recent comparison studies are found in
Refs.~\cite{Allanach:2003jw,Allanach:2004rh}.  Though
\tsc{Pythia}~\cite{Sjostrand:2000wi} also contains an internal
spectrum calculator~\cite{Mrenna:1996hu}, the resulting spectrum is
very approximate and should not be used for serious studies.

There are also a few spectrum calculators with more specialised areas
of application, such as \tsc{CPSuperH}~\cite{Lee:2003nt},
\tsc{FeynHiggs}~\cite{Heinemeyer:1998yj}, and 
\tsc{NMHDecay}~\cite{Ellwanger:2004xm}.  \tsc{NMHDecay} computes the 
entire mass spectrum in the NMSSM (and has a limit which is equivalent
to the MSSM), but couplings and decay widths are so far only
calculated for the Higgs sector, though improvements are underway.  It
is compatible with an extension of the SLHA~\cite{slha2}.  The program
\tsc{FeynHiggs} deals with the Higgs sector of the MSSM, for which it
contains higher-precision calculations than the general-purpose
programs mentioned above.  It is also able to handle both minimal
flavor violation (MFV) and CP violation, and is compatible with the
SLHA, hence can be used to e.g.\ provide a final adjustment to the
Higgs sector of a general spectrum calculated by one of the other
codes.  Finally, \tsc{CPSuperH} deals with the Higgs sector in the
MSSM with explicit CP violation and contains a number of refinements
which makes it interesting also in the CP conserving case.

\paragraph{Observables}

This includes programs that calculate one or more of the following:
cross sections, decay partial widths, dark matter relic density, and
indirect/precision observables.  Note that we here focus on
calculations relevant for hadron colliders and that matrix element and
event generators, which also calculate many of these things, are
treated separately below.

For hadron collider cross sections,
\tsc{Prospino}~\cite{Beenakker:1996ed} can be used to calculate
inclusive SUSY-NLO cross sections, both total and differential.  It
also calculates LO cross sections and gives the corresponding
K-factors.

For partial decay widths, several specialised packages exist.  For the
MSSM, \tsc{SPheno} calculates tree-level decays of all (s)particles
(soon to include RPV\footnote{RPV in SPheno is not yet public, but a
private version is available from the author}),
\tsc{SDecay}~\cite{Muhlleitner:2003vg} 
computes sparticle decay widths including NLO SUSY-QCD effects, and
both \tsc{FeynHiggs}~\cite{Heinemeyer:1998yj} and
\tsc{HDecay}~\cite{Djouadi:1998yw} compute Higgs partial widths with
higher-order corrections.  Recently, \textsc{HDecay}, \textsc{SDecay},
and \textsc{SuSpect} were combined into one package,
\textsc{Susy-Hit}~\cite{Djouadi:2006bz}.  
\tsc{NMHDecay}~\cite{Ellwanger:2004xm} computes partial widths for 
all Higgs bosons in the NMSSM.

For the density of dark matter, \tsc{DarkSUSY}~\cite{Gondolo:2004sc},
\tsc{IsaTools}~\cite{Baer:2003jb}, and 
\tsc{MicrOMEGAs}~\cite{Belanger:2001fz} represent the publically
available state-of-the-art tools.  All of these work for the MSSM,
though a special effort has been put into \tsc{MicrOMEGAs} to make it
easily extendable~\cite{micromegas2}, recently resulting in an
implementation of the NMSSM~\cite{Belanger:2005kh}, and work on CP
violation is in progress.

For precision observables, \tsc{NMHDecay} includes a check against LEP
Higgs searches, $\brm\to\s\gamma$, and can be interfaced to
\tsc{MicrOMEGAs} for the relic density.  \tsc{Isajet}/\tsc{IsaTools} 
include calculations of $\brm\to\s\gamma$, $(g-2)_\mu$,
$\SKB_\s\to\mu^+\mu^-$, $\SKB_d\to\tau^+\tau^-$, and
neutralino-nucleon scattering cross sections.  \tsc{SPheno} includes
$\brm\to\s\gamma$, $(g-2)_\mu$, as well as the SUSY contributions to
the $\rho$ parameter due to sfermions.  Finally, \tsc{SuSpect} also
includes a calculation of $\brm\to\s\gamma$.

\paragraph{Matrix Element and Event Generators}

By a matrix element generator, we mean a program that, given a set of
fields and a Lagrangian, is able to generate Feynman diagrams for any
process and square them.  Note, however, that many of the codes are
able to do quite a bit more than that.  An event generator is a
program that, given a matrix element, is able to generate a series of
random exclusive events in phase space, often including resonance
decays, parton showers, underlying event, hadronisation, and hadron
decays.

The automated tools for generating matrix elements for SUSY are
\tsc{Amegic++}~\cite{Krauss:2001iv}, \tsc{CalcHEP}~\cite{Pukhov:2004ca}, 
\tsc{CompHEP}~\cite{Pukhov:1999gg}, \tsc{Grace-SUSY}~\cite{Tanaka:1997qn}, 
\tsc{SUSY-MadGraph}~\cite{Reuter:2005us}, and 
\tsc{O'Mega}~\cite{Moretti:2001zz}.  All of these work at leading 
order, except \tsc{Grace}, and all currently only deal with the MSSM,
except \tsc{CalcHEP}, which contains an NMSSM implementation.

\tsc{CalcHEP} and \tsc{CompHEP} provide internal event generators, 
while the event generator \tsc{Sherpa}~\cite{Gleisberg:2003xi} is
built on \tsc{Amegic++}, \tsc{Gr@ppa}~\cite{Tsuno:2005ih} builds on
\tsc{Grace}, \tsc{SUSY-MadEvent}~\cite{Maltoni:2002qb} builds on 
\tsc{SUSY-MadGraph}, and \tsc{Whizard}~\cite{Kilian:2001qz} builds on 
\tsc{O'Mega}.  Of these, most are matrix-element-level event generators.
That is, they provide events consisting of just a few partons and
their four-momenta, corresponding to the given matrix element
convoluted with phase space.  These events must then be externally
interfaced~\cite{Boos:2001cv,Alwall:2006yp} e.g.\ to \tsc{Pythia} or
\tsc{Herwig} for resonance decays, parton showering, underlying event,
and hadronisation.  The exception is \tsc{Sherpa}, which contains its
own parton shower and underlying event models (similar to the
\tsc{Pythia} ones), and for which a cluster-based hadronisation model
is being developed.

In addition, both \tsc{Herwig}~\cite{Corcella:2000bw} and 
\tsc{Pythia} contain a large number of internal hardcoded leading-order 
matrix elements, including $R$-parity violating (RPV) decays in both
cases~\cite{Dreiner:1999qz,Skands:2001it,Sjostrand:2002ip}, and RPV
single sparticle production in \tsc{Herwig}~\cite{Dreiner:1999qz}.  In
\tsc{Pythia}, the parton shower off SUSY resonance decays is merged to
the real NLO jet emission matrix elements~\cite{Norrbin:2000uu}, an
interface to \tsc{CalcHEP} and \tsc{NMHDecay} exists for the
NMSSM~\cite{Pukhov:2005je}, and an implementation of the hadronisation
of $R$-hadrons is available~\cite{Kraan:2004tz,Kraan:2005ji}.

Two other event generators should be mentioned. 
\tsc{Isajet}~\cite{Baer:1993ae} also
contains a large amount of SUSY phenomenology, but its parton shower
and hadronisation machineries are much less sophisticated than those
of \tsc{Herwig}, \tsc{Pythia}, and \tsc{Sherpa}.  The active
development of \tsc{Susygen}~\cite{Ghodbane:1999va} (which among other
things includes RPV single sparticle production) is currently at a
standstill, though basic maintenance is still being carried out.

\paragraph{Fitters}

Roughly speaking, the tools described above all have one thing in
common: given a set of fundamental parameters (themselves not directly
observable) they calculate the (observable) phenomenological and
experimental consequences.  However, if SUSY is at some point
discovered, a somewhat complementary game will ensue: given a set of
observed masses, cross sections, and branching ratios, how much can we
say about the fundamental parameters?

The fitting programs \tsc{Fittino}~\cite{Bechtle:2004pc} and 
\tsc{Sfitter}~\cite{Lafaye:2004cn} attempt to address this question. 
In a spirit similar to codes like \tsc{Zfitter}~\cite{Bardin:1999yd},
they combine the above tools in an automated statistical analysis,
taking as input a set of measured observables and yielding as output a
set of fundamental parameters.

Obviously, the main difficulty does not lie in determining the actual
central values of the parameters, although this can require
significant computing resources in itself; by far the most important
aspect of these tools is a proper and thorough error analysis.
Statistical uncertainties can be treated rigorously, and are included
in both programs.  Theoretical and systematic uncertainties are
trickier.  In a conventional analysis, these uncertainties are
evaluated by careful consideration of both the experimental setup and
of the particular theoretical calculations involved.  In an automated
analysis, which has to deal simultaneously with the entire parameter
space of supersymmetry, a `correct' evaluation of these errors poses a
truly formidable challenge, one that cannot be considered fully dealt
with yet.


\subsubsection*{Acknowledgments}

The TeV4LHC series of workshops have been highly enjoyable, and I am
thankful for all the hard work put in by the organisers. This work was
supported by Universities Research Association, Inc. under Contract
No.~DE-AC02-76CH03000 with the United States Department of Energy.

\clearpage\setcounter{equation}{0}\setcounter{figure}{0}\setcounter{table}{0}
\newcommand{\misset}{$E_{T}$ }
\newcommand{\mzero}{$m_0$}
\newcommand{\mdemi}{$m_{1/2}$}
\newcommand{\Azero}{$A_0$}
\newcommand{\tanbeta}{$\tan \beta$} 
\newcommand{\signmu}{$sign \mu$}
\newcommand{\sbottom}{$\tilde{b}$}
\newcommand{\neutb}{$\tilde{\chi}_2^0$ }
\newcommand{\chara}{$\tilde{\chi}_1^{\pm}$ }
\newcommand{\neuta}{$\tilde{\chi}_1^0$ }
\newcommand{\higgs}{$h^0$}
\subsection{SFITTER}
\label{sec:sfitter}

{\em Remy Lafaye$^1$, Tilman Plehn$^2$, Dirk Zerwas$^3$\\
$^1$ CERN, Geneva, Switzerland\\
$^2$ MPI for Physics, Munich, Germany and University of Edinburgh, Scotland\\
$^3$ LAL Universit\'e de Paris-Sud, Orsay Cedex, France
}\\

{\em The impact of Tevatron measurements on the
determination of supersymmetric parameters is presented.}\\


\subsubsection{Introduction}

The supersymmetric extension~\cite{Wess:1974tw} of the Standard Model
is a well-motivated framework which can link particle physics and
astrophysics and provides us with a consistent and perturbative
description of physics up to the unification scale.  If supersymmetry
is discovered, it will be crucial to determine the fundamental
SUSY--breaking parameters at an unknown high scale from weak-scale
measurements~\cite{Blair:2002pg}.  Large production cross sections for
strongly interacting supersymmetric particles at the LHC combined with
the typical cascade decays can provide a wealth of
measurements~\cite{TDR,Aguilar-Saavedra:2001rg,Abe:2001wn,Abe:2001gc,Weiglein:2004hn}.
A precise theoretical link of masses and couplings at the high scale
and the weak scale are available, for example via Spheno, SuSpect,
SoftSUSY~\cite{Porod:2003um,Djouadi:2002ze,Allanach:2001kg}.  These
renormalization group analyses have to be combined with highly
efficient parameter extraction tools such as
Fittino~\cite{Bechtle:2004pc,Bechtle:2005vt} and
SFitter~\cite{Lafaye:2005cr,Lafaye:2004cn} to properly determine the
underlying fundamental parameters.

In the following we explore the SPS1a~\cite{Allanach:2002nj} parameter
point
(\mzero$=100$~GeV,\mdemi$=250$~GeV,\Azero$=-100$~GeV,\tanbeta$=10$ and
$\mu>0$) using the standard set of measurements as listed in
Ref.~\cite{Weiglein:2004hn}, corresponding to an integrated luminosity
of 300~fb$^{-1}$ at the LHC.  The focus of this particular study is
the determination of the expected errors on the supersymmetric
parameters using SFitter.


\subsubsection{Finding and Fitting}

Two separate tasks have to be considered for a proper determination of
supersymmetric parameters: finding the correct point in parameter
space, and determining the errors on the parameters.  For complex
parameter spaces with an increasing number of dimensions, the allowed
parameter space might not be sampled completely using a standard fit
alone, if the starting point of a fit is arbitrary.  To avoid domain
walls created by unphysical parameter regions, which can confine the
fit to a `wrong' parameter region, combining the fit with an initial
evaluation of a multi-dimensional grid offers one possible solution.
In the general MSSM, the weak-scale parameters can vastly outnumber
the collider measurements, so that a complete parameter fit is
technically not possible.  We then have to limit ourself to a subset
of parameters and carefully estimate the quantitative effect of fixing
certain parameters.  We implemented both a grid and a fit approach
which can be combined, including a general correlation matrix and the
option to exclude parameters of the model from the fit/grid by fixing
them to an arbitrary value.


\subsubsection{mSUGRA at the LHC}

The masses which could be measured in the SPS1a parameter point at the
LHC provide a sufficiently large dataset to perform a fit of
the mSUGRA parameters.  In particular, if the starting point of the
fit is far away from the true parameters (\mzero=\mdemi=1~TeV,
\tanbeta=50, \Azero=0 i.e. the central value of the entire allowed
region), the fit converges to the true values.  The sign of $\mu$ is
fixed to its true value.

\begin{table}[htb]
\begin{tabular}{|l||r|cc|}
\hline
         & SPS1a  & $\Delta$LHC$_{\rm masses}$ & $\Delta$LHC$_{\rm edges}$ \\
\hline
\mzero   & 100    & 3.9 & 1.2 \\
\mdemi   & 250    & 1.7 & 1.0 \\
\tanbeta & 10     & 1.1 & 0.9 \\
\Azero   & -100   & 33  & 20  \\
\hline
\end{tabular}
\caption{Results for mSUGRA at the LHC: all nominal values and the absolute 
errors for mass and edge measurements. Mass values are given in GeV.}
\label{Tab:Msugra}
\end{table}
Using the LHC measurements of particle masses, all SUSY breaking
parameters can be determined at the percent level
(Table~\ref{Tab:Msugra}).  It is particularly interesting to note that
the results using LHC measurements can be improved significantly when
we resort to the measured edges and thresholds instead.  Because the
mass values are extracted from the kinematic endpoints in long decay
chains, the resulting masses are strongly correlated.  In order to
restore the initial sensitivity when we extract the SUSY breaking
parameters from the measured masses we would need the full correlation
matrix.

The precision obtained with the previous fits neglects theoretical
errors.  Indeed, if we take into account reasonable theoretical
errors, such as 3~GeV~\cite{Degrassi:2002fi,Heinemeyer:2004xw} on the
lightest Higgs boson, $3\%$ on colored supersymmetric partners, $1\%$
on neutralinos and sleptons, the error on the \mzero mass increases.
Already at the LHC, the experimental precision will necessitate a
vigorous theoretical
effort~\cite{Aguilar-Saavedra:2005pw,Martin:2005ch} to fully exploit
the available experimental information.


\subsubsection{Impact of the Tevatron Data}

At the parameter point SPS1a most of the supersymmetric particles are
out of reach of the Tevatron.  However, we observe an indirect effect
from the measurement of the top mass.  The top mass and Yukawa
coupling are crucial parameters in the renormalization group analysis
and strongly influences the mass predictions of supersymmetric
particles.  Even though we use a point in the so-called bulk region of
the mSUGRA parameter space, we note that this dependence becomes
dominant for the focus point region~\cite{Drees:1995hj}.  For SPS1a a
4~GeV shift in the top mass shifts the mass of the lightest Higgs
boson by 1.5~GeV.

As a first scenario we assume one year of low-luminosity running at
the LHC, corresponding to an integrated luminosity of 10~fb$^{-1}$.
The experimental errors on the SUSY masses are scaled from the edges
measurements, even though this might be an optimistic assumption.  The
expected precision on the LHC top mass measurement of around 1~GeV
will be dominated by systematic errors and will not be available after
one year of running.  Therefore, we use the Tevatron measurement
instead.  To illustrate the influence of the Tevatron, we generate the
spectrum with a top mass of 175~GeV and fit the SUSY masses assuming a
top mass shifted by 4~GeV.  As shown in Table~\ref{Sfitter:topTeV},
the systematic effect on the extraction of \tanbeta\ and \Azero is
non-negligible: while the error of the fitted parameters is indeed
independent of the top mass we assume for the fit, the central values
shift by up to 0.7$\sigma$.
\begin{table}[t]
\begin{tabular}{|l||r|r|r|r|}
\hline
         & SPS1a & $m_t$=175     & $m_t$=179    & $m_t$=171 \\
\hline
\mzero   & 100   & $100\pm 6$    & $97.9\pm 6$  & $101\pm 6$   \\
\mdemi   & 250   & $250\pm 5$    & $250\pm 6$   & $249\pm 5$   \\
\tanbeta &  10   & $10\pm 5$     & $7.5\pm 2$   & $12.6\pm 6$      \\
\Azero   &-100   & $-100\pm 110$ & $-37\pm 140$ & $-152\pm 88$ \\
\hline
\end{tabular}
\caption{Results of the fit with a shifted top mass.  All mass 
values are given in GeV.}
\label{Sfitter:topTeV}
\end{table}

The discovery and in particular a mass measurement of the lightest
Higgs boson with only 10~fb$^{-1}$ will be challenging at the LHC, as
would several other measurements of edges and mass differences.
Therefore, a natural early--LHC scenario is the observation of only
the main decay chain $\tilde{q}_L\to q\tilde{\chi}_2\to
q\ell\tilde{\ell}_R\to q\ell\ell\tilde{\chi}_1$.  We show the
resulting mSUGRA parameters and their errors in
Table~\ref{Sfitter:LHCminimal}.

If the lightest Higgs boson is sitting at the edge of the LEP2
exclusion bound, one could expect 6 events per fb$^{-1}$ per
experiment in the $WH+ZH$ channels at the Tevatron.  A Higgs mass
determination with a precision around 4-5~GeV could be obtained.
Again, we fit the mSUGRA parameters assuming evidence of a Higgs at
the LEP limit with mass uncertainty of 4.5~GeV.  Our results are shown
in Table~\ref{Sfitter:LHCminimal}.  This hint of a light Higgs reduces
the expected error on \mzero, \tanbeta\ and \Azero\ significantly.

\begin{table}[b]
\begin{tabular}{|l||r|cc|}
\hline
         & SPS1a & $\Delta$LHC$_{\rm edges}$ & $\Delta$LHC$_{\rm
         edges}^{\rm Higgs}$ \\
\hline
\mzero   & 100  & 14   & 9 \\
\mdemi   & 250  & 10   & 9 \\
\tanbeta & 10   & 144  & 31 \\
\Azero   & -100 & 2400 & 685 \\
\hline
\end{tabular}
\caption{Main decay chain only and impact of a hint of a Higgs boson
with a mass extraction to 5~GeV at the Tevatron.  All mass values are
given in GeV.}
\label{Sfitter:LHCminimal}
\end{table}


\subsubsection{Top mass as model parameter}

\begin{table}[t]
\begin{tabular}{|l||r|ccc|}
\hline
         & SPS1a & $\Delta$LHC 
	 & $\Delta$LHC  
	 & $\Delta$LHC \\
         & & $\Delta$\mtop$=2$ & $\Delta$\mtop$=1.5$ & $\Delta$\mtop$=1$ \\ 
\hline
\mzero         & 100  & 1.28  & 1.28  & 1.26 \\
\mdemi         & 250  & 1.01  & 1.00  & 0.99 \\
\tanbeta       & 10   & 1.29  & 1.21  & 1.11 \\
\Azero         & -100 & 26.5  & 25.4  & 24.0 \\
$y_t^{\rm os}$ & 175  & 1.2   & 1.1   & 0.8 \\
\hline
\end{tabular}
\caption{Results of the fit including a top mass measurement with errors
of 2~GeV, 1.5~GeV, 1~GeV.  The errors on the top--Yukawa model
parameter $y_t^{\rm os}$ from the fit are given in the last line.  For
the SUSY masses we assume an integrated luminosity of 300~fb$^{-1}$
and an analysis based on kinematic endpoints directly.  All masses are
in GeV.}
\label{Sfitter:LHC300}
\end{table}

\begin{table}[b]
\begin{tabular}{|l||r|cc|}
\hline
         & SPS1a &  $\Delta$LHC$_{\rm edges}$
	 & $\Delta$LHC$_{\rm edges}$ \\
         &       & $\Delta$\mtop$=2$ & $\Delta$\mtop$=1$\\ 
\hline
\mzero          & 100  & 22.6  & 17.4 \\
\mdemi          & 250  & 16.1  & 12.6 \\
\tanbeta        & 10   & 253.4 & 190. \\
\Azero          & -100 & 4173  & 3108 \\
$y_t^{\rm os}$  & 175  & 2.0   & 1.0 \\
\hline
\end{tabular}
\caption{Results of the fit including Top mass measurement with errors
of 2~GeV and 1~GeV in the minimal LHC scenario with only one decay
chain measured with 10~fb$^{-1}$.  The errors on the top quark model
parameter from the fit are given in the last line.  All masses are in
GeV.}
\label{Sfitter:LHC10minimal}
\end{table}

To carefully study the impact of the top mass measurement on the
determination of supersymmetric parameters, we have to fit the
supersymmetric parameters together with the top
mass~\cite{Bechtle:2004pc,Bechtle:2005vt,Allanach:2005kz}.  The top
mass appears as an mSUGRA model parameter, just like e.g.\ \mzero.
For example, SuSpect (in agreement with the Susy Les Houches Accord
specifications~\cite{Skands:2003cj}) requires the input of an
on--shell (pole) top mass from which it computes the running top
Yukawa coupling.  We refer to this input mSUGRA model parameter as
$y_t^{\rm os}$ to differentiate it from the direct top quark mass
measurement and use 175~GeV as its central value.

At the high scale the Higgs field has not yet acquired a vacuum
expectation value, which means that all fermions are still massless
and the top model parameter should be written as the top Yukawa
coupling $y_t$.  Naturally, this Lagrangian parameter would be
renormalized in $\overline{\rm MS}$.  Note that such a replacement of
a high--scale model parameter by a weak-scale model parameter
($y_t^{\rm os}$) is nothing new to mSUGRA; the same happens with
$\tan\beta$, which as a ratio of two vacuum expectation values is by
definition a weak-scale parameter.

The top mass also appears as a measured observable at the LHC,
typically renormalized in the on--shell scheme.  In this section, the
symbol $m_t$ is reserved for this observable top mass.  The central
value we expect to extract from data is identical to our parameter
$y_t^{\rm os}$, i.e. 175~GeV.

It is instructive to start from the most precise set of measurements
(edges), corresponding to an integrated luminosity of 300~fb$^{-1}$.
We show the results in Table~\ref{Sfitter:LHC300} assuming an
experimental error on the top mass $m_t$ of 2~GeV, 1.5~GeV and 1~GeV.
Due to the high precision of supersymmetric and Higgs measurements,
the top--Yukawa model parameter is strongly constrained directly (as
the top mass) and indirectly at the same time.  Indeed, as shown in
the last line of Table~\ref{Sfitter:LHC300}, the error on the
top--Yukawa model parameter is smaller than that from direct
measurement.  The strongest impact of the top mass measurement we
observe is in the $\tan\beta$ determination via the measurement of the
lightest Higgs mass.  The precision on $\tan\beta$ improves by about
$20\%$ if the top mass measurement improves from 2~GeV to 1~GeV
accuracy.

If we perform the same fit assuming errors for an integrated
luminosity of 10~fb$^{-1}$, the improvement of the $\tan\beta$
measurement is limited to $5\%$ and the top--Yukawa model parameter
error is essentially the same as that of the direct $m_t$ measurement.

As this scenario is arguably optimistic on the LHC side, we also study
the ``minimal'' scenario described above.  The results are shown in
Table~\ref{Sfitter:LHC10minimal}.  Because of the absence of any Higgs
boson mass measurement, \tanbeta\ and \Azero\ are undetermined.  While
the top mass parameter error is not improved with respect to the
direct measurement, the errors on both \mzero\ and \mdemi\ are reduced
if the top mass precision is improved from 2~GeV to 1~GeV at the
Tevatron.  This sensitivity is essentially due to the lepton-lepton
edge measurement.  These studies show the importance of measuring
Standard Model parameters in general and the top mass in particular
with high precision.


\subsubsection{Conclusions}

If SUSY is discovered, sophisticated tools such as Fittino and SFitter
will be necessary to determine the fundamental parameters of the
theory.  Even in the absence of a discovery at the Tevatron, its top
quark measurement will impact the precision with which high--scale
SUSY breaking parameters will be measured in the early years of LHC
running.

\clearpage\setcounter{equation}{0}\setcounter{figure}{0}\setcounter{table}{0}

\section{Summary and Outlook}
\label{sec:outlook}

This report includes numerous results and new ideas relevant to the 
search for new physics at the LHC, Tevatron and elsewhere. Here we 
reiterate only a few conclusions, with the hope that the reader will 
read the detailed contributions presented in Sections~\ref{sec:exp},
\ref{sec:particle} and \ref{sec:models}.

Experimental techniques developed by the CDF and D0 collaborations 
may be used at the LHC to improve the capability to reconstruct electrons, 
photons, muons, taus, jets and missing transverse energy.
The background rates to reconstruction of some of these objects at the LHC 
may be predicted based on the Tevatron data. 

When a signal for new physics will be observed at CMS or ATLAS,
it would be useful to try to fit the signal by assuming the existence of a 
single particle beyond the ones discovered already.
We have presented several representative examples, including
$Z^\prime$ and $W^\prime$ bosons, vectorlike quarks, and
$SU(2)_w$-triplet scalars, long-lived charged particles.
If the signal cannot be convincingly explained by the existence 
of a single new particle, then one should attempt to explain it 
in models containing several new particles.

Much of model-based phenomenology presented here focuses on LHC 
signatures, or
disentangling information from a new discovery at LHC. This is not
unexpected, since many ideas for new physics lie at scales
beyond the reach of Tevatron. However, there are many cases
where Tevatron would first make a discovery, guiding searches at the LHC.
An example is Technicolor, for which there are possible hints in
the Run~I data, yet to be re-examined now in Run~II. If these hints
should turn out to be real evidence, such a discovery would refocus
the very thinly spread phenomenological preparation for LHC running
into a concentrated effort, likely helping us to be readier to analyze
LHC data in a useful way. Another possibility is supersymmetry with a
light stop. Tevatron has considerable reach yet left to explore for
such a scenario. Were a discovery made, it would similarly narrow the
LHC focus down in a highly useful way, and make preliminary SUSY
measurements which could give us strong hints for where else to look
in the LHC data for the rest of the SUSY spectrum. In some cases, the
lower-QCD background environment at Tevatron could provide a cleaner
measurement of some of the SUSY sector than at the LHC.

Given that the Tevatron produces $p\overline{p}$ collisions,
while the LHC will produce $pp$ collisions, there are physics scenarios 
where there is a complementarity between the two colliders.
If a resonance will be 
discovered at the Tevatron in a certain invariant mass distribution, 
then the observation of the same resonance at the LHC would provide
complementary information about the couplings of the new particle 
to different quark flavors.
Clearly, the synergy between the Tevatron and the LHC is a vast 
subject that requires much work beyond what has been presented 
during the TeV4LHC workshops. More generally, both the experimental and 
theoretical particle physics comunities need to intensify the preparations 
for the start-up of explorations of the energy frontier at the LHC.


\clearpage



\bibliographystyle{tev4lhc}
\bibliography{landscape-main}

\end{document}
